\documentclass[rmp,aps,amsmath,amssymb,nofootinbib,epsfig,showpacs,twocolumn,eqsecnum]{revtex4-1}

\usepackage[unicode=true,pdfusetitle,
bookmarks=false,colorlinks=true,citecolor=blue,urlcolor=blue,linkcolor=red]{hyperref}

\usepackage{amsmath,cleveref,amssymb,amsfonts,graphicx,multirow,bm,textcomp}
\usepackage{paralist}
\usepackage{srcltx}
\usepackage[all,cmtip]{xy}
\usepackage{mathrsfs}
\usepackage{threeparttable,dsfont,bbm}
\crefname{equation}{Eq.}{Eqs.}
        {\begin{enumerate}[#1]}{\end{enumerate}%
         \vspace{-.7\baselineskip}}

        {\vspace{-\baselineskip}\begin{list}{#1}{%
        \setlength{\partopsep}{0pt}%
        \setlength{\topsep}{0pt}}}
        {\end{list}\vspace{-.7\baselineskip}}

        {\begin{compactenum}[#1]}{\end{compactenum}}

\newcommand{\bee}{\begin{equation}}
\newcommand{\ee}{\end{equation}}
\newcommand{\bma}{\begin{pmatrix}}
\newcommand{\ema}{\end{pmatrix}}
\newcommand{\balig}{\begin{align}}
\newcommand{\ealig}{\end{align}}

\newcommand{\bZ}{\mathbb{Z}}
\newcommand{\ba}{\begin{align}}
\newcommand{\ea}{\end{align}}

\newcommand{\ignore}[1]{}

\newcommand{\ck}{\mathcal{K}}

\newcommand{\mK}{\mathcal{K}}

\def\sgn{\mathop{\textrm{sgn}}}

\begin{document}

\title{Classification of topological quantum matter with symmetries}

\author{Ching-Kai Chiu}
\email{chiu7@umd.edu}
\affiliation{Department of Physics and Astronomy, University of British Columbia, Vancouver, BC, Canada V6T 1Z1, Canada}
\affiliation{Condensed Matter Theory Center and Joint Quantum Institute, Department of Physics, University of Maryland, College Park, MD 20742, USA}

\author{Jeffrey C.Y. Teo}
\email{jteo@virginia.edu}
\affiliation{Department of Physics, University of Virginia, Charlottesville, VA 22904, USA}

\author{Andreas P. Schnyder}
\email{a.schnyder@fkf.mpg.de}
\affiliation{Max-Planck-Institut f\"ur Festk\"orperforschung, Heisenbergstrasse 1, D-70569 Stuttgart, Germany} 
 
\author{Shinsei Ryu}
\email{ryuu@illinois.edu}
\affiliation{Department of Physics, Institute for Condensed Matter Theory, University of Illinois at Urbana-Champaign, IL 61801, USA}

\begin{abstract}
Topological materials have become the focus of intense research in recent years, since
they exhibit fundamentally new physical phenomena with potential applications for novel
devices and quantum information technology. 
One of the hallmarks of topological materials is the existence of protected gapless surface states, which arise due to a nontrivial
topology of the bulk wave functions. This review provides a pedagogical introduction
into the field of topological quantum matter with an emphasis on classification schemes.
We consider both fully gapped and gapless topological materials and their classification
in terms of  nonspatial symmetries, such as time-reversal, as well as spatial symmetries, such
as reflection. Furthermore, we survey the classification of gapless modes localized on
topological defects. The classification of these systems is discussed by use of homotopy
groups, Clifford algebras, K-theory, and non-linear sigma models describing the Anderson (de-)localization at the surface or inside a defect of the material. Theoretical model
systems and their topological invariants are reviewed together with recent experimental
results in order to provide a unified and comprehensive perspective of the field. While
the bulk of this article is concerned with the topological properties of noninteracting or
mean-field Hamiltonians, we also provide a brief overview of recent results and open
questions concerning the topological classifications of interacting systems.
\end{abstract}

\date{\today}

\maketitle

\tableofcontents

\section{Introduction}

In the last decade since the groundbreaking discovery of topological insulators (TIs) induced by 
strong spin-orbit interactions, 
tremendous progress has been made in our understanding of topological states of quantum matter.
While many properties of condensed matter systems have an analogue in classical systems
and may be understood without referring to quantum mechanics, 
topological states and topological phenomena   
are rooted in quantum mechanics in an essential way:
They are states of matter whose quantum mechanical wave functions are topologically nontrivial and 
distinct from trivial states of matter, i.e., an atomic insulator.  
The precise meaning of the wave function topology will be elaborated below.
The best known example of a topological phase is the integer quantum Hall state, 
in which protected chiral edge states give rise to a quantized transverse Hall conductivity.
These edge states arise due to a nontrivial wave function topology, that 
can be measured in terms of a quantized topological invariant, i.e., the Chern or TKNN number
\cite{Thouless:1982rz,Kohmoto:1985zl}.
This invariant, which is proportional to the Hall conductivity, 
 remains unchanged under  adiabatic deformations of the system, as long as the bulk gap is not closed.
It was long thought that topological states and topological phenomena are rather rare in nature and 
occur only under extreme conditions.
However, with the advent of spin-orbit induced topological insulators, it became clear that topological 
quantum states are more ubiquitous than previously thought.
In fact, the study of topological aspects has become increasingly widespread in
the investigation of insulating and semi-metallic electronic structures, 
unconventional superconductors, and interacting bosonic and fermionic systems.

Another theme that emerged from spin-orbit-induced topological insulators is 
the interplay between symmetry and topology. 
Symmetries play an important role in the Landau-Ginzburg-Wilson framework of spontaneous symmetry breaking
for the classification of different states of matter
\cite{Landau,Wilson:1973jj}.
Intertwined with the topology of quantum states, symmetries serve again as an important guiding principle, but in 
a way that is drastically different from the Landau-Ginzburg-Wilson theory. 
First,  topological insulators cannot be distinguished from ordinary, topologically trivial insulators
in terms of their symmetries and
their topological nontriviality cannot be detected by a local order parameter.  
Second, in making a distinction between spin-orbit-induced topological insulators and ordinary insulators, 
time-reversal symmetry is crucial. 
That is, in the absence of time-reversal symmetry, it is possible to adiabatically deform
spin-orbit-induced topological insulators into a topologically trivial state without closing the bulk gap.
For this reason, topological insulators are called symmetry-protected topological (SPT) phases of matter. 
Roughly speaking, an SPT phase is a short-range entangled gapped phase whose topological properties 
rely on the presence of  symmetries.

\subsection{Overview of topological materials}
Let us now give a brief overview of material systems in which topology plays an important role.

First, insulating  electronic band structures can be categorized in terms of topology. 
By now, spin-orbit induced topological insulators have become classic examples of topological band insulators.
In these systems strong spin-orbit interactions open up a bulk band gap and give rise to an odd number of band inversions, 
thereby altering the wave function topology. Experimentally, this topological quantum state has been realized
in HgTe/CdTe semiconductor quantum wells~\cite{Bernevig:2006kx,MarkusKonig11022007}, 
in InAs/GaSb heterojunctions sandwiched by AlSb~\cite{LiuZhangTypeIISemiCondPRL08,knezPRL11},
in BiSb alloys~\cite{Hsieh:2008fk}, in Bi$_2$Se$_3$~\cite{Xia:2009uq,hsiehNature2009},  
and in many other systems~\cite{HasanMoore2011,andoJPSJreview13}.
The nontrivial wave function topology of these band insulators
manifests itself at the boundary 
 as an odd number of helical edge states or Dirac cone surface states, which are protected by time-reversal symmetry. 
As first shown by Kane and Mele, the topological properties of these insulators are characterized by 
a $\mathbb{Z}_2$ invariant~\cite{Kane:2005kx,Kane:2005vn,Roy2009kx,Roy2009_3D,Moore2007uq,Fu:2007fk,Fu2007uq,FuKane2006}, in a similar way as the Chern invariant characterizes the integer quantum Hall state.
Besides the exotic surface states which completely evade Anderson localization 
\cite{nomuraPRL07, Bardarson2007,Roushan2009, Alpichshev2010},
many other novel phenomena have been theoretically predicted to occur in these systems, 
including  
axion electrodynamics~\cite{Qi2008sf,essinPRL09}, 
dissipationless spin currents, and proximity-induced topological superconductivity~\cite{FuKane_SC_STI}.
These novel properties have recently attracted great interest, since they
 could potentially be used for new technical applications, ranging from spin electronic devices to quantum information technology.

In the case of spin-orbit induced topological insulators the topological nontriviality is guaranteed by time-reversal symmetry, 
a nonspatial symmetry that acts locally in position space. 
However, SPT quantum states can also arise from spatial symmetries, 
i.e., symmetries that act nonlocally in position space, such as rotation, reflection, or other space-group  symmetries~\cite{Fu_first_TCI}.
One prominent experimental realization of a topological phase with spatial symmetries is 
the  rocksalt semiconductor SnTe, whose Dirac cone surface states
are protected by reflection symmetry~\cite{Hsieh:2012fk,Tanaka:2012fk,Xu2012,Dziawa:2012uq}. 

Second, topological concepts can  be applied to unconventional superconductors and superfluids. 
In fact, there is a direct analogy between TIs and topological superconductors (SCs). Both quantum states are fully gapped in the bulk, but exhibit gapless conducting  modes on their surfaces.
In contrast to topological insulators, the surface excitations of topological superconductors are not electrons (or holes), but Bogoliubov quasiparticles, i.e., coherent superpositions of electron and hole excitations.
Due to the particle-hole symmetry of superconductors, zero-energy Bogoliubov quasiparticles
contain equal parts of electron and hole  excitations, and therefore have the properties of Majorana particles. 
While there exists an abundance of examples of topological insulators, topological superconductors are rare, since 
an unconventional pairing symmetry is required for a topologically nontrivial state.
Nevertheless, topological superconductors have become the subject of intense research,
due to their protected Majorana surface states,
which could potentially be utilized as basic building blocks of fault-tolerant quantum computers~\cite{RMP_braiding}.
Indeed, there has recently been much effort to engineer  topological superconducting states
using heterostructures with conventional superconductors~\cite{Alicea:2012em,beenakkerReview,Stanescu_Majorana_review}.
One promising proposal is to proximity induce  $p$-wave superconductivity 
in a semiconductor nanowire~\cite{Roman_SC_semi,Gil_Majorana_wire,Mourik_zero_bias}; another is to use 
Shiba bound states induced by magnetic adatoms on the surface of an $s$-wave superconductor~\cite{Nadj-Perge_Ferro_SC}.
In parallel, there has been renewed interest in the B phase of superfluid ${}^3$He, which realizes a time-reversal symmetric
topological superfluid.  
The predicted surface Majorana bound states of ${}^3$He-B
have been observed using
transverse acoustic impedance measurements
\cite{Murakawa:2009ve}.

Third, nodal systems, such as semimetals and nodal superconductors, can exhibit nontrivial band topology, 
even though the bulk gap closes at certain points in the Brillouin zone.
The Fermi surfaces (superconducting nodes) of these gapless materials are topologically protected by 
topological invariants, which are defined in terms
of an integral along a surface enclosing the gapless points.
Similar to fully gapped topological systems, the topological characteristics of 
nodal materials manifest themselves at the surface in terms of gapless boundary modes.
Depending on the symmetry properties and the dimensionality of the bulk Fermi surface,
these gapless boundary modes form Dirac  cones, Fermi arcs, or flat bands. 
Topological nodal systems can be protected by nonspatial symmetries 
(i.e., time-reversal or particle-hole symmetry) as well as spatial lattice symmetries, or a combination of the two.
Examples of gapless topological materials include, 
$d_{x^2 -y^2}$-wave superconductors
\cite{RyuHatsugaiPRL02}, 
the A phase of superfluid ${}^3$He~\cite{Volovik3HeA,Volovik:book}, 
nodal noncentrosymmetric superconductors~\cite{SchnyderRyuFlat,BrydonSchnyderTimmFlat},
 Dirac materials~\cite{Dai_predition_Na3Bi,wangCd3As2PRB13}, and Weyl semimetals~\cite{WanVishwanathSavrasovPRB11}. 
Recently, it has been experimentally shown that the Dirac semimetal is realized in 
Na$_3$Bi~\cite{Liu21022014}, while the Weyl semimetal  is realized in TaAs~\cite{Xu_Weyl_2015_first,Weyl_discovery_TaAs}.

 All of the aforementioned topological materials can be understood, at least at a phenomenological level,
in terms of noninteracting or mean-field Hamiltonians. 
While the topological properties of these single-particle theories are reasonably well understood,
less is known about the topological characteristics of strongly correlated systems.
Recently, a number of strongly correlated materials have been discussed as
interacting analogues of topological insulators. 
Among them are iridium oxide materials
\cite{Shitade2009}
transition metal oxide heterostructures \cite{okamotoNatComm2011}, and the Kondo insulator SmB$_6$~\cite{Dzero_Kondo_PRL,Dzero_Kondo_PRB,First_realization_TKI}. 
On the theory side,
the Haldane antiferromagnetic spin-1  chain  has been identified as an interacting SPT phase.
Experimentally, this phase may be realized in some quasi-one-dimensional spin-1 quantum magnets,
such as, Y$_2$BaNiO$_5$~\cite{Darriet1993409} and NENP~\cite{renardEPL87}.

\subsection{Scope and organization of the review}

A major theme of solid-state physics is the classification and characterization of different phases of matter.
Many quantum phases, such as superconductors or magnets, can be categorized within the Landau-Ginzburg-Wilson framework, i.e., by the principle of spontaneously broken symmetry.
The classification of topological quantum matter, on the other hand, 
is not based on the broken symmetry, but  the topology of the quantum mechanical wave functions 
~\cite{Thouless:1982rz,wenTopOrder1990}.
The ever-increasing number of topological materials and SPT phases,
as discussed in the previous section, 
calls for a comprehensive 
 classification scheme of topological quantum matter.
 
In this review, we survey recently developed classification schemes of 
fully gapped and gapless materials and discuss 
 new experimental developments.
Our aim is to provide a manual and reference for condensed matter theorists and experimentalists who wish to study 
the rapidly growing field of topological quantum matter.
To exemplify  the topological features we discuss
concrete model systems 
together with recent experimental findings.
While the main part of this article is concerned with the topological characteristics of  quadratic noninteracting Hamiltonians,
we will also give a brief overview of 
established results and open questions regarding the topology of interacting systems.

  The outline of the article is as follows.
After reviewing symmetries in quantum systems in Sec.~\ref{sec:Sym}, 
we start in Sec.~\ref{sec:FullyGapped} by discussing the
 topological classification
of fully gapped free fermion systems in 
terms of   nonspatial symmetries, namely,  time-reversal symmetry (TRS), particle-hole symmetry (PHS), and 
chiral symmetry, 
which define a total of ten symmetry classes~\cite{Schnyder2008,Kitaev2009,Ryu2010ten}.
This classification scheme, which is  known as the the ten-fold way, categorizes quadratic Hamiltonians 
with a given set of nonspatial symmetries into topological equivalence classes.
Assuming a full bulk gap, two Hamiltonians are defined to be topologically equivalent, if there exists
a continuous interpolation between the two that preserves the symmetries and does not close the energy gap.
Different equivalence classes for a given set of symmetries are distinguished by topological invariants, which 
measure the global phase structure of the bulk wave functions (Sec.~\ref{subsec:topinvariants}).
 We review how this classification scheme is derived using K-theory (Sec.~\ref{K-theory approach})  
and non-linear sigma models describing the Anderson (de-)localization at the surface of the material (Sec.~\ref{subsec:Anderson_delocalization}).
In Sec.~\ref{Bulk-boundary and bulk-defect correspondence} we discuss how the classification of gapless modes localized on topological defects
can be derived in a similar manner.

Recently, the ten-fold scheme has been generalized to include spatial symmetries, 
in particular reflection symmetries~\cite{Chiu_reflection,Morimoto2013,Sato_Crystalline_PRB14}, 
which is the subject of Sec.~\ref{Topological crystalline materials}.
In a topological material with spatial symmetries, only those surfaces which are invariant under
the spatial symmetry operations can support gapless boundary modes.
We  review some examples of reflection-symmetry-protected topological  systems,  in 
particular a low-energy model describing the physics of SnTe.
This is followed in Sec.~\ref{sec:gapless_materials} by a  description of the topological characteristics of gapless materials, such as
semimetals and nodal  superconductors, which can be classified 
in a similar manner as fully gapped systems~\cite{matsuuraNJP13,ZhaoWangPRL13,Sato_Crystalline_PRB14,ChiuSchnyder14}.
We discuss the topological classification of gapless materials in terms of both
nonspatial (Sec.~\ref{sectionVA}) and spatial symmetries (Sec.~\ref{sectionVB}).

In Sec.\ \ref{Effects of interactions}, 
we give a brief overview of various approaches to 
diagnose and possibly classify interacting SPT phases.
Because the field of interacting SPT phases is still rapidly growing, 
the presentation in this section is less systematic than in the other parts.  
Interactions can modify the classification in several different ways: 
(i) Two different phases which are distinct
within the free-fermion classification can merge in the presence of interactions; 
and (ii) interactions can give rise
to new topological phases which cannot exist in the absence of correlations.
As an example of case (i) we discuss in 
Sec.\ \ref{Effects of interactions}
various topological superconductors in 1, 2, and 3 spatial dimensions, 
where the interaction effects invalidate the  free fermion classification. 
Finally, we conclude in Sec.~\ref{sec:outlook}, 
where we give an outlook and mention some omitted topics, such as
symmetry-enriched topological phases, 
fractional topological insulators and Floquet topological insulators. We also  give directions
for future research.

Given the constraint of the size of this review and the large literature on topological materials, 
this article cannot provide a complete   coverage of the subject at this stage.
For further background and reviews on topological quantum matter beyond the scope of this article, 
we would like to mention in addition to the Rev.~Mod.~Phys.~articles by \onlinecite{hasan:rmp,qi:rmp},
the following works: 
\cite{Taylortheory,
mooreNatureReview2010,
andoJPSJreview13,
andoFuReviewArxiv15,
Senthil2014,
Turner2013, 
Mizushima2014, 
Volovik:book,
krempaBallentsAnnuRev2014,
BernevigHughesBook13,franzMohlenkampBook13,
shenBook2012,HasanXuNeupane2014,schnyderReviewTopNodalSCs,Bian_review_TSC_TI}.
There are also a number of reviews on the subject of Majorana fermions~\cite{Alicea:2012em,beenakkerReview,elliott_franz_review,Stanescu_Majorana_review}.

\section{Symmetries}
\label{sec:Sym}

In this section, we review how different symmetries are implemented in fermionic systems. 
Let  $\{\hat{\psi}^{\ }_I, \hat{\psi}^{\dag}_I\}^{\ }_{I=1,\ldots,N}$ 
be a set of fermion annihilation/creation operators. 
Here, we imagine for ease of notation that we have ``regularized''
the system on a lattice, and $I, J, \ldots$ are combined labels for the lattice
sites $i,j,\ldots$, and if relevant, of additional quantum numbers, such as e.g.,
a Pauli-spin quantum number  
(e.g., $I=(i,\sigma)$ with $\sigma = \pm 1/2$).
The creation and annihilation operators  satisfy the canonical anticommutation relation, 
$\{
\hat{\psi}^{\ }_{I},\hat{\psi}^\dagger_{J} 
\} = \delta_{IJ}. 
$

Let us now consider a general non-interacting 
system of fermions described by a ``second-quantized'' Hamiltonian 
$\hat{H}$. 
For a non-superconducting system,  $\hat{H}$ is given generically as
\begin{align}
\hat{H}
=  \hat{\psi}^\dagger_{I} \,  {H}^{IJ}  \, \hat{\psi}^{\ }_{J}
\equiv 
\hat{\psi}^{\dag} {H} \hat{\psi},
\label{2nd quantized H}
\end{align}
where the $N \times N$ matrix ${H}^{IJ}$ is the ``first quantized'' Hamiltonian. 
In the second expression of \eqref{2nd quantized H}
 we adopt Einstein's convention of summation on repeated indices, 
 while in the last expression in (\ref{2nd quantized H})
we use matrix notation.
(Similarly, a superconducting system is described
by a Bogoliubov-de Gennes (BdG) Hamiltonian,  
for which we use   Nambu-spinors
instead of complex fermion operators, 
and whose first quantized form is again
a matrix ${H}$ when discretized on a lattice.)

According to the symmetry representation theorem by Wigner, 
any symmetry transformation in quantum mechanics can be represented 
on the Hilbert space by an operator that is 
either linear and unitary, or antilinear and antiunitary.
We start by considering an example of a unitary symmetry,
described by a set of operators
$\{G_1, G_2, \cdots\}$ which
form a group. The Hilbert space must then be a representation of this group
with $\{ \hat{\mathscr{G}}_{1}, \hat{\mathscr{G}}_{2}, \cdots \}$ denoting the operators
acting on the Hilbert space. 
For our purposes,  it is convenient to introduce 
the symmetry transformations in terms of their action on fermionic operators.  
That is, we consider a linear transformation  
\begin{align}
 \hat{\psi}^{\ }_I \to 
 \hat{\psi}^{\prime}_I
 := 
 \hat{\mathscr{U} }
 \hat{\psi}^{\ }_I 
 \hat{\mathscr{U}}^{-1} 
 =
 U^{\ }_{I}{ }^J \hat{\psi}^{\ }_J ,
 \label{unitary symmetry}
\end{align}
where
$\hat{\mathscr{U}}$ and $\hat{\psi}^{\ }_I, \hat{\psi}^{\dag}_I$, 
are second quantized operators that act on
states in the fermionic Fock space.
$U_{I}{ }^J$, however, is ``a collection of numbers'',
i.e., not a second quantized operator. 
(More general possibilities, where a unitary symmetry operator  mixes $\hat{\psi}$ and $\hat{\psi}^{\dag}$,
will be discussed later.)
Now, the system is invariant under $\hat{\mathscr{U}}$
if the canonical anticommutation relation and $\hat{H}$ are preserved, 
$\{
\hat{\psi}^{\ }_I, \hat{\psi}^{\dag}_J\}
=
\hat{\mathscr{U}}
\{\hat{\psi}^{\ }_I, \hat{\psi}^{\dag}_J\}\hat{\mathscr{U}}^{-1}
$ 
and 
$\hat{\mathscr{U}}\hat{H} \hat{\mathscr{U}}^{-1}=\hat{H}$.
The former condition implies
that $U_{I}{ }^J$ is a unitary matrix,
while the latter leads to
$
U^*_{K}{ }^{I} {H}^{KL} U^{\ }_{L}{ }^{J}
=
{H}^{IJ} 
$, 
or 
$U^{\dag} {H} U = {H}$
in matrix notation. 

The unitary symmetry operation $\hat{\mathscr{U}}$ is called {\it spatial} ({\it nonspatial})
when it acts (does not act) on the spatial part (i.e., the lattice site labels $i, j$, $\ldots$) of the collective indices $I,J,\ldots$. 
In particular, 
when $\hat{\mathscr{U}}$ can be factorized 
as $\hat{\mathscr{U}}= \prod_i \hat{\mathscr{U}}_i$, 
i.e., when it acts on each lattice site separately, 
it is nonspatial and is called {\it on-site}. 
A similar definition also applies to antiunitary symmetry operations.
In this section, we  will focus on nonspatial symmetries, i.e., ``internal" symmetries, such as time-reversal symmetry.
Spatial symmetries will be discussed in Sec.\ \ref{Topological crystalline materials}. 

Note  that 
the unitary symmetry of the kind considered
in \eqref{unitary symmetry}
is a {\it global} (i.e., {\it non-gauge}) symmetry. 
As we will see in Sec.\ \ref{Effects of interactions},  {\it local} (i.e., {\it gauge}) symmetries will play a crucial role as a probe for SPT phases.

\subsection{Time-reversal symmetry} 
Let us now consider TRS.
Time-reversal $\hat{\mathscr{T}}$ 
is an antiunitary operator that 
acts on the fermion creation and annihilation operators as,
\begin{align}
\hat{\mathscr{T}} \hat{\psi}^{\ }_I \hat{\mathscr{T}}^{-1} 
=
   (U^{\ }_{T})_{I}{ }^{J}  \hat{\psi}_J, 
   \quad
\hat{\mathscr{T}}{i} \hat{\mathscr{T}}^{-1} = -{i}. 
\label{TR on psi}
\end{align}
(One could in principle have $\hat{\psi}^{\dag}$ appearing on the right hand side of (\ref{TR on psi}). 
But this case can be treated as a combination of
TR and PH.)
A system is TR invariant  if 
$\hat{\mathscr{T}}$ preserves the canonical anticommuator and 
if the Hamiltonian satisfies
$\hat{\mathscr{T}}\hat{H}\hat{\mathscr{T}}^{-1}=\hat{H}$. 
Note that
if a hermitian operator $\hat{O}$, built out of   fermion operators, is preserved under $\hat{\mathscr{T}}$,
then $\hat{\mathscr{T}} \hat{H}\hat{\mathscr{T}}^{-1}=\hat{H}$ implies that
$\hat{\mathscr{T}}\hat{O}(t)\hat{\mathscr{T}}^{-1}=
\hat{\mathscr{T}}e^{+{i}\hat{H}t}\hat{O}e^{-{i}\hat{H}t}\hat{\mathscr{T}}^{-1}=
\hat{O}(-t)
$. 
In non-interacting systems, 
the condition $\hat{\mathscr{T}}\hat{H}\hat{\mathscr{T}}^{-1}=\hat{H}$
leads to  
\begin{align}
\label{DEFTimeReversal}
\hat{\mathscr{T}}: 
\quad 
U^{\dag}_{T}\,   
H^*\, 
U^{\ }_{T} = + 
H. 
\end{align}

Because any given Hamiltonian has many accidental, i.e., nongeneric, symmetries,
we will consider in the following entire parameter families (i.e., ensembles) of Hamiltonians, 
whose symmetries are generic.
Such an ensemble of Hamiltonians with a given set of generic symmetries is called a {\it symmetry class}.
We now let $H$ run over all possible 
single-particle Hamiltonians of such a symmetry class with TRS. 
Applying the TRS condition  \eqref{DEFTimeReversal}  twice,
one obtains
$
(U^*_{{T}} U^{\ }_{T} )^\dagger 
H
(U^*_{{T}} U^{\ }_{T})
=
H. 
$
Since the first quantized Hamiltonian ${H}$
 runs over an irreducible representation space,
$U^{*}_{T} U^{\ }_{T}$
should be a multiple of the identity matrix $\openone$
due to  Schur's lemma, i.e.,
$U^{*}_{T} U^{\ }_{T}= e^{ i \alpha} \openone$.
Since
$U^{\ }_{T}$ is a unitary matrix, 
it follows that 
$U^{*}_{T}
=
e^{i\alpha} U^{\dag}_{{T}}$ 
$\Rightarrow$
$(U_{{T}})^T
=
e^{i\alpha} U^{\ }_{T}$. 
Hence,
we find $e^{2i \alpha}=1$, 
which leads to the two possiblities
$
U^{*}_{T} U^{\ }_{T} = \pm \openone.
$
Thus, acting on a fermion operator $\hat{\psi}_I$
with $\hat{\mathscr{T}}^2$
simply reproduces $\hat{\psi}_I$, possibly up to a sign,
$
\hat{\mathscr{T}}^2 \hat{\psi}_I \hat{\mathscr{T}}^{-2} 
=
(U^{*}_{T} U^{\ }_{T}\hat{\psi})^{\ }_I
=
\pm \hat{\psi}^{\ }_I.
$
Similarly,
for an operator consisting of $n$ fermion creation/annihilation operators,
$\hat{\mathscr{T}}^{2} \hat{O} \hat{\mathscr{T}}^{-2} = (\pm)^{n} \hat{O}$.  
To summarize, 
 TR operation $\hat{\mathscr{T}}$  
satisfies
\begin{align}
 \hat{\mathscr{T}}^2= (\pm 1)^{\hat{N}} 
 \quad
 \mbox{when}\quad U^{*}_{T} U^{\ }_{T} = \pm \openone ,
\end{align}
where 
$\hat{N}:= \sum_I \hat{\psi}^{\dag}_I \hat{\psi}^{\ }_I$ is
the total fermion number operator.
In particular, when  
$U^{*}_{T} U^{\ }_{T} = -\openone$, 
 $\hat{\mathscr{T}}$ squares to 
{\it the fermion number parity} defined by
\begin{align}
\hat{\mathscr{G}}_f:=(-1)^{\hat{N}}.
\end{align}
For systems with $\hat{\mathscr{T}}^2 = -1$ 
(i.e.,  for  systems with an odd number of fermions and $\hat{\mathscr{T}}^2=\hat{\mathscr{G}}_f$), 
 TR invariance leads to 
the Kramers degeneracy of the eigenvalues, which follows from the famous Kramers theorem.

\subsection{Particle-hole symmetry} 
\label{PH symmetry}

Particle-hole $\hat{\mathscr{C}}$ is a unitary transformation 
that mixes fermion creation and annihilation operators:
\begin{align}
\hat{\mathscr{C}} \hat{\psi}^{\ }_I \hat{\mathscr{C}}^{-1} 
=
(U^{*}_{C})^{\ }_{I}{ }^{J}    \hat{\psi}^\dagger_J.
\end{align}
$\hat{\mathscr{C}}$
is also called {\it charge-conjugation},
since in particle-number conserving systems, 
it flips the sign of the $U(1)$ charge, 
$\hat{\mathscr{C}} \hat{Q} \hat{\mathscr{C}}^{-1} = -\hat{Q}$,
where $\hat{Q}:=\hat{N}-N/2$
and $N/2$ is half the number of ``orbitals'', i.e., half the dimension of the single-particle Hilbert space.    
Requiring that the canonical anticommutation relation is invariant under 
$\hat{\mathscr{C}}$,  one finds that $U^{\ }_{C}$ is a unitary matrix. 
For the case of a non-interacting Hamiltonian $\hat{H}$, PHS leads to the condition
$
\hat{H}
=
\hat{\mathscr{C}} \hat{H} \hat{\mathscr{C}}^{-1}
=
- \hat{\psi}^{\dag} ( U^{\dag}_{{C}}{H}^T U^{\ }_{C}) \hat{\psi}^{\ } + \mathrm{Tr}\, {H} 
$,
which implies
\begin{align}
\label{DEFChargeConjugation}
\hat{\mathscr{C}}:
\quad 
U^{\dag}_{{C}}\,  {H}^T \,  U^{\ }_{C} = -H.
\end{align}
Observe that from  (\ref{DEFChargeConjugation}) it follows that $\mathrm{Tr}\, {H}={H}^{II}=0$.
Since $H$ is hermitian,
this PHS condition for  single particle 
Hamiltonians may also be written as
$- U^{\dag}_{{C}} {H}^{*}  U^{\ }_{C}={H}$.
Inspection of  Eq.\ (\ref{DEFChargeConjugation})
reveals that 
$\hat{\mathscr{C}}$
when acting on a {\it single-particle} Hilbert space, 
is not a unitary symmetry,
but  rather a reality condition on the Hamiltonian ${H}$
{\it modulo} unitary rotations.
By repeating the same arguments as in the case of TRS,  
we find that there are two kinds of PH transformations: 
\begin{align}
 \hat{\mathscr{C}}^2= (\pm 1)^{\hat{N}} 
 \quad
 \mbox{when}\quad U^{*}_{C} U^{\ }_{C} = \pm \openone.
\end{align}
In PH symmetric systems  $\hat{H}$, where
$\hat{\mathscr{C}}\hat{H}\hat{\mathscr{C}}^{-1}=\hat{H}$, 
the particle-hole reversed partner $\hat{\mathscr{C}}|\alpha\rangle$ 
of every eigenstate $|\alpha\rangle$ of $\hat{H}$
is also an eigenstate, since 
$
 \hat{\mathscr{C}}\hat{H} \hat{\mathscr{C}}^{-1} \hat{\mathscr{C}} |\alpha\rangle = E_{\alpha} \hat{\mathscr{C}}|\alpha\rangle. 
 $
Similarly, for   single-particle Hamiltonians, it follows that
for every  eigen-wave-function ${u}^A$ of ${H}$ 
with single-particle energy $\varepsilon^A$, 
$
 {H}^{IJ} {u}^A_J = \varepsilon^A {u}^A_I 
$,
its particle-hole reversed partner $U^{\dag}_{{C}} ({u}^A)^*$ 
is also an eigen-wave-function, but with energy $-\varepsilon^A$, since  
$
 U^{\dag}_{C} {H}^* U^{\ }_{C} U^{\dag}_{C} ({u}^A)^* = 
 \varepsilon^A U^{\dag}_{C} ({u}^A)^*.
 $

As an example of a PH symmetric system,
we examine the Hubbard model defined on a bipartite lattice 
\begin{align}
 \hat{H} &=  \sum^{i\neq j}_{ij} \sum_{\sigma} t_{ij} \hat{c}^{\dag}_{i\sigma} \hat{c}^{\ }_{j\sigma} 
 -\mu \sum_i \sum_{\sigma} \hat{n}_{i\sigma}
 + U \sum_i  \hat{n}_{i\uparrow} \hat{n}_{i\downarrow}, 
 \label{hubbard model}
\end{align}
where $\hat{c}^{\dag}_{i\sigma}$ is the electron creation operator at lattice site $i$  with spin $\sigma=\uparrow/\downarrow$
and $\hat{n}_{i\sigma} = \hat{c}^{\dag}_{i\sigma}\hat{c}^{\ }_{i\sigma}$. Here,   
$t_{i,j}=t^*_{ji}$, $\mu$, and $U$ denote hopping matrix element, chemical potential, and interaction strength, respectively. 
Now consider the following PH transformation:
$
 \hat{\mathscr{C}} \hat{c}^{\ }_{i\sigma} \hat{\mathscr{C}}^{-1} = (-1)^{i} \hat{c}^{\dag}_{i\sigma},
 $
 $
 \hat{\mathscr{C}} \hat{c}^{\dag}_{i\sigma} \hat{\mathscr{C}}^{-1} = (-1)^{i} \hat{c}^{\ }_{i\sigma}, 
 $
where the sign $(-1)^i$ is $+ 1$ ($-1$) for sites $i$ belonging to  sublattice  A (B).
Hamiltonian~\eqref{hubbard model} is invariant under $\hat{\mathscr{C}}$
when the  
$t_{ij}$'s connecting sites from the same (different) sublattice are imaginary (real)
and $\mu = U/2$.

\subsection{Chiral symmetry}

The combination of $\hat{\mathscr{T}}$ with $\hat{\mathscr{C}}$ leads to a third symmetry, the
so called chiral symmetry. That is, one
can have a situation where both $\hat{\mathscr{T}}$ and $\hat{\mathscr{C}}$  are broken, but 
their combination is satisfied
\begin{align}
 \hat{\mathscr{S}} = \hat{\mathscr{T}} \cdot \hat{\mathscr{C}} .
\end{align}
Chiral symmetry
$\hat{\mathscr{S}}$
acts on  fermion operators as 
\begin{align}
\hat{\mathscr{S}} \hat{\psi}_I \hat{\mathscr{S}}^{-1}
=
(U^{\ }_{C}U^{\ }_{{T}  })_{I}{ }^J  \hat{\psi}^{\dag}_{J}.
\end{align}
It follows from $\hat{\mathscr{S}} \hat{H} \hat{\mathscr{S}}^{-1}=\hat{H}$
that 
the invariance of a quadratic Hamiltonian  $H$ under $\hat{\mathscr{S}}$
is described by
\begin{align} \label{chiral_Def_eq}
\hat{\mathscr{S}}:\quad 
 U^{\dag}_{S}
 {H} U^{\ }_{{S}} = -{H} ,
\quad
\mbox{where}
\quad 
U^{\ }_{S}
=
U^{*}_{C} U^{*}_{T}.
\end{align}
Note that $\mathrm{Tr}\, {H} =0$ follows immediately from~\eqref{chiral_Def_eq}.
Applying the same reasoning that we used to derive
$\hat{\mathscr{T}}^2
=\hat{\mathscr{C}}^2 = (\pm)^{\hat{N}}$,
we find that  $U^2_{{S}} = e^{ i\alpha } \openone$.
By redefining $U_{{S}} \to e^{i\alpha/2} U_{{S}}$,
the chiral symmetry condition for  single-particle Hamiltonians
simplifies to
\begin{align}
 \hat{\mathscr{S}}:\quad 
\{ {H}, U^{\ }_{S}  \}
=0,
\quad
U^{2}_{{S}} =
U^{\dag}_{S} U^{\ }_{S}  = 
\openone. 
\end{align}
With this, one infers that the eigenvalues of the chiral operator are 
$\pm 1$. 
Additionally, one may impose the condition 
$
 \mathrm{Tr}\, U_{\mathrm{S}}  = 0 
$,
which, however, is not necessary
(see below for an example).
Chiral symmetry gives rise to a symmetric spectrum
of single-particle Hamiltonians:
if $|u\rangle$ is an eigenstate of ${H}$ 
with eigenvalue $\varepsilon$,
then $U_{{S}}|u\rangle$ 
is also an eigenstate, but with eigenvalue $-\varepsilon$. 
In the basis in which  $U_{{S}}$  is diagonal, 
the single-particle Hamiltonian $H$ is   block-off-diagonal,
\begin{align}
 {H}
 =
 \left(
 \begin{array}{cc}
 0 & D \\
 D^{\dag} & 0 
 \end{array}
 \right), 
\end{align}
where $D$ is a $N_A \times N_B$ rectangular matrix with $N_A+N_B=N$. 

As an example, 
let us consider a tight-binding Hamiltonian of spinless fermions
on a bipartite lattice:
\begin{align}
\hat{H} &= \sum_{m,n} t^{\ }_{mn} \hat{c}^{\dag}_m \hat{c}^{\ }_{n}, 
\quad
t^{\ }_{mn}=t^*_{nm} \in \mathbb{C}. 
\label{hopping model}
\end{align}
To construct a chiral symmetry we combine the PH transformation discussed in 
\eqref{hubbard model}
(but drop the spin degree of freedom $\sigma$)
with TRS for spinless fermions, which is defined as
$
\hat{\mathscr{T}}
\hat{c}^{\ }_{m}
\hat{\mathscr{T}}^{-1}
=
\hat{c}^{\ }_{m},
$
with $
\hat{\mathscr{T}}{i}\hat{\mathscr{T}}= -{i}. 
$
This leads to the symmetry condition
$
\hat{\mathscr{S}}
\hat{c}^{\ }_{m}
\hat{\mathscr{S}}^{-1}
=
(-)^m
\hat{c}^{\dag}_{m},
$
with
$
\hat{\mathscr{S}}
{i}
\hat{\mathscr{S}}^{-1}
=-{i}. 
$
Hence, $\hat{H}$ is invariant under $\hat{\mathscr{S}}$
when $t_{mn}$ is a bipartite hopping, 
i.e., when $t_{mn}$ only connects sites on different sublattices. 
Observe that in this example  $\mathrm{Tr}\, U_S = N_A - N_B$, 
where $N_{A/B}$ is the number of sites on sublattice A/B.

Besides the bipartite hopping model
\cite{Gade:1993np, GadeWegner1991},
chiral symmetry is realized in 
BdG systems with TRS and $S_z$ conservation (see below)
\cite{Foster2008}
and in QCD  
\cite{Verbaarschot1994}. 
Chiral symmetry also appears in bosonic systems 
\cite{Dyson1953, GurarieChalker2002, GurarieChalker2003,KaneLubensky2014}
and in  entanglement Hamiltonians 
\cite{Turner:2010qf, Hughes:2011uq, Entanglement_crystalline_Chang}.

\subsection{BdG systems}

Important examples of systems with PHS and 
chiral symmetry are BdG Hamiltonians, which we discuss in this section.
These BdG examples clearly demonstrate that
physically different symmetry conditions {\it at the many-body level}
may lead to the same set of constraints  on {\it single-particle} Hamiltonians.

\paragraph{Class D}
\label{Class D}

BdG Hamiltonians are defined in terms of  Nambu spinors,
\begin{align}
\hat{\Upsilon} = \left(
 \begin{array}{c}
 \hat{\psi}_1 \\
\vdots\\
 \hat{\psi}_N \\
 \hat{\psi}^{\dag}_1 \\
 \vdots \\
 \hat{\psi}^{\dag}_N \\
 \end{array}
 \right),
 \quad
 \hat{\Upsilon}^{\dag}
 =
 \big(
 \hat{\psi}^{\dag}_1,\cdots,\hat{\psi}^{\dag}_N, 
 \hat{\psi}^{\ }_1, \cdots, \hat{\psi}^{\ }_N
 \big), 
\end{align}
which satisfy 
the canonical anticommutation relation 
$
\{ \hat{\Upsilon}^{\ }_A, \hat{\Upsilon}^{\dag}_B \}
 =
 \delta_{AB}
 $
($A,B = 1,\ldots, 2N$).
It is important to note that $\hat{\Upsilon}$ and $\hat{\Upsilon}^{\dag}$ are not 
independent, but are related to each other by 
\begin{align} \label{nambu_spinor_constraint}
\big( \tau_1 \hat{\Upsilon} \big)^{T} 
=
\hat{\Upsilon}^{\dag},
\quad
\big(\hat{\Upsilon}^{\dag}\tau_1 \big)^T
=
\hat{\Upsilon},  
\end{align}
where the Pauli matrix $\tau_1$ acts on  Nambu space. 
Using Nambu spinors, the BdG Hamiltonian $\hat{H}$
is written as
\begin{align}
\hat{H} = 
 \frac{1}{2}
 \hat{\Upsilon}^{\dag}_A\,  {H}^{AB}\, \hat{\Upsilon}^{\ }_B
 =
 \frac{1}{2}\hat{\Upsilon}^{\dag}
 H
 \hat{\Upsilon}. 
\end{align}
Since $\hat{\Upsilon}$ and $\hat{\Upsilon}^{\dag}$ are not independent, 
 the single-particle Hamiltonian ${H}$ must satisfy a constraint.
 Using~\eqref{nambu_spinor_constraint}, we obtain
$
 \hat{H} 
 = ({1}/{2}) \big(\tau_1 \hat{\Upsilon}\big)^T {H} 
 \big(\hat{\Upsilon}^{\dag}\tau_1\big)^T
 =
 -({1}/{2}) \hat{\Upsilon}^{\dag} (\tau_1 {H} \tau_1)^T \hat{\Upsilon}
 +
 ({1}/{2})
 \mathrm{Tr}\, (\tau_1 {H} \tau_1),
 $
which yields
\begin{align}  
 \tau_1 {H}^T \tau_1 = - {H}. 
 \label{PH constraint}
 \end{align}
Thus, every single-particle BdG Hamiltonian satisfies 
PHS of the form (\ref{DEFChargeConjugation}).
However,   condition~\eqref{PH constraint} does not arise 
due to an imposed symmetry, but is rather
 a ``built-in" feature of BdG Hamiltonians that 
originates from  Fermi statistics.
For this reason, 
$\tau_1 {H}^T \tau_1 = - {H}$ in BdG systems 
should be called a particle-hole constraint,
or Fermi constraint 
\cite{KennedyZirnbauer2014}, and not  a symmetry.
Due to~\eqref{PH constraint}, 
any BdG Hamiltonian can be written as 
\begin{align}
{H}=
\left(
\begin{array}{cc}
\Xi & \Delta \\
 -\Delta^{*} & -\Xi^{{T}}
\end{array}
\right),
\quad
\Xi=\Xi^{\dag}, 
\quad
\Delta=-\Delta^{{T}},  
\label{eq: BdG hamiltonian}
\end{align}
where $\Xi$ represents the ``normal'' part
and 
$\Delta$ is the ``anomalous'' part (i.e., the pairing term). 

 BdG Hamiltonians can be thought of as
 single-particle Hamiltonians of Majorana fermions.
The Majorana representation of   
BdG Hamiltonians is obtained by letting
\begin{align}
\left(
\begin{array}{cc}
\hat{\lambda}_{I} \\
\hat{\lambda}_{I+N}
\end{array}
\right)
=
\left(
\begin{array}{cc}
\hat{\psi}^{\ }_I + \hat{\psi}^{\dag}_I\\
{i}\big(
\hat{\psi}^{\ }_I - \hat{\psi}^{\dag}_I\big)\\
\end{array}
\right) ,
\label{Nambu to Majorana}
\end{align}
where
$\hat{\lambda}$ are Majorana fermions satisfying 
\begin{align}
\{\hat{\lambda}^{\ }_A, \hat{\lambda}^{\ }_B\}
= 2\delta^{\ }_{AB},
\quad 
\hat{\lambda}^{\dag}_A=\hat{\lambda}^{\ }_A,
\quad 
(A,B=1,\ldots, 2N) .
\end{align}
In this Majorana basis, the BdG Hamiltonian can be written as
\begin{align}
\hat{H} &=
i 
\hat{\lambda}^{\ }_A 
{X}^{AB}
\hat{\lambda}^{\ }_B,
\quad 
{X}^{*} ={X}, \quad {X}^T = -{X}. \label{eq: class D majorana}
\end{align}
The $4N\times 4N$ matrix ${X}$ can be expressed in terms
of $\Xi$ and $\Delta$ as
\begin{align}
&
\quad 
i {X}=
\frac{1}{2}
\left(
\begin{array}{cc}
R_- + S_- &  -{i}\left(R_+ - S_+\right) \\
{i}\left(R_+ + S_+\right) &
R_- - S_-\\
\end{array}
\right),
\nonumber 
\end{align}
where
\begin{align}
&
R_{\pm} = \Xi \pm \Xi^T=\pm R^T_{\pm},\quad
S_{\pm} = \Delta \pm \Delta^*=  - S^T_{\pm}.
\end{align}
We note that the real skew-symmetric matrix $X$ can be brought into a block diagonal form by an orthogonal transformation, i.e.,
\begin{align}
X = O \Sigma O^T,\quad 
\Sigma = \left(
\begin{array}{ccccc}
0                        & \varepsilon_1   &              &  & \\
- \varepsilon_1 & 0                        &             &  & \\
                          &                           & \ddots & & \\ 
                          &                           &             &  0 & \varepsilon_N \\
                          &                           &             &  -\varepsilon_N& 0 \\
\end{array}
\right),
\label{canonical form X}
\end{align}
where $O$ is orthogonal and $\varepsilon_I \ge 0$.
In the rotated basis $\hat{\xi}:=O^T\hat{\lambda}$, 
the Hamiltonian takes the form
$\hat{H}= i \hat{\xi}^T \Sigma \hat{\xi} = 
2 \sum_{I=1}^N \varepsilon_I \hat{\xi}_{2I-1} \hat{\xi}_{2I}$.

While it is always possible to rewrite a BdG Hamiltonian in terms of Majorana operators,
it is quite rare that the Majorana operator is an eigenstate of the Hamiltonian.
That is, {\it unpaired} or {\it isolated} Majorana {\it zero-energy eigenstates}
are quite rare in BdG systems, and appear only in special occasions.
Moreover, we note that in general there is  no natural way to rewrite a given Majorana Hamiltonian 
in the form of a BdG Hamiltonian, since in general there does not exist any natural prescription
on how to form complex fermion operators out of a given set of Majorana operators.
(A necessary condition for such a prescription to be well defined, is that
the Majorana Hamiltonian must be an even-dimensional matrix.)

To summarize,
 single-particle BdG Hamiltonians are characterized by
the PH constraint (\ref{PH constraint}).
The ensemble of Hamiltonians satisfying (\ref{PH constraint})
is called symmetry class~D.
By imposing various symmetries, 
 BdG Hamiltonians can realize five other symmetry classes:
DIII, A, AIII, C, and CI,
which we will  discuss below.

\paragraph{Class DIII}

Let us start by studying how TRS with $\hat{\mathscr{T}}^2=\hat{\mathscr{G}}_f$
restricts the form of  BdG Hamiltonians.
For this purpose, we label the fermion operators by the spin index
$\sigma=\uparrow/\downarrow$, i.e., we let
$
\hat{\psi}_{I}\to \hat{\psi}_{I\sigma}
$.
We introduce TRS by the condition
\begin{align} \label{TRS_classDIIIeq}
\hat{\mathscr{T}} \hat{\psi}_{I\sigma}\hat{\mathscr{T}}^{-1} = 
 (i \sigma_2)_{\sigma\sigma'} \hat{\psi}_{I \sigma'}, 
\end{align}
where $\sigma_2$ is the second Pauli matrix acting on spin space.  
The BdG Hamiltonian then satisfies
\begin{align} \label{DIII_conditions_eq}
\tau_1 H^T \tau_1 = -H,
\; \; \textrm{and} \; \; 
\sigma_2 H^* \sigma_2 = H. 
\end{align}
As discussed before,  the PH constraint \eqref{PH constraint}
and the TRS \eqref{TRS_classDIIIeq} can be combined to yield a chiral symmetry, 
$
 \tau_1 \sigma_2 H \tau_1\sigma_2 = -H. 
$
Observe that in this realization of chiral symmetry, $\mathrm{Tr}\, U_S=0$. 
The ensemble of Hamiltonians satisfying  conditions~\eqref{DIII_conditions_eq}
is called symmetry class DIII. 
(Imposing $\hat{\mathscr{T}}^2=+1$ instead of 
$\hat{\mathscr{T}}^2=\hat{\mathscr{G}}_f$
leads to a different symmetry class, namely class BDI.)

\paragraph{Class A and AIII}
\label{Class A and AIII}
Next, we consider BdG systems 
with a $U(1)$ spin rotation symmetry around the $S_z$  axis in spin space.
This symmetry allows us to
rearrange the BdG Hamiltonian into a reduced form, i.e.,
\begin{align} \label{BdGHam_ClassAIII}
\hat{H} =
\hat{\Psi}^{\dag}_A H^{AB} \hat{\Psi}^{\ }_B,  
\end{align}
up to a constant, 
where 
$H$ is an {\it unconstrained} $2N\times 2N$ matrix and   
\begin{align}
\hat{\Psi}^{\dag}
=
\left(
\begin{array}{cc}
 \hat{\psi}^{\dag}_{I \uparrow} & 
 \hat{\psi}^{\ }_{I \downarrow} 
\end{array}
\right),
\quad 
 \hat{\Psi} =
\left(
\begin{array}{c}
 \hat{\psi}^{\ }_{I \uparrow}
 \\
 \hat{\psi}^{\dag}_{I \downarrow}
\end{array}
\right). 
\label{Psi in C and CI}
\end{align}
Observe that,
unlike for $\Upsilon, \Upsilon^{\dag}$, 
there is no constraint relating $\hat{\Psi}$ and $\hat{\Psi}^{\dag}$.
As $H$ is unconstrained, 
this Hamiltonian is a member of symmetry class A. 
Since $\hat{\Psi}$ and $\hat{\Psi}^{\dag}$ are independent
operators,   it is possible to
rename the fermion operator  $\hat{\psi}^{\dag}_{\downarrow}$ as  
$
\hat{\psi}^{\dag}_{\downarrow} 
\to
\hat{\psi}^{\ }_{\downarrow} 
$. 
With this relabelling, the BdG Hamiltonian~\eqref{BdGHam_ClassAIII} can be converted to 
an ordinary fermion system with particle number conservation. In this process,
 the  $U(1)$ spin rotation symmetry of the BdG system becomes a fictitious charge $U(1)$ symmetry. 

Let us now impose TRS on~\eqref{BdGHam_ClassAIII}, which acts on
$\hat{\Psi}$ as
\begin{align}
 \hat{\mathscr{T}}
 \hat{\Psi}
 \hat{\mathscr{T}}^{-1}
 =
 \left(
 \begin{array}{c}
 \hat{\psi}^{\ }_{\downarrow}
 \\ 
 -\hat{\psi}^{\dag}_{\uparrow}
 \end{array}
 \right)
 =
 i \rho_2 (\hat{\Psi}^{\dag})^T
 =: \hat{\Psi}^c,
 \label{TRS BdG with Sz}
\end{align}
where $\rho_{1,2,3}$ denote Pauli matrices acting on 
the particle-hole/spin components of the spinor (\ref{Psi in C and CI}).
Observe that,
if we let   
$
\hat{\psi}^{\dag}_{\uparrow} 
\to
\hat{\psi}^{\ }_{\uparrow} 
$,
then  $\hat{\mathscr{T}}$ in \eqref{TRS BdG with Sz} looks like 
a composition of $\hat{\mathscr{T}}$ and $\hat{\mathscr{C}}$, i.e.,  it represents a chiral symmetry. 
Indeed, the relationship  between chiral symmetry $\hat{\mathscr{T}}\hat{\mathscr{C}}$ and the $U(1)$ charge   $\hat{Q}$
 in particle-number conserving systems,
$
(\hat{\mathscr{T}}\hat{\mathscr{C}})
\hat{Q}
(\hat{\mathscr{T}}\hat{\mathscr{C}})^{-1} = 
\hat{Q}  
$,
is isomorphic to the relationship between TRS and $\hat{S}_z$ in BdG systems with $S_z$ conservation,
$ \hat{ \mathscr{T}} \hat{S}_z \hat{\mathscr{T}}^{-1} = \hat{S}_z$.
That is, by reinterpreting \eqref{BdGHam_ClassAIII} as a particle-number conserving system
TRS leads to an effective chiral symmetry. 
The ensemble of Hamiltonians satisfying a chiral symmetry is called symmetry class AIII.
Hence, BdG systems with $S_z$ conservation and TRS belong to symmetry class AIII.


\paragraph{Class C and CI}
We now study the constraints due to 
 $SU(2)$ spin rotation symmetries other than $S_z$ conservation. 
A spin rotation $\hat{\mathscr{U}}^{\phi}_{\boldsymbol{n}}$
by an angle $\phi$ around the rotation axis $\boldsymbol{n}$  
acts on  the doublet 
$(\hat{\psi}_{\uparrow}, \hat{\psi}_{\downarrow})^T$
as
\begin{align}
&
 \left(
 \begin{array}{c}
\hat{\psi}_{\uparrow} \\
\hat{\psi}_{\downarrow}
 \end{array}
 \right)
 \to
 \hat{\mathscr{U}}^{\phi}_{\boldsymbol{n} }
 \left(
 \begin{array}{c}
  \hat{\psi}_{\uparrow} \\
  \hat{\psi}_{\downarrow}
 \end{array}
 \right)
 \hat{\mathscr{U}}^{-\phi}_{\boldsymbol{n} }
=  
e^{- i \frac{\phi}{2} \boldsymbol{\sigma} \cdot \boldsymbol{n}} 
 \left(
 \begin{array}{c}
  \hat{\psi}_{\uparrow} \\
  \hat{\psi}_{\downarrow}
 \end{array}
 \right). 
\end{align}
That is, 
a spin rotation  by $\phi$ around the $S_{x}$ or $S_{y}$ axis
transforms $\hat{\Psi}$ into
\begin{align}
\hat{\mathscr{U}}^{\phi}_{S_x}
\hat{\Psi} 
\hat{\mathscr{U}}^{-\phi}_{S_x}
&=
 \cos(\phi/2) \hat{\Psi} - i 
 \sin( \phi/2)\hat{\Psi}^c, 
\nonumber\\
\hat{\mathscr{U}}^{\phi}_{S_y} 
\hat{\Psi} 
\hat{\mathscr{U}}^{-\phi}_{S_y} 
&=
 \cos(\phi/2) \hat{\Psi} - 
 \sin (\phi/2) 
 \hat{\Psi}^c,
\end{align}
respectively.
Thus, both $\hat{\mathscr{U}}^{\phi}_{S_x}$ and $\hat{\mathscr{U}}^{\phi}_{S_y}$
rotate $\hat{\Psi}$   smoothly into $\hat{\Psi}^c$.
In particular, a  rotation by $\pi$ around $S_x$ or $S_y$ 
acts as a discrete PH transformation, 
$
\hat{\Psi} 
\to 
- i \hat{\Psi}^c 
$
or
$
- \hat{\Psi}^c 
$.
That is, if we interpret~\eqref{BdGHam_ClassAIII} as a particle-number conserving system, then
$
\hat{\mathscr{U}}^{\pi}_{S_i}
\hat{S}^{\ }_z 
\hat{\mathscr{U}}^{-\pi}_{S_i}
=-\hat{S}_z
$ for $i=x,y$
can be viewed as a charge-conjugation 
$\hat{\mathscr{C}}\hat{Q}\hat{\mathscr{C}}^{-1}= -\hat{Q}$.
Observer that the $\pi$ rotations
 $\hat{\mathscr{U}}^{\pi}_{S_{i}}$ 
are examples of PH transformations which square to $-1$,
which is in contrast to the PH constraint of class D. 
For the single-particle Hamiltonian $H$ the 
$\pi$-rotation symmetries  $\hat{\mathscr{U}}^{\pi}_{S_{i}}$  lead 
to the condition  
\begin{align}
\rho_2  H^T \rho_2 = -H. 
\label{PH symmetry, BdG with Sz}
\end{align}
The ensemble of Hamiltonians satisfying this condition 
is called symmetry class C.
We note that for quadratic Hamiltonians the $\pi$-rotation symmetry constrains 
of $\hat{\mathscr{U}}^{\pi}_{S_{i}}$ actually correspond to a full $SU(2)$ spin rotation symmetry.
This is because 
for an arbitrary $SU(2)$ rotation around $S_{x}$ or $S_{y}$,  
the Hamiltonian $\hat{H}$ is transformed into a superposition of 
$
\hat{\Psi}^{\dag} H \hat{\Psi} 
$
and 
its conjugate 
$
\hat{\Psi}^{c \dag} H\hat{\Psi}^{c}
$
(i.e.,
$\hat{H}
\to  
\alpha \hat{\Psi}^{\dag} H \hat{\Psi} 
 + (1-\alpha) \hat{\Psi}^{c \dag} H\hat{\Psi}^{c}, 
$
for some $\alpha$),
since $
 \hat{\Psi}^{\dag}H \hat{\Psi}^c = \hat{\Psi}^{c \dag} H \hat{\Psi}=0$.
It follows from
$ \hat{\Psi}^{\dag} H \hat{\Psi} 
=\hat{\Psi}^{c \dag} H\hat{\Psi}^{c}
$
together with the $S_z$ invariance
that the BdG Hamiltonian is fully
invariant under $SU(2)$ spin rotation symmetry.

Finally, imposing TRS \eqref{TRS BdG with Sz}
in addition to $S_z$ conservation leads to  
$\hat{\Psi}^{\dag} H \hat{\Psi} \to \hat{\Psi}^T \rho_2 H^* \rho_2 (\hat{\Psi}^\dag)^T = -\hat{\Psi}^{\dag} \rho_2 H^{\dag} \rho_2 \hat{\Psi}=\hat{H}$.
I.e., $\rho_2 H^{\dag} \rho_2 = -H$. 
Combined with PHS \eqref{PH symmetry, BdG with Sz}, 
this gives  the conditions
\begin{align}
\rho_2 H^T \rho_2 = -H,
 \quad
 H^* = H ,
\end{align}
which defines symmetry class CI.

\subsection{Symmetry classes of ten-fold way}

Let us now discuss a general symmetry classification of single-particle Hamiltonians
in terms of non-unitary symmetries.
Note that unitary symmetries, which
commute with the Hamiltonian, 
allow us to bring the Hamiltonian into a block diagonal form.
Here, our aim is to classify the symmetry
properties of these irreducible blocks, which do not exhibit any unitary symmetries.
So far we have considered the following set of discrete  symmetries
\begin{align}
&
T^{-1}HT = H, \quad T = U_{T} \mathcal{K},
\quad
U_T U^*_T = \pm \openone,
\nonumber \\
&
C^{-1}HC = -H, \quad C = U_{C} \mathcal{K}, 
\quad
U_C U^*_C = \pm \openone,
\nonumber \\
&
S^{-1}HS = -H, \quad S = U_{S},
\quad
U^2_S =\openone, 
\label{symmetrysummaryeq}
\end{align}
where $\mathcal{K}$ is the complex conjugation operator. 
As it turns out, this set of symmetries is exhaustive. That is, without loss of generality 
we may assume that there is only a {\it single} TRS with operator 
${T}$ and a {\it single} PHS with operator ${C}$.
If the Hamiltonian ${H}$ was invariant under,
say, {\it two} PH operations 
${C}_1$ and ${C}_2$, 
then the composition ${C}_1 \cdot {C}_2$ of these
two symmetries would be a {\it unitary} symmetry
of the single-particle Hamiltonian $H$.
i.e., the product $U^{\ }_{{C}_1} \cdot U^*_{{C}_2}$ would {\it commute} with ${H}$. 
Hence, it would be possible to bring  ${H}$ into block form, 
such that $U^{\ }_{{C}_1} \cdot U^*_{{C}_2}$ is a constant on each block.
Thus, on each block $U^{\ }_{{C}_1}$ and $U^{\ }_{{C}_2}$ would be trivially related
to each other, and therefore it would be sufficient to consider only one of the two PH operations. 
-- The product ${T} \cdot {C}$, however, 
corresponds to a {\it unitary} symmetry operation for the single-particle Hamiltonian ${H}$.
But in this case, the unitary matrix $U^{\ }_{T} \cdot U^{*}_{C}$
does not commute, but 
{\it anti-}commutes with ${H}$.
Therefore, ${T} \cdot {C}$ does not represent an ``ordinary'' unitary symmetry of ${H}$.
This is the reason why we need to consider the product
${T} \cdot {C}$ 
[i.e., chiral symmetry $S$ in Eq.~\eqref{symmetrysummaryeq}]
as an additional crucial ingredient for the classification of the irreducible blocks,
besides TR and PH symmetries.

Now it is easy to see that there are only ten possible
ways for how a Hamiltonian $H$ can transform under the general non-unitary symmetries~\eqref{symmetrysummaryeq}.
First we observe that there are three different possibilities for 
how $H$ can transform under TRS  (${T}$): (i) $H$ is not TR invariant,
which we denote by ``$\mathrm{T}=0$" in Table~\ref{tab:classification}; (ii)
the Hamiltonian is TR invariant and the TR operator 
${T}$ squares to $+\mathbbm{1}$, in which
case we write ``$\mathrm{T}=+1$"; and (iii)
$H$ is symmetric under TR  and ${T}$
squares to $-\mathbbm{1}$,  which
we denote by ``$\mathrm{T}=-1$".
Similarly, there are three possible ways for how the Hamiltonian ${H}$
can transform under PHS  with PH operator ${C}$ (again, ${C}$
can square to $+\mathbbm{1}$ or $-\mathbbm{1}$).
For these three possibilities we write ``$\mathrm{C}=0, +1, -1$".
Hence, there are $3 \times 3 = 9$  
possibilities for how $H$ can transform under both
TRS and PHS.
These are not yet all ten cases, since it is also necessary to consider
the behavior of the Hamiltonian under the product 
${S} = {T} \cdot {C}$.
A moment's thought shows that for
8 of the 9 possibilities the presence or absence of
${S} = {T} \cdot {C}$
is fully determined by how $H$ transforms under TRS and PHS.
(We write ``$\mathrm{S}=0$" if 
 ${S}$ is not
a symmetry of the Hamiltonian, and ``$\mathrm{S}=1$" if it is.)
But in the case where both TRS and PHS are absent, there exists the extra possibility that
$S$ is still conserved, i.e.,
either $\mathrm{S}=0$ or $\mathrm{S}=1$ is possible. This then
yields $(3\times3 -1) +2 = 10$ possible behaviors of the Hamiltonian.

These ten possible behaviors of the first quantized
Hamiltonian $H$ under 
${T}$, ${C}$, and ${S}$
are listed in the first column of Table~ \ref{tab:classification}. 
These are the ten generic
symmetry classes (the ``ten-fold way'') which are the
framework within which the classification scheme of TIs and TSCs is formulated in Sec.~\ref{sec:FullyGapped}.
We note that these ten symmetry classes were originally described by 
 Altland and Zirnbauer in the context of disordered systems~\cite{Zirnbauer:1996fk,altlandZirnbauerPRB10}
and are therefore sometimes called ``Altland-Zirnbauer" (AZ) symmetry classes. 
The ten-fold way extends and completes the well known ``three-fold way'' scheme of Wigner and Dyson~\cite{dyson:1199}.

\section{Fully gapped free fermion systems and topological defects}
\label{sec:FullyGapped}

In this section we discuss the topological classification of fully gapped non-interacting fermionic systems, 
such as band insulators and fermionic quasiparticles in fully gapped SCs described by BdG Hamiltonians, 
in terms of the ten Altland-Zirnbauer (AZ) symmetry classes. 
When considering superconductors, the superconducting pairing potentials will be treated at the mean-field level, 
i.e., as a fixed background to fermionic quasiparticles.
We also discuss in this section the topological classification of zero-energy modes localized at topological defects 
in insulators and SCs.
As shown below, 
gapped topological phases and zero modes bound to topological defects can be discussed in a fully parallel and unified fashion 
\cite{Teo:2010fk} 
by introducing the parameter $\delta := d - D$, where $d$ is the space dimension 
and $D+1$ denotes the codimension of defects (see Sec.\ \ref{Topological defects} for more details).
When necessary, by taking $\delta=d$ and $D=0$,
one can easily specialize to the case of 
gapped topological systems, instead of defects of codimension greater than one.



\subsection{Ten-fold classification of gapped free fermion systems and topological defects} 
\label{secTenFoldClass}

\subsubsection{Gapped free fermion systems}

Gapped phases of quantum matter can be distinguished {\it topologically} by asking 
if they are connected in a phase diagram. 
If two gapped quantum phases can be transformed into each other through an adiabatic/a continuous path in the phase diagram 
without closing the gap (i.e., without encountering a quantum phase transition), then
they are said to be {\it topologically equivalent}. 
In particular, states which are continuously deformable to an atomic insulator,
i.e., a collection of independent atoms, are called
{\it topologically trivial}
or  
{\it trivial},
e.g., trivial band insulators.
On the other hand, those that cannot be connected to atomic insulators are called 
{\it topologically non-trivial}
or 
{\it topological}.

\begin{table}[tb]
\centering 
\begin{ruledtabular}
\begin{tabular}{c|ccc|cccccccc} 
   $\mathrm{class} \backslash \delta $ & T & C & S & 0 & 1 & 2 & 3 & 4 & 5 & 6 & 7   \\  \hline
  A & 0 & 0 &0 & $\mathbb{Z}$ & 0 & $\mathbb{Z}$ & 0 & $\mathbb{Z}$ & 0 & $\mathbb{Z}$ & 0               \\
  AIII & 0 & 0 & 1 & 0 & $\mathbb{Z}$ & 0 & $\mathbb{Z}$ & 0 & $\mathbb{Z}$ & 0 & $\mathbb{Z}$             \\  
  \hline 
  AI & $+$ & 0 & 0  & $\mathbb{Z}$ & 0 & 0 & 0 & $2\mathbb{Z}$ & 0 & $\mathbb{Z}_2$ & $\mathbb{Z}_2$       \\
  BDI    & $+$ & $+$ & 1   & $\mathbb{Z}_2$ & $\mathbb{Z}$ & 0 & 0 & 0 & $2\mathbb{Z}$ & 0 & $\mathbb{Z}_2$   \\
  D & 0 & $+$ & 0  & $\mathbb{Z}_2$ & $\mathbb{Z}_2$ & $\mathbb{Z}$ & 0 & 0 & 0 & $2\mathbb{Z}$ & 0        \\
  DIII & $-$ & $+$ & 1 & 0 & $\mathbb{Z}_2$ & $\mathbb{Z}_2$ & $\mathbb{Z}$ & 0 & 0 & 0 & $2\mathbb{Z}$      \\
  AII & $-$ & 0 & 0 & $2\mathbb{Z}$ & 0 & $\mathbb{Z}_2$ & $\mathbb{Z}_2$ & $\mathbb{Z}$ & 0 & 0 & 0       \\
  CII & $-$ & $-$ & 1  & 0 & $2\mathbb{Z}$ & 0 & $\mathbb{Z}_2$ & $\mathbb{Z}_2$ & $\mathbb{Z}$ & 0 & 0      \\
  C & 0 & $-$ & 0 & 0 & 0 & $2\mathbb{Z}$ & 0 & $\mathbb{Z}_2$ & $\mathbb{Z}_2$ & $\mathbb{Z}$ & 0         \\
  CI & $+$ & $-$ & 1& 0 & 0 & 0 & $2\mathbb{Z}$ & 0 & $\mathbb{Z}_2$ & $\mathbb{Z}_2$ & $\mathbb{Z}$    
  \\   
\end{tabular}
\end{ruledtabular}
\caption{
Periodic table of topological insulators and superconductors;
$\delta:=d-D$, where  $d$ is the space dimension and
$D+1$ is the codimension of defects;
the left-most column (A, AIII, $\ldots$, CI)
denotes the ten symmetry classes of fermionic Hamiltonians,
which are characterized by the presence/absence of
time-reversal (T), particle-hole (C), and chiral (S) symmetry of different types denoted by $\pm 1$. 
The entries ``$\mathbb{Z}$'', ``$\mathbb{Z}_2$'', ``$2\mathbb{Z}$'',
and ``$0$'' represent the presence/absence of non-trivial
topological insulators/superconductors or topological defects,
and when they exist, types of these states.
The case of $D=0$ (i.e., $\delta=d$) corresponds 
to the  tenfold classification of gapped bulk topological insulators and superconductors. 
\label{tab:classification}
}
\end{table}

Since physical systems can be characterized by the presence/absence of symmetries 
(Sec.\ \ref{sec:Sym}),
it is meaningful to discuss the topological distinction of quantum phases 
in the presence of a certain set of symmetry conditions. 
Let us then 
consider an ensemble of Hamiltonians within a given symmetry class
and for a fix spatial dimension $d$,
and ask if there is a topological distinction among 
ground states of gapped insulators and SCs 
\footnote{
More specifically, we are here not interested in  
systems with genuine {\it topological order}, 
whose existence has nothing to do with the presence/absence of 
symmetries, 
but in {\it symmetry-protected topological phases}
-- see Sec.\ \ref{Effects of interactions}.
}.
In particular, we will focus below on the classification of 
{\it topological insulators} and {\it superconductors}
in free fermion systems, described by  quadratic Bloch-BdG Hamiltonians.
Namely, we are interested in quadratic Hamiltonians  of the form
\begin{align}
\hat{H} 
=
\sum_{\mathsf{r},\mathsf{r}'} 
\hat{\psi}^{\dag}_i(\mathsf{r})\, 
{H}^{ij}(\mathsf{r},\mathsf{r}')\, 
\hat{\psi}_j(\mathsf{r}'), 
\end{align}
where 
$
\hat{\psi}_i(\mathsf{r})
$ 
is a multi-component fermion annihilation operator,
and index $\mathsf{r}$ labels a site
on a $d$-dimensional lattice.
Quadratic BdG Hamiltonians defined on a $d$-dimensional lattice
can be treated/discussed similarly. 
The single-particle Hamiltonians $H^{ij}(\mathsf{r},\mathsf{r'})$
belong to one of the ten AZ symmetry classes
and are, in general, subject to a set of symmetry constraints, see \eqref{symmetrysummaryeq}.

Assuming that the physical system has translation symmetry,
$
{H}^{ij}(\mathsf{r},\mathsf{r}') = 
{H}^{ij}(\mathsf{r}-\mathsf{r}'),
$
with periodic boundary conditions in each spatial direction, 
it is convenient to use  
the corresponding single-particle Hamiltonian in momentum space, 
$H^{ij}(\mathsf{k})$,
\begin{align}
\hat{H} = \sum_{\mathsf{k} \in \mathrm{BZ}^d} 
\hat{\psi}^{\dag}_i(\mathsf{k}) \, {H}^{ij}(\mathsf{k})\, 
\hat{\psi}_j (\mathsf{k}), 
\end{align}
where the crystal momentum $\mathsf{k}$
runs over the first Brillouin zone (BZ).
The Fourier components of the fermion operator
and the Hamiltonian are given by 
$
\hat{\psi}_i(\mathsf{r})
=
\sqrt{V}^{-1}
\sum_{\mathsf{k}\in \mathrm{BZ}^d}
e^{{i} \mathsf{k}\cdot \mathsf{r}}
\hat{\psi}_i(\mathsf{k})
$
and
$
{H}^{ij}(\mathsf{k})
=
\sum_{\mathsf{r}} 
e^{- {i} \mathsf{k}\cdot \mathsf{r}}
{H}^{ij}(\mathsf{r})
$, respectively, 
where 
$V$ is the total number of sites
\footnote{
It should however be emphasized that all TIs and TSCs in the ten AZ symmetry classes are stable against disorder,
and hence the assumption of translation invariance is not  at all necessary
(see Sec.\ \ref{subsec:Anderson_delocalization}).}. 
 TRS, PHS, and chiral symmetry act on the single-particle Hamiltonian $H(\mathsf{k})$ as 
\begin{align}
TH(\mathsf{k})T^{-1}&=H(-\mathsf{k}), 
\label{TRS_eq} \\
CH(\mathsf{k})C^{-1}&=-H(-\mathsf{k}), 
\label{PHS} \\
SH(\mathsf{k})S^{-1}&=-H(\mathsf{k}),
\label{chiral eq}
\end{align} 
where $T$, $C$, and $S$ are the antiunitary TR, PH, and unitary chiral operators, respectively. 
%
With this setup, 
we then ask, whether two gapped quadratic Hamiltonians, which belong to the same symmetry class, 
 can be continuously transformed into each other without closing the gap. That is, 
 we classify gapped Hamiltonians of a given symmetry class into different topological equivalence classes.
The result of this classification is summarized by the {\it Periodic Table} of TIs and TSCs 
\cite{Schnyder2008, Kitaev2009,Ryu2010ten,SchnyderAIP, Qi2008sf};
see Table \ref{tab:classification}. 
(The case of $D=0$ (i.e., $\delta=d$) corresponds to the  tenfold classification of gapped 
bulk TIs and TSCs.)
Systematic derivations of this classification table will be discussed later.
Here, a few comments on noticeable features of the table are in order:

\begin{figure}[t]
\centering
\includegraphics[width=0.25\textwidth]{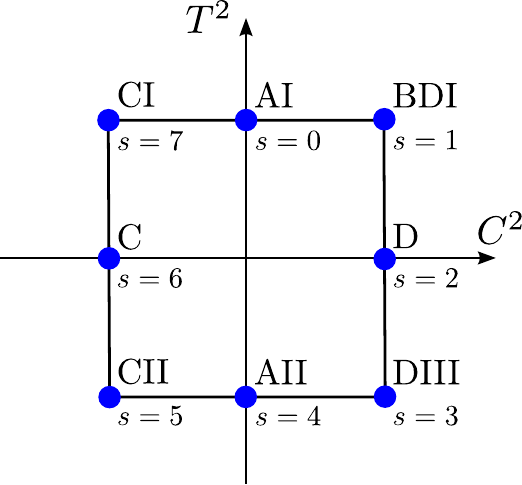}
\caption{
The 8 real symmetry classes that involve the antiunitary symmetries $T$ (time reversal) and/or $C$ (particle-hole) are specified by 
the values of $T^2 = \pm 1$ and $C^2=\pm 1$.  They can be visualized on an eight-hour ``clock". Adapted from \cite{Teo:2010fk}.
\label{fig:symmetryclock}
}
\end{figure}

-- The symmetry classes A and AIII, and the other eight classes are separately displayed. 
We will call the former ``the complex symmetry classes'',
and the latter ``the real symmetry classes''. 
The complex symmetry classes do not have TRS nor PHS. 

--
The symbols  
``$\mathbb{Z}$'',
``$\mathbb{Z}_2$'',
``$2\mathbb{Z}$'',
and 
``$0$'', 
indicate
whether or not TIs/TSCs exist 
for a give symmetry class in a given dimension, 
and 
if they exist,
what kind of topological invariant characterizes the topological phases. 
For example, 
``$2\mathbb{Z}$''
\footnote{
The label ``$2\bZ$" indicates that
the topological invariant is given by an even integer, reflecting 
the fact that there is an even number of protected gapless surface modes. 
Note, however, that the group of integers ($\bZ$) and 
the group of even integers ($2\bZ$) are isomorphic.
} 
indicates that the topological phase is characterized by an even-integer 
topological invariant,
and 
``$0$''
simply means there is no TI/TSC.
I.e.,
all states in a symmetry class in a given dimension
are adiabatically deformable. 

-- 
In the table, 
the so-called {\it weak} TIs/TSCs,  
which are non-trivial topological phases that exist in the presence of lattice translation symmetries, 
are not presented. That is, Table~\ref{tab:classification} only shows 
the {\it strong} TIs and TSCs
whose existence does not rely on translation symmetries.  
However, the presence/absence of weak TIs/TSCs in a given symmetry class 
can be deduced from 
the presence/absence of strong TIs/TSCs in lower dimensions
in the same symmetry class.

-- 
The classification table exhibits a periodicity of 2 and 8 as a function of spatial dimension,
for the complex and real symmetry classes, respectively.
(The table is only shown up to $d=7$ for this reason.)
In addition, note that the classifications for different symmetry classes are related by a dimensional shift.
For this reason it is convenient to label the eight real AZ symmetry classes by an integer $s$ running from 0 to 7, 
which can be arranged on a periodic {\em eight-hour clock}, ``the Bott clock'' (Fig.\ \ref{fig:symmetryclock}). 
Denoting 
the classification of TIs/TSCs in symmetry class $s$
and in space dimension $d$ by $K(s; d, 0)$, 
the periodic table can be summarized as Table \ref{tab:defectclassification}.

-- 
Now, let us examine the pattern in which the different kinds of topological phases appear in the table.
Along the main diagonal of the table there appear the entries for topological phases characterized by an integer
topological invariant (``$\mathbb{Z}$''). 
These topological phases will be called ``primary series''.
Just below the primary series (i.e., to the lower left),  
there are two sets of diagonal entries 
for topological phase characterized by a $\mathbb{Z}_2$ topological invariant. 
These topological phases are called ``the first descendants'' and 
``the second descendants'', 
respectively.
There is also a series of topological phases characterized by  $2\mathbb{Z}$ invariants,
i.e.,  by an even integer topological invariant.
These entries will be called ``even series''. 

\begin{table}[t!]
\centering
\begin{ruledtabular}
\begin{tabular}{c|cccccccc}
$s-\delta$&0&1&2&3&4&5&6&7\\\hline
$K(s;d,D)$&$\mathbb{Z}$&$\mathbb{Z}_2^{(1)}$&$\mathbb{Z}_2^{(2)}$&0&$2\mathbb{Z}$&0 & 0 & 0
\end{tabular}
\end{ruledtabular}
\caption{
The eightfold periodic classification of topological insulators/superconductors and topological defects with 
time-reversal and/or particle-hole symmetries;
$s$ labels the Altland-Zirnbauer symmetry classes (see Fig.\ \ref{fig:symmetryclock}); 
$\delta=d-D$ is the topological dimension; 
$\mathbb{Z}_2^{(1,2)}$ are the first and second descendant $\mathbb{Z}_2$ classifications.
\label{tab:defectclassification}
}
\end{table}

\begin{center}
***
\end{center}

To discuss an observable consequence of having a topologically non-trivial state,
let us recall that, by definition, 
topologically non-trivial and trivial states
in the phase diagram  
are always separated by a quantum phase transition,  
if the symmetry conditions are strictly enforced. 
This, in turn, implies that if a TI or TSC is in spatial proximity to a trivial phase, 
there should be a gapless state localized at the boundary between the two phases.
This gapless (i.e., critical) state can be thought of as arising due to a phase transition occurring locally in space,
where the parameters of the Hamiltonian change as a function of the direction transverse to the boundary.
Such gapless boundary modes are {\it protected} in the sense that 
they are stable against perturbations as long as the bulk gap is not destroyed  
and the symmetries are preserved.
In particular,  
gapless boundary modes are completely immune to disorder and evade Anderson localization completely
(Sec.\ \ref{subsec:Anderson_delocalization}). 
The presence of such gapless boundary states is the most salient feature of 
TIs and TSCs, and in fact, can be considered as a {\it definition} of TIs and TSCs.
This close connection between non-trivial bulk topological properties and gapless boundary modes
is known as the {\it bulk-boundary correspondence} 
(Sec.\ \ref{Bulk-boundary and bulk-defect correspondence}).

\subsubsection{Topological defects} 
\label{Topological defects}

\begin{figure}[t]
\centering\includegraphics[width=0.35\textwidth]{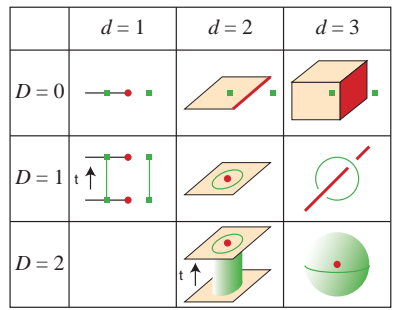}
\caption{
Topological defects characterized by a $D$ parameter family of $d$-dimensional Bloch-BdG Hamiltonians.  
Line defects correspond to $d-D=2$, while point defects correspond to $d-D=1$. 
Temporal cycles for point defects correspond to $d-D=0$. Adapted from \cite{Teo:2010fk}.}
\label{fig:defectdimension}
\end{figure}

 Boundaries 
separating bulk TIs/TSCs from trivial states of matter,
which host topologically protected gapless modes, 
are {\em codimension} one objects, 
i.e., one dimension less than the bulk. 
It is possible to discuss general higher codimension 
{\it topological defects}, such as point and line defects introduced 
in a gapped bulk system,  
and their topological classification.
Topological properties of {\it adiabatic cycles} can also be discussed in a similar manner.

Topological defects have been discussed
originally in the context of spontaneous symmetry breaking. 
For example, 
the quantum flux vortex of a type II SC~\cite{deGennesbook} 
involves the winding of the pairing order parameter, which breaks the charge conserving $U(1)$ symmetry. 
Dislocations and disclinations~\cite{ChaikinLubensky, Nelsonbook} 
are crystalline defects that associate discrete torisional and curvature fluxes in a lattice medium, 
which breaks continuous translation and rotation symmetries. 
They all involve non-trivial long length scale modulations of some order parameter around the defects. 

Topological defects in the context of topological band theories
\cite{Teo:2010fk} 
have a different origin 
in that they are not necessarily associated to
spontaneous symmetry breaking.
For example,
the mass gap that inverts between topological and trivial insulators 
does not break any symmetry. 
It is nonetheless a parameter in the band theory
that controls the topology of the bulk material, and we will refer to them as {\em band parameters} or {\em topological parameters}. 
Topological defects in insulators and SCs are therefore non-trivial long length scale windings 
of these topological parameters around the defects.

Topological defects of our interest are described by a defect Hamiltonian, which is a band Hamiltonian 
$H_{\mathsf{r}}(\mathsf{k})=H(\mathsf{k},\mathsf{r})$ 
that is slowly modulated by a parameter $\mathsf{r}$,
which includes spatial coordinates and/or a temporal parameter. 
A defect Hamiltonian describes the long length scale environment surrounding of a defect -- far away from it. 
The modulation is slow enough so that the bulk system well-separated from the defect core 
has {\em microscopic} spacetime translation symmetry, 
and hence can be characterized by momentum $\mathsf{k}$. 
More precisely, we assume 
$
\xi\left|\nabla_{\mathsf{r}}H(\mathsf{k},\mathsf{r})\right| \ll \varepsilon_g
$, 
where $\xi$ is a characteristic microscopic length scale similar to the lattice spacing, or a time scale similar to $1/\varepsilon_g$, 
where $\varepsilon_g$  is the bulk energy gap.%
\footnote{
Note, however, that the topological classification of topological defects,  which is presented in the following,
also applies to the cases where this assumption is not satisfied,
such as sharp interfaces or domain walls between different gapped bulk phases.  
}
TR, PH, and chiral symmetry act on a defect Hamiltonian as
\begin{align}
TH(\mathsf{k},\mathsf{r})T^{-1}&=H(-\mathsf{k},\mathsf{r}), \label{TRS_r}
\\
CH(\mathsf{k},\mathsf{r})C^{-1}&=-H(-\mathsf{k},\mathsf{r}), \label{PHS_r}
\\
SH(\mathsf{k},\mathsf{r})S^{-1}&=-H(\mathsf{k},\mathsf{r}), \label{chiral_r}
\end{align} 
where the spatial (temporal, when discussing adiabatic cycles) 
parameter $\mathsf{r}$ is unaltered, since the symmetries act on local microscopic degrees of freedom, 
which are independent of the slowly varying modulation.

Different defect Hamiltonians are distinguished by (i) the AZ symmetry class $s$, 
(ii) the bulk dimension $d$, and (iii) the defect codimension $d_c$
defined in terms of the dimension of the defect 
$d_{\mathrm{defect}}$
by 
$
d_c=d-d_{\mathrm{defect}}. 
$
A spatial defect of dimension $d_{\mathrm{defect}}$ 
is wrapped by a $D$-dimension sphere $S^D$, 
where $D=d_c-1=d-d_{\mathrm{defect}}-1$. 
For example,
a point defect in 3d has codimension $d_c=3-0=3$ 
and thus is surrounded by a 2d sphere. 
Adiabatic cycles are incorporated as topological defects that depend on a cyclic temporal parameter. 
In this case the defect is enclosed by a sphere $S^{D-1}$ of dimension 
$D-1=d-d_{\mathrm{defect}}-1$ in $d$-dimensional real space. 
Together with the temporal parameter that lives on $S^1$, 
the adiabatic cycle is {\em wrapped} by a $D$-dimensional manifold such as $S^{D-1}\times S^1$. 
A table of low dimensional defects is presented in Fig.\ \ref{fig:defectdimension}.

For real AZ symmetry classes, 
it was shown that the classification of topological defects depends only
on a single number~\cite{Teo:2010fk, Freedman2011} 
\begin{align}s-\delta=s-d+D\quad\mbox{modulo 8}, \end{align} 
where $\delta=d-D$ is called the {\em topological dimension} that takes the role of the usual dimension $d$ 
in the case of gapped TIs and TSCs.
For spatial defects, 
the topological dimension is related to the defect dimension by $\delta=d_{\mathrm{defect}}+1$ 
and is independent of the bulk dimension $d$.
For instance, point defects always have $\delta=1$, 
while line defects always have $\delta=2$. 
For adiabatic cycles, the extra temporal parameter 
in the $D$-component parameter $\mathsf{r}$ 
reduces the topological dimension by one. 
For example, a temporal cycle of point defects has $\delta=0$.
The classification is summarized in 
Tables \ref{tab:classification} and \ref{tab:defectclassification}.
(As in the case of gapped TIs and TSCs, 
we are interested in the highest dimension {\em strong} topologies of 
the defect that do not involve lower dimensional cycles.)

Topological defects in the two complex AZ classes A and AIII are classified in a similar manner, except that the symmetry classes 
now live on a periodic two-hour clock,
and the topological dimension $\delta=d-D$ as well as the number $d-\delta$ are now integers modulo 2. 
Topological defects in class A (class AIII) are $\mathbb{Z}$-classified when $\delta$ is even (when $\delta$ is odd). 
Otherwise they are trivially classified. 
By {\em forgetting} the antiunitary symmetries, the real AZ classes separate into the two complex classes AI,D,AII,C$\to$A and BDI,DIII,CI,CII$\to$AIII, 
where the chiral operator $S$ is given by the product of TR and PH (possibly up to a factor of $i$). 
This procedure ({\em forgetful functor} -- see Sec.\ \ref{K-theory approach}) 
relates real and complex classifications. 
For instance, the $2\mathbb{Z}$ classification for $s-\delta\equiv4$ modulo 8 in Table~\ref{tab:defectclassification} 
is normalized according to the corresponding complex $\mathbb{Z}$ classification. 
This means when forgetting the antiunitary symmetries, the topological invariants must be even for $s-\delta\equiv4$.

Like the bulk-boundary correspondence that relates bulk topology to boundary gapless excitations, 
we have a {\em bulk-defect} correspondence that guarantees gapless defect excitations from the non-trivial winding 
of bulk topological parameters around the defect. 
This framework unifies numerous TI and TSC defect systems 
(Sec.\ \ref{Bulk-boundary and bulk-defect correspondence}).

\subsection{Topological invariants}
\label{subsec:topinvariants}

In this section, we discuss 
the tenfold classification of gapped TIs/TSCs
and topological defects, 
in terms of bulk topological invariants.
A short summary of topological invariants that will be discussed is presented in Table~\ref{tab:defectinvariant}. 
Various specific examples of the topological invariants and systems characterized by the topological invariants
will be discussed, but we will mostly confine ourselves to examples taken from gapped TIs and TSCs.    
Examples of topological defects will be discussed later, in
 Sec.\ \ref{Bulk-boundary and bulk-defect correspondence}.
A systematic derivation of the periodic table 
and 
physical consequences of the non-trivial bulk topologies measured by the topological invariants,
such as gapless modes localized at boundaries and defects, 
will be discussed in Sec.\  \ref{K-theory approach} 
and in Sec.\ \ref{Bulk-boundary and bulk-defect correspondence}, respectively.

The topological invariants that will be introduced in this section
are given in terms of the eigen function of a Bloch-BdG Hamiltonian.  
We denote the $a$-th eigen function 
with energy $\varepsilon^{{a}} (\mathsf{k},\mathsf{r})$
by $|u^{{a}}(\mathsf{k}, \mathsf{r})\rangle$,
$
{H}(\mathsf{k},\mathsf{r}) |u^{{a}}(\mathsf{k},\mathsf{r}) \rangle
=
\varepsilon^{a} (\mathsf{k},\mathsf{r})|u^{a}(\mathsf{k},\mathsf{r}) \rangle.
$
By assumption, there is a spectral gap at the Fermi energy in the band structure given by 
$\varepsilon^a(\mathsf{k},\mathsf{r})$. 
We assume that there are $N_{-/+}$ bands below/above the Fermi energy.
The total number of the bands is $N_{+}+N_{-}$.
We denote the set of filled Bloch wavefunctions  by
$
\{
|u_{-}^{\alpha}(\mathsf{k},\mathsf{r}) \rangle
\}
$,
or simply
$
\{
|u^{\alpha}(\mathsf{k},\mathsf{r}) \rangle
\}
$,
where 
the Greek index $\alpha=1,\ldots, N_-$ 
labels the occupied bands only.

The Bloch wave functions are defined on 
the base manifold, $\mathrm{BZ}^d\times\mathcal{M}^D$, 
the $(d+D)$-dimensional total phase space parameterized by $(\mathsf{k},\mathsf{r})$.
Here,
the $D$-dimensional manifold $\mathcal{M}^D$ wraps around the topological defect 
(Fig.~\ref{fig:defectdimension}).
(It deformation retracts from the defect complement of spacetime.) 
For example, taking away a point defect in real $3$-space leaves behind a punctured space, 
which has the same homotopy type as the 2-sphere $S^2$. 
The complement of an infinite defect line in $3$-space can be compressed along the defect direction onto a punctured disc, 
which then can be deformation retracted to the circle $S^1$. 
The $D$-manifold $\mathcal{M}^D$ enclosing a more complicated topological defect may not be spherical. 
For instance, the one surrounding a link in $3$-space is a $2$-torus. 
The $D$-manifold of a temporal cycle must contain a non-contractible 1-cycle that corresponds to the periodic time direction. 
For the bulk of the review, we are interested in the highest dimension {\em strong} topologies of defects that do not involve lower dimensional cycles. 
For this purpose, 
we compactify the phase space into a sphere 
$
(\mathsf{k},\mathsf{r})\in \mathrm{BZ}^d\times\mathcal{M}^D\xrightarrow{\mbox{\small compactify}}S^{d+D}
$
by contracting all lower dimensional cycles.
Physically this means the defect band theory are assumed to have trivial winding around those low-dimensional cycles.

\begin{table}[tbp]
\centering
\begin{ruledtabular}
\begin{tabular}{ccc}
&non-chiral classes ($s$ even)&chiral classes ($s$ odd)\\ 
\hline 
$\mathbb{Z}$&Chern number (Ch)&winding number ($\nu$)\\
$\mathbb{Z}_2^{(1)}$&CS (CS)&Fu-Kane (FK)\\
$\mathbb{Z}_2^{(2)}$&Fu-Kane (FK)&CS ($\widetilde{\mbox{CS}}$)\\
\end{tabular}
\end{ruledtabular}
\caption{
Strong topological invariants for topological defects. 
The $\mathbb{Z}$-invariants apply to both complex and real 
Altland-Zirnbauer classes. 
\label{tab:defectinvariant}
}
\end{table}

\subsubsection{Primary series for \texorpdfstring{$s$}{s} even -- the Chern number}

For gapped topological phases and topological defects in non-chiral classes (i.e., $s$ is even), 
the $\mathbb{Z}$-classified topologies are characterized by 
{\it the Chern number} 
\begin{align}
\mbox{Ch}_n=\frac{1}{n!}\left(\frac{i}{2\pi}\right)^n
\int_{\mathrm{BZ}^d\times\mathcal{M}^D}
\mbox{Tr}\left(\mathcal{F}^n\right) ,
\label{defectChern}   
\end{align} 
where $n:=(d+D)/2$. 
The Berry curvature%
\footnote{
As in \eqref{defectChern} - \eqref{Berry curv}, we will use the differential form notation.
E.g., 
\begin{align}
\mathcal{A}^{\alpha\beta}&= A^{\alpha\beta}_I(\mathsf{s}) ds^{I}, \quad
A^{\alpha\beta}_I (\mathsf{s}):=\langle u(\mathsf{s})| \partial_I u(\mathsf{s})\rangle, 
\nonumber \\
\mathcal{F}^{\alpha\beta}&= 
d\mathcal{A}^{\alpha\beta} +
\mathcal{A}^{\alpha\gamma}
\wedge
\mathcal{A}^{\gamma\beta}
\nonumber \\
&= 
(\partial_{I} A^{\alpha\beta}_J + A^{\alpha\gamma}_I A^{\gamma\beta}_J) ds^{I}\wedge ds^J
\nonumber \\
&=
\frac{1}{2}
(\partial_{I} A_J 
-\partial_{J}A_I
+
[A_I, A_J])^{\alpha\beta} ds^{I}\wedge ds^J, 
\end{align}
where $\mathsf{s}=(\mathsf{k},\mathsf{r})$ and $I,J=1,\cdots, d+D$. 
The wedge symbol $\wedge$ is often omitted. 
When necessary, we use a subscript to indicate that a differential form $\mathcal{A}_n$ is an $n$-form. 
}
\begin{align}
 \mathcal{F} = d\mathcal{A} + \mathcal{A}^2 
 \label{Berry curv}
\end{align}
is given in terms of the non-Abelian Berry connection 
\begin{align}
 \mathcal{A}^{\alpha\beta} (\mathsf{k},\mathsf{r})
 &=
 \langle u^{\alpha}(\mathsf{k},\mathsf{r}) 
 | du^{\beta}(\mathsf{k},\mathsf{r})\rangle 
 \nonumber \\
&=
\langle u^{\alpha}(\mathsf{k},\mathsf{r})|\nabla_{\mathsf{k}} u^{\beta}(\mathsf{k},\mathsf{r})
\rangle\cdot d\mathsf{k}
\nonumber\\
&\quad 
+
\langle u^{\alpha}(\mathsf{k},\mathsf{r})|\nabla_{\mathsf{r}}u^{\beta}(\mathsf{k},\mathsf{r})
\rangle\cdot d\mathsf{r} .
\end{align}
The Chern number is well-defined only when $d+D$ is even. 
Furthermore, 
it vanishes in the presence of  TRS (or PHS) when $\delta=d-D$ is 2 (resp.~0) mod 4. 
Moreover it must be even when $s-\delta$ is 4 mod 8.

The Chern number detects an {\it obstruction}
in defining  a set of Bloch wave functions smoothly over the base space
$\mathrm{BZ}^d\times \mathcal{M}^{D}$.
Associated with each $(\mathsf{k}, \mathsf{r})$, 
we have a set of wave functions, $|u^a (\mathsf{k},\mathsf{r})\rangle$,
a collection of which can be thought of as a member of $U(N_++N_-)$.
There is however a gauge redundancy: 
$U(N_{\pm})$ rotations among unoccupied/occupied Bloch wave functions give rise to 
the same quantum ground state (the Fermi-Dirac sea) 
at given $(\mathsf{k},\mathsf{r})$. 
In other words, 
the quantum ground state at a given $(\mathsf{k},\mathsf{r})$
is a member of the coset space $U(N_+ + N_-)/U(N_-)\times U(N_+)$,
the complex Grassmannian.
The Fermi-Dirac sea at $(\mathsf{k},\mathsf{r})$
can be conveniently described  by {\it the spectral projector}: 
\begin{align}
\label{projector}
 P(\mathsf{k},\mathsf{r})  
 &= 
 \sum_{\alpha=1}^{N_-} 
 |u^{\alpha}(\mathsf{k},\mathsf{r})\rangle
 \langle u^{\alpha}(\mathsf{k},\mathsf{r})|,
\end{align}
(or $P^{ij}(\mathsf{k},\mathsf{r}) = \sum_{\alpha=1}^{N_-} u_i{ }^{\alpha}(\mathsf{k},\mathsf{r}) [u_j{ }^{\alpha}(\mathsf{k},\mathsf{r})]^*$
if indices are shown explicitly),
which specifies
a subspace of the total Hilbert space defined by the set of occupied Bloch wave functions.
The projector is gauge invariant and a member of
the complex Grassmannian: 
$
P(\mathsf{k},\mathsf{r}) \in U(N_++N_-)/U(N_-)\times U(N_+) 
$.
For what follows, it is convenient to introduce 
the ``$Q$-matrix'' by
\begin{align}
 Q(\mathsf{k},\mathsf{r}) = \openone - 2P(\mathsf{k},\mathsf{r}). 
\end{align}
The $Q$-matrix is hermitian and has the same set of eigen functions as 
$H(\mathsf{k},\mathsf{r})$, but its eigenvalues are either $\pm 1$ since 
$Q^2=\openone$. 

As we move around in the base space $\mathrm{BZ}^d\times \mathcal{M}^D$,
the set of wave functions undergo adiabatic changes.
Such wave functions thus define a fibre-bundle,
which may be ``twisted":
It may not be possible to find smooth wave functions that are well-defined everywhere over the base space.
One quick way to see when the fibre bundle is twisted 
is to note that the set of Bloch functions (or equivalently the projector)
defines a map from the base space to $U(N_++N_-)/U(N_+)\times U(N_-)$. 
Topologically distinct maps of this type can be classified by 
the homotopy group
\begin{align}
 \pi_{d+D} \left[ U(N_++N_-)/U(N_+)\times U(N_-) \right]. 
\end{align}
For large enough $N_{\pm}$ and when $d+D$ is even, 
$ \pi_{d+D} \left[ U(N_++N_-)/U(N_+)\times U(N_-) \right]=\mathbb{Z}$.
Topologically distinct maps are therefore characterized by an integer topological invariant, namely by
\begin{align}
\frac{-1}{2^{2n+1} } \frac{1}{n!} 
\left(
\frac{i}{2\pi}
\right)^n
 \int_{\mathrm{BZ}^d \times \mathcal{M}^D} 
 \mathrm{Tr}\, \left[ Q (dQ)^{2n}\right]. 
\label{Q matrix and Ch number} 
\end{align}
This, in turn, is nothing but the Chern number.

\paragraph{Example: The 2d class A quantum anomalous Hall effect}

As an illustration, let us consider band insulators with $N_{+}=N_{-}=1$
in two spatial dimensions $d=2$.  
In general,   two-band Bloch Hamiltonians can be written in 
terms of four real functions 
$R_{0,1,2,3}(\mathsf{k})$ as 
\begin{align} \label{two_band_model_classA}
 {H}(\mathsf{k}) =
 R_0(\mathsf{k}) \sigma_0 +
 \boldsymbol{R}(\mathsf{k}) \cdot \boldsymbol{\sigma},  
\end{align}
where $\boldsymbol{R} = (R_1, R_2, R_3)$. 
The energy dispersions of the bands are given by
$\varepsilon_{\pm}(\mathsf{k}) = R_0(\mathsf{k}) \pm R(\mathsf{k})$ 
with $R(\mathsf{k}):= |\boldsymbol{R}(\mathsf{k})|$. 
For band insulators, there is a spectral gap at the Fermi energy, 
which we take to be zero for convenience. 
Hence we assume 
$
R_0(\mathsf{k})+{R}(\mathsf{k})  >0> R_0(\mathsf{k})-{R}(\mathsf{k}) 
$,
which in particular implies ${R}(\mathsf{k})>0$ for all $\mathsf{k}$.

In this two-band example,
the Bloch Hamiltonian $H(\mathsf{k})$
or 
the four vector $R_{\mu=0,1,2,3}(\mathsf{k})$
defines a map from the BZ to the space of the unconstrained four vector $R_{\mu}$.
The Bloch wave functions, however,
depend only on the normalized vector 
$\boldsymbol{n}(\mathsf{k})\equiv 
\boldsymbol{R}(\mathsf{k})/{R}(\mathsf{k})$,
as seen easily from 
(i) $R_0(\mathsf{k})$ in $H(\mathsf{k})$ does not affect the wave functions,
and
(ii) $\boldsymbol{R}(\mathsf{k}) \cdot \boldsymbol{\sigma} =
{R}(\mathsf{k})
\boldsymbol{n}(\mathsf{k})\cdot \boldsymbol{\sigma}$. 
(Note that because of the presence of the spectral gap, 
${R}(\mathsf{k})>0$ for all $\mathsf{k}$, and
the normalized vector $\boldsymbol{n}(\mathsf{k})$ is always well defined). 
Thus, 
from the point of view of the Bloch wave functions, 
we consider a map from the BZ to the space of the normalized vector $\boldsymbol{n}$,
which is simply $S^2$. 
The latter is the simplest example of the complex Grassmannian,
$U(2)/U(1)\times U(1) \simeq S^2$. 

Within the two-band model, different band insulators can thus be characterized by
different maps $\boldsymbol{n}(\mathsf{k})$.
By ``compactifying'' the BZ $T^2$ to $S^2$, 
topologically distinct maps can be classified by
the second Homotopy group $\pi_2(S^2)$, which is given by 
$
 \pi_2 \left( S^2 \right) = \mathbb{Z}.   
$
For a given map $\boldsymbol{n}$, the integer topological invariant
\begin{align} 
\frac{1}{4\pi}
 \int_{\mathrm{BZ}} \boldsymbol{n} \cdot d\boldsymbol{n}\times d\boldsymbol{n}
 \in
 \mathbb{Z} 
 \label{hedgehog}
\end{align}
counts the number of times the unit vector $\boldsymbol{n}$ ``wraps'' around
$S^2$ as we go around the BZ, and hence tells us to which topological class
the map $\boldsymbol{n}(\mathsf{k})$ belongs.

Let us now construct the Bloch wave functions explicitly. 
One possible choice is  
\begin{align}
|{u}^{\pm}\rangle 
&=
\frac{1}{\sqrt{2{R}({R}\mp R_3)}}
\left(
\begin{array}{c}
R_1 -{i} R_2 \\
\pm {R} -R_3
\end{array}
\right).
\label{wfn gauge 1}
\end{align}
Observe that the occupied Bloch wave function $|u^-\rangle$
has a singularity at $\boldsymbol{R}= (0,0,-{R})$, i.e., at the
``south pole''.
When the topological invariant (\ref{hedgehog})
is non-zero, the vector $\boldsymbol{n}(\mathsf{k})$ necessarily 
maps at least one point in the BZ to the south pole, and hence one encounters a singularity,
if one insists on using the wave function (\ref{wfn gauge 1}) everywhere in the BZ.
There is an obstruction in this sense in defining 
wave functions that are smooth and well-defined globally in the BZ. 
To avoid the singularity, one can ``patch'' the BZ and use
different wave functions on different patches. 
For example, near the south pole
one can make an alternative choice,
\begin{align}
|{u}^{\pm}\rangle 
&=
\frac{1}{\sqrt{2{R}({R}\pm R_3)}}
\left(
\begin{array}{c}
\pm {R} +R_3 \\
R_1 +{i} R_2 \\
\end{array}
\right),
\label{wfn gauge 2}
\end{align}
which is smooth at the south pole, but singular at the north pole $\boldsymbol{R}=(0,0,{R})$. 
With the two patches with the wave functions 
(\ref{wfn gauge 1})
and
(\ref{wfn gauge 2}), one can cover the entire BZ.
In those
regions where the two patches overlap,
the two  wave functions are related to each other by a gauge transformation 

With the explicit form of the Bloch wave functions,
one can compute the spectral projector or the $Q$-matrix,
and check that the different gauge choices 
(\ref{wfn gauge 1})
and
(\ref{wfn gauge 2})
give rise to the same projector
(the projector is gauge invariant), 
and that it depends only on $\boldsymbol{n}(\mathsf{k})$, i.e., 
$Q(\mathsf{k}) = \boldsymbol{n}(\mathsf{k}) \cdot \boldsymbol{\sigma}$. 
From the Bloch wavefunctions, one can compute the Berry connection and then
the Chern number
$
\mathrm{Ch}  = 
({i}/{4\pi})
\int_{\mathrm{BZ}}
\mathrm{Tr}\, \mathcal{F}.  
$
The Chern number is, in fact, equal to the topological invariant
(\ref{hedgehog}),
as can been seen from (\ref{Q matrix and Ch number}), since
$
\mathrm{Tr}\, \mathcal{F}(\mathsf{k}) = 
 ({i}/{2}) \varepsilon^{ijk} 
n_i (\partial_{\mu} n_j) (\partial_{\nu} n_k) dk^{\mu}\wedge dk^{\nu}. 
$

An explicit example of the two-band model~\eqref{two_band_model_classA} with non-zero Chern number
is given in momentum space by
\begin{align}
 \boldsymbol{R}(\mathsf{k})
 =
 \left(
 \begin{array}{c}
-2  \sin k_x \\
-2  \sin k_y \\
\mu+ 2  \sum_{i=x,y} \cos k_i
 \end{array}
 \right). 
 \label{Model, Chern insulator}
\end{align}
There are four phases separated by three quantum critical points
at $\mu=0,\pm 4$, which are labeled by
the Chern number as
$\mathrm{Ch}=0$ $( |\mu| > 4)$,
$\mathrm{Ch}=-1$ $(-4 < \mu < 0)$, and
$\mathrm{Ch}=+1$ $( 0 < \mu < +4)$.
Band insulators on $d=2$ dimensional lattices having non-zero Chern number and without net magnetic field  
are commonly called {\it Chern insulators} 
and exhibit the quantum anomalous Hall effect \cite{Haldane1988,RevModPhys_AHE}, 
which generalize the integer QHE realized in the presence of a uniform magnetic field \cite{Klitzing,Thouless:1982rz,Laughlin_IQHE,Kohmoto:1985zl,QHE_book_Prange}. 
The Chern number is nothing but the quantized Hall conductance $\sigma_{xy}$.
Experimental realizations
of Chern insulators include
Cr-doped (Bi,Sb)$_2$Te$_3$ thin films \cite{Xue_QAHE,Yu_QAHE,XuZhang15}, 
InAs/GaSb and ${\mathrm{Hg}}_{1-y}{\mathrm{Mn}}_{y}\mathrm{Te}$ quantum wells \cite{WangLiu14,LiuZhang08}, 
graphene with adatoms \cite{QiaoNiu10}, 
and La$_2$MnIrO$_6$ monolayers \cite{ZhangVanderbilt14}.

\subsubsection{Primary series for \texorpdfstring{$s$}{s} odd  -- the winding number}

\paragraph{winding number}

The Chern number can be defined for Bloch-BdG Hamiltonians in 
any symmetry class as long as $d+D$ is even 
(although its allowed value depends on symmetry classes and $\delta$). 
On the other hand, there are topological invariants which can be defined 
only in the presence of symmetries. 
One example is the {\it winding number topological invariant} $\nu$,
which can be defined only in the presence of chiral symmetry,
$
 \{ {H}(\mathsf{k},\mathsf{r}), U_S \} =0
 $,
with $
U_S^2 = \openone. \label{chiral_eq}
$
For simplicity, we will focus below on the case of $\mathrm{Tr}\, U_S=0$, i.e., 
$N_+ = N_-=N$.

While in the absence of chiral symmetry the spectral projector is a member of the complex Grassmannian, in the presence of chiral symmetry
the relevant space  is the unitary group 
$U(N)$. 
This can be seen from 
the block-off-diagonal form of chiral symmetric Hamiltonians,
\begin{align}
 {H}(\mathsf{k},\mathsf{r})
 =
 \left(
 \begin{array}{cc}
 0 & D(\mathsf{k},\mathsf{r}) \\
 D^{\dag}(\mathsf{k},\mathsf{r}) & 0 
 \end{array}
 \right).  
\end{align}
Correspondingly, in this basis, the $Q$-matrix is also block-off diagonal, 
\begin{align}
 {Q}(\mathsf{k},\mathsf{r})
 =
 \left(
 \begin{array}{cc}
 0 & q(\mathsf{k},\mathsf{r}) \\
 q^{\dag}(\mathsf{k},\mathsf{r}) & 0 
 \end{array}
 \right),
 \label{block off diagonal Q}
\end{align}
where the off-diagonal block $q(\mathsf{k},\mathsf{r})$ is a unitary matrix. 
Hence, the $q$-matrix defines a map from the base space $\mathrm{BZ}^d\times\mathcal{M}^D$ to 
the space of unitary matrices $U(N)$. 
Topologically distinct  maps of this type are classified by the homotopy group $\pi_{d+D}[U(N)]$,
which is non-trivial when $d+D$ is odd, i.e.,
$\pi_{d+D}[U(N)]=\mathbb{Z}$
(for large enough $N$). 
Topologically distinct maps are characterized by the winding number, which is given by
\begin{align}
\nu_{2n+1}[q]
&=
\int_{\mathrm{BZ}^{d}\times \mathcal{M}^{D}}
\omega_{2n+1}[q],
\label{def winding number} \\
\omega_{2n+1}[q]
&=
\frac{(-1)^n n!}{(2n+1)!} 
\left(
\frac{i}{2\pi}
\right)^{n+1}
\mathrm{Tr}\,
\left[ 
(q^{-1} dq)^{2n+1}
\right], 
\nonumber 
\end{align}
where $d+D=2n+1$ is an odd integer. 
For example, 
when $(d,D)=(1,0), (3,0)$, we have
\begin{align}
\nu_1
&=
\frac{i}{2 \pi}
\int_{\mathrm{BZ}} d k \,
 \mathrm{Tr} \left[  q^{-1} \partial_k q \right], 
\label{winding number examples} \\
\nu_3 
&=
\int_{\mathrm{BZ}} \frac{d^3 \mathsf{k}}{24\pi^2} 
\epsilon^{\mu \nu \rho}
\mathrm{Tr} \left[ 
( q^{-1} \partial_\mu q ) ( q^{-1} \partial_\nu q ) ( q^{-1} \partial_\rho q )
\right],
\nonumber 
\end{align}
respectively, 
where $\partial_{\mu} = \partial_{k_{\mu}}$.

\paragraph{Chern-Simons invariant}
\label{paragraph CS invariant}

We now introduce yet another topological invariant, 
{\it the Chern-Simons invariant} (CS invariant). 
This invariant can be defined when $d+D=\mbox{odd}$,
and is not quantized in general, unlike the Chern number.
In the presence of symmetries, however, it may take discrete values.
We will use the quantized CS invariant later to characterize 
first and second descendants.
Here, we will show  that the CS invariant is also quantized in the presence of chiral symmetry. 

The CS invariant is defined in terms of the CS form $\mathcal{Q}_{2n+1}$ in $d+D=2n+1$ dimensions, where
\begin{align}
&
\mathcal{Q}_{2n+1}(\mathcal{A})
:=
\frac{1}{n!} \left(
\frac{i}{2\pi}
\right)^{n+1}
\int^1_0 dt\, \mathrm{Tr}\, (\mathcal{A} \mathcal{F}^n_t),
\label{CSformdef}
\nonumber \\
&
\mbox{with}
\quad 
\mathcal{F}_t = t d\mathcal{A} + t^2 \mathcal{A}^2
=
t\mathcal{F} + (t^2-t) \mathcal{A}^2. 
\end{align}
Integrating the CS form over the base space, yields the CS invariant
\begin{align}
\label{IntegralOfChernSimonsForm}
\mathrm{CS}_{2n+1} [\mathcal{A}] 
:=
\int_{\mathrm{BZ}^{d}\times \mathcal{M}^D}
\mathcal{Q}_{2n+1}(\mathcal{A}).
\end{align}
For example,   
for $n=0, 1, 2$, 
\begin{align}
\mathcal{Q}_{1}(\mathcal{A})
&=
\frac{i}{2\pi}
\mathrm{Tr}\, \mathcal{A}, 
\nonumber \\
\mathcal{Q}_{3}(\mathcal{A})
&=
\frac{ -1}{8\pi^2} 
\mathrm{Tr}\, 
\Big(
\mathcal{A} d \mathcal{A}
+
\frac{2}{3} \mathcal{A}^3
\Big),
\nonumber \\
\mathcal{Q}_{5}(\mathcal{A})
&=
\frac{ -i}{48\pi^3} 
\mathrm{Tr}\, 
\Big(
\mathcal{A} (d \mathcal{A})^2
+
\frac{3}{2} \mathcal{A}^3 d \mathcal{A}
+
\frac{3}{5} \mathcal{A}^5 
\Big). 
\label{CSformdef3}
\end{align}
The CS forms are not gauge invariant. 
Neither are the integrals of the CS forms.
However, 
for two different choices of gauge, 
$\mathcal{A}$ and $\mathcal{A}^g$,
which are connected by a gauge transformation $g$ as
\begin{align}
\mathcal{A}^g
:= g^{-1}\mathcal{A} g + g^{-1} d g,
\quad
\mathcal{F}^g
= g^{-1}\mathcal{F} g,
\end{align}
the difference 
$\mathcal{Q}_{2n+1}(\mathcal{A}^g) - \mathcal{Q}_{2n+1}(\mathcal{A})$
is given by the winding number density $\omega_{2n+1}[g]$ 
up to a total derivative term, 
\begin{align}
\mathcal{Q}_{2n+1}(\mathcal{A}^g ) - \mathcal{Q}_{2n+1}(\mathcal{A})
=
\omega_{2n+1}[g]
+
d \alpha_{2n+1} (\mathcal{A}, g).
\label{eq: Cartan homotopy}
\end{align}
Thus, for the integral of the CS form, 
\begin{align}
\mathrm{CS}_{2n+1}\left[\mathcal{A}^g\right]
-
\mathrm{CS}_{2n+1}\left[\mathcal{A}\right]
=
\mbox{integer}, 
\label{eq: CS invariant upto integer}
\end{align}
and hence the exponential 
\begin{align}
W_{2n+1}
: =
\exp\{
 2\pi i
\mathrm{CS}_{2n+1}
\left[\mathcal{A}\right]
\}
\end{align}
is a well-defined, gauge invariant quantity, 
although it is not necessarily quantized.

The discussion so far has been general. 
We now compute the CS invariant in the presence of chiral symmetry.
To this end, we first explicitly write down the Berry connection for chiral symmetric Hamiltonians.
For a given $q(\mathsf{k},\mathsf{r})$,  the eigen functions can explicitly be constructed as: 
\begin{align}
|u^{\alpha}_{\epsilon}(\mathsf{k},\mathsf{r})\rangle_{\mathrm{N}}
=
\frac{1}{\sqrt{2}}
\left(
\begin{array}{c}
|{n}^{\alpha} \rangle\\
\epsilon q^{\dag}(\mathsf{k},\mathsf{r}) 
|{n}^{\alpha} \rangle 
\end{array}
\right), 
\quad 
\epsilon=\pm, 
\label{u_N}
\end{align}
where $|n^{\alpha}\rangle$ are $N$ momentum independent orthonormal vectors. 
For simplicity we choose $(n^{\alpha})_{{\beta}}= \delta_{\alpha \beta}$.
These wave functions are free from any singularity. I.e., 
we have explicitly demonstrated that there is no obstruction to constructing eigen wavefuctions globally. 
The Berry connection is computed as
$
\mathcal{A}_{\mathrm{N}}
=
({1}/{2})
q(\mathsf{k},\mathsf{r})d q^{\dag}(\mathsf{k},\mathsf{r}) 
$.
In this gauge, 
the CS form $\mathcal{Q}_{2n+1}$
is shown to be one half of the winding number density, i.e.,
$
\mathcal{Q}_{2n+1}(\mathcal{A}_{\mathrm{N}})
=
\omega_{2n+1}[q^{\dag}]/2
$.
We conclude that
$
\mathrm{CS}_{2n+1}
\left[\mathcal{A}_{\mathrm{N}}\right]
=
\nu_{2n+1} [q^{\dag}]/2 
$
and hence
\begin{align}
W_{2n+1}
&=
\exp\{ \pi  {i} \, \nu_{2n+1}[q]\} = \pm 1.
\label{eq: master formula winding num and Wilson loop}
\end{align}
That is, for Hamiltonians with chiral symmetry
$W_{2n+1}$ can take on only two values, $W_{2n+1}=\pm 1$.

When $(d,D)=(1,0)$ ($n=0$),
the CS invariant $W_1$
is a $U(1)$ Wilson loop defined in the $\mathrm{BZ}^{d=1}\simeq S^1$.
The logarithm of $W_1$ represents
the electric polarization
\cite{kingSmithPRB93a,kingSmithPRB93b,restaRMP94},
which can be quantized by chiral symmetry and inversion symmetry 
\cite{RyuHatsugaiPRL02, Zak1989}.
In this context, 
the non-invariance of 
$\mathrm{CS}_{1}\left[\mathcal{A}\right]$,
(\ref{eq: CS invariant upto integer}), 
is  related to the fact that   
the displacement of electron coordinates
in periodic systems has a meaning only within a unit cell,
i.e., two coordinates
that differ by an integer multiple of the
lattice constant should be identified. 

When $(d,D)=(3,0)$ ($n=1$), 
$\mathrm{CS}_{3}$ represents the quantized magnetoelectric polarizability or ``$\theta$-angle''.
The $\theta$-angle, which is given in terms of  the Chern-Simons integral as
\begin{align}
\theta=
2\pi \int_{ \mathrm{BZ}^3}\mathcal{Q}_3(\mathsf{k})\quad\mbox{mod $2\pi$}, 
\end{align} 
appears in the electrodynamic efffective action through 
the axion term $\delta S=(\theta\alpha/4\pi)\int d^3\mathsf{r}dt\, {\bf E}\cdot{\bf B}$,
where $\alpha$ is the fine structure constant. 
The quantized magnetoelectric polarizability was first noted in the context of
3d TR symmetric TIs (in class AII)
\cite{Qi2008sf,essinPRL09,xiaoPRL09}.
Besides TRS, also chiral and inversion symmetry
quantize the CS invariant $W_3$ 
\cite{Turner:2010qf,Ryu2010ten,Ryu_CTI_SC,Deng_chiral,Deng_chiral_extended}.

\paragraph{Example: the 1d class AIII Polyacetylene}
\label{Example: Polyacetylene}

Consider the bipartite hopping model (\ref{hopping model}) on the 1d lattice, 
\begin{align}
\hat{H}  &=
t
\sum_i
(\hat{a}^{\dag}_{ i} \hat{b}^{\ }_{ i} +\mathrm{h.c.} )
-
t'
\sum_i
(\hat{b}^{\dag}_{ i} \hat{a}^{\ }_{ i+1} + \mathrm{h.c.} ),
\end{align}
where $\hat{a}_{i}$/$\hat{b}_{i}$ are  the fermion annihilation operators on sublattice  A/B
in the $i$-th unit cell.
We consider only real-valued nearest neighbor hopping amplitudes in (\ref{hopping model}), 
which we denote by $t, t'$, where we assume that $t,t'\ge 0$.
This is the Su-Schrieffer-Heeger (SSH) model describing \emph{trans}-polyacetylene \cite{heegerRMP88,suSchriefferHeegerPRB80}.
In momentum space, the Hamiltonian is written as
$\hat{H} = \sum_k \hat{\Psi}^{\dag}(k) {H}(k) \hat{\Psi}(k)$,
where
$
\hat{\Psi}(k)= 
( \hat{a}^{\ }_{k},  \hat{b}^{\ }_{k} )^T
$,
$k\in [-\pi,\pi]$, and 
\begin{align} \label{SSH_k_space}
{H}(k)
=
\boldsymbol{R}(k) \cdot \boldsymbol{\sigma},
\quad
\boldsymbol{R}(k)
=
\left(
\begin{array}{c}
t - t' \cos k \\
-t' \sin k \\
0 \\
\end{array}
\right). 
\end{align}
The energy dispersion is 
$\varepsilon(k) = \pm \sqrt{t^2 - 2 t t' \cos k + t^{\prime 2}}$.
The Hamiltonian has chiral symmetry as discussed around (\ref{hopping model}),
which in momentum space translates into the condition 
$\sigma_3 {H}(k) \sigma_3=- {H}(k)$.
With this symmetry,
the two gapped phases with $t>t'$ and $t<t'$ are topologically distinct
and are separated by a quantum critical point at $t=t'$.
Ground states in the phase $t>t'$ are adiabatically connected to
an atomic insulator (a collection of decoupled lattice sites) realized at $t'=0$.
On the other hand,
ground states in the phase $t'>t$ are topologically distinct from
topologically trivial, atomic insulators,
once chiral symmetry is imposed. 
These two phases are characterized by the winding number 
\begin{align} \label{windingNo1DAIII}
\nu [q]
&=
\frac{i}{2 \pi }
\int_{\mathrm{BZ}} d k \,  q^\dag \partial_{k} q  
=
\left\{
\begin{array}{cc}
1, & t' > t\\
0, & t' <t 
\end{array}
\right. ,
\end{align}
where the off-diagonal component of the projector is given by 
$
q(k) 
=
(t -t' e^{-ik} ) / |\varepsilon(k)|
$.
Correspondingly, the CS invariant
also takes two distinct quantized values
$\mathrm{CS}=1(0)$ for $t'>t$ and $t>t'$, respectively. 
Provided $t / t'$ is close to the critical point, the low-energy physics of the SSH model 
is captured by the continuum Dirac Hamiltonian
\begin{align}
H(k) \simeq -t' k \sigma_2 + (t - t') \sigma_1 , 
\end{align}
which is obtained from~\eqref{SSH_k_space} by expanding round $k=0$.
Note that $t-t'$ plays the role of the mass $m$.

To discuss domain walls, we first simplify the notation by letting $t\to t +m$ and $t' \to t$.
Furthermore, we make $m$ position dependent,
which defines a defect Hamiltonian in class AIII or BDI:
\begin{align}
H(k,r)=[t(1-\cos k)+m(r)]\sigma_1-t \sin k\sigma_2.
\label{defectSSH}
\end{align} 
Let us consider a spatially modulated mass gap $m(r)$ that describes a domain wall profile, i.e.,
$m(r)=\mbox{sgn}(r)\, m_0$ for $|r|\geq R_0$ with $m_0\neq0$. 
From (\ref{def winding number}),
we associate a topological invariant to this domain wall 
\begin{align}
\nu_1&=\frac{i}{2\pi}\int_{\mathrm{BZ}\times S^0}q^\dagger dq
\label{SSHinvariant1}\\
&=
\frac{i}{2\pi}\int_0^{2\pi}dk\left[
q(k,R_0)^\dagger \partial_k q(k,R_0)
\right.
\nonumber\\
&\quad\quad\quad-\left.
q(k,-R_0)^\dagger \partial_k q(k,-R_0)\right]=\pm1,
\nonumber
\end{align} where $S^0=\{R_0,-R_0\}$ is the two points that sandwich the point defect at the origin. 
The defect is also characterized by the CS integral \eqref{defectCS}, 
which in this case is the electric polarization: 
\begin{align}
\mbox{CS}_1&=\frac{i}{2\pi}\int_{\mathrm{BZ}\times S^0}\mathcal{A}
\label{SSHinvariant2}
\\&=\frac{i}{2\pi}\int_0^{2\pi}dk\left[{A}(k,R_0)-{A}(k,-R_0)\right]
=\frac{1}{2}\quad\mbox{mod $\mathbb{Z}$}. 
\nonumber
\end{align} 
The invariants \eqref{SSHinvariant1} and \eqref{SSHinvariant2} 
tell the difference between the two sides of the domain wall. 
They are well-defined even for the continuum Jackiw-Rebbi analogue~\cite{JackiwRebbi76} 
\begin{align}
H(k,r)=-t k\sigma_2+m(r)\sigma_1 ,
\label{JackiwRebbi}
\end{align} 
where the bulk topological invariants on either side do not take integer values without a regularization. 
Their difference as presented in \eqref{SSHinvariant1} and \eqref{SSHinvariant2}, 
however, are regularization independent, 
and detects the localized zero-energy mode at the domain wall.
The properties of these localized modes will be further discussed later in 
Sec.\ \ref{Bulk-boundary and bulk-defect correspondence}.

\paragraph{Example: The 3d class DIII \texorpdfstring{$^3$He-B}{He3-B}}  

Three-dimensional
TSCs in class DIII have been discussed in 
the context of the B phase of superfluid ${}^3$He 
\cite{Qi_hughes_zhang_09,Schnyder2008,Ryu2010ten,Volovik:book, 
Wada:2008ly, Murakawa:2009ve, JPSJ.80.013602, Chung:2009fk}, 
in superconducting copper dopped Bismuth Selinide 
\cite{FuBerg2010,HorCava2010,WrayCavaHasan},
and in non-centrosymmetric SCs
\cite{SchnyderRyuFlat}. 
Here, we consider the B (BW) phase of $^3$He.  
The BdG Hamiltonian that describes the B phase is 
given in terms of the Nambu spinor
$\hat{\Psi}^{\dag} = 
(\hat{\psi}^{\dag}_{\uparrow}, 
\hat{\psi}^{\dag}_{\downarrow},
\hat{\psi}^{\ }_{\uparrow},
\hat{\psi}^{\ }_{\downarrow})$
composed of 
the fermion annihilation operator for $^3$He 
$\hat{\psi}_{\uparrow,\downarrow}$
as
$\hat{H} = (1/2) \sum_{\mathsf{k}} 
\hat{\Psi}^{\dag}(\mathsf{k})
H(\mathsf{k}) 
\hat{\Psi}(\mathsf{k})
$, 
where 
\begin{align}
H(\mathsf{k}) =
\left(
\begin{array}{cc}
\xi(\mathsf{k}) & \Delta (\mathsf{k}) \\
\Delta^{\dag}(\mathsf{k}) & -\xi(\mathsf{k})
\end{array}
\right),
\quad
\left\{
\begin{array}{l}
\xi(\mathsf{k}) = k^2/2m - \mu
\\
\Delta (\mathsf{k}) = \Delta_0 i \sigma_2 \mathsf{k}\cdot \boldsymbol{\sigma}.
\end{array}
\right.
\label{BdG 3He B}
\end{align}
The BdG Hamiltonian satisfies 
$\tau_1 H(-\mathsf{k})^T \tau_1 = - H(\mathsf{k})$
and 
$\sigma_2 H(-\mathsf{k})^* \sigma_2 = H(\mathsf{k})$,
and belongs to class DIII.
From the periodic table,
class DIII in $d=3$ dimensions admits topologically non-trivial 
SCs (superfluid), which are characterized by an 
integer topological invariant, i.e., the winding number $\nu_3[q]$.
The winding number for the BdG Hamiltonian (\ref{BdG 3He B})
is given by
$\nu_3 = (1/2) (\mathrm{sgn}\, \mu +1)$.
Hence, for $\mu>0$ a topological superfluid is realized. 
When terminated by a surface, 
topological superfluids support
a topologically stable surface Andreev bound state
(Majorana cone). 
Surface acoustic impedance measurements
experimentally detected such a surface Andreev bound state
in $^3$He-B
\cite{Wada:2008ly, Murakawa:2009ve, JPSJ.80.013602}.

\subsubsection{The first \texorpdfstring{$\mathbb{Z}_2$}{Z2} descendant for \texorpdfstring{$s$}{s} even}

While for the primary series the topological phases or topological defects
are characterized by an integer-valued Chern number (or  winding number),
for the 1st and 2nd descendants the topological phases are characterized by a $\mathbb{Z}_2$ invariant.
To discuss these $\mathbb{Z}_2$ indices in  a unified framework, we will follow  two strategies:
First, we construct various $\mathbb{Z}_2$ topological invariants
by starting from the CS invariants and using symmetry conditions to restrict their possible values.
(``CS'' and ``$\widetilde{\mathrm{CS}}$'' in Table \ref{tab:defectinvariant}).
Second, we use both the Chern numbers  and CS integrals to construct 
$\mathbb{Z}_2$ invariants
(``FK'' in Table \ref{tab:defectinvariant}).

The first $\mathbb{Z}_2$ descendant topologies 
are characterized by the CS integral 
\begin{align}
\mbox{CS}_{2n-1}=\int_{\mathrm{BZ}^d\times\mathcal{M}^D}\mathcal{Q}_{2n-1}\in\frac{1}{2}\mathbb{Z} ,
\label{defectCS}
\end{align} 
for $n=(d+D+1)/2$.
The CS-invariant is well-defined only up to an integer. 
Note that under antiunitary symmetries, the CS-invariant can in general take half-integer values. The $\mathbb{Z}_2$ topology is trivial when $\mbox{CS}_{2n-1}$ is an integer; or non-trivial when $\mbox{CS}_{2n-1}$ is a half-integer.

There is a subtlety when computing 
the CS integrals \eqref{defectCS}  for a general defect Hamiltonian
(this also applies to the Fu-Kane invariant \eqref{defectFK}, which will be discussed later):
they require a set of occupied states  
defined globally on the base space, 
which is unnecessary for the definition of the Chern number \eqref{defectChern} and the winding number 
\eqref{def winding number}.
There may be a topological obstruction to such global continuous basis. 
In particular, a global valence frame does not exist whenever there are non-trivial weak topologies with non-zero Chern invariants in lower dimensions. 
In this case, one needs to include artificial Hamiltonians, i.e., $H(\mathsf{k},\mathsf{r})\to H(\mathsf{k},\mathsf{r})\oplus H_0(\mathsf{k},\mathsf{r}_0)$,
that cancel the weak topologies while at the same time doe not affect the highest dimensional strong topology
\cite{Teo:2010fk}. 
This can be achieved by a   lower dimensional Hamiltonian $H_0(\mathsf{k},\mathsf{r}_0)$,
where $\mathsf{r}_0$ lives in some proper cycles $\mathcal{N}^{D'}\subsetneq\mathcal{M}^D$ that do not wrap around the defect under consideration.
See Sec.\ \ref{sec: chiral p wave superconductors} for an example.

\paragraph{Class D in \texorpdfstring{$d=1$}{d=1}}
\label{Topological superconductor in one dimension}

A BdG Hamiltonian in Class D in $d=1$ dimensions satisfies
$
C^{-1} H(-k) C = -H(k),
$
with 
$
C = \tau_1 \mathcal{K}$,
where $k\in (-\pi, \pi]$ is the 1d momentum. 
Class D TSCs in $d=1$  are characterized by the CS integral 
(\ref{defectCS}). 
As chiral symmetry, PHS also quantizes 
$W = \exp(2\pi i \mathrm{CS}[\mathcal{A}])$
to be $\pm 1$ 
\cite{Qi2008sf, BudichArdonne2013}.
To see this, we first recall that
if $|u^{\alpha}_-({k})\rangle$ is a negative energy solution
with energy $-\varepsilon({k})$, then
$|\tau_1 u^{* \alpha}_-(-{k}) \rangle$
is a positive energy solution with energy $\varepsilon({k})$  
(Sec.\ \ref{PH symmetry}).
Consequently, the Berry connections for negative and positive energy states are related by  
\begin{align}
 A^{\alpha\beta}_{-} ({k}) &=
 \langle u^{\alpha}_- ({k}) | \partial_{k} u^{\beta}_{-} ({k})\rangle
 =
 A^{\alpha\beta}_{+} (-{k}). 
\end{align}
The 1d CS integral is then given by
\begin{align}
& \int^{+\pi}_{-\pi}dk\, \mathrm{Tr}\, A_{-}
= \int^{\pi}_{0}dk\, \mathrm{Tr}\,\left[ A_{-}+ A_{+} \right]
 \nonumber \\
 &= \int^{\pi}_{0}dk\, 
 u^{* a}_{i} \partial_{k} u^{a}_{i}
= \int^{\pi}_{0}dk\, \mathrm{Tr}\, U^{\dag} \partial_{k}U,  
\end{align}
where $a$ runs over all the bands, while $\alpha$ runs over  half of the bands 
(i.e., only the negative energy bands).
Here, we have introduced unitary matrix notation by $U^{a}_i (k):= u^{a}_i(k)$. 
By noting that
$
 \int^{\pi}_{0}dk\, \mathrm{Tr}\, U^{\dag} \partial_{k}U 
 =
 \int^{\pi}_{0}dk\, \partial_k \ln \det [U(k) ]
=
\ln \det U(\pi) - \ln \det U(0) 
$,
the CS invariant reduces to
\begin{align}
W = \left[\det U(\pi)\right]^{-1} \left[ \det U(0)\right].
\end{align}
At the PH symmetric momenta $k=0,\pi$,
the unitary matrix $U(k)$ has special properties.
This can be seen most easily by using the Majorana basis~\eqref{Nambu to Majorana}.
That is, by the basis change in Eq.\ (\ref{Nambu to Majorana}), 
we   obtain from $H(k)$ the Hamiltonian $X(k)$ in the Majorana basis.
Remember that at TR invariant momenta $\tau_1 H^*(k) \tau_1 = - H(k)$.
Hence, $X(k=0,\pi)$ is a real skew symmetric matrix, which
can be transformed into 
its canonical form by an orthogonal matrix $O(k=0,\pi)$  
[see Eq.\ \eqref{canonical form X}].
$W$ can then be written in terms of $O(k=0,\pi)$ as 
\begin{align}
W = 
\left[\mathrm{det}\, O(\pi)\right]^{-1}
\left[\mathrm{det}\, O(0)\right]. 
\end{align}
Since $O(k=0,\pi)$ are orthogonal matrices,
their determinants are either  $+1$ or $-1$, 
and so is the CS invariant, $W=\pm 1$.
Using Pfaffian of $2n$-dimensional skew symmetric matrices 
\begin{align}
\mathrm{Pf}(X)=
\frac{1}{2^nn!}\sum_{\sigma\in S_{2n}}(-1)^{|\sigma|}X_{\sigma(1)\sigma(2)}\ldots X_{\sigma(2n-1)\sigma(2n)} ,
\end{align} 
where $\sigma$ runs through permutations of $1,\ldots,2n$, 
and noting further the identities  
$\mathrm{Pf}\, (OXO^T) = \mathrm{Pf} (X) \mathrm{det}\, (O)$, 
and
$\mathrm{sgn}(\mathrm{Pf}\, [X(k)]\, \mathrm{det}[O(k)])=1$, 
$W$ can also be written as 
\begin{align}
W=
\mathrm{sgn}\, 
\big(
\mathrm{Pf}\, [X(0)] \, \mathrm{Pf}\, [X(\pi)]
\big), 
\label{pfaffian way}
\end{align}
which is manifestly gauge invariant 
(i.e., independent of the choice of wave functions).

\paragraph{Example: The class D Kitaev chain}\label{sec: The Kitaev chain}

The 1d TSC proposed by Kitaev has stimulated many studies on Majorana physics
\cite{Kitaev2001,Alicea:2012em,Sau_semiconductor_heterostructures}. 
Evidence for the existence of Majorana modes in 1d chains has been observed in
a number of recent
experiments
\cite{Roman_SC_semi,
Gil_Majorana_wire,
Mourik_zero_bias,
Das_zero_bias,
Deng_zero_bias,
Finck_zero_bias,
Churchill_zero_bias,
Lee_1DSC_semi,
Nadj-Perge_Ferro_SC,Franz_nanowire}. 
The Hamiltonian of the Kitaev chain is given by
\begin{align}
\hat{H}
&=
\frac{t}{2}
\sum_{i}
\left(
\hat{c}^{\dag}_i \hat{c}^{\ }_{i+1} 
+
\hat{c}^{\dag}_{i+1} \hat{c}^{\ }_{i} 
\right)
-\mu \sum_i \left( \hat{c}^{\dag}_{i}\hat{c}^{\ }_{i} -1/2\right) 
\nonumber \\
&
\quad 
+
\frac{1}{2}
\sum_i
\left(
\Delta^* 
\hat{c}^{\dag}_i \hat{c}^{\dag}_{i+1}  
-
\Delta \hat{c}^{\ }_{i} \hat{c}^{\ }_{i+i}  
\right). 
\label{Kitaev chain}
\end{align}
Without loss of generality, $\Delta$ can be taken as a real number, 
since the global phase of the order parameter, $\Delta=e^{i \theta}\Delta_0$,
can be removed by a simple gauge transformation 
$\hat{c}_i \rightarrow  \hat{c}_i e^{i\theta/2}$.
In momentum space $\hat{H}$ reads
\begin{align}
&
\hat{H}=
\frac{1}{2}
\sum_k 
\bma
\hat{c}^\dagger_k & \hat{c}^{\ }_{-k}
\ema
H(k)
\bma 
\hat{c}^{\ }_k \\
\hat{c}_{-k}^\dagger \\
\ema, 
\nonumber \\
&
\mbox{where}
\quad 
H(k)=(t\cos k - \mu)\tau_3 - \Delta_0 \sin k \tau_2.
\label{momentum Kitaev}
\end{align}
There are gapped phases  
for $|t|>\mu$ and $|t|<\mu$, 
which are separated by a line of critical points 
at $t=\pm \mu$. 
The Kitaev chain can be written in terms of the Majorana basis
\begin{align}
\left\{
\begin{array}{l}
\hat{\lambda}_{j} := \hat{c}_{j}^{\dag}+ \hat{c}_{j}^{\ },
\\
\hat{\lambda}'_{j} := (\hat{c}_{j}^{\ }- \hat{c}_{j}^{\dag})/{{i}}, 
\end{array}
\right.
\quad
\hat{\Lambda}_j 
:=
\left(
\begin{array}{c}
\hat{\lambda}_j \\
\hat{\lambda}'_j
\end{array}
\right), 
\label{Majorana basis Kitaev chain}
\end{align}
as
$
 \hat{H}
 =
 ({i}/{2}) \sum_k 
\hat{\Lambda}^T(k) X(k) \hat{\Lambda}(-k) 
$,
where 
\begin{align}
X(k)=-i(t \cos k -\mu )\tau_2  + i \Delta_0 \sin k \tau_1. 
\end{align}
We read off
the CS invariant as 
$W=\mp 1$ for $|\mu|<|t|$ and $|\mu|>|t|$, respectively.

Similar to the SSH model, we can also consider a domain wall by changing $\mu$ as a function of space,
which traps a localized zero-energy Majorana mode.
Properties of the localized zero-energy Majorana mode will be discussed in  
Sec.\ \ref{Bulk-boundary and bulk-defect correspondence}.

\paragraph{Class AII in \texorpdfstring{$d=3$}{d=3}}

We now discuss the topological property of TR invariant insulators in $d=3$ dimensions
\cite{Fu:2007fk,Moore2007uq,Roy2009kx}. 
The topological characteristics of these band insulators
are intimately tied to the invariance of the Hamiltonian under TRS, 
i.e., $
T^{-1}
{H}(-\mathsf{k} )
T=
{H}(\mathsf{k} )
$.
Because of this relation, 
the Bloch wave functions at $\mathsf{k}$
and those at $-\mathsf{k}$ are related. 
If $|u^{\alpha}(\mathsf{k})\rangle$ is an eigen state at $\mathsf{k}$, 
then $T|u^{\alpha}(\mathsf{k})\rangle$ is an eigen state at $-\mathsf{k}$. 
Imagine now that we can define $|u^{\alpha}(\mathsf{k})\rangle$ smoothly
for the entire BZ. 
(This is possible since TRS forces the Chern number to be zero  and, hence, there is no obstruction).
We then compare $|u^{\alpha}(-\mathsf{k})\rangle$ and 
$T|u^{\alpha}(\mathsf{k}) \rangle$.
Since both 
$|u^{\alpha}(-\mathsf{k})\rangle$ and 
$T|u^{\alpha}(\mathsf{k})\rangle$ 
are  eigen states of the same Hamiltonian ${H}(-\mathsf{k})$, 
they must be related to each other by a unitary matrix,
$
 |u^{\alpha}(-\mathsf{k})\rangle =
 [w^{\alpha\beta }(\mathsf{k})]^*
 |T u^{\beta }(\mathsf{k})\rangle
 $. 
(The complex conjugation on $w$ here is to comply with a common convention.)
Hence, the sewing matrix 
\begin{align}
w^{\alpha\beta}(\mathsf{k} )&=
\langle u^{\alpha }(-\mathsf{k} )
|
T u^{\beta }( \mathsf{k} )
\rangle,
\label{eq: def sewing matrix}
\end{align}
which is given by the overlaps between the occupied  eigenstates 
with momentum $-\mathsf{k} $ and the time reversed images
of the occupied  eigenstates with momentum 
$\mathsf{k}$, 
plays an important role in defining the $\mathbb{Z}_2$ index~\cite{Fu:2007fk}.
The matrix elements~(\ref{eq: def sewing matrix}) obey 
\begin{align}
w^{\alpha\beta}(-\mathsf{k})
=
-w^{\beta\alpha}(\mathsf{k}),
\label{eq: sewing matrix is as}
\end{align}
which follows from the fact that $T$ is antilinear and antiunitary,   
and $T^{2}=-1$.
Consequently, there is a relation between the Berry connection at 
$\mathsf{k}$ and at $-\mathsf{k}$: 
\begin{align}
A^{\ }_{\mu}(-\mathsf{k} )&=
-
w(\mathsf{k} )
A^{*}_{\mu}(\mathsf{k} )
w^{\dag}(\mathsf{k} )
-
w(\mathsf{k} )
\partial^{\ }_{\mu}
w^{\dag}(\mathsf{k} ).
\label{eq: sewing matrix as gauge trsf}
\end{align}
I.e., 
$-A_{\mu}(-\mathsf{k})$ and $A^*_{\mu}(\mathsf{k})= -A^T_{\mu}(\mathsf{k})$
are related to each other by a gauge transformation. 

With this constraint on the Berry connection, 
we now show
that the CS invariant is given in terms 
of the winding number of the sewing matrix $w$ as 
\begin{align}
 \mathrm{CS}[\mathcal{A}] =
 \frac{1}{2} \int_{\mathrm{BZ}} \omega[w]=
 \frac{1}{2} \times \mbox{integer}, 
 \label{quantization of CS}
\end{align}
and hence
$
W = \exp ( 2\pi i \mathrm{CS}[\mathcal{A}])  = \pm 1. 
$
To see this,
we change variables from $\mathsf{k}$ to $-\mathsf{k}$
in the integral $\mathrm{CS}[\mathcal{A}]$,
and use 
(\ref{eq: sewing matrix as gauge trsf}), 
$
 A_{\mu}(-\mathsf{k}) 
 = - [A^{g^*}_{\mu} (\mathsf{k})]^* 
 $
with $g=w^{\dag}$, 
to show
$
\mathrm{CS}[\mathcal{A}]
=
-
\mathrm{CS}[(\mathcal{A}^{g^*})^*]
=
-(\mathrm{CS}[\mathcal{A}^{g^*}])^*
=
-\mathrm{CS}[\mathcal{A}^{g^*}]
$,
where in the last equality we noted that $\mathrm{CS}[\mathcal{A}]$ is real.
Using Eq.\ (\ref{eq: Cartan homotopy}), 
\begin{align}
& \mathrm{CS}[\mathcal{A}] 
=
 -\mathrm{CS}[ \mathcal{A} ]
 - \int_{\mathrm{BZ}} 
 \{ 
 w[g^*]
+ d\alpha(\mathcal{A}, g^*)  
\},   
\end{align}
and 
$\int_{\mathrm{BZ}} \omega[g] =  \int_{\mathrm{BZ}} \omega[w^{\dag}] = -\int_{\mathrm{BZ}} \omega[w]$
proves the quantization of the CS invariant (\ref{quantization of CS}).

The CS invariant  can also be written 
by using the Pfaffian of the gluing matrix $w$ 
at TR invariant momenta $\mathsf{K}$ in the BZ
as
\cite{Kane:2005vn,Kane:2005kx,Fu2007uq} 
\begin{align} 
W
&=
 \prod_{\mathsf{K}}
 \frac{ \mathrm{Pf}\, \left[w(\mathsf{K} )\right] }
 { \sqrt{ \det \left[  w ( \mathsf{K} ) \right] }}.
 \label{Pfaffian formula}
\end{align}
The equivalence between the quantized CS invariant
and the Pfaffian invariant 
(\ref{Pfaffian formula})
was shown in  
\cite{WangQiZhang2010}.

\subsubsection{The second \texorpdfstring{$\mathbb{Z}_2$}{Z2} descendant for \texorpdfstring{$s$}{s} even}
The Fu-Kane invariant~\cite{FuKane2006} 
applies to the second $\mathbb{Z}_2$ descendent for non-chiral symmetry classes,
and is defined by. 
\begin{align}
\mbox{FK}_n=
\frac{1}{n!}\left(\frac{i}{2\pi}\right)^n&
\int_{\mathrm{BZ}^d_{1/2}\times\mathcal{M}^D}
\mbox{Tr}\left(\mathcal{F}^n\right)
\nonumber\\
&
\quad-\oint_{\partial \mathrm{BZ}^d_{1/2}\times\mathcal{M}^D}\mathcal{Q}_{2n-1} ,
\label{defectFK}
\end{align} 
where $n=(d+D)/2$. 
It involves an open integral of the Berry curvature over half of the Brillouin zone 
$\mathrm{BZ}^d_{1/2}$, 
where one of the momentum paramenter, say $k_1$, runs between $[0,\pi]$ so that the complement of 
$\mathrm{BZ}^d_{1/2}$ is its TR conjugate. The CS integral over 
$\partial \mathrm{BZ}^d_{1/2}$, the boundary of the half BZ where $k_1=0,\pi$, is gauge dependent and requires special attention in the choice of basis. 
For TRS systems (class AI and AII), 
the occupied states $|u^{\alpha}(\mathsf{k},\mathsf{r})\rangle$ 
that build the Berry connection $\mathcal{A}^{\alpha\beta}$
need to satisfy the gauge constraint 
\begin{align}
w^{\alpha\beta}
(\mathsf{k},\mathsf{r})=\langle u^{\alpha}(-\mathsf{k},\mathsf{r})|Tu^{\beta}(\mathsf{k},\mathsf{r})\rangle=\mbox{constant} ,
\label{FKconstraint1}
\end{align} 
for $(\mathsf{k},\mathsf{r})\in\partial \mathrm{BZ}^d_{1/2}\times\mathcal{M}^D$. 
For instance the original FK-invariant characterizing 2d class AII TIs requires 
$w(\mathsf{k},\mathsf{r})=i\sigma_2$. 
For PHS systems (class D and C), the occupied states $|u^{\alpha}(\mathsf{k},\mathsf{r})\rangle$ generate 
the unoccupied ones $|v^{\alpha}(\mathsf{k},\mathsf{r})\rangle$ by the PH operator $C$, i.e.,
 $|v^{\alpha}(\mathsf{k},\mathsf{r})\rangle=|Cu^{\alpha}(-\mathsf{k},\mathsf{r})\rangle$.
The CS form in the FK-invariant \eqref{defectFK} 
needs to be built from occupied states satisfying 
\begin{align}
\int_{\partial \mathrm{BZ}^d_{1/2}\times\mathcal{M}^D}\mbox{Tr}
\big[\left(XdX^\dagger\right)^{d+D-1}\big]=0 ,
\label{FKconstraint2}
\end{align}
where 
$
X(\mathsf{k},\mathsf{r})=\left(u^1,\ldots,u^N,v^1,\ldots,v^N\right)
$ 
is the unitary matrix formed by the eigenstates.
The gauge constraints \eqref{FKconstraint1} and \eqref{FKconstraint2} 
are essential for the FK-invariant in \eqref{defectFK}. 
Without them, the CS integral can be changed by any ineger value by a large gauge transformation of 
occupied states $|u^{\alpha}\rangle \to g_{\alpha\beta}|u^{\beta}\rangle$. 
The gauge constraints restrict such transformations so that the CS term can only be changed by an even integer. The FK-invariant therefore takes values in $\mathbb{Z}_2=\{0,1\}$.

\paragraph{Class AII in \texorpdfstring{$d=2$}{d=2}}

The topological invariant for 2d time-reversal symmetric TIs is
the Fu-Kane invariant (\ref{defectFK})
\cite{FuKane2006}.
As in the case of 3d time-reversal symmetric TIs,
this $\mathbb{Z}_2$ invariant   has an alternative expression: 
\begin{align} 
W
&=
 \prod_{\mathsf{K}}
 \frac{ \mathrm{Pf}\, \left[w(\mathsf{K} )\right] }
 { \sqrt{ \det \left[  w ( \mathsf{K} ) \right] }},
\end{align}
where $\mathsf{K}$ runs over two dimensional TR fixed momenta.
This topological invariant can also be written in a number of different ways.
For example, it can be introduced as TR invariant polarization 
\cite{FuKane2006},
which can be written as an $SU(2)$ Wilson loop in momentum space
\cite{RyuMudryObuseFurusaki2010NJPh,LeeRyu2008, YuQiBernevig2011}.
See also \onlinecite{Kane:2005kx,Kane:2005vn,Freed:2013bv,Prodan2011,FruchartCarpentier2013,SoluyanovVanderbilt2011} 
for different representations of the $\mathbb{Z}_2$ invariant.

\subsubsection{The first \texorpdfstring{$\mathbb{Z}_2$}{Z2} descendant for \texorpdfstring{$s$}{s} odd}

The first $\mathbb{Z}_2$ descendant 
for the chiral classes relates isomorphically to the second $\mathbb{Z}_2$ descendant for the non-chiral classes. 
This relation will be discussed in more detail later in Sec.\ \ref{Defect K-theory}.
The topological invariant for chiral $\mathbb{Z}_2^{(1)}$ is therefore given by the FK-invariant \eqref{defectFK} with the gauge constraint \eqref{FKconstraint1} for $s=1,5$ (class CI and DIII) or \eqref{FKconstraint2} for $s=3,7$ (class BDI and CII).

\paragraph{Class DIII in \texorpdfstring{$d=2$}{d=2}}

As in the case of time-reversal symmetric TIs in $d=2$ (AII), 
the FK invariant for time-reversal symmetric TSCs  in $d=2$ (DIII)
can be written in terms of the Pfaffian formula (\ref{Pfaffian formula}). 
The presence of TRS allows us to define the $\mathbb{Z}_2$ invariant. 
The Pfaffian formula can also be given in terms of the $Q$-matrix. 
To see this, we write  
the BdG Hamiltonian in the off-diagonal basis, i.e., in the  form
\begin{align}
H (\mathsf{k})
&=
\left(
\begin{array}{cc}
0 & D( \mathsf{k} ) \\
D^{\dag}( \mathsf{k} ) & 0
\end{array}
\right),
\quad
D( \mathsf{k} )= -D^T(- \mathsf{k} ).
\label{diii}
\end{align}
In this representation, the TR operator is given by  
${T} = U_T \mathcal{K} =     {i} \sigma_2 \otimes \openone \mathcal{K}$,
and the $Q$-matrix reads
\begin{align} \label{Qzwei}
Q( \mathsf{k} )
=
\left(
\begin{array}{cc}
0  & q( \mathsf{k} ) \\
q^{\dag}(\mathsf{k} ) & 0 
\end{array}
\right),
\quad
q( \mathsf{k} ) = - q^T (-\mathsf{k} ). 
\end{align}
To compute the $\mathbb{Z}_2$ topological number 
we choose the basis
$|u^{\alpha}_{\pm}(\mathsf{k})\rangle_{\mathrm{N}}$,
in which 
the sewing matrix is given by
$w^{\alpha\beta} ( \mathsf{k} ) 
=
-q^{\alpha \beta}(- \mathsf{k} ) 
$.
The $\mathbb{Z}_2$ topological number can thus be express as~\cite{SchnyderRyuFlat}
\begin{align} \label{eqW}
W   = 
\prod_{\mathsf{K}} 
\frac{
\mathrm{Pf}\, \left[q(\mathsf{K})\right] 
}{
\sqrt{ \det \left[ q ( \mathsf{K} ) \right] }} ,
\end{align}
where $\mathsf{K}$ denotes 
the four TR invariant momenta 
of the 2d BZ.

\subsubsection{The second \texorpdfstring{$\mathbb{Z}_2$}{Z2} descendant for \texorpdfstring{$s$}{s} odd}

The second $\mathbb{Z}_2$ descendant for chiral classes is given 
by the CS integral $\mbox{CS}_{2n-1}$ in \eqref{defectCS} for $n=(d+D+1)/2$. 
Similar to the FK-invariants, the CS form here needs to be built from occupied states that satisfy the gauge constraint 
\eqref{FKconstraint1} for class CI and DIII or \eqref{FKconstraint2} for class BDI and CII. 
Together with the antiunitary symmetry, this gauge constraint forces the Chern-Simions invariant \eqref{defectCS} to be a full integer. 
The $\mathbb{Z}_2$ topology is trivial if $\mbox{CS}_{2n-1}$ is even, and  non-trivial if $\mbox{CS}_{2n-1}$ is odd.

\paragraph{Class DIII in \texorpdfstring{$d=1$}{d=1}}

In $d=1$ 
the gauge constraint \eqref{FKconstraint1} is automatically satisfied. 
The CS integral \eqref{defectCS} becomes the ``polarization" \eqref{SSHinvariant2}, 
which takes value in full integers. 
By taking the basis where the Hamiltonian and the $Q$-matrix take the form of \eqref{Qzwei},  
the CS integral can be simplified into the following $\mathbb{Z}_2$ 
invariant 
\begin{align}
(-1)^\nu=
\frac{\mbox{Pf}\, [q(\pi)]}{\mbox{Pf}\, [q(0)]}
\frac{\sqrt{\det[q(0)]}}{\sqrt{\det[q(\pi)]}}, 
\label{PfaffianDIIId=1}
\end{align} 
that relies on information only on the fixed momenta ${k}=0,\pi$
\cite{QiHughesZhang10}.
Notice that the branch $\sqrt{\det\, [q({k})]}$ must be chosen continuously between the two fixed momenta. 
A proof of the equivalence of the 1D CS integral and \eqref{PfaffianDIIId=1} can be found in \cite{Teo:2010fk}.
As an example, let us consider the class DIII Hamiltonian in the form of \eqref{Qzwei} with 
\begin{align}
D({k})=-t\sin{k}\sigma_1-i[\Delta+u(1-\cos{k})]\sigma_2, 
\end{align} 
where ${k}\in[-\pi,\pi]$ and $u\gg |\Delta|$. 
By noting that $\det[q({k})]$ is always real and positive,
 $\mbox{Pf}\, [q(0)]/\sqrt{\det\, [q(0)]}=\mbox{sgn}\, (\Delta)$ while 
$\mbox{Pf}\, [q(\pi)]/\sqrt{\det\, [q(\pi)]}=1$. 
Hence this model is non-trivial according to \eqref{PfaffianDIIId=1} when the pairing $\Delta$ is negative.

\begin{center}
***
\end{center}

Before leaving this section,
it is worth while mentioning that the topological invariants discussed in this section
can be cast in many different forms. Moreover, they  can be extended, in certain cases,   
in a way that they are valid in the presence of disorder and interactions.  
For example, 
the Chern invariant can be written in terms of 
many-body ground state wave functions, which depend on twisting boundary conditions  
\cite{ThoulessWu1985,WangZhang2014}.
All topological invariants discussed in this section 
can be written in the language of scattering matrices 
\cite{Akhmerov2011, Fulga2011,classification_scattering}.
Topological invariants can also be written in terms of
Green's functions 
\cite{Ishikawa:1987zi, Volovik:book, Gurarie2011, WangGreenFunction2012, WangZhang2012}
and by
using $C^*$-algebra 
\cite{belissard_JMatPhys94,Prodan2014,Prodan_noncommute,Prodan_odd_Chern,loring_hastings_EPL_10,Hastings_annals_physics11}.

\subsection{K-theory approach}
\label{K-theory approach}

In this section, we derive the classification of gapped topological phases and topological defects, which
is summarized in Table~\ref{tab:classification}.
The classification can be shown by either relating to the homotopy groups of classifying spaces 
or by a K-theoretical argument
\cite{Kitaev2009}.
We also demonstrate the use of the Clifford algebra in identifying classifying spaces of symmetry-allowed Dirac mass terms.   
This method effectively allows us to translate topological problems into algebraic problems,
and makes use of a known connection between 
K-theory and Clifford algebras;
the Bott periodicity of K-theory is proved by using Clifford algebras
\cite{spingeometrybook, Hatcherbook, Atiyah19643}. 
For a more complete and precise description of K-theory, 
Clifford algebra, and Bott periodicity we refer  to the
literature in mathematics~\cite{Karoubibook, spingeometrybook, Atiyahbook, Milnorbook} as well as in physics~\cite{classification_scattering,Freed:2013bv,KennedyZirnbauer2014,Thiang2014,Stone:2011qo,Abramovici:2012zz,xiaogang_noninteract,Budich_Ktheory}.

\subsubsection{Homotopy classification of Dirac mass gaps}\label{sec:defecthomotopy}

We have seen already that many topologically 
non-trivial phases (as well as trivial phases)
have a massive Dirac Hamiltonian representative. 
One could then be interested in focusing on and classifying Dirac representatives.
One may think this is a crude approximation, 
but as it turns out, one does not lose much by narrowing one's focus in this way (see \ref{Defect K-theory}).
We thus consider the  low-energy description of  Bloch-BdG Hamiltonians
near the relevant momentum point $\mathsf{K}_0$, which generically takes the Dirac form 
\begin{align}
H(\mathsf{k},\mathsf{r})=\mathsf{k}\cdot\boldsymbol\Gamma+m\Gamma_0(\mathsf{r}), 
\label{domainwallH}
\end{align} 
where $\mathsf{k}=(k_1,\ldots,k_d)$ is the momentum deviation from $\mathsf{K}_0$, 
$\boldsymbol\Gamma=(\Gamma_1,\ldots,\Gamma_d)$ are Dirac matrices that satisfy 
the Clifford relation $\{\Gamma_\mu,\Gamma_\nu\}
=\Gamma_\mu\Gamma_\nu+\Gamma_\nu\Gamma_\mu=2\delta_{\mu\nu}$
($\mu,\nu=0,\cdots, d$).
The mass term $m\Gamma_0(\mathsf{r})$,
which depends on a $D$-dimensional spatial parameter $\mathsf{r}$, 
anticommutes with all Dirac matrices in the kinetic term,
and is responsible for a bulk energy gap. 
For a {\it stable} classification, 
which is independent of and insensitive to the addition of irrelevant trivial bands,
the dimension of the Dirac matrices (the number of bands) are taken to be sufficiently large,
$\log(\dim(\Gamma_0))\gg d+D$, the motivation of which will become clear later.  
In the presence of symmetries [\cref{TRS_r,PHS_r,chiral_r}], the Dirac matrices satisfy  
\begin{align}
\label{Dirac matrices sym}
&T\Gamma_0(\mathsf{r})T^{-1}=\Gamma_0(\mathsf{r}),  \quad T\boldsymbol\Gamma T^{-1}=-\boldsymbol\Gamma,
\\
&C\Gamma_0(\mathsf{r})C^{-1}=-\Gamma_0(\mathsf{r}),\quad C\boldsymbol\Gamma C^{-1}=\boldsymbol\Gamma,
\\
&S\Gamma_0(\mathsf{r})S^{-1}=-\Gamma_0(\mathsf{r}),\quad S\boldsymbol\Gamma S^{-1}=-\boldsymbol\Gamma.
\end{align}

For a general TI or TSC, the mass term $m\Gamma_0$ lives in some parameter space $\mathcal{R}$ 
that has the same topology (or homotopy type) as a certain {\em classifying space}
\cite{spingeometrybook, Hatcherbook, Freedman2011}, 
which will be identified shortly. 
Suppose we have a domain wall sandwiched by two bulk regions A and B.
E.g., a domain wall separating a Chern insulator and a trivial insulator in 2d 
can be topologically captured by \eqref{domainwallH}, 
where the mass term changes its sign across the interface.  
Now, pick  arbitrary points $\mathsf{r}_A$ in A and $\mathsf{r}_B$ in B. 
The domain wall is topological and carries protected interface modes, if there does not exist any continuous path in the parameter space $\mathcal{R}$ 
that connects $m\Gamma_0(\mathsf{r}_A)$ and $m\Gamma_0(\mathsf{r}_B)$. 
The topology is therefore characterized by $\pi_0(\mathcal{R})=[S^0,\mathcal{R}]$, the $0^{th}$-homotopy group of 
$\mathcal{R}$ that counts the path connected components.

For general topological defects other than domain walls, 
we first approximate the defect Hamiltonian by the Dirac Hamiltonian,
where $\mathsf{r}$ is now the modulation parameter that wraps around the defect in spacetime. 
In this case, we are interested in highest dimensional strong topologies, where $\mathsf{r}$ lives on (or deformation retracts to) the compactified sphere $S^D$.

The mass term $m\Gamma_0$ belongs to different classifying spaces $\mathcal{R}_{s-d}$ 
for different symmetry classes $s$ and bulk dimension $d$. 
As we will see, the classifying space is determined by the symmetries (\ref{Dirac matrices sym}).
Let us now demonstrate this for a few cases.

\paragraph{Class A in \texorpdfstring{$d=2$}{d=2} and \texorpdfstring{$d=1$}{d=1}}

As a first example, we will identity the classifying space that is relevant for 2d Chern insulators in class A.
To this end, let us first recall the lattice model given by \eqref{Model, Chern insulator}.
By linearizing the spectrum near $\mathsf{K}_0=0$, 
we obtain from \eqref{Model, Chern insulator}  a $d=2$ massive Dirac model:  
$
{H}(\mathsf{k}) = k_x \sigma_1 + k_y \sigma_2 + m \sigma_3
$. 
There are two distinct phases in this model
for $m>0$ and $m<0$, whose Chern number differ by one. 
(Observe here that we discuss only the relative Chern number.) 
To discuss phases with more general values of the Chern number, 
we enlarge the matrix dimension of the Hamiltonian
and consider the following $2N\times 2N$ Dirac Hamiltonian: 
\begin{align}
{H}(\mathsf{k},\mathsf{r}) = k_x \sigma_1 \otimes \openone_N + k_y \sigma_2 \otimes \openone_N +M.   
\end{align}
Since the mass $M$ should anti-commutes with the kinetic term, 
$M$ should have the form
$
 M = \sigma_3 \otimes A,
$
where $A$ is a $N\times N$ hermitian matrix.
By considering
$
 A = \mathrm{diag}\, (m_1, \cdots, m_N), 
 $
$
m_i \neq  0 
$,
we can realize band insulators with different values of the (relative) Chern number.
These are simply $N$ decoupled copies of different Dirac insulators 
with different masses. 
The magnitude of the masses does not matter for the Chern number,
while the sign $\mathrm{sgn}\, m_i$ does.
So, without loosing generality, we can consider
$ A = \Lambda_{n,N-n}$,
where $\Lambda_{n,m}=\mathrm{diag}\, (\openone_n, -\openone_{m})$.
Starting from $\Lambda_{n,N-n}$,
more generic mass terms can be generated by a unitary matrix $U$ as 
$A = U \Lambda_{n,N-n} U^{\dag}$, which share the same Chern number as $\Lambda_{n,N-n}$.
Conversely, for a given $A$, as far as its eigen values are properly normalized, 
one can diagonalize $A$ by a unitary matrix $U$ 
and write
$A = U \Lambda_{n,N-n} U^{\dag}$.
Thus, $A$ is a member of 
$U(N)/U(n)\times U(N-n)$. 
Two masses $A_1$ and $A_2$ which have the same
canonical form are unitarily related to each other.
I.e., 
$U(N)/U(n)\times U(N-n)$ is simply connected. 
However, 
two masses $A_1$ and $A_2$ which have  different canonical forms
(i.e., different $n$) are not.
Summarizing, the set of masses for a given $N$ is $ \bigcup_{0\le n \le N} U(N)/ [ U(n)\times U(N-n) ]$.

So far we have fixed $N$, but this is clearly
not enough for the purpose of realizing  
Dirac representatives for all possible phases
since for given $N$, the (relative) Chern number can be at most $N$,
whereas insulators in class A in $d=2$ can be characterized by
the Chern number which can be any integer. 
To realize insulators with arbitrary Chern number, we can take $N$ as large as possible,
and this leads us to consider: 
\begin{align}
\mathcal{C}_0 =\bigcup_{n=0}^N\frac{U(N)}{U(n)\times U(N-n)}
\xrightarrow{N\to\infty}BU\times\mathbb{Z}.
\label{Cclassifyingspace0}
\end{align}
The disconnected components of this space,
$\pi_0(C_0)$, 
is the space of topologically distinct masses,
for which it is known that 
$ \pi_0 (\mathcal{C}_0) = \mathbb{Z}$. 
This agrees with the classification of class A in $d=2$.

The fact that we take the limit of an infinite number of bands,
which can be achieved by adding as many orbitals as we want, 
is an essential ingredient of K-theory. 
In general, one would expect that
the addition of trivial atomic bands should not affect 
the non-trivial topological properties of gapped phases.
Hence, 
one is interested in general in topological properties that are stable against 
inclusion of trivial bands. 
However, 
there are topological distinctions of gapped phases that exist only 
when the number of bands is restricted to be some particular integer. 
For example, it is known that there does not exist non-trivial class A TIs in 3d  with an arbitrary number of bands.
However, if we restrict ourselves to  2-band models,   non-trivial topologies exist
as supported by the non-trivial homotopy 
$\pi_3(\mathbb{C}P^1)=\mathbb{Z}$~\cite{MooreRanWen08, KennedyZirnbauer2014,
Nittis2014a, Nittis2014b}, 
which is unstable against the addition of trivial bands.
By taking the limit of infinitely many bands, 
we eliminate in the following such unstable or accidental topologies.  
Viz.,  
we are interested in the {\it stable equivalence} of the ground states of gapped non-interacting systems. 
 
The problem of classifying possible masses can be formulated in an alternative way as follows
\cite{Kitaev2009, Abramovici:2012zz, Morimoto2013}.
First of all, the Dirac kinetic term (the part without mass), 
consists of 
gamma matrices, 
forming a {\it Clifford algebra}. 
In general, 
a complex Clifford algebra $Cl_{n}$ is
given in terms of a set of generators $\{e_i\}_{i=1,\ldots, n}$, 
which satisfy 
\begin{align}
 \{ e_i, e_j \} = 2 \delta_{ij}. 
\end{align}
``Complex'' here means we allow these generators to be 
represented by a complex matrix. 
(More formally, we are interested in 
a $2^n$-dimensional complex vector space
$\{ C_{p_1 p_2 \cdots} e^{p_1}_1 e^{p_2}_2 \cdots \}$, 
where $p_i=0,1$ and $C_{p_1 p_2 \cdots}$ is a complex number.)
For the present example of the class A TI in $d=2$, 
the Dirac matrices in the kinetic term satisfy
$
\{\sigma_i, \sigma_j\} = 2\delta_{ij}
$
($i = 1,2$).
I.e., they form $Cl_2$. 
A mass should anticommute with all Dirac matrices in the kinetic term,
$
 \{ \sigma_i, M\} =0,
 $
 $\forall i$. 
I.e., with the mass, we now have $Cl_3$. 
When considering a mass, 
we are thus {\it extending} 
the algebra from $Cl_2$ to $Cl_3$
by adding one generator (mass). 
Counting different ways to extend the algebra
is nothing but counting unitary non-equivalent masses.

In the general case, we first consider a set of symmetry operators
(and Dirac kinetic terms).
They are represented as Clifford generators. 
We then consider, in addition to these generators, possible mass terms,
which in turn extend the Clifford algebra.
That is, for a fixed representation of the symmetry 
generators, we look for possible representations of a new additional generator (= mass). 
The set of these representations  
form  a classifying space
\cite{spingeometrybook, Hatcherbook}. 
Topologically distinct states correspond 
to distinct extensions of the algebra. 

As yet another example, let us consider 
class A insulators in $d=1$ and their Dirac representatives given by
$
{H}(k) = k_x \sigma_3\otimes \openone_N + M. 
$
As before, the mass must anticommute with 
the Dirac kinetic term, $\{\sigma_3, M\}=0$.
The generic solution to this is
\begin{align}
 M = \left( 
 \begin{array}{cc}
 0 & U^{\dag} \\
 U & 0
 \end{array}
 \right),
 \quad
 U\in U(N). 
\end{align}
Since $\pi_0 (U(N))=0$,
for fixed $N$, all masses can be continuously deformed to 
each other. That is, there is no topological distinction among gapped phases. 
As before, this problem can be formulated as an extension
problem $Cl_{1}\to Cl_2$.
The space classifying the extension is
\begin{align}
\mathcal{C}_1 = U(N)
\label{Cclassifyingspace1}
\end{align}
and its homotopy group is given by $\pi_0 (\mathcal{C}_1) = 0$.

This analysis can be repeated for arbitrary $d$.
One considers the extension $Cl_{d}\to Cl_{d+1}$.
Denoting the corresponding classifying space 
$\mathcal{C}_{d}$, 
we look for $\pi_0 (\mathcal{C}_d)$. 
Because of
$ Cl_{n+2}\simeq Cl_n \otimes \mathbb{C}(2), 
$
where $\mathbb{C}(2)$ is an algebra
of $2\times 2$ complex matrices 
(which does not affect the extension problem),
we have a periodicity of classifying spaces 
\begin{align}
 \mathcal{C}_{n+2}\simeq \mathcal{C}_{n},
\end{align}
from which 
the 2-fold dimensional periodicity for the topological classification of class A 
follows.

\paragraph{Class AIII} 
As we have seen, the dimensional periodicity of the topological 
classification problem for a given symmetry class follows directly
from the Clifford algebras. Similarly, 
the dimensional shift in the classification, caused by adding a symmetry, can  also 
be understood using Clifford algebras.
As an example, let us consider 
a zero-dimensional system in symmetry class AIII, 
whose mass (i.e., the Hamiltonian itself) $H$ satisfies the chiral symmetry relation
$\{H, U_S \}=0$.
The unitary matrix $U_S$, like the gamma matrices in the Dirac kinetic term,
can be thought of as a Clifford generator. 
With a proper normalization (spectral flattening),
the zero-dimensional Hamiltonian
$H$
has eigenvalues
$\pm 1$ and
can be considered as an additional Clifford generator. 
We then consider an extension problem
$Cl_1\to Cl_2$,
whose classifying space is $\mathcal{C}_1$
and $\pi_0(\mathcal{C}_1)=0$.
Thus, the presence of symmetries can be treated by 
adding a proper number of Clifford generators, 
and has effectively the same effect as 
increasing the space dimension.

\paragraph{Class D in \texorpdfstring{$d=0,1,2$}{d=0,1,2}}

So far we have discussed the use of complex Clifford algebras  
for the classification of Dirac masses in class A and AIII. 
Real Clifford algebras are relevant for the 
classification of Dirac masses in the 8 real symmetry classes,
as we now illustrate.

We begin with the class D example in $d=0$, i.e., we consider
 the Hamiltonian $H(\mathsf{r})=m\Gamma_0(\mathsf{r})$, which anticommutes with $C=\mathcal{K}$. 
Let $\mathbf{u}_1,\ldots,\mathbf{u}_N$ be the orthonormal positive eigenvectors of $\Gamma_0$. 
By PHS, $\mathbf{u}_1^\ast,\ldots,\mathbf{u}_N^\ast$ are negative eigenvectors.
Let $\mathbf{a}_j$ and $\mathbf{b}_j$ be the real and imaginary parts of $\mathbf{u}_j$, 
$\mathbf{u}_j =(\mathbf{a}_j+i\mathbf{b}_j)/\sqrt{2}$. 
The orthonormal relation $\mathbf{u}_i^\dagger\mathbf{u}_j=\delta_{ij}$ 
and $\mathbf{u}_i^T\mathbf{u}_j=0$ translates
into $\mathbf{a}_i^T\mathbf{a}_j=\mathbf{b}_i^T\mathbf{b}_j=\delta_{ij}$ and $\mathbf{a}_i^T\mathbf{b}_j=0$.
Thus we have an $O(2N)$ matrix $A=\left(\mathbf{a}_1,\ldots,\mathbf{a}_N,\mathbf{b}_1,\ldots,\mathbf{b}_N\right)$. 
Note that the same $\Gamma_0$ can correspond to
different orthogonal matrices $A$, due to the $U(N)$ basis transformation $\mathbf{u}_j\to\mathbf{u}_j'=U_{jk}\mathbf{u}_k$. 
Hence, the class D mass term $\Gamma_0$ in $d=0$ lives in the classifying space
\begin{align}
\mathcal{R}_2=O(2N)/U(N). 
\end{align}

Moving on to 1d,
we consider $H(k,\mathsf{r})=k\Gamma_1+m\Gamma_0(\mathsf{r})$. 
By a suitable choice of basis, we can assume
that the PH operator has the form $C=\mathcal{K}$ and $\Gamma_1=\tau_3\otimes\openone_N$. 
The mass term is thus $\Gamma_0(\mathsf{r})=\tau_2\otimes\gamma_1(\mathsf{r})+\tau_1\otimes i\gamma_2(\mathsf{r})$, 
where $\gamma_1,\gamma_2$ are the real symmetric and antisymmetric components of an $N\times N$ matrix $\gamma(\mathsf{r})=\gamma_1(\mathsf{r})+\gamma_2(\mathsf{r})$. 
The normalization $\Gamma_0^2=\openone$ implies that $\gamma$ must be orthogonal. 
Thus, class D mass terms in 1d belong to the classifying space 
\begin{align}
\mathcal{R}_1=O(N).
\end{align}

Finally, we discuss the 2d case,
where $H(\mathsf{k},\mathsf{r})=k_1\Gamma_1+k_2\Gamma_2+m\Gamma_0(\mathsf{r})$. 
We choose the basis, such that $C=\mathcal{K}$, 
$\Gamma_1=\tau_1\otimes\openone_N$, and $\Gamma_2=\tau_3\otimes\openone_N$. 
The mass term must be of the form $\Gamma_0(\mathsf{r})=\tau_2\otimes\gamma(\mathsf{r})$, 
where $\gamma$ is real symmetric and $\gamma^2=1$ in order for $\Gamma_0$ to have the appropriate symmetry and square to unity. 
One can diagonalize $\gamma$ by an orthogonal matrix 
$O=\left(\mathbf{a}_1,\ldots,\mathbf{a}_n,\mathbf{a}_{n+1},\ldots,\mathbf{a}_N\right)  \in O(N)$, 
where the first $n$ vectors are positive eigenvectors of $\gamma$ and the others are negative ones. 
We observe that the same $\gamma$ can correspond to different orthogonal matrices, due to  $O(n)\times O(N-n)$ 
basis transformations that do not mix positive and negative eigenvectors. 
Thus class D mass terms in 2d belong to the classifying space 
\begin{align}
\mathcal{R}_0=\bigcup_{n=0}^N\frac{O(N)}{O(n)\times O(N-n)}
\xrightarrow{N\to\infty}BO\times\mathbb{Z}.
\end{align}

As in complex symmetry classes, 
the relevant classifying spaces can be identified through an extension problem.
Similar to complex Clifford algebras, a real Clifford algebra
$Cl_{p,q}$ is generated by a set of 
generators $\{e_i\}$,
which satisfy
\begin{align}
&
 \{ e_i, e_j \} = 0, 
 \quad
 i\neq j
 \nonumber \\
 &
 e^2_i 
 =
 \left\{
 \begin{array}{ll}
  -1 & 1\le i \le p \\
  +1 & p+1 \le i \le p+q
 \end{array}
 \right.
\end{align}
``Real'' here means we are interested in real matrices
if these generators are represented by matrices. 
For real symmetry classes, we will use the Majorana representation of quadratic Hamiltonians, 
$
 \hat{H} = 
 \hat{\Psi}^{\dag}_A
 {H}^{AB}
 \hat{\Psi}^{\ }_B
 =
 i
 \hat{\lambda}_A
 {X}^{AB}
 \hat{\lambda}_B
$.
The real antisymmetric matrix ${X}$ 
can be brought into its canonical form by 
an orthogonal transformation
$
X\to O^T {X} O
$,
which reveals the condition $X^2=-1$, the only condition in class D.
Thus, we have the extension problem: 
$Cl_{0,0}\to Cl_{1,0}$. 
Let us denote the classifying space of the extension problem
$
 Cl_{p,q} \to Cl_{p, q+1}
$
as
$
 \mathcal{R}_{p,q}
$. 
It then turns out that all other extension problems are 
described by $\mathcal{R}_{p,q}$.
First of all, since
$
 Cl_{p+1, q+1}\simeq Cl_{p,q}\otimes \mathbb{R}(2), 
 $
$\mathcal{R}_{p,q}$ depends only on $q-p$,
$\mathcal{R}_{p,q}\equiv \mathcal{R}_{q-p}$. 
Second, 
since 
$
 Cl_{p,q}\otimes \mathbb{R}(2)
 \simeq 
 Cl_{q, p+2}, 
 $
the extension problem
$
 Cl_{p,q} \to Cl_{p+1, q}
 $
is mapped to 
$
 Cl_{q, p+2} \to Cl_{q, p+3}. 
 $
Thus, the classifying space of
$
 Cl_{p,q}\to Cl_{p+1, q} 
 $
is 
$
\mathcal{R}_{p+2-q}.
$
Finally,
since 
$
 Cl_{p+8, q} \simeq 
 Cl_{p, q+8} \simeq
 Cl_{p,q} \otimes \mathbb{R}(16)
$
the Bott periodicity 
\begin{align}
 \mathcal{R}_{q+8} \simeq \mathcal{R}_{q}
\end{align}
follows. 
By using these results, 
the extension problem $Cl_{0,0}\to Cl_{1,0}$ can be mapped to 
$
 Cl_{0,2}\to Cl_{0,3}
 $
and the corresponding classifying space is
$
 \mathcal{R}_{0,2}= \mathcal{R}_{2}= O(2N)/U(N).  
$

\begin{table}[tbp]
\centering
\begin{ruledtabular}
\begin{tabular}{ccccc}
& classifying space 
& extension & 
$\pi_0(*)$ 
& AZ class \\ \hline
 $\mathcal{C}_0$ & $BU\times\mathbb{Z}$ 
& $Cl_0\to Cl_1$ & $\mathbb{Z}$ & A \\
 $\mathcal{C}_1$ & ${U}(N)$ & $Cl_1\to Cl_2$ & 0 & AIII \\ 
\hline  
 $\mathcal{R}_0$ & $BO\times\mathbb{Z}$ & $Cl_{p,p}\to Cl_{p,p+1}$ & $\mathbb{Z}$ & AI \\
 $\mathcal{R}_1$ & ${O}(N)$& $Cl_{p,p+1}\to Cl_{p,p+2}$ & $\mathbb{Z}_2$ & BDI \\
 $\mathcal{R}_2$ & ${O}(2N)/{U}(N)$& $Cl_{p,p+2}\to Cl_{p,p+3}$ & $\mathbb{Z}_2$ & D \\
 $\mathcal{R}_3$ & ${U}(N)/{Sp}(N)$& $Cl_{p, p+3}\to Cl_{p,p+4}$ & 0 & DIII \\
$\mathcal{R}_4$ & $BSp\times\mathbb{Z}$ & $Cl_{p,p+4}\to Cl_{p,p+5}$& $\mathbb{Z}$ & AII \\
$\mathcal{R}_5$ & ${Sp}(N)$& $Cl_{p,p+5}\to Cl_{p,p+6}$&   0 & CII \\
$\mathcal{R}_6$ & ${Sp}(2N)/{U}(N)$ & $Cl_{p,p+6}\to Cl_{p,p+7}$ & 0 & C \\
$\mathcal{R}_7$ & ${U}(N)/{O}(N)$  & $Cl_{p,p+7}\to Cl_{p,p+8}$ & 0 & CI 
\end{tabular}
\end{ruledtabular}
\caption{
Classifying spaces for complex ($\mathcal{C}_s$) and real ($\mathcal{R}_s$) classes.
The right most column shows the corresponding 
Altland-Zirnbauer symmetry classes for zero-dimensional systems.  
\label{tab:classifyingspace}
}
\end{table}

\paragraph{Summary}

One can repeat this process for different symmetry classes and dimensions. 
The classifying space for symmetry $s$ in $d$ dimension is given by $\mathcal{C}_{s-d}$ for the complex AZ classes, 
or $\mathcal{R}_{s-d}$ for the real cases 
(Table~\ref{tab:classifyingspace}). 
The winding of the mass terms $m\Gamma_0(\mathsf{r})$ as the spacetime parameter $\mathsf{r}$ 
wraps once around the defect is classified by the homotopy group
\cite{Freedman2011} 
$
\pi_D(\mathcal{R}_{s-d})=\left[S^D,\mathcal{R}_{s-d}\right], 
$
which counts the number of topologically distinct non-singular mass terms 
as continuous maps $m\Gamma:S^D\to\mathcal{R}_{s-d}$. 
We recall  that classifying spaces are related to each other 
by looping, i.e., $\mathcal{R}_{p+1}\simeq\Omega\mathcal{R}_{p}=\mbox{Map}(S^1,\mathcal{R}_p)$. 
This implies the following relation between homotopy groups: $\pi_n(\mathcal{R}_{p+1})=\pi_{n+1}(\mathcal{R}_p)$. 
Hence \begin{align}\pi_D(\mathcal{R}_{s-d})=\pi_0(\mathcal{R}_{s-d+D})\end{align} 
classifies topological defects in class $s$ with topological dimension $\delta=d-D$. 
This shows that the classification only depends on the combination $s-d+D$ 
and proves the classification Table~\ref{tab:defectclassification} by use of Table~\ref{tab:classifyingspace}.

As a digression,
let us  briefly mention that Table~\ref{tab:defectclassification} can also be derived
from  a stability analysis of gapless surface  Hamiltonians, 
instead of using the homotopy group  classification of  mass terms.
The first step in this approach is to write down a $(d-1)$-dimensional gapless  
Dirac Hamiltonian with minimal matrix dimension
\bee
H_{\rm{surf}}(\mathsf{k})=\sum_{j=1}^{d-1} k_j\gamma_j,
\quad
\{\gamma_i, \gamma_j\}=2\delta_{ij}\openone,
\ee
which describes the surface state of a $d$-dimensional gapped bulk system
belonging to a given symmetry class.
Note that the form of $H_{\rm{surf}}$ is restricted by the symmetries of \cref{PHS,TRS_eq,chiral eq}.
Second, we ask if there exists a symmetry allowed mass term $m\gamma_0$, which anticommutes with $H_{\rm{surf}}$. 
If so, the surface mode can be gapped, which indicates that the bulk system has trivial topology labeled by ``0" in Table~\ref{tab:defectclassification}.
On the other hand, if there does not exist any symmetry allowed mass term  $m \gamma_0$,
then the surface state is topologically stable (i.e., protected by the symmetries), which indicates 
that the bulk is topologically non-trivial. 
To distinguish between a $\bZ$ and a $\bZ_2$ classification, one needs to 
consider multiple copies of the surface Hamiltonian, e.g.,  $H_{\rm{surf}}\otimes\openone_N$. 
If the surface Dirac Hamiltonian is stable for an arbitrary number of copies (i.e., if there does not exist any symmetry allowed mass term), 
 the corresponding bulk is classified by an integer topological invariant  $\mathbb{Z}$.
If, however, the surface state is stable only for an odd number of copies, 
the bulk is classified by a $\mathbb{Z}_2$ invariant.

It is possible to derive the entire classification table in this way.
As an example,
let us consider class A, AII, and AIII in $d=3$ dimensions.
A 2d surface Dirac Hamiltonian with minimal matrix dimension can be written as
\begin{align}
H_{\rm{surf}}(\mathsf{k})=k_1 \sigma_1 + k_2 \sigma_2.
\end{align}
For class A,  the mass term $m\sigma_z$ gaps out the surface mode, leading to
the trivial classification ``0'' in Table~\ref{tab:classification}.
For class AII and AIII, however,  $m\sigma_z$, which is the only possible mass term,
breaks TRS (\ref{TRS_eq}) and chiral symmetry (\ref{chiral eq}) 
with $T=\sigma_y \mathcal{K}$ and $S=\sigma_z$, respectively. 
To further distinguish between a $\bZ_2$ and $\bZ$ classification, 
we consider $H_{\rm{surf}}\otimes \openone_2$, 
for which the symmetry operators are given by $T=\sigma_y\otimes \openone_2 \mK$ and $S=\sigma_z\otimes \openone_2$.
There exist only one mass term for this doubled Hamiltonian, namely $m\sigma_z\otimes \sigma_y$, which preserves TRS but breaks chiral symmetry. 
Thus, class AII and AIII are classified by $\bZ_2$ and $\bZ$ invariants, respectively. 

Using a similar approach it is also possible to classify TIs and TSCs in terms of crystalline symmetries, see Sec.~\ref{Topological crystalline materials}.
Furthermore, this classification strategy can also be applied
to topological semimetals and nodal SCs, see Sec.~\ref{sec:gapless_materials}.

\subsubsection{Defect \texorpdfstring{$K$}{K}-theory}
\label{Defect K-theory}
The homotopy group classification of mass terms discussed in the previous subsection seemingly depends on the fact that the defect Hamiltonian \eqref{domainwallH} is of Dirac-type. 
However, it actually applies to a general defect Hamiltonian $H(\mathsf{k},\mathsf{r})$ (i.e., not only to Dirac Hamiltonians), as long as there is a finite energy gap separating the occupied bands from unoccupied ones. 
This general classification can be presented in the language of $K$-theory 
\cite{Teo:2010fk}. 
For a fixed AZ symmetry class and dimensions $(d,D)$, 
the collection of defect Hamiltonians forms a {\em commutative monoid} 
-- an associative additive structure with an identity -- 
by considering a direct sum 
\begin{align}
H_1\oplus H_2=
\left(\begin{array}{*{20}c}H_1&0\\0&H_2
\end{array}\right),
\end{align} where direct sums of symmetry operators,
$T_1\oplus T_2$, $C_1\oplus C_2$,
are defined similarly. 
Clearly, $H_1\oplus H_2$ has the same symmetries and dimensions as its constituents. 
The identity element is the $0\times0$ empty Hamiltonian $H=\emptyset$. 
Physically, 
the direct sum operation simply means to put the two systems on top of each other without letting them couple to each other.

As in ordinary $K$-theories, 
this monoid can be promoted to a group by introducing topological equivalence and applying the Grothendieck construction, 
which will be explained below. 
Two defect Hamiltonians $H_1(\mathsf{k},\mathsf{r})$ and $H_2(\mathsf{k},\mathsf{r})$ 
with the same symmetries and spatial dimensions,
but not necessarily with the same matrix dimensions
($\mathrm{dim}\, H_1 \neq \mathrm{dim}\, H_2$), 
are {\em stably} topologically equivalent, 
\begin{align}
H_1(\mathsf{k},\mathsf{r})\simeq H_2(\mathsf{k},\mathsf{r}), 
\end{align} 
if, for large enough $M$ and $N$, 
$H_1(\mathsf{k},\mathsf{r})\oplus(\sigma_3\otimes\openone_M)$ can be continuously deformed into $H_2(\mathsf{k},\mathsf{r})\oplus(\sigma_3\otimes\openone_N)$ 
without closing the energy gap or breaking symmetries.
Here, $\sigma_3\otimes\openone_M$ is a trivial atomic $2M\times 2M$ Hamiltonian that does not depend on 
$\mathsf{k}$ and $\mathsf{r}$,
and $M-N=\mathrm{dim}\, H_2-\mathrm{dim}\, H_1$.

Stable topological equivalence defines equivalent classes of defect Hamiltonians 
\begin{align}[H]=\{H':H'\simeq H\},
\end{align} 
which is compatible with the addition structure $[H_1]\oplus[H_2]=[H_1\oplus H_2]$. 
The identity element is $0=[\emptyset]$ which consists of all topologically trivial Hamiltonians that can be deformed into $\sigma_3\otimes\openone_N$. 
Each Hamiltonian class now has an additive inverse. 
By adding trivial bands, we can always assume a Hamiltonian has an equal number of occupied and unoccupied bands. 
Consider the direct sum $H\oplus(-H)$, where in $(-H)$ the occupied states are inverted to unoccupied ones. 
This sum is topologically trivial as the states below the gap consist of both the valence and conduction states in $H$ 
and they are allowed to mix. 
This shows that $[H]\oplus[-H]=0$ and $[-H]$ is the additive inverse of $[H]$.
We now see that the collection of equivalent classes of defect Hamiltonians forms a group 
and defines a $K$-theory 
\begin{align}K(s;d,D)=\left\{[H]:
\begin{smallmatrix}
\mbox{\footnotesize $H(\mathsf{k},\mathsf{r})$, a gapped defect}\hfill\\\mbox{\footnotesize Hamiltonian of AZ class}\\\mbox{\footnotesize $s$ and dimensions $(d,D)$}\hfill
\end{smallmatrix}\right\}.
\end{align} 

We now establish group homomorphisms relating $K$-groups with different symmetries and dimensions \cite{Teo:2010fk} 
\begin{align}
\Phi_+:K(s;d,D)\longrightarrow K(s+1;d+1,D),
\label{Phi+}
\\\Phi_-:K(s;d,D)\longrightarrow K(s-1;d,D+1) .
\label{Phi-}
\end{align} 
That is, given any defect Hamiltonian $H_s(\mathsf{k},\mathsf{r})$ in symmetry class $s$, one can define a new gapped Hamiltonian 
\begin{align} \label{11periodicity}
&
H_{s\pm1}(\mathsf{k},\theta,\mathsf{r})
\\
&=
\left\{
\begin{array}{ll}
\cos\theta H_s(\mathsf{k},\mathsf{r})+\sin\theta S,
& \mbox{$s$ odd} \\
\cos\theta H_s(\mathsf{k},\mathsf{r})\otimes\sigma_3+\sin\theta\openone\otimes\sigma_{1,2}, 
& \mbox{$s$ even}
\end{array}
\right. .
\nonumber 
\end{align}
Here $\theta\in[-\pi/2,\pi/2]$ is a new variable that extends $(\mathsf{k},\mathsf{r})$, 
which lives on the sphere ${S}^{d+D}$, to the suspension $\Sigma S^{d+D}=S^{d+1+D}$. 
This is because the new Hamiltonian $H_{s\pm1}$ is independent of $(\mathsf{k},\mathsf{r})$ at the north and south poles where $\theta=\pm\pi/2$.

We first look at the case when $s$ is odd. For real symmetry classes, the chiral operator is set to be the product $S=i^{(s+1)/2}TC$ of the TR and PH operators. 
The factor of $i$ is to make $S$ hermitian and square to unity. 
The addition of the chiral operator in \eqref{11periodicity} breaks the chiral symmetry since the Hamiltonian $H_{s\pm1}$ does not anticommute with $S$ anymore. 
Depending on how the new variable $\theta$ transforms under the symmetries $\theta\to\pm\theta$, the new Hamiltonian $H_{s\pm1}$ preserves only either TRS or PHS. 
If $\theta$ is odd (even), it belongs to the symmetry class $s+1$ (resp.~$s-1$). This also applies to complex symmetry classes.

Next we consider the even $s$ cases. 
For real symmetry classes, $H_s$ has one antiunitary symmetry, say TRS. 
(The case of PHS can be argued by a similar manner.) 
The introduction of the $\sigma$ degree of freedom doubles the number of bands and the new Hamiltonian $H_{s\pm1}$ 
in \eqref{11periodicity} has a chiral symmetry $S=\openone\otimes\sigma_{2,1}$ which anticomutes with the extra term $\sin\theta\openone\otimes\sigma_{1,2}$. 
For the case when $S=\sigma_2$, there is a new PHS with the operator $C=iT\otimes\sigma_2$ that fixes the new parameter $\theta\to\theta$. 
For the other case for $S=\sigma_1$, the new PHS operator is $C=T\otimes\sigma_1$ and the new parameter flips $\theta\to-\theta$ under the symmetry. 
The new Hamiltonian then belongs to the symmetry class $s-1$ for the former case, and $s+1$ for the latter.

To summarize, equation \eqref{11periodicity} 
defines the correspondences 
\begin{align}
\Phi_\pm:[H_s(\mathsf{k},\mathsf{r})]\longrightarrow[H_{s\pm1}(\mathsf{k},\theta,\mathsf{r})].
\end{align} 
For the $+$ case, $\theta$ is odd under the symmetry and behaves like a new momentum parameter. 
It increases the dimension $d\to d+1$. 
For the $-$ case, $\theta$ is even under the symmetry. 
The extra space-like parameter then increases $D\to D+1$. $\Phi_\pm$ commutes with direct
sums $\Phi_\pm[H_1\oplus H_2]=\Phi_\pm[H_1]\oplus\Phi_\pm[H_2]$ and therefore are group homomorphisms between $K$-theories.
These homomorphisms are actually invertible and provide isomorphisms between \cite{Teo:2010fk}
\begin{align}
K(s;d,D)&\cong K(s+1;d+1,D)\nonumber\\&\cong K(s-1;d,D+1) .
\label{defectKiso}
\end{align}
To see this, we begin with an arbitrary defect Hamiltonian $H_{s\pm1}(\mathsf{k},\theta,\mathsf{r})$. 
It can be shown to be topologically equivalent to one with the particular form in
\eqref{11periodicity}. 
We  then consider the artificial action 
\begin{align}
S[\overline{H}(\mathsf{k},\theta,\mathsf{r})]=\int d\theta d^d\mathsf{k}d^D\mathsf{r}\,
\mbox{Tr}\left(\partial_\theta\overline{H}\partial_\theta\overline{H}\right)
\end{align} on the moduli space of flat band Hamiltonians $\overline{H}$ so that $\overline{H}^2=\openone$. 
By satisfying the Euler-Lagrangian equation 
\begin{align}
\left.\frac{\delta S}{\delta H}\right|_{H^2=1}=\partial_\theta^2H+H=0,
\end{align} 
Eq.\ \eqref{11periodicity} locally minimizes the action. 
The action also defines a natural minimizing flow direction that deforms an arbitrary Hamiltonian $H(\mathsf{k},\theta,\mathsf{r})$ 
to the form of 
\eqref{11periodicity}.
This shows the invertibility of $\Phi_\pm$.

The isomorphisms \eqref{defectKiso} prove that the classification of topological defects depends only on the combination $s-d+D$. 
Furthermore, the defect $K$-theory is related to the classification of TI and TSC by 
\begin{align}
K(s;d,D)\cong K(s+D;d,0) \cong K(s;d-D,0),
\label{defecttoTITSC}
\end{align} 
which classifies class $s$ topological band theories in $\delta=d-D$ dimensions. 
The equivalence \eqref{defecttoTITSC} extends characteristics of the classification of TIs and TSCs to the classification of topological defects. 

Beside \eqref{defecttoTITSC},
there are further relationships among K-groups
having different $s, d, D$.
For example, 
topological states in the 1st and 2nd descendants 
are related to their ``parent'' states in primary series,
by {\it dimensional reduction} \cite{Qi2008sf,Ryu2010ten}.
This procedure is one way to 
understand how the $\mathbb{Z}_2$ characterization 
of the 1st and 2nd descendants emerge.  
Let us consider a 
$d$-dimensional
Bloch Hamiltonian ${H}(\mathsf{k})$ 
describing a gapped topological state in the 1st descendants. 
One can then consider a $(d+1)$-dimensional 
Bloch Hamiltonian $\tilde{{H}}(\mathsf{k},k_{d+1})$ 
which belongs to the {\it same} symmetry class 
and satisfy $\tilde{{H}}(\mathsf{k},0)={H}(\mathsf{k})$. 
Furthermore, 
if there is a spectral gap in
$\tilde{{H}}(\mathsf{k},k_{d+1})$,
one can compute the topological invariants introduced above,
since $\tilde{{H}}(\mathsf{k},k_{d+1})$ belongs to the primary series.
However, as one immediately notices, 
there is no unique higher-dimensional Hamiltonian to which the original Hamiltonian can be embedded,
nor a unique value for the topological invariant.
Nevertheless, the parity of the topological invariant can be shown to be independent of the way we embed 
the Hamiltonian.
This is the origin of the $\mathbb{Z}_2$ classification of the 1st descendants. 
Similar arguments apply to the 2nd descendants. 

Summarizing, 
the first and second $\mathbb{Z}_2$ topologies are related to their parent $\mathbb{Z}$ topology of the same symmetry class by the surjections 
\begin{align}\xymatrix{\mathbb{Z}_2^{(2)}&\mathbb{Z}_2^{(1)}\ar[l]_{i^\ast}^\cong&\mathbb{Z}\ar[l]_{i^\ast}}\end{align} 
where $i^\ast:K(s;d+1,D)\to K(s;d,D)$ is the restriction homomorphism that restricts 
\begin{align}
i^\ast:H_s(\mathsf{k},k_{d+1},\mathsf{r})
\mapsto\left.H_s(\mathsf{k},\mathsf{r})\right|_{k_{d+1}=0}
\end{align} 
where $(\mathsf{k},k_{d+1},\mathsf{r})$ 
lives on the compactified $S^{d+1+D}$ and $(\mathsf{k},\mathsf{r})$ belongs to the equator $S^{d+D}$. 

As yet another relationship, 
the first $\mathbb{Z}_2$ descendant for the chiral classes relates isomorphically to the second $\mathbb{Z}_2$ descendant 
for the non-chiral classes: 
\begin{align}
\xymatrix{\mbox{chiral class ($s$ odd):}&\mathbb{Z}_2^{(1)} \ar[d]_f \ar[rd]^{\Phi_+}_\cong&\\\mbox{non-chiral class ($s+1$):}&\mathbb{Z}_2^{(2)}&\mathbb{Z}_2^{(1)} \ar[l]^{i^\ast}_\cong}
\end{align} Here the map between $K$-theories 
\begin{align}
f:K(s;d,D)\cong\mathbb{Z}_2^{(1)}\longrightarrow K(s+1;d,D)\cong\mathbb{Z}_2^{(2)}
\end{align} is the {\em forgetful functor} that ignores either TRS or PHS so that the chiral band theory now belongs to the non-chiral symmetry class $s+1$. 
It agrees with the composition $f=i^\ast\circ\Phi_+$, where $\Phi_+:K(s;d,D)\to K(s+1;d+1,D)$ is the isomorphism \eqref{11periodicity} and $i^\ast$ restricts 
the Hamiltonian $H_{s+1}(\mathsf{k},\theta,\mathsf{r})$ onto the equator where $\theta=0$. 
Since both $i^\ast$ and $\Phi_+$ are isomorphisms, so is the forgetful map $f$. The topological invariant for chiral $\mathbb{Z}_2^{(1)}$ is therefore given by the FK-invariant \eqref{defectFK} with the gauge constraint \eqref{FKconstraint1} for $s=1,5$ (class CI and DIII) or \eqref{FKconstraint2} for $s=3,7$ (class BDI and CII).

\subsection{Bulk-boundary and bulk-defect correspondence}
\label{Bulk-boundary and bulk-defect correspondence}

In this section, we will relate the bulk topological invariants discussed in Sec.~\ref{subsec:topinvariants}
to the protected gapless excitations localized at boundaries/defects. 
This will be done by introducing proper indices that ``count'' the number of 
zero modes and gapless modes localized at defects (\`a la index theorems),
and by identifying these indices as the topological invariants.   
This bulk-boundary/defect correspondence unifies numerous TI and TSC defect systems, 
which we will  demonstrate in terms of a variety of examples. 
In addition to the discussion below, we refer the reader to the 
literature of
\onlinecite{EssinGurarie2011, GrafMicheleMarcello2013},
where different approaches to establish the bulk-boundary/defect correspondence have been studied.

\subsubsection{Zero modes at point defects and index theorems}
\begin{table}[tbp]
\centering
\begin{ruledtabular}
\begin{tabular}{ccl}
Symmetry  & Topological classes & Bound States at $\varepsilon=0$\\
\hline
AIII & $\mathbb{Z}$ &  Chiral Dirac \\
BDI & $\mathbb{Z}$ &  Chiral Majorana \\
D & $\mathbb{Z}_2$ &  Majorana \\
DIII & $\mathbb{Z}_2$ &  Majorana Kramers Doublet  \\
& &  (= Dirac)\\
CII & $2\mathbb{Z}$ &   Chiral Majorana Kramers\\
& &   Doublet (=Chiral Dirac)
\end{tabular}
\end{ruledtabular}
\caption{Symmetry classes supporting non trivial point topological defects
and their associated zero-energy modes. }
\label{tab:pointdefect}
\end{table}

We start by demonstrating the protected zero-energy modes localized at topological point defects 
($\delta=d-D=1$). 
The simplest examples are given by 
the SSH and 1d Kitaev models,
or their continuum counter parts, the Jackiw-Rebbi model,  
discussed in Secs.\ \ref{Example: Polyacetylene} and \ref{sec: The Kitaev chain}.
The domain wall defects in these 1d models trap zero-energy bound states protected by chiral or PH symmetry. 
The continuum version of these models \eqref{JackiwRebbi} are given by the differential operator 
\begin{align} \label{1d_examp_bnd_state}
\mathcal{H}=-i v \sigma_2\frac{d}{dr}+m(r)\sigma_3,
\quad 
r \in (-\infty, +\infty) ,
\end{align} 
where  the mass $m(r)$, which changes sign at the origin,
describes the domain wall.  
The zero-energy bound state $|\gamma\rangle$,
which is exponentially localized at the domain wall (i.e., at the origin), is
an eigenstate of the chiral or PH operator,
$S|\gamma\rangle=\pm|\gamma\rangle$ or $C|\gamma\rangle=|\gamma\rangle$,
where $S=\sigma_1$ and $C= \sigma_1 \mathcal{K}$.
The chiral eigenvalue, called {\it chirality}, 
of the zero mode has a definite sign, depending on the sign of the winding number \eqref{SSHinvariant1}. 
The sign of the PH eigenvalue,
on the other hand, 
is unphysical, since it can be flipped by multiplying $|\gamma\rangle$ by $i$.
Hence, for zero-energy Majorana bound state (MBS) protected by PHS the PH eigenvalue
can always be assumed to be +1.

Since the 1d example~\eqref{1d_examp_bnd_state} is invariant under chiral or PH symmetry,
its energy levels must come in $\pm \varepsilon$ conjugate pairs. 
The zero mode $|\gamma\rangle$ , however, is self-conjugate, 
and therefore does not have a conjugate partner.
Hence, 
$|\gamma\rangle$ is pinned at zero energy and, as a consequence, is robust against any perturbation that does not close the bulk energy gap. 
We list in Table \ref{tab:pointdefect} the different  symmetry classes  that can support zero-energy modes at topological point defects.
Depending on the symmetry class, these zero-energy modes are of different type, as indicated by the last column in Table \ref{tab:pointdefect}.

\paragraph{Index theorems}

In general, if a point defect supports an odd number of zero-energy bound states, 
only an even number of them can be paired up and gapped out
upon inclusion of PH symmetric perturbations.
This leaves at least one unpaired zero-energy bound state. 
The even-odd parity of the number of zero modes is known as 
a $\mathbb{Z}_2$-{\em analytic index} of the differential operator $\mathcal{H}$,   
\begin{align}
\mbox{ind}^{(1)}_{\mathbb{Z}_2}[\mathcal{H}]
=
\left(
\begin{array} {*{20}l}
\mbox{\footnotesize number of zero-}\\\mbox{\footnotesize energy bound states}
\end{array}
\right)
\quad\mbox{mod 2}, 
\end{align} 
which we claim is identical to the $\mathbb{Z}_2$-{\em topological index},  
\begin{align}
\mbox{ind}^{(1)}_{\mathbb{Z}_2}[\mathcal{H}]
=2\mbox{CS}_{2d-1}[H(\mathsf{k},\mathsf{r})] ,
\label{Z2indexthm1}
\end{align} 
given by the Chern-Simons integral in \eqref{defectCS} for a point defect in $d$ dimensions.
The equality (\ref{Z2indexthm1})
is an example of the bulk-boundary correspondence.

For chiral symmetric systems, on the other hand, the chiral operator $\mathcal{S}$ defines in addition an integral quantity 
\begin{align}
\mbox{ind}_{\mathbb{Z}}[\mathcal{H}]=\mbox{Tr}\left(\mathcal{S}\right),
\label{Zanalyticindexdef}
\end{align} 
which is referred to as the chirality of the point defect. It counts 
the difference between
the number of zero modes with positive and negative chiral eigenvalues.  
This $\mathbb{Z}$-{\em analytical index} is robust against 
any chiral symmetric perturbation that does not close the bulk gap. 
This is because all conjugate pairs of energy eigen states, 
which can always be related by the chiral symmetry $|-\varepsilon\rangle=\mathcal{S}|+\varepsilon\rangle$,
do not contribute to $\mbox{Tr}(\mathcal{S})$, 
as $|+\varepsilon\rangle\pm|-\varepsilon\rangle$ must have opposite eigenvalues of $\mathcal{S}$.
For a point defect in $d$ dimensions, it is found that
the chirality is identical to the $\mathbb{Z}$-{\em topological index}, 
i.e., the winding number given in~\eqref{def winding number} 
\begin{align}
\mbox{ind}_{\mathbb{Z}}[\mathcal{H}]=\nu_{2d-1}[H(\mathsf{k},\mathsf{r})] .
\label{Zindexthmc}
\end{align} 
Moreover, \eqref{Zanalyticindexdef}  also agrees with the $\mathbb{Z}_2$-analytic index 
\begin{align}
\mbox{ind}_{\mathbb{Z}}[\mathcal{H}]=\mbox{ind}_{\mathbb{Z}_2}^{(1)}[\mathcal{H}]
\quad
\mod 2.
\end{align}

Equation \eqref{Z2indexthm1} applies to general point defects in all symmetry classes in any dimension, 
while \eqref{Zindexthmc} applies to arbitrary chiral ones. 
For instance, from the defect classification 
(Tables~\ref{tab:defectclassification} and \ref{tab:pointdefect}), 
we see that the CS-invariant for a point defect is non-vanishing only for class AIII, BDI and D. 
Equation \eqref{Z2indexthm1} then agrees with the fact that only point defects in these AZ classes can support 
an odd number of zero-energy MBS. 
All other classes either do not have a PHS, or  the zero modes must come in Kramers doublets. 
This also explains the even chirality $\mbox{ind}_{\mathbb{Z}}[\mathcal{H}]$ for class CII point defects and matches 
-- by the index theorem \eqref{Zindexthmc} -- with the $2\mathbb{Z}$ winding number $\nu_{2d-1}[H(\mathsf{k},\mathsf{r})]$.

Lastly, there is another $\mathbb{Z}_2$-analytic index associated to the second descendants.
It applies to point defects with an antiunitary symmetry $T$ or $C$ that 
squares to minus one, so that zero-energy states come in Kramers pairs:
\begin{align}
\mbox{ind}_{\mathbb{Z}_2}^{(2)}[\mathcal{H}]=
\left(
\begin{array}
{*{20}l}\mbox{\footnotesize number of zero}\\\mbox{\footnotesize energy Kramers pairs}
\end{array}
\right)
\quad\mbox{mod 2}. 
\end{align} 
This index is identical to the second descendant $\mathbb{Z}_2$-topological index 
\begin{align}
\mbox{ind}_{\mathbb{Z}_2}^{(2)}[\mathcal{H}]=\mbox{CS}_{2d-1}[H(\mathsf{k},\mathsf{r})], 
\label{Z2indexthm2}
\end{align} 
for a $d$-dimensional point defect, 
where the Chern-Simons invariant is defined in \eqref{defectCS} 
with the gauge constraint \eqref{FKconstraint1} for $T^2=-1$ or \eqref{FKconstraint2} for $C^2=-1$. 
The defect classification (Table~\ref{tab:defectclassification}) 
tells us that only point defects in class DIII support protected zero-energy Majorana Kramers pairs. 
These zero modes cannot be detected by the other indices in \eqref{Z2indexthm1} or \eqref{Zindexthmc},
since there are an even number of MBSs which necessarily carry opposite chirality, as $S$ and $T$ anticommutes.

It is worth noting that the $\mathbb{Z}$-analytic index \eqref{Zanalyticindexdef} 
and its identification to the topological index \eqref{Zindexthmc} is a rendition of the original celebrated index theorem 
in the mathematics literature~\cite{AtiyahSinger63}. 
A chiral symmetric defect Hamiltonian $\mathcal{H}$, 
in the form of a differential operator, takes the off-diagonal form 
\begin{align}
\mathcal{H}=\left(\begin{array}{*{20}c}0&\mathcal{D}^\dagger\\\mathcal{D}&0\end{array}\right),
\end{align} 
where $\mathcal{D}$ is a Dirac operator, which is {\em Fredholm}. 
Equation \eqref{Zanalyticindexdef} is identical to 
\begin{align}
\mbox{ind}_{\mathbb{Z}}[\mathcal{H}]=\dim\ker(\mathcal{D})-\dim\ker(\mathcal{D}^\dagger), 
\end{align} 
which is the original definition of the analytic index of a Dirac operator. 
The index theorem \eqref{Zindexthmc} can be proven by means of  a {\em heat kernel} method~\cite{Berlinebook, spingeometrybook}. 
Several alternative proofs have been derived in the context of both condensed matter and high energy physics~\cite{JackiwRebbi76, JackiwRossi81, Volovik:book, Nakahara:2003ve, Weinberg81, FukuiFujiwara10, Fukui10}.

In the following, 
we present some examples of zero-energy bound states at topological point defects in both 2d and 3d. 
We will focus on point defects that trap  unpaired zero-energy MBSs or Majorana Kramers doublets. 
In many cases the topological invariants can be simplified into products 
of a bulk topological invariant and a defect winding number. 
MBSs are predicted to exists in many systems, e.g., in
quantum flux vortices in chiral $p_x+ip_y$ SCs or in superfluid $^3$He-A,
in TI-SC-ferromagnet (FM) heterostructures in  2d and 3d, and so on.  
The theory of topological defects unifies the topological origin of all these examples.
For instance, the appearance of protected zero-energy MBSs is always a consequence of
$K(s;d,D)=\mathbb{Z}_2^{(1)}$ for $s=2$ (class D), 
while the presence of protected zero-energy Majorana Kramers doublets is a result of $K(s;d,D)=\mathbb{Z}_2^{(2)}$
for $s=3$ (class DIII).
For example,
the protected zero-energy MBS 
at a quantum flux vortex of a spinless chiral $p_x+ip_y$ SC
turns out to have the same topological origin as 
a MBS located at a dislocation or disclination of a non-chiral $p$-wave SC
~\cite{HughesYaoQi13, Teo_2013_disclination}.

\paragraph{Example: 2d class D \texorpdfstring{$p_x+ip_y$}{px+ipy} superconductors}\label{sec: chiral p wave superconductors}

We first look at a quantum flux vortex of a spinless chiral $p_x+ip_y$ SC
\cite{Volovik:book,Leggettbook,AndersonMorel61,BalianWerthamer63,
Leggett75,SigristUeda91,RiceSigrist95,LukeSigrist98,XiaKapitulnik06,Volovik99, ReadGreen2000, Ivanov_braiding, Kitaev2006, GurarieRadzihovsky07,TewariZhangDasSarmaNayakLee08}. 
Consider a 2d BdG Hamiltonian on the square lattice
\begin{align}
H_0(\mathsf{k})&=
\Delta(\sin k_x\tau_1+\sin k_y\tau_2)\nonumber\\&\;\;\;+[t(\cos k_x+\cos k_y)-\mu]\tau_3, 
\label{ReadGreenmodel}
\end{align} 
where $\tau_{i=1,2}$ acts on the Nambu degrees of freedom 
$(\hat{c},\hat{c}^\dagger)$, and the PH operator is $C=\tau_1\mathcal{K}$. 
When the electron hopping strength and fermi energy are arranged so that $2t>|\mu|>0$, 
this bulk 2d model has a unit Chern invariant and carries a chiral Majorana edge mode. 
In the continuum limit, a chiral $p_x+ip_y$ SC can be represented by 
\begin{align}
H_0(\mathsf{k})=\Delta(k_x\tau_1+k_y\tau_2)+\left(\frac{\hbar^2k^2}{2m}-\mu\right)\tau_3,
\end{align} 
where the fermi energy $\mu$ is positive.
A $\phi=hc/2e$ quantum flux vortex can be described by the defect Hamiltonian 
\begin{align}
H(\mathsf{k},\mathsf{r})=e^{-i\varphi(\mathsf{r}) \tau_3/2}H_0(\mathsf{k})e^{i\varphi(\mathsf{r}) \tau_3/2} ,
\label{vortexH}
\end{align} 
where the SC pairing phase $\varphi$ winds by $2\pi\times l$ ($l\in \mathbb{Z}$) around the vortex,
and can be taken as the angular parameter $\varphi(\mathsf{r})=\tan^{-1}(y/x)\times l$.
The vortex can be shown to carry a protected zero-energy Majorana bound state, so that $\mbox{ind}^{(1)}_{\mathbb{Z}_2}[\mathcal{H}]=1$. 
The index theorem \eqref{Z2indexthm1} can be verified by evaluating the Chern-Simons invariant $\mbox{CS}_{3}$.
(For the technical reason explained below Eq.~\eqref{defectCS}, 
we need to consider the modified defect Hamiltonian 
$\widetilde{H}(\mathsf{k},\mathsf{r})=H(\mathsf{k},\mathsf{r})\oplus (-H_0(\mathsf{k}))$, 
where the lower block cancels the 2d Chern invariant without contributing to extra point defect states. 
This modification is to ensure that there is a global continuous basis of occupied states for the CS-integral.) 
The CS 3-form can be simplified~\cite{Teo:2010fk} and decomposed into 
\begin{align}
\mathcal{Q}_3=\left(\frac{i}{2\pi}\right)^2\mbox{Tr}[\mathcal{F}_0(\mathsf{k})]\wedge d\varphi\label{CS=Fdvarphi},
\end{align} 
where $\mathcal{F}_0$ is the Berry curvature for the $p+ip$ SC $H_0(\mathsf{k},\mathsf{r})$ without a vortex. 
The topological index therefore is a simple product 
of the bulk Chern number and the vorticity $l$, 
\begin{align}
2\mbox{CS}_3[H(\mathsf{k},\mathsf{r})]&=\frac{i}{2\pi}\int_{\mathrm{BZ}^2}
\mbox{Tr}(\mathcal{F}_0)\oint_{S^1}d
\varphi(\mathsf{r})
\nonumber\\
&=\mbox{Ch}_1\times l. 
\label{Chm}
\end{align} 
Equations \eqref{CS=Fdvarphi} and \eqref{Chm} apply to a general defect Hamiltonian of the form of \eqref{vortexH}, 
and the parity of the number of zero energy MBSs at a flux vortex can always be read off from 
\cite{Volovik:book, Kitaev2006, StoneRoy04}
\begin{align}
\mbox{ind}^{(1)}_{\mathbb{Z}_2}[\mathcal{H}]
=\mbox{Ch}_1[H_0(\mathsf{k})]\times l.
\label{ind=Chxm}
\end{align}

\begin{figure}[t]
\includegraphics[width=0.35\textwidth]{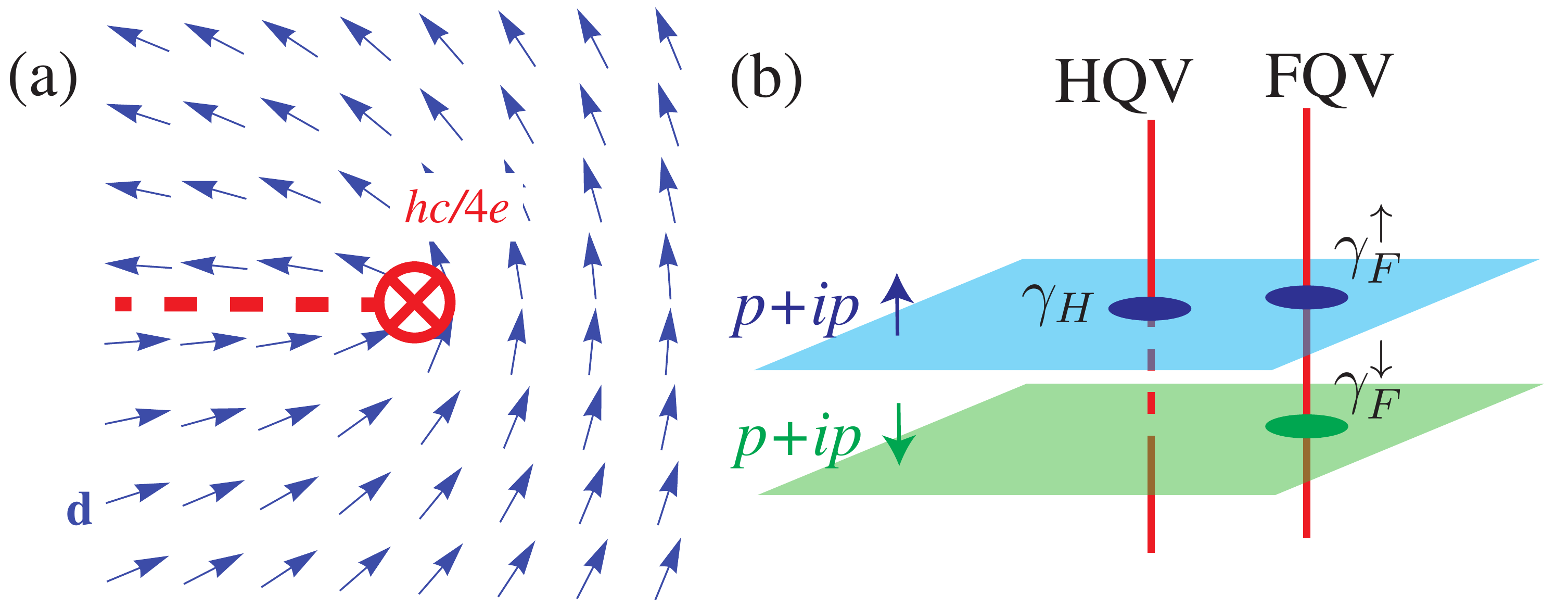}
\caption{(a) Spatial configuration of the $\mathbf{d}$-vector around a half-quantum vortex of a $p+ip$ SC. 
(b) Zero energy Majorana modes of a half-quantum vortex (HQV) and a full quantum vortex (FQV).}\label{fig:pvortex}
\end{figure}

Physical chiral $p_x+ip_y$ SCs are spinful. 
Strontium ruthenate (Sr$_2$RuO$_4$) is a plausible candidate of a spinful chiral $p$-wave SC with 
odd parity spin-triplet pairing~\cite{RiceSigrist95,LukeSigrist98,XiaKapitulnik06}, 
although its precise pairing nature is still under debate \cite{MaenoRiceSigrist2001,RaghuKapitulnikKivelson2010,maeno_review_JPSJ_12}. 
A continuum model of a 2d spinful chiral $p$-wave SC is given by   
\begin{align}
H_0(\mathsf{k})&=
\Delta({\boldsymbol\sigma}\cdot\mathbf{d})\sigma_2
(k_x\tau_1+k_y\tau_2)
\nonumber\\&\;\;\;
+\left(\frac{\hbar^2k^2}{2m}-\mu\right)
\tau_3, 
\end{align} 
where $\sigma_{i=1,2,3}$ acts on the spin degree of freedom, 
and the $\mathbf{d}$-vector specifies a special spin direction, 
say along the $xy$-plane, in the triplet pairing. 
The Nambu basis is taken to be 
$(\hat{c}_\uparrow,\hat{c}_\downarrow,\hat{c}_\downarrow^\dagger,-\hat{c}_\uparrow^\dagger)$ 
and the PH operator is $C=\sigma_2\otimes\tau_2\mathcal{K}$. 
From \eqref{ind=Chxm}, 
a full $hc/2e$ quantum vortex (FQV) carries two MBSs $\hat{\gamma}_\uparrow,\hat{\gamma}_\downarrow$,
which split by a perturbation $\delta \hat{H}=i\varepsilon\hat{\gamma}_\uparrow\hat{\gamma}_\downarrow$ 
into a $\pm\varepsilon$ pair, due to, e.g.,   spin-orbit coupling (SOC) or an in-plane magnetic field. 
On the other hand, 
a half quantum vortex (HQV) of flux $\phi=hc/4e$ consists of a $\pi$-rotation of the pairing phase as well as 
the $\mathbf{d}$-vector about the $z$-axis~\cite{SalomaaVolovik85,DasSarmaNayakTewari06,ChungBluhmKim07,Jang_SrRuO}. 
The HQV is represented by the defect Hamiltonian 
\begin{align}
H(\mathsf{k},\mathsf{r})=
e^{-i\varphi(\mathsf{r})
(\tau_3+\sigma_3)/4}H_0(\mathsf{k})e^{i\varphi(\mathsf{r}) (\tau_3+\sigma_3)/4}, 
\end{align} 
where $\varphi$ is the angular parameter around the vortex. 
The spatial configuration of the $\mathbf{d}$-vector  is shown in Fig.\ \ref{fig:pvortex}(a). 
Effectively, the HQV acts as a quantum vortex only on one of the two spin sectors where $\tau_3$ and $\sigma_3$ have the same sign. 
This gives a single protected zero energy MBS as shown in Fig.\  \ref{fig:pvortex}(b). 

\paragraph{Example: 2d class DIII \texorpdfstring{$(p+ip)\times (p-ip)$}{(p+ip)x(p-ip)} superconductors}
There exists also an unconventional spinful $p$-wave SC that preserves TRS~\cite{SchnyderAIP, Kitaev2009}. 
It involves an opposite chirality in the two spin species, 
and the pairing has a $(p_x+ip_y)\uparrow\times(p_x-ip_y)\downarrow$ structure. 
A continuum BdG Hamiltonian describing this SC is given by
\begin{align} \label{2D_examp_DIII_vortex}
H_0(\mathsf{k})=\Delta(k_x\tau_1+k_y\sigma_3\tau_2)+\left(\frac{\hbar^2k^2}{2m}
-\mu\right)\tau_3,
\end{align} 
where the Nambu basis is chosen to be 
$(\hat{c}_\uparrow, \hat{c}_\downarrow, \hat{c}_\uparrow^\dagger, \hat{c}_\downarrow^\dagger)$ 
and the PH operator is $C=\tau_1\mathcal{K}$. 
$H_0(\mathsf{k})$ has a TRS with $T=\sigma_2\tau_3\mathcal{K}$ and therefore belongs to class DIII. 
The non-trivial $\mathbb{Z}_2$ topology of $H_0(\mathsf{k})$  corresponds to a gapless helical Majorana edge mode. 
Hamiltonian~\eqref{2D_examp_DIII_vortex} is topologically equivalent -- by a basis transformation -- to the 2d $^3$He-B 
model~\cite{Volovik:book} 
\begin{align}
H_0(\mathsf{k})=\Delta(k_x\sigma_1+k_y\sigma_2)\tau_1+\left(\frac{\hbar^2k^2}{2m}-\mu\right)\tau_3 ,
\label{2d 3HeB}
\end{align} 
where the Nambu basis is now 
$(\hat{c}_\uparrow,\hat{c}_\downarrow,\hat{c}_\downarrow^\dagger,-\hat{c}_\uparrow^\dagger)$ with 
PH operator $C=\sigma_2\tau_2\mathcal{K}$ and TR operator $i\sigma_2\mathcal{K}$.
A  vortex that respects TRS
can be introduced in (\ref{2d 3HeB}) via
\begin{align}
H(\mathsf{k},\mathsf{r})=e^{-i\varphi(\mathsf{r}) \sigma_3/2}H_0(\mathsf{k})e^{i\varphi(\mathsf{r}) \sigma_3/2}, 
\end{align} 
which consists of a $2\pi$ rotation of spin once around the vortex core. 
One finds that a Majorana Kramers doublet is bound at the vortex core, as guaranteed by the second $\mathbb{Z}_2$-index 
\eqref{Z2indexthm2}.

\paragraph{Example: Dislocations and disclinations in crystalline superconductors (class D)}
\begin{figure}[t]
\centering\includegraphics[width=0.4\textwidth]{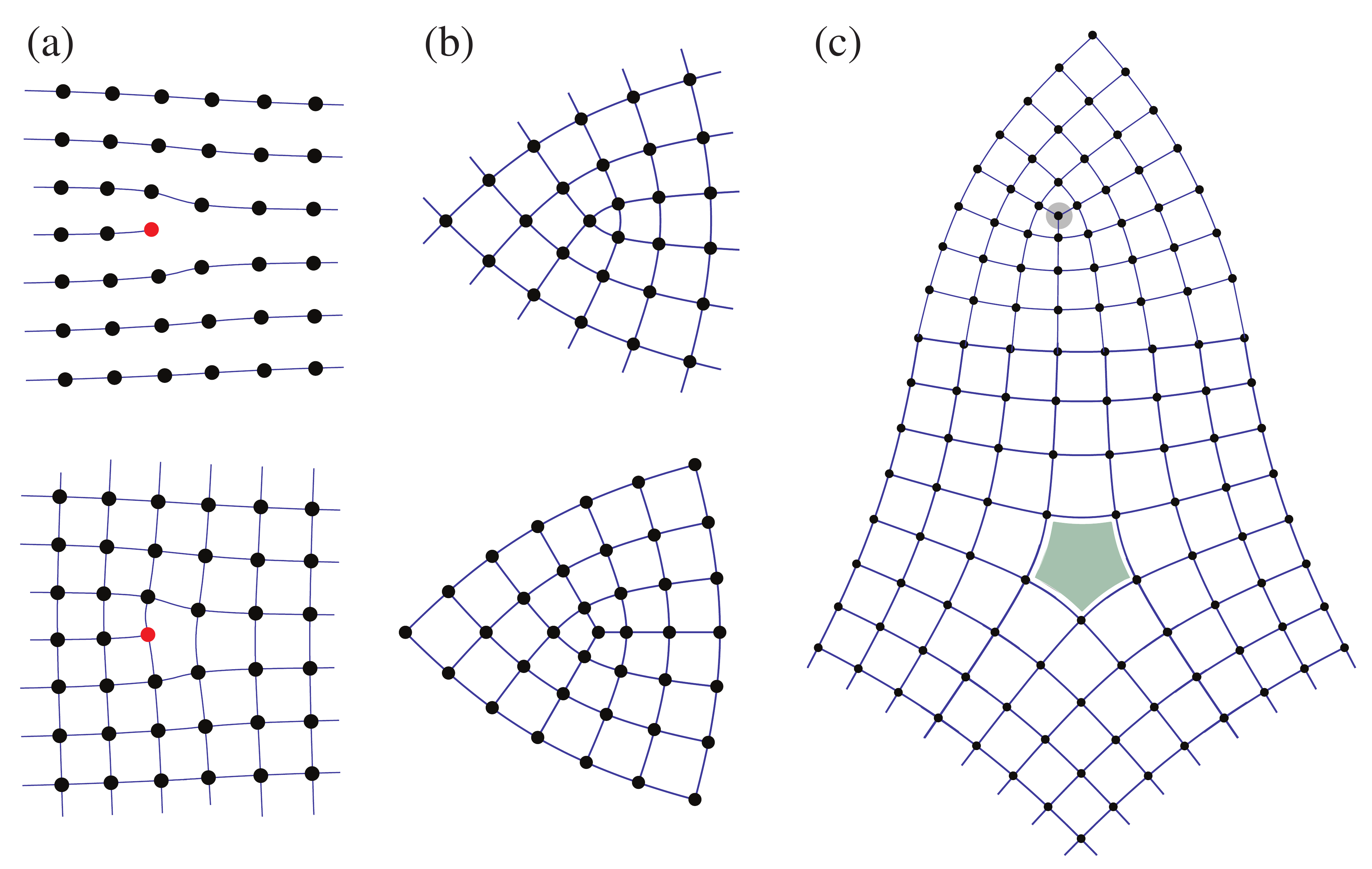}
\caption{(a) Dislocation on a square lattice. (b) Two inequivalent $\Omega=-\pi/2$ disclinations. (c) A $\Omega=\pm\pi/2$ disclination dipole.}\label{fig:dislocationdisclination}
\end{figure}

Zero-energy MBSs can also exist in non-chiral media. 
We have already seen that they appear as boundary modes in a topological 1d $p$-wave SC [see \eqref{defectSSH}]. 
This can be generalized to 2d by stacking the 1d chains into a 2d array. 
A lattice dislocation (see Fig.~\ref{fig:dislocationdisclination}) binds a zero-energy MBS 
if the 1d chains are aligned horizontally, so that the MBS is located at the end of a half-line~\cite{Teo_disclination_class, Teo_2013_disclination, HughesYaoQi13, AsahiNagaosa12, Juricic12}.
In general, a non-chiral $p$-wave SC in 2d can carry a {\em weak} $\mathbb{Z}_2$ topology. 
This is described by   weak indices, which originate
from the lower dimensional cycles of the 2d BZ, $\mathrm{BZ}^2=S^1\times S^1$. 
The weak indices characterize a homogeneous 2d SC that is topologically equivalent to an anisotropic array of $p$-wave chains. 
They can be written in the form of a $\mathbb{Z}_2$-valued reciprocal lattice vector 
\begin{align}
\mathbf{G}_{\nu }&=
\nu _1\mathbf{b}_1+\nu _2\mathbf{b}_2
\quad\mbox{with}
\quad 
\nu _i
=
\frac{i}{\pi }\oint_{\mathcal{C}_i}\text{Tr}(\mathcal{A}) \bmod 2, 
\label{eq:1_weak_invariant_2}
\end{align} 
where 
$\mathcal{C}_i=\{\pi\mathbf{b}_i+ s\epsilon_{ij}\mathbf{b}_j|s \in 
(-\pi,\pi]
\}$ 
is the cycle on the boundary of the BZ along the primitive reciprocal lattice direction $\mathbf{b}_i$. 
On a boundary normal to $\mathbf{G}_\nu$, 
the weak TSC carries a protected non-chiral Majorana edge mode, 
where the zero energy left and right moving modes are located at different PH symmetric momenta $0$ and $\pi$ 
so that back-scattering is prohibited by PH and translation symmetry.
By use of \eqref{Z2indexthm1} together with \eqref{eq:1_weak_invariant_2}, one finds the following bulk-defect correspondence, 
%
%
\begin{align} \label{examp_bulk_defect_correspond_p_wave}
\mbox{ind}^{(1)}_{\mathbb{Z}_2}=\frac{1}{2\pi}\mathbf{B}\cdot\mathbf{G}_\nu\quad\mbox{mod 2},
\end{align} 
where $\mathbf{B}$ is the Burgers vector -- the Bravais lattice vector associates to the net translation picked up by a particle going once around the dislocation~\cite{ChaikinLubensky, Nelsonbook}. The product in \eqref{examp_bulk_defect_correspond_p_wave} counts the parity of the number of zero energy MBSs located at a dislocation in a 2d {\em weak} TSC. It does not rely on a chiral $p_x+ip_y$ pairing order or a non-vanishing Chern invariant.

Discrete rotation symmetries of a crystalline SC provide further lattice symmetry protected topologies
\cite{Teo_disclination_class, Teo_2013_disclination}, see also Sec.~\ref{Topological crystalline materials}.
These {\em topological crystalline superconductors} (TCSs) possess BdG states $|u^a(\mathsf{K})\rangle$ 
that behave differently under rotation ${R}$ at different rotation fixed points $\mathsf{K}$. 
For example, the fourfold symmetric BdG model \eqref{ReadGreenmodel} 
has at the two fourfold fixed momenta $(0,0)$ and $(\pi, \pi)$
inverted occupied states, i.e.,  $|u(0,0)\rangle=\mathbf{e}_2$ and $|u(\pi,\pi)\rangle=\mathbf{e}_1$.
These two eigenstates   have distinct rotation eigenvalues ${R}=e^{i(\pi/4)\tau_3}$, since  
  $\tau_3=\pm1$ for these two BdG states.
The lattice symmetry protected bulk topologies can lead to zero-energy MBSs located at disclinations, i.e., at conical point defects. 
These disclinations correspond to singularities of the curvature that rotate the frame of an orbiting particle 
by a Frank angle $\Omega$ after one cycle. 
Examples on a square lattice are illustrated in Figs.\ \ref{fig:dislocationdisclination}(b) and \ref{fig:dislocationdisclination}(c). 
The $\mathbb{Z}_2$-index that counts the parity of the zero-energy MBSs at a disclination takes 
the form of \cite{Teo_disclination_class, Teo_2013_disclination} 
\begin{align}
\mbox{ind}^{(1)}_{\mathbb{Z}_2}=\frac{1}{2\pi}\mathbf{T}\cdot\mathbf{G}_\nu+\frac{\Omega}{2\pi}
\left(\mbox{Ch}_1+\begin{array}{*{20}l}\mbox{rotation}\\\mbox{invariant}\end{array}\right)\quad\mbox{mod 2} ,
\end{align} 
where $\mathbf{T}$ is a translation piece of the disclination similar to the Burgers vector of a dislocation. The specific form of the rotation invariant depends on the rotation symmetry and is always a combination involving the rotation eigenvalues of BdG states. Disclination MBSs are proposed to be present in the form of corner states in Sr$_2$RuO$_4$ and at grain-boundaries in superconducting graphene and silicene.

\paragraph{Example: Superconducting heterostructures (class D)}
We have seen that MBSs appear in the form of vortex states in 
chiral $(p+ip)$-SCs and as lattice defects in non-chiral $p$-wave SCs. 
Here we review 2d and 3d heterostructures that involve
$s$-waves SCs,
but still support robust zero energy MBSs~\cite{hasan:rmp, qi:rmp, FuKane_SC_STI, FuKaneJosephsoncurrent09, Teo:2010fk,Chiu:2011fk,Chiu_SC_TI_extended,Hung_TI_SC,Hosur:2011uq,SC_Proximity_Jia,Xu_TI_SC}. 

\begin{figure}[t]
\centering\includegraphics[width=0.4\textwidth]{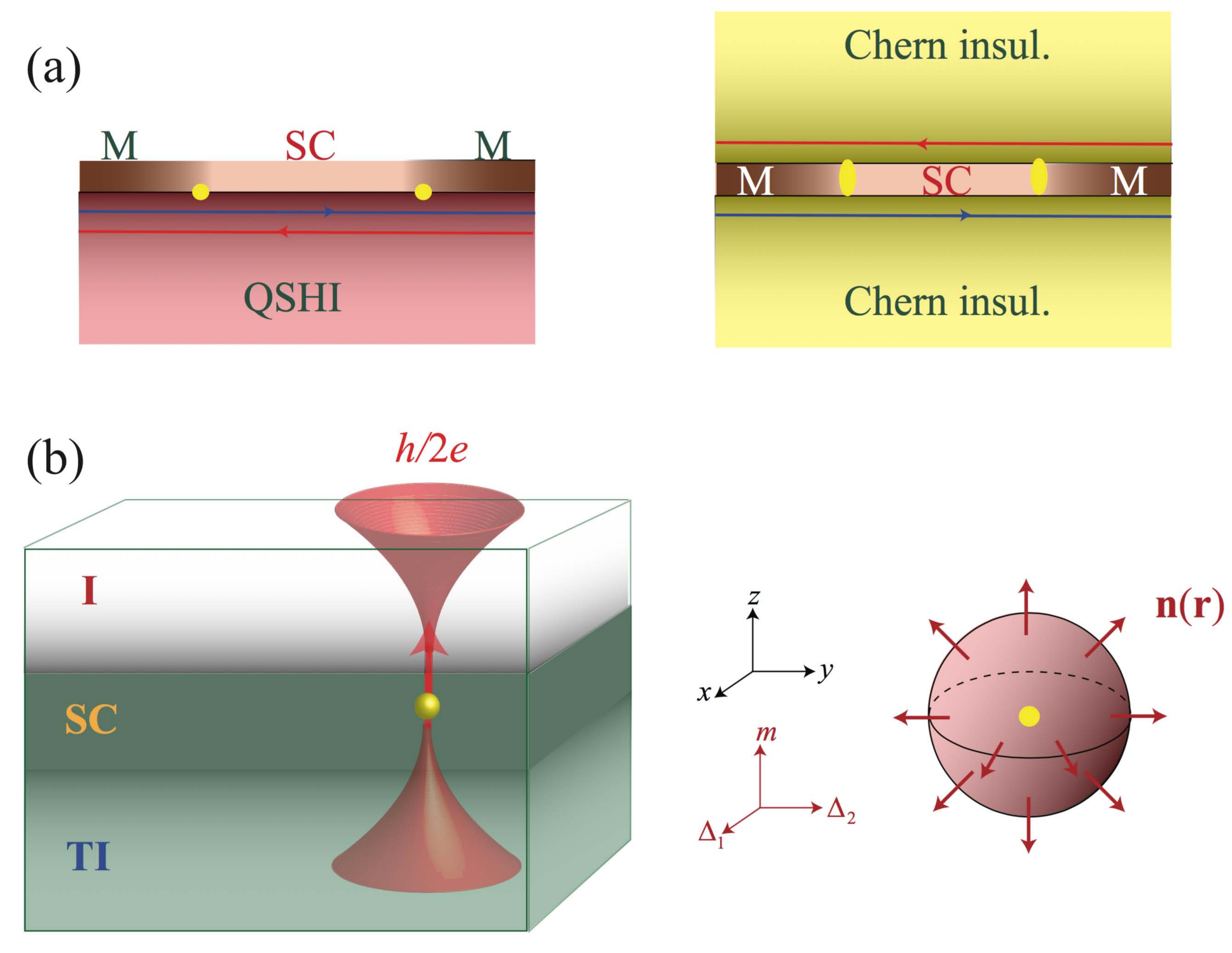}
\caption{Zero-energy MBSs (yellow dots) in heterostructures: (a) superconductor (SC) - magnet (M) domain wall along a QSH edge or a Chern insulator interface; (b) a flux vortex across a superconducting interface between a 3d topological (TI) and trivial insulator (I).}
\label{fig:TISCFM}
\end{figure}

(i) 
We first look at the gapless helical edge modes of a 2d quantum spin Hall (QSH) insulator
consisting of a pair of counter-propagating electronic states
(\ref{Gapless modes along line defects and index theorems}), 
which couple
to a TR breaking back-scattering potential $h$ and a $U(1)$ symmetry breaking SC pairing $\Delta$
[Fig.\ \ref{fig:TISCFM} (a)].
This setup can be described by the boundary BdG Hamiltonian 
\begin{align}
H_{1d}(k,r)=v_Fk\sigma_3\tau_3+h(r)\sigma_1+\Delta(r)\tau_1,
\label{QSHSCM}
\end{align} 
where $\sigma_i$ and $\tau_i$ act on spin and Nambu degrees of freedom, respectively. 
Here, the Nambu basis is chosen to be 
$(\hat{c}_\uparrow, \hat{c}_\downarrow, \hat{c}_\downarrow^\dagger,-\hat{c}_\uparrow^\dagger)$, 
so that the PH operator is $C=\sigma_2\tau_2\mathcal{K}$
and the TR operator is $T=i\sigma_2\mathcal{K}$. 
The TR-breaking mass gap $h$ can be generated by magnetic impurities, by a Zeeman field, or by proximity with a ferromagnet~(M).
The SC pairing $\Delta$, on the other hand,  can be induced by proximity coupling with an $s$-wave SC. 
The two terms $\Delta$ and $h$ commute and correspond to competing orders. 
An SC-M domain wall, where $|h(r)|-|\Delta(r)|$ changes its sign, 
traps a zero energy MBS. 
This can be seen by decomposing \eqref{QSHSCM} into $H=H^+\oplus H^-$ by the good quantum number $\sigma_1\tau_1=\pm1$, 
where \begin{align}
H_{1d}^\pm(k,r)=v_Fk\tilde\sigma_3+[h(r)\pm\Delta(r)]\tilde\sigma_2,
\end{align} 
where $\tilde\sigma$ acts on the two-dimensional subspaces. 
Assuming both $h(r)$ and $\Delta(r)$ are non-negative throughout the edge, 
$H^{+}(k,r)$ always has a gap while the mass term for $H^-(k,r)$ changes its sign across the domain wall. 
$H^-$ is exactly the Jackiw-Rebbi model \eqref{JackiwRebbi} and therefore traps a zero mode between the SC and M regions.

(ii) 
Helical modes also occur in an  interface between an $s$-wave SC and two adjacent Chern insulators that have the same unit Chern number 
[Fig.\ \ref{fig:TISCFM} (a)]. 
For example, consider the spinful band Hamiltonian \eqref{ReadGreenmodel} 
on a square lattice, where $\tau$ now acts on the spin degree of freedom. 
It supports a spin polarized chiral edge mode and has opposite polarizations on opposite edges. 
A protected MBS therefore is located at the SC-M domain wall of a weakly coupled Chern insulator interface. 
More exotic parafermionic defects are proposed in SC-M heterostructures in fractional TIs 
\cite{LindnerBergRefaelStern,ClarkeAliceaKirill,MChen2012,Vaezi2013}.

(iii) The same idea applies to semiconducting nanowires with strong SOC  in a magnetic field.
This ballistic 1d system can be modeled in the continuum limit by the following spinful Hamiltonian 
\begin{align}
H_0(k)=\frac{\hbar^2k^2}{2m}\openone_2+u_{\mbox{\tiny SO}}k\sigma_3+b\sigma_1. 
\label{SO coupled 1d wire}
\end{align} 
Hamiltonian~\eqref{SO coupled 1d wire} has an energy spectrum,  
\begin{figure}
\begin{align}
\vcenter{\hbox{\includegraphics[width=0.25\textwidth]{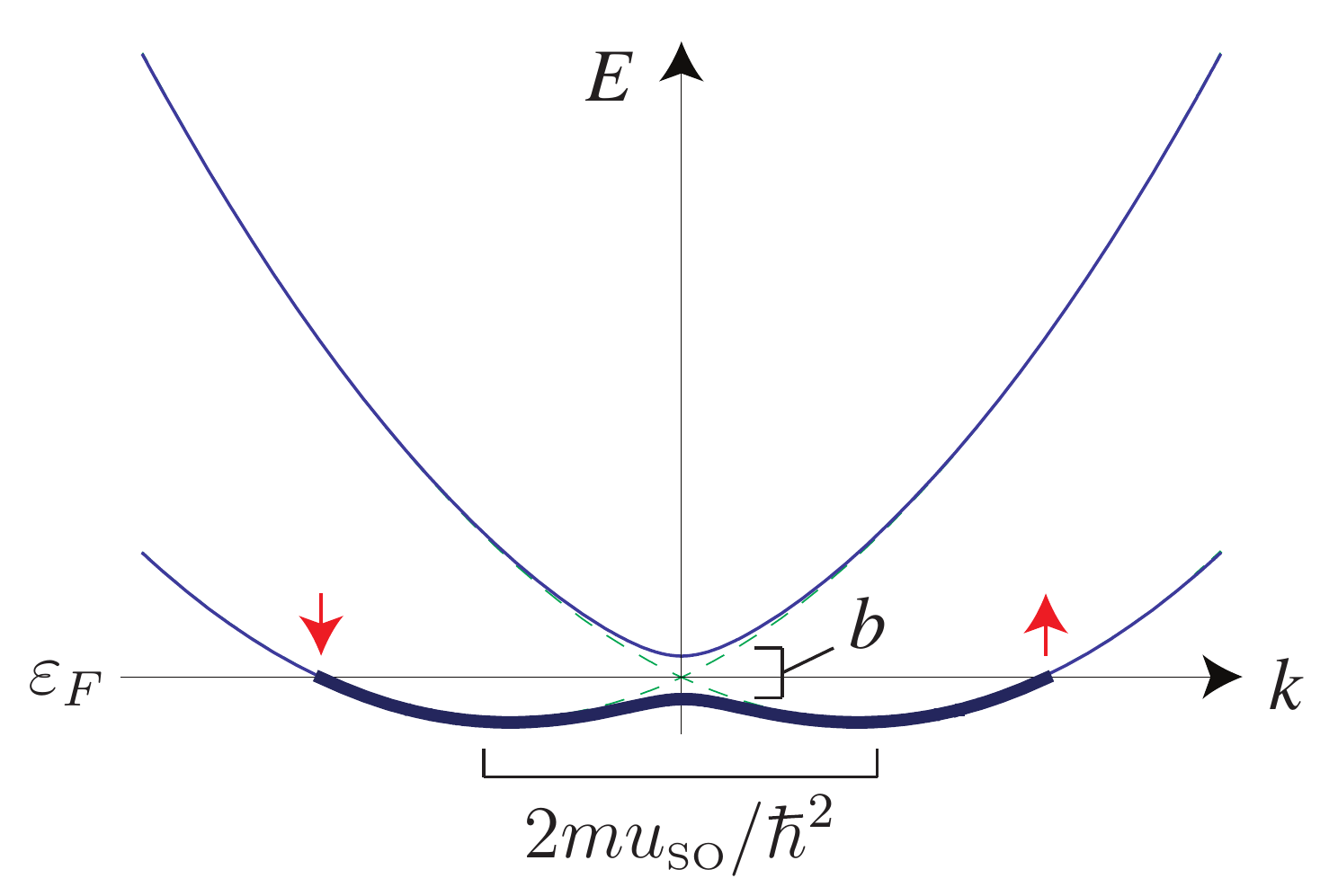}}}\nonumber
\end{align}
\caption{ \label{fig_SO_coupled_1d_wire}
Energy spectrum of the spin-orbit coupled nanowire~\eqref{SO coupled 1d wire}
in a magnetic field. 
}
\end{figure}
which consists of a spin-filtered pair of counter-propagating modes, provided that the fermi energy lies within the direct magnetic gap
(Fig.~\ref{fig_SO_coupled_1d_wire}). 
These helical modes can be gapped out by a superconducting pairing, which is proximity induced by  a bulk $s$-wave SC. 
This SC nanowire then behaves like a 1d Kitaev   $p$-wave SC and 
hosts   protected boundary MBSs~\cite{Roman_SC_semi,Sau_semiconductor_heterostructures,Gil_Majorana_wire,Alicea_Majorana_wire}. 
InSb nanowires with a low impurity density 
which are proximity-coupled to an ordinary $s$-wave SC provide an experimental realization of this 1d $p$-wave SC.
Recently, numerous transport measurements on these systems have observed zero-bias conductance peaks, which were
interpreted as  evidence of  the boundary MBSs~\cite{Mourik_zero_bias, Deng_zero_bias, Yuval_zero_bias, RokhinsonLiuFurdyna12}.

(iv) 
Going back to the SC-M domain wall along a QSH edge, 
the point defect can be equivalently described in 2d. 
The defect Hamiltonian that incorporates the 2d bulk takes 
in the continuum limit
the 8-band form of 
\begin{align}
H_{2d}(\mathsf{k},\mathsf{r})&=\left[t(k_x\sigma_1+k_y\sigma_2)\mu_1+m(\mathsf{r})\mu_3\right]\tau_3
\nonumber\\
&\;\;\;+h(\mathsf{r})\mu_2+\Delta(\mathsf{r})\tau_1 .
\label{QSHSCM2d}
\end{align} 
Here, the first line describes the transition between the QSHI and the trivial vacuum as 
the mass gap $m(\mathsf{r})$ changes its sign along the $y$-axis 
in Fig.\ \ref{fig:TISCFM}(a). 
The Pauli matrices $\sigma$ and $\mu$ act on spin and orbital degrees of freedom. 
(Notice that the $k^2$ regularization $m(\mathsf{r})\to m(\mathsf{r})-\varepsilon k^2$ is not necessary for a defect Hamiltonian, 
just like in the Jackiw-Rebbi model \eqref{JackiwRebbi}.) 
The TR breaking $h(r)\mu_2$ term 
here is actually antiferromagnetic as it also breaks the inversion  $P=\mu_3$. 
It, however, can be replaced by an ordinary ferromagnetic one, like $h(\mathsf{r})\sigma_2$. 
The magnetic and pairing orders appear only near the interface -- the $x$-axis in Fig.~\ref{fig:TISCFM}(a) -- 
where $|h(\mathsf{r})|-|\Delta(\mathsf{r})|$ changes its sign
across each domain wall point defect.
Similar to the boundary Hamiltonian \eqref{QSHSCM}, 
the 2d point defect Hamiltonian can be decomposed into $H=H^+\oplus H^-$ according to the good quantum number $\mu_2\tau_1$. 
Let us assume that $h$ and $\Delta$ are both non-negative. 
Then $H^+$ is always gapped and the defect is captured by 
\begin{align}
H_{2d}^-(\mathsf{k},\mathsf{r})&=t(k_x\sigma_1+k_y\sigma_2)\tilde\tau_3+m(\mathsf{r})\tilde\tau_1+n(\mathsf{r})\tilde\tau_2 ,
\end{align} 
where $n(\mathsf{r})=h(\mathsf{r})-\Delta(\mathsf{r})$, and $\tilde\tau$ acts 
on the  2d subspace where $\mu_2\tau_1=-1$.
This Hamiltonian is identical to the 2d Jackiw-Rossi model -- c.f.~Eq.\ \eqref{domainwallH}~\cite{JackiwRossi81, Teo:2010fk} -- 
where the mass terms in $m\Gamma_0(\mathsf{r})=m(\mathsf{r})\tilde\tau_1+n(\mathsf{r})\tilde\tau_2$ 
can be organized as a vector field $\mathbf{v}(\mathsf{r})=\left(m(\mathsf{r}),n(\mathsf{r})\right)$ that winds once around the defect. 
This winding mass term represents the non-trivial element in $\pi_1(BO)=\mathbb{Z}_2$,
which classifies class D point defects in 2d (Sec.\ \ref{sec:defecthomotopy}). 
As a consequence of the unit winding, the non-trivial topological index $\mbox{CS}_3$ in \eqref{defectCS}  guarantees a protected zero energy MBS.

(v) 
The idea generalizes even to 3d~\cite{Teo_Majorana}
[Fig.\ \ref{fig:TISCFM} (b)].
An SC interface between a bulk TI and a trivial insulator (I) in 3d 
can be described by the 8-band BdG Hamiltonian 
\begin{align}
H_{3d}(\mathsf{k},\mathsf{r})&=t(k_x\sigma_1+k_y\sigma_2+k_z\sigma_3)\mu_1\tau_3\nonumber\\
&\;\;\;+m(\mathsf{r})\mu_3\tau_3+\Delta_x(\mathsf{r})\tau_1+\Delta_y(\mathsf{r})\tau_2 ,
\label{TISCM3d}\end{align} 
where the TRS mass gap $m(\mathsf{r})$ changes its sign across the TI-I interface, 
and $\Delta=\Delta_x+i\Delta_y$ is the SC pairing. 
The model is of the same form as  Dirac Hamiltonian \eqref{domainwallH} with spatially 
modulated mass term $m\Gamma_0(\mathsf{r})=m(\mathsf{r})\mu_3\tau_3+\Delta_x(\mathsf{r})\tau_1+\Delta_y(\mathsf{r})\tau_2$. 
A quantum flux vortex brings a unit winding to the pairing phase $\Delta=|\Delta|e^{i\varphi}$. 
The coefficients in the Dirac mass term can be grouped together into a 3d vector field 
\begin{align}
\mathbf{n}(\mathsf{r})=\left(\Delta_x(\mathsf{r}),\Delta_y(\mathsf{r}),m(\mathsf{r})\right), 
\end{align} 
which looks like a ``hedgehog" around the vortex core [Fig.\ \ref{fig:TISCFM} (b)]. 
As the vector field is non-singular except at the point defect, 
the hedgehog configuration corresponds to a continuous map 
$\hat{\mathbf{n}}:S^2\to S^2$ over a 2-sphere spatially enclosing the defect. 
This map has a unit winding 
\begin{align}
\nu=\frac{1}{4\pi}\int_{S^2}
\hat{\mathbf{n}}\cdot d\hat{\mathbf{n}}\times d\hat{\mathbf{n}}=
\pm1
\end{align} 
and represents the generator in the homotopy group 
$\pi_2(S^2)=[S^2,S^2]=\mathbb{Z}$. 
It also represents the non-trivial element in 
$\pi_2(U(N)/O(N))=\mathbb{Z}_2$ -- for instance $\hat{\mathbf{n}}$ 
wraps the 2-cycle in $U(2)/O(2)\sim U(1)\times S^2$ -- that classifies class D point defects in 3d. 
The winding number $\nu$ translates into a non-trivial topological index $\mbox{CS}_5$ in 
\eqref{defectCS} and guarantees a protected zero energy MBS at the vortex core. 
The 1d, 2d, and 3d point defect models \eqref{QSHSCM}, \eqref{QSHSCM2d}, and \eqref{TISCM3d} are unified 
by the $K(s;d,D)$ classification~\cite{Teo:2010fk} 
\begin{align}
K(2;1,0)\cong K(2;2,1)\cong K(2;3,2)\cong\mathbb{Z}_2,
\end{align} 
where $s=2$ for class D.

\subsubsection{Gapless modes along line defects and index theorems}
\label{Gapless modes along line defects and index theorems}

\begin{figure}[t]
\centering\includegraphics[width=0.4\textwidth]{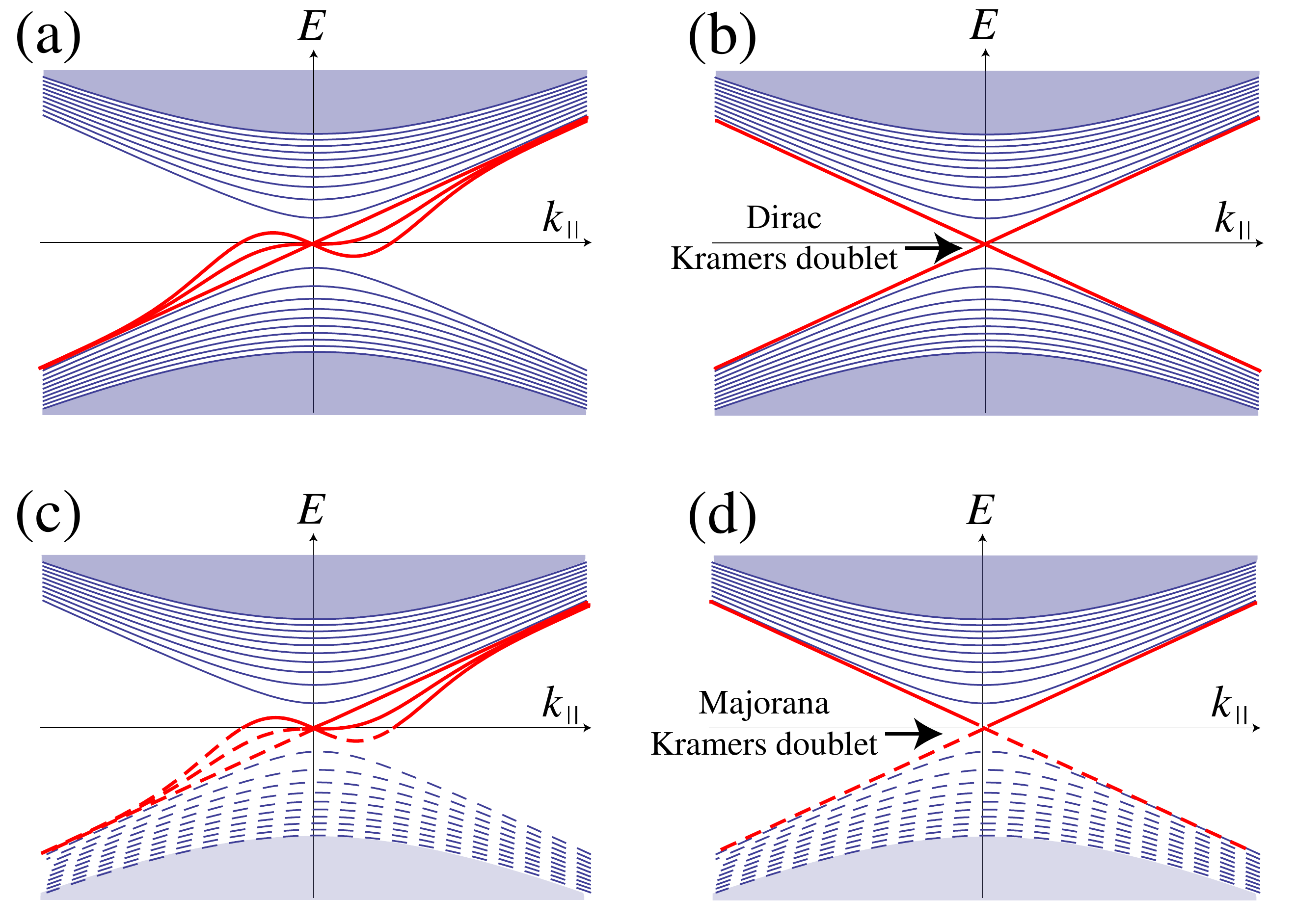}
\caption{Gapless spectra inside the bulk gap~\cite{Qi_hughes_zhang_09}: 
(a) chiral Dirac modes, 
(b)  helical Dirac mode, 
(c) chiral Majorana modes, 
and (d) helical Majorana mode.}
\label{fig:linemodes}
\end{figure}

In this section, we discuss  protected gapless modes that propagate
along topological line defects ($\delta=d-D=2$).
Relevant symmetry classes and types of gapless modes are summarized in 
Table \ref{tab:linedefect} and  Fig.\ \ref{fig:linemodes}. 
By discussing their transport properties, we introduce proper indices counting the 
degrees of freedom of the propagating gapless modes, 
which, by the bulk-boundary correspondence, will be identified with the topological invariants.

\paragraph{Edge transports}

Here, we demonstrate the appearance of protected 1d modes along topological line defects 
in terms of 1d edges of 2d bulk topological systems. 
Topological line defects in higher dimensions and their topological origin will be discussed later.

The most well-known example are the chiral edge modes [Fig.\ \ref{fig:linemodes}(a)]
along the boundary of an integer QH fluid~\cite{Laughlin_IQHE, Halperin82, Hatsugai93, Schulz00, Volovik92}. 
A chiral mode is an electronic channel that propagates in a single direction. 
For example the (spin polarized) lowest Landau level in 2d 
-- despite having a bulk cyclotron gap --
carries one conducting gapless chiral edge mode. 
At zero temperature, each chiral channel carries an electric current 
\begin{align}
I^1_e=\int_{k_{\mbox{\tiny cutoff}}}^{k_F}\frac{dk}{2\pi}ev(k)=\frac{e}{h}(\varepsilon_F-\varepsilon_{\mbox{\tiny cutoff}}) ,
\label{electriccurrent}
\end{align} 
where $e$ is the electric charge,
$\varepsilon_F$ is the fermi energy,
and 
$v(k) = \partial{\varepsilon(k)}/{\partial k}$ is the velocity. 
In a more general scenario, 
the 1d boundary may carry multiple chiral channels. 
Dropping the fermi energy independent cutoff term, 
the net electric current takes the form of 
\begin{align}
I_e\approx c_-\frac{e}{h}\varepsilon_F, \label{electriccureent}
\end{align} 
where the integer coefficient is a $\mathbb{Z}$-{\em analytic index} 
that counts the spectral flow~\cite{Volovik:book, Nakahara:2003ve} 
\begin{align}
c_-=
\left(
\begin{array}{*{20}l}\mbox{\scriptsize number of forward}\\\mbox{\scriptsize propagating Dirac modes}
\end{array}
-\begin{array}{*{20}l}\mbox{\scriptsize number of backward}\\\mbox{\scriptsize propagating Dirac modes}
\end{array}
\right).
\label{c-1}
\end{align}
The integer QHE~\cite{Klitzing} is generated by a transverse bias across the top and bottom edge of a Hall bar. 
This gives a potential difference, 
$edV_y=d\varepsilon_F=\varepsilon_F^{\mbox{\tiny top}}-\varepsilon_F^{\mbox{\tiny bottom}}$, 
between the two edges and drives a horizontal current $dI_x=I_e^{\mbox{\tiny top}}-I_e^{\mbox{\tiny bottom}}=\sigma_{xy}dV_y$, where \begin{align}\sigma_{xy}=c_-\frac{e^2}{h}\label{electricHall}\end{align} is the quantized Hall conductance.

\begin{table}[tbp]
\centering
\begin{ruledtabular}
\begin{tabular}{ccl}
Symmetry  & Topological classes & 1d gapless fermion modes\\
\hline
A & $\mathbb{Z}$ &  Chiral Dirac \\
D & $\mathbb{Z}$ &  Chiral Majorana \\
DIII & $\mathbb{Z}_2$ &  Helical Majorana \\
AII & $\mathbb{Z}_2$ &  Helical Dirac  \\
C & $2\mathbb{Z}$ &   Chiral Dirac
\end{tabular}
\end{ruledtabular}
\caption{Symmetry classes that support topologically non-trivial line defects and their associated
protected gapless modes. }
\label{tab:linedefect}
\end{table}

At small temperature $T$, 
each chiral Dirac mode also carries an energy (thermal) current~\cite{KaneFisher97, Cappelli01, Kitaev2006, Luttinger64} 
\begin{align}
I_T^1
&=\int\frac{dk}{2\pi}f(\varepsilon(k)-\mu)v(k) \varepsilon(k)
\nonumber \\
& \approx I_0+\frac{\pi^2k_B^2}{6h}T^2+O\left(T^4\right) ,
\end{align} 
where $f(\varepsilon)=(e^{\varepsilon/(k_B T)}+1)^{-1}$ is the fermi-Dirac distribution and $k_B$ is the Boltzmann constant. 
Dropping the $T$-independent contribution $I_0$, 
a general boundary with multiple chiral modes  carries the net anomalous energy current 
\begin{align}I_T\approx c_-\frac{\pi^2k_B^2}{6h}T^2.
\label{thermalcurrent}\end{align} 
The QH fluid has a thermal Hall response so that a transverse temperature difference 
$dT=T^{\mbox{\tiny top}}-T^{\mbox{\tiny bottom}}$ 
across the Hall bar drives an energy current 
$dI_T=I_T^{\mbox{\tiny top}}-I_T^{\mbox{\tiny bottom}}=\kappa_{xy}dT$, 
where 
\begin{align}
\kappa_{xy}=c_-\frac{\pi^2k_B^2}{3h}T
\label{thermalHall}
\end{align} 
is the thermal Hall conductivity, 
which can be related to the gravitational anomaly~\cite{Stone12, NomuraRyuFurusakiNagaosa12, WangQiZhang2011, RyuMooreLudwig2012, AlvarezGaumeWitten1983, Volovik1990}.
Thermal response applies to systems
that lack $U(1)$ charge conservation, like SCs. 
A chiral SC hosts chiral Majorana edge modes. 
These neutral modes do not carry electric currents, 
but they do carry energy current \eqref{thermalcurrent}. 
A chiral SC in general has no quantized electric Hall response, but exhibits a thermal Hall response \eqref{thermalHall}. 
Since a Dirac mode is decomposed into two Majorana ones as its real and imaginary components, 
$\psi=(\gamma_1+i\gamma_2)/2$, the $\mathbb{Z}$-analytical index $c_-$ in \eqref{c-1} 
translates into 
\begin{align}
c_-=\frac{1}{2}
\left(\begin{array}{*{20}l}\mbox{\scriptsize number of forward}\\\mbox{\scriptsize Majorana modes}
\end{array}-\begin{array}{*{20}l}\mbox{\scriptsize number of backward}\\\mbox{\scriptsize Majorana modes}
\end{array}\right) , 
\label{c-2}
\end{align}
 so that $c_-$ now can take half-integral values. 
For instance, $c_-=\pm1/2$ for a chiral spinless $p_x+ip_y$ SC. 
This quantity extends to many-body systems supporting fractionalization (e.g., fractional QH systems),   
where the ($1+1$)d gapless boundary can be effectively described by a conformal field theory (CFT)~\cite{FMS-CFT}. 
It corresponds to the {\em chiral central charge} $c_-=c_R-c_L$, 
the difference of the central charges between forward and backward propagating channels of the edge CFT.

Chiral modes necessarily break TRS, 
as they are not symmetric under $k\leftrightarrow-k$. 
But in the presence of TRS with $T^2=-1$ another type of robust gapless edge modes can exist:
Helical modes [Figs.\ \ref{fig:linemodes}(b) and  \ref{fig:linemodes}(d)]
are non-chiral, as they have both forward and backward channels. 
Backscattering is however forbidden by TRS, since the crossing is protected by Kramers theorem. 
Unlike chiral modes, helical modes are $\mathbb{Z}_2$-classified since a TR symmetric backscattering term can remove a pair of them. 
The $\mathbb{Z}_2$-analytical index thus counts the (non-chiral) central charge $c=c_R=c_L$ 
\begin{align}
c&=(\mbox{number of Dirac helical modes})\quad\mbox{mod 2},
\end{align} 
for $U(1)$ preserving systems, 
or 
\begin{align}
c&=
\frac{1}{2}
(\mbox{number of Majorana helical modes})
\quad \mbox{mod 1} ,
\end{align} 
for $U(1)$ breaking SCs. 
These TRS protected modes appear 
on the boundaries of 2d TIs in class AII 
(such as a QSH insulator with $c=1$) 
and TSCs in class DIII 
(such as a $(p+ip)\uparrow\times(p-ip)\downarrow$ SC with $c=1/2$). 

Along an unequilibrated edge, 
the pair of counter-propagating channels of a helical mode can have different temperatures,
or different chemical potentials, if they are of Dirac type. 
This difference can be generated by connecting two charge or heat reservoirs to the 1d boundary. 
As each Dirac chiral channel carries an electric current \eqref{electriccurrent}, 
a potential difference $edV=\varepsilon_F^R-\varepsilon_F^L$ between the forward and backward components of a Dirac helical mode drives 
a net electric current $dI_e=\sigma_{xx}dV$, where
\begin{align}\sigma_{xx}=c\frac{e^2}{h}\quad\mbox{mod }\frac{2e^2}{h}\label{electriccond}
\end{align} is the longitudinal electric conductance along a single edge. In reality the two charge reservoirs must be connected by a pair of edges -- the top and bottom boundaries of a 2d bulk -- so that the measured conductance is twice that of \eqref{electriccond}. A conductance close to $2e^2/h$ is experimentally seen across the QSHI of HgTe/CdTe quantum wells~\cite{MarkusKonig11022007}. 
On the other hand, a helical Majorana edge mode -- in a SC where $U(1)$ symmetry is broken -- responds to a thermal difference $dT=T^R-T^L$ between the counter-propagating components and gives the net energy current $dI_T=\kappa_{xx}dT$, where \begin{align}\kappa_{xx}=c\frac{\pi^2k_B^2}{3h}T\quad\mbox{mod }\frac{\pi^2k_B^2}{3h}T\label{thermalcond}\end{align} is the longitudinal thermal conductance along a single edge. Again, the measured conductance must be contributed by two edges and is double of that in \eqref{thermalcond}.

\paragraph{Index theorems}

We have seen that gapless modes along 1d boundaries of a 2d topological bulk can give rise to anomalous transport signatures. Similar signatures  also arise when a line defect in higher dimensions carries these gapless modes. Just like the bulk-boundary correspondence that relates the bulk topology to edge excitations, the gapless excitations along a line defect is guaranteed by the topology of the defect.

 The net chirality \eqref{c-1} of gapless Dirac modes along a line defect in $d$ dimensions is determined by the Chern invariant 
\begin{align}
c_-=\mbox{Ch}_{d-1}[H(\mathsf{k},\mathsf{r})]\in\mathbb{Z}\label{c-index1} ,
\end{align} 
where $\mbox{Ch}_{d-1}$ is defined in  \eqref{defectChern} and the defect Hamiltonian $H(\mathsf{k},\mathsf{r})$ 
 describes the long length scale spatial variation of the insulating band Hamiltonian around the defect. 
For instance in 2d, the chirality of a QH fluid or Chern insulator is given by the 1st Chern number. 
If the system is superconducting, the line defect  carries Majorana instead of Dirac modes. 
The net chirality is then given by
\begin{align}
c_-=\frac{1}{2}\mbox{Ch}_{d-1}[H_{\mbox{\tiny BdG}}(\mathsf{k},\mathsf{r})]\in\frac{1}{2}\mathbb{Z} ,
\label{c-index2}
\end{align} 
where $H(\mathsf{k},\mathsf{r})$ is now the BdG defect Hamiltonian. 
For example, the edge chirality of a $p+ip$ SC is half of the bulk 1st Chern invariant. 
Equations \eqref{c-index1} and \eqref{c-index2} are consistent with each other, since
the band Hamiltonian of an insulator is artificially doubled in the BdG description which has twice the original Chern number.

The defect classification (Table~\ref{tab:defectclassification}) allows non-trivial chirality only for the TR breaking symmetry classes A, D, and C. 
The PH operator for class C squares to $-1$,  $C^2=-1$. 
The Kramers theorem applies to zero energy modes at the symmetric momenta $k_\|=0,\pi$ and requires chiral modes to come in pairs. 
This agrees with the $2\mathbb{Z}$ defect classification.

The number parity of gapless helical modes along a TRS line defect in $d$ dimensions equates to a Fu-Kane invariant 
\begin{align}
c=\mbox{FK}_{d-1}[H(\mathsf{k},\mathsf{r})]\quad\mbox{mod 2} ,
\end{align} 
for Dirac systems in bulk insulators, or 
\begin{align}
c=\frac{1}{2}\mbox{FK}_{d-1}[H(\mathsf{k},\mathsf{r})]\quad\mbox{mod 1} ,
\end{align} 
for Majorana systems in bulk SCs. 
Class AII and DIII are the only TR symmetric classes that support Kramers degenerated helical modes. 
For example, 
the helical Dirac mode along the edge of a 2d TI or QSHI is protected by the original Fu-Kane $\mathbb{Z}_2$ invariant.
The helical Majorana edge mode of a 2d TSC, such as $^3$He-B, has the same topological origin.
Similarly,  the helical 1d 
mode along a dislocation line in a 3d weak TI
falls under the same classification as
the helical edge mode of a 2d QSHI~\cite{Ran_dislocation}.

\paragraph{Line defects in three dimensions}

\begin{figure}[tbp]
\centering\includegraphics[width=0.45\textwidth]{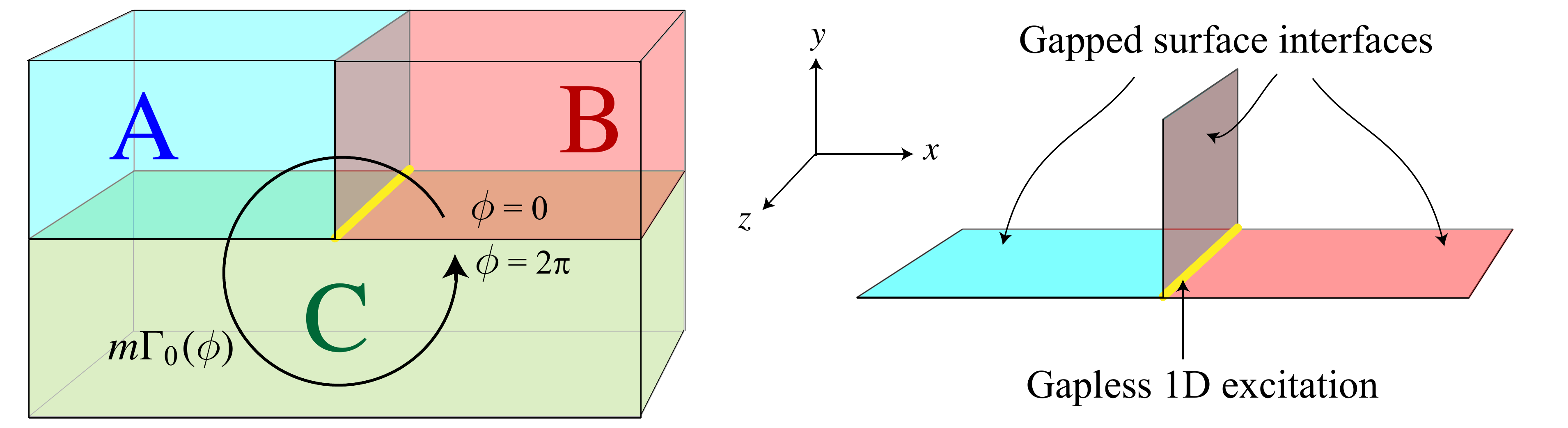}
\caption{Line defect (yellow line) at a heterostructure. A, B, and C are different bulk gapped materials put together so that there is no gapless surface modes along interfaces of any pairs. The mass term $m\Gamma_0(\phi)$ wraps non-trivially around the line interface which results in a gapless 1d excitation localized at the line defect.}\label{fig:heterostructureline}
\end{figure}

We consider various examples of line defects in 3d 
that host topologically protected gapless modes. 
The defect Hamiltonian $H(\mathsf{k},\phi)$ is slowly modulated 
by the spatial angular parameter $\phi\in[0,2\pi]$ that wraps once around the defect line. 
We begin by looking at heterostructures where the line defect is the tri-junction between three different bulk electronic materials 
(Fig.\ \ref{fig:heterostructureline}). 
A finite energy gap is required everywhere away from the tri-junction line. 
This includes the three surface interfaces that separate the three bulk materials. 
For instance, 
when the three bulk materials have non-competing orders, 
the surface interfaces can be smeared out into the 3d bulk where different orders coexist. 
The defect Hamiltonian would take the Dirac form \eqref{domainwallH} 
\begin{align}
H(\mathsf{k},\phi)=\hbar v\mathsf{k}\cdot\boldsymbol\Gamma+m\Gamma_0(\phi),
\end{align} 
where the $m\Gamma_0(\phi)$ incorporates coexisting orders as anticommuting mass terms and winds non-trivially around the defect line.

\begin{figure}[tbp]
\centering\includegraphics[width=0.45\textwidth]{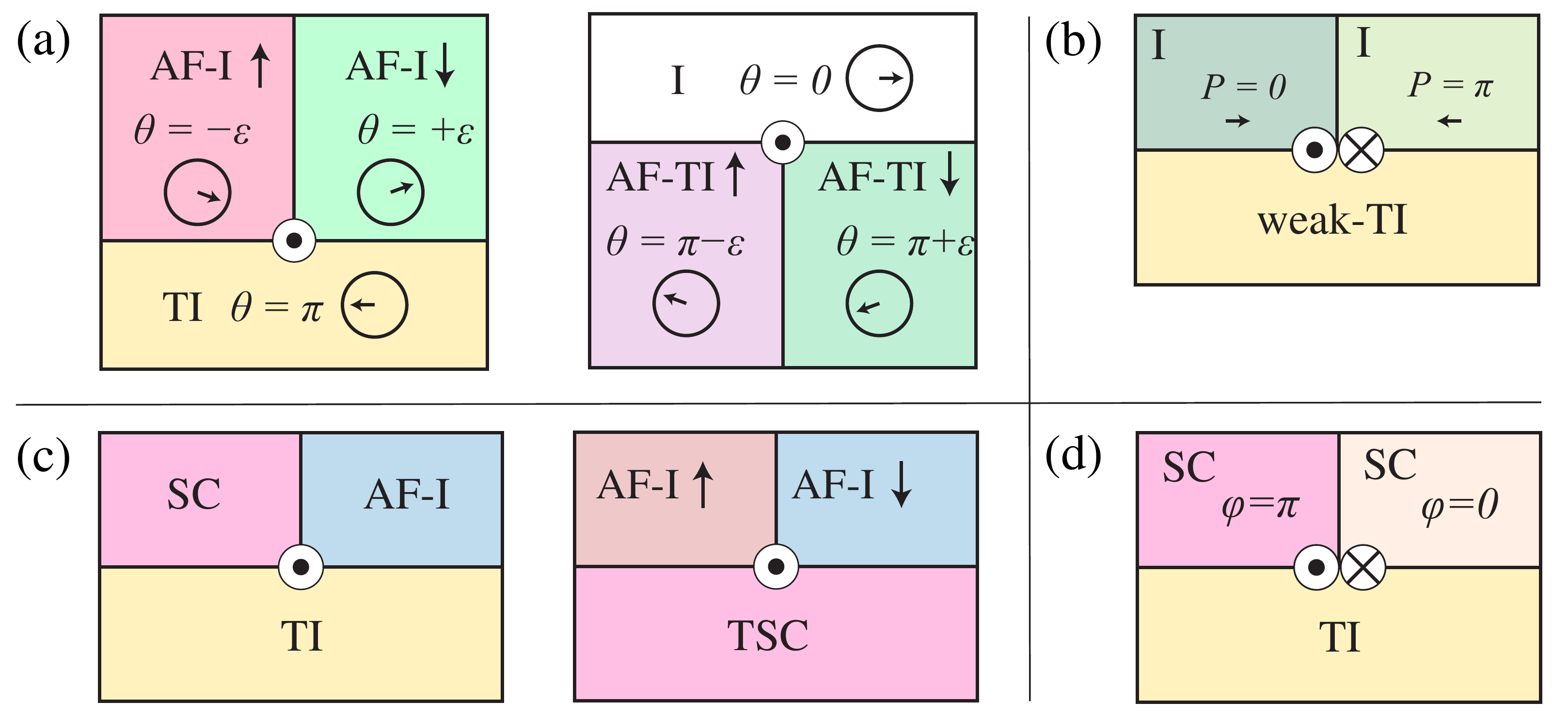}
\caption{
Heterostructure cross-section on $xy$-plane. 
AF = antiferromagnetic, I = trivial insulator, TI = topological insulator, SC = superconductor, TSC = class DIII topological superconductor. 
(a) Chiral Dirac mode $\boldsymbol\odot$ protected by winding of the magnetoelectric $\theta$-angle. 
(b) Helical Dirac mode $\boldsymbol\odot\boldsymbol\otimes$ separating opposite polarization insulating domains. 
(c) Chiral Majorana mode. 
(d) Helical Majorana mode between SC domains with TRS pairing phase $\varphi=0,\pi$.
}
\label{fig:heterostructureline1}
\end{figure}

\paragraph{Example: TI-AF heterostructure (class A)}
We now explicitly demonstrate the chiral Dirac mode bounded by the TI-AF heterostructure shown in Fig.\ \ref{fig:heterostructureline1}(a). 
The surface Dirac cone of a TI can be gapped out by a TRS breaking mass term 
\begin{align}
H_{\mbox{\tiny surface}}(k_x,k_z,x)=\hbar v(k_x\sigma_1+k_z\sigma_2)+m(x)\sigma_3, 
\label{TIAFsurface}
\end{align} 
where the surface is parallel to the $xz$-plane. 
When the mass term changes its sign $m(x\to\pm\infty)=\pm m_0$, 
there is a chiral Dirac mode running along the domain wall. 
Near the line defect the system is described by \eqref{TIAFsurface} with $k_x$ replaced by $-id/dx$,
i.e, by the differential operator 
\begin{align}
\mathcal{H}_{\mbox{\tiny surface}}(k_z)=\left[-i\hbar v\sigma_1\frac{d}{dx}+m(x)\sigma_3\right]+\hbar vk_z\sigma_2 .
\end{align} 
Notice that the operator inside the square bracket is exactly the Jackiw-Rebbi model 
\eqref{JackiwRebbi},
which traps a zero mode $|\psi_0\rangle$ for $k_z=0$. 
As $\sigma_2|\psi_0\rangle=+|\psi_0\rangle$, it has a chiral energy spectrum $\mathcal{H}_{\mbox{\tiny surface}}(k_z)|\psi_0\rangle=+\hbar vk_z|\psi_0\rangle$.
This chiral Dirac mode is topologically guaranteed by the Chern invariant \eqref{c-index1} \begin{align}c_-&=\mbox{Ch}_1[H(k_x,k_y,x)]\nonumber\\&=\frac{i}{2\pi}\int_{k_x,k_y}\mbox{Tr}\left(\left.\mathcal{F}\right|_{x>0}\right)-\mbox{Tr}\left(\left.\mathcal{F}\right|_{x<0}\right)=1 ,
\end{align} 
where the integral is taken over $k_x,k_y\in(-\infty,\infty)$. Notice that in the defect description, like in the Jackiw-Rebbi model, a $\varepsilon k^2$ regularization in \eqref{TIAFsurface} is unnecessary.

Alternatively, 
the TI-AF heterostructure can be described by a 3d defect Hamiltonian in the continuum limit
\begin{align}
H_{3d}(\mathsf{k},\phi)=\hbar v\mathsf{k}\cdot\boldsymbol\sigma\mu_1+m_1(\phi)\mu_3+m_2(\phi)\mu_2,
\label{TIAF3dc}
\end{align} 
or its discrete counter part obtained by making the replacements 
$k_i \to \sin k_i$
and 
$m_1(\phi) \to m_1(\phi) + 
m_1(\phi)+\varepsilon(3-\sum_{i=1}^3\cos k_i)
$, 
where $\sigma$ and $\mu$ are Pauli matrices acting on spin and orbital degrees of freedom. 
The Dirac mass $m\Gamma_0(\phi)=m_1(\phi)\mu_3+m_2(\phi)\mu_2$ incorporates 
(i) the TRS mass that changes its sign $m_1(y\to\pm\infty)=\pm m_0$ across the horizontal TI surface,
and (ii) the AF mass that changes its sign $m_2(x\to\pm\infty)=\pm m_0$ 
across the vertical $yz$-plane where the Neel order flips. 
Here, $m_2\mu_2$ corresponds to an AF order, as it breaks inversion symmetry $I=\mu_3$. 
It, however, can be replaced by a ferromagnetic one, i.e., $h\sigma_1$. 
The mass parameter ${\bf m}(\phi)=(m_1(\phi),m_2(\phi))$ is modulated along a circle with radius $R_0$ far away from the line defect. 
In the homogeneous limit, $m_1,m_2$ coexist and can be approximated by ${\bf m}(\phi)\approx(m_0\sin\phi,m_0\cos\phi)$. 
It winds once around the origin. 
This corresponds to the generator of the homotopy classification $\pi_1(U(N))=\mathbb{Z}$ of class A line defect in 3d, 
where $U(N)$ is the classifying space for 3d class A band Hamiltonians. 
For instance $m\Gamma_0(\phi)$ wraps around the non-trivial cycle in $U(4)$. 
This non-trivial winding matches with the 2nd Chern invariant \eqref{c-index1} and \eqref{defectChern} 
\begin{align}
c_-&=\mbox{Ch}_2\left[H_{3d}(\mathsf{k},\phi)\right]\nonumber\\
&=\frac{-1}{8\pi^2}\int_{\mathrm{BZ}^3\times S^1}
\mbox{Tr}
\left[\mathcal{F}(\mathsf{k},\phi)\wedge\mathcal{F}(\mathsf{k},\phi)\right]
\nonumber\\&=\frac{1}{2\pi}\int_{S^1}d\theta(\phi)=1, 
\end{align} 
where $\theta(\phi)
=2\pi \int_{ \mathrm{BZ}^3}\mathcal{Q}_3(\mathsf{k},\phi)
$ 
(mod $2\pi$) 
is the magnetoelectric polarizablility (theta-angle)
(Sec.\ \ref{paragraph CS invariant}),
which in this case
is slowly modulated by the spatial angle $\phi$
and  
winds once from $0$ to $2\pi$ around the origin
[Fig.~\ref{fig:heterostructureline1}(a)].

The topology of the long length scale Hamiltonian \eqref{TIAF3dc}  
corresponds to the chiral Dirac mode appearing at the heterostructure. 
Near the line defect the system is effectively described by the differential operator 
\begin{align}
&\mathcal{H}_{3d}(k_z)=
\left[-i\hbar v(\partial_x\sigma_1+\partial_y\sigma_2)\mu_1\right.
\nonumber\\
&\;
\qquad 
\left.+m_1(x,y)\mu_3+m_2(x,y)\mu_2\right]+\hbar vk_z\sigma_3\mu_1,
\end{align} 
which is obtained from  \eqref{TIAF3dc} by replacing $k_{x/y}\leftrightarrow-i\partial_{x/y}$. 
Notice that the operator inside the square bracket is exactly the 2d Jackiw-Rossi model which has a zero mode $|\psi_0\rangle$ at $k_z=0$. As the zero mode has positive chirality $S|\psi_0\rangle=+|\psi_0\rangle$ with respect to the chiral operator $S=\sigma_3\mu_1$, it gives a chiral Dirac mode with a linear energy spectrum $\mathcal{H}(k_z)|\psi_0\rangle=+\hbar vk_z|\psi_0\rangle$.

\paragraph{Example: Helical modes in heterostructures (class AII)}

Heterostructures in symmetry classes AII, D, and DIII can host helical modes. 
Figure~\ref{fig:heterostructureline1}(b) shows a helical Dirac mode on the surface of a weak TI,
which hosts a pair of Dirac cones at the two TR invariant surface momenta (TRIM) $\mathsf{K}_1$ and $\mathsf{K}_2$. 
The two cones can be gapped out by a translation breaking TRS perturbation $u$ with a finite wave vector $\mathsf{K}_1-\mathsf{K}_2$. 
This density wave $u$ introduces a polarization $P=0,\pi$ (mod $2\pi$) depending on the sign of $u$. 
A domain wall on the surface separating two regions with opposite polarization traps a protected helical Dirac mode \cite{Liu:2011fk,Nontrivial_surface_chiu}.

\paragraph{Example: Chiral Majorana modes in heterostructures (class D)}

Figure~\ref{fig:heterostructureline1}(c) shows a chiral Majorana mode realized 
in two superconducting heterostructures. 
First, the surface Dirac cone of a TI can be gapped by a TRS or $U(1)$ symmetry breaking order. 
When restricting the defect momentum to $k_z=0$~\cite{Fe_SC_TI_Tanaka,Tanaka_SC_odd_even}. 
This problem reduces to the previous 2d QSHI-FM-SC heterostructure [Fig.\ \ref{fig:TISCFM}(a) and \eqref{QSHSCM2d}].
The zero energy MBS now turns into a gapless chiral Majorana mode that disperses linearly in $k_z$ and carries the chiral central charge $c_-=1/2$. 
Instead of the SC-AF domain wall on the TI surface, 
one can also consider a domain wall in the SC phase on the TI surface, 
which  hosts a helical Majorana defect mode
[Fig.\ \ref{fig:heterostructureline1}(d)].
Second, the surface of 
a TSC in class DIII can host multiple ($=n$) copies of Majorana cones,  
with chiralities $\chi_1,\ldots,\chi_n=\pm1$. 
These surface states are sensitive to a TR breaking perturbation that opens up mass gaps $m_1,\ldots,m_n$. 
A domain wall where certain mass gaps change their sign hosts chiral Majorana modes with a chiral central charge of 
\begin{equation}
c_-=\frac{1}{2}\lim_{x\to\infty}\sum_{a=1}^n\chi_a\frac{\mbox{sgn}(m_a(x))-\mbox{sgn}(m_a(-x))}{2}.
\end{equation}

\paragraph{Example: Dislocations in weak TIs and TSCs}

Gapless modes can also appear along lattice dislocations in weak TIs and TSCs. 
The 2d weak topological indices of a 3d bulk TI and TSC 
is expressed as a reciprocal lattice vector 
$\mathbf{G}_\nu=\nu_1\mathbf{b}_1+\nu_2\mathbf{b}_2+\nu_3\mathbf{b}_3$, 
where the $i^{th}$ weak index $\nu_i$ is evaluated on the 2-cycle 
$\mathcal{C}_i=\{\mathsf{k}\in \mathrm{BZ}^3:\mathsf{k}\cdot\mathsf{a}_i=\pi\}$ 
perpendicular to $\mathsf{a}_j$ and $\mathsf{a}_k$ on the boundary of the BZ, 
where $\mathsf{a}_{i,j,k}$ are distinct primitive lattice vectors. 
Notice that the 2-cycles $\mathcal{C}_i$ are invariant under the involution $\mathsf{k}\leftrightarrow-\mathsf{k}$ 
and restricting the Hamiltonian onto these momentum planes give 2d Hamiltonians with the same symmetries. 
For example, $\nu_i$ are 1st Chern invariants for class A, D, and C, or Fu-Kane invariants (or equivalently Pfaffian invariants) for class AII and DIII. 
The topological index that characterizes the gapless modes along a dislocation line defect is the product 
\cite{Ran:dislocation}
\begin{align}\mbox{ind}=\frac{1}{2\pi}\mathbf{B}\cdot\mathbf{G}_\nu ,
\end{align} 
where $\mathbf{B}$ is the Burgers vector, the net amount of translation when a particle circles once around the dislocation line. 
This integral quantity counts the chiral central charge $c_-$ of Dirac dislocation modes in a weak 3d Chern insulator, 
or twice the chiral central charge of Majorana dislocation modes in a weak 3d class D SC. 
For weak class AII TI or class DIII TSC, 
this index becomes a $\mathbb{Z}_2$ number that counts helical Dirac or Majorana dislocation modes, respectively.

\subsection{Adiabatic pumps}
Adiabatic pumps are temporal cycles of defect systems. 
The Hamiltonian is of the form $H(\mathsf{k},\mathsf{r},t)$,
where $\mathsf{k}$ lives in the BZ, $\mathrm{BZ}^d$, 
$\mathsf{r}\in\mathcal{M}^{D-1}$ wraps the defect in real space, 
and $t$ is temporal parameter of the adiabatic cyclic.
 The topological classification is determined by the symmetry class $s$ of the Hamiltonian 
and the topological dimension $\delta=d-D$, and is given by the classification Table~\ref{tab:classification}. 
The antiunitary symmetries normally flip $(\mathsf{k},\mathsf{r},t)\to(-\mathsf{k},\mathsf{r},t)$. 
However in some cases it can also flip the temporal parameter \cite{ZhangKane14},
but this will not be the focus of this review.

The simplest pumps appear in symmetry class A in 1d, 
known as Thouless pumps \cite{Thoulesspump}, 
and are classified by an integer topological invariant, 
the first Chern invariant:
\begin{align}
\mbox{Ch}_1=\frac{i}{2\pi}\int_{\mathrm{BZ}^1\times S^1}\mbox{Tr}\left(\mathcal{F}(k,t)\right),
\end{align} 
where $\mathcal{F}$ is the Berry curvature of the occupied states and $S^1$ parametrizes the temporal cycle. 
For example, 
\begin{align}H(k,t)=t\sin k\sigma_1+u\sin t\sigma_2+m\left(\frac{3}{2}-\cos k-\cos t\right)\sigma_3
\end{align} 
realizes a non-trivial pump with $\mbox{Ch}_1=1$, where $t$ runs a cycle in $[0,2\pi]$ so that $H(k,0)=H(k,2\pi)$. 
The signature of a Thouless pump is the spectral flow of boundary modes:
The end of the 1d system does not hold protected bound modes. 
However, during the adiabatic cycle, a certain number of boundary modes appear and 
as a function time connect the occupied and unoccupied bands. 
(See Fig.\ \ref{fig:linemodes}(a), but with $k_\|$ replaced by $t$ such that the red mid-gap bands represent 
the temporal evolution of the boundary states.) 
A charge is pumped to (or away from) the boundary when a boundary state is dragged from the unoccupied bands to the occupied ones 
(resp.~occupied bands to the unoccupied ones) after a cycle. 
The index theorem relates the Chern invariant and the spectral flow: 
\begin{align}
\mbox{Ch}_1=(\mbox{charge accumulated at boundary after 1 cycle}). 
\end{align} 

General charge pumps are adiabatic cycles of point defects in $d$ dimensions. 
The class A Hamiltonian takes the form $H(\mathsf{k},\mathsf{r},t)$ where $(\mathsf{k},\mathsf{r},t)\in \mathrm{BZ}^d\times S^{d-1}\times S^1$. 
They are characterized by the $d^{th}$-Chern invariant \eqref{defectChern} defined by the Berry curvature $\mathcal{F}(\mathsf{k},\mathsf{r},t)$ of occupied states.
For example, the Laughlin argument 
(which proves that a $hc/2e$ flux quantum in an integer QH fluid carries charge $h\sigma_{xy}/e$, with $\sigma_{xy}$ the Hall conductance) 
can be rephrased as an adiabatic pump of a 2d point defect \cite{Laughlin_IQHE}.

Adiabatic pumps can also appear in superconducting class D or BDI systems. 
They are $\mathbb{Z}_2$ classified and are characterized by the Fu-Kane invariant \eqref{defectFK} with the gauge constraint \eqref{FKconstraint2}. 
The simplest example is the fermion parity pump realized along a 1d $p$-wave SC wire 
\cite{Kitaev2001,FuKaneJosephsoncurrent09,Teo:2010fk}. 
The bulk BdG Hamiltonian is of the form of 
\begin{align}
H(k,t)&=
e^{-it\sigma_3/2}[\Delta\sin k\sigma_1+(u\cos k-\mu)\sigma_3]e^{it\sigma_3/2}\nonumber\\
&=\Delta\sin k(\cos t\sigma_1+\sin t\sigma_2)+(u\cos k-\mu)\sigma_3,\end{align} 
where $t$ evolves from 0 to $2\pi$ in a cycle. 
Notice that $H(k,0)=H(k,2\pi)$. At all time $H(k,t)$ is a $p$-wave SC with a non-trivial $\mathbb{Z}_2$ index when $|u|>|\mu|$ 
and, hence, supports protected boundary Majorana zero modes. 
The SC pairing phase winds by $2\pi$ as the system goes through a cycle. 
The evolution operator $e^{-it\sigma_3/2}$, however, is not cyclic as it has a period of $4\pi$. 
Hence, since $|\gamma_t\rangle=e^{-it\sigma_3/2}|\gamma_0\rangle$,  the Majorana zero mode $\gamma$ at the wire end changes its sign
after a cycle. 

Consider a weak link along a topological $p$-wave SC wire. 
At the completely cut-off limit, there are two uncoupled Majorana zero modes 
$\hat{\gamma}_1,\hat{\gamma}_2$ sitting at the two sides of the link. 
They form a fermionic degree of freedom 
$\hat{c}=(\hat{\gamma}_1+i\hat{\gamma}_2)/2$, 
which realizes a two-level system $|0\rangle$ and $|1\rangle=\hat{c}^\dagger|0\rangle$. 
Electron tunneling across the link splits the zero modes with an energy gap proportional to the tunneling strength, 
where the ground state has now a definite fermion parity $|0\rangle$ or $|1\rangle$. 
A phase slip $\delta\varphi=\varphi_R-\varphi_L$ is a discontinuity of the SC pairing phase across the weak link, 
where $\varphi_{R/L}$ are the phases of the two disconnected SC wires on the two sides of the weak link. 
In the scenario where the phase slip adiabatically winds by $2\pi$, 
the fermion parity operator $(-1)^{\hat{N}(\delta\varphi)}=i\hat{\gamma}_1(\delta\varphi)\hat{\gamma}_2(\delta\varphi)$ 
evolves and acquires an extra sign after a cycle, i.e.~$(-1)^{\hat{N}(2\pi)}=-(-1)^{\hat{N}(0)}$. 
In other words, this flips $\hat{c}\leftrightarrow \hat{c}^\dagger$ (up to a $U(1)$-phase). 
Physically, although there is an energy gap when $\delta\varphi=0$, 
this gap has to close and re-open as the two-level system undergoes a level-crossing during the adiabatic cycle. 
The single (or in general odd number of) level-crossing cannot be removed and is protected by the non-trivial $\mathbb{Z}_2$ bulk topology. 
The fractional Josephson effect is a consequence of such a non-trivial topology 
\cite{Kitaev2001,FuKaneJosephsoncurrent09,ZhangKane14PRL},
which can also arise in TR symmetric systems 
\cite{ZhangKane14,KeselmanFuSternBerg13}.
Unconventional Josephson effects which may have a topological origin have recently been observed in certain experimental 
systems~\cite{WilliamsGoldhaber12,YamakageTanaka13,KurterHarlingen14}.

\subsection{Anderson ``delocalization'' and topological phases}
\label{subsec:Anderson_delocalization}

 So far TIs/TSCs were described from the bulk point of view
and by establishing a bulk/defect-boundary correspondence.
Here, we show that it is also possible to identify TIs/TSCs
from the boundary point of view, i.e., by studying the effects of disorder  on the boundary modes
\cite{Schnyder2008}. 

Let us recall how the bulk topological properties of TIs/TSCs manifest themselves at the boundary of the system:
TIs and TSCs are always accompanied by gapless excitations localized at their boundaries.
These boundary states are stable against perturbations which respect the symmetries of the system 
(Sec.~\ref{Bulk-boundary and bulk-defect correspondence}).
As a deformation of the system let us consider
spatially inhomogeneous perturbations, i.e., disorder. 
As it turns out,
the bulk/defect-boundary correspondence  holds even in the presence of disorder, 
and hence gapless boundary/defect excitations are stable against disorder.
That is, the boundary modes do not {\it Anderson localize} even in the presence of disorder,
as long as the symmetry conditions are preserved (enforced), and as
long as the inhomogeneous perturbations due to disorder
do not close the bulk gap. 

Adding sufficiently strong disorder in an ordinary metal 
almost always leads to an (Anderson) insulator\footnote{
There are a few exceptions to this rule, 
but even in such cases, 
homogeneous but lattice-translation symmetry breaking perturbations  
(i.e., charge density wave or dimerization) can turn the system into a band insulator.}.
In his seminal paper
\cite{Anderson1958},
Anderson showed by the so-called ``locator expansion''  that, 
if one starts from the atomic limit,  the presence
of sufficiently strong impurities
leads to the absence of electron diffusion (i.e., to {\it Anderson localization}).
If we follow Anderson's ana\-lysis, we expect Anderson localization 
in any system with sufficiently strong disorder,
as long  as Anderson's assumption applies -- i.e., that the system is connected to the atomic limit. 
Reversing this logic,
the absence of Anderson localization implies
the absence of an atomic limit,
or the impossibility of discretizing the system on a lattice. 
The absence of Anderson localization (i.e.,  ``Anderson delocalization''), can thus be used as a criterion to identify theories
that cannot be discretized on a lattice -- lattice versions of such theories can be realized only as a
boundary of some topological bulk system. 
Historically, Anderson delocalization at boundaries 
was hypothesized to be the defining property of TIs/TSCs. 
Adopting this hypothesis, it was shown that the Anderson delocalization
 approach is powerful enough to establish the ten-fold classification of TIs/TSCs
\cite{Schnyder2008}. 

In this subsection, we will review Anderson delocalization and the ten-fold classification of TIs/TSCs mainly 
by using effective field theories, i.e., non-linear sigma models (NL$\sigma$Ms)
\cite{Wegner1979,Efetov1980,Efetov1983,EversMirlin2008}
-- a convenient framework to discuss the physics of Anderson localization/delocalization
in various dimensions and in the presence of various symmetry conditions.
We will also briefly touch upon the effects of disorder on bulk TIs/TSCs.

\subsubsection{Non-linear sigma models}

The NL$\sigma$Ms for the Anderson localization problem are effective field theories 
that describe the properties of (disorder-averaged) single-particle Green's functions and products thereof.
Using the NL$\sigma$M framework, one can compute 
all essential properties of single-particle Green's functions,
and hence of single-particle Hamiltonians. 

The basic concepts that underly the framework of NL$\sigma$Ms
can be illustrated by taking a classical magnet as an example. 
The classical Heisenberg ferromagnet in $d$ space dimensions 
can be described, in the long-wave length limit, by an $O(3)$ NL$\sigma$M.
Its action is given by
\begin{align}
 S[\mathbf{n}]
 =
 \frac{1}{t} \int d^d \mathsf{r}\, \partial_{\mu} {\bf n}\cdot \partial_{\mu} {\bf n},
 \label{O(3) NLSM}
\end{align}
where  ${\bf n}$ is a three-component unit vector
and
$t$ is the coupling constant, which is proportional to the temperature, the magnitude of spin, and the magnetic exchange interaction.
The partition function is given by the functional integral 
$
Z = \int \mathcal{D}[{\bf n}] \exp( - S[{\bf n}])
$,
where the sum runs over all maps ${\bf n}(\mathsf{r})$ 
from the $d$-dimensional space to the space of the order parameter $S^2\simeq {O}(3)/{O}(2)$.
The space of the order parameter is  called the ``{\it target space}".
Here, ${O}(3) ( = G)$ is the symmetry of the classical Heisenberg ferromagnet,
and ${O}(2) (=H)$ is the residual symmetry when 
the ${O}(3)$ is spontaneously broken. 
The Nambu-Goldstone theorem tells us that $G/H = O(3)/O(2)$ is the target manifold
representing the fluctuations of the order parameter.

The Nambu-Goldstone modes that are relevant to the physics of Anderson localization
correspond to the diffusive motion of electrons or Bogoliubov quasiparticles.
These modes are called ``diffusons'' and ``Cooperons'' 
and their dynamics can be described by   NL$\sigma$Ms, whose
 action and path integral are given by
\cite{Friedan1980}
\begin{align}
S[X] &=
\frac{1}{t} \int d^d\mathsf{r}\, G_{AB} [X] \partial_{\mu} X^A \partial_{\mu}X^B ,
\nonumber \\
\quad
Z &= \int \mathcal{D}[X] \exp( -S[X]),
\label{NLSM generic}
\end{align}
respectively.
Here, $X^A (\mathsf{r}): \mathbb{R}^d \to G/H$ are coordinates on a suitably chosen target manifold $G/H$ (see below), 
which represents a map from  $d$-dimensional physical space
to the target manifold $G/H$. $G_{AB}[X]$ denotes the metric of the target space. 
In the context of Anderson localization/delocalization, the coupling constant $t$ in the NL$\sigma$Ms is inversely proportional to the conductivity.
For our discussion, 
$d$ can be either the spatial dimension of the boundary of a ($d+1$)-dimensional (topological) insulator or SC,
or can be the bulk dimension of a TIs/TSCs. 
(Technically, the NL$\sigma$Ms in Anderson localization physics 
are derived by using the replica trick to handle quenched disorder averaging.  
In the following, we will use the fermionic replica trick.) 
Two typical phases described by NL$\sigma$Ms,  
ordered and disordered phases, 
correspond,
in the context of Anderson localization,
to a metallic and an insulator phase, respectively.

In the NL$\sigma$M description of Anderson localization,
the difference between symmetry classes 
are encoded by different target manifolds. 
(See Table~\ref{TargeManifolds}, which lists the target manifolds.)
While generic NL$\sigma$Ms can have more than one coupling constant,
  the action \eqref{NLSM generic} has only one coupling constant $t$.
This is a crucial feature of NL$\sigma$Ms relevant to Anderson localization.
This fact is nothing but a reincarnation of the single parameter scaling hypothesis by the gang of four 
\cite{Abrahams1979}.
The target spaces of the NL$\sigma$Ms which allow only one coupling constant are called {\it symmetric spaces}. 
These have been fully classified by  the mathematician E.~Cartan~\cite{Helgason_book}.
Ignoring those symmetric spaces which involve exceptional Lie groups, there are only ten (families of) symmetric spaces.

To summarize, the action~\eqref{NLSM generic} depends only on 
spatial dimension, choice of target manifolds, and the conductivity.
(This fact indicates universality in the physics of Anderson localization.)
According to the scaling theory (and also the locator theory of Anderson), 
if one starts from sufficiently strong disorder, then, by the renormalization group, 
disorder will be renormalized and become stronger. In other words,  Anderson localization is inevitable.
Using the analogy to the classical magnet, this means that at infinite temperature  a ``paramagnetic phase'' is always realized. I.e.,
the NL$\sigma$Ms universally predict Anderson localization at $t=\infty$. 
We are thus led to conclude that the NL$\sigma$Ms above cannot describe the physics at the boundaries of TIs and TSCs.

\begin{table}[t]
\begin{center}
\begin{tabular}{cc}\hline
AZ class  & NL$\sigma$M target space 
           \\ \hline \hline 
A 
&
${U}(2N)/{U}(N)\times {U}(N)$ 
\\ \hline
AI 
&
${Sp}(2N)/{Sp}(N)\times {Sp}(N)$ 
\\ \hline
AII 
& 
${O}(2N)/{O}(N)\times {O}(N)$
\\ \hline \hline
AIII
& 
${U}(N)\times {U}(N)/ {U}(N)$
\\ \hline
BDI
& 
${U}(2N)/ {Sp}(N)$ 
\\ \hline
CII
& 
${U}(N)/ {O}(N)$ 
\\ \hline \hline
D 
  & ${O}(2N)/ {U}(N)$
\\ \hline
C
& ${Sp}(N)/ {U}(N)$ 
\\ \hline
DIII 
& 
${O}(N)\times {O}(N)/{O}(N)$ 
\\ \hline
CI
& 
${Sp}(N)\times {Sp}(N)/{Sp}(N)$
\\ \hline
\end{tabular}
\end{center}
\caption{\label{TargeManifolds}
This table lists the NL$\sigma$M target manifolds (in the fermionic replica approach)
for the symmetry classes of the ten-fold way.
}
\end{table}

How can Anderson delocalization possibly happen, then?
We need a mechanism that prevents Anderson localization. 
What has escaped from our attention in the above discussion
is the effects of topology of the target manifolds.
When the target manifolds have non-trivial topology,
one can add a topological term to the action of the NL$\sigma$M:
\begin{align}
 Z = \int \mathcal{D}[X]
 \exp (-S[X] - i S_{top}[X]).
\end{align}
The topological term $S_{top}[X]$ is an imaginary part of the 
action and depends only on global information of field configurations.
If there is a topological term,
there are interferences (cancellations) in the functional integral 
among different field configurations, 
and there is a possibility that different physics may emerge.

A famous example of  topological terms are the so-called {\it theta terms}.
They can appear when $\pi_d(G/H)=\mathbb{Z}$. 
Taking again an example from magnetic systems,  
consider the Haldane topological term in quantum spin chains.
Similar to the classical Heisenberg ferromagnet in 2d,
 the {\it quantum} Heisenberg {\it antiferrmagnet} in (1+1)d
can   be described at low-energies and long wave-lengths by the $O(3)$ NL$\sigma$M, Eq.\ \eqref{O(3) NLSM}
\cite{Haldane1983a, Haldane1983b}. 
However, an important twist in the quantum case is the possible presence of a topological term, 
whose presence/absence   crucially affects the structure of the low-energy spectrum 
(i.e., it leads to the presence/absence of the ``Haldane gap''). 
The theta term in this case is given by $S_{top}[{\bf n}] = \theta \times (\mbox{integer})$ with $\theta=2\pi S$, 
where $S$ is the spin magnitude,
and the $\mbox{integer}$ is a topological invariant defined for a given
texture ${\bf n}(\mathsf{r})$. 
The low-energy properties of the system are dramatically different for integer spin $S$ than for half-odd integer spin $S$.

\subsubsection{Anderson delocalization at boundaries}

For the application of NL$\sigma$Ms to the boundary physics of TIs/TSCs, 
{\it Wess-Zumino-Novikov-Witten (WZNW) terms} and {\it $\mathbb{Z}_2$ topological terms}
\cite{ryuFurusakiPRL07, Ostrovsky2007, Fendley2000}
are important,
rather than theta terms.
In contrast to theta terms,
for which the coefficient (``the theta angle'') can be tuned continuously as one changes microscopic details
\cite{Affleck:1988}, 
the coefficients of WZNW or $\mathbb{Z}_2$ topological terms are not tunable.
Furthermore, when these terms are present, 
it is expected that, 
as in the case of theta terms with $\theta=\pi \times \mbox{odd integer}$, 
systems are at their critical points. 
In the context of Anderson localization, 
critical Nambu-Goldstone bosons indicate that the localization length is diverging and hence the system delocalizes. 
Hence, in   NL$\sigma$Ms with  WZNW or $\mathbb{Z}_2$ terms, Anderson delocalization is unavoidable. 

From the mathematical point of view, 
$\mathbb{Z}_2$ topological terms 
and WZNW terms
exist 
when $\pi_d(G/H)=\mathbb{Z}_2$
and 
$\pi_{d+1}(G/H) = \mathbb{Z}$, respectively.
Thus, by merely looking at the homotopy group of the target manifolds,
one can infer if Anderson delocalization can occur or not. 
In turn, such delocalized states that cannot be Anderson localized
must be realized as a boundary state of some bulk TIs/TSCs. 
Hence, the bulk topological classification of $\mathbb{Z}_2$ or $\mathbb{Z}$ type corresponds 
to the type of topological terms ($\mathbb{Z}_2$ or WZNW) at the boundary. 
Combining these considerations all together, one derives the periodic table of TIs/TSCs. 
For Dirac models of  boundary modes of TIs/TSCs,
one can  explicitly check (i.e., one by one) the existence of these topological terms
in the NL$\sigma$M description
\cite{LeClair2002,ryuFurusakiPRL07, Ostrovsky2007,RyuMudry2012, AltlandSimonsZirnbauer2002, Bocquet2000}.

Generically, however, it is difficult to quantify the precise effects of topological terms in NL$\sigma$Ms in a controlled way when the boundary is of dimension larger than one.
Only general arguments are then available \cite{XuLudwig2011}. 
When the boundary is 0d or 1d, 
it is possible to decide in a controlled way if the boundary state is immune to disorder.
For example, 
along the 1d boundary of a TI in the symmetry class AII,
by using the Dorokov-Mello-Pereyra-Kumar (DMPK) equations for the transmission eigenvalues of quasi-1d disordered wires,
it is possible to show that the edge states contribute
a longitudinal conductance of order one in the thermodynamic limit 
\cite{Takane2004a, Takane2004b, Takane2004c}.
Historically, this problem was also studied by using the NL$\sigma$M which can be augmented by the $\mathbb{Z}_2$ topological term
\cite{Zirnbauer1992, Brouwer1996},
but the connection to bulk topology phases was only realized after the discovery of the quantum spin Hall effect.


In 2d, some Dirac fermion models in the presence of disorder can be solved exactly
\cite{LudwigFisherShankarGrinstein1994, Chamon1996, Nersesyan1994, Tsvelik1995}.
2d Dirac modes with disorder, realized on the surface of 3d TR symmetric TIs, 
can be studied numerically to demonstrate Anderson delocalization 
\cite{nomuraPRL07, Bardarson2007}.
For the latter, the complete absence of backscattering 
\cite{AndoNakanishiSaito1998}
was later confirmed in experiments on the surface of 3d TR symmetric TIs
\cite{Roushan2009, Alpichshev2010}. 
The combined effects of disorder and interactions in 2d boundaries of 3d TI/TSCs
have  also been studied in the literature
\cite{Ostrovsky2010PhRvL.105c6803O, Foster2012PhRvL.109x6801F, Foster2014PhRvB..89o5140F, Foster2015PhRvB..91b4203X}.

\subsubsection{Effects of bulk disorder}

Before concluding this section, let us briefly discuss the effects of bulk disorder in TIs/TSCs.
The effects of disorder in the most famous example of TIs, the integer QHE, manifest themselves by 
quantized plateaus of the Hall conductivity separated by a continuous phase transitions. 
If this is the case, the bulk phase diagram of TIs/TSCs can be understood qualitatively by a NL$\sigma$M augmented by a theta term, 
e.g., 
the so-called Pruisken term for the IQHE 
\cite{Pruisken1984}. 
From this NL$\sigma$M which now has two coupling constants, $t$ and the theta term,
one then expects that the phase diagram of the integer QH 
system is described in terms of
  two parameters, i.e., the longitudinal and transverse conductivities
\cite{Khmelnitski1983, Pruisken1984}.
Transitions in the presence of disorder between topologically distinct phases 
were also studied in 2d bulk TSCs
\cite{Senthil1998PhRvL..81.4704S,ReadGreen2000, Gruzberg1999PhRvL..82.4524G, Senthil1999PhRvB..60.4245S},
in 3d TIs/TSCs
\cite{Ryu2012PhRvB..85o5138R},
and 
in \mbox{(quasi-)1d} 
by scattering matrix approaches 
\cite{Brouwer1998, Brouwer2000PhRvL..84.2913B, Brouwer2000PhRvL..85.1064B, Titov2001PhRvB..63w5318T, Gruzberg2005, Akhmerov2011PhRvL.106e7001A,
RiederPhysRevB.90.205404}
and by using  NL$\sigma$Ms
\cite{Altland2014PhRvL.112t6602A, Altland2015PhRvB..91h5429A}.
For the effects of disorder on TR symmetric $\mathbb{Z}_2$ TIs, see, for example, 
\cite{
Ryu2012PhRvB..85o5138R, 
PhysRevB.76.075301,
PhysRevB.79.045321,
PhysRevLett.112.016402,
1367-2630-12-6-065008,
PhysRevLett.107.196803, PhysRevB.78.115301}
and, in particular, for topological Anderson insulators (i.e., disorder-driven transition from a trivial insulator into a TI), 
see
\cite{top_anderson_insul_Shen_PRL09,top_anderson_insul_Beenakker_PRL09, GuoPhysRevLett.105.216601,doi:10.1143/JPSJ.80.053703,
PhysRevB.87.205141}. 
Phase diagrams for disordered TIs/TSCs  
can  also be studied by using  non-commutative geometry
\cite{Prodan_noncommute,Prodan_odd_Chern,loring_hastings_EPL_10,Hastings_annals_physics11,belissard_JMatPhys94}
and by  K-theory
\cite{Morimoto2015PhRvB..91w5111M}.

%
%

\section{Topological crystalline materials}
\label{Topological crystalline materials}

We have so far focused on topological phases and topological phenomena 
protected only by non-spatial AZ symmetries. 
In this section, we introduce additional spatial symmetries 
and discuss how these modify the topological distinction of gapped phases. 
There are two possible effects upon imposing additional symmetries. 
First, additional symmetries may not change 
the topological classification,  
but lead to simplified expressions for 
the topological invariants of the ten-fold classification
\cite{KaiSun_crystalline_Kondo,Dzero_Kondo_PRL,Dzero_Kondo_PRB,Chen_chernN_GSP}. 
For example, the $\bZ_2$ invariant of 3d TR symmetric TIs in the presence of inversion symmetry 
can be computed from the parity eigenvalues at TR invariant momenta
\cite{Fu2007uq}. 
Second, additional spatial symmetries can modify the topological classification~\cite{Fu_first_TCI}. 
Example of this case include weak TIs and TSCs, whose existence relies on the presence of a lattice translation symmetry,
 see Sec.~\ref{secTenFoldClass}
\cite{Fu:2007fk,Ran_dislocation,Teo_2013_disclination,HughesYaoQi13,Ran:dislocation}. 
Besides translation symmetries,
point group symmetries, such as reflection and rotation,
can lead to new topological phases,
giving rise to an enrichment of the tenfold classification of TIs and TSCs
\cite{Ando_Fu_TCI_review}.
These TIs and TSCs which are protected by crystalline symmetries
are called {\it topological crystalline insulators and superconductors} (TCIs and TCSs). 

\subsection{Spatial symmetries}

Spatial symmetries of a crystal or a lattice are described by {\it space groups}.
Operations in space groups are composed of 
translations, including in particular lattice translations, 
and point group operations that leave at least one point in space unchanged. 
The latter includes reflection, inversion, proper and improper rotations.
By the crystallographic restriction theorem, 
only rotations with 1, 2, 3, 4, and 6-fold axes are compatible with lattice translation symmetries. 
A space group operation $G$ maps 
the $m$-th site in the unit cell at $\mathsf{r}$
to  
the $m'$-th site in the unit cell at $u_G \mathsf{r}+\mathsf{R}_m$,
where $u_G$ is a $d\times d$ orthogonal matrix 
and $\mathsf{R}_m$ is a lattice vector. 
Correspondingly, 
fermion annihilation operators in real space, $\hat{\psi}_{i}(\mathsf{r})$,
are transformed by a unitary operator $\hat{\mathscr{G}}$  acting on the electron field operator as 
\begin{align}
\hat{\mathscr{G}} 
\hat{\psi}_{i}(\mathsf{r}) 
\hat{\mathscr{G}}^{-1}=
(U_{G})_i^{\ j}\hat{\psi}_j (u_{G} \mathsf{r}+\mathsf{R}_i),
\end{align} 
where  $U_G$ is a unitary matrix, 
and $i$ and $j$ are combined indices  
labeling the sites within a unit cell as well as internal degrees of freedom, such as spin
(summation over the index $j$ is implied).
It is known that one can always choose the lattice translation operators to be diagonal
in an irreducible representation.
In other words, one can always use momentum-space Bloch functions as the basis functions 
in generating irreducible representations of a space group.
The fermion annihilation operators in momentum space transform as  
\begin{align}
\hat{\mathscr{G}} 
\hat{\psi}_{i}(\mathsf{k}) 
\hat{\mathscr{G}}^{-1}=
(U_{G}(u_G \mathsf{k}))_i^{\ j}\hat{\psi}_j (u_{G} \mathsf{k}),
\end{align} 
where $(U_G(u_G \mathsf{k}))_i{ }^j= (U_G)_i{ }^j e^{-i u_G \mathsf{k}\cdot \mathsf{R}_i}$
($i$ is not summed over).
For example, for a 1d chain with two different sublattices A and B (i.e., two atoms in the unit cell) 
reflection $\hat{\mathscr{R}}$ about the A atom in the $j=0$-th unit cell is given by  
$
\hat{\mathscr{R}}:
$
$\hat{a}_j \to  \hat{a}_{-j}$ and 
$\hat{b}_{j} \to  \hat{b}_{-j-1}$ 
(see the example discussed in Sec.~\ref{Example: Polyacetylene}).
In momentum space reflection acts as
$\hat{\mathscr{R}}:
$
$\hat{a}(k) \to  \hat{a}(-k)$ and
$\hat{b}(k) \to  e^{-ik} \hat{b}(-k)$.

In the presence of the crystalline symmetry
$\hat{\mathscr{G}} \hat{H} \hat{\mathscr{G}}^{-1}= \hat{H}$,
the Bloch-BdG Hamiltonian obeys 
\begin{align}
H(\mathsf{k})=U_{G}^\dagger(\mathsf{k}) 
H(u_G^{-1}\mathsf{k})
U^{\ }_{{G}}(\mathsf{k}).
\label{crystalline equation}
\end{align}
For crystalline symmetry operations,  
which leave at least one point fixed ($\mathsf{k}_0$, say),
we have $[H(\mathsf{k}_0),U_{G}(\mathsf{k}_0)]=0$. 
It is thus possible to define topological invariants 
at $\mathsf{k}_0$ in each eigenspace of $U_G(\mathsf{k}_0)$. 
Crystalline symmetries are either {\it symmorphic} or {\it non-symmorphic} space group symmetries. 
In the following, we mainly focus on reflection symmetry, which is symmorphic. 
Topological phases and gapless surface states protected by non-symmorphic
space group symmetries have recently been discussed in 
\onlinecite{nonsymmorphic_Liu,nonsymmorphic_Sato,New_crystalline_Fang,Roy2012_nonsymmorphic,Parameswaran2013_nonsymmorphic,2D_classification_Liu,nonsymmorphic_cone_Kane,photonic_TI_Fu}.

Let us consider a reflection symmetry 
in the $r_l$ direction,  
$
\hat{\mathscr{R}}^{\ }_l
\hat{\psi}_i(\mathsf{r})
\hat{\mathscr{R}}^{-1}_l
=(U_{R_l})_i^{\ j}
\hat{\psi}_j(\bar{\mathsf{r}}+\mathsf{R}_i),
$
where $\bar{\mathsf{r}}
=(r_1,\ldots,r_{l-1},-r_l,r_{l+1},\ldots,r_d)$.
This reflection symmetry acts on the Bloch Hamiltonian as 
\begin{align}
{H}(\mathsf{k})=
U_{R_l}^\dagger(\mathsf{k}) 
{H}(\bar{\mathsf{k}}) 
U_{R_l}^{\ }(\mathsf{k}),
\end{align}
where $\bar{\mathsf{k}}
=(k_1,\ldots,k_{l-1},-k_l,k_{l+1},\ldots,k_d)$.
For particles with spin, spatial symmetries also transform the spin degrees of freedom.
For example, reflection flips the sign of orbital angular momentum, and
hence, the sign of spin, i.e.:
$
\hat{\mathscr{R}}_x \hat{S}_{x} \hat{\mathscr{R}}^{-1}_x= \hat{S}_{x}
$ 
and
$
\hat{\mathscr{R}}_x \hat{S}_{y,z} \hat{\mathscr{R}}^{-1}_x= -\hat{S}_{y,z}
$. 
Hence, for spin-$1/2$ particles, $U_{R_l}$ is given by $U_{R_l}=i s_l$. 
The reason to include the factor of $i$ here is to ensure $U^2_R=-1$, 
since $\mathscr{R}^{-1}_l$ effectively corresponds to a spin rotation by $2 \pi$.
In general, pure spin reflection operation is often combined
with some internal symmetry operation. 
To allow for this possibility we
loosely call any 
 symmetry that 
involves $\mathsf{r}\to \bar{\mathsf{r}}$ 
 a reflection symmetry.

\subsection{Classification of topological insulators and superconductors in the presence of reflection symmetry} 
\label{secReflectClass}

\begin{figure}[tb]
 \begin{center} 
\includegraphics[clip,width=0.8\columnwidth]{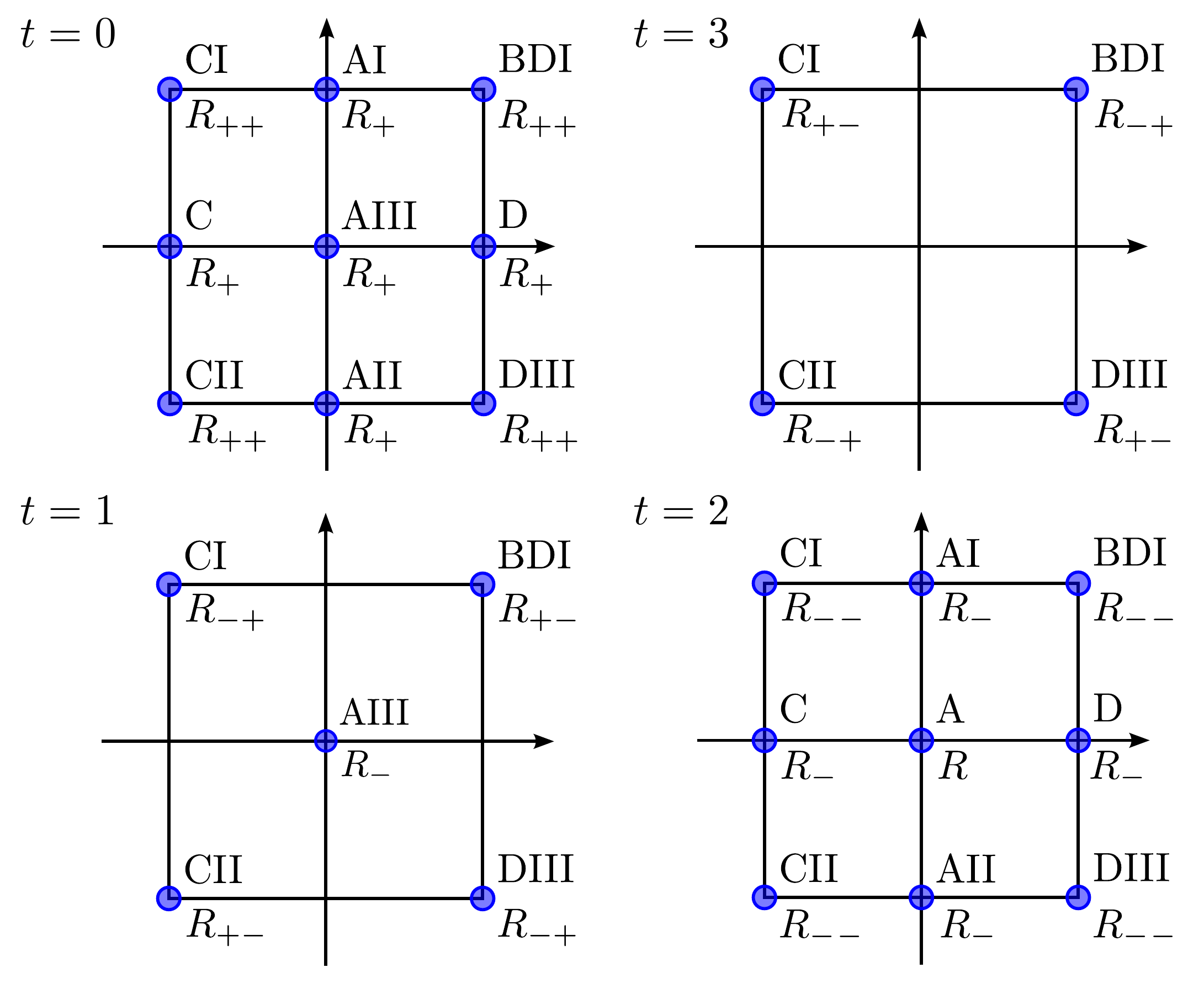}
\hfill
 \end{center}
 \caption{(Color online) 
 The 27 symmetry classes with reflection symmetry 
 can be visualized as ``the extended Bott clock''. 
 }
 \label{reflection_clock}
\end{figure}

We now discuss the classification of TCIs and TCSs in the presence of reflection symmetry. 
Consider a $d$-dimensional Bloch Hamiltonian $H(\mathsf{k})$,
which is invariant under reflection in the $r_1$ direction:
\begin{eqnarray} 
\label{refSym}
R^{-1}_1 H( -k_1, \tilde{\mathsf{k}}) R_1 = H (k_1,\tilde{\mathsf{k}} ),
\end{eqnarray}
where $\tilde{\mathsf{k}}=(k_2,\cdots, k_d)$, and 
the reflection operator $R_1$ is unitary and can depend only on $k_1$, since it is symmorphic. 
(For simplicity, we will drop the subscript ``1'' in $R_1$ henceforth.)
With a proper choice of the phase of $R$,
$R$ satisfies
on a given reflection plane, 
\bee
R^\dagger=R,
\quad
R^2 = \openone.  
\label{hermitian}
\ee 
Thus, all eigenvalues of $R$ are either $+1$ or $-1$.
The algebraic relations obeyed by $R$ and the AZ symmetry operators $T$, $C$, and $S$,
can be summarized as
\begin{align} 
\label{indicesComAnti}
SR = \eta_S RS,
\quad
TR = \eta_T RT,
\quad
CR = \eta_C RC,
\end{align}
where $\eta_{S,T,C}=\pm 1$
specify whether $R$ commutes ($+1$) or anticommutes ($-1$) with 
$S$, $T$, and $C$.
These different possibilities are labeled by $R_{\eta_T}$, $R_{\eta_{S}}$, and $R_{\eta_{C}}$ 
for the non-chiral symmetry classes AI, AII, AIII, C, and D, 
and 
by $R_{\eta_T \eta_C}$ for the chiral symmetry classes BDI, CI, CII, and DIII. 
Hence, we distinguish a total of 27 different symmetry classes 
in the presence of AZ and reflection symmetries
(Table~\ref{reflection_table_full} and Fig.\ \ref{reflection_clock}). 
(Note that the physical reflection operator always commutes with non-spatial symmetries (e.g., TRS).  
However, due to the phase convention adopted in \cref{hermitian}, $R$ may fail to commute with $T$.
For example, for spin-$1/2$ fermions
$R$ anticommutes with $T$
since, in order to fulfill \cref{hermitian},
$R$ is defined as the physical reflection operator multiplied by $-i$.
On the other hand, for spinless fermions, $R$ commutes with $T$.) 

The  classification of TCIs and TCSs in the 27 symmetry classes with reflection symmetry 
is summarized in Table~\ref{reflection_table_full}
\cite{Chiu_reflection,Morimoto2013,Sato_Crystalline_PRB14}.
In even (odd) spatial dimension $d$,  10 (17) out of the 27 symmetry classes allow 
for the existence of nontrivial TCIs/TCSs,
which are characterized and labeled by 
the following topological invariants:
(i) 
integer or $\mathbb{Z}_2$ topological invariants ($\mathbb{Z}$ or $\mathbb{Z}_2$) 
of the original 10-fold classification of TIs and TSCs without reflection symmetry;
(ii)
mirror Chern or winding numbers ($M\mathbb{Z}$)~\cite{Teo:2008fk}, or  mirror $\mathbb{Z}_2$ invariants ($M\mathbb{Z}_2$);
(iii)
$\mathbb{Z}_2$ invariants with translation symmetry ($T\mathbb{Z}_2$);
(iv) 
a combined invariant $M \mathbb{Z} \oplus \mathbb{Z}$ (or $M \mathbb{Z}_2 \oplus \mathbb{Z}_2$),
which consists of 
an integer $\mathbb{Z}$ number (or $\mathbb{Z}_2$ quantity) 
and a mirror Chern or winding number $M\mathbb{Z}$ 
(or mirror  $\mathbb{Z}_2$ quantity $M\mathbb{Z}_2$).
Let us now give a more precise description of these invariants 
and of the boundary modes that arise as a consequence.

\begin{table*}[t!]
\begin{ruledtabular}
\begin{tabular}{c|c|cccccccc}
& TCI/TCS & $d$=1 & $d$=2 & $d$=3 & $d$=4 & $d$=5 & $d$=6 & $d$=7 & $d$=8  \\
  \cline{2-10}
  $\begin{array}{c}
\mbox{ Reflection }
\end{array}$   &  
  $\begin{array}{c}
\mbox{FS1 in mirror} \\
\end{array}$
  & $p$=8 & $p$=1 & $p$=2 & $p$=3 & $p$=4 & $p$=5 & $p$=6 & $p$=7    \\
  \cline{2-10}
   &  
     $\begin{array}{c}
\mbox{FS2 in mirror} \\
\end{array}$  
  & $p$=2 & $p$=3 & $p$=4 & $p$=5 & $p$=6 & $p$=7 & $p$=8 & $p$=1     \\
\hline\hline
 $R$ &  A  & $M\bZ$ & 0 & $M\bZ$ & 0 & $M\bZ$ & 0 & $M\bZ$ & 0                \\
 $R_+$ &  AIII  & 0 & $M\bZ$ & 0 & $M\bZ$ & 0 & $M\bZ$ & 0 & $M\bZ$            \\ 
$R_-$ &  AIII   &  $M\bZ\oplus\mathbb{Z}$ & 0 & $M\bZ\oplus\mathbb{Z}$ & 0 & $M\bZ\oplus\mathbb{Z}$ & 0 & $M\bZ\oplus\mathbb{Z}$ & 0            \\ 
\hline
\multirow{8}{*}{$R_+$,$R_{++}$} & AI  & $M\bZ$ & 0 & $0^a$ & 0 & $2M\bZ^a$ & 0 & $M\bZ_2^a$ & $M\bZ_2$        \\
  & BDI  & $M\bZ_2$ & $M\bZ$ & 0 & $0^a$ & 0 & $2M\bZ^a$ & 0 & $M\bZ_2^a$        \\
  & D  & $M\bZ_2^a$ & $M\bZ_2$ & $M\bZ$ & 0 & $0^a$ & 0 & $2M\bZ^a$   & 0       \\
  & DIII  & 0 & $M\bZ_2^a$ & $M\bZ_2$ & $M\bZ$ & 0 & $0^a$ & 0  & $2M\bZ^a$       \\
  & AII   & $2M\bZ^a$ & 0 & $M\bZ_2^a$ & $M\bZ_2$ & $M\bZ$ & 0 & $0^a$   & 0  \\
  & CII  & 0 & $2M\bZ^a$ & 0 & $M\bZ_2^a$ & $M\bZ_2$ & $M\bZ$ & 0     & $0^a$   \\
  & C & $0^a$ & 0 & $2M\bZ^a$ & 0 & $M\bZ_2^a$ & $M\bZ_2$ & $M\bZ$  & 0        \\
  & CI & 0 & $0^a$ & 0 & $2M\bZ^a$ & 0 & $M\bZ_2^a$ & $M\bZ_2$ & $M\bZ$    
   \\
\hline
\multirow{8}{*}{$R_-$,$R_{--}$} & AI  & $0^a$ & 0 & $2M\bZ^a$ & 0 & $T\bZ_2^a$ &  $\bZ_2$ & $M\bZ$ & 0      \\
  & BDI   & 0  & $0^a$ & 0 & $2M\bZ^a$ & 0 & $T\bZ_2^a$ &  $\bZ_2$ & $M\bZ$        \\
  & D &  $M\bZ$   &  0  & $0^a$ & 0 & $2M\bZ^a$ & 0 & $T\bZ_2^a$ &  $\bZ_2$        \\
  & DIII  &  $\bZ_2$ & $M\bZ$   &  0  & $0^a$ & 0 & $2M\bZ^a$ & 0 & $T\bZ_2^a$       \\
  & AII  & $T\bZ_2^a$ &  $\bZ_2$ & $M\bZ$   &  0  & $0^a$ & 0 & $2M\bZ^a$ & 0    \\
  & CII  & 0 & $T\bZ_2^a$ &  $\bZ_2$ & $M\bZ$   &  0  & $0^a$ & 0 & $2M\bZ^a$        \\
  & C   & $2M\bZ^a$ & 0 & $T\bZ_2^a$ &  $\bZ_2$ & $M\bZ$   &  0  & $0^a$ & 0          \\
  & CI  & 0 & $2M\bZ^a$ & 0 & $T\bZ_2^a$ &  $\bZ_2$ & $M\bZ$   &  0  & $0^a$     
   \\
\hline
$R_{-+}$ & BDI, CII  & $2\bZ^a$ & 0 & $2M\bZ^a$ & 0 & $2\bZ^a$ & 0 & $2M\bZ^a$ & 0       \\
$R_{+-}$  & DIII, CI  & $2M\bZ^a$ & 0 & $2\bZ^a$ & 0 & $2M\bZ^a$ & 0 & $2\bZ^a$ & 0        \\
\hline
$R_{+-}$  & BDI  & $M\bZ\oplus\bZ$ & 0 & $0^a$ & 0 &  $2M\bZ\oplus 2\bZ^a$ & 0 & $M\bZ_2\oplus \bZ_2^a$ & $M\bZ_2\oplus \bZ_2$        \\
$R_{-+}$ & DIII  & $M\bZ_2\oplus \bZ_2^a$ & $M\bZ_2\oplus \bZ_2$    & $M\bZ\oplus\bZ$ & 0 & $0^a$ & 0 &  $2M\bZ\oplus 2\bZ^a$ & 0     \\
$R_{+-}$  & CII &   $2M\bZ\oplus 2\bZ^a$ & 0 & $M\bZ_2\oplus \bZ_2^a$ & $M\bZ_2\oplus \bZ_2$    & $M\bZ\oplus\bZ$ & 0 & $0^a$ & 0       \\
$R_{-+}$  & CI   & $0^a$ & 0    & $2M\bZ\oplus 2\bZ^a$ & 0 & $M\bZ_2\oplus \bZ_2^a$ & $M\bZ_2\oplus \bZ_2$    & $M\bZ\oplus\bZ$ & 0        \\
\end{tabular}
\caption{
 Classification of 
reflection-symmetry-protected
topological crystalline  insulators and superconductors (``TCI/TCS'')
as well as of stable Fermi surfaces (``FS1'' and ''FS2'') 
in terms of the spatial dimension $d$ of the TCIs/TCSs,
and the codimension $p$ of the Fermi surfaces. 
``FS1'' denotes Fermi surfaces  
that are located at high-symmetry points  within mirror planes.
``FS2'' stands for Fermi surfaces that are within mirror planes but away from high-symmetry points.
Note that for gapless topological materials the presence of translation symmetry is always assumed. 
Hence, there is no distinction between  $T\bZ_2$ and $\bZ_2$ for the classification of stable Fermi surfaces. 
Furthermore, we remark that $\bZ_2$, $M\bZ_2$, and $T\mathbb{Z}_2$ invariants can only protect Fermi surfaces of dimension zero ($d_{\mathrm{FS}}=0$) at high-symmetry points of the Brillouin zone (``FS1'').
For the entries labeled by the superscript ``$a$", 
there can exist surface states and bulk Fermi surfaces of type ``FS2" that are protected by $\bZ$ and $M\bZ$ invariants inherited from class A or AIII.  That is, in these cases TRS or PHS does not trivialize   these topological invariants.
}
\label{reflection table}
\label{reflection_table_full}
\end{ruledtabular}
\end{table*}

{\it (i) $\bZ$ and $\bZ_2$ invariants:}
For symmetry classes with at least one AZ symmetry that anticommutes with $R$, 
the topological invariants ($\mathbb{Z}$ or $\mathbb{Z}_2$) of the original ten-fold classification  
continue to exist in certain cases, even in the presence of reflection. 
These topological invariants protect gapless boundary modes, 
independent of the orientation of the boundary.

{\it (ii) $M\bZ$ and $M\bZ_2$ invariants:}
The mirror Chern numbers, the mirror winding numbers, and the mirror $\bZ_2$ invariants, 
denoted by $M\bZ$ and $M\bZ_2$, respectively, 
are defined on the hyperplanes in the BZ that are symmetric under reflection. 
For concreteness, let us consider space groups possessing the two reflection hyperplanes $k_1=0$ and $k_1=\pi$.
Since 
the Bloch Hamiltonian at $k_1=0$ and $k_1= \pi$,  
$\left. H ( \mathsf{k} ) \right|_{k_1 = 0, \pi}$, 
commutes with $R$,  
it can be block diagonalized with respect to the two eigenspaces $R=\pm 1$ of the reflection operator.
Note that each of the two blocks of $\left. H ( \mathsf{k} ) \right|_{k_1 = 0, \pi}$ is invariant under only those nonspatial symmetries that commute with the reflection operator $R$. Therefore, depending on the nonspatial symmetries of the $R = \pm 1$ blocks of 
$\left. H ( \mathsf{k} ) \right|_{k_1 = 0, \pi}$, each block can
be characterized by topological invariants of the original ten-fold classification   
in $d-1$ dimension.  
For instance, when the $R=+1$ block of $\left. H ( \mathsf{k} ) \right|_{k_1 = 0 ( \pi)} $ is characterized by the Chern  or winding number,  
$\nu_{k_1 = 0 (\pi)}$,
we introduce a mirror Chern or winding invariant by 
\cite{Chiu_reflection} 
\begin{align}
n_{M \mathbb{Z}} 
= \sgn \left( \nu_{k_1 = 0} - \nu_{k_1 = \pi} \right)
\left( \left| \nu_{k_1 = 0} \right| - \left| \nu_{k_1 = \pi} \right| \right).
\end{align}
Similarly, 
when 
the $R=+1$ block of $\left. H ( \mathsf{k} ) \right|_{k_1 = 0 ( \pi)} $ is 
characterized by a $\mathbb{Z}_2$ invariant, 
$n_{k_1 = 0 ( \pi )}=\pm 1$,
the mirror $\mathbb{Z}_2$ invariant $M\mathbb{Z}_2$ is defined by
\begin{align}
n_{M \mathbb{Z}_2} 
= 1 - \left|   n_{k_1 = 0} - n_{k_1 = \pi} \right|.
 \end{align}
A nontrivial value of these mirror indices indicates 
the appearance of protected boundary modes at reflection symmetric surfaces, 
i.e., at surfaces that are perpendicular to the reflection hyperplane $x_1=0$.
Surfaces that break reflection symmetry, however, are gapped in general.

{\it (iii) $T\bZ_2$ invariant:}
 In  symmetry classes where $R$ anticommutes with TR and PH operators
($R_-$ and $R_{--}$ in Table~\ref{reflection_table_full}),  
the second descendant $\bZ_2$ invariants are well defined only in the presence of translation symmetry. 
That is, boundary modes of these phases 
can be gapped out by density-wave type perturbations, 
which preserve reflection and AZ symmetries but break translation symmetry. 
Hence, protected TCIs/TCSs can exist when
reflection, translation, and AZ antiunitary symmetries are all there.

{\it (iv) $M\bZ \oplus \bZ$  and $M\bZ_2 \oplus \bZ_2$ invariants:}
In some cases, 
topological properties of reflection symmetric insulators (SCs) 
with chiral symmetry are described both by 
a global $\bZ$ or $\bZ_2$ invariant and by a mirror index $M\bZ$ or $M\bZ_2$,
which are independent of each other.
At boundaries which are perpendicular  to the mirror plane, 
the number of protected gapless states is
given by $\mathrm{max} \left\{ \left| n_{\bZ} \right|, \left| n_{M \bZ} \right| \right\}$
\cite{Chiu_reflection}, 
where $n_{\bZ}$ denotes the global $\bZ$ invariant, whereas 
$n_{M\bZ}$  is the mirror $\bZ$ invariant.
 
\begin{center}
 ***
 \end{center}

The classification of reflection-symmetric TIs and TSCs (Table~\ref{reflection_table_full})
can be generalized to any order-two symmetry ($\mathbb{Z}_2$ symmetry)
and, moreover, to include the presence
of topological defects (cf.~Sec.~\ref{Defect K-theory}).
 The generalized classification can be
inferred from K-groups labeled by 6 integers
$K(s, t, d, d_{\parallel}, D, D_{\parallel})$,
where $d_{\parallel}$ ($D_{\parallel}$) is the number of momentum (spatial) coordinates 
that are flipped by the $\mathbb{Z}_2$ operation, $s$ denotes the AZ symmetry class,
$t=0,1,2,3$ labels the reflection Bott clock (Fig.\ \ref{reflection_clock}), and $(d,D)$ are
the dimensions of the defect Hamiltonian.
It was shown by \onlinecite{Sato_Crystalline_PRB14} that the generalized
classification follows from the relation 
$
K(s, t, d, d_{\parallel}, D, D_{\parallel})
=
K(s-d+D, t-d_{\parallel}+ D_{\parallel}, 0,0,0,0)
$. For reflection symmetric TIs and SCs, we have $d_{\parallel}=1$, $D_{\parallel}=0$, and $D=0$, which reproduces Table~\ref{reflection table}.

\paragraph{Bulk-boundary correspondence in topological crystalline systems}
While gapless topological surface states  exist at any boundary of TIs/TSCs protected by 
non-spatial AZ symmetries (cf.~Sec.~\ref{Bulk-boundary and bulk-defect correspondence}), this is not the case for topological crystalline materials.
TCIs/TCSs exhibit gapless modes  on only those
surfaces that are left invariant by the crystal symmetries. In other words,
the absence of gapless modes  at boundaries that break the spatial symmetries
does not indicate trivial bulk topology, and therefore cannot be used to infer
the topology of TCIs/TCSs. 
However, for  topological crystalline materials one can use the midgap states in the 
 {\it the entanglement spectrum} or in  {\it the entanglement Hamiltonian} as
 a generic way to distinguish between topological trivial and nontrivial phases~\cite{RyuHatsugai2006, Entanglement_crystalline_Chang,Entanglement_real_spectrum_Fidkowski,Fang:2012kx}.
For example, 
for TCIs/TCSs protected by inversion symmetry, 
for which there is no boundary that respects the inversion,
and hence no gapless topological state at physical surfaces,
stable gapless boundary modes in the entanglement spectrum indicate the nontriviality of the bulk topology 
\cite{Hughes:2011uq,Turner:2010qf,Turner:2012bh}.

Another difference between the boundary modes of TCIs/TCSs and those of ordinary TIs/TSCs
exists with regard to disorder. While the surface modes of TIs/TSCs with AZ symmetries are 
robust to spatial disorder (Sec.~\ref{subsec:Anderson_delocalization}), the protection of the delocalized surface modes of topological crystalline materials 
relies crucially on spatial symmetries, which typically are broken by disorder.
However, 
the gapless surface modes of TCIs/TCSs may evade Anderson localization  
when the disorder  respects the spatial symmetries {\it on average}. 
This is the case, for example, for the surface states of weak TIs in class AII in $d=3$,
which can be gapped out  by charge density wave perturbations that preserve TRS but break translation symmetry.
However, inhomogeneous perturbations due to disorder which respect translation symmetry on average
do not lead to Anderson localization of the surface states
\cite{Ringel_strong_weakTI,Weak_TI_Mong,Weak_TI_Obuse,Fulga_STI,weak_TSC}.  
Similarly,
for class AII$+R_-$ in $d=3$
the surface modes remain delocalized in the presence of disorder which preserves TRS
and respects reflection symmetry on average \cite{Fu:average_symmetry}.
The quantum spin Hall effect with spin $S_z$ conservation is another similar case: When spin $S_z$ rotation symmetry is preserved only 
on average due to disorder, 
the spin Chern number remains well-defined ~\cite{Prodan_disorder_QSH} and leads to delocalized edge modes even if TRS is broken. 
Whether the surface states of TCIs/TCSs remain delocalized in the presence of 
disorder that respects the spatial symmetries only on average depends, in general,
on the symmetry class and the spatial dimension of the system.
A more detailed discussion of this topic can be found in 
\onlinecite{Morimoto_TCI_disorder,Average_reflection_Diez}.

\paragraph{Example: 3d reflection-symmetric topological crystalline insulators (class A$+R$ and class AII+$R_-$)}
\label{3Dreflection}

Using angle-resolved photoemission spectroscopy (ARPES),
SnTe, Pb$_{1-x}$Sn$_{x}$, and Pb$_{1-x}$Sn$_{x}$Te
have been experimentally identified as TCIs protected by reflection symmetry
\cite{Tanaka:2012fk,Hsieh:2012fk,Xu2012,Dziawa:2012uq}.
The topology of these materials is characterized by non-zero mirror Chern numbers,
which leads to four surface Dirac cones, that are protected by reflection symmetry (TRS in not necessary).
For example, 
on the (001) surface of SnTe,  
the low-energy Hamiltonian 
near the high-symmetry point $\bar{X}_1=(0,\pi)$ in the surface BZ is given by 
\cite{Liu_surface_TCI,PhysRevB.88.125141} 
\begin{align}
H_{\bar{X}_1}(\mathsf{k})=( v_x k_x s_2 -  v_y k_y  s_1)\tau_0+ m s_0\tau_3 + \delta s_1 \tau_2, 
\label{reflection_surface}
\end{align}
where $v_{x,y}$ are Fermi velocities,
$s_i$ and $\tau_i$ are Pauli matrices acting on spin and A/B sublattice degrees of freedom,
respectively, and $\delta,\ m$ are small parameters. 
The Hamiltonian~\eqref{reflection_surface} preserves TRS with $T=is_2 \mathcal{K}$ and reflection symmetry in the $x$ direction. 
The reflection operator in the entire surface BZ is $k$-dependent due to the rock-salt structure of SnTe, i.e.,
$
U_{R_x}=i s_1 \otimes \mathrm{diag}\, (1, e^{ -i k_x})
$.
Near $\bar{X}_1=(0,\pi)$ the reflection operator
reduces to $U_{R_x}\approx is_1\tau_0$.
One verifies that the low-energy Hamiltonian~\eqref{reflection_surface} is indeed invariant under $R_x$,  i.e.,
$
U_{R_x}^\dagger H_{\bar{X}_1}(-k_x, k_y) U_{R_x}^{}=H_{\bar{X}_1}(k_x, k_y) 
$.
It can be checked that all gap opening perturbations are forbidden by $R_{x}$. 
(Note that on the (001) surface  of SnTe there are two additional Dirac cones located near $\bar{X}_2=(\pi,0)$, which are protected by reflection in the $y$ direction.) 

The fact that the bulk Hamiltonian of SnTe is characterized by   $\mathbb{Z}$ topological invariants (i.e., mirror Chern numbers)
can be inferred by 
considering $n$ identical copies of the surface Hamiltonian, $H_{\bar{X}_1}\otimes \openone_{n}$,
and by checking that all perturbations that (partially) gap out the enlarged surface Hamiltonian 
are prohibited by reflection symmetry with the operator $U_{R_x}\otimes \openone_{n}$. 
Furthermore, one finds that TRS breaking perturbations that respect reflection symmetry 
do not remove the gapless surface states.
In the absence of TRS, the Hamiltonian for SnTe belongs to class A$+R$. 
In the presence of TRS, we redefine $U_{R_x}\rightarrow i U_{R_x}$ to make $U_{R_x}$ hermitian.
Hence $\{U_{R_x},T\}=0$, which corresponds to class AII$+R_-$. 
As shown in Table~\ref{reflection_table_full}, class A$+R$ and AII$+R_-$ in 3d are both classified by $M\bZ$. 

In the presence of TRS (i.e., class AII$+R_-$) the
surface states of SnTe are robust against 
disorder which respects reflection symmetry \emph{on average}. 
To gain some insight into this,
consider the mass perturbation $m s_3\tau_2$ in  (\ref{reflection_surface}), 
which preserves TRS but breaks reflection. 
As shown in  \cite{Liu:2011fk,Nontrivial_surface_chiu,Hsieh:2012fk}, domain walls in $m s_3\tau_2$ 
support protected helical 1d modes.
When the mass $m$ varies randomly over the surface, 
but in a way such that reflection symmetry is preserved {\it on average}, 
domain walls and their associated helical modes appear on the entire surface,
leading to a gapless (i.e., conducting) surface. 
Further interesting features of the surface states of these TCIs,
such as instabilites towards symmetry broken phases, Lifishitz transitions, and Landau level spectroscopy, etc., 
have been investigated in
\cite{Liu_surface_TCI,Surface_TCI_Safaei,Surface_TCI_Fu,Valla_TCI_surface,Rohlfing_TCI_surface,Duan_TCI_surface,Hsieh:2012fk,Dirac_node_TCI_Okada,Liu_TCI,Wojek_TCI_surface,PhysRevB.88.125141,PhysRevLett.112.046801}. 

Recently, it has been proposed that the anti-perovskite materials 
Ca$_3$PbO and Sr$_3$PbO also realize a reflection symmetric 
TCI~\cite{OgataJPSJ2011,TCI_Fu_antiperovskites}. 
Furthermore, 
it was shown that   TlBiS$_2$ turns into a TCI with mirror symmetry upon applying pressure~\cite{Zhang_TCI}.

\subsection{TCIs and TCSs protected by other point-group symmetries}

Besides reflection symmetry, other point-group symmetries can also give rise to new TCIs. 
For example, TCIs protected by $C_n$ point-group symmetries 
\cite{Fang:2012kx,liu_law_PRB_14,Chen_chernN_GSP,Fu_first_TCI}
and  $C_{nv}$ point-group symmetries \cite{alexandradinata_bernevig_PRL14}
have recently been discussed. It has been argued that
 graphene on a BN substrate is a possible candidate for a TCI protected by $C_3$ rotation symmetry~\cite{Niu_point_symmetry}. A monolayer of PbSe has been proposed to realize a TCI
  protected by a combination of mirror and $C_2$ rotation symmetry~\cite{2D_TCI}.
  Inversion symmetric TCIs have been considered by \onlinecite{LuLee2014}.
TCIs protected by magnetic symmetry groups have been investigated by \onlinecite{zhang_chao-xing_PRB_15}.  
A partial classification of TCIs protected by space group symmetries has been developed by \onlinecite{Slager:2013fk}. 
The classification of 2d gapless surfaces on 3d TCIs has been completed by \onlinecite{2D_classification_Liu} 
by investigating all 17 2d space groups.

As for TCSs, 
TCSs in 2d with discrete rotation symmetries have been discussed by 
\onlinecite{Teo_2013_disclination,Teo_disclination_class}.
TCSs protected by magnetic  symmetry groups
\cite{Magnetic_group_Fang} 
and by
$C_3$ symmetry
\cite{TSC_C3}
have also been studied.
Finally, there are also TCSs which are protected by a combination of
PHS and reflection symmetry
\cite{Sato_Mirror_SC,Kane_Mirror,Mirror_Majorana_Sato, YaoRyu2013,TSC_reflection_Kotetes},
cf.\ Table~\ref{reflection_table_full}.
Majorana gapless modes on the surfaces of these TCSs are protected by reflection.

\section{Gapless topological materials}
\label{sec:gapless_materials}


By definition, Fermi surfaces, Fermi points, and nodal lines
are sets of zeros of the energy dispersion, $\varepsilon(\mathsf{k})=\rm{const.}$, in momentum space. 
For simplicity, all these objects will be collectively called Fermi surfaces (FSs) in the following. 
When an FS exists at any energy, the (bulk) system is gapless. 
FSs are said to be  topologically stable (or simply ``stable"), 
when they cannot be fully gapped by perturbations that are {\it local} in momentum space and small, 
such that the bulk gap remains intact sufficiently far away from the FS.
(The precise meaning of ``local'' here will be elaborated shortly).
In this section, we review topological classifications of stable FSs
that appear in gapless (semi-)metals and nodal SCs 
\cite{HoravaPRL05, Sato_Crystalline_PRB14,ZhaoWangPRL13,ZhaoWangPRB14,ChiuSchnyder14,matsuuraNJP13,Volovik:book,volovikLectNotes13}.
As we will see, the classification of gapless topological materials and fully gapped TIs/TSCs can be developed along parallel lines.

 It should be stressed that in lattice systems it is only meaningful to discuss the stability for a ``single'' FS (i.e., of one FS that is
 ``isolated'' from the other FSs in the BZ).
That is, we consider FSs that are located only within a part of the BZ, but do not include all FSs in the entire BZ.
This is so since, for any lattice system, 
it is expected that FSs can be gapped pairwise by {\it nesting},
i.e., by including perturbations that connect different FSs. 
Thus, FSs are at best only locally stable in momentum space,
i.e., robust against
perturbations that are smooth in real space and slowly varying on the scale of the lattice.
This is closely related to the fermion doubling theorem 
\cite{Nielsen_Ninomiya_1981}, 
from which it follows that FSs with non-trivial topological charges in any lattice system are always accompanied by  ``partners''
with opposite topological charges.
Hence, the sum of the topological charges of all FSs in a compact BZ adds up to zero.
As a consequence of this, the topological invariants for FSs are defined 
in terms of an integral along a submanifold of the BZ, and not 
in terms of an integral over the entire BZ as in the case of TIs/TSCs.

We start by reviewing the classification of stable FSs 
protected by  non-spatial AZ symmetries,
and then describe how this classification is modified and extended
in the presence of additional crystal symmetries, such as reflection.   
The properties of theses topologically stable FSs are illustrated by selected examples.

\subsection{Ten-fold classification of gapless topological materials}
\label{sectionVA}

The topological classification of gapless materials depends on the symmetry class of the Hamiltonians 
and the codimension $p$ of the FS,  
\begin{eqnarray}
p =  d  - d_{\mathrm{FS}}, 
\end{eqnarray} 
where $d$ and $d_{\mathrm{FS}}$ denote the dimension of the BZ and the ``minimal" dimension of the FS, respectively. 
Since the dimension of the FS can be different for different Fermi energies, we define here
$d_{\mathrm{FS}}$  as the 
the dimension of the band crossing, which is independent of the Fermi energy.
In other words, $d_{\mathrm{FS}}$ is the smallest possible dimension (i.e.,  the ``minimal dimension") of the Fermi surface,
as the Fermi energy is varied.%
\footnote{If necessary, the energy bands should be adjusted without changing the topology to reach the minimal dimension of the FS. For example, although a type II Weyl node does not possess a 0d FS~\cite{Soluyanov:2015aa}, the node can be continuously deformed into a type I Weyl node. Hence, $d_{\rm{FS}}=0$.                 
}
For example, for Weyl semi-metals
$d_{\rm{FS}}=0$, since the Fermi surface is either  0d (when the Fermi energy is at the Weyl node) or 2d
(when the Fermi energy is away from the band crossing).
Furthermore, we note that $p\le d$ since $d_{\rm{FS}}$ cannot be negative. 

\begin{figure}[tb]
 \begin{center} 
\includegraphics[clip,width=0.60\columnwidth]{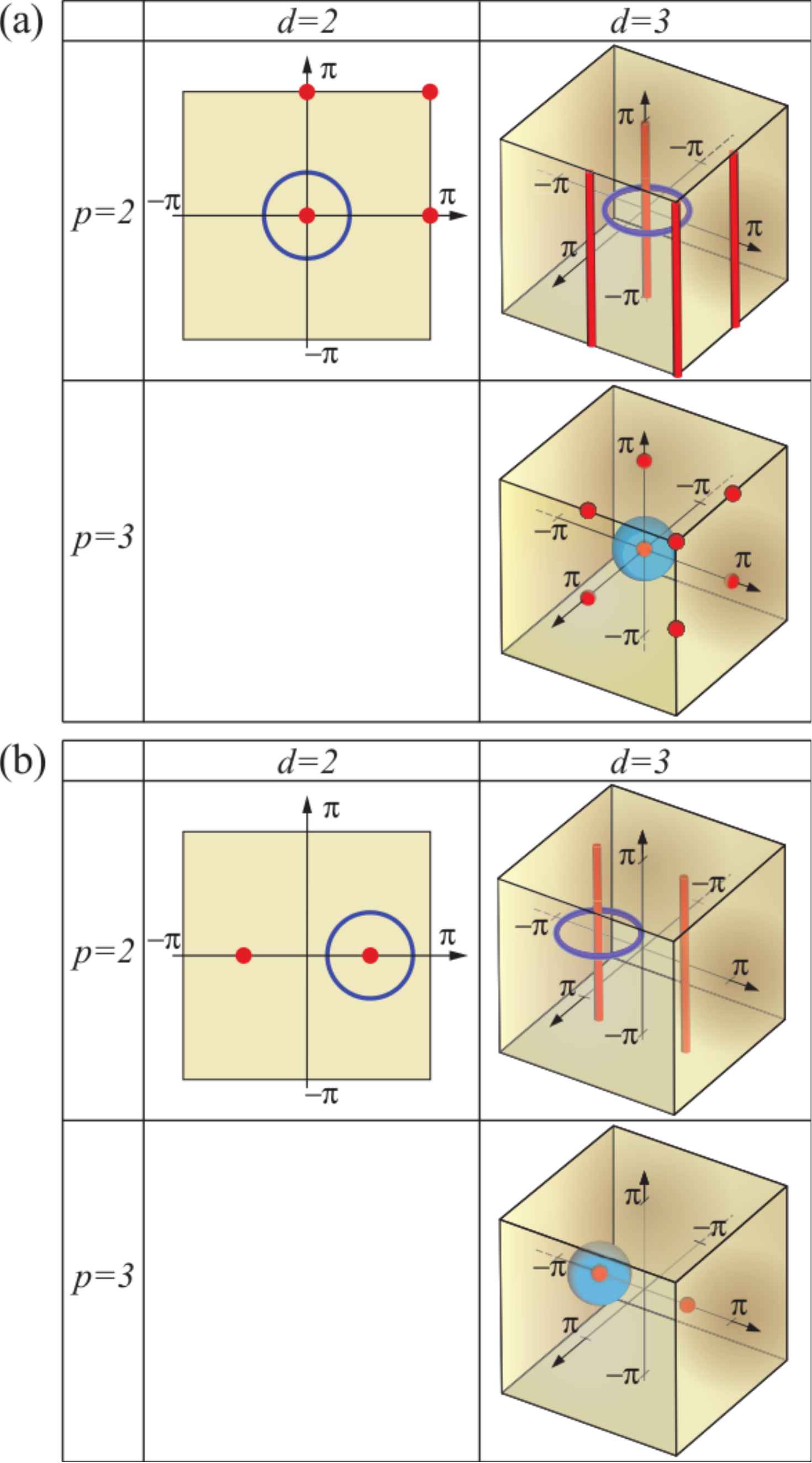}
\hfill
 \end{center}
 \vspace{-0.45cm}
 \caption{(Color online) 
The classification of stable Fermi surfaces depends on
how the Fermi surfaces transform under non-spatial antiunitary symmetries, 
and hence their location in the Brillouin zone. 
Here, $d$ denotes the spatial dimension (the dimension of the Brillouin zone) 
and $p$ is the codimension of the Fermi surface. 
The blue  circles/spheres represent the contour on which the topological invariant is defined.
(a)~Each Fermi surface (red point/line) is left invariant under non-spatial symmetries. 
(b) Different Fermi surfaces
are pairwise related by the non-spatial symmetries which map $\mathsf{k} \leftrightarrow -\mathsf{k}$.  Adapted from \cite{ChiuSchnyder14}.
}
 \label{FSwithGlobalSym}
\end{figure}

For the classification of topological FSs, we need to distinguish whether or not the
FSs are left invariant by the non-spatial AZ symmetries~\cite{matsuuraNJP13}.
I.e., two different cases have to be examined (Fig.~\ref{FSwithGlobalSym}): 
(i)~each individual FS is left invariant under anti-unitary AZ symmetries (``FS1'')
\cite{Sato_Crystalline_PRB14,ZhaoWangPRL13,ZhaoWangPRB14}, 
and (ii) different FSs are pairwise related to each other by AZ symmetries (``FS2")
\cite{ChiuSchnyder14,matsuuraNJP13}.
Note that in cease (i) the FSs must be  located at  high-symmetry points of the BZ, which are 
invariant under $\mathsf{k}\rightarrow -\mathsf{k}$.

\subsubsection{Fermi surfaces at high-symmetry points (FS1)}

The complete ten-fold classification of stable FSs 
that are located at high-symmetry points (i.e., of FSs which are left invariant under AZ symmetries)
is shown in Table~\ref{real symmetry},
where the firs row (``FS1") indicates the codimension $p$ of the FS
\cite{matsuuraNJP13,HoravaPRL05,ZhaoWangPRB14,ZhaoWangPRL13,Sato_Crystalline_PRB14,ChiuSchnyder14}.
We observe that this classification is related to the periodic table of gapped TIs and TSCs (Table~\ref{tab:classification})
by a dimensional shift.
It is important to point out that for a given symmetry class and codimension $p$, a $\bZ$-type topological invariant guarantees the stability of the FS independent of $d_{\mathrm{FS}}$. A $\bZ_2$-type topological number, on the other hand, only protects FSs with $d_{\mathrm{FS}}=0$, i.e., Fermi points. 
By the bulk-boundary correspondence,  gapless topological materials support protected boundary states, which, depending on the case, are either Dirac or Majorana cones, dispersionless flat bands, or Fermi arc surface states, etc.
(See below for examples.)

\paragraph{Example: 2d nodal SC with TRS ($p=2$, class DIII)}

As an example of stable point nodes in a SC, 
let us consider the following 2d Hamiltonian on the square lattice,
\begin{align}
\label{exampDIII}
H(\mathsf{k})=\sin k_x \sigma_1 + \sin k_y \sigma_2, 
\end{align}
which belongs to class DIII, since it preserves TRS and PHS 
with $T=\sigma_2 \mK$ and $C=\sigma_1 \mK$
($T^2=- \openone$ and $C^2 = + \openone$). 
This SC exhibits four point nodes ($d_{\mathrm{FS}}=0$, $p=2$) 
at the four TR invariant momenta $(0,0),\ (0,\pi),\ (\pi,0)$, and $(\pi,\pi)$.
According to Table~\ref{real symmetry}, these point nodes are
 protected by an integer topological invariant,
which takes the form of the winding number 
\eqref{winding number examples},
$
\nu =({i}/{2\pi})\int_{\mathcal{C}}q^*dq 
$,
where the closed contour $\mathcal{C}$ encircles one of the four nodal points
and
$q(\mathsf{k})=(\sin k_x -i \sin k_y)/\sqrt{\sin^2 k_x+\sin^2 k_y }$.
One finds that
$\nu=+1$ for the nodes at $(0,0)$ and $(\pi,\pi)$, whereas $\nu=-1$ for the nodes at $(0,\pi)$ and $(\pi,0)$.
(The contour integral is performed counterclockwise.) 
The topological nature of these point nodes results in the appearance of protected
flat-band edge states at all surfaces, except the (10) and (01) faces. 
These flat-band states connect two nodal points with opposite winding numbers  in the boundary BZ.


\begin{table}[t!]
\begin{ruledtabular}
\begin{tabular}{c|cccccccc}

                  \mbox{FS1}  &{$p$=8}  & {$p$=1} & {$p$=2} & {$p$=3} & {$p$=4} & {$p$=5} & {$p$=6} & {$p$=7}   \\
                \mbox{FS2}    &{$p$=2} & {$p$=3} & {$p$=4} & {$p$=5} & {$p$=6} & {$p$=7} & {$p$=8} & {$p$=1}  \\
                 \mbox{TI/TSC}    & {$d$=1} & {$d$=2} & {$d$=3} & {$d$=4} & {$d$=5} & {$d$=6} & {$d$=7} & {$d$=8}  \\
\hline  
     A   &   0 & $\mathbb{Z}$ & 0 & $\mathbb{Z}$ & 0 & $\mathbb{Z}$ & 0    & $\bZ$            \\
  AIII    &  $\mathbb{Z}$ & 0 & $\mathbb{Z}$ & 0 & $\mathbb{Z}$ & 0 & $\mathbb{Z}$  & 0           \\ 
\hline 
  AI     & 0 & $0^a$ & 0 & $2\mathbb{Z}$ & 0 & $\mathbb{Z}_2^{a,b}$ & $\mathbb{Z}_2^b$   &  $\mathbb{Z}$      \\
  BDI    & $\mathbb{Z}$ & 0 & $0^a$ & 0 & $2\mathbb{Z}$ & 0 & $\mathbb{Z}_2^{a,b}$   & $\mathbb{Z}_2^b$     \\
  D   & $\mathbb{Z}_2^b$ & $\mathbb{Z}$ & 0 & $0^a$ & 0 & $2\mathbb{Z}$ & 0  & $\mathbb{Z}_2^{a,b}$        \\
  DIII      & $\mathbb{Z}_2^{a,b}$ & $\mathbb{Z}_2^b$ & $\mathbb{Z}$ & 0 & $0^a$ & 0 & $2\mathbb{Z}$ & 0      \\
  AII      & 0 & $\mathbb{Z}_2^{a,b}$ & $\mathbb{Z}_2^b$ & $\mathbb{Z}$ & 0 & $0^a$ & 0   & $2\mathbb{Z}$     \\
  CII      & $2\mathbb{Z}$ & 0 & $\mathbb{Z}_2^{a,b}$ & $\mathbb{Z}_2^b$ & $\mathbb{Z}$ & 0 & $0^a$   & 0     \\
  C   &   0 & $2\mathbb{Z}$ & 0 & $\mathbb{Z}_2^{a,b}$ & $\mathbb{Z}_2^b$ & $\mathbb{Z}$ & 0  & $0^a$         \\
  CI    &  $0^a$ & 0 & $2\mathbb{Z}$ & 0 & $\mathbb{Z}_2^{a,b}$ & $\mathbb{Z}_2^b$ & $\mathbb{Z}$   & 0  \\
\end{tabular}
\end{ruledtabular}
\caption{
Classification of stable Fermi surfaces
in terms of the ten AZ symmetry classes, which are listed in 
the first column.  
The first and second rows (``FS1'' and ``FS2'') give the codimension $p =  d  - d_{\mathrm{FS}}$ 
for Fermi surfaces at high-symmetry points [Fig.~\ref{FSwithGlobalSym}(a)] and away from 
high-symmetry points of the BZ [Fig.~\ref{FSwithGlobalSym}(b)], respectively.
The classification of stable Fermi surfaces is related to the classification of gapped topological insulators and superconductors (the third row) by a simple dimensional shift.
For entries labelled by the superscript ``$a$", 
there can exist surface states and bulk Fermi surfaces of type ``FS2"
that are protected by $\bZ$ invariants inherited from class A or AIII, 
since in these cases TRS or PHS does not trivialize the $\bZ$  invariants. 
Also note that $\bZ_2$ topological invariants only protect Fermi surfaces of dimension zero at high-symmetry points. That is,
$\bZ_2$ topological numbers cannot protect  Fermi surfaces located away from high-symmetry points.
 This is indicated by the superscript ``$b$" in the table.
}
\label{real symmetry}
\label{original table}
\end{table}

\subsubsection{Fermi surfaces off high-symmetry points (FS2)}
\label{FermiSurfOffHighSym}

The classification of stable FSs that are located away from 
high-symmetry points of the BZ  is shown in Table~\ref{real symmetry},
where the second row (``FS2") gives the codimension $p$ of the FS.
We remark that
only $\bZ$ invariants can guarantee the stability of FSs away from high symmetry points.
$\bZ_2$ indices, on the other hand, cannot protect these FSs, but  
 they may lead to the appearance of zero-energy surface states at 
high-symmetry points of the boundary BZ~\cite{ChiuSchnyder14}. 
It is important to note that, in contrast to the classification of fully gapped systems, 
the label ``0" in Table~\ref{real symmetry}  does not always indicate trivial topology. 
That is, for entries with the superscript ``$a$" 
there can exist surface states and stable bulk FSs that are protected by the $\bZ$ invariants
inherited from class A and AIII. I.e., in these cases, the $\bZ$ invariants
are not required to be zero in the presence of TRS or PHS.

%

In experimental systems, the FSs are usually positioned away from the high-symmetry points of the BZ.
Indeed, there are numerous experimental examples of protected FSs off high-symmetry points, 
such as Weyl point nodes protected by a Chern number
in superfluid $^3$He A phase (class A) 
\cite{Volovik3HeA}
and in chiral ($d \pm id$)-wave  SCs~\cite{chiral_p_wave_fischer,goswami_arXiv13},
point nodes   in $d_{x^2-y^2}$-wave SCs 
protected by a winding number
\cite{RyuHatsugaiPRL02}, 
and line nodes 
in nodal noncentrosymmetric SCs
protected by a winding number
\cite{beriPRB10,SatoPRB06,SchnyderRyuFlat,BrydonSchnyderTimmFlat}. 
 In order to illustrate some of the properties of these gapless topological materials let us consider
two examples in more detail, namely, protected point nodes in Weyl semimetals
and unprotected Dirac nodes in a 3d TR symmetric semimetal.

\paragraph{Example: Weyl semimetal (\texorpdfstring{$p=3$}{p=3}, class A)} 
\label{Weyl section}

The point nodes of 3d Weyl semimetals 
are a canonical example of 
gapless topological  bulk modes located away from high-symmetry points. 
These bulk modes are linearly-dispersing Weyl fermions, 
which are robust without requiring any symmetry protection 
\cite{Murakami2007,BurkovBalentsPRB11,burkovBalenstPRL11,WanVishwanathSavrasovPRB11,Ashvin_Weyl_review}. 
The generic low-energy Hamiltonian for a Weyl node located at $\mathsf{k}^0=(k_x^0,k_y^0,k_z^0)$ 
is given by 
\begin{align}
H_{\rm{Weyl}}(\mathsf{k})=
\sum_{i,j=1,2,3}v_{ij}(k_i-k_i^0)\sigma_j, 
\label{Weyl_eq}
\end{align}
where $v_{ij}$ denotes the Fermi velocity. 
Weyl nodes cannot be gapped out, 
since there exists no ``fourth Pauli matrix" that anticommutes with $H_{\rm{Weyl}}$. 
A Weyl node is characterized by its chirality 
$
\chi_{\mathsf{k}^0}={\rm sgn}(\det(v_{ij}))=\pm 1,
$
which measures the relative handedness 
of the three momenta $\mathsf{k} - \mathsf{k}_i^0$ with respect to the Pauli matrices $\sigma_j$ in~\eqref{Weyl_eq}.

In a lattice model, Weyl nodes must come in pairs with opposite chiralities
\cite{Nielsen_Ninomiya_1981}. 
Let us demonstrate how Weyl nodes arise in a simple four-band lattice model,
and show that Weyl semimetals support {\it Fermi arc surface states}, which
connect the projected bulk Weyl nodes with opposite chiralities
in the surface BZ. 
To that end, consider the following cubic-lattice Hamiltonian describing a four-band semimetal with two Dirac points
\begin{align} \label{semimetalHam}
H(\mathsf{k})=\sin k_x \tau_1 s_1+ \sin k_y \tau_1 s_2 + M(\mathsf{k}) \tau_3 s_0,
\end{align}
where the two sets of Pauli matrices 
$\tau_\alpha$ and $s_\alpha$ operate in spin and orbital spaces, respectively,
and 
$M(\mathsf{k})=\cos k_x + \cos k_y + \cos k_z-m$. 
For concreteness, we set $m=2$. 
With this choice, the bulk Dirac points of $H(\mathsf{k})$ 
are located at
$\mathsf{k}_\pm=(0,0,\pm \pi/2)$. 
The {\it Dirac semimetal} \eqref{semimetalHam} preserves TRS and inversion symmetry
with $T=\tau_0 s_2 \ck$ and $U_I=\tau_0s_3$, respectively. When one of these two symmetries is broken, a Dirac node can be separated into two Weyl nodes. 
For example, a Zeeman term $\Delta\tau_0s_3$ (with $\Delta=1/2$ for simplicity), which breaks TRS,
separates the two Dirac cones into  
four Weyl nodes located at $\mathsf{k}^0=(0,0,\pm\pi/3)$ and $\mathsf{k}^0=(0,0,\pm2\pi/3)$.  
These Weyl points realize (anti-)hedgehog defects of the vector of the Berry curvature 
($\mathrm{Tr} (\mathcal{F}_{ij})\epsilon^{ijl} d{k}^l$) 
[see right part of Fig.~\ref{fig:TISCFM}(b)], 
and are protected by the nonzero Chern number  
\begin{align}
&\label{Weyl top inv.}
{\rm Ch}(\mathcal{N}_{{\mathsf{k}^0}})
:=\frac{i}{2\pi}
\int_{\mathcal{N}_{\mathsf{k}^0} }
\mathrm{Tr}\, (\mathcal{F})
\nonumber \\
&
\quad =\left\{ 
  \begin{array}{l l}
    +1,&  \mbox{for} \quad \mathsf{k}^0=(0,0,-\frac{\pi}{3}),(0,0,\frac{2\pi}{3}) \\
    -1, &     \mbox{for} \quad \mathsf{k}^0=(0,0,-\frac{2\pi}{3}),(0,0,\frac{\pi}{3})
  \end{array} \right.  , 
\end{align}
where the integral is over  a small closed surface $\mathcal{N}_{\mathsf{k}^0} $ surrounding the Weyl node at  $\mathsf{k}^0$.
We observe that the chiralities $\chi_{\mathsf{k}^0}$ of the Weyl nodes,
which can be computed from the low-energy description \eqref{Weyl_eq}, 
are identical to the topological invariant, i.e.,
${\rm Ch}(\mathcal{N}_{{\mathsf{k}^0}})=\chi_{\mathsf{k}^0}$, where  $\mathcal{N}_{\mathsf{k}^0} $ encloses a single Weyl point.
In general, the integral topological invariant ${\rm Ch}(\mathcal{N}_{{\mathsf{k}^0}})$
counts
the number of Weyl points 
within $\mathcal{N}_{\mathsf{k}^0}$ 
weighted by their chiralities.
Two Weyl nodes  with opposite chiralities at the same momentum in the BZ can be easily gapped out by local perturbations.
However, when the two Weyl nodes are located at different momenta,
nesting instabilities that gap out the Weyl nodes carry finite momentum, and hence 
necessarily break translation symmetry. 
Therefore, 
as long as translation symmetry is preserved, Weyl nodes are robust.
Even in the presence of disorder
which is sufficiently smooth on the scale of the lattice 
and does not induce scattering between Weyl nodes with opposite chiralities,
the Weyl points are protected and do not Anderson localize.

\begin{figure}
\begin{center}
\includegraphics[clip,width=0.85\columnwidth]{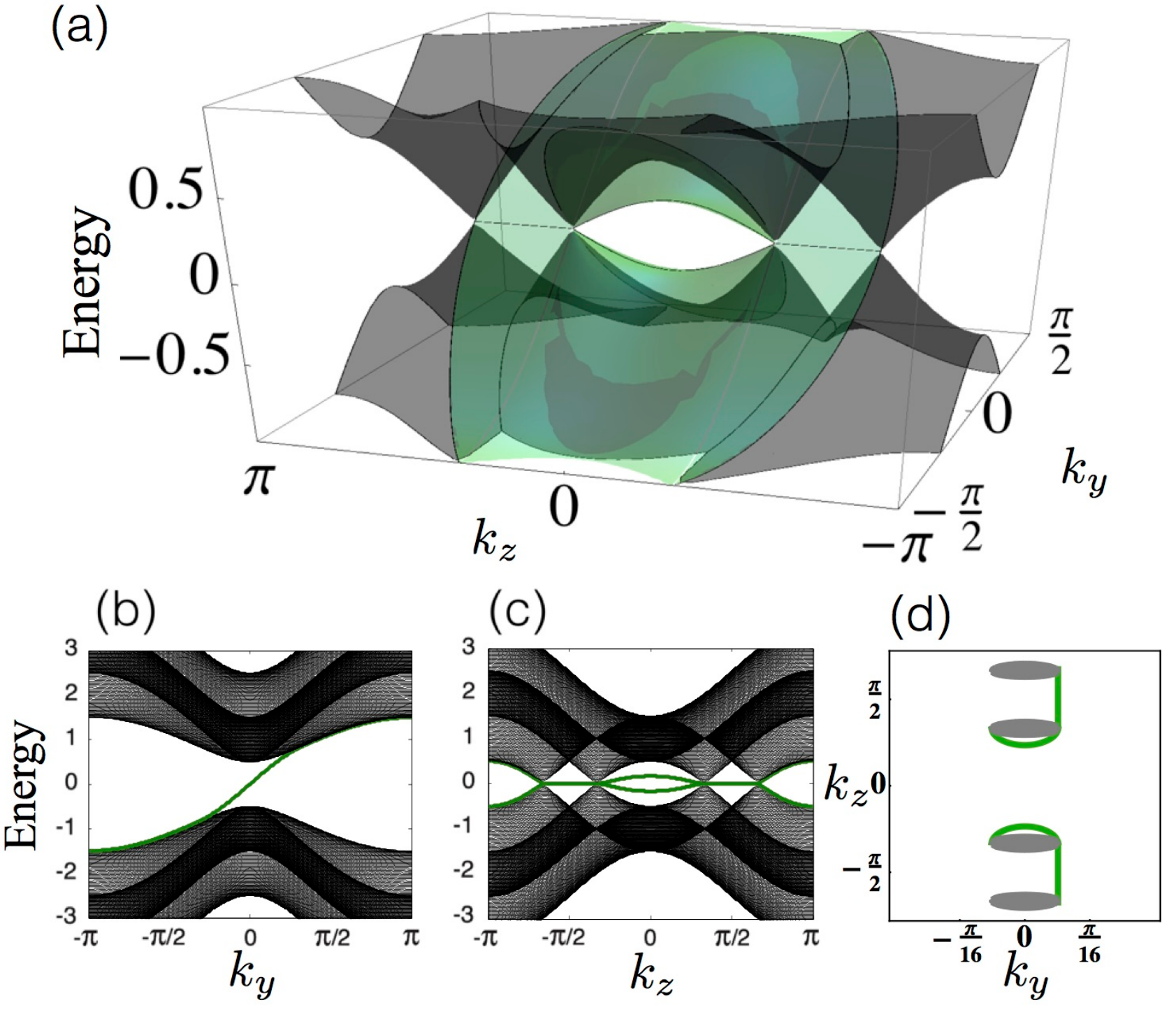}
\end{center}
  \caption{\label{surfaceWeyl} 
Surface spectrum of the Weyl semimetal~\eqref{semimetalHam}  for the (100) face
in the presence of the Zeeman term $\Delta\tau_0s_3$.  
The surface and bulk states are colored in green and gray, respectively.
(a) Surface spectrum as a function of surface momenta $(k_y,k_z)$. 
(b), (c) Surface spectrum as a function of surface momentum $k_y$ and $k_z$ with fixed $k_z=\pi/4$  
and $k_y=0$, respectively. 
(d) Bulk Fermi surface and surface Fermi arc at the energy $E=0.1$. 
The Fermi arcs located within the interval $\pi/3<|k_z|< 2\pi/3$ 
are protected by the non-zero Chern number ${\rm Ch}(k_z)=-1$, see Eq.~\eqref{WeylChernNo}.
The surface modes with $|k_z|<\pi/3$ are unstable
and can be gapped out by surface perturbations 
(e.g., by the term $ \cos(3k_z/2)\tau_1s_3$). 
}   
\end{figure}

As seen from Eq.\ (\ref{Weyl top inv.}),
Weyl nodes are sources and drains of  Berry flux, i.e., there is 
a Berry flux of $2 \pi$ flowing from one Weyl node to another along the $k_z$ direction,
which is measured by the Chern number~\eqref{Weyl top inv.}.
To exemplify this, consider a family of planes, 
$\{\mathcal{N}(k_z)\}$,
which are perpendicular to the $k_z$ axis and parameterized by $k_z$.
When $k_z$ is in between a pair of Weyl nodes with opposite chiralities, 
$\mathcal{N}(k_z)$ has a non-zero Chern number
\begin{align} \label{WeylChernNo}
&{\rm Ch}(k_z)
:=\frac{i}{2\pi}\int_{\mathcal{N}(k_z)} \mathrm{Tr}\,  \left[ \mathcal{F}(k_z) \right]
\nonumber \\
&\quad 
=\left\{ 
  \begin{array}{l l}
    -1,&  \mbox{for}\quad \pi/3< \left| k_z \right| <2 \pi/3 \\
    0,&  \mbox{for}\quad \left| k_z \right| < \pi/3\ \&\  2\pi/3 < \left| k_z \right| 
  \end{array} \right.   .
\end{align}
Each of these planes can be interpreted as a 2d fully gapped Chern insulator with a chiral edge mode. 
Hence, the surface states of the Weyl semimetal form a 1d open Fermi arc in the surface BZ, 
connecting the projected bulk Weyl nodes with opposite chiralities, see Fig.~\ref{surfaceWeyl}.
These chiral surface states give rise to a quantum anomalous Hall effect, 
with the Hall conductivity proportional to the separation of Weyl nodes with opposite chiralities in momentum space.  
A number of other exotic transport phenomena have been also discussed for Weyl semimetals, 
including negative magnetoresistance, nonlocal transport, chiral magnetic and vortical effects
\cite{Burkov_Weyl_electromagnetic_2012,Hosur_Weyl_develop,Lu_anomaly_Weyl_2013,Sid_anomaly_Weyl,Marcel_Weyl_response}.
 
An alternative way to create Weyl nodes in the Hamiltonian (\ref{semimetalHam}) is to 
break inversion symmetry by adding $\sin k_z \tau_1 s_3$ 
(which, however, preserves reflection symmetry and TRS). 
The resulting four Weyl nodes are located at $(0,0,\pm\pi/4)$ and $(0,0,\pm 3 \pi/4)$, and are robust in the absence of scattering between these nodes.
The Weyl nodes are protected by a $\bZ$ topological invariant, even though the Hamiltonian \eqref{semimetalHam} itself
belongs to class AII with $p=3$ (see footnote ``$a$" in Table~\ref{real symmetry} for more details). 
In the presence of TRS the number of Weyl nodes with chirality $\pm 1$ is always a multiple of 4 due to the vanishing Chern numbers on TR symmetric planes. 
Note that this TR symmetric Weyl semimetal exhibits besides the arc surface states
also Dirac surface states  
at $k_z=0,\ \pi$ which are protected by a $\bZ_2$ topological invariant (cf.~discussion in the example below).
Thus, this is an example of a gapless topological material with surface states that are protected by a different
invariant than the bulk nodes.

Over the last few years a number of materials with Weyl nodes in their band structure have been investigated.
For example, the transition-metal monophosphide TaAs is an experimental realization of
a TR symmetric Weyl semimetal.  
Based on first-principle calculations,
this material was theoretically identified 
to be an inversion-symmetric Weyl semimetal~\cite{TaAs_Weng,Huang_Hasan_Weyl},
which was later confirmed by ARPES experiments~\cite{Weyl_discovery_TaAs,Xu_Weyl_2015_first}.
Magnetotransport measurements on TaAs have revealed a negative megnetoresistance, which
is a signature of the chiral anomaly of Weyl semimetals~\cite{Huang_Weyl_2015,Zhang_anomaly_Weyl_2015}.
Other experimental realizations of TR symmetric Weyl semimetals are 
TaP, NbAs, and NbP \cite{TaAs_Weng,NbP_weyl_magnetotransport_arXiv1502,Weyl_NbAs_Xu}. 
A Weyl phase with broken TRS has been theoretically proposed to exist in 
 pyrochlore iridates \cite{WanVishwanathSavrasovPRB11,witczak_kim_weyl_2012,chen_hermele_weyl}, 
magnetically doped TIs, and  TI multilayers \cite{burkovBalenstPRL11}.
However, these  TRS breaking Weyl semimetals have not yet been discovered experimentally. 
A double Weyl semimetal, where the Weyl nodes have chiralities $\chi_{\mathsf{k}^0} = \pm 2$,
has been predicted to be realized
in the ferromagnetic spinel HgCr$_2$Se$_4$ \cite{HgCrSe_Weyl_2011}. 
The conditions for the existence of double Weyl nodes was recently discussed 
by \onlinecite{Fang_bernevig_multi_weyl_PRL_12}.
Furthermore, the band structure of photonic crystals can be designed in such a way that it exhibits Weyl nodes \cite{Lu_photonic_Weyl_2015,Lu_early_photonic_Weyl}.


\paragraph{Example: 3d Dirac semimetal ($p=3$, class AII)}

\begin{figure}
\begin{center}
\includegraphics[clip,width=1.0\columnwidth]{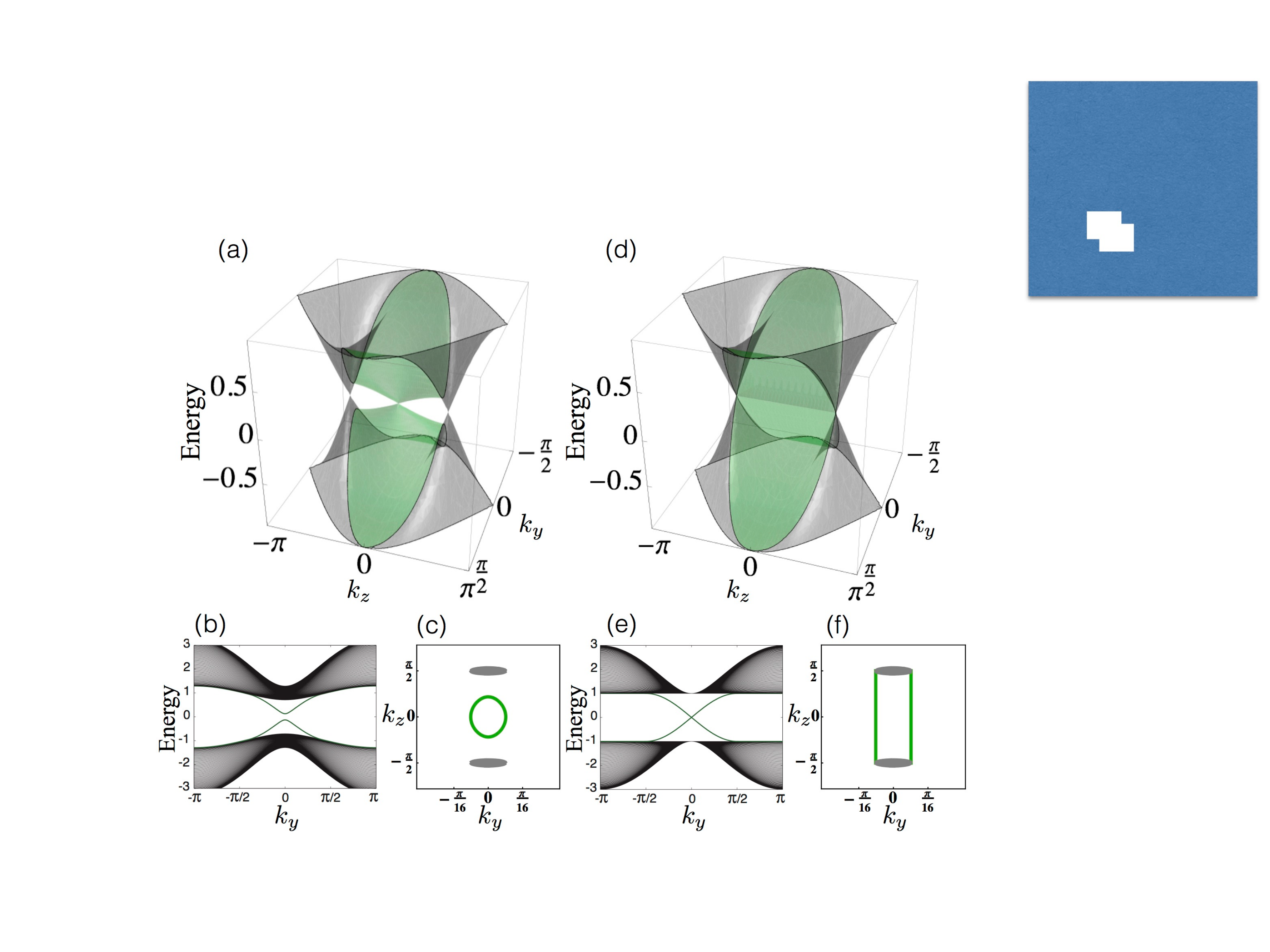}
\end{center}
  \caption{\label{surfacBandStruct2D} 
Surface spectrum of the time-reversal symmetric  (i.e., without Zeeman term) 
Dirac semimetal~\eqref{semimetalHam} for the (100) face.  
The surface and bulk states are colored in green and gray, respectively.
Panels (a)-(c) and (d)-(f) show the surface spectrum in 
the presence and absence of the surface perturbation $+g\sin k_z \tau_1 s_3$, respectively, which
breaks reflection and chiral symmetry.
Panels (b) and (e) show the surface bands as a function of surface momentum $k_y$ with fixed $k_z = \pi/4$ and $k_z = 0$, respectively. Panels (c) and (f) show the
bulk and surface states for a fixed energy $\varepsilon=0.1$.
The surface Dirac cone of panel (a) is 
protected by a $\mathbb{Z}_2$ invariant, while the surface Fermi arc 
of panel (d) is protected by the mirror winding number $\nu^+$ (see Sec.~\ref{p3 DIII}).
}   
\end{figure}

As a second example we consider  Hamiltonian \eqref{semimetalHam} with two Dirac points, 
which are located away from high-symmetry momenta in the BZ, i.e.\ at $(0,0,\pm \pi/2)$,
and impose TRS with $T=\tau_0 s_2 \ck$. 
Although a $\mathbb{Z}_2$ invariant can be defined for this case,
these Dirac points are not protected by TRS (Table \ref{original table}),
since there exists a TRS preserving mass term, namely $\sin k_z \tau_2 s_0$.
While the class AII $\mathbb{Z}_2$ invariant does not guarantee the stability of the bulk Dirac points, 
it nevertheless leads to  
 protected gapless surface states at high-symmetry momenta of the surface BZ. 
To see this, we first need to remove some accidental symmetries of  (\ref{semimetalHam})
that also give rise to protected surface states (see discussion in Sec.~\ref{p3 DIII}).
These accidental symmetries are  reflection with $R_y=\tau_3 s_2$ [cf.~Eq.~\eqref{reflectSym}] and    
 chiral symmetry with $S=\tau_1 s_3$. Both of these accidental symmetries
can be broken on the surface by adding the perturbation $+g\sin k_z \tau_1 s_3$ on the (100) and  ($\bar{1}$00) faces.
In the presence of this perturbation  
 the surface states are gapped 
except at $k_z=0$, where there exists 
a helical mode protected by TRS and the $\mathbb{Z}_2$ invariant of class AII,  see Figs.~\ref{surfacBandStruct2D}(a)-(c). 
This type of helical surface mode has been observed by ARPES in the Dirac semimetal Na$_3$Bi 
\cite{Xu18122014}. 

%

\subsection{Topological semimetals and nodal superconductors protected by reflection symmetry}
\label{sectionVB}

Let us now discuss how the classification of stable FSs is enriched by the presence of reflection symmetry \cite{ChiuSchnyder14}.
Similar to the classification of fully gapped TCIs and TCSs (cf.~Sec.~\ref{secReflectClass}), one needs to distinguish whether 
the reflection operator commutes or anticommutes with the operators of the AZ symmetries \cite{ChiuSchnyder14}.
The classification of 
reflection-symmetry-protected semimetals and nodal SCs also depends
on the codimension of the FSs, 
 $p = d - d_{\mathrm{FS}}$,
and on how the FSs transform
under reflection and AZ symmetries. 
In general, one distinguishes the following three different situations: 
(i) Each FS is left invariant by both reflection and AZ symmetries;
(ii) FSs are invariant under reflection symmetry, but are pairwise related to each other by the internal symmetries;
and 
(iii) different FSs are pairwise related to each other by both reflection and  AZ symmetries.
In cases (i) and (ii), the FSs are located within a reflection plane,
whereas in case (iii) they lie outside the reflection plane.
For brevity we focus here only on case (i) and  (ii). 
Case (iii) has been discussed extensively in Refs.~\onlinecite{morimotoFurusakiPRB14,ChiuSchnyder14}.

\subsubsection{Fermi surfaces at high-symmetry points within mirror plane (FS1 in mirror)}

First we consider case (i), where
the FSs are located within a reflection plane and at high-symmetry points in the BZ.
In this situation the classification of stable FSs with $d_{\mathrm{FS}}= 0$ can be inferred 
from the classification of TIs and TSCs protected by reflection by a dimensional reduction procedure. 
Namely, the surface states of reflection symmetric $d$-dimensional TIs/TSCs
can be viewed as reflection-symmetry-protected FSs in $d-1$ dimensions. 
It then follows that the classification of stable Fermi points ($d_{\mathrm{FS}}= 0$) is obtained
from the classification of reflection symmetric TIs/TSCs by a dimensional shift 
$d\to d -1$, see Table~\ref{reflection_table_full}.
This logic also works for FSs with $d_{\mathrm{FS}} > 0$, if their stability is guaranteed
by an $M\mathbb{Z}$ or $2 M\mathbb{Z}$ topological number.
However, $\mathbb{Z}_2$ and $M\mathbb{Z}_2$ 
topological numbers ensure only the stability of Fermi points, i.e., FSs with $d_{\mathrm{FS}} = 0$.
Derivations based on Clifford algebras and K-theory
\cite{Sato_Crystalline_PRB14,ChiuSchnyder14}
corroborate these findings.

\subsubsection{Fermi surfaces within mirror plane but off high-symmetry points (FS2 in mirror)}\label{off high in mirror}

In case (ii), the FSs transform pairwise into each other by AZ symmetries, which relate $\mathsf{k}$ and $-\mathsf{k}$.
Using an analysis based on the minimal-Dirac-Hamiltonian method
\cite{ChiuSchnyder14} it was shown  that only $M \mathbb{Z}$
and  $2 M \mathbb{Z}$ topological numbers can ensure the stability of reflection symmetric FSs off high-symmetry points.
$\mathbb{Z}_2$ and $M \mathbb{Z}_2$ invariants, on the other hand, do not give rise to stable FSs.
Nevertheless, $\mathbb{Z}_2$ or $M \mathbb{Z}_2$  invariants 
may lead to protected zero-energy surface states at TR invariant momenta of the surface BZ. 
We observe that  the classification of reflection-symmetric FSs located away from high symmetry points with codimension $p$ 
is related to the classification of reflection-symmetric TIs/TSCs with spatial dimension $d = p-1$,
see Table~\ref{reflection_table_full}.

Reflection-symmetry protected FSs in most experimental systems are of type ``FS2". Let us in the following illustrate
the properties of these FSs using two examples.

\paragraph{Example: ``FS2'' with \texorpdfstring{$p=3$}{p=3} in DIII + \texorpdfstring{$R_{--}$}{R--}}\label{p3 DIII}

We consider a topological nodal SC with point nodes, described by the Hamiltonian~\eqref{semimetalHam}. 
It preserves TRS with $T=\tau_0 s_2 \ck$
and PHS with $C = i \tau_1 s_1 \mathcal{K}$.  
In addition, it is symmetric under reflection,   
\bee
R_y^{-1}H(k_x,-k_y,k_z) R_y= H(k_x,k_y, k_z), 
\label{reflectSym}
\ee
with $R_y=\tau_3 s_2$.
The reflection operator $R_y$ anti-commutes with $T$ and $C$,
and hence the Hamiltonian~\eqref{semimetalHam} is a member 
of symmetry class DIII$+R_{--}$.
According to Table~\ref{reflection_table_full},
the Dirac nodes in \eqref{semimetalHam} 
[Fig.\ \ref{surfacBandStruct2D}(d)]
are protected by an $M\mathbb{Z}$ invariant, i.e., the mirror winding number $\nu^+$. 
The mirror invariant is defined by a 1d integral
along a contour that lies within the mirror plane $k_y=0$. 
Within the $k_y=0$ mirror plane
the Hamiltonian can be block-diagonalized with respect to $R_y$. 
For the block with mirror eigenvalue $R_y=+1$ 
by choosing a one-parameter family of contours $\mathcal{C}(k_z)$
that are parallel to the $k_x$-axis with fixed $k_z$,
the mirror winding number is given by 
$\nu^+(k_z)=-1$ for $|k_z|<\pi/2$
whereas
$\nu^+(k_z)=0$ for $|k_z|>\pi/2$.
This indicates that there exists a gapless Fermi arc state 
on the (100) surface, 
connecting 
the projection of the bulk 
Dirac nodes at $\mathsf{k}_\pm=(0,0,\pm \pi/2)$,
see Figs.~\ref{surfacBandStruct2D}(d)-(f). 
Other  types of topological nodal SCs with crystal symmetries have
 been studied by~\onlinecite{Nodal_SC_Shingo,schnyderReviewTopNodalSCs,early_nodal_SC}.

\begin{figure}[t]
 \begin{center} 
\includegraphics[width=0.42\textwidth,angle=-0]{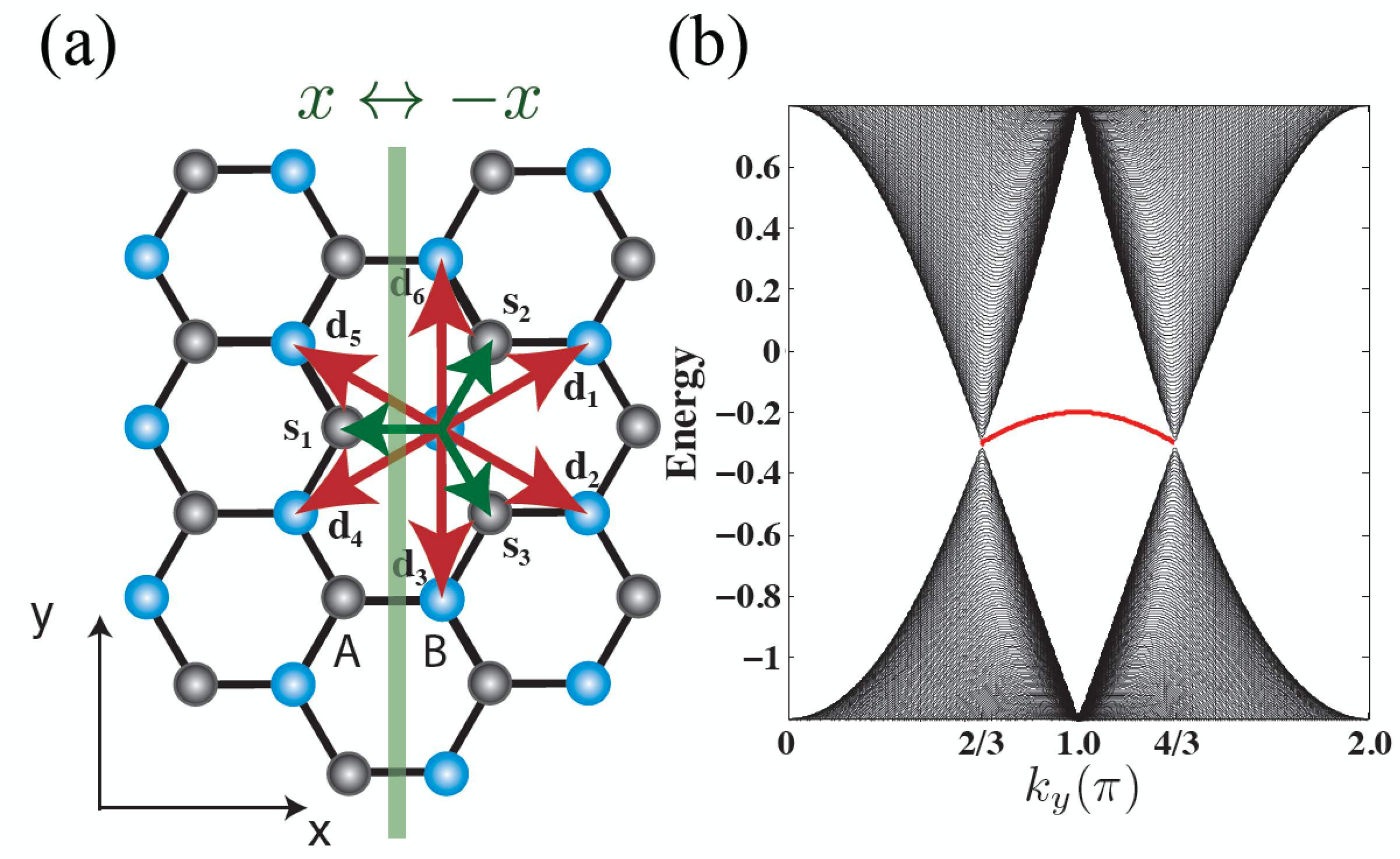}
\end{center}
\caption{
(a) 
The honeycomb lattice is a bipartite lattice composed of two interpenetrating triangular sublattices
A (black dots) and B (blue dots). 
The vectors connecting nearest-neighbor and next-nearest-neighbor sites are denoted by ${\bf s}_i$ (green) and ${\bf d}_i$ (red),
respectively,
where ${\bf s}_1=(-1,0)$, ${\bf s}_2=\frac{1}{2}(1,\sqrt{3})$, ${\bf s}_3=\frac{1}{2}(1, -\sqrt{3})$,
 and ${\bf d}_1=-{\bf d}_4=\frac{1}{2}(3,\sqrt{3})$, ${\bf d}_2=-{\bf d}_5=\frac{1}{2}(3,-\sqrt{3})$, 
${\bf d}_3=-{\bf d}_6=(0,-\sqrt{3})$. 
The mirror line $x \to - x$ is indicated by the green line.
(b) Energy spectrum of a graphene ribbon with (10) edges (i.e., zigzag edges) and $(t_1, t_2)=(1.0, 0.1)$.
A linearly dispersing edge state (red trace) connects the Dirac points at $k_y=2\pi/3$ and $k_y = 4\pi/3$ 
in the edge BZ. Adapted from \cite{ChiuSchnyder14}.
 \label{Honeycomb}
}
\end{figure}

\paragraph{Example: ``FS2'' with \texorpdfstring{$p=2$}{p=2} in class AI + \texorpdfstring{$R_{+}$}{R+} (``spinless graphene'')}

As a second example we discuss spinless fermions hopping on the honeycomb lattice.
Provided one neglects the spin degrees of freedom, this model describes the electronic properties of graphene 
\cite{castroNetoRMP09}.
The Dirac cones of spinless graphene are protected 
by TR, reflection, and translation symmetry.
(Note that the Dirac cones are also stable in the presence of inversion symmetry
instead of reflection symmetry
\cite{GuineaVozmedianoPRB07}.)
The tight-binding Hamiltonian is given by 
$\hat{H} = 
\sum_{\mathsf{k}} 
\hat{\Psi}^{\dag}_{\mathsf{k}} H( \mathsf{k} ) \hat{\Psi}^{\ }_{\mathsf{k}}$
with the spinor 
$\hat{\Psi}^{\ }_{\mathsf{k}} = 
( \hat{a}^{\ }_{\mathsf{k}},
\hat{b}^{\ }_{\mathsf{k}} )^{{T}}$ and
\begin{align} 
\label{hamGraphene}
&H ( \mathsf{k} )
=
\begin{pmatrix}
\Theta_{\mathsf{k}} & \Phi_{\mathsf{k}} \cr
\Phi^{\ast}_{\mathsf{k}} & \Theta_{\mathsf{k}} \cr
\end{pmatrix},
\quad
\left\{
\begin{array}{ll}
\Phi_{\mathsf{k}} = t_1 \sum_{i=1}^3 e^{ i {\mathsf{k}} \cdot {\bf s}_i },
\\
\Theta_{\mathsf{k}} = t_2 \sum_{i=1}^6 e^{ i {\mathsf{k}} 
\cdot {\bf d}_i },
 \end{array}
 \right.
\end{align}
where 
$\hat{a}_{\mathsf{k}}$ and $\hat{b}_{\mathsf{k}}$
denote the fermion annihilation operators with momentum ${\mathsf{k}}$ on sublattice A and B, respectively,
${\bf s}_i$ and ${\bf d}_i$ are the nearest- and second-neighbor bond vectors, respectively [Fig.~\ref{Honeycomb}(a)],
and the hopping integrals  $t_{1,2}$ are assumed to be positive. 
The Hamiltonian~\eqref{hamGraphene} is invariant under TR with $T= \sigma_0 \mathcal{K}$
and reflection $k_x \to - k_x$ with $R = \sigma_1$.
(Incidentally, the Hamiltonian \eqref{hamGraphene} 
is also symmetric under $k_y \to - k_y$,
which, however does not play any role for the protection of the Dirac points.)
Since $T^2=+\openone$ and $[R, T] = 0$,
the Hamiltonian~\eqref{hamGraphene} belongs to symmetry class AI$+R_+$.

The energy spectrum of~\eqref{hamGraphene},
$\varepsilon_{\mathsf{k}}^{\pm}
= \Theta_{\mathsf{k}} \pm |\Phi_{\mathsf{k}}|$,
exhibits two Dirac points, 
which are located on the mirror line $k_x =0$, 
i.e., at $(k_x, k_y) = (0, \pm k_0 )$ with $k_0=4 \pi /( 3 \sqrt{3})$.
These two Dirac points transform pairwise into each other under TRS.
Any gap opening term is forbidden by TRS and reflection symmetry, 
and the Dirac points are topologically stable. 
In particular, the TRS preserving mass term $\sigma_3$ is forbidden by reflection symmetry $R$.
This finding is consistent with 
the classification in Table~\ref{reflection_table_full}, 
which indicates that the stability of the Dirac points 
is guaranteed by an $M \bZ$-type invariant. 

The mirror invariant $n_{M \bZ}$ can be computed
from the eigenstates $\psi^{\pm}_{\mathsf{k}}$ of $H(  {\mathsf{k}} )$ with energy $\varepsilon^{\pm}_{\mathsf{k}}$, 
$
\psi^{\pm }_{\mathsf{k}}
=
(\pm  e^{i \varphi_{\mathsf{k}}}, 1)^T/\sqrt{2} 
$, 
where $\varphi_{\mathsf{k}} = \arg ( \Phi_{\mathsf{k}}) $.
Noting 
$e^{i \varphi_{(0, k_y) }}  
= +1(-1)$
for $|k_y|<k_0$ ($|k_y|>k_0$), 
$ \psi^{\pm}_{(0, k_y ) }$ are simultaneous eigenstates of the reflection operator 
with opposite eigenvalues ($+1$ and $-1$),
and do not hybridize.
The mirror invariant $n_{M \bZ}$ is  
given in terms of the number of states 
with energy $\varepsilon^-_{\mathsf{k}}$ and
reflection eigenvalue $R = +1$,  
$n_{\rm neg} ( k_y) $,  
as
\begin{align} \label{mirroInvGraph}
n_{M \bZ} 
=
 n_{\rm neg} ( | k_y | > k_0) - n_{\rm neg} ( | k_y | < k_0)
 = +1.
\end{align}
By the bulk-boundary correspondence, the nontrivial topology of the Dirac points leads to a linearly dispersing edge mode, 
which connects the projected Dirac points in the (10) edge BZ [Fig.~\ref{Honeycomb}(b)].

\subsection{Dirac semimetals protected by other point-group symmetries}
\label{secDiracSemimetal}

Besides reflection symmetry, 
other point group symmetries, such as rotation or inversion, can give rise
to topologically stable FSs  \cite{BiO3_Dirac_semimetal,Dai_predition_Na3Bi,wangCd3As2PRB13,Dai_Cu3PdN_ring,Kane_Cu3N_ring}.

\subsubsection{3d semimetals with $p=3$}
First, let us briefly illustrate how rotation symmetry 
can lead to protected Dirac points by using 
the Hamiltonian \eqref{semimetalHam} again as a simple example.
As discussed above, the Dirac points of Eq.~\eqref{semimetalHam}, located at $(0,0,\pm \pi/2)$,
are not protected by TRS.
However, spatial symmetries can protect these Dirac cones.
One example is chiral symmetry together with mirror symmetry~\eqref{reflectSym}, which was described above; another example is
the fourfold $C_4$ rotation symmetry along the $z$ axis, which acts on
$H ( \mathsf{k} )$ as 
\begin{eqnarray} \label{C4RotSym}
R^{-1}_{C_4}H(- k_y, k_x,k_z)R^{\ }_{C_4}=H(k_x, k_y, k_z),
\end{eqnarray}
where $R_{C_4}=\tau_3 (s_0+is_3)/\sqrt{2}$.
We find that there exist two mass terms that can gap out the Dirac nodes,
namely $f_1(k_z)\tau_2 s_0$ and $f_2(k_z)\tau_1 s_3$, 
since these are the terms that anti-commute with $H ( \mathsf{k} )$.
Here,  $f_1(k_z)$ and $f_2(k_z)$ represent $k_z$ dependent masses. 
However, these two gap opening terms break the $C_4$ rotation symmetry~\eqref{C4RotSym},
since they anti-commute with $R_{C_4}$.
However, each Dirac point can be decomposed into two Weyl nodes along the $z$ direction in the presence of the $C_4$-preserving term $\tau_0s_3$ 
(The additional inversion symmetry and TRS forbid this term). 
Thus, the gapless nature of the Hamiltonian~\eqref{semimetalHam} is protected by the $C_{4}$ rotation symmetry~\eqref{C4RotSym},
and the Dirac points are protected by the full point group $D_{6h}$.
In passing we note that similar arguments can be used to explain
the \emph{gapless} stability of the bulk Dirac points of 
 Na$_3$Bi and Cd$_3$As$_2$, which possess $C_3$ and $C_4$ rotation symmetries, respectively~\cite{Chiu_C4_Dirac,Bulk_Dirac_cone_rotation_Akira,bulk_Dirac_Yang:2014aa}.

Recently, several materials have been experimentally identified as topological semimetals protected by crystalline symmetry.
Among them are the Dirac materials Cd$_3$As$_2$~\cite{Yazdani_CdAs,neupaneDiracHasan,borisenkoPRLCd3As2,Cd3As2Chen2014,Liang_Dirac_magento}
and Na$_3$Bi~\cite{Liu21022014,Dai_predition_Na3Bi,Xu18122014}, whose gapless spectrum is protected by rotation symmetry. 
The Fermi arc states of Na$_3$Bi have recently been observed by ARPES~\cite{Xu18122014}. 
Unusual magnetoresistence has also been reported in these Dirac systems~\cite{Ando_Dirac_magneto,Liang_Dirac_magento,ZrTe_Dirac_2014}. 
Superconducting Dirac semimetals have been theoretically investigated by~\cite{superconducting_Dirac}. 

\subsubsection{3d semimetals with $p=2$} 

Topological nodal lines with $p=2$, i.e.,  1d FSs in a 3d BZ, 
have been theoretically proposed  to exist in several materials.
For semimetals with negligible spin-orbit coupling,  it has been  shown that topological nodal lines are typically protected by either reflection symmetry or the combination of TRS and inversion symmetry~\cite{Nodal_Line_Fang,Nodal_line_chan}. 
There are two different types of topological line nodes, namely, Weyl and Dirac line nodes. 
While for the stability of Weyl line nodes the presence of just a single symmetry 
(e.g., reflection or chiral symmetry) is usually sufficient~\cite{Fang_bernevig_multi_weyl_PRL_12,ChiuSchnyder14}, 
Dirac line nodes, which can be viewed as two copies of Weyl line nodes, need additional symmetries for their protection. 
For example, the compound Ca$_3$P$_2$~\cite{Cava_CaP_ring,Nodal_line_chan} possesses 
a stable Dirac line protected by reflection symmetry together with SU(2) spin-rotation symmetry. 
In Ca$_3$P$_2$ the nodal line is located at the Fermi level, which makes  
it an ideal system to study the unconventional transport properties of nodal line semimetals.
Besides Ca$_3$P$_2$, CaAgP, CaAgAs~\cite{Nodal_line_Tanaka}, and rare earth monopnictides LaX (X=N, P, As, Sb, Bi)~\cite{Fang_ring_point} 
have been proposed to possess Dirac nodal lines protected by reflection symmetry and SU(2) spin-rotation symmetry. 
Dirac line nodes also appear in the band structure of 
some orthorhombic perovskite iridates~\cite{Kim_chiral_ring}. Examples of Dirac line nodes
protected by inversion, TRS, and SU(2) spin-rotation symmetry  include
Cu$_3$N~\cite{Kane_Cu3N_ring}, Cu$_3$PbN~\cite{Dai_Cu3PdN_ring}, and 
all-carbon Mackay-Terrones crystals~\cite{MTC_nodal_line}.

Weyl line nodes protected by reflection symmetry exist in the band structure 
of  PbTaSe$_2$~\cite{Bian_nodal_line} and TlTaSe$_2$~\cite{Bian_TlTaSe2_line}. 
These materials belong to symmetry class AII$+R_-$ in Table~\ref{reflection_table_full}. 
Their Weyl lines, 
which are located away from high-symmetry points, are protected by the $M\bZ$ invariant 
that is inherited from class A$+R$ 
(compare with the discussion about Weyl nodes in Sec.~\ref{Weyl section}).

\section{Effects of interactions -- the collapse of non-interacting classifications}
\label{Effects of interactions}

\subsection{Introduction}

In this section we present a brief overview of topics that go beyond the classification of non-interacting fermionic systems.
Interactions can affect/modify topological classifications of non-interacting fermion systems in various ways.
For example,
interactions can ``destroy'' non-interacting topological phases -- 
a would-be topological state 
of a single-particle Hamiltonian,
characterized by a  topological invariant built out of single-particle wave functions,
can be adiabatically deformable to a topologically trivial state, once interactions are included. 
To describe such situations, we say the non-interacting classification ``collapses'' or ``reduces''.
Another possibility is that interactions can create new topological states which are topologically distinct from trivial states.

Examples of the latter case include, e.g.,
interaction-enabled symmetry protected topological phases in 1d
\cite{Lapa2014},
topological insulating phases in 3d that arise only in the presence of interactions
(together with topological band insulators, 
these fall into a $\mathbb{Z}_2\times \mathbb{Z}_2$ classification of 3d gapped insulating phases)
\cite{WangPotterSenthil2014}, 
and 
fractional topological insulators in (2+1)d and (3+1)d
\cite{LevinSternFTI2009,MaciejkoQiKarchZhangFTI2010,swingle_barkeshli_macgreevy_senthil,fiete_review_natPhys15,Sheng_FTI,Roy_FTI,Titus_FTI,Atma_EFT,Bernevig_Z2_FTI,Kallin_FTI,Zhao_FTI,Titus_wire_deconstruction}. 

Even when interactions do not destroy a non-interacting topological phase  
(i.e., it exists irrespective of the absence/presence of interactions),
characterizing such states without relying on the single-particle picture is often non-trivial.
Due to the rapidly developing nature of the field of strongly interacting topological phases,
we do not aim to give a complete review of this field here, 
but focus our discussion instead on the collapse of the classification of  non-interacting fermionic systems.
More specifically, we will discuss the classification of interacting TSCs (fermionic phases which lack $U(1)$ charge conservation)
with various symmetries (such as TRS, spin parity conservation, and reflection symmetry)
in one, two, and three spatial dimensions.

\subsubsection{Symmetry-protected topological phases, short-range and long-range entanglement}

Before discussing examples of interacting fermionic systems,
let us first introduce a few concepts and common terminologies,
which are useful in discussing general interacting (topological) phases.
In the previous sections, we have discussed TIs
and TSCs within non-interacting band theories, 
described by quadratic Bloch or BdG Hamiltonians.
In a broader context,
including bosonic systems, and
in particular in the presence of interactions,
the terminology {\it symmetry-protected topological (SPT) phases} is used
\cite{GuWen2009}.
In the absence of symmetry conditions these phases are trivial states of matter which are continuously deformable
to, e.g., an atomic insulator.
On the other hand, in the presence of a set of symmetry conditions,
they are topologically distinct from trivial states,
and are separated from trivial states by a quantum phase transition.

SPT phases are also called states with {\it short-range entanglement} or
{\it short-range entangled (SRE) states}.
To be more precise, SRE states are states that can be transformed, 
by applying a finite-depth local unitary quantum circuit, 
into a product state.
In contrast,
those states which cannot be disentangled into a product state by
a finite-depth local unitary quantum circuit
are called
states with {\it long-range entanglement},
or
{\it long-range entangled (LRE) states}
\cite{Xie_LRE}.
Note that in this definition,
non-interacting, integer QH states are examples of LRE states,
even though they do not have topological order as measured by the topological entanglement entropy \cite{KitaevPreskill06}
or by the  non-trivial topological ground state degeneracy \cite{WenNiu90}.
Due to the lack of topological order, SPT phases are also sometimes called symmetry-protected {\it trivial} phases
\cite{Wen2014symmetry-protected}.

There exists an alternative definition for short-range entanglement in the literature,
where SRE states are defined as
systems with gapped and non-degenerate bulk spectra,
namely as having no topological entanglement order
\cite{Kitaev_Unpublished}.
In this definition,  SRE states include SPT states as a subset.
SRE states of this kind are also called {\it invertible} or having invertible topological order
\cite{Freed2014, KongWen2014}.

While LRE states are not adiabatically deformable to trivial states
even in the absence of any symmetry,
symmetries can coexist and intertwine with topological orders,
and can lead to a distinction between states which share the same topological order.
To discuss such distinctions between topologically ordered states with symmetries,
the terminology {\it symmetry-enriched topological (SET) phases} is used
\cite{Chen2013}, while in other contexts the term {\it weak symmetry breaking} or {\it projective symmetry groups} \cite{Kitaev2006, Wen2002} is used.
In the following, we will focus on fermionic SPT phases, although some of  the
techniques/concepts that we will discuss are also applicable to SET phases.

\subsection{Example in (1+1)d: class BDI Majorana chain}
\label{Example in (1+1)d: class BDI Majorana chain}

The first example of a collapse of a non-interacting classification
was shown by Fidkowski and Kitaev
for a $(1+1)$d TSCs
\cite{Fidkowski2010, Fidkowski2011}.
To discuss this example we use as our starting point
 the Kitaev chain defined in Eq.\ (\ref{Kitaev chain}) in terms of spinless fermions.
The Kitaev chain is a member of symmetry class D
and its different phases are classified by the $\mathbb{Z}_2$ topological index
discussed in Sec.\ \ref{Topological superconductor in one dimension}.
To impose on this 1d model TRS, we recall that TRS acts on spinless fermions as
\begin{align}
&
\hat{\mathscr{T}} \hat{c}_j \hat{\mathscr{T}}^{-1}  = \hat{c}_j,
\quad
\hat{\mathscr{T}} \hat{c}^{\dag}_j \hat{\mathscr{T}}^{-1}  = \hat{c}^{\dag}_j,
\quad
\hat{\mathscr{T}}^2 = 1.
\label{trs kitaev chain}
\end{align}
(In the Majorana fermion basis (\ref{Majorana basis Kitaev chain}),
TRS acts as
$\hat{\mathscr{T}}\hat{\lambda}_{j} \hat{\mathscr{T}}^{-1} = -\hat{\lambda}_{j}$
and
$
\hat{\mathscr{T}}\hat{\lambda}'_{j} \hat{\mathscr{T}}^{-1} = \hat{\lambda}'_{j}$.)
While particle number conservation is broken  in BdG systems,
the fermion number parity $\hat{\mathscr{G}}_f$ remains conserved.
$\hat{\mathscr{G}}_f$ acts on the fermion operators as
\begin{align}
 \hat{\mathscr{G}}^{\ }_f \hat{c}^{\ }_j \hat{\mathscr{G}}^{-1}_f = - \hat{c}^{\ }_j,
 \quad
 \hat{\mathscr{G}}^{\ }_f \hat{c}^{\dag}_j \hat{\mathscr{G}}^{-1}_f = - \hat{c}^{\dag}_j.
\label{fermion pairty kitaev chain}
\end{align}
The symmetry operations $\hat{\mathscr{T}}$ and $\hat{\mathscr{G}}_f$ constitute
the full symmetry group of the example at hand.
These operators satisfy $\hat{\mathscr{T}} \hat{\mathscr{G}}_f = \hat{\mathscr{G}}_f \hat{\mathscr{T}}$ and $\hat{\mathscr{T}}^2 =\hat{\mathscr{G}}^2_f=1$.
Hence, since  $\hat{\mathscr{T}}^2 =1$, the relevant symmetry class is BDI, whose
 topologically distinct ground states in 1d are distinguished
 by a winding number $\nu$, see Sec.\ \ref{Example: Polyacetylene}.
For~\eqref{Kitaev chain} we find that
$\nu=0$ for $|t|<|\mu|$
whereas
$\nu=1$ for $|t|>|\mu|$.
In the topologically non-trivial phases with  $\nu \ne 0 $
there appear $\nu$ isolated Majorana zero modes
localized at the end.
These Majorana end states are stable against quadratic perturbations which preserve the symmetries.
Phases with higher winding number $\nu=N_f$ can be realized by
taking $N_f$ identical copies of the Majorana chain,
$\sum_{a=1}^{N_f}\hat{H}_0(\hat{c}^{a\dag}, \hat{c}^a)$,
where $\hat{H}_0(\hat{c}^{a \dag}, \hat{c}^a)$ is the quadratic Hamiltonian of the Kitaev chain for the $a$-th copy (flavor),
and the fermion creation/annihilation operators for different copies are denoted by
$\hat{c}^{a \dag}_j, \hat{c}^a_j$ with $a=1,\ldots, N_f$.

Fidkowski and Kitaev demonstrated
that when $N_f=0$ ($\mbox{mod}\, 8$), 
the non-interacting topological phase with the winding number $\nu = N_f$
can be adiabatically connected to the topologically trivial phase,
once interactions are included
\cite{Fidkowski2010, Fidkowski2011}.
Specifically, they considered 
the following interacting Hamiltonian for the case of $N_f=8$
\begin{align}
\hat{H}=  \sum_{a=1}^{N_f}\hat{H}_0(\hat{c}^{a\dag}, \hat{c}^a) + w
 \sum_j \left[
\hat{W}(\hat{\lambda}^{a}_{j}) + \hat{W}(\hat{\lambda}^{\prime a}_{j})
\right],
\label{Kitaev chain w/ interactions}
\end{align}
where $\hat{W}(\hat{\lambda}^{a})$ can be 
given, conveniently and suggestively, 
in terms of two species of spin-full complex fermion operators, 
$\hat{c}^{\ }_{1\uparrow}= (\hat{\lambda}^1 + i \hat{\lambda}^2)/2$, 
$\hat{c}^{\dag}_{1\downarrow}= (\hat{\lambda}^3 + i\hat{\lambda}^4)/2$,
$\hat{c}^{\ }_{2\downarrow}=(\hat{\lambda}^5 + i \hat{\lambda}^6)/2$,
$\hat{c}^{\dag}_{2\uparrow}=(\hat{\lambda}^7+i\hat{\lambda}^8)/2$, 
as 
$
\hat{W}
=
16 \hat{\boldsymbol{S}}_1 \cdot \hat{\boldsymbol{S}}_2
+
2 (\hat{n}_1-1)^2
+
2 (\hat{n}_2-1)^2 -2
$,
where 
$\hat{\boldsymbol{S}}_i= c^{\dag}_{i \alpha} ( \boldsymbol{\sigma}_{\alpha \beta}  / 2 ) c^{\ }_{i \beta}$ and
$\hat{n}_i =\hat{n}_{i \uparrow}+\hat{n}_{i \downarrow}$.
This interaction preserves an $SO(7)$ subgroup of the $SO(8)$ acting on the flavor index. 
Since the Hamiltonian now depends on three parameters, i.e. on $(t,\mu,w)$ (we set $\Delta_0=t$ for simplicity),
it is possible to construct a path that connects the non-interacting topological phase
($|t|>|\mu|$ and $w=0$)
to the non-interacting trivial phase
($|t|<|\mu|$ and $w=0$) via 
the   interacting phase ($w \ne 0$)
without gap closing.
To explicitly construct this path,
we start from
$(t,\mu,0)$ with $|t|>|\mu|$
and switch off $\mu$,
$(t,\mu,0)\to (t,0,0)$.
Along this deformation, we stay in the topological phase.
At the point $(t,0,0)$, the system is a collection of decoupled dimers.
We then switch on $w$ and let $t \to 0$, $(t,0,0)\to (0,0,w)$.
The interaction term $\hat{W}$ is designed so that
the system remains gapped throughout this path.
Finally, we switch on $\mu$ and let $w \to 0$, $(0,0,w)\to (0,\mu,0)$,
which brings us to the non-interacting trivial phase without closing the gap.
This completes the construction of a path in the phase diagram connecting the non-interacting topological phase
to the trivial phase, and
proves the triviality of the $\nu=0$ (mod 8) phase.
Thus, the non-interacting classification  reduces from $\mathbb{Z}$ to $\mathbb{Z}_8$.
Similar interaction effects on other 1d fermionic topological phases have been studied in  \cite{Tang2012,Rosch2012,Lapa2014,ning_liu_PRB_15}.
Proposals on how to realize  1d interaction enable topological 
phases in experiments have been discussed in \cite{Chiu_interaction_enabled,Chiu_strong_inter_Majorana}.

\subsubsection{Projective representation analysis}

More insight into the underlying ``mechanism'' of the collapse of the classification can be gained
by considering the symmetry properties of the boundary Majorana fermion modes of the Kitaev chain.
When $\nu=N_f$, there are $N_f$  zero-energy Majorana bound states at the end of the Kitaev chain, which are
described by $N_f$ dangling Majorana fermion operators,
$
\hat{\eta}_1, \hat{\eta}_2, \cdots, \hat{\eta}_{N_f}
$.
As emphasized in Sec.\ \ref{Class D},
these bound states are {\it unpaired} (i.e., {\it isolated}) Majorana zero-energy modes, which are different from
the ones that appear in the bulk BdG Hamiltonian, i.e., $\hat{\lambda}$ and $\hat{\lambda}'$, which always come in pairs.
While 
the symmetry operators $\hat{\mathscr{T}}$ and  $\hat{\mathscr{G}}_f$  
act on the full Hilbert space of fermion operators $\hat{c}^{\dag}_j, \hat{c}^{\ }_j$
in a way such that
the standard group multiplication laws,
$\hat{\mathscr{T}} \hat{\mathscr{G}}_f = \hat{\mathscr{G}}_f \hat{\mathscr{T}}$
and $\hat{\mathscr{T}}^2 =\hat{\mathscr{G}}^2_f=1$, are satisfied,
these symmetries  act on the Hilbert space of the dangling Majorana fermions $\hat{\eta}_i$  in a way such that
the group composition/multiplication law is respected only up to a phase.
That is, the symmetries in the Hilbert space of the dangling Majorana fermions 
are realized only {\it projectively} or {\it anomalously}.
The group structure of the symmetry generators $\hat{T}$ and $\hat{G}_f$
acting on the Hilbert space of the dangling Majorana fermions was calculated in
\cite{Fidkowski2011, turnerPollmannPRB11}.
The result of this calculation is summarized here:
\begin{align}
\begin{array}{c|cccccccc}
 \nu \, (\mbox{mod 8}) & 0& 1 & 2 & 3 & 4 & 5 & 6 &7  \\
 \hline
 a & +1 & +1& -1 &-1& +1 & +1 & -1 &-1  \\
 \hat{T}^2 & +1 &+1 & +1 &-1 & -1 & -1 & -1 & +1
 \end{array} ,
 \label{tab FK}
\end{align}
where $a$ specifies the (anti-)commutation relation between
$\hat{T}$ and $\hat{G}_f$ as
$\hat{T}\hat{G}_f \hat{T}^{-1} = a \hat{G}_f$.
From the 8-fold periodicity of Table \ref{tab FK}, 
we see that the non-interacting classification
 collapses from  $\mathbb{Z}$ to $\mathbb{Z}_8$.
Note that this result can 
 also be derived in terms of  
Green's functions 
\cite{Gurarie2011, BenTov2014} 
and in terms of non-linear sigma models 
\cite{YouXu2014}. 

\paragraph{Matrix Product States (MPSs)}
\label{MPS}
The above analysis of the projective symmetry group realized at the boundary of the Kitaev chain
can be generalized to arbitrary SPT phases in (1+1)d.
Besides the interacting Kitaev chain,
another well-known example of a 1d interacting SPT phase is
the Haldane antiferromagnetic spin-1 chain  \cite{Haldane1983a, Haldane1983b},
which has an $SO(3)$ spin-rotation symmetry. The Haldane spin-1 chain exhibits
dangling spin-1/2 moments at its ends, which
transform according to a half-integer projective representations of the $SO(3)$ group.

A convenient and unifying way to describe generic SPT phases in (1+1)d
is provided by the {\it matrix product state (MPS)} representation of ground states of (1+1)d systems
\cite{Chen2011, Pollmann2010, Pollmann2012, Schuch2011}.
In the MPS representation,
a quantum state $|\Psi\rangle$ defined on a 1d lattice is written as
\begin{align}
 |\Psi\rangle
 &=
 \sum_{s_1, s_2, \cdots}
 A^{s_1}_{ij} A^{s_2}_{jk}\cdots | s_1 s_2 \cdots \rangle
 \nonumber \\
 &=
 \sum_{s_1, s_2, \cdots}
 \mathrm{Tr}_{\chi}\,
 \left[
 A^{s_1} A^{s_2}\cdots \right] | s_1 s_2 \cdots \rangle ,
\end{align}
where $|s_1 s_2 \cdots\rangle$ is a basis ket of the many-body Hilbert space,
which is composed of the basis kets $|s_j\rangle$ at each site $j$ of the 1d lattice, e.g., $|s_j\rangle = |\uparrow\rangle, |\downarrow\rangle$ for a spin 1/2 chain.
The  $A^s_{ij}$'s are $\chi\times \chi$ matrices on site $s$, with $i,j,k,\ldots = 1,\ldots, \chi$, 
and $\chi$  the bond dimension of the MPS. 
For simplicity, periodic boundary conditions are assumed.
By suitably choosing the matrix elements of the  $A^s_{ij}$'s (using a variational approach, say)
and by making the bond dimension $\chi$ large enough, an MPS is in many cases
a good approximation to the true ground state. 
In fact, it has been shown that  
the ground state of any gapped (local) 1d Hamiltonian can efficiently and faithfully be represented by an MPS
with sufficiently large, but finite, bond dimension $\chi$.
\cite{Hastings2007, Schuch2008, Gottesman2010}.

In order to describe SPT phases using MPSs, one needs to examine
how the symmetries act on the matrices $A^s_{ij}$ that constitute the MPS.
To that end, it is crucial to distinguish between the ``physical" indices $s_i$
and the ``auxiliary" indices $i,j$.
Physical indices represent physical degrees of freedom.
The way they transform under symmetries is fully determined by the microscopic
physical laws.
The symmetry transformations of the 
auxiliary indicies (or the auxiliary Hilbert space), on the other hand, 
are not entirely fixed by the symmetries of the physical system.
Instead,  MPSs representing different phases  with the same physical symmetries
may transform differently under the symmetries.
More precisely, the symmetries may be realized {\it projectively}
within the auxiliary Hilbert space of  the MPS.

To make this more explicit, let us consider a system with the symmetry group $G=\{g, h,\cdots \}$.
For simplicity, we consider only unitary and on-site symmetry operations here.
For the physical degrees of freedom there exists a unitary representation of $G$ with unitary operators
$\hat{U}(g)$, that acts on the local physical degrees of freedom as $|s\rangle \to \hat{U}(g)_{s}^{s'} |s' \rangle$.
Now, since the quantum state $|\Psi\rangle$ is left invariant by the symmetries $G$ up to an overall phase $\theta_g$,
we find that 
the symmetry transformation $\hat{U}(g)$ induces a corresponding transformation on the auxiliary space as
\begin{align}
\hat{U}(g)^{s'}_s A^{s} = \hat{V}^{-1}(g)A^{s'} \hat{V}(g) e^{i\theta_{g}},
\label{def_V_fun_for_MPS}
\end{align}
where $\hat{V}(g)$ operates on the auxiliary space indices, $i, j$.
While the transformations on the physical index $s$ form a  linear representation of the group $G$,
i.e., $\hat{U}(g)\hat{U}(h) = \hat{U}(gh)$,
the operations $\hat{V}(g)$ form, in general,  
a {\it projective} representation of $G$, i.e.,
\begin{align} \label{MPS_projective_reps}
\hat{V}(g)\hat{V}(h) = e^{i \alpha(g, h) } \hat{V}(g h).
\end{align}
The phase $\alpha(g,h)$ distinguishes between different projective representations of $G$
which, as it turns out, correspond to different SPT phases. 
In particular, when $e^{i \alpha(g, h) } \ne 1$ the corresponding SPT phase
is topologically non-trivial.

\subsection{Examples in (2+1)d: TSCs with \texorpdfstring{$\mathbb{Z}_2$}{Z2} and reflection symmetry}

In this section, we presents two examples of 2d TSCs, for which the non-interacting
classification collapses due to interactions.
Furthermore, we show that the collapse of these classifications
can be inferred from (i) the absence of a global gravitational anomaly
and (ii) the braiding statistics of the quasiparticles of the SPT phase
with gauged global symmetry.


\paragraph{Example: \texorpdfstring{$\mathbb{Z}_2$}{Z2} symmetric TSC}
\label{Example: Z2 symmetric TSC}

The first example is a  2d TSC 
with $N_f$ left- and right-moving Majorana edge modes,
protected by a $\mathbb{Z}_2$
symmetry in addition to the fermion number parity conservation
\cite{Qi2013, RyuZhang2012}.
To introduce this TSC we first consider a
spin-1/2 systems with two conserved $U(1)$ charges,
given by the total fermion number $N_{\uparrow}+N_{\downarrow}$
and the total spin $S_z$
quantum number 
$N_{\uparrow}-N_{\downarrow}$,
respectively.
By introducing an SC pair potential,
we break the electromagnetic $U(1)$ symmetry down to $\mathbb{Z}_2$, such that only
the fermion number parity $(-1)^{N_{\uparrow}+N_{\downarrow}}$ is conserved. 
To generate a second $\mathbb{Z}_2$ symmetry, we relax the conservation of total $S_z$,
and demand that only the parity $(-1)^{N_{\uparrow}}$ [and consequently $(-1)^{N_{\downarrow}}$] is conserved.
%
%
Observe that in the presence of these two $\mathbb{Z}_2$ symmetries
it is possible to block diagonalize 
the single-particle BdG Hamiltonian into a spin up and a spin down block, since
the $\mathbb{Z}_2 \times \mathbb{Z}_2$ symmetry
does not allow any spin flip terms,
i.e., any bilinears connecting the spin up and spin down sectors.
These two sub blocks belong to symmetry class A (cf.~Sec.~\ref{Class A and AIII})
and their topological properties are characterized by Chern numbers, i.e., by
$\mathrm{Ch}_{\uparrow}$ and  $\mathrm{Ch}_{\downarrow}$ for the spin-up and
spin-down blocks, respectively.
When $\mathrm{Ch}_{\uparrow} + \mathrm{Ch}_{\downarrow} \neq 0$,
TRS is necessarily broken,
which corresponds to a class D TSC with
 $\mathrm{Ch}_{\textrm{tot}}:=\mathrm{Ch}_{\uparrow} + \mathrm{Ch}_{\downarrow}$
{\it chiral} Majorana edge modes.
The class D TSC is robust against interactions as well as disorder for any $\mathrm{Ch}_{\textrm{tot}}$.

Here, however, we are interested in the case where
the total Chern number is vanishing, 
$\textrm{Ch}_{\textrm{tot}} = 0$,
but the spin Chern number is non zero,
$\textrm{Ch}_s :=
( \textrm{Ch}_{\uparrow} - \textrm{Ch}_{\downarrow})/2 \neq 0$.
A lattice model that realizes this situation
can  be constructed, by combining two
copies of chiral $p$-wave SCs
with opposite chiralities.
This TSC supports  $\mathrm{Ch}_s = N_f $ 
{\it non-chiral}  (i.e., {\it helical}) edge modes,
which are described  by
\begin{align}
\hat{H}
&=
\int dx\,
\sum^{N_f}_{a=1}
\big[
\hat{\psi}^a_L {i}v \partial_x \hat{\psi}^a_L
-
\hat{\psi}^a_R {i} v \partial_x \hat{\psi}^a_R
\big],
\label{edge theory N=8}
\end{align}
where
$x$ is the spatial coordinate along the edge of the TSC,
$\hat{\psi}^a_L$ ($\hat{\psi}^a_R$) denote the left- (right-) moving
(1+1)d Majorana fermions with flavor index $a$, and $v$ is the Fermi velocity.
The generators of the $\mathbb{Z}_2\times \mathbb{Z}_2$ symmetry of the bulk TSC are realized
within the edge theory \eqref{edge theory N=8} as
$
\hat{\mathscr{G}}_L = (-)^{\hat{N}_L}
$
and
$
\hat{\mathscr{G}}_R = (-)^{\hat{N}_R},
$
where $\hat{N}_L(=\hat{N}_{\uparrow})$
[$\hat{N}_R(=\hat{N}_{\downarrow})$] is the
total left-moving (right-moving) fermion number
at the edge.
The $\mathbb{Z}_2\times\mathbb{Z}_2$ symmetry
prohibits all mass terms
$\hat{\psi}^a_{L} \hat{\psi}^b_{R}$ at the edge,
since they are odd under the 
left- or right-$\mathbb{Z}_2$ parity ($\hat{\mathscr{G}}_L$ or~$\hat{\mathscr{G}}_R$).
Hence, this non-interacting TCS is classified by a $\mathbb{Z}$ invariant, which is
simply the number of  flavors of the (non-chiral) modes $N_f$.


Now, to study the effects of interactions
we consider quartic interaction terms of the form
$\hat{\psi}^a_{L} \hat{\psi}^b_{L}\hat{\psi}^c_{R}\hat{\psi}^d_{R}$
that preserve the $\mathbb{Z}_2\times \mathbb{Z}_2$ symmetries.
As it turns out,
when $N_f\equiv 0$ mod 8,
one can construct an interaction of this form
that destabilizes the edge; i.e., that
gaps out the edge without breaking the symmetries
(neither explicitly nor spontaneously).
This interaction term takes the form of the $SO(7)$ Gross-Neveu interaction, 
and is given essentially by the continuum-limit version of the interaction $\hat{W}$ in Eq.\ (\ref{Kitaev chain w/ interactions}).
We note that this  
interaction  can also be constructed in terms of
  {\it twist operators},
which twist the boundary conditions of the Majorana fermion fields when inserted in the path integral
(see Sec.~\ref{Braiding statistics approach}). To conclude, in the presence of interactions the classification of 
the $\mathbb{Z} \times \mathbb{Z}$ symmetric TSC collapses from $\mathbb{Z}$ to $\mathbb{Z}_8$.


\paragraph{Example: TCS in DIII+R$_{--}$}

The second example is a 2d topological crystalline superconductor
belonging to class DIII+R$_{--}$ \cite{YaoRyu2013}, see Sec. \ref{secReflectClass}.
(Note that this example and the $\mathbb{Z}_2$ symmetric TSCs discussed above
are related by the CPT theorem \cite{Hsieh2014b}.)
According to the non-interacting classification of Table~\ref{reflection_table_full},
TCSs in this symmetry class are characterized by an integer topological invariant, i.e, the mirror winding number.
Hence, in the absence of interactions this TCS supports an integer number of
stable gapless non-chiral edge states, provided that the edge is symmetric under reflection.
These edge states are described by Hamiltonian (\ref{edge theory N=8}).
Time-reversal $\hat{\mathscr{T}}$ and reflection $\hat{\mathscr{R}}$
act on the Majorana fields in the edge theory (\ref{edge theory N=8}) as
\begin{align}
&
\hat{\mathscr{T}} \hat{\psi}^a_L(x) \hat{\mathscr{T}}^{-1} = \hat{\psi}^a_R(x),
\,\,
\hat{\mathscr{T}} \hat{\psi}^a_R(x) \hat{\mathscr{T}}^{-1} = -\hat{\psi}^a_L(x),
\nonumber \\
&
 \hat{\mathscr{R}} \hat{\psi}^a_L(x) \hat{\mathscr{R}}^{-1} = \hat{\psi}^a_R(-x),
\,\,
\hat{ \mathscr{R}} \hat{\psi}^a_R(x) \hat{\mathscr{R}}^{-1} = -\hat{\psi}^a_L(-x),
 \nonumber \\
 &
\hat{\mathscr{T}}^{2} = \hat{ \mathscr{R}}^{2} = \hat{\mathscr{G}}_f.
\end{align}
One can check that in the presence of both TRS and reflection symmetry, 
there exists no gap opening quadratic mass term 
 within the edge theory (\ref{edge theory N=8}) for any $N_f$.
On the other hand, quartic interaction terms can fully gap
out the edge states of   phases with $N_f=0$ mod 8.
These quartic interactions are of the same form
as those  of the $\mathbb{Z}_2$ symmetric
TSC, see above.
Thus, in the presence of interactions the classification of the 2d TCS in class DIII+R$_{--}$
reduces from 
 from $\mathbb{Z}$ to $\mathbb{Z}_8$.
\begin{center}
***
\end{center}
The approach that we took in the above two examples can be summarized as follows:
For a topological bulk state with a given set of symmetries,
we first obtain representative edge theories
(and many copies thereof when necessary), describing the gapless edge
modes.
As a second step, we derive interaction terms within the edge theory 
which gap out the edge modes and which do not break the symmetries, 
neither explicitly nor spontaneously.
Such a microscopic stability analysis of  edge theories is quite powerful in (2+1)d,
and has been applied to many SPT as well as SET phases, such as bosonic SPT phases and fractional TIs
\cite{LuVishwanath2012, LuVishwanath2013, LevinStern2012,LevinSternFTI2009, Titus_FTI,HungWen2014}.

As in the (1+1)d example of Sec.~\ref{Example in (1+1)d: class BDI Majorana chain}, we now 
present alternative derivations of the collapse of the free-fermion classification, which will give us a deeper insight into why certain edge theories are stable while
others are not.
To that end,  we will introduce three important concepts:
{\it twisting/gauging (i.e., orbifolding) symmetries},
{\it quantum anomalies}, and {\it braiding statistics}.

\subsubsection{Twisting and gauging symmetries}

SPT phases, by definition, are topologically trivial in the absence of symmetries.
In order to determine whether a given SPT phase is topological or not, it is thus necessary to
probe the phase in a way that takes into account the symmetries.
This can be done by many different means, as we will describe below.

First of all, quantum systems with symmetries can be probed by coupling them to an external (source) gauge field
corresponding to the symmetry.
This is most commonly done for  unitary on-site (i.e., non-spatial) symmetries (e.g., continuous $U(1)$ symmetries)
in the spirit of  linear response theory. 
While for discrete symmetries (e.g., non-spatial unitary $\mathbb{Z}_2$ symmetries) 
linear-response functions cannot be defined,
the coupling to external gauge fields 
is in this case still a useful probe for SPT phases. 
The partition functions of SPT phases in the presence of external gauge fields,
typically given in terms of topological terms of gauge theories, 
can be used to distinguish and even classify different SPT phases
\cite{Wen2014symmetry-protected,HungWen2014, ChengZu2014, Wang2014c,SPT_invariant_Juven}.

A second possibility to probe the topology of an SPT phase 
is to {\it twist} the boundary conditions in space and time by elements of the symmetry group $G$. 
(Note that twisted boundary conditions can be turned into   untwisted ones,
by introducing   background gauge fields and by applying suitable gauge transformations.)
This approach, which can be applied in the presence of
both interactions and disorder, is commonly used to define and compute many-body Chern numbers.
Specifically, this is done by twisting the spatial boundary conditions by a $U(1)$ symmetry~\cite{Laughlin_IQHE,ThoulessWu1985,WangZhang2014}.

Making a step further, 
one can promote symmetries in SPT phases to  gauge symmetries, 
by making the external gauge field dynamical.  
This ``gauging"  of symmetries was proposed and shown to be a useful method
to diagnose and distinguish different SPT phases
\cite{Levin2012}. 
A similar procedure is the so-called \emph{orbi\-folding}  (as known from conformal field theories),
where one introduces twisted boundary conditions in space and time, and then considers
the sum (average) over all possible twisted boundary conditions~\cite{RyuZhang2012, Sule2013}.
Gauging and orbifolding have a similar effect in
that both procedures remove states in the original theory that are not singlets under the symmetry group $G$. 
I.e., the theory is projected onto the gauge singlet sector.
Another effect of gauging/orbifolding is to introduce (i.e., ``deconfine'') additional topological excitations (quasiparticles). 

Orbifolding and gauging can be applied
not only to SPT phases with unitary non-spatial symmetries, but   also 
to phases with unitary \emph{spatial} symmetries, such as reflection.
For example, twisting the boundary conditions by reflection leads to theories that are  defined on   non-orientied
manifolds, e.g.,  Klein bottles, which has recently been discussed for SPT phases  
 in (2+1)d  and (3+1)d~\cite{Hsieh2014a,ChoHsiehMorimotoRyu2015, HsiehChoRyu2015}.
Interestingly, this twisting procedure provides a  link
between SPT phases and so-called ``orientifold field theories'', i.e., field theories discussed
in the context of unoriented superstring theory.

\subsubsection{Quantum anomalies}
\label{global gravitational anomaly}

Another diagnostic for topological phases with symmetries are {\it quantum anomalies}.
A quantum anomaly is the breaking of a symmetry of the classical action by quantum effects. 
That is, an anomalous symmetry is a symmetry of the action, but not of
the quantum mechanical partition function. 
The presence of quantum anomalies can be used for diagnosing, 
defying and perhaps even classifying SPT phases. Quantum anomalies 
give us a deeper insight into the properties of the edge theory of a topological phase.

%

For example, the edge theory of the QHE suffers from a $U(1)$ {\it gauge anomaly},
i.e., the $U(1)$ charge is not conserved  by the edge theory due to quantum mechanical effects.
The presence of this anomaly is directly related to the nontrivial topology of the bulk: Charge conservation
is broken at the boundary, 
since current can \emph{leak} 
into the bulk due to non-zero Hall conductance, and hence due to the QHE.
Besides the $U(1)$ charge, also energy is not conserved at the edge of the QH system.
This is caused by the {\it gravitational anomaly}, i.e., by the fact that the chiral edge
theory of the QH state is not invariant under infinitesimal coordinate transformations~\cite{AlvarezGaumeWitten1983}.
The breaking of energy conservation at the edge signals that the bulk is topologically
non-trivial, which allows leaking of energy-momentum into the bulk due to the
non-zero thermal Hall conductance $\kappa_{xy}$
\cite{ Volovik1990,ReadGreen2000,Cappelli01}. 

The $U(1)$ and gravitational anomalies that we have discussed so far are examples of {\it perturbative anomalies}.
That is, the edge theory is not invariant under infinitesimal gauge/general coordinate transformations
that can be reached by successive infinitesimal transformations from the identity.
On the other hand, edge theories may also possess {\it global anomalies}, in which case
the quantum theory is not invariant under large gauge or large coordinate transformations that are 
preserved in the classical theory. 
Here, the term ``large" (or  ``global") refers to a transformation that cannot be continuously connected to the identity.
\emph{Global gauge} and \emph{global gravitational anomalies} lead to
anomalous phases picked up by the partition function
of quantum field theories under large gauge and coordinate transformations, respectively
\cite{Witten1982, Witten:1985xe}.
Note that Laughlin's gauge argument for the robustness of the QHE against disorder and interactions~\cite{Laughlin_IQHE},
is based on the global $U(1)$ gauge anomaly. The presence of such a global anomaly can be used as 
a powerful diagnostic for TR breaking interacting topological phases with conserved particle number.


It has been shown in numerous works that quantum anomalies generically appear in 
the boundary theories of SPT phases~\cite{RyuMooreLudwig2012, Xiaogang_anomalies, Wang2014a, Wang2014b, WangWen2013, Wang2014c,Ringel2013,
Koch-Janusz2014, Cappelli2013, Cappelli2014}. Due to the presence
of various types of quantum anomalies, the $d$-dimensional boundary theory of these SPT phases in $(d+1)$ dimensions
cannot be realize in isolation, i.e., there exists an ``obstruction" to discretize the boundary theory
on a $d$-dimensional lattice.


\paragraph{Global gravitational anomaly and orbifolds of a $\mathbb{Z}_2$ symmetric TSC}
Let us now discuss how the collapse of the non-interacting classification of the 
$\mathbb{Z}_2$ TSCs of Sec.~\ref{Example: Z2 symmetric TSC} 
can be inferred from the presence or absence of global gravitational anomalies. 
To this end, we put the edge theory~\eqref{edge theory N=8} on a flat space-time torus $T^2=S^1\times S^1$
with periodic spatial and imaginary time coordinates. 
The geometry of the flat torus $T^2$ is specified by two real parameters (so-called moduli), 
which can be arranged into a single complex parameter
$\tau = \omega_2 /\omega_1$,
namely the ratio of the two periods $\omega_i$ of the torus
($\mathrm{Im}\,\tau > 0$).
Two different modular parameters
$\tau$ and $\tau'$
describe the same toroidal geometry
if they are related by an integer linear transformation with unit determinant,
$
\tau\to \tau'=(a \tau+b)/(c\tau + d)
$
with 
$
a,b,c,d\in \mathbb{Z},
$
and
$
ad-bc=1.
$
These are large coordinate transformations on the torus $T^2$ and are referred to as modular transformations,
which form a group.
In general, any conformal field theory on $T^2$ that describes the continuum limit
of an isolated (1+1)d lattice  system is required to be invariant under modular transformations,
and hence anomaly free~\cite{Cardy1986}. For an edge theory, however, modular invariance
is not necessarily required. That is,   the inability to construct a modular-invariant partition function
signals that the  theory cannot be realized as an isolated (1 + 1)d system and must be realized as 
an edge theory of a  (2 + 1)d topological bulk state.

For the edge theory~\eqref{edge theory N=8}  we find that  
 the partition function  is modular invariant in the absence of the $\mathbb{Z}_2 \times \mathbb{Z}_2$ symmetry. 
In the presence of this symmetry, however, modular invariance cannot always be achieved. 
To see this,
we need to examine the orbifolded partition function of ~\eqref{edge theory N=8}, 
i.e., the partition function summed over all possible twisted boundary conditions
\begin{align} \label{orbifolded_part_fun}
 Z(\tau,\bar{\tau})
 =
|G|^{-1}
 \sum_{g,h\in G}
 \epsilon(g,h) 
 Z^{g}_{h}(\tau, \bar{\tau}), 
\end{align}
where
the group elements $g, h\in G=\mathbb{Z}_2\times \mathbb{Z}_2$ specify the boundary conditions
for the  partition function $Z^{g}_{h}$ in time and space directions, respectively.
That is, $g$ ($h$) specify if 
the left-moving/right-moving fermions obey
periodic or antiperiodic  temporal (spatial) boundary conditions. 
The weights $\epsilon(g,h)$ in the superposition \eqref{orbifolded_part_fun} are   constant phases with $|\epsilon(g,h)|=1$.  
Now the question is whether the orbifolded partition function $Z(\tau, \bar{\tau})$ 
can be made modular invariant, $Z(\tau, \bar{\tau}) = Z(\tau', \bar{\tau}')$,
(i.e., free from global gravitational anomalies)
by a suitable choice of $\epsilon(g,h)$. 
One can show that this is possible only when the number of Majorana fermion flavors is $N_f=0$ mod 8 \cite{Sule2013},
which indicates that the non-interacting classification collapses from $\mathbb{Z} \to \mathbb{Z}_8$, thereby
confirming the microscopic stability analysis of the edge theory, see Sec.~\ref{Example: Z2 symmetric TSC}.

\begin{center}
***
\end{center}

The discussed approach of studying modular invariance of orbifolded partition functions of edge theories to
determine the topological character of the bulk, has been successfully applied to 
other models, for example, 2d SPT phases
\cite{Sule2013} and 2d electron systems without any symmetries
\cite{Levin2013}.
For the examples considered in \onlinecite{Sule2013},
it was shown that the orbifolded partition functions can be made modular invariant, whenever
 the symmetry group acts on left- and right-moving sectors of the edge theory (i.e, the holomorphic and anti-holomorphic sectors of the edge CFT) in a symmetric fashion. On the other hand, 
 modular invariance can no longer be achieved,
if the symmetry group acts in an asymmetric manner on the left- and right-moving sectors.
In that case the corresponding orbifolded partition function is referred to as an ``asymmetric orbifold".
It turns out that many non-trivial SPT phases are directly related to asymmetric orbifolds.


\subsubsection{Braiding statistics approach}
\label{Braiding statistics approach}

By promoting the symmetry group $G$ of an SPT phase to a gauge symmetry 
one can associate a topologically
ordered phase  to each SPT phase.
As shown by \onlinecite{Levin2012}
the topological properties of the original SPT phase can then be inferred
by constructing the excitations of the gauged theory and 
by examining their quasiparticle braiding statistics.
This provides a way to distinguish between different SPT phases:
If two gauged  theories have different quasiparticle statistics,
then the corresponding ``ungauged" SPT phases must be distinct and cannot
be continuously connected without breaking the symmetries. 
Moreover, using this so-called ``braiding statistics approach" one can infer the stability 
of the edge theory. That is, 
for the cases where the gauged theories are Abelian topological phases
[i.e., phases that do not allow non-Abelian statistics but only Abelian (fractional) statistics],
the stability of the edge theories can be diagnosed from the
braiding statistics of the gauge theories.
This ``braiding statistics approach" has recently been used to show
that the non-interacting classification of the $\mathbb{Z}_2$ symmetric TSCs
(Sec.~\ref{Example: Z2 symmetric TSC}) 
 collapses from $\mathbb{Z}\to \mathbb{Z}_8$
\cite{Gu2014}, thereby confirming the microscopic stability analysis
(see also \onlinecite{Cheng2015}).

As discussed above, gauging and orbifolding are similar in that  both procedures project the theory onto
the gauge singlet ($G$-invariant)  sector.
(Although gauging means in general that the singlet condition is imposed locally (e.g., at each site of a lattice), while
orbifolding enforces the projection only globally.)
To make this connection between orbifolding and gauging more explicit,
let us consider edge theories with symmetry group $G$.
As in any quantum field theory we can use the global symmetries $g \in G$
to twist the boundary conditions.
This leads to a ``$g$-twisted" sector in the edge theory, which has twisted boundary conditions and whose ground state $|g\rangle$
satisfies
$
[ \hat{\Phi}(x+\ell) - U_g \cdot \hat{\Phi}(x)
]|g\rangle
=0.
$
Here, 
$\hat{\Phi}(x)$ denotes a field operator that is composed of, e.g., left- and right-moving Majorana fermions
 $\hat{\psi}^a_L$ and $\hat{\psi}^a_R$.
$U_g \hat{\Phi}$ is the field operator $\hat{\Phi}$ transformed by $g$
and $\ell$ is the circumference of the edge. All the states in this 
 $g$-twisted sector can be constructed from the ground state $|g\rangle$.
 Now, by using the state-operator correspondence, we can also construct a corresponding operator, the so-called twist operator
$\hat{\sigma}_g$, which implements this twisting. That is, by dragging the field operator 
 $\hat{\Phi}$ around the twist operator $\hat{\sigma}_g$ in spacetime,
$\hat{\Phi}$ gets  twisted by $g$, i.e., $\hat{\Phi}\to U_g \cdot \hat{\Phi}$. 
By use of the bulk-boundary correspondence, we find that corresponding to this there
exists  a  bulk excitation (i.e., an ``anyon'').
The bulk statistical properties of the gauged theory can then be read off
from the operator product expansions and fusion rules obeyed by
the twist operators $\hat{\sigma}_g$.
Hence, by the braiding-statistics approach it follows that different  ungauged SPT phases
can be distinguished by studying 
the statistical (i.e, braiding) properties of the corresponding twist operators $\hat{\sigma}_g$, 
I.e., two ÒungaugedÓ SPT phases must be distinct if their corresponding twist operators 
have different statistical properties.

It is known that for Abelian edge theories
(e.g., multicomponent chiral/nonchiral bosons compactified on a lattice),
the braiding statistics approach 
and the principle of the
modular invariance of the orbifolded (gauged) edge theory
give the same stability criterion for the edge theories.
This follows, for example, from
a self-dual condition together with an even-lattice condition that guarantee
that modular invariance is achieved~\cite{Sule2013}.
Alternatively, this result can be derived from arguments based on braiding statistics 
\cite{Levin2013}.

In closing, we note that the braiding statistics approach has recently been extended
to (3+1)d SPT phases, in which case one needs to examine the
 statistics among loop excitations
\cite{WangLevin2014, Jiang2014, WangWen2014,Jian2014}.
Gauging symmetries of (3+1)d SPT phases has been studied in
\cite{Chen2014, Cho2014, HsiehChoRyu2015,ChoHsiehMorimotoRyu2015}.

\subsection{Example in (3+1)d: class DIII TSCs}
\label{sec: 3+1 D}

To illustrate the collapse of a non-interacting classification in (3+1)d, let
us now consider TR symmetric SCs in class DIII.

\paragraph{Example: class DIII TSCs}

At the non-interacting level, 3d TR symmetric SCs with 
$\hat{\mathscr{T}}^{2}=\hat{\mathscr{G}}_f$ (i.e., 3d SCs of class DIII)
are classified by the 3d winding number $\nu$ (\ref{winding number examples}),
which counts the number of gapless surface Majorana cones.
One example of a class DIII TSC is the
 B-phase of $^3\mathrm{He}$, described by 
(\ref{BdG 3He B}). This topological superfluid has
$\nu=1$ and supports  at its surface a single Majorana cone
which the low-energy Hamiltonian
\begin{align} \label{surfaceHamDIII_int}
 \hat{H} = \int dx\, dy\, \hat{\psi}^T (-i \partial_x \sigma_3 -i \partial_y \sigma_1)\hat{\psi}, 
\end{align}
where
$\hat{\psi}$ denotes a two-component real fermionic field satisfying $\hat{\psi}^{\dag}=\hat{\psi}$.
The surface Hamiltonian is invariant under TRS, which acts on $\hat{\psi}$  as
$
\hat{ \mathscr{T}} \hat{\psi} \hat{\mathscr{T}}^{-1}  =  i \sigma_2 \hat{\psi}.
$
For TSCs with $\nu=N_f$, the
surface modes are described by $N_f$ copies of Hamiltonian~\eqref{surfaceHamDIII_int}. 

One can verify that in the absence of interactions this surface theory is robust against perturbations for any value of $\nu=N_f$.
In the presence of interactions, however, the surface theory~\eqref{surfaceHamDIII_int} is unstable 
when $\nu=0$ mod 16, leading to a collapse of the non-interacting classification from $\mathbb{Z}$
to $\mathbb{Z}_{16}$ \cite{Fidkowski2013, Metlitski2014, Wang2014, Senthil2014}.
This result has been obtained by a number of different approaches.
Among them are the,
so-called ``vortex condensation approach'' and a method based on
symmetry-preserving surface topological order,
which we will review below (see also \onlinecite{Kitaev_Unpublished, YouXu2014, Kapustin2014e}).
Note that in recent works a similar collapse of non-interacting classifications has been derived 
for (3+1)d crystalline TIs and TSCs
\cite{HsiehChoRyu2015, Isobe2015}.

\subsubsection{Vortex condensation approach and symmetry-preserving surface topological order}

The vortex condensation approach was first developed in the context of bosonic TIs
\cite{Vishwanath2013}, but can also be applied to fermionic SPT phases~\cite{Metlitski2014,Wang2014,you_bentov_xu_arXiv_14}. 
A crucial observation used in this approach is that the gapped surface theory of a trivial insulator
is dual to a quantum disordered superfluid, which is similar to the duality between the superfluid and the Mott insulator phases
of the (2+1)d Bose-Hubbard model.
This approach hence applies most directly to SPT phases,
whose symmetry group contains a $U(1)$ symmetry,
for example, 
an $S_z$ spin-rotation symmetry or an electromagnetic charge conservation.
One then imagines driving the surface of the SPT phase into a ``superfluid" phase, which spontaneously 
breaks the $U(1)$ symmetry and leads to a gapped surface. 
The non-trivial topology of the symmetry-broken surface state can then be inferred 
from the properties of the topological defects of the order parameter, i.e., from
the vortices of the superfluid. 

One possibility is that
quantum disordering the superfluid by proliferating (condensing) the vortices restores the $U(1)$ symmetry, leading
to a topologically trivial gapped surface that respects all symmetries. This then indicates that the bulk phase
is topologically trivial.
However, this is only possible if the vortices do not have any abnormal properties.
For example, if the vortices transform abnormally under the symmetries or if they have exotic exchange statistics,
it may not be possible to condense the vortices, such that the surface becomes a gapped trivial state respecting
all the symmetries.

Another possibility is that, while vortices may be anomalous in the above sense, 
vortices with vorticity $>1$ (i.e., multi-vortices) may behave in an ordinary way.
If this is the case, it might be possible to condense these multi-vortices to form
a gapped surface state that respects all symmetries.
This surface state, however, inevitably exhibits an intrinsic topological order
\cite{Balents1999, Senthil2000}, thereby signaling that the bulk phase is nontrivial.
This surface topological order is anomalous, since it cannot be  realized
in an isolated (2+1)d system while preserving the symmetries.
The existence of symmetry preserving surface topological order may in fact be used as
a non-perturbative definition of 3d SPT phases.
Surface states with symmetry preserving topological order
have recently been constructed for  fermionic TIs
\cite{Metlitski2013b, Bonderson2013, ChenFidkowskiVishwanath2014, WangPotterSenthilgapTI2013,WangPotterSenthil2014}, 
as well as for bosonic TIs \cite{Vishwanath2013,Metlitski2013a,WangSenthil2013,Burnell2013},


\paragraph{Application to class DIII TSC}

Let us now discuss how the vortex condensation approach works 
for the 3d class DIII TSC with an even number $\nu$ of Majorana surface
cones~\cite{Metlitski2014,Wang2014}. Since $\nu$ is even, we can construct
an artificial ``flavor'' $U(1)$ symmetry by combing Majorana cones pairwise.
We then drive the surface state into a superfluid phase
where this artificial $U(1)$ symmetry is spontaneously broken and the surface Majorana cones are gapped. 
Next, we imagine quantum disordering the superfluid by condensing the vortices.
However, it turns out that for general $\nu$ the vortices are nontrivial: 
An elementary vortex (with vorticity $1$) binds $\nu/2$ chiral Majorana modes.
Hence the vortex core resembles the edge of a 1D TSC in class BDI.
As discussed in Sec.~\ref{Example in (1+1)d: class BDI Majorana chain},
 interactions can gap out these Majorana modes without breaking the symmetries when they come in multiples of 8.
 Thus, for  $\nu =16$ the vortices on the surface of the class DIII TSC are trivial.  
Hence, by condensing these vortices the $U(1)$ symmetry can be restored, which gives
rise to a topologically trivial  gapped surface state which respects all symmetries of class DIII. 
However, for smaller even $\nu$, the elementary vortices  are non-trivial and cannot condense without breaking the symmetries.
This confirms the collapse of the non-interacting classification from
$\mathbb{Z}\to \mathbb{Z}_{16}$, discussed above.
In \onlinecite{Fidkowski2013}  symmetry-preserving gapped surface states with intrinsic topological order
have been constructed explicitly for this 3d TSC.

\subsection{Proposed classification scheme of SPT phases}
\label{proposedclassificationschemeSPT}

So far we have introduced  various approaches 
to {\it diagnose} the properties of a given interacting SPT phase.  
More generically, one would like to obtain a comprehensive and exhaustive {\it classification} all possible SPT phases.
Here, we present two approaches to this problem:
the group cohomology method and the cobordism approach.
For other related and complementary approaches, see also 
\cite{Freed2014, WenNLSMs2014}. 

\subsubsection{Group cohomology approach}

The idea of using MPSs to diagnose and distinguish SPT phases 
discussed in Sec.\ \ref{MPS}, 
can be used
for ground states of generic gapped Hamiltonians in $(1+1)$d, and
in fact, provides a complete classification of SPT phases in $(1+1)$d
\cite{Pollmann2010, Chen2011, Pollmann2012, Schuch2011,Chen2011b}.
Recall from Sec.~\ref{MPS} that the phases 
$\alpha(g,h)$  of \eqref{MPS_projective_reps} distinguish
between different projective representations of the symmetry group $G$
and hence between different SPT phases. 
(Note that the set of phase functions $\alpha(g,h)$ are called 2-cocycles, since they must satisfy 
the so-called 2-cocycle condition, 
$\alpha(h,k)+\alpha(g, hk)= \alpha(gh,k)+\alpha(h,k)$,
which follows from associativity of the symmetry group.)
Since $\hat{V}(g)$  in \eqref{MPS_projective_reps} is defined only up to a phase, 
one finds that two different projective representations with the phase functions $\alpha_1(g,h)$ and $\alpha_2(g,h)$
are equivalent, if they are related by  
$\alpha_2(g,h) = \alpha_1(g,h) + \beta(gh) -\beta(g) - \beta(h)$.
(Here, the function $\beta$ is called a coboundary).
This relation defines equivalent classes, which form a group the so-called
second cohomology group of $G$ over $U(1)$ denoted by $H^2 (G, {U} (1))$.
Different gapped (1+1)d phases with symmetry $G$ are then classified by 
$H^2 (G, {U} (1))$. 
In higher dimensions $d > 1$, 
a large class of bosonic SPT phases can be systematically constructed using
the tensor network method and path integrals on discrete spacetime
using elements in $H^{d+1}(G, U(1))$ 
\cite{Chen2012, Chen2013, DijkgraafWitten1990}. 
For fermionic systems, group supercohomology theory has been used to classify SPT phases
\cite{Gu2012}.

\subsubsection{Cobordism approach}

While the group cohomology approach is one of the most systematic and general methods to classify SPT phases,
it was shown that it does not describe all possible phases~\cite{Kapustin2014b,Vishwanath2013,WangSenthil2013}.
Among these is a bosonic SPT phase in (3+1)d
\cite{Vishwanath2013}.
An alternative approach to classify SPT phases,  based on the cobordism,
was proposed 
\cite{Kapustin2014a, Kapustin2014b, Kapustin2014c, Kapustin2014d, Kapustin2014e}.
Assuming that the low-energy effective action of the SPT phase
is cobordism-invariant, SPT phases with finite symmetry group $G$ have been classified by use of
the cobordism groups of the classifying spaces corresponding to $G$.
As an example, the result of this classification for fermionic SPT phases with various symmetries is shown in Table \ref{tab:cobordism}. 
Note that the classification shown in this table agrees with the results presented in the previous sections.


\begin{table}[tb]
\centering 
\begin{ruledtabular}
\begin{tabular}{c|cccccccccc} 
symmetry   $\backslash d $ & 0 & 1 & 2 & 3 & 4 & 5 & 6    
\\  \hline
no symmetry  (D) & $\mathbb{Z}_2$ & $\mathbb{Z}_2$ &$\mathbb{Z}$ & 0 & 0 & 0 & $\mathbb{Z}^2$  
\\
$\hat{\mathscr{T}}^2=1$ (BDI)  & $\mathbb{Z}_2$ & $\mathbb{Z}_8$ & 0 & 0 & 0 & $\mathbb{Z}_{16}$ & 0 
\\  
$\hat{\mathscr{T}}^2=\hat{\mathscr{G}}_f$ (DIII)  & 0 & $\mathbb{Z}_2$ & $\mathbb{Z}_2$  & $\mathbb{Z}_{16}$ & 0 & 0 & 0      
\\
unitary  $\mathbb{Z}_2$ & $\mathbb{Z}^2_2$ & $\mathbb{Z}_2^2$ & $\mathbb{Z}_8\times \mathbb{Z}$   & 0 & 0 & 0 & $\mathbb{Z}_{12}\times \mathbb{Z}^2$  
\end{tabular}
\end{ruledtabular}
\caption{
Classification of interacting fermionic SPT phases as a function of spatial dimension $d$, as derived from the cobordism approach 
\cite{Kapustin2014e}. 
\label{tab:cobordism}
}
\end{table}

\section{Outlook and future directions}
\label{sec:outlook}

The discovery of spin-orbit induced topological insulators has taught us that
topological effects, which were long thought to occur only under extreme conditions, can profoundly affect the properties of seemingly normal materials, such as band insulators, even under ordinary conditions.
Over the last few years, research on topological materials has made impressive progress starting from 
the experimental realizations of the quantum spin Hall  
 and  quantum anomalous Hall effects to the construction and classification of interacting SPT phases.
While the basic properties of noninteracting topological systems are relatively well understood theoretically,
there is much work to be done to find, design, and improve material systems that
realize the theoretical models and allow to experimentally verify the theoretical predictions. 

For further progress on the theoretical front, 
a deeper understanding of fractional topological phases and 
correlated SPT phases in $d>1$ is important.
Other outstanding problems include realistic material predictions for
interacting SPTs, fractional TIs, and TSCs,
and the development of effective field theory descriptions. 
Furthermore, the full classification of noninteracting Hamiltonians
in terms of all (magnetic) space group symmetries, in particular nonsymmorphic symmetries, remains as 
an important open issue for future research.

On the experimental side, perhaps the most important task is the engineering of topological electronic states.
An attractive possibility is to realize topological phases in heterostructures, involving for example
iridates or other materials with strong SOC~\cite{okamotoNatComm2011}, since this allows for a fine control
of the interface properties and therefore of the topological state. There is already a large number of experimental studies, that investigate interfaces between TIs and $s$-wave~\cite{wang_Xue_science_12} or $d_{x^2-y^2}$-wave superconductors~\cite{burch_natComm12,wang_d_wave_Bi2Se3_NatPhys13}.
We expect that this remains a major research direction for the foreseeable future.
Another important task is the perfection of existing materials, 
in particular the growth of topological materials with sufficiently high purity, such that the bulk is truly insulating.

There are numerous topics and developments which we could not mention in this review due to space limitations.
These include 
topological field theories describing the electromagnetic, thermal, or gravitational responses of topological phases
~\cite{Qi2008sf,Furusaki:responses,RyuMooreLudwig2012,Atma_EFT},
Floquet topological insulators~\cite{lindnerFloquetTI_NatPhys11,Takuya_floquet_TI,Ezawa_floquet_TI,Wang453}, 
topological phases of ultracold atoms~\cite{zoller_PRL_11,goldman_PRL12,sun_NatPhys_12,AMO_topo_review}, 
photonic topological insulators~\cite{Rechtsman_photonic_TI,Khanikaev_photonic_TI}, 
topological states in quasicrystals~\cite{Kraus_topo_quasicrystal,Verbin_Topo_quasicrystal}, 
and quantum phase transitions without gap closing in the presence of interactions~\cite{transition_nogap}.
Other interesting topics that we left out are
symmetry-enriched topological phases
\cite{
2014arXiv1410.4540B,
2015arXiv150306812T, 
symmetry_enriched_ying_ran_PRB_13,LuVishwanath2012,essin_hermele_PRB_13,
cho_moore_PRB_2012,Chen2013,wang_essin_hermele_motrunich_PRB_15,Hung_SET}, 
and experimental realizations of interacting SPT phases~\cite{lu_lee_PRB_14}.

We also did not have space to discuss potential applications that utilize the conducting edge states of topological materials.
Possible avenues for technological uses are low-power-consumption electronic devices based on the dissipationless
edge currents of TIs. 
Furthermore, TSCs or heterostructures between TIs and SCs might lead
to new architectures of quantum computing devices.
An important first step in order to realization such devices is to control and manipulate the topological currents using, e.g., electric fields~\cite{ezawa_siliceneNJP12,ezawaPRL15,wray_hasan_JPCS_13}, magnetic fields~\cite{garate_franz_2010,schnyder_PRL13,linder_tanaka_PRL_10}, or mechanical strain~\cite{wenliang_PRB_11,winterfeld_PRB_13}. 
\acknowledgments 

We would like to thank our colleagues and collaborators.
We are grateful to the community for many valuable comments and encouragements.
S.R., A.P.S., and J.C.Y.T. wish to thank the ESI (Vienna) for its hospitality.
C.K.C. and A.P.S. acknowledge
the support of the Max-Planck-UBC Centre for Quantum Materials.  
The work by S.R. has been supported by the NSF under Grant
No. DMR-1455296 and the Alfred P. Sloan foundation. C.K.C. was also supported by Microsoft and LPS-MPO-CMTC during the resubmission stage of this work at the University of Maryland.

\bibliography{TOPO}

\begin{thebibliography}{568}%
\makeatletter
\providecommand \@ifxundefined [1]{%
 \@ifx{#1\undefined}
}%
\providecommand \@ifnum [1]{%
 \ifnum #1\expandafter \@firstoftwo
 \else \expandafter \@secondoftwo
 \fi
}%
\providecommand \@ifx [1]{%
 \ifx #1\expandafter \@firstoftwo
 \else \expandafter \@secondoftwo
 \fi
}%
\providecommand \natexlab [1]{#1}%
\providecommand \enquote  [1]{``#1''}%
\providecommand \bibnamefont  [1]{#1}%
\providecommand \bibfnamefont [1]{#1}%
\providecommand \citenamefont [1]{#1}%
\providecommand \href@noop [0]{\@secondoftwo}%
\providecommand \href [0]{\begingroup \@sanitize@url \@href}%
\providecommand \@href[1]{\@@startlink{#1}\@@href}%
\providecommand \@@href[1]{\endgroup#1\@@endlink}%
\providecommand \@sanitize@url [0]{\catcode `\\12\catcode `\$12\catcode
  `\&12\catcode `\#12\catcode `\^12\catcode `\_12\catcode `\%12\relax}%
\providecommand \@@startlink[1]{}%
\providecommand \@@endlink[0]{}%
\providecommand \url  [0]{\begingroup\@sanitize@url \@url }%
\providecommand \@url [1]{\endgroup\@href {#1}{\urlprefix }}%
\providecommand \urlprefix  [0]{URL }%
\providecommand \Eprint [0]{\href }%
\providecommand \doibase [0]{http://dx.doi.org/}%
\providecommand \selectlanguage [0]{\@gobble}%
\providecommand \bibinfo  [0]{\@secondoftwo}%
\providecommand \bibfield  [0]{\@secondoftwo}%
\providecommand \translation [1]{[#1]}%
\providecommand \BibitemOpen [0]{}%
\providecommand \bibitemStop [0]{}%
\providecommand \bibitemNoStop [0]{.\EOS\space}%
\providecommand \EOS [0]{\spacefactor3000\relax}%
\providecommand \BibitemShut  [1]{\csname bibitem#1\endcsname}%
\let\auto@bib@innerbib\@empty
\bibitem [{\citenamefont {Abrahams}\ \emph {et~al.}(1979)\citenamefont
  {Abrahams}, \citenamefont {Anderson}, \citenamefont {Licciardello},\ and\
  \citenamefont {Ramakrishnan}}]{Abrahams1979}%
  \BibitemOpen
  \bibfield  {author} {\bibinfo {author} {\bibnamefont {Abrahams},
  \bibfnamefont {E.}}, \bibinfo {author} {\bibfnamefont {P.~W.}\ \bibnamefont
  {Anderson}}, \bibinfo {author} {\bibfnamefont {D.~C.}\ \bibnamefont
  {Licciardello}}, \ and\ \bibinfo {author} {\bibfnamefont {T.~V.}\
  \bibnamefont {Ramakrishnan}}} (\bibinfo {year} {1979}),\ \href {\doibase
  10.1103/PhysRevLett.42.673} {\bibfield  {journal} {\bibinfo  {journal} {Phys.
  Rev. Lett.}\ }\textbf {\bibinfo {volume} {42}},\ \bibinfo {pages}
  {673}}\BibitemShut {NoStop}%
\bibitem [{\citenamefont {Abramovici}\ and\ \citenamefont
  {Kalugin}(2012)}]{Abramovici:2012zz}%
  \BibitemOpen
  \bibfield  {author} {\bibinfo {author} {\bibnamefont {Abramovici},
  \bibfnamefont {G.}}, \ and\ \bibinfo {author} {\bibfnamefont
  {P.}~\bibnamefont {Kalugin}}} (\bibinfo {year} {2012}),\ \href {\doibase
  10.1142/S0219887812500235} {\bibfield  {journal} {\bibinfo  {journal} {Int.
  J. Geom. Meth. Mod. Phys.}\ }\textbf {\bibinfo {volume} {9}},\ \bibinfo
  {pages} {1250023}}\BibitemShut {NoStop}%
\bibitem [{\citenamefont {Affleck}(1988)}]{Affleck:1988}%
  \BibitemOpen
  \bibfield  {author} {\bibinfo {author} {\bibnamefont {Affleck}, \bibfnamefont
  {I.}}} (\bibinfo {year} {1988}),\ in\ \href@noop {} {\emph {\bibinfo
  {booktitle} {Les Houches, Session XLIX}}},\ \bibinfo {editor} {edited by\
  \bibinfo {editor} {\bibfnamefont {E.}~\bibnamefont {Brezin}}\ and\ \bibinfo
  {editor} {\bibfnamefont {J.}~\bibnamefont {Zinn-Justin}}}\ (\bibinfo
  {publisher} {North-Holland},\ \bibinfo {address} {Amsterdam})\ pp.\ \bibinfo
  {pages} {563--640}\BibitemShut {NoStop}%
\bibitem [{\citenamefont {{Akhmerov}}\ \emph
  {et~al.}(2011{\natexlab{a}})\citenamefont {{Akhmerov}}, \citenamefont
  {{Dahlhaus}}, \citenamefont {{Hassler}}, \citenamefont {{Wimmer}},\ and\
  \citenamefont {{Beenakker}}}]{Akhmerov2011}%
  \BibitemOpen
  \bibfield  {author} {\bibinfo {author} {\bibnamefont {{Akhmerov}},
  \bibfnamefont {A.~R.}}, \bibinfo {author} {\bibfnamefont {J.~P.}\
  \bibnamefont {{Dahlhaus}}}, \bibinfo {author} {\bibfnamefont
  {F.}~\bibnamefont {{Hassler}}}, \bibinfo {author} {\bibfnamefont
  {M.}~\bibnamefont {{Wimmer}}}, \ and\ \bibinfo {author} {\bibfnamefont
  {C.~W.~J.}\ \bibnamefont {{Beenakker}}}} (\bibinfo {year}
  {2011}{\natexlab{a}}),\ \href {\doibase 10.1103/PhysRevLett.106.057001}
  {\bibfield  {journal} {\bibinfo  {journal} {Phys. Rev. Lett.}\ }\textbf
  {\bibinfo {volume} {106}}~(\bibinfo {number} {5}),\ \bibinfo {eid}
  {057001}}\BibitemShut {NoStop}%
\bibitem [{\citenamefont {{Akhmerov}}\ \emph
  {et~al.}(2011{\natexlab{b}})\citenamefont {{Akhmerov}}, \citenamefont
  {{Dahlhaus}}, \citenamefont {{Hassler}}, \citenamefont {{Wimmer}},\ and\
  \citenamefont {{Beenakker}}}]{Akhmerov2011PhRvL.106e7001A}%
  \BibitemOpen
  \bibfield  {author} {\bibinfo {author} {\bibnamefont {{Akhmerov}},
  \bibfnamefont {A.~R.}}, \bibinfo {author} {\bibfnamefont {J.~P.}\
  \bibnamefont {{Dahlhaus}}}, \bibinfo {author} {\bibfnamefont
  {F.}~\bibnamefont {{Hassler}}}, \bibinfo {author} {\bibfnamefont
  {M.}~\bibnamefont {{Wimmer}}}, \ and\ \bibinfo {author} {\bibfnamefont
  {C.~W.~J.}\ \bibnamefont {{Beenakker}}}} (\bibinfo {year}
  {2011}{\natexlab{b}}),\ \href {\doibase 10.1103/PhysRevLett.106.057001}
  {\bibfield  {journal} {\bibinfo  {journal} {Physical Review Letters}\
  }\textbf {\bibinfo {volume} {106}}~(\bibinfo {number} {5}),\ \bibinfo {eid}
  {057001}}\BibitemShut {NoStop}%
\bibitem [{\citenamefont {Alexandradinata}\ \emph {et~al.}(2014)\citenamefont
  {Alexandradinata}, \citenamefont {Fang}, \citenamefont {Gilbert},\ and\
  \citenamefont {Bernevig}}]{alexandradinata_bernevig_PRL14}%
  \BibitemOpen
  \bibfield  {author} {\bibinfo {author} {\bibnamefont {Alexandradinata},
  \bibfnamefont {A.}}, \bibinfo {author} {\bibfnamefont {C.}~\bibnamefont
  {Fang}}, \bibinfo {author} {\bibfnamefont {M.~J.}\ \bibnamefont {Gilbert}}, \
  and\ \bibinfo {author} {\bibfnamefont {B.~A.}\ \bibnamefont {Bernevig}}}
  (\bibinfo {year} {2014}),\ \href {\doibase 10.1103/PhysRevLett.113.116403}
  {\bibfield  {journal} {\bibinfo  {journal} {Phys. Rev. Lett.}\ }\textbf
  {\bibinfo {volume} {113}},\ \bibinfo {pages} {116403}}\BibitemShut {NoStop}%
\bibitem [{\citenamefont {Alicea}(2012)}]{Alicea:2012em}%
  \BibitemOpen
  \bibfield  {author} {\bibinfo {author} {\bibnamefont {Alicea}, \bibfnamefont
  {J.}}} (\bibinfo {year} {2012}),\ \href
  {http://stacks.iop.org/0034-4885/75/i=7/a=076501} {\bibfield  {journal}
  {\bibinfo  {journal} {Reports on Progress in Physics}\ }\textbf {\bibinfo
  {volume} {75}},\ \bibinfo {pages} {076501}}\BibitemShut {NoStop}%
\bibitem [{\citenamefont {Alicea}\ \emph {et~al.}(2011)\citenamefont {Alicea},
  \citenamefont {Oreg}, \citenamefont {Refael}, \citenamefont {von Oppen},\
  and\ \citenamefont {Fisher}}]{Alicea_Majorana_wire}%
  \BibitemOpen
  \bibfield  {author} {\bibinfo {author} {\bibnamefont {Alicea}, \bibfnamefont
  {J.}}, \bibinfo {author} {\bibfnamefont {Y.}~\bibnamefont {Oreg}}, \bibinfo
  {author} {\bibfnamefont {G.}~\bibnamefont {Refael}}, \bibinfo {author}
  {\bibfnamefont {F.}~\bibnamefont {von Oppen}}, \ and\ \bibinfo {author}
  {\bibfnamefont {M.~P.~A.}\ \bibnamefont {Fisher}}} (\bibinfo {year} {2011}),\
  \href {\doibase 10.1038/nphys1915} {\bibfield  {journal} {\bibinfo  {journal}
  {Nature Physics}\ }\textbf {\bibinfo {volume} {7}},\ \bibinfo {pages}
  {412}}\BibitemShut {NoStop}%
\bibitem [{\citenamefont {{Alpichshev}}\ \emph {et~al.}(2010)\citenamefont
  {{Alpichshev}}, \citenamefont {{Analytis}}, \citenamefont {{Chu}},
  \citenamefont {{Fisher}}, \citenamefont {{Chen}}, \citenamefont {{Shen}},
  \citenamefont {{Fang}},\ and\ \citenamefont {{Kapitulnik}}}]{Alpichshev2010}%
  \BibitemOpen
  \bibfield  {author} {\bibinfo {author} {\bibnamefont {{Alpichshev}},
  \bibfnamefont {Z.}}, \bibinfo {author} {\bibfnamefont {J.~G.}\ \bibnamefont
  {{Analytis}}}, \bibinfo {author} {\bibfnamefont {J.-H.}\ \bibnamefont
  {{Chu}}}, \bibinfo {author} {\bibfnamefont {I.~R.}\ \bibnamefont {{Fisher}}},
  \bibinfo {author} {\bibfnamefont {Y.~L.}\ \bibnamefont {{Chen}}}, \bibinfo
  {author} {\bibfnamefont {Z.~X.}\ \bibnamefont {{Shen}}}, \bibinfo {author}
  {\bibfnamefont {A.}~\bibnamefont {{Fang}}}, \ and\ \bibinfo {author}
  {\bibfnamefont {A.}~\bibnamefont {{Kapitulnik}}}} (\bibinfo {year} {2010}),\
  \href {\doibase 10.1103/PhysRevLett.104.016401} {\bibfield  {journal}
  {\bibinfo  {journal} {Phys. Rev. Lett.}\ }\textbf {\bibinfo {volume}
  {104}}~(\bibinfo {number} {1}),\ \bibinfo {eid} {016401}}\BibitemShut
  {NoStop}%
\bibitem [{\citenamefont {{Altland}}\ \emph {et~al.}(2014)\citenamefont
  {{Altland}}, \citenamefont {{Bagrets}}, \citenamefont {{Fritz}},
  \citenamefont {{Kamenev}},\ and\ \citenamefont
  {{Schmiedt}}}]{Altland2014PhRvL.112t6602A}%
  \BibitemOpen
  \bibfield  {author} {\bibinfo {author} {\bibnamefont {{Altland}},
  \bibfnamefont {A.}}, \bibinfo {author} {\bibfnamefont {D.}~\bibnamefont
  {{Bagrets}}}, \bibinfo {author} {\bibfnamefont {L.}~\bibnamefont {{Fritz}}},
  \bibinfo {author} {\bibfnamefont {A.}~\bibnamefont {{Kamenev}}}, \ and\
  \bibinfo {author} {\bibfnamefont {H.}~\bibnamefont {{Schmiedt}}}} (\bibinfo
  {year} {2014}),\ \href {\doibase 10.1103/PhysRevLett.112.206602} {\bibfield
  {journal} {\bibinfo  {journal} {Physical Review Letters}\ }\textbf {\bibinfo
  {volume} {112}}~(\bibinfo {number} {20}),\ \bibinfo {eid}
  {206602}}\BibitemShut {NoStop}%
\bibitem [{\citenamefont {{Altland}}\ \emph {et~al.}(2015)\citenamefont
  {{Altland}}, \citenamefont {{Bagrets}},\ and\ \citenamefont
  {{Kamenev}}}]{Altland2015PhRvB..91h5429A}%
  \BibitemOpen
  \bibfield  {author} {\bibinfo {author} {\bibnamefont {{Altland}},
  \bibfnamefont {A.}}, \bibinfo {author} {\bibfnamefont {D.}~\bibnamefont
  {{Bagrets}}}, \ and\ \bibinfo {author} {\bibfnamefont {A.}~\bibnamefont
  {{Kamenev}}}} (\bibinfo {year} {2015}),\ \href {\doibase
  10.1103/PhysRevB.91.085429} {\bibfield  {journal} {\bibinfo  {journal}
  {\prb}\ }\textbf {\bibinfo {volume} {91}}~(\bibinfo {number} {8}),\ \bibinfo
  {eid} {085429}}\BibitemShut {NoStop}%
\bibitem [{\citenamefont {{Altland}}\ \emph {et~al.}(2002)\citenamefont
  {{Altland}}, \citenamefont {{Simons}},\ and\ \citenamefont
  {{Zirnbauer}}}]{AltlandSimonsZirnbauer2002}%
  \BibitemOpen
  \bibfield  {author} {\bibinfo {author} {\bibnamefont {{Altland}},
  \bibfnamefont {A.}}, \bibinfo {author} {\bibfnamefont {B.~D.}\ \bibnamefont
  {{Simons}}}, \ and\ \bibinfo {author} {\bibfnamefont {M.~R.}\ \bibnamefont
  {{Zirnbauer}}}} (\bibinfo {year} {2002}),\ \href {\doibase
  10.1016/S0370-1573(01)00065-5} {\bibfield  {journal} {\bibinfo  {journal}
  {Physics Report}\ }\textbf {\bibinfo {volume} {359}},\ \bibinfo {pages}
  {283}}\BibitemShut {NoStop}%
\bibitem [{\citenamefont {Altland}\ and\ \citenamefont
  {Zirnbauer}(1997)}]{altlandZirnbauerPRB10}%
  \BibitemOpen
  \bibfield  {author} {\bibinfo {author} {\bibnamefont {Altland}, \bibfnamefont
  {A.}}, \ and\ \bibinfo {author} {\bibfnamefont {M.~R.}\ \bibnamefont
  {Zirnbauer}}} (\bibinfo {year} {1997}),\ \href {\doibase
  10.1103/PhysRevB.55.1142} {\bibfield  {journal} {\bibinfo  {journal} {Phys.
  Rev. B}\ }\textbf {\bibinfo {volume} {55}},\ \bibinfo {pages}
  {1142}}\BibitemShut {NoStop}%
\bibitem [{\citenamefont {Alvarez-Gaume}\ and\ \citenamefont
  {Witten}(1984)}]{AlvarezGaumeWitten1983}%
  \BibitemOpen
  \bibfield  {author} {\bibinfo {author} {\bibnamefont {Alvarez-Gaume},
  \bibfnamefont {L.}}, \ and\ \bibinfo {author} {\bibfnamefont
  {E.}~\bibnamefont {Witten}}} (\bibinfo {year} {1984}),\ \href {\doibase
  10.1016/0550-3213(84)90066-X} {\bibfield  {journal} {\bibinfo  {journal}
  {Nucl. Phys. B}\ }\textbf {\bibinfo {volume} {234}},\ \bibinfo {pages}
  {269}}\BibitemShut {NoStop}%
\bibitem [{\citenamefont {Amaricci}\ \emph {et~al.}(2015)\citenamefont
  {Amaricci}, \citenamefont {Budich}, \citenamefont {Capone}, \citenamefont
  {Trauzettel},\ and\ \citenamefont {Sangiovanni}}]{transition_nogap}%
  \BibitemOpen
  \bibfield  {author} {\bibinfo {author} {\bibnamefont {Amaricci},
  \bibfnamefont {A.}}, \bibinfo {author} {\bibfnamefont {J.~C.}\ \bibnamefont
  {Budich}}, \bibinfo {author} {\bibfnamefont {M.}~\bibnamefont {Capone}},
  \bibinfo {author} {\bibfnamefont {B.}~\bibnamefont {Trauzettel}}, \ and\
  \bibinfo {author} {\bibfnamefont {G.}~\bibnamefont {Sangiovanni}}} (\bibinfo
  {year} {2015}),\ \href {\doibase 10.1103/PhysRevLett.114.185701} {\bibfield
  {journal} {\bibinfo  {journal} {Phys. Rev. Lett.}\ }\textbf {\bibinfo
  {volume} {114}},\ \bibinfo {pages} {185701}}\BibitemShut {NoStop}%
\bibitem [{\citenamefont {Anderson}(1958)}]{Anderson1958}%
  \BibitemOpen
  \bibfield  {author} {\bibinfo {author} {\bibnamefont {Anderson},
  \bibfnamefont {P.~W.}}} (\bibinfo {year} {1958}),\ \href {\doibase
  10.1103/PhysRev.109.1492} {\bibfield  {journal} {\bibinfo  {journal} {Phys.
  Rev.}\ }\textbf {\bibinfo {volume} {109}},\ \bibinfo {pages}
  {1492}}\BibitemShut {NoStop}%
\bibitem [{\citenamefont {Anderson}\ and\ \citenamefont
  {Morel}(1961)}]{AndersonMorel61}%
  \BibitemOpen
  \bibfield  {author} {\bibinfo {author} {\bibnamefont {Anderson},
  \bibfnamefont {P.~W.}}, \ and\ \bibinfo {author} {\bibfnamefont
  {P.}~\bibnamefont {Morel}}} (\bibinfo {year} {1961}),\ \href {\doibase
  10.1103/PhysRev.123.1911} {\bibfield  {journal} {\bibinfo  {journal} {Phys.
  Rev.}\ }\textbf {\bibinfo {volume} {123}},\ \bibinfo {pages}
  {1911}}\BibitemShut {NoStop}%
\bibitem [{\citenamefont {Ando}\ \emph {et~al.}(1998)\citenamefont {Ando},
  \citenamefont {Nakanishi},\ and\ \citenamefont
  {Saito}}]{AndoNakanishiSaito1998}%
  \BibitemOpen
  \bibfield  {author} {\bibinfo {author} {\bibnamefont {Ando}, \bibfnamefont
  {T.}}, \bibinfo {author} {\bibfnamefont {T.}~\bibnamefont {Nakanishi}}, \
  and\ \bibinfo {author} {\bibfnamefont {R.}~\bibnamefont {Saito}}} (\bibinfo
  {year} {1998}),\ \href {\doibase 10.1143/JPSJ.67.2857} {\bibfield  {journal}
  {\bibinfo  {journal} {J. Phys. Soc. Jpn.}\ }\textbf {\bibinfo {volume}
  {67}}~(\bibinfo {number} {8}),\ \bibinfo {pages} {2857}}\BibitemShut
  {NoStop}%
\bibitem [{\citenamefont {Ando}(2013)}]{andoJPSJreview13}%
  \BibitemOpen
  \bibfield  {author} {\bibinfo {author} {\bibnamefont {Ando}, \bibfnamefont
  {Y.}}} (\bibinfo {year} {2013}),\ \href@noop {} {\bibfield  {journal}
  {\bibinfo  {journal} {J. Phys. Soc. Jpn.}\ }\textbf {\bibinfo {volume}
  {82}}~(\bibinfo {number} {10}),\ \bibinfo {pages} {102001}}\BibitemShut
  {NoStop}%
\bibitem [{\citenamefont {{Ando}}\ and\ \citenamefont
  {{Fu}}(2015)}]{andoFuReviewArxiv15}%
  \BibitemOpen
  \bibfield  {author} {\bibinfo {author} {\bibnamefont {{Ando}}, \bibfnamefont
  {Y.}}, \ and\ \bibinfo {author} {\bibfnamefont {L.}~\bibnamefont {{Fu}}}}
  (\bibinfo {year} {2015}),\ \href@noop {} {\ }\Eprint
  {http://arxiv.org/abs/1501.00531} {arXiv:1501.00531} \BibitemShut {NoStop}%
\bibitem [{\citenamefont {Ando}\ and\ \citenamefont
  {Fu}(2015)}]{Ando_Fu_TCI_review}%
  \BibitemOpen
  \bibfield  {author} {\bibinfo {author} {\bibnamefont {Ando}, \bibfnamefont
  {Y.}}, \ and\ \bibinfo {author} {\bibfnamefont {L.}~\bibnamefont {Fu}}}
  (\bibinfo {year} {2015}),\ \href {\doibase
  10.1146/annurev-conmatphys-031214-014501} {\bibfield  {journal} {\bibinfo
  {journal} {Annual Review of Condensed Matter Physics}\ }\textbf {\bibinfo
  {volume} {6}}~(\bibinfo {number} {1}),\ \bibinfo {pages} {361}}\BibitemShut
  {NoStop}%
\bibitem [{\citenamefont {Asahi}\ and\ \citenamefont
  {Nagaosa}(2012)}]{AsahiNagaosa12}%
  \BibitemOpen
  \bibfield  {author} {\bibinfo {author} {\bibnamefont {Asahi}, \bibfnamefont
  {D.}}, \ and\ \bibinfo {author} {\bibfnamefont {N.}~\bibnamefont {Nagaosa}}}
  (\bibinfo {year} {2012}),\ \href {\doibase 10.1103/PhysRevB.86.100504}
  {\bibfield  {journal} {\bibinfo  {journal} {Phys. Rev. B}\ }\textbf {\bibinfo
  {volume} {86}},\ \bibinfo {pages} {100504}}\BibitemShut {NoStop}%
\bibitem [{\citenamefont {Atiyah}(1994)}]{Atiyahbook}%
  \BibitemOpen
  \bibfield  {author} {\bibinfo {author} {\bibnamefont {Atiyah}, \bibfnamefont
  {M.}}} (\bibinfo {year} {1994}),\ \href@noop {} {\emph {\bibinfo {title}
  {K-Theory}}}\ (\bibinfo  {publisher} {Westview Press})\BibitemShut {NoStop}%
\bibitem [{\citenamefont {Atiyah}\ \emph {et~al.}(1964)\citenamefont {Atiyah},
  \citenamefont {Bott},\ and\ \citenamefont {Shapiro}}]{Atiyah19643}%
  \BibitemOpen
  \bibfield  {author} {\bibinfo {author} {\bibnamefont {Atiyah}, \bibfnamefont
  {M.}}, \bibinfo {author} {\bibfnamefont {R.}~\bibnamefont {Bott}}, \ and\
  \bibinfo {author} {\bibfnamefont {A.}~\bibnamefont {Shapiro}}} (\bibinfo
  {year} {1964}),\ \href {\doibase
  http://dx.doi.org/10.1016/0040-9383(64)90003-5} {\bibfield  {journal}
  {\bibinfo  {journal} {Topology}\ }\textbf {\bibinfo {volume} {3, Supplement
  1}}~(\bibinfo {number} {0}),\ \bibinfo {pages} {3 }}\BibitemShut {NoStop}%
\bibitem [{\citenamefont {Atiyah}\ and\ \citenamefont
  {Singer}(1963)}]{AtiyahSinger63}%
  \BibitemOpen
  \bibfield  {author} {\bibinfo {author} {\bibnamefont {Atiyah}, \bibfnamefont
  {M.~F.}}, \ and\ \bibinfo {author} {\bibfnamefont {I.~M.}\ \bibnamefont
  {Singer}}} (\bibinfo {year} {1963}),\ \href
  {http://www.ams.org/journals/bull/1963-69-03/S0002-9904-1963-10957-X/home.html}
  {\bibfield  {journal} {\bibinfo  {journal} {Bull. Amer. Math. Soc.}\ }\textbf
  {\bibinfo {volume} {69}},\ \bibinfo {pages} {422}}\BibitemShut {NoStop}%
\bibitem [{\citenamefont {{Balents}}\ \emph {et~al.}(1999)\citenamefont
  {{Balents}}, \citenamefont {{Fisher}},\ and\ \citenamefont
  {{Nayak}}}]{Balents1999}%
  \BibitemOpen
  \bibfield  {author} {\bibinfo {author} {\bibnamefont {{Balents}},
  \bibfnamefont {L.}}, \bibinfo {author} {\bibfnamefont {M.~P.~A.}\
  \bibnamefont {{Fisher}}}, \ and\ \bibinfo {author} {\bibfnamefont
  {C.}~\bibnamefont {{Nayak}}}} (\bibinfo {year} {1999}),\ \href {\doibase
  10.1103/PhysRevB.60.1654} {\bibfield  {journal} {\bibinfo  {journal} {\prb}\
  }\textbf {\bibinfo {volume} {60}},\ \bibinfo {pages} {1654}}\BibitemShut
  {NoStop}%
\bibitem [{\citenamefont {Balian}\ and\ \citenamefont
  {Werthamer}(1963)}]{BalianWerthamer63}%
  \BibitemOpen
  \bibfield  {author} {\bibinfo {author} {\bibnamefont {Balian}, \bibfnamefont
  {R.}}, \ and\ \bibinfo {author} {\bibfnamefont {N.~R.}\ \bibnamefont
  {Werthamer}}} (\bibinfo {year} {1963}),\ \href {\doibase
  10.1103/PhysRev.131.1553} {\bibfield  {journal} {\bibinfo  {journal} {Phys.
  Rev.}\ }\textbf {\bibinfo {volume} {131}},\ \bibinfo {pages}
  {1553}}\BibitemShut {NoStop}%
\bibitem [{\citenamefont {Bardarson}\ \emph {et~al.}(2007)\citenamefont
  {Bardarson}, \citenamefont {Tworzyd\l{}o}, \citenamefont {Brouwer},\ and\
  \citenamefont {Beenakker}}]{Bardarson2007}%
  \BibitemOpen
  \bibfield  {author} {\bibinfo {author} {\bibnamefont {Bardarson},
  \bibfnamefont {J.~H.}}, \bibinfo {author} {\bibfnamefont {J.}~\bibnamefont
  {Tworzyd\l{}o}}, \bibinfo {author} {\bibfnamefont {P.~W.}\ \bibnamefont
  {Brouwer}}, \ and\ \bibinfo {author} {\bibfnamefont {C.~W.~J.}\ \bibnamefont
  {Beenakker}}} (\bibinfo {year} {2007}),\ \href {\doibase
  10.1103/PhysRevLett.99.106801} {\bibfield  {journal} {\bibinfo  {journal}
  {Phys. Rev. Lett.}\ }\textbf {\bibinfo {volume} {99}},\ \bibinfo {pages}
  {106801}}\BibitemShut {NoStop}%
\bibitem [{\citenamefont {{Barkeshli}}\ \emph {et~al.}(2014)\citenamefont
  {{Barkeshli}}, \citenamefont {{Bonderson}}, \citenamefont {{Cheng}},\ and\
  \citenamefont {{Wang}}}]{2014arXiv1410.4540B}%
  \BibitemOpen
  \bibfield  {author} {\bibinfo {author} {\bibnamefont {{Barkeshli}},
  \bibfnamefont {M.}}, \bibinfo {author} {\bibfnamefont {P.}~\bibnamefont
  {{Bonderson}}}, \bibinfo {author} {\bibfnamefont {M.}~\bibnamefont
  {{Cheng}}}, \ and\ \bibinfo {author} {\bibfnamefont {Z.}~\bibnamefont
  {{Wang}}}} (\bibinfo {year} {2014}),\ \href@noop {} {\bibfield  {journal}
  {\bibinfo  {journal} {ArXiv e-prints}\ }}\Eprint
  {http://arxiv.org/abs/1410.4540} {arXiv:1410.4540 [cond-mat.str-el]}
  \BibitemShut {NoStop}%
\bibitem [{\citenamefont {Beenakker}(2013)}]{beenakkerReview}%
  \BibitemOpen
  \bibfield  {author} {\bibinfo {author} {\bibnamefont {Beenakker},
  \bibfnamefont {C.}}} (\bibinfo {year} {2013}),\ \href@noop {} {\bibfield
  {journal} {\bibinfo  {journal} {Annual Review of Condensed Matter Physics}\
  }\textbf {\bibinfo {volume} {4}}~(\bibinfo {number} {1}),\ \bibinfo {pages}
  {113}}\BibitemShut {NoStop}%
\bibitem [{\citenamefont {Bellissard}\ \emph {et~al.}(1994)\citenamefont
  {Bellissard}, \citenamefont {van Elst},\ and\ \citenamefont
  {Schulz‐~Baldes}}]{belissard_JMatPhys94}%
  \BibitemOpen
  \bibfield  {author} {\bibinfo {author} {\bibnamefont {Bellissard},
  \bibfnamefont {J.}}, \bibinfo {author} {\bibfnamefont {A.}~\bibnamefont {van
  Elst}}, \ and\ \bibinfo {author} {\bibfnamefont {H.}~\bibnamefont
  {Schulz‐~Baldes}}} (\bibinfo {year} {1994}),\ \href {\doibase
  http://dx.doi.org/10.1063/1.530758} {\bibfield  {journal} {\bibinfo
  {journal} {Journal of Mathematical Physics}\ }\textbf {\bibinfo {volume}
  {35}}~(\bibinfo {number} {10}),\ \bibinfo {pages} {5373}}\BibitemShut
  {NoStop}%
\bibitem [{\citenamefont {Benalcazar}\ \emph {et~al.}(2014)\citenamefont
  {Benalcazar}, \citenamefont {Teo},\ and\ \citenamefont
  {Hughes}}]{Teo_disclination_class}%
  \BibitemOpen
  \bibfield  {author} {\bibinfo {author} {\bibnamefont {Benalcazar},
  \bibfnamefont {W.~A.}}, \bibinfo {author} {\bibfnamefont {J.~C.~Y.}\
  \bibnamefont {Teo}}, \ and\ \bibinfo {author} {\bibfnamefont {T.~L.}\
  \bibnamefont {Hughes}}} (\bibinfo {year} {2014}),\ \href {\doibase
  10.1103/PhysRevB.89.224503} {\bibfield  {journal} {\bibinfo  {journal} {Phys.
  Rev. B}\ }\textbf {\bibinfo {volume} {89}},\ \bibinfo {pages}
  {224503}}\BibitemShut {NoStop}%
\bibitem [{\citenamefont {BenTov}(2015)}]{BenTov2014}%
  \BibitemOpen
  \bibfield  {author} {\bibinfo {author} {\bibnamefont {BenTov}, \bibfnamefont
  {Y.}}} (\bibinfo {year} {2015}),\ \bibfield  {booktitle} {\emph {\bibinfo
  {booktitle} {Journal of High Energy Physics}},\ }\href {\doibase
  10.1007/JHEP07(2015)034} {\ \textbf {\bibinfo {volume} {2015}}~(\bibinfo
  {number} {7}),\ \bibinfo {pages} {1}}\BibitemShut {NoStop}%
\bibitem [{\citenamefont {Bergholtz}\ and\ \citenamefont
  {Liu}(2013)}]{Zhao_FTI}%
  \BibitemOpen
  \bibfield  {author} {\bibinfo {author} {\bibnamefont {Bergholtz},
  \bibfnamefont {E.~J.}}, \ and\ \bibinfo {author} {\bibfnamefont
  {Z.}~\bibnamefont {Liu}}} (\bibinfo {year} {2013}),\ \href {\doibase
  10.1142/S021797921330017X} {\bibfield  {journal} {\bibinfo  {journal}
  {International Journal of Modern Physics B}\ }\textbf {\bibinfo {volume}
  {27}}~(\bibinfo {number} {24}),\ \bibinfo {pages} {1330017}}\BibitemShut
  {NoStop}%
\bibitem [{\citenamefont {B\'eri}(2010)}]{beriPRB10}%
  \BibitemOpen
  \bibfield  {author} {\bibinfo {author} {\bibnamefont {B\'eri}, \bibfnamefont
  {B.}}} (\bibinfo {year} {2010}),\ \href {\doibase 10.1103/PhysRevB.81.134515}
  {\bibfield  {journal} {\bibinfo  {journal} {Phys. Rev. B}\ }\textbf {\bibinfo
  {volume} {81}},\ \bibinfo {pages} {134515}}\BibitemShut {NoStop}%
\bibitem [{\citenamefont {Berline}\ \emph {et~al.}(1992)\citenamefont
  {Berline}, \citenamefont {Getzler},\ and\ \citenamefont
  {Vergne}}]{Berlinebook}%
  \BibitemOpen
  \bibfield  {author} {\bibinfo {author} {\bibnamefont {Berline}, \bibfnamefont
  {N.}}, \bibinfo {author} {\bibfnamefont {E.}~\bibnamefont {Getzler}}, \ and\
  \bibinfo {author} {\bibfnamefont {M.}~\bibnamefont {Vergne}}} (\bibinfo
  {year} {1992}),\ \href@noop {} {\emph {\bibinfo {title} {Heat Kernels and
  Dirac Operators}}}\ (\bibinfo  {publisher} {Springer})\BibitemShut {NoStop}%
\bibitem [{\citenamefont {{Bernard}}\ and\ \citenamefont
  {{LeClair}}(2002)}]{LeClair2002}%
  \BibitemOpen
  \bibfield  {author} {\bibinfo {author} {\bibnamefont {{Bernard}},
  \bibfnamefont {D.}}, \ and\ \bibinfo {author} {\bibfnamefont
  {A.}~\bibnamefont {{LeClair}}}} (\bibinfo {year} {2002}),\ \href {\doibase
  10.1088/0305-4470/35/11/303} {\bibfield  {journal} {\bibinfo  {journal} {J.
  Phys. Soc. Jpn.}\ }\textbf {\bibinfo {volume} {35}},\ \bibinfo {pages}
  {2555}}\BibitemShut {NoStop}%
\bibitem [{\citenamefont {Bernevig}\ and\ \citenamefont
  {Hughes}(2013)}]{BernevigHughesBook13}%
  \BibitemOpen
  \bibfield  {author} {\bibinfo {author} {\bibnamefont {Bernevig},
  \bibfnamefont {A.~B.}}, \ and\ \bibinfo {author} {\bibfnamefont {T.~L.}\
  \bibnamefont {Hughes}}} (\bibinfo {year} {2013}),\ \href@noop {} {\emph
  {\bibinfo {title} {Topological Insulators and Topological Superconductors}}}\
  (\bibinfo  {publisher} {Princeton University Press},\ \bibinfo {address}
  {Princeton, NJ})\BibitemShut {NoStop}%
\bibitem [{\citenamefont {Bernevig}\ \emph {et~al.}(2006)\citenamefont
  {Bernevig}, \citenamefont {Hughes},\ and\ \citenamefont
  {Zhang}}]{Bernevig:2006kx}%
  \BibitemOpen
  \bibfield  {author} {\bibinfo {author} {\bibnamefont {Bernevig},
  \bibfnamefont {B.~A.}}, \bibinfo {author} {\bibfnamefont {T.~L.}\
  \bibnamefont {Hughes}}, \ and\ \bibinfo {author} {\bibfnamefont {S.-C.}\
  \bibnamefont {Zhang}}} (\bibinfo {year} {2006}),\ \href@noop {} {\bibfield
  {journal} {\bibinfo  {journal} {Science}\ }\textbf {\bibinfo {volume}
  {314}},\ \bibinfo {pages} {1757}}\BibitemShut {NoStop}%
\bibitem [{\citenamefont {Bian}\ \emph {et~al.}(2016)\citenamefont {Bian},
  \citenamefont {Chang}, \citenamefont {Sankar}, \citenamefont {Xu},
  \citenamefont {Zheng}, \citenamefont {Neupert}, \citenamefont {Chiu},
  \citenamefont {Huang}, \citenamefont {Chang}, \citenamefont {Belopolski},
  \citenamefont {Sanchez}, \citenamefont {Neupane}, \citenamefont {Alidoust},
  \citenamefont {Liu}, \citenamefont {Wang}, \citenamefont {Lee}, \citenamefont
  {Jeng}, \citenamefont {Zhang}, \citenamefont {Yuan}, \citenamefont {Jia},
  \citenamefont {Bansil}, \citenamefont {Chou}, \citenamefont {Lin},\ and\
  \citenamefont {Hasan}}]{Bian_nodal_line}%
  \BibitemOpen
  \bibfield  {author} {\bibinfo {author} {\bibnamefont {Bian}, \bibfnamefont
  {G.}}, \bibinfo {author} {\bibfnamefont {T.-R.}\ \bibnamefont {Chang}},
  \bibinfo {author} {\bibfnamefont {R.}~\bibnamefont {Sankar}}, \bibinfo
  {author} {\bibfnamefont {S.-Y.}\ \bibnamefont {Xu}}, \bibinfo {author}
  {\bibfnamefont {H.}~\bibnamefont {Zheng}}, \bibinfo {author} {\bibfnamefont
  {T.}~\bibnamefont {Neupert}}, \bibinfo {author} {\bibfnamefont {C.-K.}\
  \bibnamefont {Chiu}}, \bibinfo {author} {\bibfnamefont {S.-M.}\ \bibnamefont
  {Huang}}, \bibinfo {author} {\bibfnamefont {G.}~\bibnamefont {Chang}},
  \bibinfo {author} {\bibfnamefont {I.}~\bibnamefont {Belopolski}}, \bibinfo
  {author} {\bibfnamefont {D.~S.}\ \bibnamefont {Sanchez}}, \bibinfo {author}
  {\bibfnamefont {M.}~\bibnamefont {Neupane}}, \bibinfo {author} {\bibfnamefont
  {N.}~\bibnamefont {Alidoust}}, \bibinfo {author} {\bibfnamefont
  {C.}~\bibnamefont {Liu}}, \bibinfo {author} {\bibfnamefont {B.}~\bibnamefont
  {Wang}}, \bibinfo {author} {\bibfnamefont {C.-C.}\ \bibnamefont {Lee}},
  \bibinfo {author} {\bibfnamefont {H.-T.}\ \bibnamefont {Jeng}}, \bibinfo
  {author} {\bibfnamefont {C.}~\bibnamefont {Zhang}}, \bibinfo {author}
  {\bibfnamefont {Z.}~\bibnamefont {Yuan}}, \bibinfo {author} {\bibfnamefont
  {S.}~\bibnamefont {Jia}}, \bibinfo {author} {\bibfnamefont {A.}~\bibnamefont
  {Bansil}}, \bibinfo {author} {\bibfnamefont {F.}~\bibnamefont {Chou}},
  \bibinfo {author} {\bibfnamefont {H.}~\bibnamefont {Lin}}, \ and\ \bibinfo
  {author} {\bibfnamefont {M.~Z.}\ \bibnamefont {Hasan}}} (\bibinfo {year}
  {2016}),\ \href {http://dx.doi.org/10.1038/ncomms10556} {\bibfield  {journal}
  {\bibinfo  {journal} {Nat Commun}\ }\textbf {\bibinfo {volume}
  {7}}}\BibitemShut {NoStop}%
\bibitem [{\citenamefont {{Bian}}\ \emph {et~al.}(2015)\citenamefont {{Bian}},
  \citenamefont {{Chang}}, \citenamefont {{Zheng}}, \citenamefont {{Velury}},
  \citenamefont {{Xu}}, \citenamefont {{Neupert}}, \citenamefont {{Chiu}},
  \citenamefont {{Sanchez}}, \citenamefont {{Belopolski}}, \citenamefont
  {{Alidoust}}, \citenamefont {{Chen}}, \citenamefont {{Chang}}, \citenamefont
  {{Bansil}}, \citenamefont {{Jeng}}, \citenamefont {{Lin}},\ and\
  \citenamefont {{Zahid Hasan}}}]{Bian_TlTaSe2_line}%
  \BibitemOpen
  \bibfield  {author} {\bibinfo {author} {\bibnamefont {{Bian}}, \bibfnamefont
  {G.}}, \bibinfo {author} {\bibfnamefont {T.-R.}\ \bibnamefont {{Chang}}},
  \bibinfo {author} {\bibfnamefont {H.}~\bibnamefont {{Zheng}}}, \bibinfo
  {author} {\bibfnamefont {S.}~\bibnamefont {{Velury}}}, \bibinfo {author}
  {\bibfnamefont {S.-Y.}\ \bibnamefont {{Xu}}}, \bibinfo {author}
  {\bibfnamefont {T.}~\bibnamefont {{Neupert}}}, \bibinfo {author}
  {\bibfnamefont {C.-K.}\ \bibnamefont {{Chiu}}}, \bibinfo {author}
  {\bibfnamefont {D.~S.}\ \bibnamefont {{Sanchez}}}, \bibinfo {author}
  {\bibfnamefont {I.}~\bibnamefont {{Belopolski}}}, \bibinfo {author}
  {\bibfnamefont {N.}~\bibnamefont {{Alidoust}}}, \bibinfo {author}
  {\bibfnamefont {P.-J.}\ \bibnamefont {{Chen}}}, \bibinfo {author}
  {\bibfnamefont {G.}~\bibnamefont {{Chang}}}, \bibinfo {author} {\bibfnamefont
  {A.}~\bibnamefont {{Bansil}}}, \bibinfo {author} {\bibfnamefont {H.-T.}\
  \bibnamefont {{Jeng}}}, \bibinfo {author} {\bibfnamefont {H.}~\bibnamefont
  {{Lin}}}, \ and\ \bibinfo {author} {\bibfnamefont {M.}~\bibnamefont {{Zahid
  Hasan}}}} (\bibinfo {year} {2015}),\ \href@noop {} {\bibfield  {journal}
  {\bibinfo  {journal} {ArXiv e-prints}\ }}\Eprint
  {http://arxiv.org/abs/1508.07521} {arXiv:1508.07521} \BibitemShut {NoStop}%
\bibitem [{\citenamefont {{Bocquet}}\ \emph {et~al.}(2000)\citenamefont
  {{Bocquet}}, \citenamefont {{Serban}},\ and\ \citenamefont
  {{Zirnbauer}}}]{Bocquet2000}%
  \BibitemOpen
  \bibfield  {author} {\bibinfo {author} {\bibnamefont {{Bocquet}},
  \bibfnamefont {M.}}, \bibinfo {author} {\bibfnamefont {D.}~\bibnamefont
  {{Serban}}}, \ and\ \bibinfo {author} {\bibfnamefont {M.~R.}\ \bibnamefont
  {{Zirnbauer}}}} (\bibinfo {year} {2000}),\ \href {\doibase
  10.1016/S0550-3213(00)00208-X} {\bibfield  {journal} {\bibinfo  {journal}
  {Nuclear Physics B}\ }\textbf {\bibinfo {volume} {578}},\ \bibinfo {pages}
  {628}}\BibitemShut {NoStop}%
\bibitem [{\citenamefont {{Bonderson}}\ \emph {et~al.}(2013)\citenamefont
  {{Bonderson}}, \citenamefont {{Nayak}},\ and\ \citenamefont
  {{Qi}}}]{Bonderson2013}%
  \BibitemOpen
  \bibfield  {author} {\bibinfo {author} {\bibnamefont {{Bonderson}},
  \bibfnamefont {P.}}, \bibinfo {author} {\bibfnamefont {C.}~\bibnamefont
  {{Nayak}}}, \ and\ \bibinfo {author} {\bibfnamefont {X.-L.}\ \bibnamefont
  {{Qi}}}} (\bibinfo {year} {2013}),\ \href {\doibase
  10.1088/1742-5468/2013/09/P09016} {\bibfield  {journal} {\bibinfo  {journal}
  {Journal of Statistical Mechanics: Theory and Experiment}\ }\textbf {\bibinfo
  {volume} {9}},\ \bibinfo {eid} {09016}}\BibitemShut {NoStop}%
\bibitem [{\citenamefont {Borisenko}\ \emph {et~al.}(2014)\citenamefont
  {Borisenko}, \citenamefont {Gibson}, \citenamefont {Evtushinsky},
  \citenamefont {Zabolotnyy}, \citenamefont {B\"uchner},\ and\ \citenamefont
  {Cava}}]{borisenkoPRLCd3As2}%
  \BibitemOpen
  \bibfield  {author} {\bibinfo {author} {\bibnamefont {Borisenko},
  \bibfnamefont {S.}}, \bibinfo {author} {\bibfnamefont {Q.}~\bibnamefont
  {Gibson}}, \bibinfo {author} {\bibfnamefont {D.}~\bibnamefont {Evtushinsky}},
  \bibinfo {author} {\bibfnamefont {V.}~\bibnamefont {Zabolotnyy}}, \bibinfo
  {author} {\bibfnamefont {B.}~\bibnamefont {B\"uchner}}, \ and\ \bibinfo
  {author} {\bibfnamefont {R.~J.}\ \bibnamefont {Cava}}} (\bibinfo {year}
  {2014}),\ \href {\doibase 10.1103/PhysRevLett.113.027603} {\bibfield
  {journal} {\bibinfo  {journal} {Phys. Rev. Lett.}\ }\textbf {\bibinfo
  {volume} {113}},\ \bibinfo {pages} {027603}}\BibitemShut {NoStop}%
\bibitem [{\citenamefont {{Brouwer}}\ and\ \citenamefont
  {{Frahm}}(1996)}]{Brouwer1996}%
  \BibitemOpen
  \bibfield  {author} {\bibinfo {author} {\bibnamefont {{Brouwer}},
  \bibfnamefont {P.~W.}}, \ and\ \bibinfo {author} {\bibfnamefont
  {K.}~\bibnamefont {{Frahm}}}} (\bibinfo {year} {1996}),\ \href {\doibase
  10.1103/PhysRevB.53.1490} {\bibfield  {journal} {\bibinfo  {journal} {\prb}\
  }\textbf {\bibinfo {volume} {53}},\ \bibinfo {pages} {1490}}\BibitemShut
  {NoStop}%
\bibitem [{\citenamefont {{Brouwer}}\ \emph
  {et~al.}(2000{\natexlab{a}})\citenamefont {{Brouwer}}, \citenamefont
  {{Furusaki}}, \citenamefont {{Gruzberg}},\ and\ \citenamefont
  {{Mudry}}}]{Brouwer2000PhRvL..85.1064B}%
  \BibitemOpen
  \bibfield  {author} {\bibinfo {author} {\bibnamefont {{Brouwer}},
  \bibfnamefont {P.~W.}}, \bibinfo {author} {\bibfnamefont {A.}~\bibnamefont
  {{Furusaki}}}, \bibinfo {author} {\bibfnamefont {I.~A.}\ \bibnamefont
  {{Gruzberg}}}, \ and\ \bibinfo {author} {\bibfnamefont {C.}~\bibnamefont
  {{Mudry}}}} (\bibinfo {year} {2000}{\natexlab{a}}),\ \href {\doibase
  10.1103/PhysRevLett.85.1064} {\bibfield  {journal} {\bibinfo  {journal}
  {Physical Review Letters}\ }\textbf {\bibinfo {volume} {85}},\ \bibinfo
  {pages} {1064}}\BibitemShut {NoStop}%
\bibitem [{\citenamefont {{Brouwer}}\ \emph
  {et~al.}(2000{\natexlab{b}})\citenamefont {{Brouwer}}, \citenamefont
  {{Mudry}},\ and\ \citenamefont {{Furusaki}}}]{Brouwer2000PhRvL..84.2913B}%
  \BibitemOpen
  \bibfield  {author} {\bibinfo {author} {\bibnamefont {{Brouwer}},
  \bibfnamefont {P.~W.}}, \bibinfo {author} {\bibfnamefont {C.}~\bibnamefont
  {{Mudry}}}, \ and\ \bibinfo {author} {\bibfnamefont {A.}~\bibnamefont
  {{Furusaki}}}} (\bibinfo {year} {2000}{\natexlab{b}}),\ \href {\doibase
  10.1103/PhysRevLett.84.2913} {\bibfield  {journal} {\bibinfo  {journal}
  {Physical Review Letters}\ }\textbf {\bibinfo {volume} {84}},\ \bibinfo
  {pages} {2913}}\BibitemShut {NoStop}%
\bibitem [{\citenamefont {{Brouwer}}\ \emph {et~al.}(1998)\citenamefont
  {{Brouwer}}, \citenamefont {{Mudry}}, \citenamefont {{Simons}},\ and\
  \citenamefont {{Altland}}}]{Brouwer1998}%
  \BibitemOpen
  \bibfield  {author} {\bibinfo {author} {\bibnamefont {{Brouwer}},
  \bibfnamefont {P.~W.}}, \bibinfo {author} {\bibfnamefont {C.}~\bibnamefont
  {{Mudry}}}, \bibinfo {author} {\bibfnamefont {B.~D.}\ \bibnamefont
  {{Simons}}}, \ and\ \bibinfo {author} {\bibfnamefont {A.}~\bibnamefont
  {{Altland}}}} (\bibinfo {year} {1998}),\ \href {\doibase
  10.1103/PhysRevLett.81.862} {\bibfield  {journal} {\bibinfo  {journal}
  {Physical Review Letters}\ }\textbf {\bibinfo {volume} {81}},\ \bibinfo
  {pages} {862}},\ \Eprint {http://arxiv.org/abs/cond-mat/9807189}
  {cond-mat/9807189} \BibitemShut {NoStop}%
\bibitem [{\citenamefont {Brydon}\ \emph {et~al.}(2011)\citenamefont {Brydon},
  \citenamefont {Schnyder},\ and\ \citenamefont
  {Timm}}]{BrydonSchnyderTimmFlat}%
  \BibitemOpen
  \bibfield  {author} {\bibinfo {author} {\bibnamefont {Brydon}, \bibfnamefont
  {P.~M.~R.}}, \bibinfo {author} {\bibfnamefont {A.~P.}\ \bibnamefont
  {Schnyder}}, \ and\ \bibinfo {author} {\bibfnamefont {C.}~\bibnamefont
  {Timm}}} (\bibinfo {year} {2011}),\ \href {\doibase
  10.1103/PhysRevB.84.020501} {\bibfield  {journal} {\bibinfo  {journal} {Phys.
  Rev. B}\ }\textbf {\bibinfo {volume} {84}},\ \bibinfo {pages}
  {020501}}\BibitemShut {NoStop}%
\bibitem [{\citenamefont {{Budich}}\ and\ \citenamefont
  {{Ardonne}}(2013)}]{BudichArdonne2013}%
  \BibitemOpen
  \bibfield  {author} {\bibinfo {author} {\bibnamefont {{Budich}},
  \bibfnamefont {J.~C.}}, \ and\ \bibinfo {author} {\bibfnamefont
  {E.}~\bibnamefont {{Ardonne}}}} (\bibinfo {year} {2013}),\ \href {\doibase
  10.1103/PhysRevB.88.075419} {\bibfield  {journal} {\bibinfo  {journal}
  {\prb}\ }\textbf {\bibinfo {volume} {88}}~(\bibinfo {number} {7}),\ \bibinfo
  {eid} {075419}}\BibitemShut {NoStop}%
\bibitem [{\citenamefont {Budich}\ and\ \citenamefont
  {Trauzettel}(2013)}]{Budich_Ktheory}%
  \BibitemOpen
  \bibfield  {author} {\bibinfo {author} {\bibnamefont {Budich}, \bibfnamefont
  {J.~C.}}, \ and\ \bibinfo {author} {\bibfnamefont {B.}~\bibnamefont
  {Trauzettel}}} (\bibinfo {year} {2013}),\ \href {\doibase
  10.1002/pssr.201206416} {\bibfield  {journal} {\bibinfo  {journal} {physica
  status solidi (RRL) -- Rapid Research Letters}\ }\textbf {\bibinfo {volume}
  {7}}~(\bibinfo {number} {1-2}),\ \bibinfo {pages} {109}}\BibitemShut
  {NoStop}%
\bibitem [{\citenamefont {Burkov}\ and\ \citenamefont
  {Balents}(2011)}]{burkovBalenstPRL11}%
  \BibitemOpen
  \bibfield  {author} {\bibinfo {author} {\bibnamefont {Burkov}, \bibfnamefont
  {A.~A.}}, \ and\ \bibinfo {author} {\bibfnamefont {L.}~\bibnamefont
  {Balents}}} (\bibinfo {year} {2011}),\ \href {\doibase
  10.1103/PhysRevLett.107.127205} {\bibfield  {journal} {\bibinfo  {journal}
  {Phys. Rev. Lett.}\ }\textbf {\bibinfo {volume} {107}},\ \bibinfo {pages}
  {127205}}\BibitemShut {NoStop}%
\bibitem [{\citenamefont {Burkov}\ \emph {et~al.}(2011)\citenamefont {Burkov},
  \citenamefont {Hook},\ and\ \citenamefont {Balents}}]{BurkovBalentsPRB11}%
  \BibitemOpen
  \bibfield  {author} {\bibinfo {author} {\bibnamefont {Burkov}, \bibfnamefont
  {A.~A.}}, \bibinfo {author} {\bibfnamefont {M.~D.}\ \bibnamefont {Hook}}, \
  and\ \bibinfo {author} {\bibfnamefont {L.}~\bibnamefont {Balents}}} (\bibinfo
  {year} {2011}),\ \href {\doibase 10.1103/PhysRevB.84.235126} {\bibfield
  {journal} {\bibinfo  {journal} {Phys. Rev. B}\ }\textbf {\bibinfo {volume}
  {84}},\ \bibinfo {pages} {235126}}\BibitemShut {NoStop}%
\bibitem [{\citenamefont {Burnell}\ \emph {et~al.}(2014)\citenamefont
  {Burnell}, \citenamefont {Chen}, \citenamefont {Fidkowski},\ and\
  \citenamefont {Vishwanath}}]{Burnell2013}%
  \BibitemOpen
  \bibfield  {author} {\bibinfo {author} {\bibnamefont {Burnell}, \bibfnamefont
  {F.~J.}}, \bibinfo {author} {\bibfnamefont {X.}~\bibnamefont {Chen}},
  \bibinfo {author} {\bibfnamefont {L.}~\bibnamefont {Fidkowski}}, \ and\
  \bibinfo {author} {\bibfnamefont {A.}~\bibnamefont {Vishwanath}}} (\bibinfo
  {year} {2014}),\ \href {\doibase 10.1103/PhysRevB.90.245122} {\bibfield
  {journal} {\bibinfo  {journal} {Phys. Rev. B}\ }\textbf {\bibinfo {volume}
  {90}},\ \bibinfo {pages} {245122}}\BibitemShut {NoStop}%
\bibitem [{\citenamefont {Cappelli}\ \emph {et~al.}(2002)\citenamefont
  {Cappelli}, \citenamefont {Huerta},\ and\ \citenamefont
  {Zemba}}]{Cappelli01}%
  \BibitemOpen
  \bibfield  {author} {\bibinfo {author} {\bibnamefont {Cappelli},
  \bibfnamefont {A.}}, \bibinfo {author} {\bibfnamefont {M.}~\bibnamefont
  {Huerta}}, \ and\ \bibinfo {author} {\bibfnamefont {G.~R.}\ \bibnamefont
  {Zemba}}} (\bibinfo {year} {2002}),\ \href {\doibase
  http://dx.doi.org/10.1016/S0550-3213(02)00340-1} {\bibfield  {journal}
  {\bibinfo  {journal} {Nuclear Physics B}\ }\textbf {\bibinfo {volume}
  {636}}~(\bibinfo {number} {3}),\ \bibinfo {pages} {568 }}\BibitemShut
  {NoStop}%
\bibitem [{\citenamefont {{Cappelli}}\ and\ \citenamefont
  {{Randellini}}(2013)}]{Cappelli2013}%
  \BibitemOpen
  \bibfield  {author} {\bibinfo {author} {\bibnamefont {{Cappelli}},
  \bibfnamefont {A.}}, \ and\ \bibinfo {author} {\bibfnamefont
  {E.}~\bibnamefont {{Randellini}}}} (\bibinfo {year} {2013}),\ \href {\doibase
  10.1007/JHEP12(2013)101} {\bibfield  {journal} {\bibinfo  {journal} {Journal
  of High Energy Physics}\ }\textbf {\bibinfo {volume} {12}},\ \bibinfo {eid}
  {101}}\BibitemShut {NoStop}%
\bibitem [{\citenamefont {Cappelli}\ and\ \citenamefont
  {Randellini}(2015)}]{Cappelli2014}%
  \BibitemOpen
  \bibfield  {author} {\bibinfo {author} {\bibnamefont {Cappelli},
  \bibfnamefont {A.}}, \ and\ \bibinfo {author} {\bibfnamefont
  {E.}~\bibnamefont {Randellini}}} (\bibinfo {year} {2015}),\ \href
  {http://stacks.iop.org/1751-8121/48/i=10/a=105404} {\bibfield  {journal}
  {\bibinfo  {journal} {Journal of Physics A: Mathematical and Theoretical}\
  }\textbf {\bibinfo {volume} {48}}~(\bibinfo {number} {10}),\ \bibinfo {pages}
  {105404}}\BibitemShut {NoStop}%
\bibitem [{\citenamefont {{Cardy}}(1986)}]{Cardy1986}%
  \BibitemOpen
  \bibfield  {author} {\bibinfo {author} {\bibnamefont {{Cardy}}, \bibfnamefont
  {J.~L.}}} (\bibinfo {year} {1986}),\ \href {\doibase
  10.1016/0550-3213(86)90552-3} {\bibfield  {journal} {\bibinfo  {journal}
  {Nuclear Physics B}\ }\textbf {\bibinfo {volume} {270}},\ \bibinfo {pages}
  {186}}\BibitemShut {NoStop}%
\bibitem [{\citenamefont {Castro~Neto}\ \emph {et~al.}(2009)\citenamefont
  {Castro~Neto}, \citenamefont {Guinea}, \citenamefont {Peres}, \citenamefont
  {Novoselov},\ and\ \citenamefont {Geim}}]{castroNetoRMP09}%
  \BibitemOpen
  \bibfield  {author} {\bibinfo {author} {\bibnamefont {Castro~Neto},
  \bibfnamefont {A.~H.}}, \bibinfo {author} {\bibfnamefont {F.}~\bibnamefont
  {Guinea}}, \bibinfo {author} {\bibfnamefont {N.~M.~R.}\ \bibnamefont
  {Peres}}, \bibinfo {author} {\bibfnamefont {K.~S.}\ \bibnamefont
  {Novoselov}}, \ and\ \bibinfo {author} {\bibfnamefont {A.~K.}\ \bibnamefont
  {Geim}}} (\bibinfo {year} {2009}),\ \href {\doibase
  10.1103/RevModPhys.81.109} {\bibfield  {journal} {\bibinfo  {journal} {Rev.
  Mod. Phys.}\ }\textbf {\bibinfo {volume} {81}},\ \bibinfo {pages}
  {109}}\BibitemShut {NoStop}%
\bibitem [{\citenamefont {Juri\ifmmode \check{c}\else
  \v{c}\fi{}i\ifmmode~\acute{c}\else \'{c}\fi{}}\ \emph
  {et~al.}(2012)\citenamefont {Juri\ifmmode \check{c}\else
  \v{c}\fi{}i\ifmmode~\acute{c}\else \'{c}\fi{}}, \citenamefont {Mesaros},
  \citenamefont {Slager},\ and\ \citenamefont {Zaanen}}]{Juricic12}%
  \BibitemOpen
  \bibfield  {author} {\bibinfo {author} {\bibnamefont {Juri\ifmmode
  \check{c}\else \v{c}\fi{}i\ifmmode~\acute{c}\else \'{c}\fi{}}, \bibfnamefont
  {V.}}, \bibinfo {author} {\bibfnamefont {A.}~\bibnamefont {Mesaros}},
  \bibinfo {author} {\bibfnamefont {R.-J.}\ \bibnamefont {Slager}}, \ and\
  \bibinfo {author} {\bibfnamefont {J.}~\bibnamefont {Zaanen}}} (\bibinfo
  {year} {2012}),\ \href {\doibase 10.1103/PhysRevLett.108.106403} {\bibfield
  {journal} {\bibinfo  {journal} {Phys. Rev. Lett.}\ }\textbf {\bibinfo
  {volume} {108}},\ \bibinfo {pages} {106403}}\BibitemShut {NoStop}%
\bibitem [{\citenamefont {Chaikin}\ and\ \citenamefont
  {Lubensky}(2000)}]{ChaikinLubensky}%
  \BibitemOpen
  \bibfield  {author} {\bibinfo {author} {\bibnamefont {Chaikin}, \bibfnamefont
  {P.~M.}}, \ and\ \bibinfo {author} {\bibfnamefont {T.~C.}\ \bibnamefont
  {Lubensky}}} (\bibinfo {year} {2000}),\ \href@noop {} {\emph {\bibinfo
  {title} {Principles of Condensed Matter Physics}}}\ (\bibinfo  {publisher}
  {Cambridge University Press})\BibitemShut {NoStop}%
\bibitem [{\citenamefont {Chan}\ \emph {et~al.}(2013)\citenamefont {Chan},
  \citenamefont {Hughes}, \citenamefont {Ryu},\ and\ \citenamefont
  {Fradkin}}]{Atma_EFT}%
  \BibitemOpen
  \bibfield  {author} {\bibinfo {author} {\bibnamefont {Chan}, \bibfnamefont
  {A.}}, \bibinfo {author} {\bibfnamefont {T.~L.}\ \bibnamefont {Hughes}},
  \bibinfo {author} {\bibfnamefont {S.}~\bibnamefont {Ryu}}, \ and\ \bibinfo
  {author} {\bibfnamefont {E.}~\bibnamefont {Fradkin}}} (\bibinfo {year}
  {2013}),\ \href {\doibase 10.1103/PhysRevB.87.085132} {\bibfield  {journal}
  {\bibinfo  {journal} {Phys. Rev. B}\ }\textbf {\bibinfo {volume} {87}},\
  \bibinfo {pages} {085132}}\BibitemShut {NoStop}%
\bibitem [{\citenamefont {{Chan}}\ \emph {et~al.}(2015)\citenamefont {{Chan}},
  \citenamefont {{Chiu}}, \citenamefont {{Chou}},\ and\ \citenamefont
  {{Schnyder}}}]{Nodal_line_chan}%
  \BibitemOpen
  \bibfield  {author} {\bibinfo {author} {\bibnamefont {{Chan}}, \bibfnamefont
  {Y.-H.}}, \bibinfo {author} {\bibfnamefont {C.-K.}\ \bibnamefont {{Chiu}}},
  \bibinfo {author} {\bibfnamefont {M.~Y.}\ \bibnamefont {{Chou}}}, \ and\
  \bibinfo {author} {\bibfnamefont {A.~P.}\ \bibnamefont {{Schnyder}}}}
  (\bibinfo {year} {2015}),\ \href@noop {} {\bibfield  {journal} {\bibinfo
  {journal} {ArXiv e-prints}\ }}\Eprint {http://arxiv.org/abs/1510.02759}
  {arXiv:1510.02759 [cond-mat.mes-hall]} \BibitemShut {NoStop}%
\bibitem [{\citenamefont {Chang}\ \emph {et~al.}(2013)\citenamefont {Chang},
  \citenamefont {Zhang}, \citenamefont {Feng}, \citenamefont {Shen},
  \citenamefont {Zhang}, \citenamefont {Guo}, \citenamefont {Li}, \citenamefont
  {Ou}, \citenamefont {Wei}, \citenamefont {Wang}, \citenamefont {Ji},
  \citenamefont {Feng}, \citenamefont {Ji}, \citenamefont {Chen}, \citenamefont
  {Jia}, \citenamefont {Dai}, \citenamefont {Fang}, \citenamefont {Zhang},
  \citenamefont {He}, \citenamefont {Wang}, \citenamefont {Lu}, \citenamefont
  {Ma},\ and\ \citenamefont {Xue}}]{Xue_QAHE}%
  \BibitemOpen
  \bibfield  {author} {\bibinfo {author} {\bibnamefont {Chang}, \bibfnamefont
  {C.-Z.}}, \bibinfo {author} {\bibfnamefont {J.}~\bibnamefont {Zhang}},
  \bibinfo {author} {\bibfnamefont {X.}~\bibnamefont {Feng}}, \bibinfo {author}
  {\bibfnamefont {J.}~\bibnamefont {Shen}}, \bibinfo {author} {\bibfnamefont
  {Z.}~\bibnamefont {Zhang}}, \bibinfo {author} {\bibfnamefont
  {M.}~\bibnamefont {Guo}}, \bibinfo {author} {\bibfnamefont {K.}~\bibnamefont
  {Li}}, \bibinfo {author} {\bibfnamefont {Y.}~\bibnamefont {Ou}}, \bibinfo
  {author} {\bibfnamefont {P.}~\bibnamefont {Wei}}, \bibinfo {author}
  {\bibfnamefont {L.-L.}\ \bibnamefont {Wang}}, \bibinfo {author}
  {\bibfnamefont {Z.-Q.}\ \bibnamefont {Ji}}, \bibinfo {author} {\bibfnamefont
  {Y.}~\bibnamefont {Feng}}, \bibinfo {author} {\bibfnamefont {S.}~\bibnamefont
  {Ji}}, \bibinfo {author} {\bibfnamefont {X.}~\bibnamefont {Chen}}, \bibinfo
  {author} {\bibfnamefont {J.}~\bibnamefont {Jia}}, \bibinfo {author}
  {\bibfnamefont {X.}~\bibnamefont {Dai}}, \bibinfo {author} {\bibfnamefont
  {Z.}~\bibnamefont {Fang}}, \bibinfo {author} {\bibfnamefont {S.-C.}\
  \bibnamefont {Zhang}}, \bibinfo {author} {\bibfnamefont {K.}~\bibnamefont
  {He}}, \bibinfo {author} {\bibfnamefont {Y.}~\bibnamefont {Wang}}, \bibinfo
  {author} {\bibfnamefont {L.}~\bibnamefont {Lu}}, \bibinfo {author}
  {\bibfnamefont {X.-C.}\ \bibnamefont {Ma}}, \ and\ \bibinfo {author}
  {\bibfnamefont {Q.-K.}\ \bibnamefont {Xue}}} (\bibinfo {year} {2013}),\ \href
  {\doibase 10.1126/science.1234414} {\bibfield  {journal} {\bibinfo  {journal}
  {Science}\ }\textbf {\bibinfo {volume} {340}},\ \bibinfo {pages}
  {167}}\BibitemShut {NoStop}%
\bibitem [{\citenamefont {Chang}\ \emph {et~al.}(2014)\citenamefont {Chang},
  \citenamefont {Mudry},\ and\ \citenamefont
  {Ryu}}]{Entanglement_crystalline_Chang}%
  \BibitemOpen
  \bibfield  {author} {\bibinfo {author} {\bibnamefont {Chang}, \bibfnamefont
  {P.-Y.}}, \bibinfo {author} {\bibfnamefont {C.}~\bibnamefont {Mudry}}, \ and\
  \bibinfo {author} {\bibfnamefont {S.}~\bibnamefont {Ryu}}} (\bibinfo {year}
  {2014}),\ \href {http://stacks.iop.org/1742-5468/2014/i=9/a=P09014}
  {\bibfield  {journal} {\bibinfo  {journal} {Journal of Statistical Mechanics:
  Theory and Experiment}\ }\textbf {\bibinfo {volume} {2014}}~(\bibinfo
  {number} {9}),\ \bibinfo {pages} {P09014}}\BibitemShut {NoStop}%
\bibitem [{\citenamefont {Chen}\ and\ \citenamefont
  {Hermele}(2012)}]{chen_hermele_weyl}%
  \BibitemOpen
  \bibfield  {author} {\bibinfo {author} {\bibnamefont {Chen}, \bibfnamefont
  {G.}}, \ and\ \bibinfo {author} {\bibfnamefont {M.}~\bibnamefont {Hermele}}}
  (\bibinfo {year} {2012}),\ \href {\doibase 10.1103/PhysRevB.86.235129}
  {\bibfield  {journal} {\bibinfo  {journal} {Phys. Rev. B}\ }\textbf {\bibinfo
  {volume} {86}},\ \bibinfo {pages} {235129}}\BibitemShut {NoStop}%
\bibitem [{\citenamefont {Chen}\ \emph
  {et~al.}(2015{\natexlab{a}})\citenamefont {Chen}, \citenamefont {Burnell},
  \citenamefont {Vishwanath},\ and\ \citenamefont {Fidkowski}}]{Chen2014}%
  \BibitemOpen
  \bibfield  {author} {\bibinfo {author} {\bibnamefont {Chen}, \bibfnamefont
  {X.}}, \bibinfo {author} {\bibfnamefont {F.~J.}\ \bibnamefont {Burnell}},
  \bibinfo {author} {\bibfnamefont {A.}~\bibnamefont {Vishwanath}}, \ and\
  \bibinfo {author} {\bibfnamefont {L.}~\bibnamefont {Fidkowski}}} (\bibinfo
  {year} {2015}{\natexlab{a}}),\ \href {\doibase 10.1103/PhysRevX.5.041013}
  {\bibfield  {journal} {\bibinfo  {journal} {Phys. Rev. X}\ }\textbf {\bibinfo
  {volume} {5}},\ \bibinfo {pages} {041013}}\BibitemShut {NoStop}%
\bibitem [{\citenamefont {{Chen}}\ \emph {et~al.}(2014)\citenamefont {{Chen}},
  \citenamefont {{Fidkowski}},\ and\ \citenamefont
  {{Vishwanath}}}]{ChenFidkowskiVishwanath2014}%
  \BibitemOpen
  \bibfield  {author} {\bibinfo {author} {\bibnamefont {{Chen}}, \bibfnamefont
  {X.}}, \bibinfo {author} {\bibfnamefont {L.}~\bibnamefont {{Fidkowski}}}, \
  and\ \bibinfo {author} {\bibfnamefont {A.}~\bibnamefont {{Vishwanath}}}}
  (\bibinfo {year} {2014}),\ \href {\doibase 10.1103/PhysRevB.89.165132}
  {\bibfield  {journal} {\bibinfo  {journal} {\prb}\ }\textbf {\bibinfo
  {volume} {89}},\ \bibinfo {eid} {165132}}\BibitemShut {NoStop}%
\bibitem [{\citenamefont {Chen}\ \emph {et~al.}(2012)\citenamefont {Chen},
  \citenamefont {Gu}, \citenamefont {Liu},\ and\ \citenamefont
  {Wen}}]{Chen2012}%
  \BibitemOpen
  \bibfield  {author} {\bibinfo {author} {\bibnamefont {Chen}, \bibfnamefont
  {X.}}, \bibinfo {author} {\bibfnamefont {Z.-C.}\ \bibnamefont {Gu}}, \bibinfo
  {author} {\bibfnamefont {Z.-X.}\ \bibnamefont {Liu}}, \ and\ \bibinfo
  {author} {\bibfnamefont {X.-G.}\ \bibnamefont {Wen}}} (\bibinfo {year}
  {2012}),\ \href@noop {} {\bibfield  {journal} {\bibinfo  {journal} {Science}\
  }\textbf {\bibinfo {volume} {338}},\ \bibinfo {pages} {1604}}\BibitemShut
  {NoStop}%
\bibitem [{\citenamefont {{Chen}}\ \emph {et~al.}(2013)\citenamefont {{Chen}},
  \citenamefont {{Gu}}, \citenamefont {{Liu}},\ and\ \citenamefont
  {{Wen}}}]{Chen2013}%
  \BibitemOpen
  \bibfield  {author} {\bibinfo {author} {\bibnamefont {{Chen}}, \bibfnamefont
  {X.}}, \bibinfo {author} {\bibfnamefont {Z.-C.}\ \bibnamefont {{Gu}}},
  \bibinfo {author} {\bibfnamefont {Z.-X.}\ \bibnamefont {{Liu}}}, \ and\
  \bibinfo {author} {\bibfnamefont {X.-G.}\ \bibnamefont {{Wen}}}} (\bibinfo
  {year} {2013}),\ \href {\doibase 10.1103/PhysRevB.87.155114} {\bibfield
  {journal} {\bibinfo  {journal} {\prb}\ }\textbf {\bibinfo {volume} {87}},\
  \bibinfo {eid} {155114}}\BibitemShut {NoStop}%
\bibitem [{\citenamefont {Chen}\ \emph {et~al.}(2010)\citenamefont {Chen},
  \citenamefont {Gu},\ and\ \citenamefont {Wen}}]{Xie_LRE}%
  \BibitemOpen
  \bibfield  {author} {\bibinfo {author} {\bibnamefont {Chen}, \bibfnamefont
  {X.}}, \bibinfo {author} {\bibfnamefont {Z.-C.}\ \bibnamefont {Gu}}, \ and\
  \bibinfo {author} {\bibfnamefont {X.-G.}\ \bibnamefont {Wen}}} (\bibinfo
  {year} {2010}),\ \href {\doibase 10.1103/PhysRevB.82.155138} {\bibfield
  {journal} {\bibinfo  {journal} {Phys. Rev. B}\ }\textbf {\bibinfo {volume}
  {82}},\ \bibinfo {pages} {155138}}\BibitemShut {NoStop}%
\bibitem [{\citenamefont {{Chen}}\ \emph
  {et~al.}(2011{\natexlab{a}})\citenamefont {{Chen}}, \citenamefont {{Gu}},\
  and\ \citenamefont {{Wen}}}]{Chen2011}%
  \BibitemOpen
  \bibfield  {author} {\bibinfo {author} {\bibnamefont {{Chen}}, \bibfnamefont
  {X.}}, \bibinfo {author} {\bibfnamefont {Z.-C.}\ \bibnamefont {{Gu}}}, \ and\
  \bibinfo {author} {\bibfnamefont {X.-G.}\ \bibnamefont {{Wen}}}} (\bibinfo
  {year} {2011}{\natexlab{a}}),\ \href {\doibase 10.1103/PhysRevB.83.035107}
  {\bibfield  {journal} {\bibinfo  {journal} {\prb}\ }\textbf {\bibinfo
  {volume} {83}},\ \bibinfo {eid} {035107}}\BibitemShut {NoStop}%
\bibitem [{\citenamefont {{Chen}}\ \emph
  {et~al.}(2011{\natexlab{b}})\citenamefont {{Chen}}, \citenamefont {{Gu}},\
  and\ \citenamefont {{Wen}}}]{Chen2011b}%
  \BibitemOpen
  \bibfield  {author} {\bibinfo {author} {\bibnamefont {{Chen}}, \bibfnamefont
  {X.}}, \bibinfo {author} {\bibfnamefont {Z.-C.}\ \bibnamefont {{Gu}}}, \ and\
  \bibinfo {author} {\bibfnamefont {X.-G.}\ \bibnamefont {{Wen}}}} (\bibinfo
  {year} {2011}{\natexlab{b}}),\ \href {\doibase 10.1103/PhysRevB.84.235128}
  {\bibfield  {journal} {\bibinfo  {journal} {\prb}\ }\textbf {\bibinfo
  {volume} {84}},\ \bibinfo {eid} {235128}}\BibitemShut {NoStop}%
\bibitem [{\citenamefont {Chen}\ \emph
  {et~al.}(2015{\natexlab{b}})\citenamefont {Chen}, \citenamefont {Lu},\ and\
  \citenamefont {Kee}}]{Kim_chiral_ring}%
  \BibitemOpen
  \bibfield  {author} {\bibinfo {author} {\bibnamefont {Chen}, \bibfnamefont
  {Y.}}, \bibinfo {author} {\bibfnamefont {Y.-M.}\ \bibnamefont {Lu}}, \ and\
  \bibinfo {author} {\bibfnamefont {H.-Y.}\ \bibnamefont {Kee}}} (\bibinfo
  {year} {2015}{\natexlab{b}}),\ \href {http://dx.doi.org/10.1038/ncomms7593}
  {\bibfield  {journal} {\bibinfo  {journal} {Nat Commun}\ }\textbf {\bibinfo
  {volume} {6}}}\BibitemShut {NoStop}%
\bibitem [{\citenamefont {Cheng}(2012)}]{MChen2012}%
  \BibitemOpen
  \bibfield  {author} {\bibinfo {author} {\bibnamefont {Cheng}, \bibfnamefont
  {M.}}} (\bibinfo {year} {2012}),\ \href {\doibase 10.1103/PhysRevB.86.195126}
  {\bibfield  {journal} {\bibinfo  {journal} {Phys. Rev. B}\ }\textbf {\bibinfo
  {volume} {86}},\ \bibinfo {pages} {195126}}\BibitemShut {NoStop}%
\bibitem [{\citenamefont {{Cheng}}\ \emph {et~al.}(2015)\citenamefont
  {{Cheng}}, \citenamefont {{Bi}}, \citenamefont {{You}},\ and\ \citenamefont
  {{Gu}}}]{Cheng2015}%
  \BibitemOpen
  \bibfield  {author} {\bibinfo {author} {\bibnamefont {{Cheng}}, \bibfnamefont
  {M.}}, \bibinfo {author} {\bibfnamefont {Z.}~\bibnamefont {{Bi}}}, \bibinfo
  {author} {\bibfnamefont {Y.-Z.}\ \bibnamefont {{You}}}, \ and\ \bibinfo
  {author} {\bibfnamefont {Z.-C.}\ \bibnamefont {{Gu}}}} (\bibinfo {year}
  {2015}),\ \href@noop {} {\bibfield  {journal} {\bibinfo  {journal} {ArXiv
  e-prints}\ }}\Eprint {http://arxiv.org/abs/1501.01313} {arXiv:1501.01313
  [cond-mat.str-el]} \BibitemShut {NoStop}%
\bibitem [{\citenamefont {Cheng}\ and\ \citenamefont {Gu}(2014)}]{ChengZu2014}%
  \BibitemOpen
  \bibfield  {author} {\bibinfo {author} {\bibnamefont {Cheng}, \bibfnamefont
  {M.}}, \ and\ \bibinfo {author} {\bibfnamefont {Z.-C.}\ \bibnamefont {Gu}}}
  (\bibinfo {year} {2014}),\ \href {\doibase 10.1103/PhysRevLett.112.141602}
  {\bibfield  {journal} {\bibinfo  {journal} {Phys. Rev. Lett.}\ }\textbf
  {\bibinfo {volume} {112}},\ \bibinfo {pages} {141602}}\BibitemShut {NoStop}%
\bibitem [{\citenamefont {{Chiu}}(2014)}]{Nontrivial_surface_chiu}%
  \BibitemOpen
  \bibfield  {author} {\bibinfo {author} {\bibnamefont {{Chiu}}, \bibfnamefont
  {C.-K.}}} (\bibinfo {year} {2014}),\ \href@noop {} {\ }\Eprint
  {http://arxiv.org/abs/1410.1117} {arXiv:1410.1117} \BibitemShut {NoStop}%
\bibitem [{\citenamefont {Chiu}\ \emph {et~al.}(2012)\citenamefont {Chiu},
  \citenamefont {Ghaemi},\ and\ \citenamefont {Hughes}}]{Chiu_SC_TI_extended}%
  \BibitemOpen
  \bibfield  {author} {\bibinfo {author} {\bibnamefont {Chiu}, \bibfnamefont
  {C.-K.}}, \bibinfo {author} {\bibfnamefont {P.}~\bibnamefont {Ghaemi}}, \
  and\ \bibinfo {author} {\bibfnamefont {T.~L.}\ \bibnamefont {Hughes}}}
  (\bibinfo {year} {2012}),\ \href {\doibase 10.1103/PhysRevLett.109.237009}
  {\bibfield  {journal} {\bibinfo  {journal} {Phys. Rev. Lett.}\ }\textbf
  {\bibinfo {volume} {109}},\ \bibinfo {pages} {237009}}\BibitemShut {NoStop}%
\bibitem [{\citenamefont {Chiu}\ \emph {et~al.}(2011)\citenamefont {Chiu},
  \citenamefont {Gilbert},\ and\ \citenamefont {Hughes}}]{Chiu:2011fk}%
  \BibitemOpen
  \bibfield  {author} {\bibinfo {author} {\bibnamefont {Chiu}, \bibfnamefont
  {C.-K.}}, \bibinfo {author} {\bibfnamefont {M.~J.}\ \bibnamefont {Gilbert}},
  \ and\ \bibinfo {author} {\bibfnamefont {T.~L.}\ \bibnamefont {Hughes}}}
  (\bibinfo {year} {2011}),\ \href
  {http://link.aps.org/doi/10.1103/PhysRevB.84.144507} {\bibfield  {journal}
  {\bibinfo  {journal} {Phys. Rev. B}\ }\textbf {\bibinfo {volume}
  {84}}~(\bibinfo {number} {14}),\ \bibinfo {pages} {144507}}\BibitemShut
  {NoStop}%
\bibitem [{\citenamefont {{Chiu}}\ \emph {et~al.}(2015)\citenamefont {{Chiu}},
  \citenamefont {{Pikulin}},\ and\ \citenamefont
  {{Franz}}}]{Chiu_interaction_enabled}%
  \BibitemOpen
  \bibfield  {author} {\bibinfo {author} {\bibnamefont {{Chiu}}, \bibfnamefont
  {C.-K.}}, \bibinfo {author} {\bibfnamefont {D.~I.}\ \bibnamefont
  {{Pikulin}}}, \ and\ \bibinfo {author} {\bibfnamefont {M.}~\bibnamefont
  {{Franz}}}} (\bibinfo {year} {2015}),\ \href@noop {} {\ }\Eprint
  {http://arxiv.org/abs/1502.03432} {arXiv:1502.03432} \BibitemShut {NoStop}%
\bibitem [{\citenamefont {Chiu}\ \emph {et~al.}(2015)\citenamefont {Chiu},
  \citenamefont {Pikulin},\ and\ \citenamefont
  {Franz}}]{Chiu_strong_inter_Majorana}%
  \BibitemOpen
  \bibfield  {author} {\bibinfo {author} {\bibnamefont {Chiu}, \bibfnamefont
  {C.-K.}}, \bibinfo {author} {\bibfnamefont {D.~I.}\ \bibnamefont {Pikulin}},
  \ and\ \bibinfo {author} {\bibfnamefont {M.}~\bibnamefont {Franz}}} (\bibinfo
  {year} {2015}),\ \href {\doibase 10.1103/PhysRevB.91.165402} {\bibfield
  {journal} {\bibinfo  {journal} {Phys. Rev. B}\ }\textbf {\bibinfo {volume}
  {91}},\ \bibinfo {pages} {165402}}\BibitemShut {NoStop}%
\bibitem [{\citenamefont {Chiu}\ and\ \citenamefont
  {Schnyder}(2014)}]{ChiuSchnyder14}%
  \BibitemOpen
  \bibfield  {author} {\bibinfo {author} {\bibnamefont {Chiu}, \bibfnamefont
  {C.-K.}}, \ and\ \bibinfo {author} {\bibfnamefont {A.~P.}\ \bibnamefont
  {Schnyder}}} (\bibinfo {year} {2014}),\ \href {\doibase
  10.1103/PhysRevB.90.205136} {\bibfield  {journal} {\bibinfo  {journal} {Phys.
  Rev. B}\ }\textbf {\bibinfo {volume} {90}},\ \bibinfo {pages}
  {205136}}\BibitemShut {NoStop}%
\bibitem [{\citenamefont {Chiu}\ and\ \citenamefont
  {Schnyder}(2015)}]{Chiu_C4_Dirac}%
  \BibitemOpen
  \bibfield  {author} {\bibinfo {author} {\bibnamefont {Chiu}, \bibfnamefont
  {C.-K.}}, \ and\ \bibinfo {author} {\bibfnamefont {A.~P.}\ \bibnamefont
  {Schnyder}}} (\bibinfo {year} {2015}),\ \href
  {http://stacks.iop.org/1742-6596/603/i=1/a=012002} {\bibfield  {journal}
  {\bibinfo  {journal} {Journal of Physics: Conference Series}\ }\textbf
  {\bibinfo {volume} {603}}~(\bibinfo {number} {1}),\ \bibinfo {pages}
  {012002}}\BibitemShut {NoStop}%
\bibitem [{\citenamefont {Chiu}\ \emph {et~al.}(2013)\citenamefont {Chiu},
  \citenamefont {Yao},\ and\ \citenamefont {Ryu}}]{Chiu_reflection}%
  \BibitemOpen
  \bibfield  {author} {\bibinfo {author} {\bibnamefont {Chiu}, \bibfnamefont
  {C.-K.}}, \bibinfo {author} {\bibfnamefont {H.}~\bibnamefont {Yao}}, \ and\
  \bibinfo {author} {\bibfnamefont {S.}~\bibnamefont {Ryu}}} (\bibinfo {year}
  {2013}),\ \href {\doibase 10.1103/PhysRevB.88.075142} {\bibfield  {journal}
  {\bibinfo  {journal} {Phys. Rev. B}\ }\textbf {\bibinfo {volume} {88}},\
  \bibinfo {pages} {075142}}\BibitemShut {NoStop}%
\bibitem [{\citenamefont {Cho}\ \emph {et~al.}(2015)\citenamefont {Cho},
  \citenamefont {Hsieh}, \citenamefont {Morimoto},\ and\ \citenamefont
  {Ryu}}]{ChoHsiehMorimotoRyu2015}%
  \BibitemOpen
  \bibfield  {author} {\bibinfo {author} {\bibnamefont {Cho}, \bibfnamefont
  {G.~Y.}}, \bibinfo {author} {\bibfnamefont {C.-T.}\ \bibnamefont {Hsieh}},
  \bibinfo {author} {\bibfnamefont {T.}~\bibnamefont {Morimoto}}, \ and\
  \bibinfo {author} {\bibfnamefont {S.}~\bibnamefont {Ryu}}} (\bibinfo {year}
  {2015}),\ \href {\doibase 10.1103/PhysRevB.91.195142} {\bibfield  {journal}
  {\bibinfo  {journal} {Phys. Rev. B}\ }\textbf {\bibinfo {volume} {91}},\
  \bibinfo {pages} {195142}}\BibitemShut {NoStop}%
\bibitem [{\citenamefont {Cho}\ \emph {et~al.}(2012)\citenamefont {Cho},
  \citenamefont {Lu},\ and\ \citenamefont {Moore}}]{cho_moore_PRB_2012}%
  \BibitemOpen
  \bibfield  {author} {\bibinfo {author} {\bibnamefont {Cho}, \bibfnamefont
  {G.~Y.}}, \bibinfo {author} {\bibfnamefont {Y.-M.}\ \bibnamefont {Lu}}, \
  and\ \bibinfo {author} {\bibfnamefont {J.~E.}\ \bibnamefont {Moore}}}
  (\bibinfo {year} {2012}),\ \href {\doibase 10.1103/PhysRevB.86.125101}
  {\bibfield  {journal} {\bibinfo  {journal} {Phys. Rev. B}\ }\textbf {\bibinfo
  {volume} {86}},\ \bibinfo {pages} {125101}}\BibitemShut {NoStop}%
\bibitem [{\citenamefont {{Cho}}\ \emph {et~al.}(2014)\citenamefont {{Cho}},
  \citenamefont {{Teo}},\ and\ \citenamefont {{Ryu}}}]{Cho2014}%
  \BibitemOpen
  \bibfield  {author} {\bibinfo {author} {\bibnamefont {{Cho}}, \bibfnamefont
  {G.~Y.}}, \bibinfo {author} {\bibfnamefont {J.~C.~Y.}\ \bibnamefont {{Teo}}},
  \ and\ \bibinfo {author} {\bibfnamefont {S.}~\bibnamefont {{Ryu}}}} (\bibinfo
  {year} {2014}),\ \href {\doibase 10.1103/PhysRevB.89.235103} {\bibfield
  {journal} {\bibinfo  {journal} {\prb}\ }\textbf {\bibinfo {volume} {89}},\
  \bibinfo {eid} {235103}}\BibitemShut {NoStop}%
\bibitem [{\citenamefont {Chung}\ \emph {et~al.}(2007)\citenamefont {Chung},
  \citenamefont {Bluhm},\ and\ \citenamefont {Kim}}]{ChungBluhmKim07}%
  \BibitemOpen
  \bibfield  {author} {\bibinfo {author} {\bibnamefont {Chung}, \bibfnamefont
  {S.~B.}}, \bibinfo {author} {\bibfnamefont {H.}~\bibnamefont {Bluhm}}, \ and\
  \bibinfo {author} {\bibfnamefont {E.-A.}\ \bibnamefont {Kim}}} (\bibinfo
  {year} {2007}),\ \href {\doibase 10.1103/PhysRevLett.99.197002} {\bibfield
  {journal} {\bibinfo  {journal} {Phys. Rev. Lett.}\ }\textbf {\bibinfo
  {volume} {99}},\ \bibinfo {pages} {197002}}\BibitemShut {NoStop}%
\bibitem [{\citenamefont {Chung}\ and\ \citenamefont
  {Zhang}(2009)}]{Chung:2009fk}%
  \BibitemOpen
  \bibfield  {author} {\bibinfo {author} {\bibnamefont {Chung}, \bibfnamefont
  {S.~B.}}, \ and\ \bibinfo {author} {\bibfnamefont {S.-C.}\ \bibnamefont
  {Zhang}}} (\bibinfo {year} {2009}),\ \href@noop {} {\bibfield  {journal}
  {\bibinfo  {journal} {Phys. Rev. Lett.}\ }\textbf {\bibinfo {volume} {103}},\
  \bibinfo {pages} {235301}}\BibitemShut {NoStop}%
\bibitem [{\citenamefont {Churchill}\ \emph {et~al.}(2013)\citenamefont
  {Churchill}, \citenamefont {Fatemi}, \citenamefont {Grove-Rasmussen},
  \citenamefont {Deng}, \citenamefont {Caroff}, \citenamefont {Xu},\ and\
  \citenamefont {Marcus}}]{Churchill_zero_bias}%
  \BibitemOpen
  \bibfield  {author} {\bibinfo {author} {\bibnamefont {Churchill},
  \bibfnamefont {H.~O.~H.}}, \bibinfo {author} {\bibfnamefont {V.}~\bibnamefont
  {Fatemi}}, \bibinfo {author} {\bibfnamefont {K.}~\bibnamefont
  {Grove-Rasmussen}}, \bibinfo {author} {\bibfnamefont {M.~T.}\ \bibnamefont
  {Deng}}, \bibinfo {author} {\bibfnamefont {P.}~\bibnamefont {Caroff}},
  \bibinfo {author} {\bibfnamefont {H.~Q.}\ \bibnamefont {Xu}}, \ and\ \bibinfo
  {author} {\bibfnamefont {C.~M.}\ \bibnamefont {Marcus}}} (\bibinfo {year}
  {2013}),\ \href {\doibase 10.1103/PhysRevB.87.241401} {\bibfield  {journal}
  {\bibinfo  {journal} {Phys. Rev. B}\ }\textbf {\bibinfo {volume} {87}},\
  \bibinfo {pages} {241401}}\BibitemShut {NoStop}%
\bibitem [{\citenamefont {Clarke}\ \emph {et~al.}(2012)\citenamefont {Clarke},
  \citenamefont {Alicea},\ and\ \citenamefont {Shtengel}}]{ClarkeAliceaKirill}%
  \BibitemOpen
  \bibfield  {author} {\bibinfo {author} {\bibnamefont {Clarke}, \bibfnamefont
  {D.~J.}}, \bibinfo {author} {\bibfnamefont {J.}~\bibnamefont {Alicea}}, \
  and\ \bibinfo {author} {\bibfnamefont {K.}~\bibnamefont {Shtengel}}}
  (\bibinfo {year} {2012}),\ \href {\doibase 10.1038/ncomms2340} {\bibfield
  {journal} {\bibinfo  {journal} {Nature Commun.}\ }\textbf {\bibinfo {volume}
  {4}},\ \bibinfo {pages} {1348}}\BibitemShut {NoStop}%
\bibitem [{\citenamefont {Cook}\ and\ \citenamefont
  {Franz}(2011)}]{Franz_nanowire}%
  \BibitemOpen
  \bibfield  {author} {\bibinfo {author} {\bibnamefont {Cook}, \bibfnamefont
  {A.}}, \ and\ \bibinfo {author} {\bibfnamefont {M.}~\bibnamefont {Franz}}}
  (\bibinfo {year} {2011}),\ \href {\doibase 10.1103/PhysRevB.84.201105}
  {\bibfield  {journal} {\bibinfo  {journal} {Phys. Rev. B}\ }\textbf {\bibinfo
  {volume} {84}},\ \bibinfo {pages} {201105}}\BibitemShut {NoStop}%
\bibitem [{\citenamefont {Darriet}\ and\ \citenamefont
  {Regnault}(1993)}]{Darriet1993409}%
  \BibitemOpen
  \bibfield  {author} {\bibinfo {author} {\bibnamefont {Darriet}, \bibfnamefont
  {J.}}, \ and\ \bibinfo {author} {\bibfnamefont {L.}~\bibnamefont {Regnault}}}
  (\bibinfo {year} {1993}),\ \href {\doibase
  http://dx.doi.org/10.1016/0038-1098(93)90455-V} {\bibfield  {journal}
  {\bibinfo  {journal} {Solid State Communications}\ }\textbf {\bibinfo
  {volume} {86}}~(\bibinfo {number} {7}),\ \bibinfo {pages} {409 }}\BibitemShut
  {NoStop}%
\bibitem [{\citenamefont {Das}\ \emph {et~al.}(2012{\natexlab{a}})\citenamefont
  {Das}, \citenamefont {Ronen}, \citenamefont {Most}, \citenamefont {Oreg},
  \citenamefont {Heiblum},\ and\ \citenamefont {Shtrikman}}]{Das_zero_bias}%
  \BibitemOpen
  \bibfield  {author} {\bibinfo {author} {\bibnamefont {Das}, \bibfnamefont
  {A.}}, \bibinfo {author} {\bibfnamefont {Y.}~\bibnamefont {Ronen}}, \bibinfo
  {author} {\bibfnamefont {Y.}~\bibnamefont {Most}}, \bibinfo {author}
  {\bibfnamefont {Y.}~\bibnamefont {Oreg}}, \bibinfo {author} {\bibfnamefont
  {M.}~\bibnamefont {Heiblum}}, \ and\ \bibinfo {author} {\bibfnamefont
  {H.}~\bibnamefont {Shtrikman}}} (\bibinfo {year} {2012}{\natexlab{a}}),\
  \href@noop {} {\bibfield  {journal} {\bibinfo  {journal} {Nat. Phys.}\
  }\textbf {\bibinfo {volume} {8}},\ \bibinfo {pages} {887}}\BibitemShut
  {NoStop}%
\bibitem [{\citenamefont {Das}\ \emph {et~al.}(2012{\natexlab{b}})\citenamefont
  {Das}, \citenamefont {Ronen}, \citenamefont {Most}, \citenamefont {Oreg},
  \citenamefont {Heiblum},\ and\ \citenamefont {Shtrikman}}]{Yuval_zero_bias}%
  \BibitemOpen
  \bibfield  {author} {\bibinfo {author} {\bibnamefont {Das}, \bibfnamefont
  {A.}}, \bibinfo {author} {\bibfnamefont {Y.}~\bibnamefont {Ronen}}, \bibinfo
  {author} {\bibfnamefont {Y.}~\bibnamefont {Most}}, \bibinfo {author}
  {\bibfnamefont {Y.}~\bibnamefont {Oreg}}, \bibinfo {author} {\bibfnamefont
  {M.}~\bibnamefont {Heiblum}}, \ and\ \bibinfo {author} {\bibfnamefont
  {H.}~\bibnamefont {Shtrikman}}} (\bibinfo {year} {2012}{\natexlab{b}}),\
  \href@noop {} {\bibfield  {journal} {\bibinfo  {journal} {Nat. Phys.}\
  }\textbf {\bibinfo {volume} {8}},\ \bibinfo {pages} {887}}\BibitemShut
  {NoStop}%
\bibitem [{\citenamefont {Das~Sarma}\ \emph {et~al.}(2006)\citenamefont
  {Das~Sarma}, \citenamefont {Nayak},\ and\ \citenamefont
  {Tewari}}]{DasSarmaNayakTewari06}%
  \BibitemOpen
  \bibfield  {author} {\bibinfo {author} {\bibnamefont {Das~Sarma},
  \bibfnamefont {S.}}, \bibinfo {author} {\bibfnamefont {C.}~\bibnamefont
  {Nayak}}, \ and\ \bibinfo {author} {\bibfnamefont {S.}~\bibnamefont
  {Tewari}}} (\bibinfo {year} {2006}),\ \href {\doibase
  10.1103/PhysRevB.73.220502} {\bibfield  {journal} {\bibinfo  {journal} {Phys.
  Rev. B}\ }\textbf {\bibinfo {volume} {73}},\ \bibinfo {pages}
  {220502}}\BibitemShut {NoStop}%
\bibitem [{\citenamefont {{De Nittis}}\ and\ \citenamefont
  {{Gomi}}(2014)}]{Nittis2014a}%
  \BibitemOpen
  \bibfield  {author} {\bibinfo {author} {\bibnamefont {{De Nittis}},
  \bibfnamefont {G.}}, \ and\ \bibinfo {author} {\bibfnamefont
  {K.}~\bibnamefont {{Gomi}}}} (\bibinfo {year} {2014}),\ \href {\doibase
  10.1016/j.geomphys.2014.07.036} {\bibfield  {journal} {\bibinfo  {journal}
  {Journal of Geometry and Physics}\ }\textbf {\bibinfo {volume} {86}},\
  \bibinfo {pages} {303}}\BibitemShut {NoStop}%
\bibitem [{\citenamefont {De~Nittis}\ and\ \citenamefont
  {Gomi}(2015)}]{Nittis2014b}%
  \BibitemOpen
  \bibfield  {author} {\bibinfo {author} {\bibnamefont {De~Nittis},
  \bibfnamefont {G.}}, \ and\ \bibinfo {author} {\bibfnamefont
  {K.}~\bibnamefont {Gomi}}} (\bibinfo {year} {2015}),\ \bibfield  {booktitle}
  {\emph {\bibinfo {booktitle} {Communications in Mathematical Physics}},\
  }\href {\doibase 10.1007/s00220-015-2390-0} {\ \textbf {\bibinfo {volume}
  {339}}~(\bibinfo {number} {1}),\ \bibinfo {pages} {1}}\BibitemShut {NoStop}%
\bibitem [{\citenamefont {Deng}\ \emph {et~al.}(2014)\citenamefont {Deng},
  \citenamefont {Wang},\ and\ \citenamefont {Duan}}]{Deng_chiral_extended}%
  \BibitemOpen
  \bibfield  {author} {\bibinfo {author} {\bibnamefont {Deng}, \bibfnamefont
  {D.-L.}}, \bibinfo {author} {\bibfnamefont {S.-T.}\ \bibnamefont {Wang}}, \
  and\ \bibinfo {author} {\bibfnamefont {L.-M.}\ \bibnamefont {Duan}}}
  (\bibinfo {year} {2014}),\ \href {\doibase 10.1103/PhysRevB.89.075126}
  {\bibfield  {journal} {\bibinfo  {journal} {Phys. Rev. B}\ }\textbf {\bibinfo
  {volume} {89}},\ \bibinfo {pages} {075126}}\BibitemShut {NoStop}%
\bibitem [{\citenamefont {Deng}\ \emph {et~al.}(2012)\citenamefont {Deng},
  \citenamefont {Yu}, \citenamefont {Huang}, \citenamefont {Larsson},
  \citenamefont {Caroff},\ and\ \citenamefont {Xu}}]{Deng_zero_bias}%
  \BibitemOpen
  \bibfield  {author} {\bibinfo {author} {\bibnamefont {Deng}, \bibfnamefont
  {M.~T.}}, \bibinfo {author} {\bibfnamefont {C.~L.}\ \bibnamefont {Yu}},
  \bibinfo {author} {\bibfnamefont {G.~Y.}\ \bibnamefont {Huang}}, \bibinfo
  {author} {\bibfnamefont {M.}~\bibnamefont {Larsson}}, \bibinfo {author}
  {\bibfnamefont {P.}~\bibnamefont {Caroff}}, \ and\ \bibinfo {author}
  {\bibfnamefont {H.~Q.}\ \bibnamefont {Xu}}} (\bibinfo {year} {2012}),\ \href
  {\doibase 10.1021/nl303758w} {\bibfield  {journal} {\bibinfo  {journal} {Nano
  Letters}\ }\textbf {\bibinfo {volume} {12}},\ \bibinfo {pages}
  {6414}}\BibitemShut {NoStop}%
\bibitem [{\citenamefont {Diez}\ \emph {et~al.}(2014)\citenamefont {Diez},
  \citenamefont {Fulga}, \citenamefont {Pikulin}, \citenamefont
  {Tworzyd{\l}o},\ and\ \citenamefont {Beenakker}}]{weak_TSC}%
  \BibitemOpen
  \bibfield  {author} {\bibinfo {author} {\bibnamefont {Diez}, \bibfnamefont
  {M.}}, \bibinfo {author} {\bibfnamefont {I.~C.}\ \bibnamefont {Fulga}},
  \bibinfo {author} {\bibfnamefont {D.~I.}\ \bibnamefont {Pikulin}}, \bibinfo
  {author} {\bibfnamefont {J.}~\bibnamefont {Tworzyd{\l}o}}, \ and\ \bibinfo
  {author} {\bibfnamefont {C.~W.~J.}\ \bibnamefont {Beenakker}}} (\bibinfo
  {year} {2014}),\ \href {http://stacks.iop.org/1367-2630/16/i=6/a=063049}
  {\bibfield  {journal} {\bibinfo  {journal} {New Journal of Physics}\ }\textbf
  {\bibinfo {volume} {16}}~(\bibinfo {number} {6}),\ \bibinfo {pages}
  {063049}}\BibitemShut {NoStop}%
\bibitem [{\citenamefont {Diez}\ \emph {et~al.}(2015)\citenamefont {Diez},
  \citenamefont {Pikulin}, \citenamefont {Fulga},\ and\ \citenamefont
  {Tworzyd{\l}o}}]{Average_reflection_Diez}%
  \BibitemOpen
  \bibfield  {author} {\bibinfo {author} {\bibnamefont {Diez}, \bibfnamefont
  {M.}}, \bibinfo {author} {\bibfnamefont {D.~I.}\ \bibnamefont {Pikulin}},
  \bibinfo {author} {\bibfnamefont {I.~C.}\ \bibnamefont {Fulga}}, \ and\
  \bibinfo {author} {\bibfnamefont {J.}~\bibnamefont {Tworzyd{\l}o}}} (\bibinfo
  {year} {2015}),\ \href {http://stacks.iop.org/1367-2630/17/i=4/a=043014}
  {\bibfield  {journal} {\bibinfo  {journal} {New Journal of Physics}\ }\textbf
  {\bibinfo {volume} {17}}~(\bibinfo {number} {4}),\ \bibinfo {pages}
  {043014}}\BibitemShut {NoStop}%
\bibitem [{\citenamefont {Dijkgraaf}\ and\ \citenamefont
  {Witten}(1990)}]{DijkgraafWitten1990}%
  \BibitemOpen
  \bibfield  {author} {\bibinfo {author} {\bibnamefont {Dijkgraaf},
  \bibfnamefont {R.}}, \ and\ \bibinfo {author} {\bibfnamefont
  {E.}~\bibnamefont {Witten}}} (\bibinfo {year} {1990}),\ \href {\doibase
  10.1007/BF02096988} {\bibfield  {journal} {\bibinfo  {journal}
  {Commun.Math.Phys.}\ }\textbf {\bibinfo {volume} {129}},\ \bibinfo {pages}
  {393}}\BibitemShut {NoStop}%
\bibitem [{\citenamefont {Dong}\ and\ \citenamefont
  {Liu}(2016)}]{2D_classification_Liu}%
  \BibitemOpen
  \bibfield  {author} {\bibinfo {author} {\bibnamefont {Dong}, \bibfnamefont
  {X.-Y.}}, \ and\ \bibinfo {author} {\bibfnamefont {C.-X.}\ \bibnamefont
  {Liu}}} (\bibinfo {year} {2016}),\ \href {\doibase
  10.1103/PhysRevB.93.045429} {\bibfield  {journal} {\bibinfo  {journal} {Phys.
  Rev. B}\ }\textbf {\bibinfo {volume} {93}},\ \bibinfo {pages}
  {045429}}\BibitemShut {NoStop}%
\bibitem [{\citenamefont {Dr\"uppel}\ \emph {et~al.}(2014)\citenamefont
  {Dr\"uppel}, \citenamefont {Kr\"uger},\ and\ \citenamefont
  {Rohlfing}}]{Rohlfing_TCI_surface}%
  \BibitemOpen
  \bibfield  {author} {\bibinfo {author} {\bibnamefont {Dr\"uppel},
  \bibfnamefont {M.}}, \bibinfo {author} {\bibfnamefont {P.}~\bibnamefont
  {Kr\"uger}}, \ and\ \bibinfo {author} {\bibfnamefont {M.}~\bibnamefont
  {Rohlfing}}} (\bibinfo {year} {2014}),\ \href {\doibase
  10.1103/PhysRevB.90.155312} {\bibfield  {journal} {\bibinfo  {journal} {Phys.
  Rev. B}\ }\textbf {\bibinfo {volume} {90}},\ \bibinfo {pages}
  {155312}}\BibitemShut {NoStop}%
\bibitem [{\citenamefont {Dyson}(1953)}]{Dyson1953}%
  \BibitemOpen
  \bibfield  {author} {\bibinfo {author} {\bibnamefont {Dyson}, \bibfnamefont
  {F.~J.}}} (\bibinfo {year} {1953}),\ \href {\doibase 10.1103/PhysRev.92.1331}
  {\bibfield  {journal} {\bibinfo  {journal} {Phys. Rev.}\ }\textbf {\bibinfo
  {volume} {92}},\ \bibinfo {pages} {1331}}\BibitemShut {NoStop}%
\bibitem [{\citenamefont {Dyson}(1962)}]{dyson:1199}%
  \BibitemOpen
  \bibfield  {author} {\bibinfo {author} {\bibnamefont {Dyson}, \bibfnamefont
  {F.~J.}}} (\bibinfo {year} {1962}),\ \href {\doibase 10.1063/1.1703863}
  {\bibfield  {journal} {\bibinfo  {journal} {Journal of Mathematical Physics}\
  }\textbf {\bibinfo {volume} {3}},\ \bibinfo {pages} {1199}}\BibitemShut
  {NoStop}%
\bibitem [{\citenamefont {Dzero}\ \emph {et~al.}(2012)\citenamefont {Dzero},
  \citenamefont {Sun}, \citenamefont {Coleman},\ and\ \citenamefont
  {Galitski}}]{Dzero_Kondo_PRB}%
  \BibitemOpen
  \bibfield  {author} {\bibinfo {author} {\bibnamefont {Dzero}, \bibfnamefont
  {M.}}, \bibinfo {author} {\bibfnamefont {K.}~\bibnamefont {Sun}}, \bibinfo
  {author} {\bibfnamefont {P.}~\bibnamefont {Coleman}}, \ and\ \bibinfo
  {author} {\bibfnamefont {V.}~\bibnamefont {Galitski}}} (\bibinfo {year}
  {2012}),\ \href {\doibase 10.1103/PhysRevB.85.045130} {\bibfield  {journal}
  {\bibinfo  {journal} {Phys. Rev. B}\ }\textbf {\bibinfo {volume} {85}},\
  \bibinfo {pages} {045130}}\BibitemShut {NoStop}%
\bibitem [{\citenamefont {Dzero}\ \emph {et~al.}(2010)\citenamefont {Dzero},
  \citenamefont {Sun}, \citenamefont {Galitski},\ and\ \citenamefont
  {Coleman}}]{Dzero_Kondo_PRL}%
  \BibitemOpen
  \bibfield  {author} {\bibinfo {author} {\bibnamefont {Dzero}, \bibfnamefont
  {M.}}, \bibinfo {author} {\bibfnamefont {K.}~\bibnamefont {Sun}}, \bibinfo
  {author} {\bibfnamefont {V.}~\bibnamefont {Galitski}}, \ and\ \bibinfo
  {author} {\bibfnamefont {P.}~\bibnamefont {Coleman}}} (\bibinfo {year}
  {2010}),\ \href {\doibase 10.1103/PhysRevLett.104.106408} {\bibfield
  {journal} {\bibinfo  {journal} {Phys. Rev. Lett.}\ }\textbf {\bibinfo
  {volume} {104}},\ \bibinfo {pages} {106408}}\BibitemShut {NoStop}%
\bibitem [{\citenamefont {Dziawa}\ \emph {et~al.}(2012)\citenamefont {Dziawa},
  \citenamefont {Kowalski}, \citenamefont {Dybko}, \citenamefont {Buczko},
  \citenamefont {Szczerbakow}, \citenamefont {Szot}, \citenamefont
  {{\L}usakowska}, \citenamefont {Balasubramanian}, \citenamefont {Wojek},
  \citenamefont {Berntsen}, \citenamefont {Tjernberg},\ and\ \citenamefont
  {Story}}]{Dziawa:2012uq}%
  \BibitemOpen
  \bibfield  {author} {\bibinfo {author} {\bibnamefont {Dziawa}, \bibfnamefont
  {P.}}, \bibinfo {author} {\bibfnamefont {B.~J.}\ \bibnamefont {Kowalski}},
  \bibinfo {author} {\bibfnamefont {K.}~\bibnamefont {Dybko}}, \bibinfo
  {author} {\bibfnamefont {R.}~\bibnamefont {Buczko}}, \bibinfo {author}
  {\bibfnamefont {A.}~\bibnamefont {Szczerbakow}}, \bibinfo {author}
  {\bibfnamefont {M.}~\bibnamefont {Szot}}, \bibinfo {author} {\bibfnamefont
  {E.}~\bibnamefont {{\L}usakowska}}, \bibinfo {author} {\bibfnamefont
  {T.}~\bibnamefont {Balasubramanian}}, \bibinfo {author} {\bibfnamefont
  {B.~M.}\ \bibnamefont {Wojek}}, \bibinfo {author} {\bibfnamefont {M.~H.}\
  \bibnamefont {Berntsen}}, \bibinfo {author} {\bibfnamefont {O.}~\bibnamefont
  {Tjernberg}}, \ and\ \bibinfo {author} {\bibfnamefont {T.}~\bibnamefont
  {Story}}} (\bibinfo {year} {2012}),\ \href@noop {} {\bibfield  {journal}
  {\bibinfo  {journal} {Nat. Mater.}\ }\textbf {\bibinfo {volume} {11}},\
  \bibinfo {pages} {1023}}\BibitemShut {NoStop}%
\bibitem [{\citenamefont {{Efetov}}(1983)}]{Efetov1983}%
  \BibitemOpen
  \bibfield  {author} {\bibinfo {author} {\bibnamefont {{Efetov}},
  \bibfnamefont {K.~B.}}} (\bibinfo {year} {1983}),\ \href {\doibase
  10.1080/00018738300101531} {\bibfield  {journal} {\bibinfo  {journal}
  {Advances in Physics}\ }\textbf {\bibinfo {volume} {32}},\ \bibinfo {pages}
  {53}}\BibitemShut {NoStop}%
\bibitem [{\citenamefont {{Efetov}}\ \emph {et~al.}(1980)\citenamefont
  {{Efetov}}, \citenamefont {{Larkin}},\ and\ \citenamefont {{Kheml'Nitski{\v
  \i}}}}]{Efetov1980}%
  \BibitemOpen
  \bibfield  {author} {\bibinfo {author} {\bibnamefont {{Efetov}},
  \bibfnamefont {K.~B.}}, \bibinfo {author} {\bibfnamefont {A.~I.}\
  \bibnamefont {{Larkin}}}, \ and\ \bibinfo {author} {\bibfnamefont {D.~E.}\
  \bibnamefont {{Kheml'Nitski{\v \i}}}}} (\bibinfo {year} {1980}),\ \href@noop
  {} {\bibfield  {journal} {\bibinfo  {journal} {Soviet Journal of Experimental
  and Theoretical Physics}\ }\textbf {\bibinfo {volume} {52}},\ \bibinfo
  {pages} {568}}\BibitemShut {NoStop}%
\bibitem [{\citenamefont {Elliott}\ and\ \citenamefont
  {Franz}(2015)}]{elliott_franz_review}%
  \BibitemOpen
  \bibfield  {author} {\bibinfo {author} {\bibnamefont {Elliott}, \bibfnamefont
  {S.~R.}}, \ and\ \bibinfo {author} {\bibfnamefont {M.}~\bibnamefont {Franz}}}
  (\bibinfo {year} {2015}),\ \href {\doibase 10.1103/RevModPhys.87.137}
  {\bibfield  {journal} {\bibinfo  {journal} {Rev. Mod. Phys.}\ }\textbf
  {\bibinfo {volume} {87}},\ \bibinfo {pages} {137}}\BibitemShut {NoStop}%
\bibitem [{\citenamefont {{Essin}}\ and\ \citenamefont
  {{Gurarie}}(2011)}]{EssinGurarie2011}%
  \BibitemOpen
  \bibfield  {author} {\bibinfo {author} {\bibnamefont {{Essin}}, \bibfnamefont
  {A.~M.}}, \ and\ \bibinfo {author} {\bibfnamefont {V.}~\bibnamefont
  {{Gurarie}}}} (\bibinfo {year} {2011}),\ \href {\doibase
  10.1103/PhysRevB.84.125132} {\bibfield  {journal} {\bibinfo  {journal}
  {\prb}\ }\textbf {\bibinfo {volume} {84}}~(\bibinfo {number} {12}),\ \bibinfo
  {eid} {125132}}\BibitemShut {NoStop}%
\bibitem [{\citenamefont {Essin}\ and\ \citenamefont
  {Hermele}(2013)}]{essin_hermele_PRB_13}%
  \BibitemOpen
  \bibfield  {author} {\bibinfo {author} {\bibnamefont {Essin}, \bibfnamefont
  {A.~M.}}, \ and\ \bibinfo {author} {\bibfnamefont {M.}~\bibnamefont
  {Hermele}}} (\bibinfo {year} {2013}),\ \href {\doibase
  10.1103/PhysRevB.87.104406} {\bibfield  {journal} {\bibinfo  {journal} {Phys.
  Rev. B}\ }\textbf {\bibinfo {volume} {87}},\ \bibinfo {pages}
  {104406}}\BibitemShut {NoStop}%
\bibitem [{\citenamefont {Essin}\ \emph {et~al.}(2009)\citenamefont {Essin},
  \citenamefont {Moore},\ and\ \citenamefont {Vanderbilt}}]{essinPRL09}%
  \BibitemOpen
  \bibfield  {author} {\bibinfo {author} {\bibnamefont {Essin}, \bibfnamefont
  {A.~M.}}, \bibinfo {author} {\bibfnamefont {J.~E.}\ \bibnamefont {Moore}}, \
  and\ \bibinfo {author} {\bibfnamefont {D.}~\bibnamefont {Vanderbilt}}}
  (\bibinfo {year} {2009}),\ \href {\doibase 10.1103/PhysRevLett.102.146805}
  {\bibfield  {journal} {\bibinfo  {journal} {Phys. Rev. Lett.}\ }\textbf
  {\bibinfo {volume} {102}},\ \bibinfo {pages} {146805}}\BibitemShut {NoStop}%
\bibitem [{\citenamefont {{Evers}}\ and\ \citenamefont
  {{Mirlin}}(2008)}]{EversMirlin2008}%
  \BibitemOpen
  \bibfield  {author} {\bibinfo {author} {\bibnamefont {{Evers}}, \bibfnamefont
  {F.}}, \ and\ \bibinfo {author} {\bibfnamefont {A.~D.}\ \bibnamefont
  {{Mirlin}}}} (\bibinfo {year} {2008}),\ \href {\doibase
  10.1103/RevModPhys.80.1355} {\bibfield  {journal} {\bibinfo  {journal}
  {Reviews of Modern Physics}\ }\textbf {\bibinfo {volume} {80}},\ \bibinfo
  {pages} {1355}}\BibitemShut {NoStop}%
\bibitem [{\citenamefont {Ezawa}(2012)}]{ezawa_siliceneNJP12}%
  \BibitemOpen
  \bibfield  {author} {\bibinfo {author} {\bibnamefont {Ezawa}, \bibfnamefont
  {M.}}} (\bibinfo {year} {2012}),\ \href
  {http://stacks.iop.org/1367-2630/14/i=3/a=033003} {\bibfield  {journal}
  {\bibinfo  {journal} {New Journal of Physics}\ }\textbf {\bibinfo {volume}
  {14}}~(\bibinfo {number} {3}),\ \bibinfo {pages} {033003}}\BibitemShut
  {NoStop}%
\bibitem [{\citenamefont {Ezawa}(2013)}]{Ezawa_floquet_TI}%
  \BibitemOpen
  \bibfield  {author} {\bibinfo {author} {\bibnamefont {Ezawa}, \bibfnamefont
  {M.}}} (\bibinfo {year} {2013}),\ \href {\doibase
  10.1103/PhysRevLett.110.026603} {\bibfield  {journal} {\bibinfo  {journal}
  {Phys. Rev. Lett.}\ }\textbf {\bibinfo {volume} {110}},\ \bibinfo {pages}
  {026603}}\BibitemShut {NoStop}%
\bibitem [{\citenamefont {Ezawa}(2015)}]{ezawaPRL15}%
  \BibitemOpen
  \bibfield  {author} {\bibinfo {author} {\bibnamefont {Ezawa}, \bibfnamefont
  {M.}}} (\bibinfo {year} {2015}),\ \href {\doibase
  10.1103/PhysRevLett.114.056403} {\bibfield  {journal} {\bibinfo  {journal}
  {Phys. Rev. Lett.}\ }\textbf {\bibinfo {volume} {114}},\ \bibinfo {pages}
  {056403}}\BibitemShut {NoStop}%
\bibitem [{\citenamefont {Fang}\ \emph {et~al.}(2015)\citenamefont {Fang},
  \citenamefont {Chen}, \citenamefont {Kee},\ and\ \citenamefont
  {Fu}}]{Nodal_Line_Fang}%
  \BibitemOpen
  \bibfield  {author} {\bibinfo {author} {\bibnamefont {Fang}, \bibfnamefont
  {C.}}, \bibinfo {author} {\bibfnamefont {Y.}~\bibnamefont {Chen}}, \bibinfo
  {author} {\bibfnamefont {H.-Y.}\ \bibnamefont {Kee}}, \ and\ \bibinfo
  {author} {\bibfnamefont {L.}~\bibnamefont {Fu}}} (\bibinfo {year} {2015}),\
  \href {\doibase 10.1103/PhysRevB.92.081201} {\bibfield  {journal} {\bibinfo
  {journal} {Phys. Rev. B}\ }\textbf {\bibinfo {volume} {92}},\ \bibinfo
  {pages} {081201}}\BibitemShut {NoStop}%
\bibitem [{\citenamefont {Fang}\ and\ \citenamefont
  {Fu}(2015)}]{New_crystalline_Fang}%
  \BibitemOpen
  \bibfield  {author} {\bibinfo {author} {\bibnamefont {Fang}, \bibfnamefont
  {C.}}, \ and\ \bibinfo {author} {\bibfnamefont {L.}~\bibnamefont {Fu}}}
  (\bibinfo {year} {2015}),\ \href {\doibase 10.1103/PhysRevB.91.161105}
  {\bibfield  {journal} {\bibinfo  {journal} {Phys. Rev. B}\ }\textbf {\bibinfo
  {volume} {91}},\ \bibinfo {pages} {161105}}\BibitemShut {NoStop}%
\bibitem [{\citenamefont {Fang}\ \emph
  {et~al.}(2012{\natexlab{a}})\citenamefont {Fang}, \citenamefont {Gilbert},\
  and\ \citenamefont {Bernevig}}]{Chen_chernN_GSP}%
  \BibitemOpen
  \bibfield  {author} {\bibinfo {author} {\bibnamefont {Fang}, \bibfnamefont
  {C.}}, \bibinfo {author} {\bibfnamefont {M.~J.}\ \bibnamefont {Gilbert}}, \
  and\ \bibinfo {author} {\bibfnamefont {B.~A.}\ \bibnamefont {Bernevig}}}
  (\bibinfo {year} {2012}{\natexlab{a}}),\ \href {\doibase
  10.1103/PhysRevB.86.115112} {\bibfield  {journal} {\bibinfo  {journal} {Phys.
  Rev. B}\ }\textbf {\bibinfo {volume} {86}},\ \bibinfo {pages}
  {115112}}\BibitemShut {NoStop}%
\bibitem [{\citenamefont {Fang}\ \emph
  {et~al.}(2013{\natexlab{a}})\citenamefont {Fang}, \citenamefont {Gilbert},\
  and\ \citenamefont {Bernevig}}]{Fang:2012kx}%
  \BibitemOpen
  \bibfield  {author} {\bibinfo {author} {\bibnamefont {Fang}, \bibfnamefont
  {C.}}, \bibinfo {author} {\bibfnamefont {M.~J.}\ \bibnamefont {Gilbert}}, \
  and\ \bibinfo {author} {\bibfnamefont {B.~A.}\ \bibnamefont {Bernevig}}}
  (\bibinfo {year} {2013}{\natexlab{a}}),\ \href {\doibase
  10.1103/PhysRevB.87.035119} {\bibfield  {journal} {\bibinfo  {journal} {Phys.
  Rev. B}\ }\textbf {\bibinfo {volume} {87}},\ \bibinfo {pages}
  {035119}}\BibitemShut {NoStop}%
\bibitem [{\citenamefont {Fang}\ \emph
  {et~al.}(2014{\natexlab{a}})\citenamefont {Fang}, \citenamefont {Gilbert},\
  and\ \citenamefont {Bernevig}}]{PhysRevLett.112.046801}%
  \BibitemOpen
  \bibfield  {author} {\bibinfo {author} {\bibnamefont {Fang}, \bibfnamefont
  {C.}}, \bibinfo {author} {\bibfnamefont {M.~J.}\ \bibnamefont {Gilbert}}, \
  and\ \bibinfo {author} {\bibfnamefont {B.~A.}\ \bibnamefont {Bernevig}}}
  (\bibinfo {year} {2014}{\natexlab{a}}),\ \href {\doibase
  10.1103/PhysRevLett.112.046801} {\bibfield  {journal} {\bibinfo  {journal}
  {Phys. Rev. Lett.}\ }\textbf {\bibinfo {volume} {112}},\ \bibinfo {pages}
  {046801}}\BibitemShut {NoStop}%
\bibitem [{\citenamefont {Fang}\ \emph
  {et~al.}(2014{\natexlab{b}})\citenamefont {Fang}, \citenamefont {Gilbert},\
  and\ \citenamefont {Bernevig}}]{Magnetic_group_Fang}%
  \BibitemOpen
  \bibfield  {author} {\bibinfo {author} {\bibnamefont {Fang}, \bibfnamefont
  {C.}}, \bibinfo {author} {\bibfnamefont {M.~J.}\ \bibnamefont {Gilbert}}, \
  and\ \bibinfo {author} {\bibfnamefont {B.~A.}\ \bibnamefont {Bernevig}}}
  (\bibinfo {year} {2014}{\natexlab{b}}),\ \href {\doibase
  10.1103/PhysRevLett.112.106401} {\bibfield  {journal} {\bibinfo  {journal}
  {Phys. Rev. Lett.}\ }\textbf {\bibinfo {volume} {112}},\ \bibinfo {pages}
  {106401}}\BibitemShut {NoStop}%
\bibitem [{\citenamefont {Fang}\ \emph
  {et~al.}(2012{\natexlab{b}})\citenamefont {Fang}, \citenamefont {Gilbert},
  \citenamefont {Dai},\ and\ \citenamefont
  {Bernevig}}]{Fang_bernevig_multi_weyl_PRL_12}%
  \BibitemOpen
  \bibfield  {author} {\bibinfo {author} {\bibnamefont {Fang}, \bibfnamefont
  {C.}}, \bibinfo {author} {\bibfnamefont {M.~J.}\ \bibnamefont {Gilbert}},
  \bibinfo {author} {\bibfnamefont {X.}~\bibnamefont {Dai}}, \ and\ \bibinfo
  {author} {\bibfnamefont {B.~A.}\ \bibnamefont {Bernevig}}} (\bibinfo {year}
  {2012}{\natexlab{b}}),\ \href {\doibase 10.1103/PhysRevLett.108.266802}
  {\bibfield  {journal} {\bibinfo  {journal} {Phys. Rev. Lett.}\ }\textbf
  {\bibinfo {volume} {108}},\ \bibinfo {pages} {266802}}\BibitemShut {NoStop}%
\bibitem [{\citenamefont {Fang}\ \emph
  {et~al.}(2013{\natexlab{b}})\citenamefont {Fang}, \citenamefont {Gilbert},
  \citenamefont {Xu}, \citenamefont {Bernevig},\ and\ \citenamefont
  {Hasan}}]{PhysRevB.88.125141}%
  \BibitemOpen
  \bibfield  {author} {\bibinfo {author} {\bibnamefont {Fang}, \bibfnamefont
  {C.}}, \bibinfo {author} {\bibfnamefont {M.~J.}\ \bibnamefont {Gilbert}},
  \bibinfo {author} {\bibfnamefont {S.-Y.}\ \bibnamefont {Xu}}, \bibinfo
  {author} {\bibfnamefont {B.~A.}\ \bibnamefont {Bernevig}}, \ and\ \bibinfo
  {author} {\bibfnamefont {M.~Z.}\ \bibnamefont {Hasan}}} (\bibinfo {year}
  {2013}{\natexlab{b}}),\ \href {\doibase 10.1103/PhysRevB.88.125141}
  {\bibfield  {journal} {\bibinfo  {journal} {Phys. Rev. B}\ }\textbf {\bibinfo
  {volume} {88}},\ \bibinfo {pages} {125141}}\BibitemShut {NoStop}%
\bibitem [{\citenamefont {{Fendley}}(2000)}]{Fendley2000}%
  \BibitemOpen
  \bibfield  {author} {\bibinfo {author} {\bibnamefont {{Fendley}},
  \bibfnamefont {P.}}} (\bibinfo {year} {2000}),\ \href@noop {} {\bibinfo
  {journal} {arXiv:cond-mat/0006360}\ }\BibitemShut {NoStop}%
\bibitem [{\citenamefont
  {Fidkowski}(2010)}]{Entanglement_real_spectrum_Fidkowski}%
  \BibitemOpen
\bibfield  {journal} {  }\bibfield  {author} {\bibinfo {author} {\bibnamefont
  {Fidkowski}, \bibfnamefont {L.}}} (\bibinfo {year} {2010}),\ \href {\doibase
  10.1103/PhysRevLett.104.130502} {\bibfield  {journal} {\bibinfo  {journal}
  {Phys. Rev. Lett.}\ }\textbf {\bibinfo {volume} {104}},\ \bibinfo {pages}
  {130502}}\BibitemShut {NoStop}%
\bibitem [{\citenamefont {{Fidkowski}}\ \emph {et~al.}(2013)\citenamefont
  {{Fidkowski}}, \citenamefont {{Chen}},\ and\ \citenamefont
  {{Vishwanath}}}]{Fidkowski2013}%
  \BibitemOpen
  \bibfield  {author} {\bibinfo {author} {\bibnamefont {{Fidkowski}},
  \bibfnamefont {L.}}, \bibinfo {author} {\bibfnamefont {X.}~\bibnamefont
  {{Chen}}}, \ and\ \bibinfo {author} {\bibfnamefont {A.}~\bibnamefont
  {{Vishwanath}}}} (\bibinfo {year} {2013}),\ \href {\doibase
  10.1103/PhysRevX.3.041016} {\bibfield  {journal} {\bibinfo  {journal}
  {Physical Review X}\ }\textbf {\bibinfo {volume} {3}},\ \bibinfo {eid}
  {041016}}\BibitemShut {NoStop}%
\bibitem [{\citenamefont {Fidkowski}\ and\ \citenamefont
  {Kitaev}(2010)}]{Fidkowski2010}%
  \BibitemOpen
  \bibfield  {author} {\bibinfo {author} {\bibnamefont {Fidkowski},
  \bibfnamefont {L.}}, \ and\ \bibinfo {author} {\bibfnamefont
  {A.}~\bibnamefont {Kitaev}}} (\bibinfo {year} {2010}),\ \href@noop {}
  {\bibfield  {journal} {\bibinfo  {journal} {Phys. Rev. B}\ }\textbf {\bibinfo
  {volume} {81}},\ \bibinfo {pages} {134509}}\BibitemShut {NoStop}%
\bibitem [{\citenamefont {Fidkowski}\ and\ \citenamefont
  {Kitaev}(2011)}]{Fidkowski2011}%
  \BibitemOpen
  \bibfield  {author} {\bibinfo {author} {\bibnamefont {Fidkowski},
  \bibfnamefont {L.}}, \ and\ \bibinfo {author} {\bibfnamefont
  {A.}~\bibnamefont {Kitaev}}} (\bibinfo {year} {2011}),\ \href@noop {}
  {\bibfield  {journal} {\bibinfo  {journal} {Phys. Rev. B}\ }\textbf {\bibinfo
  {volume} {83}},\ \bibinfo {pages} {075103}}\BibitemShut {NoStop}%
\bibitem [{\citenamefont {Finck}\ \emph {et~al.}(2013)\citenamefont {Finck},
  \citenamefont {Van~Harlingen}, \citenamefont {Mohseni}, \citenamefont
  {Jung},\ and\ \citenamefont {Li}}]{Finck_zero_bias}%
  \BibitemOpen
  \bibfield  {author} {\bibinfo {author} {\bibnamefont {Finck}, \bibfnamefont
  {A.~D.~K.}}, \bibinfo {author} {\bibfnamefont {D.~J.}\ \bibnamefont
  {Van~Harlingen}}, \bibinfo {author} {\bibfnamefont {P.~K.}\ \bibnamefont
  {Mohseni}}, \bibinfo {author} {\bibfnamefont {K.}~\bibnamefont {Jung}}, \
  and\ \bibinfo {author} {\bibfnamefont {X.}~\bibnamefont {Li}}} (\bibinfo
  {year} {2013}),\ \href {\doibase 10.1103/PhysRevLett.110.126406} {\bibfield
  {journal} {\bibinfo  {journal} {Phys. Rev. Lett.}\ }\textbf {\bibinfo
  {volume} {110}},\ \bibinfo {pages} {126406}}\BibitemShut {NoStop}%
\bibitem [{\citenamefont {Fischer}\ \emph {et~al.}(2014)\citenamefont
  {Fischer}, \citenamefont {Neupert}, \citenamefont {Platt}, \citenamefont
  {Schnyder}, \citenamefont {Hanke}, \citenamefont {Goryo}, \citenamefont
  {Thomale},\ and\ \citenamefont {Sigrist}}]{chiral_p_wave_fischer}%
  \BibitemOpen
  \bibfield  {author} {\bibinfo {author} {\bibnamefont {Fischer}, \bibfnamefont
  {M.~H.}}, \bibinfo {author} {\bibfnamefont {T.}~\bibnamefont {Neupert}},
  \bibinfo {author} {\bibfnamefont {C.}~\bibnamefont {Platt}}, \bibinfo
  {author} {\bibfnamefont {A.~P.}\ \bibnamefont {Schnyder}}, \bibinfo {author}
  {\bibfnamefont {W.}~\bibnamefont {Hanke}}, \bibinfo {author} {\bibfnamefont
  {J.}~\bibnamefont {Goryo}}, \bibinfo {author} {\bibfnamefont
  {R.}~\bibnamefont {Thomale}}, \ and\ \bibinfo {author} {\bibfnamefont
  {M.}~\bibnamefont {Sigrist}}} (\bibinfo {year} {2014}),\ \href {\doibase
  10.1103/PhysRevB.89.020509} {\bibfield  {journal} {\bibinfo  {journal} {Phys.
  Rev. B}\ }\textbf {\bibinfo {volume} {89}},\ \bibinfo {pages}
  {020509}}\BibitemShut {NoStop}%
\bibitem [{\citenamefont {Foster}\ and\ \citenamefont
  {Ludwig}(2008)}]{Foster2008}%
  \BibitemOpen
  \bibfield  {author} {\bibinfo {author} {\bibnamefont {Foster}, \bibfnamefont
  {M.~S.}}, \ and\ \bibinfo {author} {\bibfnamefont {A.~W.~W.}\ \bibnamefont
  {Ludwig}}} (\bibinfo {year} {2008}),\ \href {\doibase
  10.1103/PhysRevB.77.165108} {\bibfield  {journal} {\bibinfo  {journal} {Phys.
  Rev. B}\ }\textbf {\bibinfo {volume} {77}},\ \bibinfo {pages}
  {165108}}\BibitemShut {NoStop}%
\bibitem [{\citenamefont {{Foster}}\ \emph {et~al.}(2014)\citenamefont
  {{Foster}}, \citenamefont {{Xie}},\ and\ \citenamefont
  {{Chou}}}]{Foster2014PhRvB..89o5140F}%
  \BibitemOpen
  \bibfield  {author} {\bibinfo {author} {\bibnamefont {{Foster}},
  \bibfnamefont {M.~S.}}, \bibinfo {author} {\bibfnamefont {H.-Y.}\
  \bibnamefont {{Xie}}}, \ and\ \bibinfo {author} {\bibfnamefont {Y.-Z.}\
  \bibnamefont {{Chou}}}} (\bibinfo {year} {2014}),\ \href {\doibase
  10.1103/PhysRevB.89.155140} {\bibfield  {journal} {\bibinfo  {journal}
  {\prb}\ }\textbf {\bibinfo {volume} {89}}~(\bibinfo {number} {15}),\ \bibinfo
  {pages} {155140}},\ \Eprint {http://arxiv.org/abs/1403.6502} {1403.6502}
  \BibitemShut {NoStop}%
\bibitem [{\citenamefont {{Foster}}\ and\ \citenamefont
  {{Yuzbashyan}}(2012)}]{Foster2012PhRvL.109x6801F}%
  \BibitemOpen
  \bibfield  {author} {\bibinfo {author} {\bibnamefont {{Foster}},
  \bibfnamefont {M.~S.}}, \ and\ \bibinfo {author} {\bibfnamefont {E.~A.}\
  \bibnamefont {{Yuzbashyan}}}} (\bibinfo {year} {2012}),\ \href {\doibase
  10.1103/PhysRevLett.109.246801} {\bibfield  {journal} {\bibinfo  {journal}
  {Physical Review Letters}\ }\textbf {\bibinfo {volume} {109}}~(\bibinfo
  {number} {24}),\ \bibinfo {eid} {246801}}\BibitemShut {NoStop}%
\bibitem [{\citenamefont {Francesco}\ \emph {et~al.}(1997)\citenamefont
  {Francesco}, \citenamefont {Mathieu},\ and\ \citenamefont
  {Senechal}}]{FMS-CFT}%
  \BibitemOpen
  \bibfield  {author} {\bibinfo {author} {\bibnamefont {Francesco},
  \bibfnamefont {P.~D.}}, \bibinfo {author} {\bibfnamefont {P.}~\bibnamefont
  {Mathieu}}, \ and\ \bibinfo {author} {\bibfnamefont {D.}~\bibnamefont
  {Senechal}}} (\bibinfo {year} {1997}),\ \href@noop {} {\emph {\bibinfo
  {title} {{Conformal Field Theory}}}}\ (\bibinfo  {publisher} {Springer-Verlag
  New York})\BibitemShut {NoStop}%
\bibitem [{\citenamefont {Franz}\ and\ \citenamefont
  {Molenkamp}(2013)}]{franzMohlenkampBook13}%
  \BibitemOpen
  \bibfield  {author} {\bibinfo {author} {\bibnamefont {Franz}, \bibfnamefont
  {M.}}, \ and\ \bibinfo {author} {\bibfnamefont {L.}~\bibnamefont
  {Molenkamp}}} (\bibinfo {year} {2013}),\ \href@noop {} {\emph {\bibinfo
  {title} {Topological Insulators}}},\ Contemporary Concepts of Condensed
  Matter Science\ (\bibinfo  {publisher} {Elsevier})\BibitemShut {NoStop}%
\bibitem [{\citenamefont {{Freed}}(2014)}]{Freed2014}%
  \BibitemOpen
  \bibfield  {author} {\bibinfo {author} {\bibnamefont {{Freed}}, \bibfnamefont
  {D.~S.}}} (\bibinfo {year} {2014}),\ \href@noop {} {\ }\Eprint
  {http://arxiv.org/abs/1406.7278} {arXiv:1406.7278} \BibitemShut {NoStop}%
\bibitem [{\citenamefont {Freed}\ and\ \citenamefont
  {Moore}(2013)}]{Freed:2013bv}%
  \BibitemOpen
  \bibfield  {author} {\bibinfo {author} {\bibnamefont {Freed}, \bibfnamefont
  {D.~S.}}, \ and\ \bibinfo {author} {\bibfnamefont {G.~W.}\ \bibnamefont
  {Moore}}} (\bibinfo {year} {2013}),\ \href {\doibase
  10.1007/s00023-013-0236-x} {\emph {\bibinfo {title} {Annales Henri
  Poincar{\'e}}}},\ Vol.~\bibinfo {volume} {14}\ (\bibinfo  {publisher}
  {Springer Basel})\BibitemShut {NoStop}%
\bibitem [{\citenamefont {Freedman}\ \emph {et~al.}(2011)\citenamefont
  {Freedman}, \citenamefont {Hastings}, \citenamefont {Nayak}, \citenamefont
  {Qi}, \citenamefont {Walker},\ and\ \citenamefont {Wang}}]{Freedman2011}%
  \BibitemOpen
  \bibfield  {author} {\bibinfo {author} {\bibnamefont {Freedman},
  \bibfnamefont {M.}}, \bibinfo {author} {\bibfnamefont {M.~B.}\ \bibnamefont
  {Hastings}}, \bibinfo {author} {\bibfnamefont {C.}~\bibnamefont {Nayak}},
  \bibinfo {author} {\bibfnamefont {X.-L.}\ \bibnamefont {Qi}}, \bibinfo
  {author} {\bibfnamefont {K.}~\bibnamefont {Walker}}, \ and\ \bibinfo {author}
  {\bibfnamefont {Z.}~\bibnamefont {Wang}}} (\bibinfo {year} {2011}),\ \href
  {\doibase 10.1103/PhysRevB.83.115132} {\bibfield  {journal} {\bibinfo
  {journal} {Phys. Rev. B}\ }\textbf {\bibinfo {volume} {83}},\ \bibinfo
  {pages} {115132}}\BibitemShut {NoStop}%
\bibitem [{\citenamefont {Friedan}(1985)}]{Friedan1980}%
  \BibitemOpen
  \bibfield  {author} {\bibinfo {author} {\bibnamefont {Friedan}, \bibfnamefont
  {D.~H.}}} (\bibinfo {year} {1985}),\ \href {\doibase
  10.1016/0003-4916(85)90384-7} {\bibfield  {journal} {\bibinfo  {journal}
  {Annals Phys.}\ }\textbf {\bibinfo {volume} {163}},\ \bibinfo {pages}
  {318}}\BibitemShut {NoStop}%
\bibitem [{\citenamefont {Fruchart}\ and\ \citenamefont
  {Carpentier}(2013)}]{FruchartCarpentier2013}%
  \BibitemOpen
  \bibfield  {author} {\bibinfo {author} {\bibnamefont {Fruchart},
  \bibfnamefont {M.}}, \ and\ \bibinfo {author} {\bibfnamefont
  {D.}~\bibnamefont {Carpentier}}} (\bibinfo {year} {2013}),\ \href {\doibase
  http://dx.doi.org/10.1016/j.crhy.2013.09.013} {\bibfield  {journal} {\bibinfo
   {journal} {Comptes Rendus Physique}\ }\textbf {\bibinfo {volume}
  {14}}~(\bibinfo {number} {9--10}),\ \bibinfo {pages} {779 }}\BibitemShut
  {NoStop}%
\bibitem [{\citenamefont {Fu}(2011)}]{Fu_first_TCI}%
  \BibitemOpen
  \bibfield  {author} {\bibinfo {author} {\bibnamefont {Fu}, \bibfnamefont
  {L.}}} (\bibinfo {year} {2011}),\ \href@noop {} {\bibfield  {journal}
  {\bibinfo  {journal} {Phys. Rev. Lett.}\ }\textbf {\bibinfo {volume}
  {106}}~(\bibinfo {number} {10}),\ \bibinfo {pages} {106802}}\BibitemShut
  {NoStop}%
\bibitem [{\citenamefont {Fu}\ and\ \citenamefont {Berg}(2010)}]{FuBerg2010}%
  \BibitemOpen
  \bibfield  {author} {\bibinfo {author} {\bibnamefont {Fu}, \bibfnamefont
  {L.}}, \ and\ \bibinfo {author} {\bibfnamefont {E.}~\bibnamefont {Berg}}}
  (\bibinfo {year} {2010}),\ \href {\doibase 10.1103/PhysRevLett.105.097001}
  {\bibfield  {journal} {\bibinfo  {journal} {Phys. Rev. Lett.}\ }\textbf
  {\bibinfo {volume} {105}},\ \bibinfo {pages} {097001}}\BibitemShut {NoStop}%
\bibitem [{\citenamefont {Fu}\ and\ \citenamefont {Kane}(2006)}]{FuKane2006}%
  \BibitemOpen
  \bibfield  {author} {\bibinfo {author} {\bibnamefont {Fu}, \bibfnamefont
  {L.}}, \ and\ \bibinfo {author} {\bibfnamefont {C.}~\bibnamefont {Kane}}}
  (\bibinfo {year} {2006}),\ \href@noop {} {\bibfield  {journal} {\bibinfo
  {journal} {Phys. Rev. B}\ }\textbf {\bibinfo {volume} {74}},\ \bibinfo
  {pages} {195312}}\BibitemShut {NoStop}%
\bibitem [{\citenamefont {Fu}\ and\ \citenamefont {Kane}(2007)}]{Fu2007uq}%
  \BibitemOpen
  \bibfield  {author} {\bibinfo {author} {\bibnamefont {Fu}, \bibfnamefont
  {L.}}, \ and\ \bibinfo {author} {\bibfnamefont {C.~L.}\ \bibnamefont {Kane}}}
  (\bibinfo {year} {2007}),\ \href@noop {} {\bibfield  {journal} {\bibinfo
  {journal} {Phys. Rev. B}\ }\textbf {\bibinfo {volume} {76}},\ \bibinfo
  {pages} {045302}}\BibitemShut {NoStop}%
\bibitem [{\citenamefont {Fu}\ and\ \citenamefont
  {Kane}(2008)}]{FuKane_SC_STI}%
  \BibitemOpen
  \bibfield  {author} {\bibinfo {author} {\bibnamefont {Fu}, \bibfnamefont
  {L.}}, \ and\ \bibinfo {author} {\bibfnamefont {C.~L.}\ \bibnamefont {Kane}}}
  (\bibinfo {year} {2008}),\ \href {\doibase 10.1103/PhysRevLett.100.096407}
  {\bibfield  {journal} {\bibinfo  {journal} {Phys. Rev. Lett.}\ }\textbf
  {\bibinfo {volume} {100}},\ \bibinfo {pages} {096407}}\BibitemShut {NoStop}%
\bibitem [{\citenamefont {Fu}\ and\ \citenamefont
  {Kane}(2009)}]{FuKaneJosephsoncurrent09}%
  \BibitemOpen
  \bibfield  {author} {\bibinfo {author} {\bibnamefont {Fu}, \bibfnamefont
  {L.}}, \ and\ \bibinfo {author} {\bibfnamefont {C.~L.}\ \bibnamefont {Kane}}}
  (\bibinfo {year} {2009}),\ \href {\doibase 10.1103/PhysRevB.79.161408}
  {\bibfield  {journal} {\bibinfo  {journal} {Phys. Rev. B}\ }\textbf {\bibinfo
  {volume} {79}},\ \bibinfo {pages} {161408}}\BibitemShut {NoStop}%
\bibitem [{\citenamefont {Fu}\ and\ \citenamefont
  {Kane}(2012)}]{Fu:average_symmetry}%
  \BibitemOpen
  \bibfield  {author} {\bibinfo {author} {\bibnamefont {Fu}, \bibfnamefont
  {L.}}, \ and\ \bibinfo {author} {\bibfnamefont {C.~L.}\ \bibnamefont {Kane}}}
  (\bibinfo {year} {2012}),\ \href {\doibase 10.1103/PhysRevLett.109.246605}
  {\bibfield  {journal} {\bibinfo  {journal} {Phys. Rev. Lett.}\ }\textbf
  {\bibinfo {volume} {109}},\ \bibinfo {pages} {246605}}\BibitemShut {NoStop}%
\bibitem [{\citenamefont {Fu}\ \emph {et~al.}(2007)\citenamefont {Fu},
  \citenamefont {Kane},\ and\ \citenamefont {Mele}}]{Fu:2007fk}%
  \BibitemOpen
  \bibfield  {author} {\bibinfo {author} {\bibnamefont {Fu}, \bibfnamefont
  {L.}}, \bibinfo {author} {\bibfnamefont {C.~L.}\ \bibnamefont {Kane}}, \ and\
  \bibinfo {author} {\bibfnamefont {E.~J.}\ \bibnamefont {Mele}}} (\bibinfo
  {year} {2007}),\ \href@noop {} {\bibfield  {journal} {\bibinfo  {journal}
  {Phys. Rev. Lett.}\ }\textbf {\bibinfo {volume} {98}},\ \bibinfo {pages}
  {106803}}\BibitemShut {NoStop}%
\bibitem [{\citenamefont {Fukui}(2010)}]{Fukui10}%
  \BibitemOpen
  \bibfield  {author} {\bibinfo {author} {\bibnamefont {Fukui}, \bibfnamefont
  {T.}}} (\bibinfo {year} {2010}),\ \href {\doibase 10.1103/PhysRevB.81.214516}
  {\bibfield  {journal} {\bibinfo  {journal} {Phys. Rev. B}\ }\textbf {\bibinfo
  {volume} {81}},\ \bibinfo {pages} {214516}}\BibitemShut {NoStop}%
\bibitem [{\citenamefont {Fukui}\ and\ \citenamefont
  {Fujiwara}(2010)}]{FukuiFujiwara10}%
  \BibitemOpen
  \bibfield  {author} {\bibinfo {author} {\bibnamefont {Fukui}, \bibfnamefont
  {T.}}, \ and\ \bibinfo {author} {\bibfnamefont {T.}~\bibnamefont {Fujiwara}}}
  (\bibinfo {year} {2010}),\ \href {\doibase 10.1143/JPSJ.79.033701} {\bibfield
   {journal} {\bibinfo  {journal} {J. Phys. Soc. Jpn.}\ }\textbf {\bibinfo
  {volume} {79}}~(\bibinfo {number} {3}),\ \bibinfo {pages}
  {033701}}\BibitemShut {NoStop}%
\bibitem [{\citenamefont {Fulga}\ \emph {et~al.}(2012)\citenamefont {Fulga},
  \citenamefont {Hassler},\ and\ \citenamefont
  {Akhmerov}}]{classification_scattering}%
  \BibitemOpen
  \bibfield  {author} {\bibinfo {author} {\bibnamefont {Fulga}, \bibfnamefont
  {I.~C.}}, \bibinfo {author} {\bibfnamefont {F.}~\bibnamefont {Hassler}}, \
  and\ \bibinfo {author} {\bibfnamefont {A.~R.}\ \bibnamefont {Akhmerov}}}
  (\bibinfo {year} {2012}),\ \href {\doibase 10.1103/PhysRevB.85.165409}
  {\bibfield  {journal} {\bibinfo  {journal} {Phys. Rev. B}\ }\textbf {\bibinfo
  {volume} {85}},\ \bibinfo {pages} {165409}}\BibitemShut {NoStop}%
\bibitem [{\citenamefont {{Fulga}}\ \emph {et~al.}(2011)\citenamefont
  {{Fulga}}, \citenamefont {{Hassler}}, \citenamefont {{Akhmerov}},\ and\
  \citenamefont {{Beenakker}}}]{Fulga2011}%
  \BibitemOpen
  \bibfield  {author} {\bibinfo {author} {\bibnamefont {{Fulga}}, \bibfnamefont
  {I.~C.}}, \bibinfo {author} {\bibfnamefont {F.}~\bibnamefont {{Hassler}}},
  \bibinfo {author} {\bibfnamefont {A.~R.}\ \bibnamefont {{Akhmerov}}}, \ and\
  \bibinfo {author} {\bibfnamefont {C.~W.~J.}\ \bibnamefont {{Beenakker}}}}
  (\bibinfo {year} {2011}),\ \href {\doibase 10.1103/PhysRevB.83.155429}
  {\bibfield  {journal} {\bibinfo  {journal} {\prb}\ }\textbf {\bibinfo
  {volume} {83}}~(\bibinfo {number} {15}),\ \bibinfo {eid}
  {155429}}\BibitemShut {NoStop}%
\bibitem [{\citenamefont {Fulga}\ \emph {et~al.}(2014)\citenamefont {Fulga},
  \citenamefont {van Heck}, \citenamefont {Edge},\ and\ \citenamefont
  {Akhmerov}}]{Fulga_STI}%
  \BibitemOpen
  \bibfield  {author} {\bibinfo {author} {\bibnamefont {Fulga}, \bibfnamefont
  {I.~C.}}, \bibinfo {author} {\bibfnamefont {B.}~\bibnamefont {van Heck}},
  \bibinfo {author} {\bibfnamefont {J.~M.}\ \bibnamefont {Edge}}, \ and\
  \bibinfo {author} {\bibfnamefont {A.~R.}\ \bibnamefont {Akhmerov}}} (\bibinfo
  {year} {2014}),\ \href {\doibase 10.1103/PhysRevB.89.155424} {\bibfield
  {journal} {\bibinfo  {journal} {Phys. Rev. B}\ }\textbf {\bibinfo {volume}
  {89}},\ \bibinfo {pages} {155424}}\BibitemShut {NoStop}%
\bibitem [{\citenamefont {Furusaki}\ \emph {et~al.}(2013)\citenamefont
  {Furusaki}, \citenamefont {Nagaosa}, \citenamefont {Nomura}, \citenamefont
  {Ryu},\ and\ \citenamefont {Takayanagi}}]{Furusaki:responses}%
  \BibitemOpen
  \bibfield  {author} {\bibinfo {author} {\bibnamefont {Furusaki},
  \bibfnamefont {A.}}, \bibinfo {author} {\bibfnamefont {N.}~\bibnamefont
  {Nagaosa}}, \bibinfo {author} {\bibfnamefont {K.}~\bibnamefont {Nomura}},
  \bibinfo {author} {\bibfnamefont {S.}~\bibnamefont {Ryu}}, \ and\ \bibinfo
  {author} {\bibfnamefont {T.}~\bibnamefont {Takayanagi}}} (\bibinfo {year}
  {2013}),\ \href {\doibase http://dx.doi.org/10.1016/j.crhy.2013.03.002}
  {\bibfield  {journal} {\bibinfo  {journal} {Comptes Rendus Physique}\
  }\textbf {\bibinfo {volume} {14}},\ \bibinfo {pages} {871}}\BibitemShut
  {NoStop}%
\bibitem [{\citenamefont {Gade}(1993)}]{Gade:1993np}%
  \BibitemOpen
  \bibfield  {author} {\bibinfo {author} {\bibnamefont {Gade}, \bibfnamefont
  {R.}}} (\bibinfo {year} {1993}),\ \href {\doibase
  http://dx.doi.org/10.1016/0550-3213(93)90601-K} {\bibfield  {journal}
  {\bibinfo  {journal} {Nuclear Physics B}\ }\textbf {\bibinfo {volume}
  {398}}~(\bibinfo {number} {3}),\ \bibinfo {pages} {499 }}\BibitemShut
  {NoStop}%
\bibitem [{\citenamefont {Gade}\ and\ \citenamefont
  {Wegner}(1991)}]{GadeWegner1991}%
  \BibitemOpen
  \bibfield  {author} {\bibinfo {author} {\bibnamefont {Gade}, \bibfnamefont
  {R.}}, \ and\ \bibinfo {author} {\bibfnamefont {F.}~\bibnamefont {Wegner}}}
  (\bibinfo {year} {1991}),\ \href {\doibase
  http://dx.doi.org/10.1016/0550-3213(91)90401-I} {\bibfield  {journal}
  {\bibinfo  {journal} {Nuclear Physics B}\ }\textbf {\bibinfo {volume}
  {360}}~(\bibinfo {number} {2--3}),\ \bibinfo {pages} {213 }}\BibitemShut
  {NoStop}%
\bibitem [{\citenamefont {Garate}\ and\ \citenamefont
  {Franz}(2010)}]{garate_franz_2010}%
  \BibitemOpen
  \bibfield  {author} {\bibinfo {author} {\bibnamefont {Garate}, \bibfnamefont
  {I.}}, \ and\ \bibinfo {author} {\bibfnamefont {M.}~\bibnamefont {Franz}}}
  (\bibinfo {year} {2010}),\ \href {\doibase 10.1103/PhysRevLett.104.146802}
  {\bibfield  {journal} {\bibinfo  {journal} {Phys. Rev. Lett.}\ }\textbf
  {\bibinfo {volume} {104}},\ \bibinfo {pages} {146802}}\BibitemShut {NoStop}%
\bibitem [{\citenamefont {de~Gennes}(1999)}]{deGennesbook}%
  \BibitemOpen
  \bibfield  {author} {\bibinfo {author} {\bibnamefont {de~Gennes},
  \bibfnamefont {P.~G.}}} (\bibinfo {year} {1999}),\ \href@noop {} {\emph
  {\bibinfo {title} {Superconductivity Of Metals And Alloys}}}\ (\bibinfo
  {publisher} {Westview Press})\BibitemShut {NoStop}%
\bibitem [{\citenamefont {Goldman}\ \emph {et~al.}(2014)\citenamefont
  {Goldman}, \citenamefont {Juzeli{\=u}nas}, \citenamefont {{\"O}hberg},\ and\
  \citenamefont {Spielman}}]{AMO_topo_review}%
  \BibitemOpen
  \bibfield  {author} {\bibinfo {author} {\bibnamefont {Goldman}, \bibfnamefont
  {N.}}, \bibinfo {author} {\bibfnamefont {G.}~\bibnamefont {Juzeli{\=u}nas}},
  \bibinfo {author} {\bibfnamefont {P.}~\bibnamefont {{\"O}hberg}}, \ and\
  \bibinfo {author} {\bibfnamefont {I.~B.}\ \bibnamefont {Spielman}}} (\bibinfo
  {year} {2014}),\ \href {http://stacks.iop.org/0034-4885/77/i=12/a=126401}
  {\bibfield  {journal} {\bibinfo  {journal} {Reports on Progress in Physics}\
  }\textbf {\bibinfo {volume} {77}}~(\bibinfo {number} {12}),\ \bibinfo {pages}
  {126401}}\BibitemShut {NoStop}%
\bibitem [{\citenamefont {Goldman}\ \emph {et~al.}(2010)\citenamefont
  {Goldman}, \citenamefont {Satija}, \citenamefont {Nikolic}, \citenamefont
  {Bermudez}, \citenamefont {Martin-Delgado}, \citenamefont {Lewenstein},\ and\
  \citenamefont {Spielman}}]{goldman_PRL12}%
  \BibitemOpen
  \bibfield  {author} {\bibinfo {author} {\bibnamefont {Goldman}, \bibfnamefont
  {N.}}, \bibinfo {author} {\bibfnamefont {I.}~\bibnamefont {Satija}}, \bibinfo
  {author} {\bibfnamefont {P.}~\bibnamefont {Nikolic}}, \bibinfo {author}
  {\bibfnamefont {A.}~\bibnamefont {Bermudez}}, \bibinfo {author}
  {\bibfnamefont {M.~A.}\ \bibnamefont {Martin-Delgado}}, \bibinfo {author}
  {\bibfnamefont {M.}~\bibnamefont {Lewenstein}}, \ and\ \bibinfo {author}
  {\bibfnamefont {I.~B.}\ \bibnamefont {Spielman}}} (\bibinfo {year} {2010}),\
  \href {\doibase 10.1103/PhysRevLett.105.255302} {\bibfield  {journal}
  {\bibinfo  {journal} {Phys. Rev. Lett.}\ }\textbf {\bibinfo {volume} {105}},\
  \bibinfo {pages} {255302}}\BibitemShut {NoStop}%
\bibitem [{\citenamefont {{Goswami}}\ and\ \citenamefont
  {{Balicas}}(2013)}]{goswami_arXiv13}%
  \BibitemOpen
  \bibfield  {author} {\bibinfo {author} {\bibnamefont {{Goswami}},
  \bibfnamefont {P.}}, \ and\ \bibinfo {author} {\bibfnamefont
  {L.}~\bibnamefont {{Balicas}}}} (\bibinfo {year} {2013}),\ \href@noop {} {\
  }\Eprint {http://arxiv.org/abs/1312.3632} {arXiv:1312.3632} \BibitemShut
  {NoStop}%
\bibitem [{\citenamefont {Goswami}\ and\ \citenamefont
  {Chakravarty}(2011)}]{PhysRevLett.107.196803}%
  \BibitemOpen
  \bibfield  {author} {\bibinfo {author} {\bibnamefont {Goswami}, \bibfnamefont
  {P.}}, \ and\ \bibinfo {author} {\bibfnamefont {S.}~\bibnamefont
  {Chakravarty}}} (\bibinfo {year} {2011}),\ \href {\doibase
  10.1103/PhysRevLett.107.196803} {\bibfield  {journal} {\bibinfo  {journal}
  {Phys. Rev. Lett.}\ }\textbf {\bibinfo {volume} {107}},\ \bibinfo {pages}
  {196803}}\BibitemShut {NoStop}%
\bibitem [{\citenamefont {{Gottesman}}\ and\ \citenamefont
  {{Hastings}}(2010)}]{Gottesman2010}%
  \BibitemOpen
  \bibfield  {author} {\bibinfo {author} {\bibnamefont {{Gottesman}},
  \bibfnamefont {D.}}, \ and\ \bibinfo {author} {\bibfnamefont {M.~B.}\
  \bibnamefont {{Hastings}}}} (\bibinfo {year} {2010}),\ \href {\doibase
  10.1088/1367-2630/12/2/025002} {\bibfield  {journal} {\bibinfo  {journal}
  {New Journal of Physics}\ }\textbf {\bibinfo {volume} {12}},\ \bibinfo {eid}
  {025002}}\BibitemShut {NoStop}%
\bibitem [{\citenamefont {Graf}\ and\ \citenamefont
  {Porta}(2013)}]{GrafMicheleMarcello2013}%
  \BibitemOpen
  \bibfield  {author} {\bibinfo {author} {\bibnamefont {Graf}, \bibfnamefont
  {G.~M.}}, \ and\ \bibinfo {author} {\bibfnamefont {M.}~\bibnamefont {Porta}}}
  (\bibinfo {year} {2013}),\ \href {\doibase 10.1007/s00220-013-1819-6}
  {\bibfield  {journal} {\bibinfo  {journal} {Communications in Mathematical
  Physics}\ }\textbf {\bibinfo {volume} {324}}~(\bibinfo {number} {3}),\
  \bibinfo {pages} {851}}\BibitemShut {NoStop}%
\bibitem [{\citenamefont {Groth}\ \emph {et~al.}(2009)\citenamefont {Groth},
  \citenamefont {Wimmer}, \citenamefont {Akhmerov}, \citenamefont
  {Tworzyd\l{}o},\ and\ \citenamefont
  {Beenakker}}]{top_anderson_insul_Beenakker_PRL09}%
  \BibitemOpen
  \bibfield  {author} {\bibinfo {author} {\bibnamefont {Groth}, \bibfnamefont
  {C.~W.}}, \bibinfo {author} {\bibfnamefont {M.}~\bibnamefont {Wimmer}},
  \bibinfo {author} {\bibfnamefont {A.~R.}\ \bibnamefont {Akhmerov}}, \bibinfo
  {author} {\bibfnamefont {J.}~\bibnamefont {Tworzyd\l{}o}}, \ and\ \bibinfo
  {author} {\bibfnamefont {C.~W.~J.}\ \bibnamefont {Beenakker}}} (\bibinfo
  {year} {2009}),\ \href {\doibase 10.1103/PhysRevLett.103.196805} {\bibfield
  {journal} {\bibinfo  {journal} {Phys. Rev. Lett.}\ }\textbf {\bibinfo
  {volume} {103}},\ \bibinfo {pages} {196805}}\BibitemShut {NoStop}%
\bibitem [{\citenamefont {{Gruzberg}}\ \emph {et~al.}(1999)\citenamefont
  {{Gruzberg}}, \citenamefont {{Ludwig}},\ and\ \citenamefont
  {{Read}}}]{Gruzberg1999PhRvL..82.4524G}%
  \BibitemOpen
  \bibfield  {author} {\bibinfo {author} {\bibnamefont {{Gruzberg}},
  \bibfnamefont {I.~A.}}, \bibinfo {author} {\bibfnamefont {A.~W.~W.}\
  \bibnamefont {{Ludwig}}}, \ and\ \bibinfo {author} {\bibfnamefont
  {N.}~\bibnamefont {{Read}}}} (\bibinfo {year} {1999}),\ \href {\doibase
  10.1103/PhysRevLett.82.4524} {\bibfield  {journal} {\bibinfo  {journal}
  {Physical Review Letters}\ }\textbf {\bibinfo {volume} {82}},\ \bibinfo
  {pages} {4524}}\BibitemShut {NoStop}%
\bibitem [{\citenamefont {{Gruzberg}}\ \emph {et~al.}(2005)\citenamefont
  {{Gruzberg}}, \citenamefont {{Read}},\ and\ \citenamefont
  {{Vishveshwara}}}]{Gruzberg2005}%
  \BibitemOpen
  \bibfield  {author} {\bibinfo {author} {\bibnamefont {{Gruzberg}},
  \bibfnamefont {I.~A.}}, \bibinfo {author} {\bibfnamefont {N.}~\bibnamefont
  {{Read}}}, \ and\ \bibinfo {author} {\bibfnamefont {S.}~\bibnamefont
  {{Vishveshwara}}}} (\bibinfo {year} {2005}),\ \href {\doibase
  10.1103/PhysRevB.71.245124} {\bibfield  {journal} {\bibinfo  {journal}
  {\prb}\ }\textbf {\bibinfo {volume} {71}}~(\bibinfo {number} {24}),\ \bibinfo
  {eid} {245124}}\BibitemShut {NoStop}%
\bibitem [{\citenamefont {{Gu}}\ and\ \citenamefont {{Levin}}(2014)}]{Gu2014}%
  \BibitemOpen
  \bibfield  {author} {\bibinfo {author} {\bibnamefont {{Gu}}, \bibfnamefont
  {Z.-C.}}, \ and\ \bibinfo {author} {\bibfnamefont {M.}~\bibnamefont
  {{Levin}}}} (\bibinfo {year} {2014}),\ \href {\doibase
  10.1103/PhysRevB.89.201113} {\bibfield  {journal} {\bibinfo  {journal}
  {\prb}\ }\textbf {\bibinfo {volume} {89}},\ \bibinfo {eid}
  {201113}}\BibitemShut {NoStop}%
\bibitem [{\citenamefont {{Gu}}\ \emph {et~al.}(2015)\citenamefont {{Gu}},
  \citenamefont {{Wang}},\ and\ \citenamefont {{Wen}}}]{SPT_invariant_Juven}%
  \BibitemOpen
  \bibfield  {author} {\bibinfo {author} {\bibnamefont {{Gu}}, \bibfnamefont
  {Z.-C.}}, \bibinfo {author} {\bibfnamefont {J.~C.}\ \bibnamefont {{Wang}}}, \
  and\ \bibinfo {author} {\bibfnamefont {X.-G.}\ \bibnamefont {{Wen}}}}
  (\bibinfo {year} {2015}),\ \href@noop {} {\bibfield  {journal} {\bibinfo
  {journal} {ArXiv e-prints}\ }}\Eprint {http://arxiv.org/abs/1503.01768}
  {arXiv:1503.01768} \BibitemShut {NoStop}%
\bibitem [{\citenamefont {{Gu}}\ and\ \citenamefont {{Wen}}(2009)}]{GuWen2009}%
  \BibitemOpen
  \bibfield  {author} {\bibinfo {author} {\bibnamefont {{Gu}}, \bibfnamefont
  {Z.-C.}}, \ and\ \bibinfo {author} {\bibfnamefont {X.-G.}\ \bibnamefont
  {{Wen}}}} (\bibinfo {year} {2009}),\ \href {\doibase
  10.1103/PhysRevB.80.155131} {\bibfield  {journal} {\bibinfo  {journal}
  {\prb}\ }\textbf {\bibinfo {volume} {80}}~(\bibinfo {number} {15}),\ \bibinfo
  {eid} {155131}}\BibitemShut {NoStop}%
\bibitem [{\citenamefont {Gu}\ and\ \citenamefont {Wen}(2014)}]{Gu2012}%
  \BibitemOpen
  \bibfield  {author} {\bibinfo {author} {\bibnamefont {Gu}, \bibfnamefont
  {Z.-C.}}, \ and\ \bibinfo {author} {\bibfnamefont {X.-G.}\ \bibnamefont
  {Wen}}} (\bibinfo {year} {2014}),\ \href {\doibase
  10.1103/PhysRevB.90.115141} {\bibfield  {journal} {\bibinfo  {journal} {Phys.
  Rev. B}\ }\textbf {\bibinfo {volume} {90}},\ \bibinfo {pages}
  {115141}}\BibitemShut {NoStop}%
\bibitem [{\citenamefont {Guo}\ \emph {et~al.}(2010)\citenamefont {Guo},
  \citenamefont {Rosenberg}, \citenamefont {Refael},\ and\ \citenamefont
  {Franz}}]{GuoPhysRevLett.105.216601}%
  \BibitemOpen
  \bibfield  {author} {\bibinfo {author} {\bibnamefont {Guo}, \bibfnamefont
  {H.-M.}}, \bibinfo {author} {\bibfnamefont {G.}~\bibnamefont {Rosenberg}},
  \bibinfo {author} {\bibfnamefont {G.}~\bibnamefont {Refael}}, \ and\ \bibinfo
  {author} {\bibfnamefont {M.}~\bibnamefont {Franz}}} (\bibinfo {year}
  {2010}),\ \href {\doibase 10.1103/PhysRevLett.105.216601} {\bibfield
  {journal} {\bibinfo  {journal} {Phys. Rev. Lett.}\ }\textbf {\bibinfo
  {volume} {105}},\ \bibinfo {pages} {216601}}\BibitemShut {NoStop}%
\bibitem [{\citenamefont {{Gurarie}}(2011)}]{Gurarie2011}%
  \BibitemOpen
  \bibfield  {author} {\bibinfo {author} {\bibnamefont {{Gurarie}},
  \bibfnamefont {V.}}} (\bibinfo {year} {2011}),\ \href {\doibase
  10.1103/PhysRevB.83.085426} {\bibfield  {journal} {\bibinfo  {journal}
  {\prb}\ }\textbf {\bibinfo {volume} {83}}~(\bibinfo {number} {8}),\ \bibinfo
  {eid} {085426}}\BibitemShut {NoStop}%
\bibitem [{\citenamefont {{Gurarie}}\ and\ \citenamefont
  {{Chalker}}(2002)}]{GurarieChalker2002}%
  \BibitemOpen
  \bibfield  {author} {\bibinfo {author} {\bibnamefont {{Gurarie}},
  \bibfnamefont {V.}}, \ and\ \bibinfo {author} {\bibfnamefont {J.~T.}\
  \bibnamefont {{Chalker}}}} (\bibinfo {year} {2002}),\ \href {\doibase
  10.1103/PhysRevLett.89.136801} {\bibfield  {journal} {\bibinfo  {journal}
  {Phys. Rev. Lett.}\ }\textbf {\bibinfo {volume} {89}}~(\bibinfo {number}
  {13}),\ \bibinfo {eid} {136801}}\BibitemShut {NoStop}%
\bibitem [{\citenamefont {{Gurarie}}\ and\ \citenamefont
  {{Chalker}}(2003)}]{GurarieChalker2003}%
  \BibitemOpen
  \bibfield  {author} {\bibinfo {author} {\bibnamefont {{Gurarie}},
  \bibfnamefont {V.}}, \ and\ \bibinfo {author} {\bibfnamefont {J.~T.}\
  \bibnamefont {{Chalker}}}} (\bibinfo {year} {2003}),\ \href {\doibase
  10.1103/PhysRevB.68.134207} {\bibfield  {journal} {\bibinfo  {journal}
  {\prb}\ }\textbf {\bibinfo {volume} {68}}~(\bibinfo {number} {13}),\ \bibinfo
  {eid} {134207}}\BibitemShut {NoStop}%
\bibitem [{\citenamefont {Gurarie}\ and\ \citenamefont
  {Radzihovsky}(2007)}]{GurarieRadzihovsky07}%
  \BibitemOpen
  \bibfield  {author} {\bibinfo {author} {\bibnamefont {Gurarie}, \bibfnamefont
  {V.}}, \ and\ \bibinfo {author} {\bibfnamefont {L.}~\bibnamefont
  {Radzihovsky}}} (\bibinfo {year} {2007}),\ \href {\doibase
  10.1103/PhysRevB.75.212509} {\bibfield  {journal} {\bibinfo  {journal} {Phys.
  Rev. B}\ }\textbf {\bibinfo {volume} {75}},\ \bibinfo {pages}
  {212509}}\BibitemShut {NoStop}%
\bibitem [{\citenamefont {Haldane}(1983{\natexlab{a}})}]{Haldane1983a}%
  \BibitemOpen
  \bibfield  {author} {\bibinfo {author} {\bibnamefont {Haldane}, \bibfnamefont
  {F.}}} (\bibinfo {year} {1983}{\natexlab{a}}),\ \href {\doibase
  http://dx.doi.org/10.1016/0375-9601(83)90631-X} {\bibfield  {journal}
  {\bibinfo  {journal} {Physics Letters A}\ }\textbf {\bibinfo {volume}
  {93}}~(\bibinfo {number} {9}),\ \bibinfo {pages} {464 }}\BibitemShut
  {NoStop}%
\bibitem [{\citenamefont {Haldane}(1983{\natexlab{b}})}]{Haldane1983b}%
  \BibitemOpen
  \bibfield  {author} {\bibinfo {author} {\bibnamefont {Haldane}, \bibfnamefont
  {F.~D.~M.}}} (\bibinfo {year} {1983}{\natexlab{b}}),\ \href {\doibase
  10.1103/PhysRevLett.50.1153} {\bibfield  {journal} {\bibinfo  {journal}
  {Phys. Rev. Lett.}\ }\textbf {\bibinfo {volume} {50}},\ \bibinfo {pages}
  {1153}}\BibitemShut {NoStop}%
\bibitem [{\citenamefont {Haldane}(1988)}]{Haldane1988}%
  \BibitemOpen
  \bibfield  {author} {\bibinfo {author} {\bibnamefont {Haldane}, \bibfnamefont
  {F.~D.~M.}}} (\bibinfo {year} {1988}),\ \href@noop {} {\bibfield  {journal}
  {\bibinfo  {journal} {Phys. Rev. Lett.}\ }\textbf {\bibinfo {volume} {61}},\
  \bibinfo {pages} {2015}}\BibitemShut {NoStop}%
\bibitem [{\citenamefont {Halperin}(1982)}]{Halperin82}%
  \BibitemOpen
  \bibfield  {author} {\bibinfo {author} {\bibnamefont {Halperin},
  \bibfnamefont {B.~I.}}} (\bibinfo {year} {1982}),\ \href {\doibase
  10.1103/PhysRevB.25.2185} {\bibfield  {journal} {\bibinfo  {journal} {Phys.
  Rev. B}\ }\textbf {\bibinfo {volume} {25}},\ \bibinfo {pages}
  {2185}}\BibitemShut {NoStop}%
\bibitem [{\citenamefont {Hasan}\ and\ \citenamefont {Kane}(2010)}]{hasan:rmp}%
  \BibitemOpen
  \bibfield  {author} {\bibinfo {author} {\bibnamefont {Hasan}, \bibfnamefont
  {M.~Z.}}, \ and\ \bibinfo {author} {\bibfnamefont {C.~L.}\ \bibnamefont
  {Kane}}} (\bibinfo {year} {2010}),\ \href@noop {} {\bibfield  {journal}
  {\bibinfo  {journal} {Rev. Mod. Phys.}\ }\textbf {\bibinfo {volume} {82}},\
  \bibinfo {pages} {3045}}\BibitemShut {NoStop}%
\bibitem [{\citenamefont {Hasan}\ and\ \citenamefont
  {Moore}(2011)}]{HasanMoore2011}%
  \BibitemOpen
  \bibfield  {author} {\bibinfo {author} {\bibnamefont {Hasan}, \bibfnamefont
  {M.~Z.}}, \ and\ \bibinfo {author} {\bibfnamefont {J.~E.}\ \bibnamefont
  {Moore}}} (\bibinfo {year} {2011}),\ \href@noop {} {\bibfield  {journal}
  {\bibinfo  {journal} {Annu. Rev. Condens. Matter Phys.}\ }\textbf {\bibinfo
  {volume} {2}},\ \bibinfo {pages} {55}}\BibitemShut {NoStop}%
\bibitem [{\citenamefont {Hasan}\ \emph {et~al.}(2015)\citenamefont {Hasan},
  \citenamefont {Xu},\ and\ \citenamefont {Bian}}]{Bian_review_TSC_TI}%
  \BibitemOpen
  \bibfield  {author} {\bibinfo {author} {\bibnamefont {Hasan}, \bibfnamefont
  {M.~Z.}}, \bibinfo {author} {\bibfnamefont {S.-Y.}\ \bibnamefont {Xu}}, \
  and\ \bibinfo {author} {\bibfnamefont {G.}~\bibnamefont {Bian}}} (\bibinfo
  {year} {2015}),\ \href {http://stacks.iop.org/1402-4896/2015/i=T164/a=014001}
  {\bibfield  {journal} {\bibinfo  {journal} {Physica Scripta}\ }\textbf
  {\bibinfo {volume} {2015}}~(\bibinfo {number} {T164}),\ \bibinfo {pages}
  {014001}}\BibitemShut {NoStop}%
\bibitem [{\citenamefont {{Hastings}}(2007)}]{Hastings2007}%
  \BibitemOpen
  \bibfield  {author} {\bibinfo {author} {\bibnamefont {{Hastings}},
  \bibfnamefont {M.~B.}}} (\bibinfo {year} {2007}),\ \href {\doibase
  10.1088/1742-5468/2007/08/P08024} {\bibfield  {journal} {\bibinfo  {journal}
  {J. Stat. Mech. Theor. Exp.}\ }\textbf {\bibinfo {volume} {8}},\ \bibinfo
  {pages} {24}}\BibitemShut {NoStop}%
\bibitem [{\citenamefont {Hastings}\ and\ \citenamefont
  {Loring}(2011)}]{Hastings_annals_physics11}%
  \BibitemOpen
  \bibfield  {author} {\bibinfo {author} {\bibnamefont {Hastings},
  \bibfnamefont {M.~B.}}, \ and\ \bibinfo {author} {\bibfnamefont {T.~A.}\
  \bibnamefont {Loring}}} (\bibinfo {year} {2011}),\ \href {\doibase
  http://dx.doi.org/10.1016/j.aop.2010.12.013} {\bibfield  {journal} {\bibinfo
  {journal} {Annals of Physics}\ }\textbf {\bibinfo {volume} {326}}~(\bibinfo
  {number} {7}),\ \bibinfo {pages} {1699 }},\ \bibinfo {note} {july 2011
  Special Issue}\BibitemShut {NoStop}%
\bibitem [{\citenamefont {Hatcher}(2001)}]{Hatcherbook}%
  \BibitemOpen
  \bibfield  {author} {\bibinfo {author} {\bibnamefont {Hatcher}, \bibfnamefont
  {A.}}} (\bibinfo {year} {2001}),\ \href
  {http://www.math.cornell.edu/~hatcher/} {\emph {\bibinfo {title} {Algebraic
  Topology}}}\ (\bibinfo  {publisher} {Cambridge University Press})\BibitemShut
  {NoStop}%
\bibitem [{\citenamefont {Hatsugai}(1993)}]{Hatsugai93}%
  \BibitemOpen
  \bibfield  {author} {\bibinfo {author} {\bibnamefont {Hatsugai},
  \bibfnamefont {Y.}}} (\bibinfo {year} {1993}),\ \href {\doibase
  10.1103/PhysRevLett.71.3697} {\bibfield  {journal} {\bibinfo  {journal}
  {Phys. Rev. Lett.}\ }\textbf {\bibinfo {volume} {71}},\ \bibinfo {pages}
  {3697}}\BibitemShut {NoStop}%
\bibitem [{\citenamefont {Heeger}\ \emph {et~al.}(1988)\citenamefont {Heeger},
  \citenamefont {Kivelson}, \citenamefont {Schrieffer},\ and\ \citenamefont
  {Su}}]{heegerRMP88}%
  \BibitemOpen
  \bibfield  {author} {\bibinfo {author} {\bibnamefont {Heeger}, \bibfnamefont
  {A.~J.}}, \bibinfo {author} {\bibfnamefont {S.}~\bibnamefont {Kivelson}},
  \bibinfo {author} {\bibfnamefont {J.~R.}\ \bibnamefont {Schrieffer}}, \ and\
  \bibinfo {author} {\bibfnamefont {W.~P.}\ \bibnamefont {Su}}} (\bibinfo
  {year} {1988}),\ \href {\doibase 10.1103/RevModPhys.60.781} {\bibfield
  {journal} {\bibinfo  {journal} {Rev. Mod. Phys.}\ }\textbf {\bibinfo {volume}
  {60}},\ \bibinfo {pages} {781}}\BibitemShut {NoStop}%
\bibitem [{\citenamefont {Helgason}(1978)}]{Helgason_book}%
  \BibitemOpen
  \bibfield  {author} {\bibinfo {author} {\bibnamefont {Helgason},
  \bibfnamefont {S.}}} (\bibinfo {year} {1978}),\ \href@noop {} {\emph
  {\bibinfo {title} {Differential geometry, Lie groups, and symmetric
  spaces}}}\ (\bibinfo  {publisher} {American Mathematical Society, Providence,
  RI})\BibitemShut {NoStop}%
\bibitem [{\citenamefont {Hor}\ \emph {et~al.}(2010)\citenamefont {Hor},
  \citenamefont {Williams}, \citenamefont {Checkelsky}, \citenamefont
  {Roushan}, \citenamefont {Seo}, \citenamefont {Xu}, \citenamefont
  {Zandbergen}, \citenamefont {Yazdani}, \citenamefont {Ong},\ and\
  \citenamefont {Cava}}]{HorCava2010}%
  \BibitemOpen
  \bibfield  {author} {\bibinfo {author} {\bibnamefont {Hor}, \bibfnamefont
  {Y.~S.}}, \bibinfo {author} {\bibfnamefont {A.~J.}\ \bibnamefont {Williams}},
  \bibinfo {author} {\bibfnamefont {J.~G.}\ \bibnamefont {Checkelsky}},
  \bibinfo {author} {\bibfnamefont {P.}~\bibnamefont {Roushan}}, \bibinfo
  {author} {\bibfnamefont {J.}~\bibnamefont {Seo}}, \bibinfo {author}
  {\bibfnamefont {Q.}~\bibnamefont {Xu}}, \bibinfo {author} {\bibfnamefont
  {H.~W.}\ \bibnamefont {Zandbergen}}, \bibinfo {author} {\bibfnamefont
  {A.}~\bibnamefont {Yazdani}}, \bibinfo {author} {\bibfnamefont {N.~P.}\
  \bibnamefont {Ong}}, \ and\ \bibinfo {author} {\bibfnamefont {R.~J.}\
  \bibnamefont {Cava}}} (\bibinfo {year} {2010}),\ \href {\doibase
  10.1103/PhysRevLett.104.057001} {\bibfield  {journal} {\bibinfo  {journal}
  {Phys. Rev. Lett.}\ }\textbf {\bibinfo {volume} {104}},\ \bibinfo {pages}
  {057001}}\BibitemShut {NoStop}%
\bibitem [{\citenamefont {Hosur}\ \emph {et~al.}(2011)\citenamefont {Hosur},
  \citenamefont {Ghaemi}, \citenamefont {Mong},\ and\ \citenamefont
  {Vishwanath}}]{Hosur:2011uq}%
  \BibitemOpen
  \bibfield  {author} {\bibinfo {author} {\bibnamefont {Hosur}, \bibfnamefont
  {P.}}, \bibinfo {author} {\bibfnamefont {P.}~\bibnamefont {Ghaemi}}, \bibinfo
  {author} {\bibfnamefont {R.~S.~K.}\ \bibnamefont {Mong}}, \ and\ \bibinfo
  {author} {\bibfnamefont {A.}~\bibnamefont {Vishwanath}}} (\bibinfo {year}
  {2011}),\ \href {http://link.aps.org/doi/10.1103/PhysRevLett.107.097001}
  {\bibfield  {journal} {\bibinfo  {journal} {Phys. Rev. Lett.}\ }\textbf
  {\bibinfo {volume} {107}}~(\bibinfo {number} {9}),\ \bibinfo {pages}
  {097001}}\BibitemShut {NoStop}%
\bibitem [{\citenamefont {Hosur}\ and\ \citenamefont
  {Qi}(2013)}]{Hosur_Weyl_develop}%
  \BibitemOpen
  \bibfield  {author} {\bibinfo {author} {\bibnamefont {Hosur}, \bibfnamefont
  {P.}}, \ and\ \bibinfo {author} {\bibfnamefont {X.}~\bibnamefont {Qi}}}
  (\bibinfo {year} {2013}),\ \href {\doibase
  http://dx.doi.org/10.1016/j.crhy.2013.10.010} {\bibfield  {journal} {\bibinfo
   {journal} {Comptes Rendus Physique}\ }\textbf {\bibinfo {volume}
  {14}}~(\bibinfo {number} {9--10}),\ \bibinfo {pages} {857 }}\BibitemShut
  {NoStop}%
\bibitem [{\citenamefont {Hosur}\ \emph {et~al.}(2010)\citenamefont {Hosur},
  \citenamefont {Ryu},\ and\ \citenamefont {Vishwanath}}]{Ryu_CTI_SC}%
  \BibitemOpen
  \bibfield  {author} {\bibinfo {author} {\bibnamefont {Hosur}, \bibfnamefont
  {P.}}, \bibinfo {author} {\bibfnamefont {S.}~\bibnamefont {Ryu}}, \ and\
  \bibinfo {author} {\bibfnamefont {A.}~\bibnamefont {Vishwanath}}} (\bibinfo
  {year} {2010}),\ \href {\doibase 10.1103/PhysRevB.81.045120} {\bibfield
  {journal} {\bibinfo  {journal} {Phys. Rev. B}\ }\textbf {\bibinfo {volume}
  {81}},\ \bibinfo {pages} {045120}}\BibitemShut {NoStop}%
\bibitem [{\citenamefont {Hsieh}\ \emph {et~al.}(2016)\citenamefont {Hsieh},
  \citenamefont {Cho},\ and\ \citenamefont {Ryu}}]{HsiehChoRyu2015}%
  \BibitemOpen
  \bibfield  {author} {\bibinfo {author} {\bibnamefont {Hsieh}, \bibfnamefont
  {C.-T.}}, \bibinfo {author} {\bibfnamefont {G.~Y.}\ \bibnamefont {Cho}}, \
  and\ \bibinfo {author} {\bibfnamefont {S.}~\bibnamefont {Ryu}}} (\bibinfo
  {year} {2016}),\ \href {\doibase 10.1103/PhysRevB.93.075135} {\bibfield
  {journal} {\bibinfo  {journal} {Phys. Rev. B}\ }\textbf {\bibinfo {volume}
  {93}},\ \bibinfo {pages} {075135}}\BibitemShut {NoStop}%
\bibitem [{\citenamefont {Hsieh}\ \emph
  {et~al.}(2014{\natexlab{a}})\citenamefont {Hsieh}, \citenamefont {Morimoto},\
  and\ \citenamefont {Ryu}}]{Hsieh2014b}%
  \BibitemOpen
  \bibfield  {author} {\bibinfo {author} {\bibnamefont {Hsieh}, \bibfnamefont
  {C.-T.}}, \bibinfo {author} {\bibfnamefont {T.}~\bibnamefont {Morimoto}}, \
  and\ \bibinfo {author} {\bibfnamefont {S.}~\bibnamefont {Ryu}}} (\bibinfo
  {year} {2014}{\natexlab{a}}),\ \href {\doibase 10.1103/PhysRevB.90.245111}
  {\bibfield  {journal} {\bibinfo  {journal} {Phys. Rev. B}\ }\textbf {\bibinfo
  {volume} {90}},\ \bibinfo {pages} {245111}}\BibitemShut {NoStop}%
\bibitem [{\citenamefont {Hsieh}\ \emph
  {et~al.}(2014{\natexlab{b}})\citenamefont {Hsieh}, \citenamefont {Sule},
  \citenamefont {Cho}, \citenamefont {Ryu},\ and\ \citenamefont
  {Leigh}}]{Hsieh2014a}%
  \BibitemOpen
  \bibfield  {author} {\bibinfo {author} {\bibnamefont {Hsieh}, \bibfnamefont
  {C.-T.}}, \bibinfo {author} {\bibfnamefont {O.~M.}\ \bibnamefont {Sule}},
  \bibinfo {author} {\bibfnamefont {G.~Y.}\ \bibnamefont {Cho}}, \bibinfo
  {author} {\bibfnamefont {S.}~\bibnamefont {Ryu}}, \ and\ \bibinfo {author}
  {\bibfnamefont {R.~G.}\ \bibnamefont {Leigh}}} (\bibinfo {year}
  {2014}{\natexlab{b}}),\ \href {\doibase 10.1103/PhysRevB.90.165134}
  {\bibfield  {journal} {\bibinfo  {journal} {Phys. Rev. B}\ }\textbf {\bibinfo
  {volume} {90}},\ \bibinfo {pages} {165134}}\BibitemShut {NoStop}%
\bibitem [{\citenamefont {Hsieh}\ \emph {et~al.}(2008)\citenamefont {Hsieh},
  \citenamefont {Qian}, \citenamefont {Wray}, \citenamefont {Xia},
  \citenamefont {Hor}, \citenamefont {Cava},\ and\ \citenamefont
  {Hasan}}]{Hsieh:2008fk}%
  \BibitemOpen
  \bibfield  {author} {\bibinfo {author} {\bibnamefont {Hsieh}, \bibfnamefont
  {D.}}, \bibinfo {author} {\bibfnamefont {D.}~\bibnamefont {Qian}}, \bibinfo
  {author} {\bibfnamefont {L.}~\bibnamefont {Wray}}, \bibinfo {author}
  {\bibfnamefont {Y.}~\bibnamefont {Xia}}, \bibinfo {author} {\bibfnamefont
  {Y.~S.}\ \bibnamefont {Hor}}, \bibinfo {author} {\bibfnamefont {R.~J.}\
  \bibnamefont {Cava}}, \ and\ \bibinfo {author} {\bibfnamefont {M.~Z.}\
  \bibnamefont {Hasan}}} (\bibinfo {year} {2008}),\ \href@noop {} {\bibfield
  {journal} {\bibinfo  {journal} {Nature (London)}\ }\textbf {\bibinfo {volume}
  {452}},\ \bibinfo {pages} {970}}\BibitemShut {NoStop}%
\bibitem [{\citenamefont {Hsieh}\ \emph {et~al.}(2009)\citenamefont {Hsieh},
  \citenamefont {Xia}, \citenamefont {Qian}, \citenamefont {Wray},
  \citenamefont {Dil}, \citenamefont {Meier}, \citenamefont {Osterwalder},
  \citenamefont {Patthey}, \citenamefont {Checkelsky}, \citenamefont {Ong},
  \citenamefont {Fedorov}, \citenamefont {Lin}, \citenamefont {Bansil},
  \citenamefont {Grauer}, \citenamefont {Hor}, \citenamefont {Cava},\ and\
  \citenamefont {Hasan}}]{hsiehNature2009}%
  \BibitemOpen
  \bibfield  {author} {\bibinfo {author} {\bibnamefont {Hsieh}, \bibfnamefont
  {D.}}, \bibinfo {author} {\bibfnamefont {Y.}~\bibnamefont {Xia}}, \bibinfo
  {author} {\bibfnamefont {D.}~\bibnamefont {Qian}}, \bibinfo {author}
  {\bibfnamefont {L.}~\bibnamefont {Wray}}, \bibinfo {author} {\bibfnamefont
  {J.~H.}\ \bibnamefont {Dil}}, \bibinfo {author} {\bibfnamefont
  {F.}~\bibnamefont {Meier}}, \bibinfo {author} {\bibfnamefont
  {J.}~\bibnamefont {Osterwalder}}, \bibinfo {author} {\bibfnamefont
  {L.}~\bibnamefont {Patthey}}, \bibinfo {author} {\bibfnamefont {J.~G.}\
  \bibnamefont {Checkelsky}}, \bibinfo {author} {\bibfnamefont {N.~P.}\
  \bibnamefont {Ong}}, \bibinfo {author} {\bibfnamefont {A.~V.}\ \bibnamefont
  {Fedorov}}, \bibinfo {author} {\bibfnamefont {H.}~\bibnamefont {Lin}},
  \bibinfo {author} {\bibfnamefont {A.}~\bibnamefont {Bansil}}, \bibinfo
  {author} {\bibfnamefont {D.}~\bibnamefont {Grauer}}, \bibinfo {author}
  {\bibfnamefont {Y.~S.}\ \bibnamefont {Hor}}, \bibinfo {author} {\bibfnamefont
  {R.~J.}\ \bibnamefont {Cava}}, \ and\ \bibinfo {author} {\bibfnamefont
  {M.~Z.}\ \bibnamefont {Hasan}}} (\bibinfo {year} {2009}),\ \href@noop {}
  {\bibfield  {journal} {\bibinfo  {journal} {Nature}\ }\textbf {\bibinfo
  {volume} {460}}~(\bibinfo {number} {7259}),\ \bibinfo {pages}
  {1101}}\BibitemShut {NoStop}%
\bibitem [{\citenamefont {Hsieh}\ \emph {et~al.}(2012)\citenamefont {Hsieh},
  \citenamefont {Lin}, \citenamefont {Liu}, \citenamefont {Duan}, \citenamefont
  {Bansil},\ and\ \citenamefont {Fu}}]{Hsieh:2012fk}%
  \BibitemOpen
  \bibfield  {author} {\bibinfo {author} {\bibnamefont {Hsieh}, \bibfnamefont
  {T.~H.}}, \bibinfo {author} {\bibfnamefont {H.}~\bibnamefont {Lin}}, \bibinfo
  {author} {\bibfnamefont {J.}~\bibnamefont {Liu}}, \bibinfo {author}
  {\bibfnamefont {W.}~\bibnamefont {Duan}}, \bibinfo {author} {\bibfnamefont
  {A.}~\bibnamefont {Bansil}}, \ and\ \bibinfo {author} {\bibfnamefont
  {L.}~\bibnamefont {Fu}}} (\bibinfo {year} {2012}),\ \href
  {http://dx.doi.org/10.1038/ncomms1969} {\bibfield  {journal} {\bibinfo
  {journal} {Nat. Commun.}\ }\textbf {\bibinfo {volume} {3}},\ \bibinfo {pages}
  {982}}\BibitemShut {NoStop}%
\bibitem [{\citenamefont {Hsieh}\ \emph
  {et~al.}(2014{\natexlab{c}})\citenamefont {Hsieh}, \citenamefont {Liu},\ and\
  \citenamefont {Fu}}]{TCI_Fu_antiperovskites}%
  \BibitemOpen
  \bibfield  {author} {\bibinfo {author} {\bibnamefont {Hsieh}, \bibfnamefont
  {T.~H.}}, \bibinfo {author} {\bibfnamefont {J.}~\bibnamefont {Liu}}, \ and\
  \bibinfo {author} {\bibfnamefont {L.}~\bibnamefont {Fu}}} (\bibinfo {year}
  {2014}{\natexlab{c}}),\ \href {\doibase 10.1103/PhysRevB.90.081112}
  {\bibfield  {journal} {\bibinfo  {journal} {Phys. Rev. B}\ }\textbf {\bibinfo
  {volume} {90}},\ \bibinfo {pages} {081112}}\BibitemShut {NoStop}%
\bibitem [{\citenamefont {Huang}\ \emph
  {et~al.}(2015{\natexlab{a}})\citenamefont {Huang}, \citenamefont {Xu},
  \citenamefont {Belopolski}, \citenamefont {Lee}, \citenamefont {Chang},
  \citenamefont {Wang}, \citenamefont {Alidoust}, \citenamefont {Bian},
  \citenamefont {Neupane}, \citenamefont {Zhang}, \citenamefont {Jia},
  \citenamefont {Bansil}, \citenamefont {Lin},\ and\ \citenamefont
  {Hasan}}]{Huang_Hasan_Weyl}%
  \BibitemOpen
  \bibfield  {author} {\bibinfo {author} {\bibnamefont {Huang}, \bibfnamefont
  {S.-M.}}, \bibinfo {author} {\bibfnamefont {S.-Y.}\ \bibnamefont {Xu}},
  \bibinfo {author} {\bibfnamefont {I.}~\bibnamefont {Belopolski}}, \bibinfo
  {author} {\bibfnamefont {C.-C.}\ \bibnamefont {Lee}}, \bibinfo {author}
  {\bibfnamefont {G.}~\bibnamefont {Chang}}, \bibinfo {author} {\bibfnamefont
  {B.}~\bibnamefont {Wang}}, \bibinfo {author} {\bibfnamefont {N.}~\bibnamefont
  {Alidoust}}, \bibinfo {author} {\bibfnamefont {G.}~\bibnamefont {Bian}},
  \bibinfo {author} {\bibfnamefont {M.}~\bibnamefont {Neupane}}, \bibinfo
  {author} {\bibfnamefont {C.}~\bibnamefont {Zhang}}, \bibinfo {author}
  {\bibfnamefont {S.}~\bibnamefont {Jia}}, \bibinfo {author} {\bibfnamefont
  {A.}~\bibnamefont {Bansil}}, \bibinfo {author} {\bibfnamefont
  {H.}~\bibnamefont {Lin}}, \ and\ \bibinfo {author} {\bibfnamefont {M.~Z.}\
  \bibnamefont {Hasan}}} (\bibinfo {year} {2015}{\natexlab{a}}),\ \href
  {http://dx.doi.org/10.1038/ncomms8373} {\bibfield  {journal} {\bibinfo
  {journal} {Nat Commun}\ }\textbf {\bibinfo {volume} {6}}}\BibitemShut
  {NoStop}%
\bibitem [{\citenamefont {Huang}\ \emph
  {et~al.}(2015{\natexlab{b}})\citenamefont {Huang}, \citenamefont {Zhao},
  \citenamefont {Long}, \citenamefont {Wang}, \citenamefont {Chen},
  \citenamefont {Yang}, \citenamefont {Liang}, \citenamefont {Xue},
  \citenamefont {Weng}, \citenamefont {Fang}, \citenamefont {Dai},\ and\
  \citenamefont {Chen}}]{Huang_Weyl_2015}%
  \BibitemOpen
  \bibfield  {author} {\bibinfo {author} {\bibnamefont {Huang}, \bibfnamefont
  {X.}}, \bibinfo {author} {\bibfnamefont {L.}~\bibnamefont {Zhao}}, \bibinfo
  {author} {\bibfnamefont {Y.}~\bibnamefont {Long}}, \bibinfo {author}
  {\bibfnamefont {P.}~\bibnamefont {Wang}}, \bibinfo {author} {\bibfnamefont
  {D.}~\bibnamefont {Chen}}, \bibinfo {author} {\bibfnamefont {Z.}~\bibnamefont
  {Yang}}, \bibinfo {author} {\bibfnamefont {H.}~\bibnamefont {Liang}},
  \bibinfo {author} {\bibfnamefont {M.}~\bibnamefont {Xue}}, \bibinfo {author}
  {\bibfnamefont {H.}~\bibnamefont {Weng}}, \bibinfo {author} {\bibfnamefont
  {Z.}~\bibnamefont {Fang}}, \bibinfo {author} {\bibfnamefont {X.}~\bibnamefont
  {Dai}}, \ and\ \bibinfo {author} {\bibfnamefont {G.}~\bibnamefont {Chen}}}
  (\bibinfo {year} {2015}{\natexlab{b}}),\ \href {\doibase
  10.1103/PhysRevX.5.031023} {\bibfield  {journal} {\bibinfo  {journal} {Phys.
  Rev. X}\ }\textbf {\bibinfo {volume} {5}},\ \bibinfo {pages}
  {031023}}\BibitemShut {NoStop}%
\bibitem [{\citenamefont {Hughes}\ \emph {et~al.}(2011)\citenamefont {Hughes},
  \citenamefont {Prodan},\ and\ \citenamefont {Bernevig}}]{Hughes:2011uq}%
  \BibitemOpen
  \bibfield  {author} {\bibinfo {author} {\bibnamefont {Hughes}, \bibfnamefont
  {T.~L.}}, \bibinfo {author} {\bibfnamefont {E.}~\bibnamefont {Prodan}}, \
  and\ \bibinfo {author} {\bibfnamefont {B.~A.}\ \bibnamefont {Bernevig}}}
  (\bibinfo {year} {2011}),\ \href@noop {} {\bibfield  {journal} {\bibinfo
  {journal} {Phys. Rev. B}\ }\textbf {\bibinfo {volume} {83}},\ \bibinfo
  {pages} {245132}}\BibitemShut {NoStop}%
\bibitem [{\citenamefont {Hughes}\ \emph {et~al.}(2014)\citenamefont {Hughes},
  \citenamefont {Yao},\ and\ \citenamefont {Qi}}]{HughesYaoQi13}%
  \BibitemOpen
  \bibfield  {author} {\bibinfo {author} {\bibnamefont {Hughes}, \bibfnamefont
  {T.~L.}}, \bibinfo {author} {\bibfnamefont {H.}~\bibnamefont {Yao}}, \ and\
  \bibinfo {author} {\bibfnamefont {X.-L.}\ \bibnamefont {Qi}}} (\bibinfo
  {year} {2014}),\ \href {\doibase 10.1103/PhysRevB.90.235123} {\bibfield
  {journal} {\bibinfo  {journal} {Phys. Rev. B}\ }\textbf {\bibinfo {volume}
  {90}},\ \bibinfo {pages} {235123}}\BibitemShut {NoStop}%
\bibitem [{\citenamefont {Hung}\ \emph {et~al.}(2013)\citenamefont {Hung},
  \citenamefont {Ghaemi}, \citenamefont {Hughes},\ and\ \citenamefont
  {Gilbert}}]{Hung_TI_SC}%
  \BibitemOpen
  \bibfield  {author} {\bibinfo {author} {\bibnamefont {Hung}, \bibfnamefont
  {H.-H.}}, \bibinfo {author} {\bibfnamefont {P.}~\bibnamefont {Ghaemi}},
  \bibinfo {author} {\bibfnamefont {T.~L.}\ \bibnamefont {Hughes}}, \ and\
  \bibinfo {author} {\bibfnamefont {M.~J.}\ \bibnamefont {Gilbert}}} (\bibinfo
  {year} {2013}),\ \href {\doibase 10.1103/PhysRevB.87.035401} {\bibfield
  {journal} {\bibinfo  {journal} {Phys. Rev. B}\ }\textbf {\bibinfo {volume}
  {87}},\ \bibinfo {pages} {035401}}\BibitemShut {NoStop}%
\bibitem [{\citenamefont {Hung}\ and\ \citenamefont {Wen}(2013)}]{Hung_SET}%
  \BibitemOpen
  \bibfield  {author} {\bibinfo {author} {\bibnamefont {Hung}, \bibfnamefont
  {L.-Y.}}, \ and\ \bibinfo {author} {\bibfnamefont {X.-G.}\ \bibnamefont
  {Wen}}} (\bibinfo {year} {2013}),\ \href {\doibase
  10.1103/PhysRevB.87.165107} {\bibfield  {journal} {\bibinfo  {journal} {Phys.
  Rev. B}\ }\textbf {\bibinfo {volume} {87}},\ \bibinfo {pages}
  {165107}}\BibitemShut {NoStop}%
\bibitem [{\citenamefont {{Hung}}\ and\ \citenamefont
  {{Wen}}(2014)}]{HungWen2014}%
  \BibitemOpen
  \bibfield  {author} {\bibinfo {author} {\bibnamefont {{Hung}}, \bibfnamefont
  {L.-Y.}}, \ and\ \bibinfo {author} {\bibfnamefont {X.-G.}\ \bibnamefont
  {{Wen}}}} (\bibinfo {year} {2014}),\ \href {\doibase
  10.1103/PhysRevB.89.075121} {\bibfield  {journal} {\bibinfo  {journal}
  {\prb}\ }\textbf {\bibinfo {volume} {89}}~(\bibinfo {number} {7}),\ \bibinfo
  {eid} {075121}}\BibitemShut {NoStop}%
\bibitem [{\citenamefont {Ishikawa}\ and\ \citenamefont
  {Matsuyama}(1987)}]{Ishikawa:1987zi}%
  \BibitemOpen
  \bibfield  {author} {\bibinfo {author} {\bibnamefont {Ishikawa},
  \bibfnamefont {K.}}, \ and\ \bibinfo {author} {\bibfnamefont
  {T.}~\bibnamefont {Matsuyama}}} (\bibinfo {year} {1987}),\ \href {\doibase
  10.1016/0550-3213(87)90160-X} {\bibfield  {journal} {\bibinfo  {journal}
  {Nucl. Phys.}\ }\textbf {\bibinfo {volume} {B280}},\ \bibinfo {pages}
  {523}}\BibitemShut {NoStop}%
\bibitem [{\citenamefont {Isobe}\ and\ \citenamefont {Fu}(2015)}]{Isobe2015}%
  \BibitemOpen
  \bibfield  {author} {\bibinfo {author} {\bibnamefont {Isobe}, \bibfnamefont
  {H.}}, \ and\ \bibinfo {author} {\bibfnamefont {L.}~\bibnamefont {Fu}}}
  (\bibinfo {year} {2015}),\ \href {\doibase 10.1103/PhysRevB.92.081304}
  {\bibfield  {journal} {\bibinfo  {journal} {Phys. Rev. B}\ }\textbf {\bibinfo
  {volume} {92}},\ \bibinfo {pages} {081304}}\BibitemShut {NoStop}%
\bibitem [{\citenamefont {Ivanov}(2001)}]{Ivanov_braiding}%
  \BibitemOpen
  \bibfield  {author} {\bibinfo {author} {\bibnamefont {Ivanov}, \bibfnamefont
  {D.~A.}}} (\bibinfo {year} {2001}),\ \href {\doibase
  10.1103/PhysRevLett.86.268} {\bibfield  {journal} {\bibinfo  {journal} {Phys.
  Rev. Lett.}\ }\textbf {\bibinfo {volume} {86}},\ \bibinfo {pages}
  {268}}\BibitemShut {NoStop}%
\bibitem [{\citenamefont {Jackiw}\ and\ \citenamefont
  {Rebbi}(1976)}]{JackiwRebbi76}%
  \BibitemOpen
  \bibfield  {author} {\bibinfo {author} {\bibnamefont {Jackiw}, \bibfnamefont
  {R.}}, \ and\ \bibinfo {author} {\bibfnamefont {C.}~\bibnamefont {Rebbi}}}
  (\bibinfo {year} {1976}),\ \href {\doibase 10.1103/PhysRevD.13.3398}
  {\bibfield  {journal} {\bibinfo  {journal} {Phys. Rev. D}\ }\textbf {\bibinfo
  {volume} {13}},\ \bibinfo {pages} {3398}}\BibitemShut {NoStop}%
\bibitem [{\citenamefont {Jackiw}\ and\ \citenamefont
  {Rossi}(1981)}]{JackiwRossi81}%
  \BibitemOpen
  \bibfield  {author} {\bibinfo {author} {\bibnamefont {Jackiw}, \bibfnamefont
  {R.}}, \ and\ \bibinfo {author} {\bibfnamefont {P.}~\bibnamefont {Rossi}}}
  (\bibinfo {year} {1981}),\ \href {\doibase 10.1016/0550-3213(81)90044-4}
  {\bibfield  {journal} {\bibinfo  {journal} {Nucl. Phys. B}\ }\textbf
  {\bibinfo {volume} {190}},\ \bibinfo {pages} {681}}\BibitemShut {NoStop}%
\bibitem [{\citenamefont {Jadaun}\ \emph {et~al.}(2013)\citenamefont {Jadaun},
  \citenamefont {Xiao}, \citenamefont {Niu},\ and\ \citenamefont
  {Banerjee}}]{Niu_point_symmetry}%
  \BibitemOpen
  \bibfield  {author} {\bibinfo {author} {\bibnamefont {Jadaun}, \bibfnamefont
  {P.}}, \bibinfo {author} {\bibfnamefont {D.}~\bibnamefont {Xiao}}, \bibinfo
  {author} {\bibfnamefont {Q.}~\bibnamefont {Niu}}, \ and\ \bibinfo {author}
  {\bibfnamefont {S.~K.}\ \bibnamefont {Banerjee}}} (\bibinfo {year} {2013}),\
  \href {\doibase 10.1103/PhysRevB.88.085110} {\bibfield  {journal} {\bibinfo
  {journal} {Phys. Rev. B}\ }\textbf {\bibinfo {volume} {88}},\ \bibinfo
  {pages} {085110}}\BibitemShut {NoStop}%
\bibitem [{\citenamefont {Jang}\ \emph {et~al.}(2011)\citenamefont {Jang},
  \citenamefont {Ferguson}, \citenamefont {Vakaryuk}, \citenamefont {Budakian},
  \citenamefont {Chung}, \citenamefont {Goldbart},\ and\ \citenamefont
  {Maeno}}]{Jang_SrRuO}%
  \BibitemOpen
  \bibfield  {author} {\bibinfo {author} {\bibnamefont {Jang}, \bibfnamefont
  {J.}}, \bibinfo {author} {\bibfnamefont {D.~G.}\ \bibnamefont {Ferguson}},
  \bibinfo {author} {\bibfnamefont {V.}~\bibnamefont {Vakaryuk}}, \bibinfo
  {author} {\bibfnamefont {R.}~\bibnamefont {Budakian}}, \bibinfo {author}
  {\bibfnamefont {S.~B.}\ \bibnamefont {Chung}}, \bibinfo {author}
  {\bibfnamefont {P.~M.}\ \bibnamefont {Goldbart}}, \ and\ \bibinfo {author}
  {\bibfnamefont {Y.}~\bibnamefont {Maeno}}} (\bibinfo {year} {2011}),\ \href
  {\doibase 10.1126/science.1193839} {\bibfield  {journal} {\bibinfo  {journal}
  {Science}\ }\textbf {\bibinfo {volume} {331}}~(\bibinfo {number} {6014}),\
  \bibinfo {pages} {186}}\BibitemShut {NoStop}%
\bibitem [{\citenamefont {Jeon}\ \emph {et~al.}(2014)\citenamefont {Jeon},
  \citenamefont {Zhou}, \citenamefont {Gyenis}, \citenamefont {Feldman},
  \citenamefont {Kimchi}, \citenamefont {Potter}, \citenamefont {Gibson},
  \citenamefont {Cava}, \citenamefont {Vishwanath},\ and\ \citenamefont
  {Yazdani}}]{Yazdani_CdAs}%
  \BibitemOpen
  \bibfield  {author} {\bibinfo {author} {\bibnamefont {Jeon}, \bibfnamefont
  {S.}}, \bibinfo {author} {\bibfnamefont {B.~B.}\ \bibnamefont {Zhou}},
  \bibinfo {author} {\bibfnamefont {A.}~\bibnamefont {Gyenis}}, \bibinfo
  {author} {\bibfnamefont {B.~E.}\ \bibnamefont {Feldman}}, \bibinfo {author}
  {\bibfnamefont {I.}~\bibnamefont {Kimchi}}, \bibinfo {author} {\bibfnamefont
  {A.~C.}\ \bibnamefont {Potter}}, \bibinfo {author} {\bibfnamefont {Q.~D.}\
  \bibnamefont {Gibson}}, \bibinfo {author} {\bibfnamefont {R.~J.}\
  \bibnamefont {Cava}}, \bibinfo {author} {\bibfnamefont {A.}~\bibnamefont
  {Vishwanath}}, \ and\ \bibinfo {author} {\bibfnamefont {A.}~\bibnamefont
  {Yazdani}}} (\bibinfo {year} {2014}),\ \href@noop {} {\bibfield  {journal}
  {\bibinfo  {journal} {Nat. Mater.}\ }\textbf {\bibinfo {volume} {13}},\
  \bibinfo {pages} {851}}\BibitemShut {NoStop}%
\bibitem [{\citenamefont {Jian}\ and\ \citenamefont {Qi}(2014)}]{Jian2014}%
  \BibitemOpen
  \bibfield  {author} {\bibinfo {author} {\bibnamefont {Jian}, \bibfnamefont
  {C.-M.}}, \ and\ \bibinfo {author} {\bibfnamefont {X.-L.}\ \bibnamefont
  {Qi}}} (\bibinfo {year} {2014}),\ \href {\doibase 10.1103/PhysRevX.4.041043}
  {\bibfield  {journal} {\bibinfo  {journal} {Phys. Rev. X}\ }\textbf {\bibinfo
  {volume} {4}},\ \bibinfo {pages} {041043}}\BibitemShut {NoStop}%
\bibitem [{\citenamefont {Jiang}\ \emph {et~al.}(2011)\citenamefont {Jiang},
  \citenamefont {Kitagawa}, \citenamefont {Alicea}, \citenamefont {Akhmerov},
  \citenamefont {Pekker}, \citenamefont {Refael}, \citenamefont {Cirac},
  \citenamefont {Demler}, \citenamefont {Lukin},\ and\ \citenamefont
  {Zoller}}]{zoller_PRL_11}%
  \BibitemOpen
  \bibfield  {author} {\bibinfo {author} {\bibnamefont {Jiang}, \bibfnamefont
  {L.}}, \bibinfo {author} {\bibfnamefont {T.}~\bibnamefont {Kitagawa}},
  \bibinfo {author} {\bibfnamefont {J.}~\bibnamefont {Alicea}}, \bibinfo
  {author} {\bibfnamefont {A.~R.}\ \bibnamefont {Akhmerov}}, \bibinfo {author}
  {\bibfnamefont {D.}~\bibnamefont {Pekker}}, \bibinfo {author} {\bibfnamefont
  {G.}~\bibnamefont {Refael}}, \bibinfo {author} {\bibfnamefont {J.~I.}\
  \bibnamefont {Cirac}}, \bibinfo {author} {\bibfnamefont {E.}~\bibnamefont
  {Demler}}, \bibinfo {author} {\bibfnamefont {M.~D.}\ \bibnamefont {Lukin}}, \
  and\ \bibinfo {author} {\bibfnamefont {P.}~\bibnamefont {Zoller}}} (\bibinfo
  {year} {2011}),\ \href {\doibase 10.1103/PhysRevLett.106.220402} {\bibfield
  {journal} {\bibinfo  {journal} {Phys. Rev. Lett.}\ }\textbf {\bibinfo
  {volume} {106}},\ \bibinfo {pages} {220402}}\BibitemShut {NoStop}%
\bibitem [{\citenamefont {Jiang}\ \emph {et~al.}(2014)\citenamefont {Jiang},
  \citenamefont {Mesaros},\ and\ \citenamefont {Ran}}]{Jiang2014}%
  \BibitemOpen
  \bibfield  {author} {\bibinfo {author} {\bibnamefont {Jiang}, \bibfnamefont
  {S.}}, \bibinfo {author} {\bibfnamefont {A.}~\bibnamefont {Mesaros}}, \ and\
  \bibinfo {author} {\bibfnamefont {Y.}~\bibnamefont {Ran}}} (\bibinfo {year}
  {2014}),\ \href {\doibase 10.1103/PhysRevX.4.031048} {\bibfield  {journal}
  {\bibinfo  {journal} {Phys. Rev. X}\ }\textbf {\bibinfo {volume} {4}},\
  \bibinfo {pages} {031048}}\BibitemShut {NoStop}%
\bibitem [{\citenamefont {Kane}\ and\ \citenamefont
  {Fisher}(1997)}]{KaneFisher97}%
  \BibitemOpen
  \bibfield  {author} {\bibinfo {author} {\bibnamefont {Kane}, \bibfnamefont
  {C.~L.}}, \ and\ \bibinfo {author} {\bibfnamefont {M.~P.~A.}\ \bibnamefont
  {Fisher}}} (\bibinfo {year} {1997}),\ \href {\doibase
  10.1103/PhysRevB.55.15832} {\bibfield  {journal} {\bibinfo  {journal} {Phys.
  Rev. B}\ }\textbf {\bibinfo {volume} {55}},\ \bibinfo {pages}
  {15832}}\BibitemShut {NoStop}%
\bibitem [{\citenamefont {{Kane}}\ and\ \citenamefont
  {{Lubensky}}(2014)}]{KaneLubensky2014}%
  \BibitemOpen
  \bibfield  {author} {\bibinfo {author} {\bibnamefont {{Kane}}, \bibfnamefont
  {C.~L.}}, \ and\ \bibinfo {author} {\bibfnamefont {T.~C.}\ \bibnamefont
  {{Lubensky}}}} (\bibinfo {year} {2014}),\ \href {\doibase 10.1038/nphys2835}
  {\bibfield  {journal} {\bibinfo  {journal} {Nature Physics}\ }\textbf
  {\bibinfo {volume} {10}},\ \bibinfo {pages} {39}}\BibitemShut {NoStop}%
\bibitem [{\citenamefont {Kane}\ and\ \citenamefont
  {Mele}(2005{\natexlab{a}})}]{Kane:2005vn}%
  \BibitemOpen
  \bibfield  {author} {\bibinfo {author} {\bibnamefont {Kane}, \bibfnamefont
  {C.~L.}}, \ and\ \bibinfo {author} {\bibfnamefont {E.~J.}\ \bibnamefont
  {Mele}}} (\bibinfo {year} {2005}{\natexlab{a}}),\ \href@noop {} {\bibfield
  {journal} {\bibinfo  {journal} {Phys. Rev. Lett.}\ }\textbf {\bibinfo
  {volume} {95}},\ \bibinfo {pages} {226801}}\BibitemShut {NoStop}%
\bibitem [{\citenamefont {Kane}\ and\ \citenamefont
  {Mele}(2005{\natexlab{b}})}]{Kane:2005kx}%
  \BibitemOpen
  \bibfield  {author} {\bibinfo {author} {\bibnamefont {Kane}, \bibfnamefont
  {C.~L.}}, \ and\ \bibinfo {author} {\bibfnamefont {E.~J.}\ \bibnamefont
  {Mele}}} (\bibinfo {year} {2005}{\natexlab{b}}),\ \href@noop {} {\bibfield
  {journal} {\bibinfo  {journal} {Phys. Rev. Lett.}\ }\textbf {\bibinfo
  {volume} {95}},\ \bibinfo {pages} {146802}}\BibitemShut {NoStop}%
\bibitem [{\citenamefont {{Kapustin}}(2014{\natexlab{a}})}]{Kapustin2014d}%
  \BibitemOpen
  \bibfield  {author} {\bibinfo {author} {\bibnamefont {{Kapustin}},
  \bibfnamefont {A.}}} (\bibinfo {year} {2014}{\natexlab{a}}),\ \href@noop {}
  {\ }\Eprint {http://arxiv.org/abs/1404.6659} {arXiv:1404.6659} \BibitemShut
  {NoStop}%
\bibitem [{\citenamefont {{Kapustin}}(2014{\natexlab{b}})}]{Kapustin2014b}%
  \BibitemOpen
  \bibfield  {author} {\bibinfo {author} {\bibnamefont {{Kapustin}},
  \bibfnamefont {A.}}} (\bibinfo {year} {2014}{\natexlab{b}}),\ \href@noop {}
  {\ }\Eprint {http://arxiv.org/abs/1403.1467} {arXiv:1403.1467} \BibitemShut
  {NoStop}%
\bibitem [{\citenamefont {{Kapustin}}\ and\ \citenamefont
  {{Thorngren}}(2014{\natexlab{a}})}]{Kapustin2014c}%
  \BibitemOpen
  \bibfield  {author} {\bibinfo {author} {\bibnamefont {{Kapustin}},
  \bibfnamefont {A.}}, \ and\ \bibinfo {author} {\bibfnamefont
  {R.}~\bibnamefont {{Thorngren}}}} (\bibinfo {year} {2014}{\natexlab{a}}),\
  \href@noop {} {\ }\Eprint {http://arxiv.org/abs/1404.3230} {arXiv:1404.3230}
  \BibitemShut {NoStop}%
\bibitem [{\citenamefont {{Kapustin}}\ and\ \citenamefont
  {{Thorngren}}(2014{\natexlab{b}})}]{Kapustin2014a}%
  \BibitemOpen
  \bibfield  {author} {\bibinfo {author} {\bibnamefont {{Kapustin}},
  \bibfnamefont {A.}}, \ and\ \bibinfo {author} {\bibfnamefont
  {R.}~\bibnamefont {{Thorngren}}}} (\bibinfo {year} {2014}{\natexlab{b}}),\
  \href {\doibase 10.1103/PhysRevLett.112.231602} {\bibfield  {journal}
  {\bibinfo  {journal} {Phys. Rev. Lett.}\ }\textbf {\bibinfo {volume} {112}},\
  \bibinfo {eid} {231602}}\BibitemShut {NoStop}%
\bibitem [{\citenamefont {Kapustin}\ \emph {et~al.}(2015)\citenamefont
  {Kapustin}, \citenamefont {Thorngren}, \citenamefont {Turzillo},\ and\
  \citenamefont {Wang}}]{Kapustin2014e}%
  \BibitemOpen
  \bibfield  {author} {\bibinfo {author} {\bibnamefont {Kapustin},
  \bibfnamefont {A.}}, \bibinfo {author} {\bibfnamefont {R.}~\bibnamefont
  {Thorngren}}, \bibinfo {author} {\bibfnamefont {A.}~\bibnamefont {Turzillo}},
  \ and\ \bibinfo {author} {\bibfnamefont {Z.}~\bibnamefont {Wang}}} (\bibinfo
  {year} {2015}),\ \href {\doibase 10.1007/JHEP12(2015)052} {\bibfield
  {journal} {\bibinfo  {journal} {Journal of High Energy Physics}\ }\textbf
  {\bibinfo {volume} {2015}}~(\bibinfo {number} {12}),\ \bibinfo {pages}
  {1}}\BibitemShut {NoStop}%
\bibitem [{\citenamefont {Kariyado}\ and\ \citenamefont
  {Ogata}(2011)}]{OgataJPSJ2011}%
  \BibitemOpen
  \bibfield  {author} {\bibinfo {author} {\bibnamefont {Kariyado},
  \bibfnamefont {T.}}, \ and\ \bibinfo {author} {\bibfnamefont
  {M.}~\bibnamefont {Ogata}}} (\bibinfo {year} {2011}),\ \href@noop {}
  {\bibfield  {journal} {\bibinfo  {journal} {J. Phys. Soc. Jpn.}\ }\textbf
  {\bibinfo {volume} {80}},\ \bibinfo {pages} {083704}}\BibitemShut {NoStop}%
\bibitem [{\citenamefont {Karoubi}(1978)}]{Karoubibook}%
  \BibitemOpen
  \bibfield  {author} {\bibinfo {author} {\bibnamefont {Karoubi}, \bibfnamefont
  {M.}}} (\bibinfo {year} {1978}),\ \href@noop {} {\emph {\bibinfo {title}
  {K-Theory: An Introduction}}}\ (\bibinfo  {publisher} {Springer})\BibitemShut
  {NoStop}%
\bibitem [{\citenamefont {Kennedy}\ and\ \citenamefont
  {Zirnbauer}(2015)}]{KennedyZirnbauer2014}%
  \BibitemOpen
  \bibfield  {author} {\bibinfo {author} {\bibnamefont {Kennedy}, \bibfnamefont
  {R.}}, \ and\ \bibinfo {author} {\bibfnamefont {M.~R.}\ \bibnamefont
  {Zirnbauer}}} (\bibinfo {year} {2015}),\ \href {\doibase
  10.1007/s00220-015-2512-8} {\bibfield  {journal} {\bibinfo  {journal}
  {Communications in Mathematical Physics}\ }\textbf {\bibinfo {volume}
  {342}}~(\bibinfo {number} {3}),\ \bibinfo {pages} {909}}\BibitemShut
  {NoStop}%
\bibitem [{\citenamefont {Keselman}\ \emph {et~al.}(2013)\citenamefont
  {Keselman}, \citenamefont {Fu}, \citenamefont {Stern},\ and\ \citenamefont
  {Berg}}]{KeselmanFuSternBerg13}%
  \BibitemOpen
  \bibfield  {author} {\bibinfo {author} {\bibnamefont {Keselman},
  \bibfnamefont {A.}}, \bibinfo {author} {\bibfnamefont {L.}~\bibnamefont
  {Fu}}, \bibinfo {author} {\bibfnamefont {A.}~\bibnamefont {Stern}}, \ and\
  \bibinfo {author} {\bibfnamefont {E.}~\bibnamefont {Berg}}} (\bibinfo {year}
  {2013}),\ \href {\doibase 10.1103/PhysRevLett.111.116402} {\bibfield
  {journal} {\bibinfo  {journal} {Phys. Rev. Lett.}\ }\textbf {\bibinfo
  {volume} {111}},\ \bibinfo {pages} {116402}}\BibitemShut {NoStop}%
\bibitem [{\citenamefont {Khanikaev}\ \emph {et~al.}(2013)\citenamefont
  {Khanikaev}, \citenamefont {Hossein~Mousavi}, \citenamefont {Tse},
  \citenamefont {Kargarian}, \citenamefont {MacDonald},\ and\ \citenamefont
  {Shvets}}]{Khanikaev_photonic_TI}%
  \BibitemOpen
  \bibfield  {author} {\bibinfo {author} {\bibnamefont {Khanikaev},
  \bibfnamefont {A.~B.}}, \bibinfo {author} {\bibfnamefont {S.}~\bibnamefont
  {Hossein~Mousavi}}, \bibinfo {author} {\bibfnamefont {W.-K.}\ \bibnamefont
  {Tse}}, \bibinfo {author} {\bibfnamefont {M.}~\bibnamefont {Kargarian}},
  \bibinfo {author} {\bibfnamefont {A.~H.}\ \bibnamefont {MacDonald}}, \ and\
  \bibinfo {author} {\bibfnamefont {G.}~\bibnamefont {Shvets}}} (\bibinfo
  {year} {2013}),\ \href {http://dx.doi.org/10.1038/nmat3520} {\bibfield
  {journal} {\bibinfo  {journal} {Nat Mater}\ }\textbf {\bibinfo {volume}
  {12}}~(\bibinfo {number} {3}),\ \bibinfo {pages} {233}}\BibitemShut {NoStop}%
\bibitem [{\citenamefont {{Khmel'Nitski{\v \i}}}(1983)}]{Khmelnitski1983}%
  \BibitemOpen
  \bibfield  {author} {\bibinfo {author} {\bibnamefont {{Khmel'Nitski{\v \i}}},
  \bibfnamefont {D.~E.}}} (\bibinfo {year} {1983}),\ \href@noop {} {\bibfield
  {journal} {\bibinfo  {journal} {Soviet Journal of Experimental and
  Theoretical Physics Letters}\ }\textbf {\bibinfo {volume} {38}},\ \bibinfo
  {pages} {552}}\BibitemShut {NoStop}%
\bibitem [{\citenamefont {Kim}\ \emph {et~al.}(2015)\citenamefont {Kim},
  \citenamefont {Wieder}, \citenamefont {Kane},\ and\ \citenamefont
  {Rappe}}]{Kane_Cu3N_ring}%
  \BibitemOpen
  \bibfield  {author} {\bibinfo {author} {\bibnamefont {Kim}, \bibfnamefont
  {Y.}}, \bibinfo {author} {\bibfnamefont {B.~J.}\ \bibnamefont {Wieder}},
  \bibinfo {author} {\bibfnamefont {C.~L.}\ \bibnamefont {Kane}}, \ and\
  \bibinfo {author} {\bibfnamefont {A.~M.}\ \bibnamefont {Rappe}}} (\bibinfo
  {year} {2015}),\ \href {\doibase 10.1103/PhysRevLett.115.036806} {\bibfield
  {journal} {\bibinfo  {journal} {Phys. Rev. Lett.}\ }\textbf {\bibinfo
  {volume} {115}},\ \bibinfo {pages} {036806}}\BibitemShut {NoStop}%
\bibitem [{\citenamefont {King-Smith}\ and\ \citenamefont
  {Vanderbilt}(1993)}]{kingSmithPRB93a}%
  \BibitemOpen
  \bibfield  {author} {\bibinfo {author} {\bibnamefont {King-Smith},
  \bibfnamefont {R.~D.}}, \ and\ \bibinfo {author} {\bibfnamefont
  {D.}~\bibnamefont {Vanderbilt}}} (\bibinfo {year} {1993}),\ \href {\doibase
  10.1103/PhysRevB.47.1651} {\bibfield  {journal} {\bibinfo  {journal} {Phys.
  Rev. B}\ }\textbf {\bibinfo {volume} {47}},\ \bibinfo {pages}
  {1651}}\BibitemShut {NoStop}%
\bibitem [{\citenamefont {{Kitaev}}(2001)}]{Kitaev2001}%
  \BibitemOpen
  \bibfield  {author} {\bibinfo {author} {\bibnamefont {{Kitaev}},
  \bibfnamefont {A.}}} (\bibinfo {year} {2001}),\ \href {\doibase
  10.1070/1063-7869/44/10S/S29} {\bibfield  {journal} {\bibinfo  {journal}
  {Physics Uspekhi}\ }\textbf {\bibinfo {volume} {44}},\ \bibinfo {pages}
  {131}}\BibitemShut {NoStop}%
\bibitem [{\citenamefont {{Kitaev}}(2006)}]{Kitaev2006}%
  \BibitemOpen
  \bibfield  {author} {\bibinfo {author} {\bibnamefont {{Kitaev}},
  \bibfnamefont {A.}}} (\bibinfo {year} {2006}),\ \href {\doibase
  10.1016/j.aop.2005.10.005} {\bibfield  {journal} {\bibinfo  {journal} {Annals
  of Physics}\ }\textbf {\bibinfo {volume} {321}},\ \bibinfo {pages}
  {2}}\BibitemShut {NoStop}%
\bibitem [{\citenamefont {Kitaev}(2009)}]{Kitaev2009}%
  \BibitemOpen
  \bibfield  {author} {\bibinfo {author} {\bibnamefont {Kitaev}, \bibfnamefont
  {A.}}} (\bibinfo {year} {2009}),\ \href@noop {} {\bibfield  {journal}
  {\bibinfo  {journal} {AIP Conf. Proc.}\ }\textbf {\bibinfo {volume} {1134}},\
  \bibinfo {pages} {22}}\BibitemShut {NoStop}%
\bibitem [{\citenamefont {Kitaev}(2015)}]{Kitaev_Unpublished}%
  \BibitemOpen
  \bibfield  {author} {\bibinfo {author} {\bibnamefont {Kitaev}, \bibfnamefont
  {A.}}} (\bibinfo {year} {2015}),\ \href
  {http://www.federalreserve.gov/boarddocs/speeches/1996/19961205.htm}
  {}\bibinfo {note} {Talk at the IPAM workshop "Symmetry and Topology in
  Quantum Matter"}\BibitemShut {NoStop}%
\bibitem [{\citenamefont {Kitaev}\ and\ \citenamefont
  {Preskill}(2006)}]{KitaevPreskill06}%
  \BibitemOpen
  \bibfield  {author} {\bibinfo {author} {\bibnamefont {Kitaev}, \bibfnamefont
  {A.}}, \ and\ \bibinfo {author} {\bibfnamefont {J.}~\bibnamefont {Preskill}}}
  (\bibinfo {year} {2006}),\ \href {\doibase 10.1103/PhysRevLett.96.110404}
  {\bibfield  {journal} {\bibinfo  {journal} {Phys. Rev. Lett.}\ }\textbf
  {\bibinfo {volume} {96}},\ \bibinfo {pages} {110404}}\BibitemShut {NoStop}%
\bibitem [{\citenamefont {Kitagawa}\ \emph {et~al.}(2010)\citenamefont
  {Kitagawa}, \citenamefont {Berg}, \citenamefont {Rudner},\ and\ \citenamefont
  {Demler}}]{Takuya_floquet_TI}%
  \BibitemOpen
  \bibfield  {author} {\bibinfo {author} {\bibnamefont {Kitagawa},
  \bibfnamefont {T.}}, \bibinfo {author} {\bibfnamefont {E.}~\bibnamefont
  {Berg}}, \bibinfo {author} {\bibfnamefont {M.}~\bibnamefont {Rudner}}, \ and\
  \bibinfo {author} {\bibfnamefont {E.}~\bibnamefont {Demler}}} (\bibinfo
  {year} {2010}),\ \href {\doibase 10.1103/PhysRevB.82.235114} {\bibfield
  {journal} {\bibinfo  {journal} {Phys. Rev. B}\ }\textbf {\bibinfo {volume}
  {82}},\ \bibinfo {pages} {235114}}\BibitemShut {NoStop}%
\bibitem [{\citenamefont {Klitzing}\ \emph {et~al.}(1980)\citenamefont
  {Klitzing}, \citenamefont {Dorda},\ and\ \citenamefont {Pepper}}]{Klitzing}%
  \BibitemOpen
  \bibfield  {author} {\bibinfo {author} {\bibnamefont {Klitzing},
  \bibfnamefont {K.~v.}}, \bibinfo {author} {\bibfnamefont {G.}~\bibnamefont
  {Dorda}}, \ and\ \bibinfo {author} {\bibfnamefont {M.}~\bibnamefont
  {Pepper}}} (\bibinfo {year} {1980}),\ \href@noop {} {\bibfield  {journal}
  {\bibinfo  {journal} {Phys. Rev. Lett.}\ }\textbf {\bibinfo {volume} {45}},\
  \bibinfo {pages} {494}}\BibitemShut {NoStop}%
\bibitem [{\citenamefont {Knez}\ \emph {et~al.}(2011)\citenamefont {Knez},
  \citenamefont {Du},\ and\ \citenamefont {Sullivan}}]{knezPRL11}%
  \BibitemOpen
  \bibfield  {author} {\bibinfo {author} {\bibnamefont {Knez}, \bibfnamefont
  {I.}}, \bibinfo {author} {\bibfnamefont {R.-R.}\ \bibnamefont {Du}}, \ and\
  \bibinfo {author} {\bibfnamefont {G.}~\bibnamefont {Sullivan}}} (\bibinfo
  {year} {2011}),\ \href {\doibase 10.1103/PhysRevLett.107.136603} {\bibfield
  {journal} {\bibinfo  {journal} {Phys. Rev. Lett.}\ }\textbf {\bibinfo
  {volume} {107}},\ \bibinfo {pages} {136603}}\BibitemShut {NoStop}%
\bibitem [{\citenamefont {Kobayashi}\ \emph
  {et~al.}(2014{\natexlab{a}})\citenamefont {Kobayashi}, \citenamefont
  {Ohtsuki}, \citenamefont {Imura},\ and\ \citenamefont
  {Herbut}}]{PhysRevLett.112.016402}%
  \BibitemOpen
  \bibfield  {author} {\bibinfo {author} {\bibnamefont {Kobayashi},
  \bibfnamefont {K.}}, \bibinfo {author} {\bibfnamefont {T.}~\bibnamefont
  {Ohtsuki}}, \bibinfo {author} {\bibfnamefont {K.-I.}\ \bibnamefont {Imura}},
  \ and\ \bibinfo {author} {\bibfnamefont {I.~F.}\ \bibnamefont {Herbut}}}
  (\bibinfo {year} {2014}{\natexlab{a}}),\ \href {\doibase
  10.1103/PhysRevLett.112.016402} {\bibfield  {journal} {\bibinfo  {journal}
  {Phys. Rev. Lett.}\ }\textbf {\bibinfo {volume} {112}},\ \bibinfo {pages}
  {016402}}\BibitemShut {NoStop}%
\bibitem [{\citenamefont {{Kobayashi}}\ and\ \citenamefont
  {{Sato}}(2015)}]{superconducting_Dirac}%
  \BibitemOpen
  \bibfield  {author} {\bibinfo {author} {\bibnamefont {{Kobayashi}},
  \bibfnamefont {S.}}, \ and\ \bibinfo {author} {\bibfnamefont
  {M.}~\bibnamefont {{Sato}}}} (\bibinfo {year} {2015}),\ \href@noop {}
  {\bibfield  {journal} {\bibinfo  {journal} {ArXiv e-prints}\ }}\Eprint
  {http://arxiv.org/abs/1504.07408} {arXiv:1504.07408} \BibitemShut {NoStop}%
\bibitem [{\citenamefont {Kobayashi}\ \emph
  {et~al.}(2014{\natexlab{b}})\citenamefont {Kobayashi}, \citenamefont
  {Shiozaki}, \citenamefont {Tanaka},\ and\ \citenamefont
  {Sato}}]{Nodal_SC_Shingo}%
  \BibitemOpen
  \bibfield  {author} {\bibinfo {author} {\bibnamefont {Kobayashi},
  \bibfnamefont {S.}}, \bibinfo {author} {\bibfnamefont {K.}~\bibnamefont
  {Shiozaki}}, \bibinfo {author} {\bibfnamefont {Y.}~\bibnamefont {Tanaka}}, \
  and\ \bibinfo {author} {\bibfnamefont {M.}~\bibnamefont {Sato}}} (\bibinfo
  {year} {2014}{\natexlab{b}}),\ \href {\doibase 10.1103/PhysRevB.90.024516}
  {\bibfield  {journal} {\bibinfo  {journal} {Phys. Rev. B}\ }\textbf {\bibinfo
  {volume} {90}},\ \bibinfo {pages} {024516}}\BibitemShut {NoStop}%
\bibitem [{\citenamefont {{Koch-Janusz}}\ and\ \citenamefont
  {{Ringel}}(2014)}]{Koch-Janusz2014}%
  \BibitemOpen
  \bibfield  {author} {\bibinfo {author} {\bibnamefont {{Koch-Janusz}},
  \bibfnamefont {M.}}, \ and\ \bibinfo {author} {\bibfnamefont
  {Z.}~\bibnamefont {{Ringel}}}} (\bibinfo {year} {2014}),\ \href {\doibase
  10.1103/PhysRevB.89.075137} {\bibfield  {journal} {\bibinfo  {journal}
  {\prb}\ }\textbf {\bibinfo {volume} {89}},\ \bibinfo {eid}
  {075137}}\BibitemShut {NoStop}%
\bibitem [{\citenamefont {Kohmoto}(1985)}]{Kohmoto:1985zl}%
  \BibitemOpen
  \bibfield  {author} {\bibinfo {author} {\bibnamefont {Kohmoto}, \bibfnamefont
  {M.}}} (\bibinfo {year} {1985}),\ \href
  {http://www.sciencedirect.com/science/article/B6WB1-4DF4YV5-XT/2/42d97844cf9edf15cdd00ec50647fb57}
  {\bibfield  {journal} {\bibinfo  {journal} {Annals of Physics}\ }\textbf
  {\bibinfo {volume} {160}}~(\bibinfo {number} {2}),\ \bibinfo {pages}
  {343}}\BibitemShut {NoStop}%
\bibitem [{\citenamefont {{Kong}}\ and\ \citenamefont
  {{Wen}}(2014)}]{KongWen2014}%
  \BibitemOpen
  \bibfield  {author} {\bibinfo {author} {\bibnamefont {{Kong}}, \bibfnamefont
  {L.}}, \ and\ \bibinfo {author} {\bibfnamefont {X.-G.}\ \bibnamefont
  {{Wen}}}} (\bibinfo {year} {2014}),\ \href@noop {} {\ }\Eprint
  {http://arxiv.org/abs/1405.5858} {arXiv:1405.5858} \BibitemShut {NoStop}%
\bibitem [{\citenamefont {K\"{o}nig}\ \emph {et~al.}(2008)\citenamefont
  {K\"{o}nig}, \citenamefont {Buhmann}, \citenamefont {Molenkamp},
  \citenamefont {Hughes}, \citenamefont {Liu}, \citenamefont {Qi},\ and\
  \citenamefont {Zhang}}]{Taylortheory}%
  \BibitemOpen
  \bibfield  {author} {\bibinfo {author} {\bibnamefont {K\"{o}nig},
  \bibfnamefont {M.}}, \bibinfo {author} {\bibfnamefont {H.}~\bibnamefont
  {Buhmann}}, \bibinfo {author} {\bibfnamefont {L.~W.}\ \bibnamefont
  {Molenkamp}}, \bibinfo {author} {\bibfnamefont {T.}~\bibnamefont {Hughes}},
  \bibinfo {author} {\bibfnamefont {C.-X.}\ \bibnamefont {Liu}}, \bibinfo
  {author} {\bibfnamefont {X.-L.}\ \bibnamefont {Qi}}, \ and\ \bibinfo {author}
  {\bibfnamefont {S.-C.}\ \bibnamefont {Zhang}}} (\bibinfo {year} {2008}),\
  \href {\doibase 10.1143/JPSJ.77.031007} {\bibfield  {journal} {\bibinfo
  {journal} {J. Phys. Soc. Jpn.}\ }\textbf {\bibinfo {volume} {77}}~(\bibinfo
  {number} {3}),\ \bibinfo {pages} {031007}}\BibitemShut {NoStop}%
\bibitem [{\citenamefont {Konig}\ \emph {et~al.}(2007)\citenamefont {Konig},
  \citenamefont {Wiedmann}, \citenamefont {Brune}, \citenamefont {Roth},
  \citenamefont {Buhmann}, \citenamefont {Molenkamp}, \citenamefont {Qi},\ and\
  \citenamefont {Zhang}}]{MarkusKonig11022007}%
  \BibitemOpen
  \bibfield  {author} {\bibinfo {author} {\bibnamefont {Konig}, \bibfnamefont
  {M.}}, \bibinfo {author} {\bibfnamefont {S.}~\bibnamefont {Wiedmann}},
  \bibinfo {author} {\bibfnamefont {C.}~\bibnamefont {Brune}}, \bibinfo
  {author} {\bibfnamefont {A.}~\bibnamefont {Roth}}, \bibinfo {author}
  {\bibfnamefont {H.}~\bibnamefont {Buhmann}}, \bibinfo {author} {\bibfnamefont
  {L.~W.}\ \bibnamefont {Molenkamp}}, \bibinfo {author} {\bibfnamefont {X.-L.}\
  \bibnamefont {Qi}}, \ and\ \bibinfo {author} {\bibfnamefont {S.-C.}\
  \bibnamefont {Zhang}}} (\bibinfo {year} {2007}),\ \href {\doibase
  10.1126/science.1148047} {\bibfield  {journal} {\bibinfo  {journal}
  {Science}\ }\textbf {\bibinfo {volume} {318}},\ \bibinfo {pages}
  {766}}\BibitemShut {NoStop}%
\bibitem [{\citenamefont {Kotetes}(2013)}]{TSC_reflection_Kotetes}%
  \BibitemOpen
  \bibfield  {author} {\bibinfo {author} {\bibnamefont {Kotetes}, \bibfnamefont
  {P.}}} (\bibinfo {year} {2013}),\ \href
  {http://stacks.iop.org/1367-2630/15/i=10/a=105027} {\bibfield  {journal}
  {\bibinfo  {journal} {New Journal of Physics}\ }\textbf {\bibinfo {volume}
  {15}}~(\bibinfo {number} {10}),\ \bibinfo {pages} {105027}}\BibitemShut
  {NoStop}%
\bibitem [{\citenamefont {Kraus}\ \emph {et~al.}(2012)\citenamefont {Kraus},
  \citenamefont {Lahini}, \citenamefont {Ringel}, \citenamefont {Verbin},\ and\
  \citenamefont {Zilberberg}}]{Kraus_topo_quasicrystal}%
  \BibitemOpen
  \bibfield  {author} {\bibinfo {author} {\bibnamefont {Kraus}, \bibfnamefont
  {Y.~E.}}, \bibinfo {author} {\bibfnamefont {Y.}~\bibnamefont {Lahini}},
  \bibinfo {author} {\bibfnamefont {Z.}~\bibnamefont {Ringel}}, \bibinfo
  {author} {\bibfnamefont {M.}~\bibnamefont {Verbin}}, \ and\ \bibinfo {author}
  {\bibfnamefont {O.}~\bibnamefont {Zilberberg}}} (\bibinfo {year} {2012}),\
  \href {\doibase 10.1103/PhysRevLett.109.106402} {\bibfield  {journal}
  {\bibinfo  {journal} {Phys. Rev. Lett.}\ }\textbf {\bibinfo {volume} {109}},\
  \bibinfo {pages} {106402}}\BibitemShut {NoStop}%
\bibitem [{\citenamefont {Kurter}\ \emph {et~al.}(2014)\citenamefont {Kurter},
  \citenamefont {Finck}, \citenamefont {Ghaemi}, \citenamefont {Hor},\ and\
  \citenamefont {Van~Harlingen}}]{KurterHarlingen14}%
  \BibitemOpen
  \bibfield  {author} {\bibinfo {author} {\bibnamefont {Kurter}, \bibfnamefont
  {C.}}, \bibinfo {author} {\bibfnamefont {A.~D.~K.}\ \bibnamefont {Finck}},
  \bibinfo {author} {\bibfnamefont {P.}~\bibnamefont {Ghaemi}}, \bibinfo
  {author} {\bibfnamefont {Y.~S.}\ \bibnamefont {Hor}}, \ and\ \bibinfo
  {author} {\bibfnamefont {D.~J.}\ \bibnamefont {Van~Harlingen}}} (\bibinfo
  {year} {2014}),\ \href {\doibase 10.1103/PhysRevB.90.014501} {\bibfield
  {journal} {\bibinfo  {journal} {Phys. Rev. B}\ }\textbf {\bibinfo {volume}
  {90}},\ \bibinfo {pages} {014501}}\BibitemShut {NoStop}%
\bibitem [{\citenamefont {Landau}\ \emph {et~al.}(1999)\citenamefont {Landau},
  \citenamefont {Lifshitz},\ and\ \citenamefont {Pitaevskii}}]{Landau}%
  \BibitemOpen
  \bibfield  {author} {\bibinfo {author} {\bibnamefont {Landau}, \bibfnamefont
  {L.~D.}}, \bibinfo {author} {\bibfnamefont {E.~M.}\ \bibnamefont {Lifshitz}},
  \ and\ \bibinfo {author} {\bibfnamefont {M.}~\bibnamefont {Pitaevskii}}}
  (\bibinfo {year} {1999}),\ \href@noop {} {\emph {\bibinfo {title}
  {Statistical Physics}}}\ (\bibinfo  {publisher} {Butterworth-Heinemann, New
  York})\BibitemShut {NoStop}%
\bibitem [{\citenamefont {Lapa}\ \emph {et~al.}(2016)\citenamefont {Lapa},
  \citenamefont {Teo},\ and\ \citenamefont {Hughes}}]{Lapa2014}%
  \BibitemOpen
  \bibfield  {author} {\bibinfo {author} {\bibnamefont {Lapa}, \bibfnamefont
  {M.~F.}}, \bibinfo {author} {\bibfnamefont {J.~C.~Y.}\ \bibnamefont {Teo}}, \
  and\ \bibinfo {author} {\bibfnamefont {T.~L.}\ \bibnamefont {Hughes}}}
  (\bibinfo {year} {2016}),\ \href {\doibase 10.1103/PhysRevB.93.115131}
  {\bibfield  {journal} {\bibinfo  {journal} {Phys. Rev. B}\ }\textbf {\bibinfo
  {volume} {93}},\ \bibinfo {pages} {115131}}\BibitemShut {NoStop}%
\bibitem [{\citenamefont {Laughlin}(1981)}]{Laughlin_IQHE}%
  \BibitemOpen
  \bibfield  {author} {\bibinfo {author} {\bibnamefont {Laughlin},
  \bibfnamefont {R.~B.}}} (\bibinfo {year} {1981}),\ \href {\doibase
  10.1103/PhysRevB.23.5632} {\bibfield  {journal} {\bibinfo  {journal} {Phys.
  Rev. B}\ }\textbf {\bibinfo {volume} {23}},\ \bibinfo {pages}
  {5632}}\BibitemShut {NoStop}%
\bibitem [{\citenamefont {Lawson}\ and\ \citenamefont
  {Michelsohn}(1990)}]{spingeometrybook}%
  \BibitemOpen
  \bibfield  {author} {\bibinfo {author} {\bibnamefont {Lawson}, \bibfnamefont
  {H.~B.}}, \ and\ \bibinfo {author} {\bibfnamefont {M.-L.}\ \bibnamefont
  {Michelsohn}}} (\bibinfo {year} {1990}),\ \href@noop {} {\emph {\bibinfo
  {title} {Spin Geometry}}}\ (\bibinfo  {publisher} {Princeton University
  Press})\BibitemShut {NoStop}%
\bibitem [{\citenamefont {Lee}\ \emph {et~al.}(2014)\citenamefont {Lee},
  \citenamefont {Jiang}, \citenamefont {Houzet}, \citenamefont {Aguado},
  \citenamefont {Lieber},\ and\ \citenamefont {De~Franceschi}}]{Lee_1DSC_semi}%
  \BibitemOpen
  \bibfield  {author} {\bibinfo {author} {\bibnamefont {Lee}, \bibfnamefont
  {E.~J.~H.}}, \bibinfo {author} {\bibfnamefont {X.}~\bibnamefont {Jiang}},
  \bibinfo {author} {\bibfnamefont {M.}~\bibnamefont {Houzet}}, \bibinfo
  {author} {\bibfnamefont {R.}~\bibnamefont {Aguado}}, \bibinfo {author}
  {\bibfnamefont {C.~M.}\ \bibnamefont {Lieber}}, \ and\ \bibinfo {author}
  {\bibfnamefont {S.}~\bibnamefont {De~Franceschi}}} (\bibinfo {year} {2014}),\
  \href@noop {} {\bibfield  {journal} {\bibinfo  {journal} {Nat. Nanotechnol.}\
  }\textbf {\bibinfo {volume} {9}},\ \bibinfo {pages} {79}}\BibitemShut
  {NoStop}%
\bibitem [{\citenamefont {Lee}\ and\ \citenamefont {Ryu}(2008)}]{LeeRyu2008}%
  \BibitemOpen
  \bibfield  {author} {\bibinfo {author} {\bibnamefont {Lee}, \bibfnamefont
  {S.-S.}}, \ and\ \bibinfo {author} {\bibfnamefont {S.}~\bibnamefont {Ryu}}}
  (\bibinfo {year} {2008}),\ \href {\doibase 10.1103/PhysRevLett.100.186807}
  {\bibfield  {journal} {\bibinfo  {journal} {Phys. Rev. Lett.}\ }\textbf
  {\bibinfo {volume} {100}},\ \bibinfo {pages} {186807}}\BibitemShut {NoStop}%
\bibitem [{\citenamefont {Leggett}(1975)}]{Leggett75}%
  \BibitemOpen
  \bibfield  {author} {\bibinfo {author} {\bibnamefont {Leggett}, \bibfnamefont
  {A.~J.}}} (\bibinfo {year} {1975}),\ \href {\doibase
  10.1103/RevModPhys.47.331} {\bibfield  {journal} {\bibinfo  {journal} {Rev.
  Mod. Phys.}\ }\textbf {\bibinfo {volume} {47}},\ \bibinfo {pages}
  {331}}\BibitemShut {NoStop}%
\bibitem [{\citenamefont {Leggett}(2006)}]{Leggettbook}%
  \BibitemOpen
  \bibfield  {author} {\bibinfo {author} {\bibnamefont {Leggett}, \bibfnamefont
  {A.~J.}}} (\bibinfo {year} {2006}),\ \href@noop {} {\emph {\bibinfo {title}
  {Quantum Liquids: Bose Condensation and Cooper Pairing in Condensed-Matter
  Systems}}}\ (\bibinfo  {publisher} {Oxford University Press})\BibitemShut
  {NoStop}%
\bibitem [{\citenamefont {{Levin}}(2013)}]{Levin2013}%
  \BibitemOpen
  \bibfield  {author} {\bibinfo {author} {\bibnamefont {{Levin}}, \bibfnamefont
  {M.}}} (\bibinfo {year} {2013}),\ \href {\doibase 10.1103/PhysRevX.3.021009}
  {\bibfield  {journal} {\bibinfo  {journal} {Phys. Rev. X}\ }\textbf {\bibinfo
  {volume} {3}},\ \bibinfo {eid} {021009}}\BibitemShut {NoStop}%
\bibitem [{\citenamefont {{Levin}}\ and\ \citenamefont
  {{Gu}}(2012)}]{Levin2012}%
  \BibitemOpen
  \bibfield  {author} {\bibinfo {author} {\bibnamefont {{Levin}}, \bibfnamefont
  {M.}}, \ and\ \bibinfo {author} {\bibfnamefont {Z.-C.}\ \bibnamefont {{Gu}}}}
  (\bibinfo {year} {2012}),\ \href {\doibase 10.1103/PhysRevB.86.115109}
  {\bibfield  {journal} {\bibinfo  {journal} {\prb}\ }\textbf {\bibinfo
  {volume} {86}},\ \bibinfo {eid} {115109}}\BibitemShut {NoStop}%
\bibitem [{\citenamefont {Levin}\ and\ \citenamefont
  {Stern}(2009)}]{LevinSternFTI2009}%
  \BibitemOpen
  \bibfield  {author} {\bibinfo {author} {\bibnamefont {Levin}, \bibfnamefont
  {M.}}, \ and\ \bibinfo {author} {\bibfnamefont {A.}~\bibnamefont {Stern}}}
  (\bibinfo {year} {2009}),\ \href {\doibase 10.1103/PhysRevLett.103.196803}
  {\bibfield  {journal} {\bibinfo  {journal} {Phys. Rev. Lett.}\ }\textbf
  {\bibinfo {volume} {103}},\ \bibinfo {pages} {196803}}\BibitemShut {NoStop}%
\bibitem [{\citenamefont {{Levin}}\ and\ \citenamefont
  {{Stern}}(2012)}]{LevinStern2012}%
  \BibitemOpen
  \bibfield  {author} {\bibinfo {author} {\bibnamefont {{Levin}}, \bibfnamefont
  {M.}}, \ and\ \bibinfo {author} {\bibfnamefont {A.}~\bibnamefont {{Stern}}}}
  (\bibinfo {year} {2012}),\ \href {\doibase 10.1103/PhysRevB.86.115131}
  {\bibfield  {journal} {\bibinfo  {journal} {\prb}\ }\textbf {\bibinfo
  {volume} {86}},\ \bibinfo {eid} {115131}}\BibitemShut {NoStop}%
\bibitem [{\citenamefont {Li}\ \emph {et~al.}(2009)\citenamefont {Li},
  \citenamefont {Chu}, \citenamefont {Jain},\ and\ \citenamefont
  {Shen}}]{top_anderson_insul_Shen_PRL09}%
  \BibitemOpen
  \bibfield  {author} {\bibinfo {author} {\bibnamefont {Li}, \bibfnamefont
  {J.}}, \bibinfo {author} {\bibfnamefont {R.-L.}\ \bibnamefont {Chu}},
  \bibinfo {author} {\bibfnamefont {J.~K.}\ \bibnamefont {Jain}}, \ and\
  \bibinfo {author} {\bibfnamefont {S.-Q.}\ \bibnamefont {Shen}}} (\bibinfo
  {year} {2009}),\ \href {\doibase 10.1103/PhysRevLett.102.136806} {\bibfield
  {journal} {\bibinfo  {journal} {Phys. Rev. Lett.}\ }\textbf {\bibinfo
  {volume} {102}},\ \bibinfo {pages} {136806}}\BibitemShut {NoStop}%
\bibitem [{\citenamefont {Li}\ \emph {et~al.}(2016)\citenamefont {Li},
  \citenamefont {Kharzeev}, \citenamefont {Zhang}, \citenamefont {Huang},
  \citenamefont {Pletikosic}, \citenamefont {Fedorov}, \citenamefont {Zhong},
  \citenamefont {Schneeloch}, \citenamefont {Gu},\ and\ \citenamefont
  {Valla}}]{ZrTe_Dirac_2014}%
  \BibitemOpen
  \bibfield  {author} {\bibinfo {author} {\bibnamefont {Li}, \bibfnamefont
  {Q.}}, \bibinfo {author} {\bibfnamefont {D.~E.}\ \bibnamefont {Kharzeev}},
  \bibinfo {author} {\bibfnamefont {C.}~\bibnamefont {Zhang}}, \bibinfo
  {author} {\bibfnamefont {Y.}~\bibnamefont {Huang}}, \bibinfo {author}
  {\bibfnamefont {I.}~\bibnamefont {Pletikosic}}, \bibinfo {author}
  {\bibfnamefont {A.~V.}\ \bibnamefont {Fedorov}}, \bibinfo {author}
  {\bibfnamefont {R.~D.}\ \bibnamefont {Zhong}}, \bibinfo {author}
  {\bibfnamefont {J.~A.}\ \bibnamefont {Schneeloch}}, \bibinfo {author}
  {\bibfnamefont {G.~D.}\ \bibnamefont {Gu}}, \ and\ \bibinfo {author}
  {\bibfnamefont {T.}~\bibnamefont {Valla}}} (\bibinfo {year} {2016}),\ \href
  {http://dx.doi.org/10.1038/nphys3648} {\bibfield  {journal} {\bibinfo
  {journal} {Nat Phys}\ }\textbf {\bibinfo {volume} {advance online
  publication}},\ }\BibitemShut {NoStop}%
\bibitem [{\citenamefont {Liang}\ \emph {et~al.}(2015)\citenamefont {Liang},
  \citenamefont {Gibson}, \citenamefont {Ali}, \citenamefont {Liu},
  \citenamefont {Cava},\ and\ \citenamefont {Ong}}]{Liang_Dirac_magento}%
  \BibitemOpen
  \bibfield  {author} {\bibinfo {author} {\bibnamefont {Liang}, \bibfnamefont
  {T.}}, \bibinfo {author} {\bibfnamefont {Q.}~\bibnamefont {Gibson}}, \bibinfo
  {author} {\bibfnamefont {M.~N.}\ \bibnamefont {Ali}}, \bibinfo {author}
  {\bibfnamefont {M.}~\bibnamefont {Liu}}, \bibinfo {author} {\bibfnamefont
  {R.~J.}\ \bibnamefont {Cava}}, \ and\ \bibinfo {author} {\bibfnamefont
  {N.~P.}\ \bibnamefont {Ong}}} (\bibinfo {year} {2015}),\ \href
  {http://dx.doi.org/10.1038/nmat4143} {\bibfield  {journal} {\bibinfo
  {journal} {Nat Mater}\ }\textbf {\bibinfo {volume} {14}}~(\bibinfo {number}
  {3}),\ \bibinfo {pages} {280}}\BibitemShut {NoStop}%
\bibitem [{\citenamefont {Linder}\ \emph {et~al.}(2010)\citenamefont {Linder},
  \citenamefont {Tanaka}, \citenamefont {Yokoyama}, \citenamefont {Sudb\o{}},\
  and\ \citenamefont {Nagaosa}}]{linder_tanaka_PRL_10}%
  \BibitemOpen
  \bibfield  {author} {\bibinfo {author} {\bibnamefont {Linder}, \bibfnamefont
  {J.}}, \bibinfo {author} {\bibfnamefont {Y.}~\bibnamefont {Tanaka}}, \bibinfo
  {author} {\bibfnamefont {T.}~\bibnamefont {Yokoyama}}, \bibinfo {author}
  {\bibfnamefont {A.}~\bibnamefont {Sudb\o{}}}, \ and\ \bibinfo {author}
  {\bibfnamefont {N.}~\bibnamefont {Nagaosa}}} (\bibinfo {year} {2010}),\ \href
  {\doibase 10.1103/PhysRevLett.104.067001} {\bibfield  {journal} {\bibinfo
  {journal} {Phys. Rev. Lett.}\ }\textbf {\bibinfo {volume} {104}},\ \bibinfo
  {pages} {067001}}\BibitemShut {NoStop}%
\bibitem [{\citenamefont {Lindner}\ \emph {et~al.}(2012)\citenamefont
  {Lindner}, \citenamefont {Berg}, \citenamefont {Refael},\ and\ \citenamefont
  {Stern}}]{LindnerBergRefaelStern}%
  \BibitemOpen
  \bibfield  {author} {\bibinfo {author} {\bibnamefont {Lindner}, \bibfnamefont
  {N.~H.}}, \bibinfo {author} {\bibfnamefont {E.}~\bibnamefont {Berg}},
  \bibinfo {author} {\bibfnamefont {G.}~\bibnamefont {Refael}}, \ and\ \bibinfo
  {author} {\bibfnamefont {A.}~\bibnamefont {Stern}}} (\bibinfo {year}
  {2012}),\ \href {\doibase 10.1103/PhysRevX.2.041002} {\bibfield  {journal}
  {\bibinfo  {journal} {Phys. Rev. X}\ }\textbf {\bibinfo {volume} {2}},\
  \bibinfo {pages} {041002}}\BibitemShut {NoStop}%
\bibitem [{\citenamefont {Lindner}\ \emph {et~al.}(2011)\citenamefont
  {Lindner}, \citenamefont {Refael},\ and\ \citenamefont
  {Galitski}}]{lindnerFloquetTI_NatPhys11}%
  \BibitemOpen
  \bibfield  {author} {\bibinfo {author} {\bibnamefont {Lindner}, \bibfnamefont
  {N.~H.}}, \bibinfo {author} {\bibfnamefont {G.}~\bibnamefont {Refael}}, \
  and\ \bibinfo {author} {\bibfnamefont {V.}~\bibnamefont {Galitski}}}
  (\bibinfo {year} {2011}),\ \href {http://dx.doi.org/10.1038/nphys1926}
  {\bibfield  {journal} {\bibinfo  {journal} {Nat Phys}\ }\textbf {\bibinfo
  {volume} {7}}~(\bibinfo {number} {6}),\ \bibinfo {pages} {490}}\BibitemShut
  {NoStop}%
\bibitem [{\citenamefont {Liu}\ \emph {et~al.}(2008{\natexlab{a}})\citenamefont
  {Liu}, \citenamefont {Hughes}, \citenamefont {Qi}, \citenamefont {Wang},\
  and\ \citenamefont {Zhang}}]{LiuZhangTypeIISemiCondPRL08}%
  \BibitemOpen
  \bibfield  {author} {\bibinfo {author} {\bibnamefont {Liu}, \bibfnamefont
  {C.}}, \bibinfo {author} {\bibfnamefont {T.~L.}\ \bibnamefont {Hughes}},
  \bibinfo {author} {\bibfnamefont {X.-L.}\ \bibnamefont {Qi}}, \bibinfo
  {author} {\bibfnamefont {K.}~\bibnamefont {Wang}}, \ and\ \bibinfo {author}
  {\bibfnamefont {S.-C.}\ \bibnamefont {Zhang}}} (\bibinfo {year}
  {2008}{\natexlab{a}}),\ \href {\doibase 10.1103/PhysRevLett.100.236601}
  {\bibfield  {journal} {\bibinfo  {journal} {Phys. Rev. Lett.}\ }\textbf
  {\bibinfo {volume} {100}},\ \bibinfo {pages} {236601}}\BibitemShut {NoStop}%
\bibitem [{\citenamefont {Liu}\ \emph {et~al.}(2008{\natexlab{b}})\citenamefont
  {Liu}, \citenamefont {Qi}, \citenamefont {Dai}, \citenamefont {Fang},\ and\
  \citenamefont {Zhang}}]{LiuZhang08}%
  \BibitemOpen
  \bibfield  {author} {\bibinfo {author} {\bibnamefont {Liu}, \bibfnamefont
  {C.-X.}}, \bibinfo {author} {\bibfnamefont {X.-L.}\ \bibnamefont {Qi}},
  \bibinfo {author} {\bibfnamefont {X.}~\bibnamefont {Dai}}, \bibinfo {author}
  {\bibfnamefont {Z.}~\bibnamefont {Fang}}, \ and\ \bibinfo {author}
  {\bibfnamefont {S.-C.}\ \bibnamefont {Zhang}}} (\bibinfo {year}
  {2008}{\natexlab{b}}),\ \href {\doibase 10.1103/PhysRevLett.101.146802}
  {\bibfield  {journal} {\bibinfo  {journal} {Phys. Rev. Lett.}\ }\textbf
  {\bibinfo {volume} {101}},\ \bibinfo {pages} {146802}}\BibitemShut {NoStop}%
\bibitem [{\citenamefont {Liu}\ \emph {et~al.}(2012)\citenamefont {Liu},
  \citenamefont {Qi},\ and\ \citenamefont {Zhang}}]{Liu:2011fk}%
  \BibitemOpen
  \bibfield  {author} {\bibinfo {author} {\bibnamefont {Liu}, \bibfnamefont
  {C.-X.}}, \bibinfo {author} {\bibfnamefont {X.-L.}\ \bibnamefont {Qi}}, \
  and\ \bibinfo {author} {\bibfnamefont {S.-C.}\ \bibnamefont {Zhang}}}
  (\bibinfo {year} {2012}),\ \href {\doibase
  http://dx.doi.org/10.1016/j.physe.2011.11.005} {\bibfield  {journal}
  {\bibinfo  {journal} {Physica E}\ }\textbf {\bibinfo {volume} {44}},\
  \bibinfo {pages} {906 }}\BibitemShut {NoStop}%
\bibitem [{\citenamefont {Liu}\ \emph {et~al.}(2013{\natexlab{a}})\citenamefont
  {Liu}, \citenamefont {Ye},\ and\ \citenamefont {Qi}}]{Lu_anomaly_Weyl_2013}%
  \BibitemOpen
  \bibfield  {author} {\bibinfo {author} {\bibnamefont {Liu}, \bibfnamefont
  {C.-X.}}, \bibinfo {author} {\bibfnamefont {P.}~\bibnamefont {Ye}}, \ and\
  \bibinfo {author} {\bibfnamefont {X.-L.}\ \bibnamefont {Qi}}} (\bibinfo
  {year} {2013}{\natexlab{a}}),\ \href {\doibase 10.1103/PhysRevB.87.235306}
  {\bibfield  {journal} {\bibinfo  {journal} {Phys. Rev. B}\ }\textbf {\bibinfo
  {volume} {87}},\ \bibinfo {pages} {235306}}\BibitemShut {NoStop}%
\bibitem [{\citenamefont {Liu}\ \emph {et~al.}(2014{\natexlab{a}})\citenamefont
  {Liu}, \citenamefont {Zhang},\ and\ \citenamefont
  {VanLeeuwen}}]{nonsymmorphic_Liu}%
  \BibitemOpen
  \bibfield  {author} {\bibinfo {author} {\bibnamefont {Liu}, \bibfnamefont
  {C.-X.}}, \bibinfo {author} {\bibfnamefont {R.-X.}\ \bibnamefont {Zhang}}, \
  and\ \bibinfo {author} {\bibfnamefont {B.~K.}\ \bibnamefont {VanLeeuwen}}}
  (\bibinfo {year} {2014}{\natexlab{a}}),\ \href {\doibase
  10.1103/PhysRevB.90.085304} {\bibfield  {journal} {\bibinfo  {journal} {Phys.
  Rev. B}\ }\textbf {\bibinfo {volume} {90}},\ \bibinfo {pages}
  {085304}}\BibitemShut {NoStop}%
\bibitem [{\citenamefont {Liu}\ \emph {et~al.}(2013{\natexlab{b}})\citenamefont
  {Liu}, \citenamefont {Duan},\ and\ \citenamefont {Fu}}]{Liu_surface_TCI}%
  \BibitemOpen
  \bibfield  {author} {\bibinfo {author} {\bibnamefont {Liu}, \bibfnamefont
  {J.}}, \bibinfo {author} {\bibfnamefont {W.}~\bibnamefont {Duan}}, \ and\
  \bibinfo {author} {\bibfnamefont {L.}~\bibnamefont {Fu}}} (\bibinfo {year}
  {2013}{\natexlab{b}}),\ \href {\doibase 10.1103/PhysRevB.88.241303}
  {\bibfield  {journal} {\bibinfo  {journal} {Phys. Rev. B}\ }\textbf {\bibinfo
  {volume} {88}},\ \bibinfo {pages} {241303}}\BibitemShut {NoStop}%
\bibitem [{\citenamefont {Liu}\ \emph {et~al.}(2014{\natexlab{b}})\citenamefont
  {Liu}, \citenamefont {Hsieh}, \citenamefont {Wei}, \citenamefont {Duan},
  \citenamefont {Moodera},\ and\ \citenamefont {Fu}}]{Liu_TCI}%
  \BibitemOpen
  \bibfield  {author} {\bibinfo {author} {\bibnamefont {Liu}, \bibfnamefont
  {J.}}, \bibinfo {author} {\bibfnamefont {T.~H.}\ \bibnamefont {Hsieh}},
  \bibinfo {author} {\bibfnamefont {P.}~\bibnamefont {Wei}}, \bibinfo {author}
  {\bibfnamefont {W.}~\bibnamefont {Duan}}, \bibinfo {author} {\bibfnamefont
  {J.}~\bibnamefont {Moodera}}, \ and\ \bibinfo {author} {\bibfnamefont
  {L.}~\bibnamefont {Fu}}} (\bibinfo {year} {2014}{\natexlab{b}}),\ \href
  {http://dx.doi.org/10.1038/nmat3828} {\bibfield  {journal} {\bibinfo
  {journal} {Nat Mater}\ }\textbf {\bibinfo {volume} {13}}~(\bibinfo {number}
  {2}),\ \bibinfo {pages} {178}}\BibitemShut {NoStop}%
\bibitem [{\citenamefont {Liu}\ \emph {et~al.}(2011)\citenamefont {Liu},
  \citenamefont {Peng}, \citenamefont {Tang}, \citenamefont {Sun},
  \citenamefont {Zhang},\ and\ \citenamefont {Zhong}}]{wenliang_PRB_11}%
  \BibitemOpen
  \bibfield  {author} {\bibinfo {author} {\bibnamefont {Liu}, \bibfnamefont
  {W.}}, \bibinfo {author} {\bibfnamefont {X.}~\bibnamefont {Peng}}, \bibinfo
  {author} {\bibfnamefont {C.}~\bibnamefont {Tang}}, \bibinfo {author}
  {\bibfnamefont {L.}~\bibnamefont {Sun}}, \bibinfo {author} {\bibfnamefont
  {K.}~\bibnamefont {Zhang}}, \ and\ \bibinfo {author} {\bibfnamefont
  {J.}~\bibnamefont {Zhong}}} (\bibinfo {year} {2011}),\ \href {\doibase
  10.1103/PhysRevB.84.245105} {\bibfield  {journal} {\bibinfo  {journal} {Phys.
  Rev. B}\ }\textbf {\bibinfo {volume} {84}},\ \bibinfo {pages}
  {245105}}\BibitemShut {NoStop}%
\bibitem [{\citenamefont {Liu}\ \emph {et~al.}(2014{\natexlab{c}})\citenamefont
  {Liu}, \citenamefont {He},\ and\ \citenamefont {Law}}]{liu_law_PRB_14}%
  \BibitemOpen
  \bibfield  {author} {\bibinfo {author} {\bibnamefont {Liu}, \bibfnamefont
  {X.-J.}}, \bibinfo {author} {\bibfnamefont {J.~J.}\ \bibnamefont {He}}, \
  and\ \bibinfo {author} {\bibfnamefont {K.~T.}\ \bibnamefont {Law}}} (\bibinfo
  {year} {2014}{\natexlab{c}}),\ \href {\doibase 10.1103/PhysRevB.90.235141}
  {\bibfield  {journal} {\bibinfo  {journal} {Phys. Rev. B}\ }\textbf {\bibinfo
  {volume} {90}},\ \bibinfo {pages} {235141}}\BibitemShut {NoStop}%
\bibitem [{\citenamefont {Liu}\ \emph {et~al.}(2014{\natexlab{d}})\citenamefont
  {Liu}, \citenamefont {Jiang}, \citenamefont {Zhou}, \citenamefont {Wang},
  \citenamefont {Zhang}, \citenamefont {Weng}, \citenamefont {Prabhakaran},
  \citenamefont {Mo}, \citenamefont {Peng}, \citenamefont {Dudin},
  \citenamefont {Kim}, \citenamefont {Hoesch}, \citenamefont {Fang},
  \citenamefont {Dai}, \citenamefont {Shen}, \citenamefont {Feng},
  \citenamefont {Hussain},\ and\ \citenamefont {Chen}}]{Cd3As2Chen2014}%
  \BibitemOpen
  \bibfield  {author} {\bibinfo {author} {\bibnamefont {Liu}, \bibfnamefont
  {Z.~K.}}, \bibinfo {author} {\bibfnamefont {J.}~\bibnamefont {Jiang}},
  \bibinfo {author} {\bibfnamefont {B.}~\bibnamefont {Zhou}}, \bibinfo {author}
  {\bibfnamefont {Z.~J.}\ \bibnamefont {Wang}}, \bibinfo {author}
  {\bibfnamefont {Y.}~\bibnamefont {Zhang}}, \bibinfo {author} {\bibfnamefont
  {H.~M.}\ \bibnamefont {Weng}}, \bibinfo {author} {\bibfnamefont
  {D.}~\bibnamefont {Prabhakaran}}, \bibinfo {author} {\bibfnamefont {S.-K.}\
  \bibnamefont {Mo}}, \bibinfo {author} {\bibfnamefont {H.}~\bibnamefont
  {Peng}}, \bibinfo {author} {\bibfnamefont {P.}~\bibnamefont {Dudin}},
  \bibinfo {author} {\bibfnamefont {T.}~\bibnamefont {Kim}}, \bibinfo {author}
  {\bibfnamefont {M.}~\bibnamefont {Hoesch}}, \bibinfo {author} {\bibfnamefont
  {Z.}~\bibnamefont {Fang}}, \bibinfo {author} {\bibfnamefont {X.}~\bibnamefont
  {Dai}}, \bibinfo {author} {\bibfnamefont {Z.~X.}\ \bibnamefont {Shen}},
  \bibinfo {author} {\bibfnamefont {D.~L.}\ \bibnamefont {Feng}}, \bibinfo
  {author} {\bibfnamefont {Z.}~\bibnamefont {Hussain}}, \ and\ \bibinfo
  {author} {\bibfnamefont {Y.~L.}\ \bibnamefont {Chen}}} (\bibinfo {year}
  {2014}{\natexlab{d}}),\ \href@noop {} {\bibfield  {journal} {\bibinfo
  {journal} {Nat. Mater.}\ }\textbf {\bibinfo {volume} {13}},\ \bibinfo {pages}
  {677}}\BibitemShut {NoStop}%
\bibitem [{\citenamefont {Liu}\ \emph {et~al.}(2014{\natexlab{e}})\citenamefont
  {Liu}, \citenamefont {Zhou}, \citenamefont {Zhang}, \citenamefont {Wang},
  \citenamefont {Weng}, \citenamefont {Prabhakaran}, \citenamefont {Mo},
  \citenamefont {Shen}, \citenamefont {Fang}, \citenamefont {Dai},
  \citenamefont {Hussain},\ and\ \citenamefont {Chen}}]{Liu21022014}%
  \BibitemOpen
  \bibfield  {author} {\bibinfo {author} {\bibnamefont {Liu}, \bibfnamefont
  {Z.~K.}}, \bibinfo {author} {\bibfnamefont {B.}~\bibnamefont {Zhou}},
  \bibinfo {author} {\bibfnamefont {Y.}~\bibnamefont {Zhang}}, \bibinfo
  {author} {\bibfnamefont {Z.~J.}\ \bibnamefont {Wang}}, \bibinfo {author}
  {\bibfnamefont {H.~M.}\ \bibnamefont {Weng}}, \bibinfo {author}
  {\bibfnamefont {D.}~\bibnamefont {Prabhakaran}}, \bibinfo {author}
  {\bibfnamefont {S.-K.}\ \bibnamefont {Mo}}, \bibinfo {author} {\bibfnamefont
  {Z.~X.}\ \bibnamefont {Shen}}, \bibinfo {author} {\bibfnamefont
  {Z.}~\bibnamefont {Fang}}, \bibinfo {author} {\bibfnamefont {X.}~\bibnamefont
  {Dai}}, \bibinfo {author} {\bibfnamefont {Z.}~\bibnamefont {Hussain}}, \ and\
  \bibinfo {author} {\bibfnamefont {Y.~L.}\ \bibnamefont {Chen}}} (\bibinfo
  {year} {2014}{\natexlab{e}}),\ \href {\doibase 10.1126/science.1245085}
  {\bibfield  {journal} {\bibinfo  {journal} {Science}\ }\textbf {\bibinfo
  {volume} {343}}~(\bibinfo {number} {6173}),\ \bibinfo {pages}
  {864}}\BibitemShut {NoStop}%
\bibitem [{\citenamefont {Loring}\ and\ \citenamefont
  {Hastings}(2010)}]{loring_hastings_EPL_10}%
  \BibitemOpen
  \bibfield  {author} {\bibinfo {author} {\bibnamefont {Loring}, \bibfnamefont
  {T.~A.}}, \ and\ \bibinfo {author} {\bibfnamefont {M.~B.}\ \bibnamefont
  {Hastings}}} (\bibinfo {year} {2010}),\ \href
  {http://stacks.iop.org/0295-5075/92/i=6/a=67004} {\bibfield  {journal}
  {\bibinfo  {journal} {EPL (Europhysics Letters)}\ }\textbf {\bibinfo {volume}
  {92}}~(\bibinfo {number} {6}),\ \bibinfo {pages} {67004}}\BibitemShut
  {NoStop}%
\bibitem [{\citenamefont {Lu}\ \emph {et~al.}(2016)\citenamefont {Lu},
  \citenamefont {Fang}, \citenamefont {Fu}, \citenamefont {Johnson},
  \citenamefont {Joannopoulos},\ and\ \citenamefont
  {Soljacic}}]{photonic_TI_Fu}%
  \BibitemOpen
  \bibfield  {author} {\bibinfo {author} {\bibnamefont {Lu}, \bibfnamefont
  {L.}}, \bibinfo {author} {\bibfnamefont {C.}~\bibnamefont {Fang}}, \bibinfo
  {author} {\bibfnamefont {L.}~\bibnamefont {Fu}}, \bibinfo {author}
  {\bibfnamefont {S.~G.}\ \bibnamefont {Johnson}}, \bibinfo {author}
  {\bibfnamefont {J.~D.}\ \bibnamefont {Joannopoulos}}, \ and\ \bibinfo
  {author} {\bibfnamefont {M.}~\bibnamefont {Soljacic}}} (\bibinfo {year}
  {2016}),\ \href {http://dx.doi.org/10.1038/nphys3611} {\bibfield  {journal}
  {\bibinfo  {journal} {Nat Phys}\ }\textbf {\bibinfo {volume} {advance online
  publication}},\ }\BibitemShut {NoStop}%
\bibitem [{\citenamefont {Lu}\ \emph {et~al.}(2013)\citenamefont {Lu},
  \citenamefont {Fu}, \citenamefont {Joannopoulos},\ and\ \citenamefont
  {Soljacic}}]{Lu_early_photonic_Weyl}%
  \BibitemOpen
  \bibfield  {author} {\bibinfo {author} {\bibnamefont {Lu}, \bibfnamefont
  {L.}}, \bibinfo {author} {\bibfnamefont {L.}~\bibnamefont {Fu}}, \bibinfo
  {author} {\bibfnamefont {J.~D.}\ \bibnamefont {Joannopoulos}}, \ and\
  \bibinfo {author} {\bibfnamefont {M.}~\bibnamefont {Soljacic}}} (\bibinfo
  {year} {2013}),\ \href {http://dx.doi.org/10.1038/nphoton.2013.42} {\bibfield
   {journal} {\bibinfo  {journal} {Nat Photon}\ }\textbf {\bibinfo {volume}
  {7}}~(\bibinfo {number} {4}),\ \bibinfo {pages} {294}}\BibitemShut {NoStop}%
\bibitem [{\citenamefont {Lu}\ \emph {et~al.}(2015)\citenamefont {Lu},
  \citenamefont {Wang}, \citenamefont {Ye}, \citenamefont {Ran}, \citenamefont
  {Fu}, \citenamefont {Joannopoulos},\ and\ \citenamefont {Solja{\v
  c}i{\'c}}}]{Lu_photonic_Weyl_2015}%
  \BibitemOpen
  \bibfield  {author} {\bibinfo {author} {\bibnamefont {Lu}, \bibfnamefont
  {L.}}, \bibinfo {author} {\bibfnamefont {Z.}~\bibnamefont {Wang}}, \bibinfo
  {author} {\bibfnamefont {D.}~\bibnamefont {Ye}}, \bibinfo {author}
  {\bibfnamefont {L.}~\bibnamefont {Ran}}, \bibinfo {author} {\bibfnamefont
  {L.}~\bibnamefont {Fu}}, \bibinfo {author} {\bibfnamefont {J.~D.}\
  \bibnamefont {Joannopoulos}}, \ and\ \bibinfo {author} {\bibfnamefont
  {M.}~\bibnamefont {Solja{\v c}i{\'c}}}} (\bibinfo {year} {2015}),\ \href
  {\doibase 10.1126/science.aaa9273} {\bibfield  {journal} {\bibinfo  {journal}
  {Science}\ }\textbf {\bibinfo {volume} {349}}~(\bibinfo {number} {6248}),\
  \bibinfo {pages} {622}}\BibitemShut {NoStop}%
\bibitem [{\citenamefont {{Lu}}\ and\ \citenamefont {{Lee}}(2014)}]{LuLee2014}%
  \BibitemOpen
  \bibfield  {author} {\bibinfo {author} {\bibnamefont {{Lu}}, \bibfnamefont
  {Y.-M.}}, \ and\ \bibinfo {author} {\bibfnamefont {D.-H.}\ \bibnamefont
  {{Lee}}}} (\bibinfo {year} {2014}),\ \href@noop {} {\ }\Eprint
  {http://arxiv.org/abs/1403.5558} {arXiv:1403.5558} \BibitemShut {NoStop}%
\bibitem [{\citenamefont {Lu}\ and\ \citenamefont {Lee}(2014)}]{lu_lee_PRB_14}%
  \BibitemOpen
  \bibfield  {author} {\bibinfo {author} {\bibnamefont {Lu}, \bibfnamefont
  {Y.-M.}}, \ and\ \bibinfo {author} {\bibfnamefont {D.-H.}\ \bibnamefont
  {Lee}}} (\bibinfo {year} {2014}),\ \href {\doibase
  10.1103/PhysRevB.89.184417} {\bibfield  {journal} {\bibinfo  {journal} {Phys.
  Rev. B}\ }\textbf {\bibinfo {volume} {89}},\ \bibinfo {pages}
  {184417}}\BibitemShut {NoStop}%
\bibitem [{\citenamefont {{Lu}}\ and\ \citenamefont
  {{Vishwanath}}(2012)}]{LuVishwanath2012}%
  \BibitemOpen
  \bibfield  {author} {\bibinfo {author} {\bibnamefont {{Lu}}, \bibfnamefont
  {Y.-M.}}, \ and\ \bibinfo {author} {\bibfnamefont {A.}~\bibnamefont
  {{Vishwanath}}}} (\bibinfo {year} {2012}),\ \href {\doibase
  10.1103/PhysRevB.86.125119} {\bibfield  {journal} {\bibinfo  {journal}
  {\prb}\ }\textbf {\bibinfo {volume} {86}},\ \bibinfo {eid}
  {125119}}\BibitemShut {NoStop}%
\bibitem [{\citenamefont {{Lu}}\ and\ \citenamefont
  {{Vishwanath}}(2013)}]{LuVishwanath2013}%
  \BibitemOpen
  \bibfield  {author} {\bibinfo {author} {\bibnamefont {{Lu}}, \bibfnamefont
  {Y.-M.}}, \ and\ \bibinfo {author} {\bibfnamefont {A.}~\bibnamefont
  {{Vishwanath}}}} (\bibinfo {year} {2013}),\ \href@noop {} {\ }\Eprint
  {http://arxiv.org/abs/1302.2634} {arXiv:1302.2634} \BibitemShut {NoStop}%
\bibitem [{\citenamefont {Ludwig}\ \emph {et~al.}(1994)\citenamefont {Ludwig},
  \citenamefont {Fisher}, \citenamefont {Shankar},\ and\ \citenamefont
  {Grinstein}}]{LudwigFisherShankarGrinstein1994}%
  \BibitemOpen
  \bibfield  {author} {\bibinfo {author} {\bibnamefont {Ludwig}, \bibfnamefont
  {A.~W.~W.}}, \bibinfo {author} {\bibfnamefont {M.~P.~A.}\ \bibnamefont
  {Fisher}}, \bibinfo {author} {\bibfnamefont {R.}~\bibnamefont {Shankar}}, \
  and\ \bibinfo {author} {\bibfnamefont {G.}~\bibnamefont {Grinstein}}}
  (\bibinfo {year} {1994}),\ \href {\doibase 10.1103/PhysRevB.50.7526}
  {\bibfield  {journal} {\bibinfo  {journal} {Phys. Rev. B}\ }\textbf {\bibinfo
  {volume} {50}},\ \bibinfo {pages} {7526}}\BibitemShut {NoStop}%
\bibitem [{\citenamefont {Luke}\ \emph {et~al.}(1998)\citenamefont {Luke},
  \citenamefont {Fudamoto}, \citenamefont {Kojima}, \citenamefont {Larkin},
  \citenamefont {Merrin}, \citenamefont {Nachumi}, \citenamefont {Uemura},
  \citenamefont {Maeno}, \citenamefont {Mao}, \citenamefont {Mori},
  \citenamefont {Nakamura},\ and\ \citenamefont {Sigrist}}]{LukeSigrist98}%
  \BibitemOpen
  \bibfield  {author} {\bibinfo {author} {\bibnamefont {Luke}, \bibfnamefont
  {G.~M.}}, \bibinfo {author} {\bibfnamefont {Y.}~\bibnamefont {Fudamoto}},
  \bibinfo {author} {\bibfnamefont {K.~M.}\ \bibnamefont {Kojima}}, \bibinfo
  {author} {\bibfnamefont {M.~I.}\ \bibnamefont {Larkin}}, \bibinfo {author}
  {\bibfnamefont {J.}~\bibnamefont {Merrin}}, \bibinfo {author} {\bibfnamefont
  {B.}~\bibnamefont {Nachumi}}, \bibinfo {author} {\bibfnamefont {Y.~J.}\
  \bibnamefont {Uemura}}, \bibinfo {author} {\bibfnamefont {Y.}~\bibnamefont
  {Maeno}}, \bibinfo {author} {\bibfnamefont {Z.~Q.}\ \bibnamefont {Mao}},
  \bibinfo {author} {\bibfnamefont {Y.}~\bibnamefont {Mori}}, \bibinfo {author}
  {\bibfnamefont {H.}~\bibnamefont {Nakamura}}, \ and\ \bibinfo {author}
  {\bibfnamefont {M.}~\bibnamefont {Sigrist}}} (\bibinfo {year} {1998}),\ \href
  {\doibase 10.1038/29038} {\bibfield  {journal} {\bibinfo  {journal} {Nature}\
  }\textbf {\bibinfo {volume} {394}},\ \bibinfo {pages} {558}}\BibitemShut
  {NoStop}%
\bibitem [{\citenamefont {Lutchyn}\ \emph {et~al.}(2010)\citenamefont
  {Lutchyn}, \citenamefont {Sau},\ and\ \citenamefont
  {Das~Sarma}}]{Roman_SC_semi}%
  \BibitemOpen
  \bibfield  {author} {\bibinfo {author} {\bibnamefont {Lutchyn}, \bibfnamefont
  {R.~M.}}, \bibinfo {author} {\bibfnamefont {J.~D.}\ \bibnamefont {Sau}}, \
  and\ \bibinfo {author} {\bibfnamefont {S.}~\bibnamefont {Das~Sarma}}}
  (\bibinfo {year} {2010}),\ \href {\doibase 10.1103/PhysRevLett.105.077001}
  {\bibfield  {journal} {\bibinfo  {journal} {Phys. Rev. Lett.}\ }\textbf
  {\bibinfo {volume} {105}},\ \bibinfo {pages} {077001}}\BibitemShut {NoStop}%
\bibitem [{\citenamefont {Luttinger}(1964)}]{Luttinger64}%
  \BibitemOpen
  \bibfield  {author} {\bibinfo {author} {\bibnamefont {Luttinger},
  \bibfnamefont {J.~M.}}} (\bibinfo {year} {1964}),\ \href {\doibase
  10.1103/PhysRev.135.A1505} {\bibfield  {journal} {\bibinfo  {journal} {Phys.
  Rev.}\ }\textbf {\bibinfo {volume} {135}},\ \bibinfo {pages}
  {A1505}}\BibitemShut {NoStop}%
\bibitem [{\citenamefont {Lv}\ \emph {et~al.}(2015)\citenamefont {Lv},
  \citenamefont {Weng}, \citenamefont {Fu}, \citenamefont {Wang}, \citenamefont
  {Miao}, \citenamefont {Ma}, \citenamefont {Richard}, \citenamefont {Huang},
  \citenamefont {Zhao}, \citenamefont {Chen}, \citenamefont {Fang},
  \citenamefont {Dai}, \citenamefont {Qian},\ and\ \citenamefont
  {Ding}}]{Weyl_discovery_TaAs}%
  \BibitemOpen
  \bibfield  {author} {\bibinfo {author} {\bibnamefont {Lv}, \bibfnamefont
  {B.~Q.}}, \bibinfo {author} {\bibfnamefont {H.~M.}\ \bibnamefont {Weng}},
  \bibinfo {author} {\bibfnamefont {B.~B.}\ \bibnamefont {Fu}}, \bibinfo
  {author} {\bibfnamefont {X.~P.}\ \bibnamefont {Wang}}, \bibinfo {author}
  {\bibfnamefont {H.}~\bibnamefont {Miao}}, \bibinfo {author} {\bibfnamefont
  {J.}~\bibnamefont {Ma}}, \bibinfo {author} {\bibfnamefont {P.}~\bibnamefont
  {Richard}}, \bibinfo {author} {\bibfnamefont {X.~C.}\ \bibnamefont {Huang}},
  \bibinfo {author} {\bibfnamefont {L.~X.}\ \bibnamefont {Zhao}}, \bibinfo
  {author} {\bibfnamefont {G.~F.}\ \bibnamefont {Chen}}, \bibinfo {author}
  {\bibfnamefont {Z.}~\bibnamefont {Fang}}, \bibinfo {author} {\bibfnamefont
  {X.}~\bibnamefont {Dai}}, \bibinfo {author} {\bibfnamefont {T.}~\bibnamefont
  {Qian}}, \ and\ \bibinfo {author} {\bibfnamefont {H.}~\bibnamefont {Ding}}}
  (\bibinfo {year} {2015}),\ \href {\doibase 10.1103/PhysRevX.5.031013}
  {\bibfield  {journal} {\bibinfo  {journal} {Phys. Rev. X}\ }\textbf {\bibinfo
  {volume} {5}},\ \bibinfo {pages} {031013}}\BibitemShut {NoStop}%
\bibitem [{\citenamefont {Ma\~nes}\ \emph {et~al.}(2007)\citenamefont
  {Ma\~nes}, \citenamefont {Guinea},\ and\ \citenamefont
  {Vozmediano}}]{GuineaVozmedianoPRB07}%
  \BibitemOpen
  \bibfield  {author} {\bibinfo {author} {\bibnamefont {Ma\~nes}, \bibfnamefont
  {J.~L.}}, \bibinfo {author} {\bibfnamefont {F.}~\bibnamefont {Guinea}}, \
  and\ \bibinfo {author} {\bibfnamefont {M.~A.~H.}\ \bibnamefont {Vozmediano}}}
  (\bibinfo {year} {2007}),\ \href {\doibase 10.1103/PhysRevB.75.155424}
  {\bibfield  {journal} {\bibinfo  {journal} {Phys. Rev. B}\ }\textbf {\bibinfo
  {volume} {75}},\ \bibinfo {pages} {155424}}\BibitemShut {NoStop}%
\bibitem [{\citenamefont {Maciejko}\ and\ \citenamefont
  {Fiete}(2015)}]{fiete_review_natPhys15}%
  \BibitemOpen
  \bibfield  {author} {\bibinfo {author} {\bibnamefont {Maciejko},
  \bibfnamefont {J.}}, \ and\ \bibinfo {author} {\bibfnamefont {G.~A.}\
  \bibnamefont {Fiete}}} (\bibinfo {year} {2015}),\ \href
  {http://dx.doi.org/10.1038/nphys3311} {\bibfield  {journal} {\bibinfo
  {journal} {Nat Phys}\ }\textbf {\bibinfo {volume} {11}}~(\bibinfo {number}
  {5}),\ \bibinfo {pages} {385}}\BibitemShut {NoStop}%
\bibitem [{\citenamefont {Maciejko}\ \emph {et~al.}(2010)\citenamefont
  {Maciejko}, \citenamefont {Qi}, \citenamefont {Karch},\ and\ \citenamefont
  {Zhang}}]{MaciejkoQiKarchZhangFTI2010}%
  \BibitemOpen
  \bibfield  {author} {\bibinfo {author} {\bibnamefont {Maciejko},
  \bibfnamefont {J.}}, \bibinfo {author} {\bibfnamefont {X.-L.}\ \bibnamefont
  {Qi}}, \bibinfo {author} {\bibfnamefont {A.}~\bibnamefont {Karch}}, \ and\
  \bibinfo {author} {\bibfnamefont {S.-C.}\ \bibnamefont {Zhang}}} (\bibinfo
  {year} {2010}),\ \href {\doibase 10.1103/PhysRevLett.105.246809} {\bibfield
  {journal} {\bibinfo  {journal} {Phys. Rev. Lett.}\ }\textbf {\bibinfo
  {volume} {105}},\ \bibinfo {pages} {246809}}\BibitemShut {NoStop}%
\bibitem [{\citenamefont {Maeno}\ \emph {et~al.}(2012)\citenamefont {Maeno},
  \citenamefont {Kittaka}, \citenamefont {Nomura}, \citenamefont {Yonezawa},\
  and\ \citenamefont {Ishida}}]{maeno_review_JPSJ_12}%
  \BibitemOpen
  \bibfield  {author} {\bibinfo {author} {\bibnamefont {Maeno}, \bibfnamefont
  {Y.}}, \bibinfo {author} {\bibfnamefont {S.}~\bibnamefont {Kittaka}},
  \bibinfo {author} {\bibfnamefont {T.}~\bibnamefont {Nomura}}, \bibinfo
  {author} {\bibfnamefont {S.}~\bibnamefont {Yonezawa}}, \ and\ \bibinfo
  {author} {\bibfnamefont {K.}~\bibnamefont {Ishida}}} (\bibinfo {year}
  {2012}),\ \href {\doibase 10.1143/JPSJ.81.011009} {\bibfield  {journal}
  {\bibinfo  {journal} {Journal of the Physical Society of Japan}\ }\textbf
  {\bibinfo {volume} {81}}~(\bibinfo {number} {1}),\ \bibinfo {pages}
  {011009}}\BibitemShut {NoStop}%
\bibitem [{\citenamefont {Maeno}\ \emph {et~al.}(2001)\citenamefont {Maeno},
  \citenamefont {Rice},\ and\ \citenamefont {Sigrist}}]{MaenoRiceSigrist2001}%
  \BibitemOpen
  \bibfield  {author} {\bibinfo {author} {\bibnamefont {Maeno}, \bibfnamefont
  {Y.}}, \bibinfo {author} {\bibfnamefont {T.~M.}\ \bibnamefont {Rice}}, \ and\
  \bibinfo {author} {\bibfnamefont {M.}~\bibnamefont {Sigrist}}} (\bibinfo
  {year} {2001}),\ \href {\doibase 10.1063/1.1349611} {\bibfield  {journal}
  {\bibinfo  {journal} {Physics Today}\ }\textbf {\bibinfo {volume} {54}},\
  \bibinfo {pages} {42}}\BibitemShut {NoStop}%
\bibitem [{\citenamefont {Matsuura}\ \emph {et~al.}(2013)\citenamefont
  {Matsuura}, \citenamefont {Chang}, \citenamefont {Schnyder},\ and\
  \citenamefont {Ryu}}]{matsuuraNJP13}%
  \BibitemOpen
  \bibfield  {author} {\bibinfo {author} {\bibnamefont {Matsuura},
  \bibfnamefont {S.}}, \bibinfo {author} {\bibfnamefont {P.-Y.}\ \bibnamefont
  {Chang}}, \bibinfo {author} {\bibfnamefont {A.~P.}\ \bibnamefont {Schnyder}},
  \ and\ \bibinfo {author} {\bibfnamefont {S.}~\bibnamefont {Ryu}}} (\bibinfo
  {year} {2013}),\ \href@noop {} {\bibfield  {journal} {\bibinfo  {journal}
  {New J. Phys.}\ }\textbf {\bibinfo {volume} {15}},\ \bibinfo {pages}
  {065001}}\BibitemShut {NoStop}%
\bibitem [{\citenamefont {Mendler}\ \emph {et~al.}(2015)\citenamefont
  {Mendler}, \citenamefont {Kotetes},\ and\ \citenamefont {Sch\"on}}]{TSC_C3}%
  \BibitemOpen
  \bibfield  {author} {\bibinfo {author} {\bibnamefont {Mendler}, \bibfnamefont
  {D.}}, \bibinfo {author} {\bibfnamefont {P.}~\bibnamefont {Kotetes}}, \ and\
  \bibinfo {author} {\bibfnamefont {G.}~\bibnamefont {Sch\"on}}} (\bibinfo
  {year} {2015}),\ \href {\doibase 10.1103/PhysRevB.91.155405} {\bibfield
  {journal} {\bibinfo  {journal} {Phys. Rev. B}\ }\textbf {\bibinfo {volume}
  {91}},\ \bibinfo {pages} {155405}}\BibitemShut {NoStop}%
\bibitem [{\citenamefont {Mesaros}\ and\ \citenamefont
  {Ran}(2013)}]{symmetry_enriched_ying_ran_PRB_13}%
  \BibitemOpen
  \bibfield  {author} {\bibinfo {author} {\bibnamefont {Mesaros}, \bibfnamefont
  {A.}}, \ and\ \bibinfo {author} {\bibfnamefont {Y.}~\bibnamefont {Ran}}}
  (\bibinfo {year} {2013}),\ \href {\doibase 10.1103/PhysRevB.87.155115}
  {\bibfield  {journal} {\bibinfo  {journal} {Phys. Rev. B}\ }\textbf {\bibinfo
  {volume} {87}},\ \bibinfo {pages} {155115}}\BibitemShut {NoStop}%
\bibitem [{\citenamefont {{Metlitski}}\ \emph {et~al.}(2014)\citenamefont
  {{Metlitski}}, \citenamefont {{Fidkowski}}, \citenamefont {{Chen}},\ and\
  \citenamefont {{Vishwanath}}}]{Metlitski2014}%
  \BibitemOpen
  \bibfield  {author} {\bibinfo {author} {\bibnamefont {{Metlitski}},
  \bibfnamefont {M.~A.}}, \bibinfo {author} {\bibfnamefont {L.}~\bibnamefont
  {{Fidkowski}}}, \bibinfo {author} {\bibfnamefont {X.}~\bibnamefont {{Chen}}},
  \ and\ \bibinfo {author} {\bibfnamefont {A.}~\bibnamefont {{Vishwanath}}}}
  (\bibinfo {year} {2014}),\ \href@noop {} {\ }\Eprint
  {http://arxiv.org/abs/1406.3032} {arXiv:1406.3032} \BibitemShut {NoStop}%
\bibitem [{\citenamefont {{Metlitski}}\ \emph {et~al.}(2013)\citenamefont
  {{Metlitski}}, \citenamefont {{Kane}},\ and\ \citenamefont
  {{Fisher}}}]{Metlitski2013a}%
  \BibitemOpen
  \bibfield  {author} {\bibinfo {author} {\bibnamefont {{Metlitski}},
  \bibfnamefont {M.~A.}}, \bibinfo {author} {\bibfnamefont {C.~L.}\
  \bibnamefont {{Kane}}}, \ and\ \bibinfo {author} {\bibfnamefont {M.~P.~A.}\
  \bibnamefont {{Fisher}}}} (\bibinfo {year} {2013}),\ \href {\doibase
  10.1103/PhysRevB.88.035131} {\bibfield  {journal} {\bibinfo  {journal}
  {\prb}\ }\textbf {\bibinfo {volume} {88}},\ \bibinfo {eid}
  {035131}}\BibitemShut {NoStop}%
\bibitem [{\citenamefont {Metlitski}\ \emph {et~al.}(2015)\citenamefont
  {Metlitski}, \citenamefont {Kane},\ and\ \citenamefont
  {Fisher}}]{Metlitski2013b}%
  \BibitemOpen
  \bibfield  {author} {\bibinfo {author} {\bibnamefont {Metlitski},
  \bibfnamefont {M.~A.}}, \bibinfo {author} {\bibfnamefont {C.~L.}\
  \bibnamefont {Kane}}, \ and\ \bibinfo {author} {\bibfnamefont {M.~P.~A.}\
  \bibnamefont {Fisher}}} (\bibinfo {year} {2015}),\ \href {\doibase
  10.1103/PhysRevB.92.125111} {\bibfield  {journal} {\bibinfo  {journal} {Phys.
  Rev. B}\ }\textbf {\bibinfo {volume} {92}},\ \bibinfo {pages}
  {125111}}\BibitemShut {NoStop}%
\bibitem [{\citenamefont {Milnor}(1963)}]{Milnorbook}%
  \BibitemOpen
  \bibfield  {author} {\bibinfo {author} {\bibnamefont {Milnor}, \bibfnamefont
  {J.}}} (\bibinfo {year} {1963}),\ \href@noop {} {\emph {\bibinfo {title}
  {Morse Theory}}}\ (\bibinfo  {publisher} {Princeton University
  Press})\BibitemShut {NoStop}%
\bibitem [{\citenamefont {Mizushima}\ \emph {et~al.}(2015)\citenamefont
  {Mizushima}, \citenamefont {Tsutsumi}, \citenamefont {Sato},\ and\
  \citenamefont {Machida}}]{Mizushima2014}%
  \BibitemOpen
  \bibfield  {author} {\bibinfo {author} {\bibnamefont {Mizushima},
  \bibfnamefont {T.}}, \bibinfo {author} {\bibfnamefont {Y.}~\bibnamefont
  {Tsutsumi}}, \bibinfo {author} {\bibfnamefont {M.}~\bibnamefont {Sato}}, \
  and\ \bibinfo {author} {\bibfnamefont {K.}~\bibnamefont {Machida}}} (\bibinfo
  {year} {2015}),\ \href {http://stacks.iop.org/0953-8984/27/i=11/a=113203}
  {\bibfield  {journal} {\bibinfo  {journal} {Journal of Physics: Condensed
  Matter}\ }\textbf {\bibinfo {volume} {27}}~(\bibinfo {number} {11}),\
  \bibinfo {pages} {113203}}\BibitemShut {NoStop}%
\bibitem [{\citenamefont {Mong}\ \emph {et~al.}(2012)\citenamefont {Mong},
  \citenamefont {Bardarson},\ and\ \citenamefont {Moore}}]{Weak_TI_Mong}%
  \BibitemOpen
  \bibfield  {author} {\bibinfo {author} {\bibnamefont {Mong}, \bibfnamefont
  {R.~S.~K.}}, \bibinfo {author} {\bibfnamefont {J.~H.}\ \bibnamefont
  {Bardarson}}, \ and\ \bibinfo {author} {\bibfnamefont {J.~E.}\ \bibnamefont
  {Moore}}} (\bibinfo {year} {2012}),\ \href {\doibase
  10.1103/PhysRevLett.108.076804} {\bibfield  {journal} {\bibinfo  {journal}
  {Phys. Rev. Lett.}\ }\textbf {\bibinfo {volume} {108}},\ \bibinfo {pages}
  {076804}}\BibitemShut {NoStop}%
\bibitem [{\citenamefont {Moore}(2010)}]{mooreNatureReview2010}%
  \BibitemOpen
  \bibfield  {author} {\bibinfo {author} {\bibnamefont {Moore}, \bibfnamefont
  {J.~E.}}} (\bibinfo {year} {2010}),\ \href@noop {} {\bibfield  {journal}
  {\bibinfo  {journal} {Nature}\ }\textbf {\bibinfo {volume} {464}}~(\bibinfo
  {number} {7286}),\ \bibinfo {pages} {194}}\BibitemShut {NoStop}%
\bibitem [{\citenamefont {Moore}\ and\ \citenamefont
  {Balents}(2007)}]{Moore2007uq}%
  \BibitemOpen
  \bibfield  {author} {\bibinfo {author} {\bibnamefont {Moore}, \bibfnamefont
  {J.~E.}}, \ and\ \bibinfo {author} {\bibfnamefont {L.}~\bibnamefont
  {Balents}}} (\bibinfo {year} {2007}),\ \href@noop {} {\bibfield  {journal}
  {\bibinfo  {journal} {Phys. Rev. B}\ }\textbf {\bibinfo {volume} {75}},\
  \bibinfo {pages} {121306}}\BibitemShut {NoStop}%
\bibitem [{\citenamefont {Moore}\ \emph {et~al.}(2008)\citenamefont {Moore},
  \citenamefont {Ran},\ and\ \citenamefont {Wen}}]{MooreRanWen08}%
  \BibitemOpen
  \bibfield  {author} {\bibinfo {author} {\bibnamefont {Moore}, \bibfnamefont
  {J.~E.}}, \bibinfo {author} {\bibfnamefont {Y.}~\bibnamefont {Ran}}, \ and\
  \bibinfo {author} {\bibfnamefont {X.-G.}\ \bibnamefont {Wen}}} (\bibinfo
  {year} {2008}),\ \href {\doibase 10.1103/PhysRevLett.101.186805} {\bibfield
  {journal} {\bibinfo  {journal} {Phys. Rev. Lett.}\ }\textbf {\bibinfo
  {volume} {101}},\ \bibinfo {pages} {186805}}\BibitemShut {NoStop}%
\bibitem [{\citenamefont {{Morimoto}}\ and\ \citenamefont
  {{Furusaki}}(2013)}]{Morimoto2013}%
  \BibitemOpen
  \bibfield  {author} {\bibinfo {author} {\bibnamefont {{Morimoto}},
  \bibfnamefont {T.}}, \ and\ \bibinfo {author} {\bibfnamefont
  {A.}~\bibnamefont {{Furusaki}}}} (\bibinfo {year} {2013}),\ \href {\doibase
  10.1103/PhysRevB.88.125129} {\bibfield  {journal} {\bibinfo  {journal}
  {\prb}\ }\textbf {\bibinfo {volume} {88}},\ \bibinfo {eid}
  {125129}}\BibitemShut {NoStop}%
\bibitem [{\citenamefont {Morimoto}\ and\ \citenamefont
  {Furusaki}(2014)}]{morimotoFurusakiPRB14}%
  \BibitemOpen
  \bibfield  {author} {\bibinfo {author} {\bibnamefont {Morimoto},
  \bibfnamefont {T.}}, \ and\ \bibinfo {author} {\bibfnamefont
  {A.}~\bibnamefont {Furusaki}}} (\bibinfo {year} {2014}),\ \href {\doibase
  10.1103/PhysRevB.89.235127} {\bibfield  {journal} {\bibinfo  {journal} {Phys.
  Rev. B}\ }\textbf {\bibinfo {volume} {89}},\ \bibinfo {pages}
  {235127}}\BibitemShut {NoStop}%
\bibitem [{\citenamefont {{Morimoto}}\ \emph {et~al.}(2015)\citenamefont
  {{Morimoto}}, \citenamefont {{Furusaki}},\ and\ \citenamefont
  {{Mudry}}}]{Morimoto2015PhRvB..91w5111M}%
  \BibitemOpen
  \bibfield  {author} {\bibinfo {author} {\bibnamefont {{Morimoto}},
  \bibfnamefont {T.}}, \bibinfo {author} {\bibfnamefont {A.}~\bibnamefont
  {{Furusaki}}}, \ and\ \bibinfo {author} {\bibfnamefont {C.}~\bibnamefont
  {{Mudry}}}} (\bibinfo {year} {2015}),\ \href {\doibase
  10.1103/PhysRevB.91.235111} {\bibfield  {journal} {\bibinfo  {journal}
  {\prb}\ }\textbf {\bibinfo {volume} {91}}~(\bibinfo {number} {23}),\ \bibinfo
  {eid} {235111}}\BibitemShut {NoStop}%
\bibitem [{\citenamefont {Morimoto}\ \emph {et~al.}(2015)\citenamefont
  {Morimoto}, \citenamefont {Furusaki},\ and\ \citenamefont
  {Mudry}}]{Morimoto_TCI_disorder}%
  \BibitemOpen
  \bibfield  {author} {\bibinfo {author} {\bibnamefont {Morimoto},
  \bibfnamefont {T.}}, \bibinfo {author} {\bibfnamefont {A.}~\bibnamefont
  {Furusaki}}, \ and\ \bibinfo {author} {\bibfnamefont {C.}~\bibnamefont
  {Mudry}}} (\bibinfo {year} {2015}),\ \href {\doibase
  10.1103/PhysRevB.91.235111} {\bibfield  {journal} {\bibinfo  {journal} {Phys.
  Rev. B}\ }\textbf {\bibinfo {volume} {91}},\ \bibinfo {pages}
  {235111}}\BibitemShut {NoStop}%
\bibitem [{\citenamefont {Mourik}\ \emph {et~al.}(2012)\citenamefont {Mourik},
  \citenamefont {Zuo}, \citenamefont {Frolov}, \citenamefont {Plissard},
  \citenamefont {Bakkers},\ and\ \citenamefont
  {Kouwenhoven}}]{Mourik_zero_bias}%
  \BibitemOpen
  \bibfield  {author} {\bibinfo {author} {\bibnamefont {Mourik}, \bibfnamefont
  {V.}}, \bibinfo {author} {\bibfnamefont {K.}~\bibnamefont {Zuo}}, \bibinfo
  {author} {\bibfnamefont {S.~M.}\ \bibnamefont {Frolov}}, \bibinfo {author}
  {\bibfnamefont {S.~R.}\ \bibnamefont {Plissard}}, \bibinfo {author}
  {\bibfnamefont {E.~P. A.~M.}\ \bibnamefont {Bakkers}}, \ and\ \bibinfo
  {author} {\bibfnamefont {L.~P.}\ \bibnamefont {Kouwenhoven}}} (\bibinfo
  {year} {2012}),\ \href {\doibase 10.1126/science.1222360} {\bibfield
  {journal} {\bibinfo  {journal} {Science}\ }\textbf {\bibinfo {volume}
  {336}},\ \bibinfo {pages} {1003}}\BibitemShut {NoStop}%
\bibitem [{\citenamefont {{Mudry}}\ \emph {et~al.}(1996)\citenamefont
  {{Mudry}}, \citenamefont {{Chamon}},\ and\ \citenamefont
  {{Wen}}}]{Chamon1996}%
  \BibitemOpen
  \bibfield  {author} {\bibinfo {author} {\bibnamefont {{Mudry}}, \bibfnamefont
  {C.}}, \bibinfo {author} {\bibfnamefont {C.}~\bibnamefont {{Chamon}}}, \ and\
  \bibinfo {author} {\bibfnamefont {X.-G.}\ \bibnamefont {{Wen}}}} (\bibinfo
  {year} {1996}),\ \href {\doibase 10.1016/0550-3213(96)00128-9} {\bibfield
  {journal} {\bibinfo  {journal} {Nuclear Physics B}\ }\textbf {\bibinfo
  {volume} {466}},\ \bibinfo {pages} {383}}\BibitemShut {NoStop}%
\bibitem [{\citenamefont {Murakami}(2007)}]{Murakami2007}%
  \BibitemOpen
  \bibfield  {author} {\bibinfo {author} {\bibnamefont {Murakami},
  \bibfnamefont {S.}}} (\bibinfo {year} {2007}),\ \href {\doibase
  10.1088/1367-2630/9/9/356} {\bibfield  {journal} {\bibinfo  {journal} {New J.
  Phys.}\ }\textbf {\bibinfo {volume} {9}},\ \bibinfo {pages}
  {356}}\BibitemShut {NoStop}%
\bibitem [{\citenamefont {Murakawa}\ \emph {et~al.}(2009)\citenamefont
  {Murakawa}, \citenamefont {Tamura}, \citenamefont {Wada}, \citenamefont
  {Wasai}, \citenamefont {Saitoh}, \citenamefont {Aoki}, \citenamefont
  {Nomura}, \citenamefont {Okuda}, \citenamefont {Nagato}, \citenamefont
  {Yamamoto}, \citenamefont {Higashitani},\ and\ \citenamefont
  {Nagai}}]{Murakawa:2009ve}%
  \BibitemOpen
  \bibfield  {author} {\bibinfo {author} {\bibnamefont {Murakawa},
  \bibfnamefont {S.}}, \bibinfo {author} {\bibfnamefont {Y.}~\bibnamefont
  {Tamura}}, \bibinfo {author} {\bibfnamefont {Y.}~\bibnamefont {Wada}},
  \bibinfo {author} {\bibfnamefont {M.}~\bibnamefont {Wasai}}, \bibinfo
  {author} {\bibfnamefont {M.}~\bibnamefont {Saitoh}}, \bibinfo {author}
  {\bibfnamefont {Y.}~\bibnamefont {Aoki}}, \bibinfo {author} {\bibfnamefont
  {R.}~\bibnamefont {Nomura}}, \bibinfo {author} {\bibfnamefont
  {Y.}~\bibnamefont {Okuda}}, \bibinfo {author} {\bibfnamefont
  {Y.}~\bibnamefont {Nagato}}, \bibinfo {author} {\bibfnamefont
  {M.}~\bibnamefont {Yamamoto}}, \bibinfo {author} {\bibfnamefont
  {S.}~\bibnamefont {Higashitani}}, \ and\ \bibinfo {author} {\bibfnamefont
  {K.}~\bibnamefont {Nagai}}} (\bibinfo {year} {2009}),\ \href {\doibase
  10.1103/PhysRevLett.103.155301} {\bibfield  {journal} {\bibinfo  {journal}
  {Phys. Rev. Lett.}\ }\textbf {\bibinfo {volume} {103}},\ \bibinfo {pages}
  {155301}}\BibitemShut {NoStop}%
\bibitem [{\citenamefont {Murakawa}\ \emph {et~al.}(2011)\citenamefont
  {Murakawa}, \citenamefont {Wada}, \citenamefont {Tamura}, \citenamefont
  {Wasai}, \citenamefont {Saitoh}, \citenamefont {Aoki}, \citenamefont
  {Nomura}, \citenamefont {Okuda}, \citenamefont {Nagato}, \citenamefont
  {Yamamoto}, \citenamefont {Higashitani},\ and\ \citenamefont
  {Nagai}}]{JPSJ.80.013602}%
  \BibitemOpen
  \bibfield  {author} {\bibinfo {author} {\bibnamefont {Murakawa},
  \bibfnamefont {S.}}, \bibinfo {author} {\bibfnamefont {Y.}~\bibnamefont
  {Wada}}, \bibinfo {author} {\bibfnamefont {Y.}~\bibnamefont {Tamura}},
  \bibinfo {author} {\bibfnamefont {M.}~\bibnamefont {Wasai}}, \bibinfo
  {author} {\bibfnamefont {M.}~\bibnamefont {Saitoh}}, \bibinfo {author}
  {\bibfnamefont {Y.}~\bibnamefont {Aoki}}, \bibinfo {author} {\bibfnamefont
  {R.}~\bibnamefont {Nomura}}, \bibinfo {author} {\bibfnamefont
  {Y.}~\bibnamefont {Okuda}}, \bibinfo {author} {\bibfnamefont
  {Y.}~\bibnamefont {Nagato}}, \bibinfo {author} {\bibfnamefont
  {M.}~\bibnamefont {Yamamoto}}, \bibinfo {author} {\bibfnamefont
  {S.}~\bibnamefont {Higashitani}}, \ and\ \bibinfo {author} {\bibfnamefont
  {K.}~\bibnamefont {Nagai}}} (\bibinfo {year} {2011}),\ \href {\doibase
  10.1143/JPSJ.80.013602} {\bibfield  {journal} {\bibinfo  {journal} {J. Phys.
  Soc. Jpn.}\ }\textbf {\bibinfo {volume} {80}},\ \bibinfo {pages}
  {013602}}\BibitemShut {NoStop}%
\bibitem [{\citenamefont {Nadj-Perge}\ \emph {et~al.}(2014)\citenamefont
  {Nadj-Perge}, \citenamefont {Drozdov}, \citenamefont {Li}, \citenamefont
  {Chen}, \citenamefont {Jeon}, \citenamefont {Seo}, \citenamefont {MacDonald},
  \citenamefont {Bernevig},\ and\ \citenamefont
  {Yazdani}}]{Nadj-Perge_Ferro_SC}%
  \BibitemOpen
  \bibfield  {author} {\bibinfo {author} {\bibnamefont {Nadj-Perge},
  \bibfnamefont {S.}}, \bibinfo {author} {\bibfnamefont {I.~K.}\ \bibnamefont
  {Drozdov}}, \bibinfo {author} {\bibfnamefont {J.}~\bibnamefont {Li}},
  \bibinfo {author} {\bibfnamefont {H.}~\bibnamefont {Chen}}, \bibinfo {author}
  {\bibfnamefont {S.}~\bibnamefont {Jeon}}, \bibinfo {author} {\bibfnamefont
  {J.}~\bibnamefont {Seo}}, \bibinfo {author} {\bibfnamefont {A.~H.}\
  \bibnamefont {MacDonald}}, \bibinfo {author} {\bibfnamefont {B.~A.}\
  \bibnamefont {Bernevig}}, \ and\ \bibinfo {author} {\bibfnamefont
  {A.}~\bibnamefont {Yazdani}}} (\bibinfo {year} {2014}),\ \href {\doibase
  10.1126/science.1259327} {\bibfield  {journal} {\bibinfo  {journal}
  {Science}\ }\textbf {\bibinfo {volume} {346}},\ \bibinfo {pages}
  {602}}\BibitemShut {NoStop}%
\bibitem [{\citenamefont {Nagaosa}\ \emph {et~al.}(2010)\citenamefont
  {Nagaosa}, \citenamefont {Sinova}, \citenamefont {Onoda}, \citenamefont
  {MacDonald},\ and\ \citenamefont {Ong}}]{RevModPhys_AHE}%
  \BibitemOpen
  \bibfield  {author} {\bibinfo {author} {\bibnamefont {Nagaosa}, \bibfnamefont
  {N.}}, \bibinfo {author} {\bibfnamefont {J.}~\bibnamefont {Sinova}}, \bibinfo
  {author} {\bibfnamefont {S.}~\bibnamefont {Onoda}}, \bibinfo {author}
  {\bibfnamefont {A.~H.}\ \bibnamefont {MacDonald}}, \ and\ \bibinfo {author}
  {\bibfnamefont {N.~P.}\ \bibnamefont {Ong}}} (\bibinfo {year} {2010}),\ \href
  {\doibase 10.1103/RevModPhys.82.1539} {\bibfield  {journal} {\bibinfo
  {journal} {Rev. Mod. Phys.}\ }\textbf {\bibinfo {volume} {82}},\ \bibinfo
  {pages} {1539}}\BibitemShut {NoStop}%
\bibitem [{\citenamefont {Nakahara}(2003)}]{Nakahara:2003ve}%
  \BibitemOpen
  \bibfield  {author} {\bibinfo {author} {\bibnamefont {Nakahara},
  \bibfnamefont {M.}}} (\bibinfo {year} {2003}),\ \href
  {http://www.amazon.ca/exec/obidos/redirect?tag=citeulike09-20&amp;path=ASIN/0750306068}
  {\emph {\bibinfo {title} {Geometry, Topology and Physics, Second Edition
  (Graduate Student Series in Physics)}}}\ (\bibinfo  {publisher} {Taylor \&
  Francis})\BibitemShut {NoStop}%
\bibitem [{\citenamefont {Nayak}\ \emph {et~al.}(2008)\citenamefont {Nayak},
  \citenamefont {Simon}, \citenamefont {Stern}, \citenamefont {Freedman},\ and\
  \citenamefont {Das~Sarma}}]{RMP_braiding}%
  \BibitemOpen
  \bibfield  {author} {\bibinfo {author} {\bibnamefont {Nayak}, \bibfnamefont
  {C.}}, \bibinfo {author} {\bibfnamefont {S.~H.}\ \bibnamefont {Simon}},
  \bibinfo {author} {\bibfnamefont {A.}~\bibnamefont {Stern}}, \bibinfo
  {author} {\bibfnamefont {M.}~\bibnamefont {Freedman}}, \ and\ \bibinfo
  {author} {\bibfnamefont {S.}~\bibnamefont {Das~Sarma}}} (\bibinfo {year}
  {2008}),\ \href {\doibase 10.1103/RevModPhys.80.1083} {\bibfield  {journal}
  {\bibinfo  {journal} {Rev. Mod. Phys.}\ }\textbf {\bibinfo {volume} {80}},\
  \bibinfo {pages} {1083}}\BibitemShut {NoStop}%
\bibitem [{\citenamefont {Nelson}(2002)}]{Nelsonbook}%
  \BibitemOpen
  \bibfield  {author} {\bibinfo {author} {\bibnamefont {Nelson}, \bibfnamefont
  {D.~R.}}} (\bibinfo {year} {2002}),\ \href@noop {} {\emph {\bibinfo {title}
  {Defects and Geometry in Condensed Matter Physics}}}\ (\bibinfo  {publisher}
  {Cambridge University Press})\BibitemShut {NoStop}%
\bibitem [{\citenamefont {{Nersesyan}}\ \emph {et~al.}(1994)\citenamefont
  {{Nersesyan}}, \citenamefont {{Tsvelik}},\ and\ \citenamefont
  {{Wenger}}}]{Nersesyan1994}%
  \BibitemOpen
  \bibfield  {author} {\bibinfo {author} {\bibnamefont {{Nersesyan}},
  \bibfnamefont {A.~A.}}, \bibinfo {author} {\bibfnamefont {A.~M.}\
  \bibnamefont {{Tsvelik}}}, \ and\ \bibinfo {author} {\bibfnamefont
  {F.}~\bibnamefont {{Wenger}}}} (\bibinfo {year} {1994}),\ \href {\doibase
  10.1103/PhysRevLett.72.2628} {\bibfield  {journal} {\bibinfo  {journal}
  {Phys. Rev. Lett.}\ }\textbf {\bibinfo {volume} {72}},\ \bibinfo {pages}
  {2628}}\BibitemShut {NoStop}%
\bibitem [{\citenamefont {Neupane}\ \emph {et~al.}(2014)\citenamefont
  {Neupane}, \citenamefont {Xu}, \citenamefont {Sankar}, \citenamefont
  {Alidoust}, \citenamefont {Bian}, \citenamefont {Liu}, \citenamefont
  {Belopolski}, \citenamefont {Chang}, \citenamefont {Jeng}, \citenamefont
  {Lin}, \citenamefont {Bansil}, \citenamefont {Chou},\ and\ \citenamefont
  {Hasan}}]{neupaneDiracHasan}%
  \BibitemOpen
  \bibfield  {author} {\bibinfo {author} {\bibnamefont {Neupane}, \bibfnamefont
  {M.}}, \bibinfo {author} {\bibfnamefont {S.-Y.}\ \bibnamefont {Xu}}, \bibinfo
  {author} {\bibfnamefont {R.}~\bibnamefont {Sankar}}, \bibinfo {author}
  {\bibfnamefont {N.}~\bibnamefont {Alidoust}}, \bibinfo {author}
  {\bibfnamefont {G.}~\bibnamefont {Bian}}, \bibinfo {author} {\bibfnamefont
  {C.}~\bibnamefont {Liu}}, \bibinfo {author} {\bibfnamefont {I.}~\bibnamefont
  {Belopolski}}, \bibinfo {author} {\bibfnamefont {T.-R.}\ \bibnamefont
  {Chang}}, \bibinfo {author} {\bibfnamefont {H.-T.}\ \bibnamefont {Jeng}},
  \bibinfo {author} {\bibfnamefont {H.}~\bibnamefont {Lin}}, \bibinfo {author}
  {\bibfnamefont {A.}~\bibnamefont {Bansil}}, \bibinfo {author} {\bibfnamefont
  {F.}~\bibnamefont {Chou}}, \ and\ \bibinfo {author} {\bibfnamefont {M.~Z.}\
  \bibnamefont {Hasan}}} (\bibinfo {year} {2014}),\ \href@noop {} {\bibfield
  {journal} {\bibinfo  {journal} {Nat. Commun.}\ }\textbf {\bibinfo {volume}
  {5}},\ \bibinfo {pages} {3786}}\BibitemShut {NoStop}%
\bibitem [{\citenamefont {Neupert}\ \emph {et~al.}(2014)\citenamefont
  {Neupert}, \citenamefont {Chamon}, \citenamefont {Mudry},\ and\ \citenamefont
  {Thomale}}]{Titus_wire_deconstruction}%
  \BibitemOpen
  \bibfield  {author} {\bibinfo {author} {\bibnamefont {Neupert}, \bibfnamefont
  {T.}}, \bibinfo {author} {\bibfnamefont {C.}~\bibnamefont {Chamon}}, \bibinfo
  {author} {\bibfnamefont {C.}~\bibnamefont {Mudry}}, \ and\ \bibinfo {author}
  {\bibfnamefont {R.}~\bibnamefont {Thomale}}} (\bibinfo {year} {2014}),\ \href
  {\doibase 10.1103/PhysRevB.90.205101} {\bibfield  {journal} {\bibinfo
  {journal} {Phys. Rev. B}\ }\textbf {\bibinfo {volume} {90}},\ \bibinfo
  {pages} {205101}}\BibitemShut {NoStop}%
\bibitem [{\citenamefont {Neupert}\ \emph {et~al.}(2011)\citenamefont
  {Neupert}, \citenamefont {Santos}, \citenamefont {Ryu}, \citenamefont
  {Chamon},\ and\ \citenamefont {Mudry}}]{Titus_FTI}%
  \BibitemOpen
  \bibfield  {author} {\bibinfo {author} {\bibnamefont {Neupert}, \bibfnamefont
  {T.}}, \bibinfo {author} {\bibfnamefont {L.}~\bibnamefont {Santos}}, \bibinfo
  {author} {\bibfnamefont {S.}~\bibnamefont {Ryu}}, \bibinfo {author}
  {\bibfnamefont {C.}~\bibnamefont {Chamon}}, \ and\ \bibinfo {author}
  {\bibfnamefont {C.}~\bibnamefont {Mudry}}} (\bibinfo {year} {2011}),\ \href
  {\doibase 10.1103/PhysRevB.84.165107} {\bibfield  {journal} {\bibinfo
  {journal} {Phys. Rev. B}\ }\textbf {\bibinfo {volume} {84}},\ \bibinfo
  {pages} {165107}}\BibitemShut {NoStop}%
\bibitem [{\citenamefont {Nielsen}\ and\ \citenamefont
  {Ninomiya}(1981)}]{Nielsen_Ninomiya_1981}%
  \BibitemOpen
  \bibfield  {author} {\bibinfo {author} {\bibnamefont {Nielsen}, \bibfnamefont
  {H.}}, \ and\ \bibinfo {author} {\bibfnamefont {M.}~\bibnamefont {Ninomiya}}}
  (\bibinfo {year} {1981}),\ \href@noop {} {\bibfield  {journal} {\bibinfo
  {journal} {Nucl. Phys. B}\ }\textbf {\bibinfo {volume} {185}},\ \bibinfo
  {pages} {20}}\BibitemShut {NoStop}%
\bibitem [{\citenamefont {Ning}\ \emph {et~al.}(2015)\citenamefont {Ning},
  \citenamefont {Jiang},\ and\ \citenamefont {Liu}}]{ning_liu_PRB_15}%
  \BibitemOpen
  \bibfield  {author} {\bibinfo {author} {\bibnamefont {Ning}, \bibfnamefont
  {S.-Q.}}, \bibinfo {author} {\bibfnamefont {H.-C.}\ \bibnamefont {Jiang}}, \
  and\ \bibinfo {author} {\bibfnamefont {Z.-X.}\ \bibnamefont {Liu}}} (\bibinfo
  {year} {2015}),\ \href {\doibase 10.1103/PhysRevB.91.241105} {\bibfield
  {journal} {\bibinfo  {journal} {Phys. Rev. B}\ }\textbf {\bibinfo {volume}
  {91}},\ \bibinfo {pages} {241105}}\BibitemShut {NoStop}%
\bibitem [{\citenamefont {Niu}\ \emph {et~al.}(1985)\citenamefont {Niu},
  \citenamefont {Thouless},\ and\ \citenamefont {Wu}}]{ThoulessWu1985}%
  \BibitemOpen
  \bibfield  {author} {\bibinfo {author} {\bibnamefont {Niu}, \bibfnamefont
  {Q.}}, \bibinfo {author} {\bibfnamefont {D.~J.}\ \bibnamefont {Thouless}}, \
  and\ \bibinfo {author} {\bibfnamefont {Y.-S.}\ \bibnamefont {Wu}}} (\bibinfo
  {year} {1985}),\ \href {\doibase 10.1103/PhysRevB.31.3372} {\bibfield
  {journal} {\bibinfo  {journal} {Phys. Rev. B}\ }\textbf {\bibinfo {volume}
  {31}},\ \bibinfo {pages} {3372}}\BibitemShut {NoStop}%
\bibitem [{\citenamefont {Nomura}\ \emph {et~al.}(2007)\citenamefont {Nomura},
  \citenamefont {Koshino},\ and\ \citenamefont {Ryu}}]{nomuraPRL07}%
  \BibitemOpen
  \bibfield  {author} {\bibinfo {author} {\bibnamefont {Nomura}, \bibfnamefont
  {K.}}, \bibinfo {author} {\bibfnamefont {M.}~\bibnamefont {Koshino}}, \ and\
  \bibinfo {author} {\bibfnamefont {S.}~\bibnamefont {Ryu}}} (\bibinfo {year}
  {2007}),\ \href {\doibase 10.1103/PhysRevLett.99.146806} {\bibfield
  {journal} {\bibinfo  {journal} {Phys. Rev. Lett.}\ }\textbf {\bibinfo
  {volume} {99}},\ \bibinfo {pages} {146806}}\BibitemShut {NoStop}%
\bibitem [{\citenamefont {Nomura}\ \emph {et~al.}(2012)\citenamefont {Nomura},
  \citenamefont {Ryu}, \citenamefont {Furusaki},\ and\ \citenamefont
  {Nagaosa}}]{NomuraRyuFurusakiNagaosa12}%
  \BibitemOpen
  \bibfield  {author} {\bibinfo {author} {\bibnamefont {Nomura}, \bibfnamefont
  {K.}}, \bibinfo {author} {\bibfnamefont {S.}~\bibnamefont {Ryu}}, \bibinfo
  {author} {\bibfnamefont {A.}~\bibnamefont {Furusaki}}, \ and\ \bibinfo
  {author} {\bibfnamefont {N.}~\bibnamefont {Nagaosa}}} (\bibinfo {year}
  {2012}),\ \href {\doibase 10.1103/PhysRevLett.108.026802} {\bibfield
  {journal} {\bibinfo  {journal} {Phys. Rev. Lett.}\ }\textbf {\bibinfo
  {volume} {108}},\ \bibinfo {pages} {026802}}\BibitemShut {NoStop}%
\bibitem [{\citenamefont {Novak}\ \emph {et~al.}(2015)\citenamefont {Novak},
  \citenamefont {Sasaki}, \citenamefont {Segawa},\ and\ \citenamefont
  {Ando}}]{Ando_Dirac_magneto}%
  \BibitemOpen
  \bibfield  {author} {\bibinfo {author} {\bibnamefont {Novak}, \bibfnamefont
  {M.}}, \bibinfo {author} {\bibfnamefont {S.}~\bibnamefont {Sasaki}}, \bibinfo
  {author} {\bibfnamefont {K.}~\bibnamefont {Segawa}}, \ and\ \bibinfo {author}
  {\bibfnamefont {Y.}~\bibnamefont {Ando}}} (\bibinfo {year} {2015}),\ \href
  {\doibase 10.1103/PhysRevB.91.041203} {\bibfield  {journal} {\bibinfo
  {journal} {Phys. Rev. B}\ }\textbf {\bibinfo {volume} {91}},\ \bibinfo
  {pages} {041203}}\BibitemShut {NoStop}%
\bibitem [{\citenamefont {Obuse}\ \emph {et~al.}(2007)\citenamefont {Obuse},
  \citenamefont {Furusaki}, \citenamefont {Ryu},\ and\ \citenamefont
  {Mudry}}]{PhysRevB.76.075301}%
  \BibitemOpen
  \bibfield  {author} {\bibinfo {author} {\bibnamefont {Obuse}, \bibfnamefont
  {H.}}, \bibinfo {author} {\bibfnamefont {A.}~\bibnamefont {Furusaki}},
  \bibinfo {author} {\bibfnamefont {S.}~\bibnamefont {Ryu}}, \ and\ \bibinfo
  {author} {\bibfnamefont {C.}~\bibnamefont {Mudry}}} (\bibinfo {year}
  {2007}),\ \href {\doibase 10.1103/PhysRevB.76.075301} {\bibfield  {journal}
  {\bibinfo  {journal} {Phys. Rev. B}\ }\textbf {\bibinfo {volume} {76}},\
  \bibinfo {pages} {075301}}\BibitemShut {NoStop}%
\bibitem [{\citenamefont {Obuse}\ \emph {et~al.}(2008)\citenamefont {Obuse},
  \citenamefont {Furusaki}, \citenamefont {Ryu},\ and\ \citenamefont
  {Mudry}}]{PhysRevB.78.115301}%
  \BibitemOpen
  \bibfield  {author} {\bibinfo {author} {\bibnamefont {Obuse}, \bibfnamefont
  {H.}}, \bibinfo {author} {\bibfnamefont {A.}~\bibnamefont {Furusaki}},
  \bibinfo {author} {\bibfnamefont {S.}~\bibnamefont {Ryu}}, \ and\ \bibinfo
  {author} {\bibfnamefont {C.}~\bibnamefont {Mudry}}} (\bibinfo {year}
  {2008}),\ \href {\doibase 10.1103/PhysRevB.78.115301} {\bibfield  {journal}
  {\bibinfo  {journal} {Phys. Rev. B}\ }\textbf {\bibinfo {volume} {78}},\
  \bibinfo {pages} {115301}}\BibitemShut {NoStop}%
\bibitem [{\citenamefont {Obuse}\ \emph {et~al.}(2014)\citenamefont {Obuse},
  \citenamefont {Ryu}, \citenamefont {Furusaki},\ and\ \citenamefont
  {Mudry}}]{Weak_TI_Obuse}%
  \BibitemOpen
  \bibfield  {author} {\bibinfo {author} {\bibnamefont {Obuse}, \bibfnamefont
  {H.}}, \bibinfo {author} {\bibfnamefont {S.}~\bibnamefont {Ryu}}, \bibinfo
  {author} {\bibfnamefont {A.}~\bibnamefont {Furusaki}}, \ and\ \bibinfo
  {author} {\bibfnamefont {C.}~\bibnamefont {Mudry}}} (\bibinfo {year}
  {2014}),\ \href {\doibase 10.1103/PhysRevB.89.155315} {\bibfield  {journal}
  {\bibinfo  {journal} {Phys. Rev. B}\ }\textbf {\bibinfo {volume} {89}},\
  \bibinfo {pages} {155315}}\BibitemShut {NoStop}%
\bibitem [{\citenamefont {Okada}\ \emph {et~al.}(2013)\citenamefont {Okada},
  \citenamefont {Serbyn}, \citenamefont {Lin}, \citenamefont {Walkup},
  \citenamefont {Zhou}, \citenamefont {Dhital}, \citenamefont {Neupane},
  \citenamefont {Xu}, \citenamefont {Wang}, \citenamefont {Sankar},
  \citenamefont {Chou}, \citenamefont {Bansil}, \citenamefont {Hasan},
  \citenamefont {Wilson}, \citenamefont {Fu},\ and\ \citenamefont
  {Madhavan}}]{Dirac_node_TCI_Okada}%
  \BibitemOpen
  \bibfield  {author} {\bibinfo {author} {\bibnamefont {Okada}, \bibfnamefont
  {Y.}}, \bibinfo {author} {\bibfnamefont {M.}~\bibnamefont {Serbyn}}, \bibinfo
  {author} {\bibfnamefont {H.}~\bibnamefont {Lin}}, \bibinfo {author}
  {\bibfnamefont {D.}~\bibnamefont {Walkup}}, \bibinfo {author} {\bibfnamefont
  {W.}~\bibnamefont {Zhou}}, \bibinfo {author} {\bibfnamefont {C.}~\bibnamefont
  {Dhital}}, \bibinfo {author} {\bibfnamefont {M.}~\bibnamefont {Neupane}},
  \bibinfo {author} {\bibfnamefont {S.}~\bibnamefont {Xu}}, \bibinfo {author}
  {\bibfnamefont {Y.~J.}\ \bibnamefont {Wang}}, \bibinfo {author}
  {\bibfnamefont {R.}~\bibnamefont {Sankar}}, \bibinfo {author} {\bibfnamefont
  {F.}~\bibnamefont {Chou}}, \bibinfo {author} {\bibfnamefont {A.}~\bibnamefont
  {Bansil}}, \bibinfo {author} {\bibfnamefont {M.~Z.}\ \bibnamefont {Hasan}},
  \bibinfo {author} {\bibfnamefont {S.~D.}\ \bibnamefont {Wilson}}, \bibinfo
  {author} {\bibfnamefont {L.}~\bibnamefont {Fu}}, \ and\ \bibinfo {author}
  {\bibfnamefont {V.}~\bibnamefont {Madhavan}}} (\bibinfo {year} {2013}),\
  \href {\doibase 10.1126/science.1239451} {\bibfield  {journal} {\bibinfo
  {journal} {Science}\ }\textbf {\bibinfo {volume} {341}}~(\bibinfo {number}
  {6153}),\ \bibinfo {pages} {1496}}\BibitemShut {NoStop}%
\bibitem [{\citenamefont {Oreg}\ \emph {et~al.}(2010)\citenamefont {Oreg},
  \citenamefont {Refael},\ and\ \citenamefont {von Oppen}}]{Gil_Majorana_wire}%
  \BibitemOpen
  \bibfield  {author} {\bibinfo {author} {\bibnamefont {Oreg}, \bibfnamefont
  {Y.}}, \bibinfo {author} {\bibfnamefont {G.}~\bibnamefont {Refael}}, \ and\
  \bibinfo {author} {\bibfnamefont {F.}~\bibnamefont {von Oppen}}} (\bibinfo
  {year} {2010}),\ \href {\doibase 10.1103/PhysRevLett.105.177002} {\bibfield
  {journal} {\bibinfo  {journal} {Phys. Rev. Lett.}\ }\textbf {\bibinfo
  {volume} {105}},\ \bibinfo {pages} {177002}}\BibitemShut {NoStop}%
\bibitem [{\citenamefont {{Ostrovsky}}\ \emph {et~al.}(2007)\citenamefont
  {{Ostrovsky}}, \citenamefont {{Gornyi}},\ and\ \citenamefont
  {{Mirlin}}}]{Ostrovsky2007}%
  \BibitemOpen
  \bibfield  {author} {\bibinfo {author} {\bibnamefont {{Ostrovsky}},
  \bibfnamefont {P.~M.}}, \bibinfo {author} {\bibfnamefont {I.~V.}\
  \bibnamefont {{Gornyi}}}, \ and\ \bibinfo {author} {\bibfnamefont {A.~D.}\
  \bibnamefont {{Mirlin}}}} (\bibinfo {year} {2007}),\ \href {\doibase
  10.1103/PhysRevLett.98.256801} {\bibfield  {journal} {\bibinfo  {journal}
  {Phys. Rev. Lett.}\ }\textbf {\bibinfo {volume} {98}}~(\bibinfo {number}
  {25}),\ \bibinfo {eid} {256801}}\BibitemShut {NoStop}%
\bibitem [{\citenamefont {{Ostrovsky}}\ \emph {et~al.}(2010)\citenamefont
  {{Ostrovsky}}, \citenamefont {{Gornyi}},\ and\ \citenamefont
  {{Mirlin}}}]{Ostrovsky2010PhRvL.105c6803O}%
  \BibitemOpen
  \bibfield  {author} {\bibinfo {author} {\bibnamefont {{Ostrovsky}},
  \bibfnamefont {P.~M.}}, \bibinfo {author} {\bibfnamefont {I.~V.}\
  \bibnamefont {{Gornyi}}}, \ and\ \bibinfo {author} {\bibfnamefont {A.~D.}\
  \bibnamefont {{Mirlin}}}} (\bibinfo {year} {2010}),\ \href {\doibase
  10.1103/PhysRevLett.105.036803} {\bibfield  {journal} {\bibinfo  {journal}
  {Physical Review Letters}\ }\textbf {\bibinfo {volume} {105}}~(\bibinfo
  {number} {3}),\ \bibinfo {eid} {036803}}\BibitemShut {NoStop}%
\bibitem [{\citenamefont {Parameswaran}\ \emph {et~al.}(2014)\citenamefont
  {Parameswaran}, \citenamefont {Grover}, \citenamefont {Abanin}, \citenamefont
  {Pesin},\ and\ \citenamefont {Vishwanath}}]{Sid_anomaly_Weyl}%
  \BibitemOpen
  \bibfield  {author} {\bibinfo {author} {\bibnamefont {Parameswaran},
  \bibfnamefont {S.~A.}}, \bibinfo {author} {\bibfnamefont {T.}~\bibnamefont
  {Grover}}, \bibinfo {author} {\bibfnamefont {D.~A.}\ \bibnamefont {Abanin}},
  \bibinfo {author} {\bibfnamefont {D.~A.}\ \bibnamefont {Pesin}}, \ and\
  \bibinfo {author} {\bibfnamefont {A.}~\bibnamefont {Vishwanath}}} (\bibinfo
  {year} {2014}),\ \href {http://link.aps.org/doi/10.1103/PhysRevX.4.031035}
  {\bibfield  {journal} {\bibinfo  {journal} {Physical Review X}\ }\textbf
  {\bibinfo {volume} {4}}~(\bibinfo {number} {3}),\ \bibinfo {pages}
  {031035}}\BibitemShut {NoStop}%
\bibitem [{\citenamefont {Parameswaran}\ \emph {et~al.}(2012)\citenamefont
  {Parameswaran}, \citenamefont {Roy},\ and\ \citenamefont {Sondhi}}]{Roy_FTI}%
  \BibitemOpen
  \bibfield  {author} {\bibinfo {author} {\bibnamefont {Parameswaran},
  \bibfnamefont {S.~A.}}, \bibinfo {author} {\bibfnamefont {R.}~\bibnamefont
  {Roy}}, \ and\ \bibinfo {author} {\bibfnamefont {S.~L.}\ \bibnamefont
  {Sondhi}}} (\bibinfo {year} {2012}),\ \href {\doibase
  10.1103/PhysRevB.85.241308} {\bibfield  {journal} {\bibinfo  {journal} {Phys.
  Rev. B}\ }\textbf {\bibinfo {volume} {85}},\ \bibinfo {pages}
  {241308}}\BibitemShut {NoStop}%
\bibitem [{\citenamefont {Parameswaran}\ \emph {et~al.}(2013)\citenamefont
  {Parameswaran}, \citenamefont {Turner}, \citenamefont {Arovas},\ and\
  \citenamefont {Vishwanath}}]{Parameswaran2013_nonsymmorphic}%
  \BibitemOpen
  \bibfield  {author} {\bibinfo {author} {\bibnamefont {Parameswaran},
  \bibfnamefont {S.~A.}}, \bibinfo {author} {\bibfnamefont {A.~M.}\
  \bibnamefont {Turner}}, \bibinfo {author} {\bibfnamefont {D.~P.}\
  \bibnamefont {Arovas}}, \ and\ \bibinfo {author} {\bibfnamefont
  {A.}~\bibnamefont {Vishwanath}}} (\bibinfo {year} {2013}),\ \href
  {http://dx.doi.org/10.1038/nphys2600} {\bibfield  {journal} {\bibinfo
  {journal} {Nat Phys}\ }\textbf {\bibinfo {volume} {9}}~(\bibinfo {number}
  {5}),\ \bibinfo {pages} {299}}\BibitemShut {NoStop}%
\bibitem [{\citenamefont {{Pletikosi{\'c}}}\ \emph {et~al.}(2014)\citenamefont
  {{Pletikosi{\'c}}}, \citenamefont {{Gu}},\ and\ \citenamefont
  {{Valla}}}]{Valla_TCI_surface}%
  \BibitemOpen
  \bibfield  {author} {\bibinfo {author} {\bibnamefont {{Pletikosi{\'c}}},
  \bibfnamefont {I.}}, \bibinfo {author} {\bibfnamefont {G.~D.}\ \bibnamefont
  {{Gu}}}, \ and\ \bibinfo {author} {\bibfnamefont {T.}~\bibnamefont
  {{Valla}}}} (\bibinfo {year} {2014}),\ \href {\doibase
  10.1103/PhysRevLett.112.146403} {\bibfield  {journal} {\bibinfo  {journal}
  {Phys. Rev. Lett.}\ }\textbf {\bibinfo {volume} {112}},\ \bibinfo {pages}
  {146403}}\BibitemShut {NoStop}%
\bibitem [{\citenamefont {{Pollmann}}\ \emph {et~al.}(2012)\citenamefont
  {{Pollmann}}, \citenamefont {{Berg}}, \citenamefont {{Turner}},\ and\
  \citenamefont {{Oshikawa}}}]{Pollmann2012}%
  \BibitemOpen
  \bibfield  {author} {\bibinfo {author} {\bibnamefont {{Pollmann}},
  \bibfnamefont {F.}}, \bibinfo {author} {\bibfnamefont {E.}~\bibnamefont
  {{Berg}}}, \bibinfo {author} {\bibfnamefont {A.~M.}\ \bibnamefont
  {{Turner}}}, \ and\ \bibinfo {author} {\bibfnamefont {M.}~\bibnamefont
  {{Oshikawa}}}} (\bibinfo {year} {2012}),\ \href {\doibase
  10.1103/PhysRevB.85.075125} {\bibfield  {journal} {\bibinfo  {journal}
  {\prb}\ }\textbf {\bibinfo {volume} {85}},\ \bibinfo {eid}
  {075125}}\BibitemShut {NoStop}%
\bibitem [{\citenamefont {Pollmann}\ \emph {et~al.}(2010)\citenamefont
  {Pollmann}, \citenamefont {Turner}, \citenamefont {Berg},\ and\ \citenamefont
  {Oshikawa}}]{Pollmann2010}%
  \BibitemOpen
  \bibfield  {author} {\bibinfo {author} {\bibnamefont {Pollmann},
  \bibfnamefont {F.}}, \bibinfo {author} {\bibfnamefont {A.~M.}\ \bibnamefont
  {Turner}}, \bibinfo {author} {\bibfnamefont {E.}~\bibnamefont {Berg}}, \ and\
  \bibinfo {author} {\bibfnamefont {M.}~\bibnamefont {Oshikawa}}} (\bibinfo
  {year} {2010}),\ \href@noop {} {\bibfield  {journal} {\bibinfo  {journal}
  {Phys. Rev. B}\ }\textbf {\bibinfo {volume} {81}},\ \bibinfo {pages}
  {064439}}\BibitemShut {NoStop}%
\bibitem [{\citenamefont {Prange}\ and\ \citenamefont
  {Girvin}(1990)}]{QHE_book_Prange}%
  \BibitemOpen
  \bibinfo {editor} {\bibnamefont {Prange}, \bibfnamefont {R.~E.}}, \ and\
  \bibinfo {editor} {\bibfnamefont {e.}~\bibnamefont {Girvin}, \bibfnamefont
  {Steven~M.}},\ Eds. (\bibinfo {year} {1990}),\ \href
  {http://www.worldcat.org/isbn/9780387971773} {\emph {\bibinfo {title} {{The
  Quantum Hall effect}}}}\ (\bibinfo  {publisher}
  {Springer-Verlag})\BibitemShut {NoStop}%
\bibitem [{\citenamefont {Prodan}(2009)}]{Prodan_disorder_QSH}%
  \BibitemOpen
  \bibfield  {author} {\bibinfo {author} {\bibnamefont {Prodan}, \bibfnamefont
  {E.}}} (\bibinfo {year} {2009}),\ \href {\doibase 10.1103/PhysRevB.80.125327}
  {\bibfield  {journal} {\bibinfo  {journal} {Phys. Rev. B}\ }\textbf {\bibinfo
  {volume} {80}},\ \bibinfo {pages} {125327}}\BibitemShut {NoStop}%
\bibitem [{\citenamefont {Prodan}(2011)}]{Prodan2011}%
  \BibitemOpen
  \bibfield  {author} {\bibinfo {author} {\bibnamefont {Prodan}, \bibfnamefont
  {E.}}} (\bibinfo {year} {2011}),\ \href {\doibase 10.1103/PhysRevB.83.235115}
  {\bibfield  {journal} {\bibinfo  {journal} {Phys. Rev. B}\ }\textbf {\bibinfo
  {volume} {83}},\ \bibinfo {pages} {235115}}\BibitemShut {NoStop}%
\bibitem [{\citenamefont {Prodan}(2014)}]{Prodan2014}%
  \BibitemOpen
  \bibfield  {author} {\bibinfo {author} {\bibnamefont {Prodan}, \bibfnamefont
  {E.}}} (\bibinfo {year} {2014}),\ \href@noop {} {\bibfield  {journal}
  {\bibinfo  {journal} {Topological Quantum Matter}\ }\textbf {\bibinfo
  {volume} {1}}~(\bibinfo {number} {1})}\BibitemShut {NoStop}%
\bibitem [{\citenamefont {Prodan}\ \emph {et~al.}(2013)\citenamefont {Prodan},
  \citenamefont {Leung},\ and\ \citenamefont {Bellissard}}]{Prodan_noncommute}%
  \BibitemOpen
  \bibfield  {author} {\bibinfo {author} {\bibnamefont {Prodan}, \bibfnamefont
  {E.}}, \bibinfo {author} {\bibfnamefont {B.}~\bibnamefont {Leung}}, \ and\
  \bibinfo {author} {\bibfnamefont {J.}~\bibnamefont {Bellissard}}} (\bibinfo
  {year} {2013}),\ \href {http://stacks.iop.org/1751-8121/46/i=48/a=485202}
  {\bibfield  {journal} {\bibinfo  {journal} {J. Phys. A: Math. Theor.}\
  }\textbf {\bibinfo {volume} {46}}~(\bibinfo {number} {48}),\ \bibinfo {pages}
  {485202}}\BibitemShut {NoStop}%
\bibitem [{\citenamefont {{Prodan}}\ and\ \citenamefont
  {{Schulz-Baldes}}(2014)}]{Prodan_odd_Chern}%
  \BibitemOpen
  \bibfield  {author} {\bibinfo {author} {\bibnamefont {{Prodan}},
  \bibfnamefont {E.}}, \ and\ \bibinfo {author} {\bibfnamefont
  {H.}~\bibnamefont {{Schulz-Baldes}}}} (\bibinfo {year} {2014}),\ \href@noop
  {} {\ }\Eprint {http://arxiv.org/abs/1402.5002} {arXiv:1402.5002}
  \BibitemShut {NoStop}%
\bibitem [{\citenamefont {{Pruisken}}(1984)}]{Pruisken1984}%
  \BibitemOpen
  \bibfield  {author} {\bibinfo {author} {\bibnamefont {{Pruisken}},
  \bibfnamefont {A.~M.~M.}}} (\bibinfo {year} {1984}),\ \href {\doibase
  10.1016/0550-3213(84)90101-9} {\bibfield  {journal} {\bibinfo  {journal}
  {Nuclear Physics B}\ }\textbf {\bibinfo {volume} {235}},\ \bibinfo {pages}
  {277}}\BibitemShut {NoStop}%
\bibitem [{\citenamefont {Qi}(2013)}]{Qi2013}%
  \BibitemOpen
  \bibfield  {author} {\bibinfo {author} {\bibnamefont {Qi}, \bibfnamefont
  {X.-L.}}} (\bibinfo {year} {2013}),\ \href@noop {} {\bibfield  {journal}
  {\bibinfo  {journal} {New J. Phys.}\ }\textbf {\bibinfo {volume} {15}},\
  \bibinfo {pages} {065002}}\BibitemShut {NoStop}%
\bibitem [{\citenamefont {Qi}\ \emph {et~al.}(2009)\citenamefont {Qi},
  \citenamefont {Hughes}, \citenamefont {Raghu},\ and\ \citenamefont
  {Zhang}}]{Qi_hughes_zhang_09}%
  \BibitemOpen
  \bibfield  {author} {\bibinfo {author} {\bibnamefont {Qi}, \bibfnamefont
  {X.-L.}}, \bibinfo {author} {\bibfnamefont {T.~L.}\ \bibnamefont {Hughes}},
  \bibinfo {author} {\bibfnamefont {S.}~\bibnamefont {Raghu}}, \ and\ \bibinfo
  {author} {\bibfnamefont {S.-C.}\ \bibnamefont {Zhang}}} (\bibinfo {year}
  {2009}),\ \href {\doibase 10.1103/PhysRevLett.102.187001} {\bibfield
  {journal} {\bibinfo  {journal} {Phys. Rev. Lett.}\ }\textbf {\bibinfo
  {volume} {102}},\ \bibinfo {pages} {187001}}\BibitemShut {NoStop}%
\bibitem [{\citenamefont {Qi}\ \emph {et~al.}(2008)\citenamefont {Qi},
  \citenamefont {Hughes},\ and\ \citenamefont {Zhang}}]{Qi2008sf}%
  \BibitemOpen
  \bibfield  {author} {\bibinfo {author} {\bibnamefont {Qi}, \bibfnamefont
  {X.-L.}}, \bibinfo {author} {\bibfnamefont {T.~L.}\ \bibnamefont {Hughes}}, \
  and\ \bibinfo {author} {\bibfnamefont {S.-C.}\ \bibnamefont {Zhang}}}
  (\bibinfo {year} {2008}),\ \href@noop {} {\bibfield  {journal} {\bibinfo
  {journal} {Phys. Rev. B}\ }\textbf {\bibinfo {volume} {78}},\ \bibinfo
  {pages} {195424}}\BibitemShut {NoStop}%
\bibitem [{\citenamefont {Qi}\ \emph {et~al.}(2010)\citenamefont {Qi},
  \citenamefont {Hughes},\ and\ \citenamefont {Zhang}}]{QiHughesZhang10}%
  \BibitemOpen
  \bibfield  {author} {\bibinfo {author} {\bibnamefont {Qi}, \bibfnamefont
  {X.-L.}}, \bibinfo {author} {\bibfnamefont {T.~L.}\ \bibnamefont {Hughes}}, \
  and\ \bibinfo {author} {\bibfnamefont {S.-C.}\ \bibnamefont {Zhang}}}
  (\bibinfo {year} {2010}),\ \href {\doibase 10.1103/PhysRevB.81.134508}
  {\bibfield  {journal} {\bibinfo  {journal} {Phys. Rev. B}\ }\textbf {\bibinfo
  {volume} {81}},\ \bibinfo {pages} {134508}}\BibitemShut {NoStop}%
\bibitem [{\citenamefont {Qi}\ and\ \citenamefont {Zhang}(2011)}]{qi:rmp}%
  \BibitemOpen
  \bibfield  {author} {\bibinfo {author} {\bibnamefont {Qi}, \bibfnamefont
  {X.-L.}}, \ and\ \bibinfo {author} {\bibfnamefont {S.-C.}\ \bibnamefont
  {Zhang}}} (\bibinfo {year} {2011}),\ \href {\doibase
  10.1103/RevModPhys.83.1057} {\bibfield  {journal} {\bibinfo  {journal} {Rev.
  Mod. Phys.}\ }\textbf {\bibinfo {volume} {83}},\ \bibinfo {pages}
  {1057}}\BibitemShut {NoStop}%
\bibitem [{\citenamefont {Qiao}\ \emph {et~al.}(2010)\citenamefont {Qiao},
  \citenamefont {Yang}, \citenamefont {Feng}, \citenamefont {Tse},
  \citenamefont {Ding}, \citenamefont {Yao}, \citenamefont {Wang},\ and\
  \citenamefont {Niu}}]{QiaoNiu10}%
  \BibitemOpen
  \bibfield  {author} {\bibinfo {author} {\bibnamefont {Qiao}, \bibfnamefont
  {Z.}}, \bibinfo {author} {\bibfnamefont {S.~A.}\ \bibnamefont {Yang}},
  \bibinfo {author} {\bibfnamefont {W.}~\bibnamefont {Feng}}, \bibinfo {author}
  {\bibfnamefont {W.-K.}\ \bibnamefont {Tse}}, \bibinfo {author} {\bibfnamefont
  {J.}~\bibnamefont {Ding}}, \bibinfo {author} {\bibfnamefont {Y.}~\bibnamefont
  {Yao}}, \bibinfo {author} {\bibfnamefont {J.}~\bibnamefont {Wang}}, \ and\
  \bibinfo {author} {\bibfnamefont {Q.}~\bibnamefont {Niu}}} (\bibinfo {year}
  {2010}),\ \href {\doibase 10.1103/PhysRevB.82.161414} {\bibfield  {journal}
  {\bibinfo  {journal} {Phys. Rev. B}\ }\textbf {\bibinfo {volume} {82}},\
  \bibinfo {pages} {161414}}\BibitemShut {NoStop}%
\bibitem [{\citenamefont {Raghu}\ \emph {et~al.}(2010)\citenamefont {Raghu},
  \citenamefont {Kapitulnik},\ and\ \citenamefont
  {Kivelson}}]{RaghuKapitulnikKivelson2010}%
  \BibitemOpen
  \bibfield  {author} {\bibinfo {author} {\bibnamefont {Raghu}, \bibfnamefont
  {S.}}, \bibinfo {author} {\bibfnamefont {A.}~\bibnamefont {Kapitulnik}}, \
  and\ \bibinfo {author} {\bibfnamefont {S.~A.}\ \bibnamefont {Kivelson}}}
  (\bibinfo {year} {2010}),\ \href {\doibase 10.1103/PhysRevLett.105.136401}
  {\bibfield  {journal} {\bibinfo  {journal} {Phys. Rev. Lett.}\ }\textbf
  {\bibinfo {volume} {105}},\ \bibinfo {pages} {136401}}\BibitemShut {NoStop}%
\bibitem [{\citenamefont {{Ran}}(2010)}]{Ran_dislocation}%
  \BibitemOpen
  \bibfield  {author} {\bibinfo {author} {\bibnamefont {{Ran}}, \bibfnamefont
  {Y.}}} (\bibinfo {year} {2010}),\ \href@noop {} {\ }\Eprint
  {http://arxiv.org/abs/1006.5454} {arXiv:1006.5454} \BibitemShut {NoStop}%
\bibitem [{\citenamefont {Ran}\ \emph {et~al.}(2009)\citenamefont {Ran},
  \citenamefont {Zhang},\ and\ \citenamefont {Vishwanath}}]{Ran:dislocation}%
  \BibitemOpen
  \bibfield  {author} {\bibinfo {author} {\bibnamefont {Ran}, \bibfnamefont
  {Y.}}, \bibinfo {author} {\bibfnamefont {Y.}~\bibnamefont {Zhang}}, \ and\
  \bibinfo {author} {\bibfnamefont {A.}~\bibnamefont {Vishwanath}}} (\bibinfo
  {year} {2009}),\ \href@noop {} {\bibfield  {journal} {\bibinfo  {journal}
  {Nat Phys}\ }\textbf {\bibinfo {volume} {5}}~(\bibinfo {number} {4}),\
  \bibinfo {pages} {298}}\BibitemShut {NoStop}%
\bibitem [{\citenamefont {{Read}}\ and\ \citenamefont
  {{Green}}(2000)}]{ReadGreen2000}%
  \BibitemOpen
  \bibfield  {author} {\bibinfo {author} {\bibnamefont {{Read}}, \bibfnamefont
  {N.}}, \ and\ \bibinfo {author} {\bibfnamefont {D.}~\bibnamefont {{Green}}}}
  (\bibinfo {year} {2000}),\ \href {\doibase 10.1103/PhysRevB.61.10267}
  {\bibfield  {journal} {\bibinfo  {journal} {\prb}\ }\textbf {\bibinfo
  {volume} {61}},\ \bibinfo {pages} {10267}}\BibitemShut {NoStop}%
\bibitem [{\citenamefont {Rechtsman}\ \emph {et~al.}(2013)\citenamefont
  {Rechtsman}, \citenamefont {Zeuner}, \citenamefont {Plotnik}, \citenamefont
  {Lumer}, \citenamefont {Podolsky}, \citenamefont {Dreisow}, \citenamefont
  {Nolte}, \citenamefont {Segev},\ and\ \citenamefont
  {Szameit}}]{Rechtsman_photonic_TI}%
  \BibitemOpen
  \bibfield  {author} {\bibinfo {author} {\bibnamefont {Rechtsman},
  \bibfnamefont {M.~C.}}, \bibinfo {author} {\bibfnamefont {J.~M.}\
  \bibnamefont {Zeuner}}, \bibinfo {author} {\bibfnamefont {Y.}~\bibnamefont
  {Plotnik}}, \bibinfo {author} {\bibfnamefont {Y.}~\bibnamefont {Lumer}},
  \bibinfo {author} {\bibfnamefont {D.}~\bibnamefont {Podolsky}}, \bibinfo
  {author} {\bibfnamefont {F.}~\bibnamefont {Dreisow}}, \bibinfo {author}
  {\bibfnamefont {S.}~\bibnamefont {Nolte}}, \bibinfo {author} {\bibfnamefont
  {M.}~\bibnamefont {Segev}}, \ and\ \bibinfo {author} {\bibfnamefont
  {A.}~\bibnamefont {Szameit}}} (\bibinfo {year} {2013}),\ \href
  {http://dx.doi.org/10.1038/nature12066} {\bibfield  {journal} {\bibinfo
  {journal} {Nature}\ }\textbf {\bibinfo {volume} {496}}~(\bibinfo {number}
  {7444}),\ \bibinfo {pages} {196}}\BibitemShut {NoStop}%
\bibitem [{\citenamefont {Renard}\ \emph {et~al.}(1987)\citenamefont {Renard},
  \citenamefont {Verdaguer}, \citenamefont {Regnault}, \citenamefont
  {Erkelens}, \citenamefont {Rossat-Mignod},\ and\ \citenamefont
  {Stirling}}]{renardEPL87}%
  \BibitemOpen
  \bibfield  {author} {\bibinfo {author} {\bibnamefont {Renard}, \bibfnamefont
  {J.~P.}}, \bibinfo {author} {\bibfnamefont {M.}~\bibnamefont {Verdaguer}},
  \bibinfo {author} {\bibfnamefont {L.~P.}\ \bibnamefont {Regnault}}, \bibinfo
  {author} {\bibfnamefont {W.~A.~C.}\ \bibnamefont {Erkelens}}, \bibinfo
  {author} {\bibfnamefont {J.}~\bibnamefont {Rossat-Mignod}}, \ and\ \bibinfo
  {author} {\bibfnamefont {W.~G.}\ \bibnamefont {Stirling}}} (\bibinfo {year}
  {1987}),\ \href@noop {} {\bibfield  {journal} {\bibinfo  {journal} {Europhys.
  Lett.}\ }\textbf {\bibinfo {volume} {3}}~(\bibinfo {number} {8}),\ \bibinfo
  {pages} {945}}\BibitemShut {NoStop}%
\bibitem [{\citenamefont {Repellin}\ \emph {et~al.}(2014)\citenamefont
  {Repellin}, \citenamefont {Bernevig},\ and\ \citenamefont
  {Regnault}}]{Bernevig_Z2_FTI}%
  \BibitemOpen
  \bibfield  {author} {\bibinfo {author} {\bibnamefont {Repellin},
  \bibfnamefont {C.}}, \bibinfo {author} {\bibfnamefont {B.~A.}\ \bibnamefont
  {Bernevig}}, \ and\ \bibinfo {author} {\bibfnamefont {N.}~\bibnamefont
  {Regnault}}} (\bibinfo {year} {2014}),\ \href {\doibase
  10.1103/PhysRevB.90.245401} {\bibfield  {journal} {\bibinfo  {journal} {Phys.
  Rev. B}\ }\textbf {\bibinfo {volume} {90}},\ \bibinfo {pages}
  {245401}}\BibitemShut {NoStop}%
\bibitem [{\citenamefont {Resta}(1994)}]{restaRMP94}%
  \BibitemOpen
  \bibfield  {author} {\bibinfo {author} {\bibnamefont {Resta}, \bibfnamefont
  {R.}}} (\bibinfo {year} {1994}),\ \href {\doibase 10.1103/RevModPhys.66.899}
  {\bibfield  {journal} {\bibinfo  {journal} {Rev. Mod. Phys.}\ }\textbf
  {\bibinfo {volume} {66}},\ \bibinfo {pages} {899}}\BibitemShut {NoStop}%
\bibitem [{\citenamefont {Rice}\ and\ \citenamefont
  {Sigrist}(1995)}]{RiceSigrist95}%
  \BibitemOpen
  \bibfield  {author} {\bibinfo {author} {\bibnamefont {Rice}, \bibfnamefont
  {T.~M.}}, \ and\ \bibinfo {author} {\bibfnamefont {M.}~\bibnamefont
  {Sigrist}}} (\bibinfo {year} {1995}),\ \href {\doibase
  10.1088/0953-8984/7/47/002} {\bibfield  {journal} {\bibinfo  {journal} {J.
  Phys.: Condens. Matter}\ }\textbf {\bibinfo {volume} {7}},\ \bibinfo {pages}
  {L643}}\BibitemShut {NoStop}%
\bibitem [{\citenamefont {Rieder}\ and\ \citenamefont
  {Brouwer}(2014)}]{RiederPhysRevB.90.205404}%
  \BibitemOpen
  \bibfield  {author} {\bibinfo {author} {\bibnamefont {Rieder}, \bibfnamefont
  {M.-T.}}, \ and\ \bibinfo {author} {\bibfnamefont {P.~W.}\ \bibnamefont
  {Brouwer}}} (\bibinfo {year} {2014}),\ \href {\doibase
  10.1103/PhysRevB.90.205404} {\bibfield  {journal} {\bibinfo  {journal} {Phys.
  Rev. B}\ }\textbf {\bibinfo {volume} {90}},\ \bibinfo {pages}
  {205404}}\BibitemShut {NoStop}%
\bibitem [{\citenamefont {Ringel}\ \emph {et~al.}(2012)\citenamefont {Ringel},
  \citenamefont {Kraus},\ and\ \citenamefont {Stern}}]{Ringel_strong_weakTI}%
  \BibitemOpen
  \bibfield  {author} {\bibinfo {author} {\bibnamefont {Ringel}, \bibfnamefont
  {Z.}}, \bibinfo {author} {\bibfnamefont {Y.~E.}\ \bibnamefont {Kraus}}, \
  and\ \bibinfo {author} {\bibfnamefont {A.}~\bibnamefont {Stern}}} (\bibinfo
  {year} {2012}),\ \href {\doibase 10.1103/PhysRevB.86.045102} {\bibfield
  {journal} {\bibinfo  {journal} {Phys. Rev. B}\ }\textbf {\bibinfo {volume}
  {86}},\ \bibinfo {pages} {045102}}\BibitemShut {NoStop}%
\bibitem [{\citenamefont {{Ringel}}\ and\ \citenamefont
  {{Stern}}(2013)}]{Ringel2013}%
  \BibitemOpen
  \bibfield  {author} {\bibinfo {author} {\bibnamefont {{Ringel}},
  \bibfnamefont {Z.}}, \ and\ \bibinfo {author} {\bibfnamefont
  {A.}~\bibnamefont {{Stern}}}} (\bibinfo {year} {2013}),\ \href {\doibase
  10.1103/PhysRevB.88.115307} {\bibfield  {journal} {\bibinfo  {journal}
  {\prb}\ }\textbf {\bibinfo {volume} {88}},\ \bibinfo {eid}
  {115307}}\BibitemShut {NoStop}%
\bibitem [{\citenamefont {Rokhinson}\ \emph {et~al.}(2012)\citenamefont
  {Rokhinson}, \citenamefont {Liu},\ and\ \citenamefont
  {Furdyna}}]{RokhinsonLiuFurdyna12}%
  \BibitemOpen
  \bibfield  {author} {\bibinfo {author} {\bibnamefont {Rokhinson},
  \bibfnamefont {L.~P.}}, \bibinfo {author} {\bibfnamefont {X.}~\bibnamefont
  {Liu}}, \ and\ \bibinfo {author} {\bibfnamefont {J.~K.}\ \bibnamefont
  {Furdyna}}} (\bibinfo {year} {2012}),\ \href {\doibase 10.1038/nphys2429}
  {\bibfield  {journal} {\bibinfo  {journal} {Nat. Phys.}\ }\textbf {\bibinfo
  {volume} {8}},\ \bibinfo {pages} {795}}\BibitemShut {NoStop}%
\bibitem [{\citenamefont {{Rosch}}(2012)}]{Rosch2012}%
  \BibitemOpen
  \bibfield  {author} {\bibinfo {author} {\bibnamefont {{Rosch}}, \bibfnamefont
  {A.}}} (\bibinfo {year} {2012}),\ \href {\doibase 10.1103/PhysRevB.86.125120}
  {\bibfield  {journal} {\bibinfo  {journal} {\prb}\ }\textbf {\bibinfo
  {volume} {86}},\ \bibinfo {eid} {125120}}\BibitemShut {NoStop}%
\bibitem [{\citenamefont {{Roushan}}\ \emph {et~al.}(2009)\citenamefont
  {{Roushan}}, \citenamefont {{Seo}}, \citenamefont {{Parker}}, \citenamefont
  {{Hor}}, \citenamefont {{Hsieh}}, \citenamefont {{Qian}}, \citenamefont
  {{Richardella}}, \citenamefont {{Hasan}}, \citenamefont {{Cava}},\ and\
  \citenamefont {{Yazdani}}}]{Roushan2009}%
  \BibitemOpen
  \bibfield  {author} {\bibinfo {author} {\bibnamefont {{Roushan}},
  \bibfnamefont {P.}}, \bibinfo {author} {\bibfnamefont {J.}~\bibnamefont
  {{Seo}}}, \bibinfo {author} {\bibfnamefont {C.~V.}\ \bibnamefont {{Parker}}},
  \bibinfo {author} {\bibfnamefont {Y.~S.}\ \bibnamefont {{Hor}}}, \bibinfo
  {author} {\bibfnamefont {D.}~\bibnamefont {{Hsieh}}}, \bibinfo {author}
  {\bibfnamefont {D.}~\bibnamefont {{Qian}}}, \bibinfo {author} {\bibfnamefont
  {A.}~\bibnamefont {{Richardella}}}, \bibinfo {author} {\bibfnamefont {M.~Z.}\
  \bibnamefont {{Hasan}}}, \bibinfo {author} {\bibfnamefont {R.~J.}\
  \bibnamefont {{Cava}}}, \ and\ \bibinfo {author} {\bibfnamefont
  {A.}~\bibnamefont {{Yazdani}}}} (\bibinfo {year} {2009}),\ \href {\doibase
  10.1038/nature08308} {\bibfield  {journal} {\bibinfo  {journal} {\nat}\
  }\textbf {\bibinfo {volume} {460}},\ \bibinfo {pages} {1106}}\BibitemShut
  {NoStop}%
\bibitem [{\citenamefont {Roy}(2009{\natexlab{a}})}]{Roy2009_3D}%
  \BibitemOpen
  \bibfield  {author} {\bibinfo {author} {\bibnamefont {Roy}, \bibfnamefont
  {R.}}} (\bibinfo {year} {2009}{\natexlab{a}}),\ \href {\doibase
  10.1103/PhysRevB.79.195322} {\bibfield  {journal} {\bibinfo  {journal} {Phys.
  Rev. B}\ }\textbf {\bibinfo {volume} {79}},\ \bibinfo {pages}
  {195322}}\BibitemShut {NoStop}%
\bibitem [{\citenamefont {Roy}(2009{\natexlab{b}})}]{Roy2009kx}%
  \BibitemOpen
  \bibfield  {author} {\bibinfo {author} {\bibnamefont {Roy}, \bibfnamefont
  {R.}}} (\bibinfo {year} {2009}{\natexlab{b}}),\ \href@noop {} {\bibfield
  {journal} {\bibinfo  {journal} {Phys. Rev. B}\ }\textbf {\bibinfo {volume}
  {79}},\ \bibinfo {pages} {195321}}\BibitemShut {NoStop}%
\bibitem [{\citenamefont {{Roy}}(2012)}]{Roy2012_nonsymmorphic}%
  \BibitemOpen
  \bibfield  {author} {\bibinfo {author} {\bibnamefont {{Roy}}, \bibfnamefont
  {R.}}} (\bibinfo {year} {2012}),\ \href@noop {} {\ }\Eprint
  {http://arxiv.org/abs/1212.2944} {arXiv:1212.2944} \BibitemShut {NoStop}%
\bibitem [{\citenamefont {Ryu}\ and\ \citenamefont
  {Hatsugai}(2002)}]{RyuHatsugaiPRL02}%
  \BibitemOpen
  \bibfield  {author} {\bibinfo {author} {\bibnamefont {Ryu}, \bibfnamefont
  {S.}}, \ and\ \bibinfo {author} {\bibfnamefont {Y.}~\bibnamefont {Hatsugai}}}
  (\bibinfo {year} {2002}),\ \href {\doibase 10.1103/PhysRevLett.89.077002}
  {\bibfield  {journal} {\bibinfo  {journal} {Phys. Rev. Lett.}\ }\textbf
  {\bibinfo {volume} {89}},\ \bibinfo {pages} {077002}}\BibitemShut {NoStop}%
\bibitem [{\citenamefont {{Ryu}}\ and\ \citenamefont
  {{Hatsugai}}(2006)}]{RyuHatsugai2006}%
  \BibitemOpen
  \bibfield  {author} {\bibinfo {author} {\bibnamefont {{Ryu}}, \bibfnamefont
  {S.}}, \ and\ \bibinfo {author} {\bibfnamefont {Y.}~\bibnamefont
  {{Hatsugai}}}} (\bibinfo {year} {2006}),\ \href {\doibase
  10.1103/PhysRevB.73.245115} {\bibfield  {journal} {\bibinfo  {journal}
  {\prb}\ }\textbf {\bibinfo {volume} {73}}~(\bibinfo {number} {24}),\ \bibinfo
  {eid} {245115}}\BibitemShut {NoStop}%
\bibitem [{\citenamefont {{Ryu}}\ \emph
  {et~al.}(2012{\natexlab{a}})\citenamefont {{Ryu}}, \citenamefont {{Moore}},\
  and\ \citenamefont {{Ludwig}}}]{RyuMooreLudwig2012}%
  \BibitemOpen
  \bibfield  {author} {\bibinfo {author} {\bibnamefont {{Ryu}}, \bibfnamefont
  {S.}}, \bibinfo {author} {\bibfnamefont {J.~E.}\ \bibnamefont {{Moore}}}, \
  and\ \bibinfo {author} {\bibfnamefont {A.~W.~W.}\ \bibnamefont {{Ludwig}}}}
  (\bibinfo {year} {2012}{\natexlab{a}}),\ \href {\doibase
  10.1103/PhysRevB.85.045104} {\bibfield  {journal} {\bibinfo  {journal}
  {\prb}\ }\textbf {\bibinfo {volume} {85}},\ \bibinfo {eid}
  {045104}}\BibitemShut {NoStop}%
\bibitem [{\citenamefont {{Ryu}}\ \emph
  {et~al.}(2012{\natexlab{b}})\citenamefont {{Ryu}}, \citenamefont {{Mudry}},
  \citenamefont {{Ludwig}},\ and\ \citenamefont {{Furusaki}}}]{RyuMudry2012}%
  \BibitemOpen
  \bibfield  {author} {\bibinfo {author} {\bibnamefont {{Ryu}}, \bibfnamefont
  {S.}}, \bibinfo {author} {\bibfnamefont {C.}~\bibnamefont {{Mudry}}},
  \bibinfo {author} {\bibfnamefont {A.~W.~W.}\ \bibnamefont {{Ludwig}}}, \ and\
  \bibinfo {author} {\bibfnamefont {A.}~\bibnamefont {{Furusaki}}}} (\bibinfo
  {year} {2012}{\natexlab{b}}),\ \href {\doibase 10.1103/PhysRevB.85.235115}
  {\bibfield  {journal} {\bibinfo  {journal} {\prb}\ }\textbf {\bibinfo
  {volume} {85}}~(\bibinfo {number} {23}),\ \bibinfo {eid}
  {235115}}\BibitemShut {NoStop}%
\bibitem [{\citenamefont {Ryu}\ \emph {et~al.}(2007)\citenamefont {Ryu},
  \citenamefont {Mudry}, \citenamefont {Obuse},\ and\ \citenamefont
  {Furusaki}}]{ryuFurusakiPRL07}%
  \BibitemOpen
  \bibfield  {author} {\bibinfo {author} {\bibnamefont {Ryu}, \bibfnamefont
  {S.}}, \bibinfo {author} {\bibfnamefont {C.}~\bibnamefont {Mudry}}, \bibinfo
  {author} {\bibfnamefont {H.}~\bibnamefont {Obuse}}, \ and\ \bibinfo {author}
  {\bibfnamefont {A.}~\bibnamefont {Furusaki}}} (\bibinfo {year} {2007}),\
  \href {\doibase 10.1103/PhysRevLett.99.116601} {\bibfield  {journal}
  {\bibinfo  {journal} {Phys. Rev. Lett.}\ }\textbf {\bibinfo {volume} {99}},\
  \bibinfo {pages} {116601}}\BibitemShut {NoStop}%
\bibitem [{\citenamefont {Ryu}\ \emph {et~al.}(2010{\natexlab{a}})\citenamefont
  {Ryu}, \citenamefont {Mudry}, \citenamefont {Obuse},\ and\ \citenamefont
  {Furusaki}}]{RyuMudryObuseFurusaki2010NJPh}%
  \BibitemOpen
  \bibfield  {author} {\bibinfo {author} {\bibnamefont {Ryu}, \bibfnamefont
  {S.}}, \bibinfo {author} {\bibfnamefont {C.}~\bibnamefont {Mudry}}, \bibinfo
  {author} {\bibfnamefont {H.}~\bibnamefont {Obuse}}, \ and\ \bibinfo {author}
  {\bibfnamefont {A.}~\bibnamefont {Furusaki}}} (\bibinfo {year}
  {2010}{\natexlab{a}}),\ \href@noop {} {\bibfield  {journal} {\bibinfo
  {journal} {New Journal of Physics}\ }\textbf {\bibinfo {volume}
  {12}}~(\bibinfo {number} {6}),\ \bibinfo {pages} {065005}}\BibitemShut
  {NoStop}%
\bibitem [{\citenamefont {{Ryu}}\ and\ \citenamefont
  {{Nomura}}(2012)}]{Ryu2012PhRvB..85o5138R}%
  \BibitemOpen
  \bibfield  {author} {\bibinfo {author} {\bibnamefont {{Ryu}}, \bibfnamefont
  {S.}}, \ and\ \bibinfo {author} {\bibfnamefont {K.}~\bibnamefont {{Nomura}}}}
  (\bibinfo {year} {2012}),\ \href {\doibase 10.1103/PhysRevB.85.155138}
  {\bibfield  {journal} {\bibinfo  {journal} {\prb}\ }\textbf {\bibinfo
  {volume} {85}}~(\bibinfo {number} {15}),\ \bibinfo {eid}
  {155138}}\BibitemShut {NoStop}%
\bibitem [{\citenamefont {Ryu}\ \emph {et~al.}(2010{\natexlab{b}})\citenamefont
  {Ryu}, \citenamefont {Schnyder}, \citenamefont {Furusaki},\ and\
  \citenamefont {Ludwig}}]{Ryu2010ten}%
  \BibitemOpen
  \bibfield  {author} {\bibinfo {author} {\bibnamefont {Ryu}, \bibfnamefont
  {S.}}, \bibinfo {author} {\bibfnamefont {A.~P.}\ \bibnamefont {Schnyder}},
  \bibinfo {author} {\bibfnamefont {A.}~\bibnamefont {Furusaki}}, \ and\
  \bibinfo {author} {\bibfnamefont {A.~W.~W.}\ \bibnamefont {Ludwig}}}
  (\bibinfo {year} {2010}{\natexlab{b}}),\ \href@noop {} {\bibfield  {journal}
  {\bibinfo  {journal} {New J. Phys.}\ }\textbf {\bibinfo {volume} {12}},\
  \bibinfo {pages} {065010}}\BibitemShut {NoStop}%
\bibitem [{\citenamefont {{Ryu}}\ and\ \citenamefont
  {{Zhang}}(2012)}]{RyuZhang2012}%
  \BibitemOpen
  \bibfield  {author} {\bibinfo {author} {\bibnamefont {{Ryu}}, \bibfnamefont
  {S.}}, \ and\ \bibinfo {author} {\bibfnamefont {S.-C.}\ \bibnamefont
  {{Zhang}}}} (\bibinfo {year} {2012}),\ \href {\doibase
  10.1103/PhysRevB.85.245132} {\bibfield  {journal} {\bibinfo  {journal}
  {\prb}\ }\textbf {\bibinfo {volume} {85}}~(\bibinfo {number} {24}),\ \bibinfo
  {eid} {245132}}\BibitemShut {NoStop}%
\bibitem [{\citenamefont {Safaei}\ \emph {et~al.}(2013)\citenamefont {Safaei},
  \citenamefont {Kacman},\ and\ \citenamefont {Buczko}}]{Surface_TCI_Safaei}%
  \BibitemOpen
  \bibfield  {author} {\bibinfo {author} {\bibnamefont {Safaei}, \bibfnamefont
  {S.}}, \bibinfo {author} {\bibfnamefont {P.}~\bibnamefont {Kacman}}, \ and\
  \bibinfo {author} {\bibfnamefont {R.}~\bibnamefont {Buczko}}} (\bibinfo
  {year} {2013}),\ \href {\doibase 10.1103/PhysRevB.88.045305} {\bibfield
  {journal} {\bibinfo  {journal} {Phys. Rev. B}\ }\textbf {\bibinfo {volume}
  {88}},\ \bibinfo {pages} {045305}}\BibitemShut {NoStop}%
\bibitem [{\citenamefont {Salomaa}\ and\ \citenamefont
  {Volovik}(1985)}]{SalomaaVolovik85}%
  \BibitemOpen
  \bibfield  {author} {\bibinfo {author} {\bibnamefont {Salomaa}, \bibfnamefont
  {M.~M.}}, \ and\ \bibinfo {author} {\bibfnamefont {G.~E.}\ \bibnamefont
  {Volovik}}} (\bibinfo {year} {1985}),\ \href {\doibase
  10.1103/PhysRevLett.55.1184} {\bibfield  {journal} {\bibinfo  {journal}
  {Phys. Rev. Lett.}\ }\textbf {\bibinfo {volume} {55}},\ \bibinfo {pages}
  {1184}}\BibitemShut {NoStop}%
\bibitem [{\citenamefont {{Santos}}\ and\ \citenamefont
  {{Wang}}(2014)}]{Wang2014a}%
  \BibitemOpen
  \bibfield  {author} {\bibinfo {author} {\bibnamefont {{Santos}},
  \bibfnamefont {L.~H.}}, \ and\ \bibinfo {author} {\bibfnamefont
  {J.}~\bibnamefont {{Wang}}}} (\bibinfo {year} {2014}),\ \href {\doibase
  10.1103/PhysRevB.89.195122} {\bibfield  {journal} {\bibinfo  {journal}
  {\prb}\ }\textbf {\bibinfo {volume} {89}},\ \bibinfo {eid}
  {195122}}\BibitemShut {NoStop}%
\bibitem [{\citenamefont {Sato}(2006)}]{SatoPRB06}%
  \BibitemOpen
  \bibfield  {author} {\bibinfo {author} {\bibnamefont {Sato}, \bibfnamefont
  {M.}}} (\bibinfo {year} {2006}),\ \href {\doibase 10.1103/PhysRevB.73.214502}
  {\bibfield  {journal} {\bibinfo  {journal} {Phys. Rev. B}\ }\textbf {\bibinfo
  {volume} {73}},\ \bibinfo {pages} {214502}}\BibitemShut {NoStop}%
\bibitem [{\citenamefont {Sato}\ \emph {et~al.}(2014)\citenamefont {Sato},
  \citenamefont {Yamakage},\ and\ \citenamefont
  {Mizushima}}]{Mirror_Majorana_Sato}%
  \BibitemOpen
  \bibfield  {author} {\bibinfo {author} {\bibnamefont {Sato}, \bibfnamefont
  {M.}}, \bibinfo {author} {\bibfnamefont {A.}~\bibnamefont {Yamakage}}, \ and\
  \bibinfo {author} {\bibfnamefont {T.}~\bibnamefont {Mizushima}}} (\bibinfo
  {year} {2014}),\ \href {\doibase
  http://dx.doi.org/10.1016/j.physe.2013.07.011} {\bibfield  {journal}
  {\bibinfo  {journal} {Physica E: Low-dimensional Systems and Nanostructures}\
  }\textbf {\bibinfo {volume} {55}}~(\bibinfo {number} {0}),\ \bibinfo {pages}
  {20 }},\ \bibinfo {note} {topological Objects}\BibitemShut {NoStop}%
\bibitem [{\citenamefont {Sau}\ \emph {et~al.}(2010)\citenamefont {Sau},
  \citenamefont {Lutchyn}, \citenamefont {Tewari},\ and\ \citenamefont
  {Das~Sarma}}]{Sau_semiconductor_heterostructures}%
  \BibitemOpen
  \bibfield  {author} {\bibinfo {author} {\bibnamefont {Sau}, \bibfnamefont
  {J.~D.}}, \bibinfo {author} {\bibfnamefont {R.~M.}\ \bibnamefont {Lutchyn}},
  \bibinfo {author} {\bibfnamefont {S.}~\bibnamefont {Tewari}}, \ and\ \bibinfo
  {author} {\bibfnamefont {S.}~\bibnamefont {Das~Sarma}}} (\bibinfo {year}
  {2010}),\ \href {\doibase 10.1103/PhysRevLett.104.040502} {\bibfield
  {journal} {\bibinfo  {journal} {Phys. Rev. Lett.}\ }\textbf {\bibinfo
  {volume} {104}},\ \bibinfo {pages} {040502}}\BibitemShut {NoStop}%
\bibitem [{\citenamefont {Schnyder}\ and\ \citenamefont
  {Brydon}(2015)}]{schnyderReviewTopNodalSCs}%
  \BibitemOpen
  \bibfield  {author} {\bibinfo {author} {\bibnamefont {Schnyder},
  \bibfnamefont {A.~P.}}, \ and\ \bibinfo {author} {\bibfnamefont {P.~M.~R.}\
  \bibnamefont {Brydon}}} (\bibinfo {year} {2015}),\ \href
  {http://stacks.iop.org/0953-8984/27/i=24/a=243201} {\bibfield  {journal}
  {\bibinfo  {journal} {Journal of Physics: Condensed Matter}\ }\textbf
  {\bibinfo {volume} {27}}~(\bibinfo {number} {24}),\ \bibinfo {pages}
  {243201}}\BibitemShut {NoStop}%
\bibitem [{\citenamefont {Schnyder}\ \emph {et~al.}(2012)\citenamefont
  {Schnyder}, \citenamefont {Brydon},\ and\ \citenamefont
  {Timm}}]{early_nodal_SC}%
  \BibitemOpen
  \bibfield  {author} {\bibinfo {author} {\bibnamefont {Schnyder},
  \bibfnamefont {A.~P.}}, \bibinfo {author} {\bibfnamefont {P.~M.~R.}\
  \bibnamefont {Brydon}}, \ and\ \bibinfo {author} {\bibfnamefont
  {C.}~\bibnamefont {Timm}}} (\bibinfo {year} {2012}),\ \href {\doibase
  10.1103/PhysRevB.85.024522} {\bibfield  {journal} {\bibinfo  {journal} {Phys.
  Rev. B}\ }\textbf {\bibinfo {volume} {85}},\ \bibinfo {pages}
  {024522}}\BibitemShut {NoStop}%
\bibitem [{\citenamefont {Schnyder}\ and\ \citenamefont
  {Ryu}(2011)}]{SchnyderRyuFlat}%
  \BibitemOpen
  \bibfield  {author} {\bibinfo {author} {\bibnamefont {Schnyder},
  \bibfnamefont {A.~P.}}, \ and\ \bibinfo {author} {\bibfnamefont
  {S.}~\bibnamefont {Ryu}}} (\bibinfo {year} {2011}),\ \href {\doibase
  10.1103/PhysRevB.84.060504} {\bibfield  {journal} {\bibinfo  {journal} {Phys.
  Rev. B}\ }\textbf {\bibinfo {volume} {84}},\ \bibinfo {pages}
  {060504}}\BibitemShut {NoStop}%
\bibitem [{\citenamefont {{Schnyder}}\ \emph {et~al.}(2008)\citenamefont
  {{Schnyder}}, \citenamefont {{Ryu}}, \citenamefont {{Furusaki}},\ and\
  \citenamefont {{Ludwig}}}]{Schnyder2008}%
  \BibitemOpen
  \bibfield  {author} {\bibinfo {author} {\bibnamefont {{Schnyder}},
  \bibfnamefont {A.~P.}}, \bibinfo {author} {\bibfnamefont {S.}~\bibnamefont
  {{Ryu}}}, \bibinfo {author} {\bibfnamefont {A.}~\bibnamefont {{Furusaki}}}, \
  and\ \bibinfo {author} {\bibfnamefont {A.~W.~W.}\ \bibnamefont {{Ludwig}}}}
  (\bibinfo {year} {2008}),\ \href {\doibase 10.1103/PhysRevB.78.195125}
  {\bibfield  {journal} {\bibinfo  {journal} {\prb}\ }\textbf {\bibinfo
  {volume} {78}},\ \bibinfo {eid} {195125}}\BibitemShut {NoStop}%
\bibitem [{\citenamefont {Schnyder}\ \emph {et~al.}(2009)\citenamefont
  {Schnyder}, \citenamefont {Ryu}, \citenamefont {Furusaki},\ and\
  \citenamefont {Ludwig}}]{SchnyderAIP}%
  \BibitemOpen
  \bibfield  {author} {\bibinfo {author} {\bibnamefont {Schnyder},
  \bibfnamefont {A.~P.}}, \bibinfo {author} {\bibfnamefont {S.}~\bibnamefont
  {Ryu}}, \bibinfo {author} {\bibfnamefont {A.}~\bibnamefont {Furusaki}}, \
  and\ \bibinfo {author} {\bibfnamefont {A.~W.~W.}\ \bibnamefont {Ludwig}}}
  (\bibinfo {year} {2009}),\ \href@noop {} {\bibfield  {journal} {\bibinfo
  {journal} {AIP Conf. Proc.}\ }\textbf {\bibinfo {volume} {1134}},\ \bibinfo
  {pages} {22}}\BibitemShut {NoStop}%
\bibitem [{\citenamefont {Schnyder}\ \emph {et~al.}(2013)\citenamefont
  {Schnyder}, \citenamefont {Timm},\ and\ \citenamefont
  {Brydon}}]{schnyder_PRL13}%
  \BibitemOpen
  \bibfield  {author} {\bibinfo {author} {\bibnamefont {Schnyder},
  \bibfnamefont {A.~P.}}, \bibinfo {author} {\bibfnamefont {C.}~\bibnamefont
  {Timm}}, \ and\ \bibinfo {author} {\bibfnamefont {P.~M.~R.}\ \bibnamefont
  {Brydon}}} (\bibinfo {year} {2013}),\ \href {\doibase
  10.1103/PhysRevLett.111.077001} {\bibfield  {journal} {\bibinfo  {journal}
  {Phys. Rev. Lett.}\ }\textbf {\bibinfo {volume} {111}},\ \bibinfo {pages}
  {077001}}\BibitemShut {NoStop}%
\bibitem [{\citenamefont {Schuch}\ \emph {et~al.}(2011)\citenamefont {Schuch},
  \citenamefont {Perez-Garcia},\ and\ \citenamefont {Cirac}}]{Schuch2011}%
  \BibitemOpen
  \bibfield  {author} {\bibinfo {author} {\bibnamefont {Schuch}, \bibfnamefont
  {N.}}, \bibinfo {author} {\bibfnamefont {D.}~\bibnamefont {Perez-Garcia}}, \
  and\ \bibinfo {author} {\bibfnamefont {I.}~\bibnamefont {Cirac}}} (\bibinfo
  {year} {2011}),\ \href@noop {} {\bibfield  {journal} {\bibinfo  {journal}
  {Phys. Rev. B}\ }\textbf {\bibinfo {volume} {84}},\ \bibinfo {pages}
  {165139}}\BibitemShut {NoStop}%
\bibitem [{\citenamefont {{Schuch}}\ \emph {et~al.}(2008)\citenamefont
  {{Schuch}}, \citenamefont {{Wolf}}, \citenamefont {{Verstraete}},\ and\
  \citenamefont {{Cirac}}}]{Schuch2008}%
  \BibitemOpen
  \bibfield  {author} {\bibinfo {author} {\bibnamefont {{Schuch}},
  \bibfnamefont {N.}}, \bibinfo {author} {\bibfnamefont {M.~M.}\ \bibnamefont
  {{Wolf}}}, \bibinfo {author} {\bibfnamefont {F.}~\bibnamefont
  {{Verstraete}}}, \ and\ \bibinfo {author} {\bibfnamefont {J.~I.}\
  \bibnamefont {{Cirac}}}} (\bibinfo {year} {2008}),\ \href {\doibase
  10.1103/PhysRevLett.100.030504} {\bibfield  {journal} {\bibinfo  {journal}
  {Phys. Rev. Lett.}\ }\textbf {\bibinfo {volume} {100}},\ \bibinfo {eid}
  {030504}}\BibitemShut {NoStop}%
\bibitem [{\citenamefont {Schulz-Baldes}\ \emph {et~al.}(2000)\citenamefont
  {Schulz-Baldes}, \citenamefont {Kellendonk},\ and\ \citenamefont
  {Richter}}]{Schulz00}%
  \BibitemOpen
  \bibfield  {author} {\bibinfo {author} {\bibnamefont {Schulz-Baldes},
  \bibfnamefont {H.}}, \bibinfo {author} {\bibfnamefont {J.}~\bibnamefont
  {Kellendonk}}, \ and\ \bibinfo {author} {\bibfnamefont {T.}~\bibnamefont
  {Richter}}} (\bibinfo {year} {2000}),\ \href
  {http://stacks.iop.org/0305-4470/33/i=2/a=102} {\bibfield  {journal}
  {\bibinfo  {journal} {Journal of Physics A: Mathematical and General}\
  }\textbf {\bibinfo {volume} {33}}~(\bibinfo {number} {2}),\ \bibinfo {pages}
  {L27}}\BibitemShut {NoStop}%
\bibitem [{\citenamefont {Senthil}(2015)}]{Senthil2014}%
  \BibitemOpen
  \bibfield  {author} {\bibinfo {author} {\bibnamefont {Senthil}, \bibfnamefont
  {T.}}} (\bibinfo {year} {2015}),\ \href {\doibase
  10.1146/annurev-conmatphys-031214-014740} {\bibfield  {journal} {\bibinfo
  {journal} {Annual Review of Condensed Matter Physics}\ }\textbf {\bibinfo
  {volume} {6}}~(\bibinfo {number} {1}),\ \bibinfo {pages} {299}}\BibitemShut
  {NoStop}%
\bibitem [{\citenamefont {{Senthil}}\ and\ \citenamefont
  {{Fisher}}(2000)}]{Senthil2000}%
  \BibitemOpen
  \bibfield  {author} {\bibinfo {author} {\bibnamefont {{Senthil}},
  \bibfnamefont {T.}}, \ and\ \bibinfo {author} {\bibfnamefont {M.~P.~A.}\
  \bibnamefont {{Fisher}}}} (\bibinfo {year} {2000}),\ \href {\doibase
  10.1103/PhysRevB.62.7850} {\bibfield  {journal} {\bibinfo  {journal} {\prb}\
  }\textbf {\bibinfo {volume} {62}},\ \bibinfo {pages} {7850}}\BibitemShut
  {NoStop}%
\bibitem [{\citenamefont {{Senthil}}\ \emph {et~al.}(1998)\citenamefont
  {{Senthil}}, \citenamefont {{Fisher}}, \citenamefont {{Balents}},\ and\
  \citenamefont {{Nayak}}}]{Senthil1998PhRvL..81.4704S}%
  \BibitemOpen
  \bibfield  {author} {\bibinfo {author} {\bibnamefont {{Senthil}},
  \bibfnamefont {T.}}, \bibinfo {author} {\bibfnamefont {M.~P.~A.}\
  \bibnamefont {{Fisher}}}, \bibinfo {author} {\bibfnamefont {L.}~\bibnamefont
  {{Balents}}}, \ and\ \bibinfo {author} {\bibfnamefont {C.}~\bibnamefont
  {{Nayak}}}} (\bibinfo {year} {1998}),\ \href {\doibase
  10.1103/PhysRevLett.81.4704} {\bibfield  {journal} {\bibinfo  {journal}
  {Physical Review Letters}\ }\textbf {\bibinfo {volume} {81}},\ \bibinfo
  {pages} {4704}}\BibitemShut {NoStop}%
\bibitem [{\citenamefont {{Senthil}}\ \emph {et~al.}(1999)\citenamefont
  {{Senthil}}, \citenamefont {{Marston}},\ and\ \citenamefont
  {{Fisher}}}]{Senthil1999PhRvB..60.4245S}%
  \BibitemOpen
  \bibfield  {author} {\bibinfo {author} {\bibnamefont {{Senthil}},
  \bibfnamefont {T.}}, \bibinfo {author} {\bibfnamefont {J.~B.}\ \bibnamefont
  {{Marston}}}, \ and\ \bibinfo {author} {\bibfnamefont {M.~P.~A.}\
  \bibnamefont {{Fisher}}}} (\bibinfo {year} {1999}),\ \href {\doibase
  10.1103/PhysRevB.60.4245} {\bibfield  {journal} {\bibinfo  {journal} {\prb}\
  }\textbf {\bibinfo {volume} {60}},\ \bibinfo {pages} {4245}}\BibitemShut
  {NoStop}%
\bibitem [{\citenamefont {Serbyn}\ and\ \citenamefont
  {Fu}(2014)}]{Surface_TCI_Fu}%
  \BibitemOpen
  \bibfield  {author} {\bibinfo {author} {\bibnamefont {Serbyn}, \bibfnamefont
  {M.}}, \ and\ \bibinfo {author} {\bibfnamefont {L.}~\bibnamefont {Fu}}}
  (\bibinfo {year} {2014}),\ \href {\doibase 10.1103/PhysRevB.90.035402}
  {\bibfield  {journal} {\bibinfo  {journal} {Phys. Rev. B}\ }\textbf {\bibinfo
  {volume} {90}},\ \bibinfo {pages} {035402}}\BibitemShut {NoStop}%
\bibitem [{\citenamefont {Shekhar}\ \emph {et~al.}(2015)\citenamefont
  {Shekhar}, \citenamefont {Nayak}, \citenamefont {Sun}, \citenamefont
  {Schmidt}, \citenamefont {Nicklas}, \citenamefont {Leermakers}, \citenamefont
  {Zeitler}, \citenamefont {Skourski}, \citenamefont {Wosnitza}, \citenamefont
  {Liu}, \citenamefont {Chen}, \citenamefont {Schnelle}, \citenamefont
  {Borrmann}, \citenamefont {Grin}, \citenamefont {Felser},\ and\ \citenamefont
  {Yan}}]{NbP_weyl_magnetotransport_arXiv1502}%
  \BibitemOpen
  \bibfield  {author} {\bibinfo {author} {\bibnamefont {Shekhar}, \bibfnamefont
  {C.}}, \bibinfo {author} {\bibfnamefont {A.~K.}\ \bibnamefont {Nayak}},
  \bibinfo {author} {\bibfnamefont {Y.}~\bibnamefont {Sun}}, \bibinfo {author}
  {\bibfnamefont {M.}~\bibnamefont {Schmidt}}, \bibinfo {author} {\bibfnamefont
  {M.}~\bibnamefont {Nicklas}}, \bibinfo {author} {\bibfnamefont
  {I.}~\bibnamefont {Leermakers}}, \bibinfo {author} {\bibfnamefont
  {U.}~\bibnamefont {Zeitler}}, \bibinfo {author} {\bibfnamefont
  {Y.}~\bibnamefont {Skourski}}, \bibinfo {author} {\bibfnamefont
  {J.}~\bibnamefont {Wosnitza}}, \bibinfo {author} {\bibfnamefont
  {Z.}~\bibnamefont {Liu}}, \bibinfo {author} {\bibfnamefont {Y.}~\bibnamefont
  {Chen}}, \bibinfo {author} {\bibfnamefont {W.}~\bibnamefont {Schnelle}},
  \bibinfo {author} {\bibfnamefont {H.}~\bibnamefont {Borrmann}}, \bibinfo
  {author} {\bibfnamefont {Y.}~\bibnamefont {Grin}}, \bibinfo {author}
  {\bibfnamefont {C.}~\bibnamefont {Felser}}, \ and\ \bibinfo {author}
  {\bibfnamefont {B.}~\bibnamefont {Yan}}} (\bibinfo {year} {2015}),\ \href
  {http://dx.doi.org/10.1038/nphys3372} {\bibfield  {journal} {\bibinfo
  {journal} {Nat Phys}\ }\textbf {\bibinfo {volume} {11}}~(\bibinfo {number}
  {8}),\ \bibinfo {pages} {645}}\BibitemShut {NoStop}%
\bibitem [{\citenamefont {Shen}(2012)}]{shenBook2012}%
  \BibitemOpen
  \bibfield  {author} {\bibinfo {author} {\bibnamefont {Shen}, \bibfnamefont
  {S.-Q.}}} (\bibinfo {year} {2012}),\ \href@noop {} {\emph {\bibinfo {title}
  {Topological insulators}}},\ Springer Series in Solid-State Sciences, Volume
  174\ (\bibinfo  {publisher} {Springer})\BibitemShut {NoStop}%
\bibitem [{\citenamefont {Sheng}\ \emph {et~al.}(2011)\citenamefont {Sheng},
  \citenamefont {Gu}, \citenamefont {Sun},\ and\ \citenamefont
  {Sheng}}]{Sheng_FTI}%
  \BibitemOpen
  \bibfield  {author} {\bibinfo {author} {\bibnamefont {Sheng}, \bibfnamefont
  {D.~N.}}, \bibinfo {author} {\bibfnamefont {Z.-C.}\ \bibnamefont {Gu}},
  \bibinfo {author} {\bibfnamefont {K.}~\bibnamefont {Sun}}, \ and\ \bibinfo
  {author} {\bibfnamefont {L.}~\bibnamefont {Sheng}}} (\bibinfo {year}
  {2011}),\ \href {http://dx.doi.org/10.1038/ncomms1380} {\bibfield  {journal}
  {\bibinfo  {journal} {Nat Commun}\ }\textbf {\bibinfo {volume} {2}},\
  \bibinfo {pages} {389}}\BibitemShut {NoStop}%
\bibitem [{\citenamefont {Shindou}\ and\ \citenamefont
  {Murakami}(2009)}]{PhysRevB.79.045321}%
  \BibitemOpen
  \bibfield  {author} {\bibinfo {author} {\bibnamefont {Shindou}, \bibfnamefont
  {R.}}, \ and\ \bibinfo {author} {\bibfnamefont {S.}~\bibnamefont {Murakami}}}
  (\bibinfo {year} {2009}),\ \href {\doibase 10.1103/PhysRevB.79.045321}
  {\bibfield  {journal} {\bibinfo  {journal} {Phys. Rev. B}\ }\textbf {\bibinfo
  {volume} {79}},\ \bibinfo {pages} {045321}}\BibitemShut {NoStop}%
\bibitem [{\citenamefont {Shindou}\ \emph {et~al.}(2010)\citenamefont
  {Shindou}, \citenamefont {Nakai},\ and\ \citenamefont
  {Murakami}}]{1367-2630-12-6-065008}%
  \BibitemOpen
  \bibfield  {author} {\bibinfo {author} {\bibnamefont {Shindou}, \bibfnamefont
  {R.}}, \bibinfo {author} {\bibfnamefont {R.}~\bibnamefont {Nakai}}, \ and\
  \bibinfo {author} {\bibfnamefont {S.}~\bibnamefont {Murakami}}} (\bibinfo
  {year} {2010}),\ \href {http://stacks.iop.org/1367-2630/12/i=6/a=065008}
  {\bibfield  {journal} {\bibinfo  {journal} {New Journal of Physics}\ }\textbf
  {\bibinfo {volume} {12}}~(\bibinfo {number} {6}),\ \bibinfo {pages}
  {065008}}\BibitemShut {NoStop}%
\bibitem [{\citenamefont {Shiozaki}\ and\ \citenamefont
  {Sato}(2014)}]{Sato_Crystalline_PRB14}%
  \BibitemOpen
  \bibfield  {author} {\bibinfo {author} {\bibnamefont {Shiozaki},
  \bibfnamefont {K.}}, \ and\ \bibinfo {author} {\bibfnamefont
  {M.}~\bibnamefont {Sato}}} (\bibinfo {year} {2014}),\ \href {\doibase
  10.1103/PhysRevB.90.165114} {\bibfield  {journal} {\bibinfo  {journal} {Phys.
  Rev. B}\ }\textbf {\bibinfo {volume} {90}},\ \bibinfo {pages}
  {165114}}\BibitemShut {NoStop}%
\bibitem [{\citenamefont {Shiozaki}\ \emph {et~al.}(2015)\citenamefont
  {Shiozaki}, \citenamefont {Sato},\ and\ \citenamefont
  {Gomi}}]{nonsymmorphic_Sato}%
  \BibitemOpen
  \bibfield  {author} {\bibinfo {author} {\bibnamefont {Shiozaki},
  \bibfnamefont {K.}}, \bibinfo {author} {\bibfnamefont {M.}~\bibnamefont
  {Sato}}, \ and\ \bibinfo {author} {\bibfnamefont {K.}~\bibnamefont {Gomi}}}
  (\bibinfo {year} {2015}),\ \href {\doibase 10.1103/PhysRevB.91.155120}
  {\bibfield  {journal} {\bibinfo  {journal} {Phys. Rev. B}\ }\textbf {\bibinfo
  {volume} {91}},\ \bibinfo {pages} {155120}}\BibitemShut {NoStop}%
\bibitem [{\citenamefont {{Shitade}}\ \emph {et~al.}(2009)\citenamefont
  {{Shitade}}, \citenamefont {{Katsura}}, \citenamefont {{Kune{\v s}}},
  \citenamefont {{Qi}}, \citenamefont {{Zhang}},\ and\ \citenamefont
  {{Nagaosa}}}]{Shitade2009}%
  \BibitemOpen
  \bibfield  {author} {\bibinfo {author} {\bibnamefont {{Shitade}},
  \bibfnamefont {A.}}, \bibinfo {author} {\bibfnamefont {H.}~\bibnamefont
  {{Katsura}}}, \bibinfo {author} {\bibfnamefont {J.}~\bibnamefont {{Kune{\v
  s}}}}, \bibinfo {author} {\bibfnamefont {X.-L.}\ \bibnamefont {{Qi}}},
  \bibinfo {author} {\bibfnamefont {S.-C.}\ \bibnamefont {{Zhang}}}, \ and\
  \bibinfo {author} {\bibfnamefont {N.}~\bibnamefont {{Nagaosa}}}} (\bibinfo
  {year} {2009}),\ \href {\doibase 10.1103/PhysRevLett.102.256403} {\bibfield
  {journal} {\bibinfo  {journal} {Phys. Rev. Lett.}\ }\textbf {\bibinfo
  {volume} {102}}~(\bibinfo {number} {25}),\ \bibinfo {eid}
  {256403}}\BibitemShut {NoStop}%
\bibitem [{\citenamefont {Sigrist}\ and\ \citenamefont
  {Ueda}(1991)}]{SigristUeda91}%
  \BibitemOpen
  \bibfield  {author} {\bibinfo {author} {\bibnamefont {Sigrist}, \bibfnamefont
  {M.}}, \ and\ \bibinfo {author} {\bibfnamefont {K.}~\bibnamefont {Ueda}}}
  (\bibinfo {year} {1991}),\ \href {\doibase 10.1103/RevModPhys.63.239}
  {\bibfield  {journal} {\bibinfo  {journal} {Rev. Mod. Phys.}\ }\textbf
  {\bibinfo {volume} {63}},\ \bibinfo {pages} {239}}\BibitemShut {NoStop}%
\bibitem [{\citenamefont {Slager}\ \emph {et~al.}(2013)\citenamefont {Slager},
  \citenamefont {Mesaros}, \citenamefont {Juricic},\ and\ \citenamefont
  {Zaanen}}]{Slager:2013fk}%
  \BibitemOpen
  \bibfield  {author} {\bibinfo {author} {\bibnamefont {Slager}, \bibfnamefont
  {R.-J.}}, \bibinfo {author} {\bibfnamefont {A.}~\bibnamefont {Mesaros}},
  \bibinfo {author} {\bibfnamefont {V.}~\bibnamefont {Juricic}}, \ and\
  \bibinfo {author} {\bibfnamefont {J.}~\bibnamefont {Zaanen}}} (\bibinfo
  {year} {2013}),\ \href@noop {} {\bibfield  {journal} {\bibinfo  {journal}
  {Nat Phys}\ }\textbf {\bibinfo {volume} {9}}~(\bibinfo {number} {2}),\
  \bibinfo {pages} {98}}\BibitemShut {NoStop}%
\bibitem [{\citenamefont {Soluyanov}\ \emph {et~al.}(2015)\citenamefont
  {Soluyanov}, \citenamefont {Gresch}, \citenamefont {Wang}, \citenamefont
  {Wu}, \citenamefont {Troyer}, \citenamefont {Dai},\ and\ \citenamefont
  {Bernevig}}]{Soluyanov:2015aa}%
  \BibitemOpen
  \bibfield  {author} {\bibinfo {author} {\bibnamefont {Soluyanov},
  \bibfnamefont {A.~A.}}, \bibinfo {author} {\bibfnamefont {D.}~\bibnamefont
  {Gresch}}, \bibinfo {author} {\bibfnamefont {Z.}~\bibnamefont {Wang}},
  \bibinfo {author} {\bibfnamefont {Q.}~\bibnamefont {Wu}}, \bibinfo {author}
  {\bibfnamefont {M.}~\bibnamefont {Troyer}}, \bibinfo {author} {\bibfnamefont
  {X.}~\bibnamefont {Dai}}, \ and\ \bibinfo {author} {\bibfnamefont {B.~A.}\
  \bibnamefont {Bernevig}}} (\bibinfo {year} {2015}),\ \href@noop {} {\bibfield
   {journal} {\bibinfo  {journal} {Nature}\ }\textbf {\bibinfo {volume}
  {527}}~(\bibinfo {number} {7579}),\ \bibinfo {pages} {495}}\BibitemShut
  {NoStop}%
\bibitem [{\citenamefont {Soluyanov}\ and\ \citenamefont
  {Vanderbilt}(2011)}]{SoluyanovVanderbilt2011}%
  \BibitemOpen
  \bibfield  {author} {\bibinfo {author} {\bibnamefont {Soluyanov},
  \bibfnamefont {A.~A.}}, \ and\ \bibinfo {author} {\bibfnamefont
  {D.}~\bibnamefont {Vanderbilt}}} (\bibinfo {year} {2011}),\ \href {\doibase
  10.1103/PhysRevB.83.035108} {\bibfield  {journal} {\bibinfo  {journal} {Phys.
  Rev. B}\ }\textbf {\bibinfo {volume} {83}},\ \bibinfo {pages}
  {035108}}\BibitemShut {NoStop}%
\bibitem [{\citenamefont {Stanescu}\ and\ \citenamefont
  {Tewari}(2013)}]{Stanescu_Majorana_review}%
  \BibitemOpen
  \bibfield  {author} {\bibinfo {author} {\bibnamefont {Stanescu},
  \bibfnamefont {T.~D.}}, \ and\ \bibinfo {author} {\bibfnamefont
  {S.}~\bibnamefont {Tewari}}} (\bibinfo {year} {2013}),\ \href
  {http://stacks.iop.org/0953-8984/25/i=23/a=233201} {\bibfield  {journal}
  {\bibinfo  {journal} {J. Phys. Condens. Matter}\ }\textbf {\bibinfo {volume}
  {25}},\ \bibinfo {pages} {233201}}\BibitemShut {NoStop}%
\bibitem [{\citenamefont {Stone}(2012)}]{Stone12}%
  \BibitemOpen
  \bibfield  {author} {\bibinfo {author} {\bibnamefont {Stone}, \bibfnamefont
  {M.}}} (\bibinfo {year} {2012}),\ \href {\doibase 10.1103/PhysRevB.85.184503}
  {\bibfield  {journal} {\bibinfo  {journal} {Phys. Rev. B}\ }\textbf {\bibinfo
  {volume} {85}},\ \bibinfo {pages} {184503}}\BibitemShut {NoStop}%
\bibitem [{\citenamefont {Stone}\ \emph {et~al.}(2011)\citenamefont {Stone},
  \citenamefont {Chiu},\ and\ \citenamefont {Roy}}]{Stone:2011qo}%
  \BibitemOpen
  \bibfield  {author} {\bibinfo {author} {\bibnamefont {Stone}, \bibfnamefont
  {M.}}, \bibinfo {author} {\bibfnamefont {C.-K.}\ \bibnamefont {Chiu}}, \ and\
  \bibinfo {author} {\bibfnamefont {A.}~\bibnamefont {Roy}}} (\bibinfo {year}
  {2011}),\ \href@noop {} {\bibfield  {journal} {\bibinfo  {journal} {J. Phys.
  A: Math. Theor.}\ }\textbf {\bibinfo {volume} {44}},\ \bibinfo {pages}
  {045001}}\BibitemShut {NoStop}%
\bibitem [{\citenamefont {Stone}\ and\ \citenamefont {Roy}(2004)}]{StoneRoy04}%
  \BibitemOpen
  \bibfield  {author} {\bibinfo {author} {\bibnamefont {Stone}, \bibfnamefont
  {M.}}, \ and\ \bibinfo {author} {\bibfnamefont {R.}~\bibnamefont {Roy}}}
  (\bibinfo {year} {2004}),\ \href {\doibase 10.1103/PhysRevB.69.184511}
  {\bibfield  {journal} {\bibinfo  {journal} {Phys. Rev. B}\ }\textbf {\bibinfo
  {volume} {69}},\ \bibinfo {pages}
  {\href{http://link.aps.org/doi/10.1103/PhysRevB.69.184511}{184511}}}\BibitemShut
  {NoStop}%
\bibitem [{\citenamefont {Su}\ \emph {et~al.}(1980)\citenamefont {Su},
  \citenamefont {Schrieffer},\ and\ \citenamefont
  {Heeger}}]{suSchriefferHeegerPRB80}%
  \BibitemOpen
  \bibfield  {author} {\bibinfo {author} {\bibnamefont {Su}, \bibfnamefont
  {W.~P.}}, \bibinfo {author} {\bibfnamefont {J.~R.}\ \bibnamefont
  {Schrieffer}}, \ and\ \bibinfo {author} {\bibfnamefont {A.~J.}\ \bibnamefont
  {Heeger}}} (\bibinfo {year} {1980}),\ \href {\doibase
  10.1103/PhysRevB.22.2099} {\bibfield  {journal} {\bibinfo  {journal} {Phys.
  Rev. B}\ }\textbf {\bibinfo {volume} {22}},\ \bibinfo {pages}
  {2099}}\BibitemShut {NoStop}%
\bibitem [{\citenamefont {{Sule}}\ \emph {et~al.}(2013)\citenamefont {{Sule}},
  \citenamefont {{Chen}},\ and\ \citenamefont {{Ryu}}}]{Sule2013}%
  \BibitemOpen
  \bibfield  {author} {\bibinfo {author} {\bibnamefont {{Sule}}, \bibfnamefont
  {O.~M.}}, \bibinfo {author} {\bibfnamefont {X.}~\bibnamefont {{Chen}}}, \
  and\ \bibinfo {author} {\bibfnamefont {S.}~\bibnamefont {{Ryu}}}} (\bibinfo
  {year} {2013}),\ \href {\doibase 10.1103/PhysRevB.88.075125} {\bibfield
  {journal} {\bibinfo  {journal} {\prb}\ }\textbf {\bibinfo {volume} {88}},\
  \bibinfo {eid} {075125}}\BibitemShut {NoStop}%
\bibitem [{\citenamefont {Sun}\ \emph {et~al.}(2012)\citenamefont {Sun},
  \citenamefont {Liu}, \citenamefont {Hemmerich},\ and\ \citenamefont
  {Das~Sarma}}]{sun_NatPhys_12}%
  \BibitemOpen
  \bibfield  {author} {\bibinfo {author} {\bibnamefont {Sun}, \bibfnamefont
  {K.}}, \bibinfo {author} {\bibfnamefont {W.~V.}\ \bibnamefont {Liu}},
  \bibinfo {author} {\bibfnamefont {A.}~\bibnamefont {Hemmerich}}, \ and\
  \bibinfo {author} {\bibfnamefont {S.}~\bibnamefont {Das~Sarma}}} (\bibinfo
  {year} {2012}),\ \href {http://dx.doi.org/10.1038/nphys2134} {\bibfield
  {journal} {\bibinfo  {journal} {Nat Phys}\ }\textbf {\bibinfo {volume}
  {8}}~(\bibinfo {number} {1}),\ \bibinfo {pages} {67}}\BibitemShut {NoStop}%
\bibitem [{\citenamefont {Swingle}\ \emph {et~al.}(2011)\citenamefont
  {Swingle}, \citenamefont {Barkeshli}, \citenamefont {McGreevy},\ and\
  \citenamefont {Senthil}}]{swingle_barkeshli_macgreevy_senthil}%
  \BibitemOpen
  \bibfield  {author} {\bibinfo {author} {\bibnamefont {Swingle}, \bibfnamefont
  {B.}}, \bibinfo {author} {\bibfnamefont {M.}~\bibnamefont {Barkeshli}},
  \bibinfo {author} {\bibfnamefont {J.}~\bibnamefont {McGreevy}}, \ and\
  \bibinfo {author} {\bibfnamefont {T.}~\bibnamefont {Senthil}}} (\bibinfo
  {year} {2011}),\ \href {\doibase 10.1103/PhysRevB.83.195139} {\bibfield
  {journal} {\bibinfo  {journal} {Phys. Rev. B}\ }\textbf {\bibinfo {volume}
  {83}},\ \bibinfo {pages} {195139}}\BibitemShut {NoStop}%
\bibitem [{\citenamefont {{Takane}}(2004{\natexlab{a}})}]{Takane2004c}%
  \BibitemOpen
  \bibfield  {author} {\bibinfo {author} {\bibnamefont {{Takane}},
  \bibfnamefont {Y.}}} (\bibinfo {year} {2004}{\natexlab{a}}),\ \href {\doibase
  10.1143/JPSJ.73.2366} {\bibfield  {journal} {\bibinfo  {journal} {J. Phys.
  Soc. Jpn.}\ }\textbf {\bibinfo {volume} {73}},\ \bibinfo {pages}
  {2366}}\BibitemShut {NoStop}%
\bibitem [{\citenamefont {{Takane}}(2004{\natexlab{b}})}]{Takane2004a}%
  \BibitemOpen
  \bibfield  {author} {\bibinfo {author} {\bibnamefont {{Takane}},
  \bibfnamefont {Y.}}} (\bibinfo {year} {2004}{\natexlab{b}}),\ \href {\doibase
  10.1143/JPSJ.73.9} {\bibfield  {journal} {\bibinfo  {journal} {J. Phys. Soc.
  Jpn.}\ }\textbf {\bibinfo {volume} {73}},\ \bibinfo {pages} {9}}\BibitemShut
  {NoStop}%
\bibitem [{\citenamefont {{Takane}}(2004{\natexlab{c}})}]{Takane2004b}%
  \BibitemOpen
  \bibfield  {author} {\bibinfo {author} {\bibnamefont {{Takane}},
  \bibfnamefont {Y.}}} (\bibinfo {year} {2004}{\natexlab{c}}),\ \href {\doibase
  10.1143/JPSJ.73.1430} {\bibfield  {journal} {\bibinfo  {journal} {J. Phys.
  Soc. Jpn.}\ }\textbf {\bibinfo {volume} {73}},\ \bibinfo {pages}
  {1430}}\BibitemShut {NoStop}%
\bibitem [{\citenamefont {Tanaka}\ \emph
  {et~al.}(2012{\natexlab{a}})\citenamefont {Tanaka}, \citenamefont {Ren},
  \citenamefont {Sato}, \citenamefont {Nakayama}, \citenamefont {Souma},
  \citenamefont {Takahashi}, \citenamefont {Segawa},\ and\ \citenamefont
  {Ando}}]{Tanaka:2012fk}%
  \BibitemOpen
  \bibfield  {author} {\bibinfo {author} {\bibnamefont {Tanaka}, \bibfnamefont
  {Y.}}, \bibinfo {author} {\bibfnamefont {Z.}~\bibnamefont {Ren}}, \bibinfo
  {author} {\bibfnamefont {T.}~\bibnamefont {Sato}}, \bibinfo {author}
  {\bibfnamefont {K.}~\bibnamefont {Nakayama}}, \bibinfo {author}
  {\bibfnamefont {S.}~\bibnamefont {Souma}}, \bibinfo {author} {\bibfnamefont
  {T.}~\bibnamefont {Takahashi}}, \bibinfo {author} {\bibfnamefont
  {K.}~\bibnamefont {Segawa}}, \ and\ \bibinfo {author} {\bibfnamefont
  {Y.}~\bibnamefont {Ando}}} (\bibinfo {year} {2012}{\natexlab{a}}),\
  \href@noop {} {\bibfield  {journal} {\bibinfo  {journal} {Nat. Phys.}\
  }\textbf {\bibinfo {volume} {8}},\ \bibinfo {pages} {800}}\BibitemShut
  {NoStop}%
\bibitem [{\citenamefont {Tanaka}\ \emph
  {et~al.}(2012{\natexlab{b}})\citenamefont {Tanaka}, \citenamefont {Sato},\
  and\ \citenamefont {Nagaosa}}]{Tanaka_SC_odd_even}%
  \BibitemOpen
  \bibfield  {author} {\bibinfo {author} {\bibnamefont {Tanaka}, \bibfnamefont
  {Y.}}, \bibinfo {author} {\bibfnamefont {M.}~\bibnamefont {Sato}}, \ and\
  \bibinfo {author} {\bibfnamefont {N.}~\bibnamefont {Nagaosa}}} (\bibinfo
  {year} {2012}{\natexlab{b}}),\ \href {\doibase 10.1143/JPSJ.81.011013}
  {\bibfield  {journal} {\bibinfo  {journal} {Journal of the Physical Society
  of Japan}\ }\textbf {\bibinfo {volume} {81}}~(\bibinfo {number} {1}),\
  \bibinfo {pages} {011013}}\BibitemShut {NoStop}%
\bibitem [{\citenamefont {Tanaka}\ \emph {et~al.}(2009)\citenamefont {Tanaka},
  \citenamefont {Yokoyama},\ and\ \citenamefont {Nagaosa}}]{Fe_SC_TI_Tanaka}%
  \BibitemOpen
  \bibfield  {author} {\bibinfo {author} {\bibnamefont {Tanaka}, \bibfnamefont
  {Y.}}, \bibinfo {author} {\bibfnamefont {T.}~\bibnamefont {Yokoyama}}, \ and\
  \bibinfo {author} {\bibfnamefont {N.}~\bibnamefont {Nagaosa}}} (\bibinfo
  {year} {2009}),\ \href {\doibase 10.1103/PhysRevLett.103.107002} {\bibfield
  {journal} {\bibinfo  {journal} {Phys. Rev. Lett.}\ }\textbf {\bibinfo
  {volume} {103}},\ \bibinfo {pages} {107002}}\BibitemShut {NoStop}%
\bibitem [{\citenamefont {{Tang}}\ and\ \citenamefont
  {{Wen}}(2012)}]{Tang2012}%
  \BibitemOpen
  \bibfield  {author} {\bibinfo {author} {\bibnamefont {{Tang}}, \bibfnamefont
  {E.}}, \ and\ \bibinfo {author} {\bibfnamefont {X.-G.}\ \bibnamefont
  {{Wen}}}} (\bibinfo {year} {2012}),\ \href {\doibase
  10.1103/PhysRevLett.109.096403} {\bibfield  {journal} {\bibinfo  {journal}
  {Phys. Rev. Lett.}\ }\textbf {\bibinfo {volume} {109}},\ \bibinfo {eid}
  {096403}}\BibitemShut {NoStop}%
\bibitem [{\citenamefont {Teo}\ \emph {et~al.}(2008)\citenamefont {Teo},
  \citenamefont {Fu},\ and\ \citenamefont {Kane}}]{Teo:2008fk}%
  \BibitemOpen
  \bibfield  {author} {\bibinfo {author} {\bibnamefont {Teo}, \bibfnamefont
  {J.~C.~Y.}}, \bibinfo {author} {\bibfnamefont {L.}~\bibnamefont {Fu}}, \ and\
  \bibinfo {author} {\bibfnamefont {C.~L.}\ \bibnamefont {Kane}}} (\bibinfo
  {year} {2008}),\ \href@noop {} {\bibfield  {journal} {\bibinfo  {journal}
  {Phys. Rev. B}\ }\textbf {\bibinfo {volume} {78}},\ \bibinfo {pages}
  {045426}}\BibitemShut {NoStop}%
\bibitem [{\citenamefont {Teo}\ and\ \citenamefont
  {Hughes}(2013)}]{Teo_2013_disclination}%
  \BibitemOpen
  \bibfield  {author} {\bibinfo {author} {\bibnamefont {Teo}, \bibfnamefont
  {J.~C.~Y.}}, \ and\ \bibinfo {author} {\bibfnamefont {T.~L.}\ \bibnamefont
  {Hughes}}} (\bibinfo {year} {2013}),\ \href {\doibase
  10.1103/PhysRevLett.111.047006} {\bibfield  {journal} {\bibinfo  {journal}
  {Phys. Rev. Lett.}\ }\textbf {\bibinfo {volume} {111}},\ \bibinfo {pages}
  {047006}}\BibitemShut {NoStop}%
\bibitem [{\citenamefont {Teo}\ \emph {et~al.}(2015)\citenamefont {Teo},
  \citenamefont {Hughes},\ and\ \citenamefont {Fradkin}}]{2015arXiv150306812T}%
  \BibitemOpen
  \bibfield  {author} {\bibinfo {author} {\bibnamefont {Teo}, \bibfnamefont
  {J.~C.~Y.}}, \bibinfo {author} {\bibfnamefont {T.~L.}\ \bibnamefont
  {Hughes}}, \ and\ \bibinfo {author} {\bibfnamefont {E.}~\bibnamefont
  {Fradkin}}} (\bibinfo {year} {2015}),\ \href {\doibase
  http://dx.doi.org/10.1016/j.aop.2015.05.012} {\bibfield  {journal} {\bibinfo
  {journal} {Annals of Physics}\ }\textbf {\bibinfo {volume} {360}},\ \bibinfo
  {pages} {349}}\BibitemShut {NoStop}%
\bibitem [{\citenamefont {Teo}\ and\ \citenamefont
  {Kane}(2010{\natexlab{a}})}]{Teo_Majorana}%
  \BibitemOpen
  \bibfield  {author} {\bibinfo {author} {\bibnamefont {Teo}, \bibfnamefont
  {J.~C.~Y.}}, \ and\ \bibinfo {author} {\bibfnamefont {C.~L.}\ \bibnamefont
  {Kane}}} (\bibinfo {year} {2010}{\natexlab{a}}),\ \href {\doibase
  10.1103/PhysRevLett.104.046401} {\bibfield  {journal} {\bibinfo  {journal}
  {Phys. Rev. Lett.}\ }\textbf {\bibinfo {volume} {104}},\ \bibinfo {pages}
  {046401}}\BibitemShut {NoStop}%
\bibitem [{\citenamefont {Teo}\ and\ \citenamefont
  {Kane}(2010{\natexlab{b}})}]{Teo:2010fk}%
  \BibitemOpen
  \bibfield  {author} {\bibinfo {author} {\bibnamefont {Teo}, \bibfnamefont
  {J.~C.~Y.}}, \ and\ \bibinfo {author} {\bibfnamefont {C.~L.}\ \bibnamefont
  {Kane}}} (\bibinfo {year} {2010}{\natexlab{b}}),\ \href@noop {} {\bibfield
  {journal} {\bibinfo  {journal} {Phys. Rev. B}\ }\textbf {\bibinfo {volume}
  {82}}~(\bibinfo {number} {11}),\ \bibinfo {pages} {115120}}\BibitemShut
  {NoStop}%
\bibitem [{\citenamefont {Tewari}\ \emph {et~al.}(2008)\citenamefont {Tewari},
  \citenamefont {Zhang}, \citenamefont {Das~Sarma}, \citenamefont {Nayak},\
  and\ \citenamefont {Lee}}]{TewariZhangDasSarmaNayakLee08}%
  \BibitemOpen
  \bibfield  {author} {\bibinfo {author} {\bibnamefont {Tewari}, \bibfnamefont
  {S.}}, \bibinfo {author} {\bibfnamefont {C.}~\bibnamefont {Zhang}}, \bibinfo
  {author} {\bibfnamefont {S.}~\bibnamefont {Das~Sarma}}, \bibinfo {author}
  {\bibfnamefont {C.}~\bibnamefont {Nayak}}, \ and\ \bibinfo {author}
  {\bibfnamefont {D.-H.}\ \bibnamefont {Lee}}} (\bibinfo {year} {2008}),\ \href
  {\doibase 10.1103/PhysRevLett.100.027001} {\bibfield  {journal} {\bibinfo
  {journal} {Phys. Rev. Lett.}\ }\textbf {\bibinfo {volume} {100}},\ \bibinfo
  {pages} {027001}}\BibitemShut {NoStop}%
\bibitem [{\citenamefont {Thiang}(2015)}]{Thiang2014}%
  \BibitemOpen
  \bibfield  {author} {\bibinfo {author} {\bibnamefont {Thiang}, \bibfnamefont
  {G.~C.}}} (\bibinfo {year} {2015}),\ \href {\doibase
  10.1007/s00023-015-0418-9} {\bibfield  {journal} {\bibinfo  {journal}
  {Annales Henri Poincar{\'e}}\ }\textbf {\bibinfo {volume} {17}}~(\bibinfo
  {number} {4}),\ \bibinfo {pages} {757}}\BibitemShut {NoStop}%
\bibitem [{\citenamefont {Thouless}(1983)}]{Thoulesspump}%
  \BibitemOpen
  \bibfield  {author} {\bibinfo {author} {\bibnamefont {Thouless},
  \bibfnamefont {D.~J.}}} (\bibinfo {year} {1983}),\ \href {\doibase
  10.1103/PhysRevB.27.6083} {\bibfield  {journal} {\bibinfo  {journal} {Phys.
  Rev. B}\ }\textbf {\bibinfo {volume} {27}},\ \bibinfo {pages}
  {6083}}\BibitemShut {NoStop}%
\bibitem [{\citenamefont {Thouless}\ \emph {et~al.}(1982)\citenamefont
  {Thouless}, \citenamefont {Kohmoto}, \citenamefont {Nightingale},\ and\
  \citenamefont {den Nijs}}]{Thouless:1982rz}%
  \BibitemOpen
  \bibfield  {author} {\bibinfo {author} {\bibnamefont {Thouless},
  \bibfnamefont {D.~J.}}, \bibinfo {author} {\bibfnamefont {M.}~\bibnamefont
  {Kohmoto}}, \bibinfo {author} {\bibfnamefont {M.~P.}\ \bibnamefont
  {Nightingale}}, \ and\ \bibinfo {author} {\bibfnamefont {M.}~\bibnamefont
  {den Nijs}}} (\bibinfo {year} {1982}),\ \href@noop {} {\bibfield  {journal}
  {\bibinfo  {journal} {Phys. Rev. Lett.}\ }\textbf {\bibinfo {volume} {49}},\
  \bibinfo {pages} {405}}\BibitemShut {NoStop}%
\bibitem [{\citenamefont {{Titov}}\ \emph {et~al.}(2001)\citenamefont
  {{Titov}}, \citenamefont {{Brouwer}}, \citenamefont {{Furusaki}},\ and\
  \citenamefont {{Mudry}}}]{Titov2001PhRvB..63w5318T}%
  \BibitemOpen
  \bibfield  {author} {\bibinfo {author} {\bibnamefont {{Titov}}, \bibfnamefont
  {M.}}, \bibinfo {author} {\bibfnamefont {P.~W.}\ \bibnamefont {{Brouwer}}},
  \bibinfo {author} {\bibfnamefont {A.}~\bibnamefont {{Furusaki}}}, \ and\
  \bibinfo {author} {\bibfnamefont {C.}~\bibnamefont {{Mudry}}}} (\bibinfo
  {year} {2001}),\ \href {\doibase 10.1103/PhysRevB.63.235318} {\bibfield
  {journal} {\bibinfo  {journal} {\prb}\ }\textbf {\bibinfo {volume}
  {63}}~(\bibinfo {number} {23}),\ \bibinfo {eid} {235318}}\BibitemShut
  {NoStop}%
\bibitem [{\citenamefont {Tsvelik}(1995)}]{Tsvelik1995}%
  \BibitemOpen
  \bibfield  {author} {\bibinfo {author} {\bibnamefont {Tsvelik}, \bibfnamefont
  {A.~M.}}} (\bibinfo {year} {1995}),\ \href {\doibase
  10.1103/PhysRevB.51.9449} {\bibfield  {journal} {\bibinfo  {journal} {Phys.
  Rev. B}\ }\textbf {\bibinfo {volume} {51}},\ \bibinfo {pages}
  {9449}}\BibitemShut {NoStop}%
\bibitem [{\citenamefont {Turner}\ \emph {et~al.}(2011)\citenamefont {Turner},
  \citenamefont {Pollmann},\ and\ \citenamefont {Berg}}]{turnerPollmannPRB11}%
  \BibitemOpen
  \bibfield  {author} {\bibinfo {author} {\bibnamefont {Turner}, \bibfnamefont
  {A.~M.}}, \bibinfo {author} {\bibfnamefont {F.}~\bibnamefont {Pollmann}}, \
  and\ \bibinfo {author} {\bibfnamefont {E.}~\bibnamefont {Berg}}} (\bibinfo
  {year} {2011}),\ \href {\doibase 10.1103/PhysRevB.83.075102} {\bibfield
  {journal} {\bibinfo  {journal} {Phys. Rev. B}\ }\textbf {\bibinfo {volume}
  {83}},\ \bibinfo {pages} {075102}}\BibitemShut {NoStop}%
\bibitem [{\citenamefont {{Turner}}\ and\ \citenamefont
  {{Vishwanath}}(2013)}]{Turner2013}%
  \BibitemOpen
  \bibfield  {author} {\bibinfo {author} {\bibnamefont {{Turner}},
  \bibfnamefont {A.~M.}}, \ and\ \bibinfo {author} {\bibfnamefont
  {A.}~\bibnamefont {{Vishwanath}}}} (\bibinfo {year} {2013}),\ \href@noop {}
  {\ }\Eprint {http://arxiv.org/abs/1301.0330} {arXiv:1301.0330} \BibitemShut
  {NoStop}%
\bibitem [{\citenamefont {Turner}\ \emph {et~al.}(2012)\citenamefont {Turner},
  \citenamefont {Zhang}, \citenamefont {Mong},\ and\ \citenamefont
  {Vishwanath}}]{Turner:2012bh}%
  \BibitemOpen
  \bibfield  {author} {\bibinfo {author} {\bibnamefont {Turner}, \bibfnamefont
  {A.~M.}}, \bibinfo {author} {\bibfnamefont {Y.}~\bibnamefont {Zhang}},
  \bibinfo {author} {\bibfnamefont {R.~S.~K.}\ \bibnamefont {Mong}}, \ and\
  \bibinfo {author} {\bibfnamefont {A.}~\bibnamefont {Vishwanath}}} (\bibinfo
  {year} {2012}),\ \href@noop {} {\bibfield  {journal} {\bibinfo  {journal}
  {Phys. Rev. B}\ }\textbf {\bibinfo {volume} {85}},\ \bibinfo {pages}
  {165120}}\BibitemShut {NoStop}%
\bibitem [{\citenamefont {Turner}\ \emph {et~al.}(2010)\citenamefont {Turner},
  \citenamefont {Zhang},\ and\ \citenamefont {Vishwanath}}]{Turner:2010qf}%
  \BibitemOpen
  \bibfield  {author} {\bibinfo {author} {\bibnamefont {Turner}, \bibfnamefont
  {A.~M.}}, \bibinfo {author} {\bibfnamefont {Y.}~\bibnamefont {Zhang}}, \ and\
  \bibinfo {author} {\bibfnamefont {A.}~\bibnamefont {Vishwanath}}} (\bibinfo
  {year} {2010}),\ \href@noop {} {\bibfield  {journal} {\bibinfo  {journal}
  {Phys. Rev. B}\ }\textbf {\bibinfo {volume} {82}}~(\bibinfo {number} {24}),\
  \bibinfo {pages} {241102}}\BibitemShut {NoStop}%
\bibitem [{\citenamefont {Ueno}\ \emph {et~al.}(2013)\citenamefont {Ueno},
  \citenamefont {Yamakage}, \citenamefont {Tanaka},\ and\ \citenamefont
  {Sato}}]{Sato_Mirror_SC}%
  \BibitemOpen
  \bibfield  {author} {\bibinfo {author} {\bibnamefont {Ueno}, \bibfnamefont
  {Y.}}, \bibinfo {author} {\bibfnamefont {A.}~\bibnamefont {Yamakage}},
  \bibinfo {author} {\bibfnamefont {Y.}~\bibnamefont {Tanaka}}, \ and\ \bibinfo
  {author} {\bibfnamefont {M.}~\bibnamefont {Sato}}} (\bibinfo {year} {2013}),\
  \href {\doibase 10.1103/PhysRevLett.111.087002} {\bibfield  {journal}
  {\bibinfo  {journal} {Phys. Rev. Lett.}\ }\textbf {\bibinfo {volume} {111}},\
  \bibinfo {pages} {087002}}\BibitemShut {NoStop}%
\bibitem [{\citenamefont {Vaezi}(2013)}]{Vaezi2013}%
  \BibitemOpen
  \bibfield  {author} {\bibinfo {author} {\bibnamefont {Vaezi}, \bibfnamefont
  {A.}}} (\bibinfo {year} {2013}),\ \href {\doibase 10.1103/PhysRevB.87.035132}
  {\bibfield  {journal} {\bibinfo  {journal} {Phys. Rev. B}\ }\textbf {\bibinfo
  {volume} {87}},\ \bibinfo {pages} {035132}}\BibitemShut {NoStop}%
\bibitem [{\citenamefont {Vafek}\ and\ \citenamefont
  {Vishwanath}(2014)}]{Ashvin_Weyl_review}%
  \BibitemOpen
  \bibfield  {author} {\bibinfo {author} {\bibnamefont {Vafek}, \bibfnamefont
  {O.}}, \ and\ \bibinfo {author} {\bibfnamefont {A.}~\bibnamefont
  {Vishwanath}}} (\bibinfo {year} {2014}),\ \href {\doibase
  10.1146/annurev-conmatphys-031113-133841} {\bibfield  {journal} {\bibinfo
  {journal} {Annual Review of Condensed Matter Physics}\ }\textbf {\bibinfo
  {volume} {5}}~(\bibinfo {number} {1}),\ \bibinfo {pages} {83}}\BibitemShut
  {NoStop}%
\bibitem [{\citenamefont {Vanderbilt}\ and\ \citenamefont
  {King-Smith}(1993)}]{kingSmithPRB93b}%
  \BibitemOpen
  \bibfield  {author} {\bibinfo {author} {\bibnamefont {Vanderbilt},
  \bibfnamefont {D.}}, \ and\ \bibinfo {author} {\bibfnamefont {R.~D.}\
  \bibnamefont {King-Smith}}} (\bibinfo {year} {1993}),\ \href {\doibase
  10.1103/PhysRevB.48.4442} {\bibfield  {journal} {\bibinfo  {journal} {Phys.
  Rev. B}\ }\textbf {\bibinfo {volume} {48}},\ \bibinfo {pages}
  {4442}}\BibitemShut {NoStop}%
\bibitem [{\citenamefont {Vazifeh}\ and\ \citenamefont
  {Franz}(2013)}]{Marcel_Weyl_response}%
  \BibitemOpen
  \bibfield  {author} {\bibinfo {author} {\bibnamefont {Vazifeh}, \bibfnamefont
  {M.~M.}}, \ and\ \bibinfo {author} {\bibfnamefont {M.}~\bibnamefont {Franz}}}
  (\bibinfo {year} {2013}),\ \href {\doibase 10.1103/PhysRevLett.111.027201}
  {\bibfield  {journal} {\bibinfo  {journal} {Phys. Rev. Lett.}\ }\textbf
  {\bibinfo {volume} {111}},\ \bibinfo {pages} {027201}}\BibitemShut {NoStop}%
\bibitem [{\citenamefont {{Verbaarschot}}(1994)}]{Verbaarschot1994}%
  \BibitemOpen
  \bibfield  {author} {\bibinfo {author} {\bibnamefont {{Verbaarschot}},
  \bibfnamefont {J.}}} (\bibinfo {year} {1994}),\ \href {\doibase
  10.1103/PhysRevLett.72.2531} {\bibfield  {journal} {\bibinfo  {journal}
  {Phys. Rev. Lett.}\ }\textbf {\bibinfo {volume} {72}},\ \bibinfo {pages}
  {2531}}\BibitemShut {NoStop}%
\bibitem [{\citenamefont {Verbin}\ \emph {et~al.}(2013)\citenamefont {Verbin},
  \citenamefont {Zilberberg}, \citenamefont {Kraus}, \citenamefont {Lahini},\
  and\ \citenamefont {Silberberg}}]{Verbin_Topo_quasicrystal}%
  \BibitemOpen
  \bibfield  {author} {\bibinfo {author} {\bibnamefont {Verbin}, \bibfnamefont
  {M.}}, \bibinfo {author} {\bibfnamefont {O.}~\bibnamefont {Zilberberg}},
  \bibinfo {author} {\bibfnamefont {Y.~E.}\ \bibnamefont {Kraus}}, \bibinfo
  {author} {\bibfnamefont {Y.}~\bibnamefont {Lahini}}, \ and\ \bibinfo {author}
  {\bibfnamefont {Y.}~\bibnamefont {Silberberg}}} (\bibinfo {year} {2013}),\
  \href {\doibase 10.1103/PhysRevLett.110.076403} {\bibfield  {journal}
  {\bibinfo  {journal} {Phys. Rev. Lett.}\ }\textbf {\bibinfo {volume} {110}},\
  \bibinfo {pages} {076403}}\BibitemShut {NoStop}%
\bibitem [{\citenamefont {{Vishwanath}}\ and\ \citenamefont
  {{Senthil}}(2013)}]{Vishwanath2013}%
  \BibitemOpen
  \bibfield  {author} {\bibinfo {author} {\bibnamefont {{Vishwanath}},
  \bibfnamefont {A.}}, \ and\ \bibinfo {author} {\bibfnamefont
  {T.}~\bibnamefont {{Senthil}}}} (\bibinfo {year} {2013}),\ \href {\doibase
  10.1103/PhysRevX.3.011016} {\bibfield  {journal} {\bibinfo  {journal} {Phys.
  Rev. X}\ }\textbf {\bibinfo {volume} {3}},\ \bibinfo {eid}
  {011016}}\BibitemShut {NoStop}%
\bibitem [{\citenamefont {{Volovik}}(1990)}]{Volovik1990}%
  \BibitemOpen
  \bibfield  {author} {\bibinfo {author} {\bibnamefont {{Volovik}},
  \bibfnamefont {G.~E.}}} (\bibinfo {year} {1990}),\ \href@noop {} {\bibfield
  {journal} {\bibinfo  {journal} {Sov. J. Exp. Theor. Phys. Lett.}\ }\textbf
  {\bibinfo {volume} {51}},\ \bibinfo {pages} {125}}\BibitemShut {NoStop}%
\bibitem [{\citenamefont {Volovik}(1992)}]{Volovik92}%
  \BibitemOpen
  \bibfield  {author} {\bibinfo {author} {\bibnamefont {Volovik}, \bibfnamefont
  {G.~E.}}} (\bibinfo {year} {1992}),\ \href@noop {} {\bibfield  {journal}
  {\bibinfo  {journal} {JETP Letters}\ }\textbf {\bibinfo {volume} {55}},\
  \bibinfo {pages} {368}}\BibitemShut {NoStop}%
\bibitem [{\citenamefont {Volovik}(1999)}]{Volovik99}%
  \BibitemOpen
  \bibfield  {author} {\bibinfo {author} {\bibnamefont {Volovik}, \bibfnamefont
  {G.~E.}}} (\bibinfo {year} {1999}),\ \href {\doibase 10.1134/1.568223}
  {\bibfield  {journal} {\bibinfo  {journal} {Pisma Zh. Eksp. Teor. Fiz.}\
  }\textbf {\bibinfo {volume} {70}},\ \bibinfo {pages} {601}}\BibitemShut
  {NoStop}%
\bibitem [{\citenamefont {Volovik}(2003)}]{Volovik:book}%
  \BibitemOpen
  \bibfield  {author} {\bibinfo {author} {\bibnamefont {Volovik}, \bibfnamefont
  {G.~E.}}} (\bibinfo {year} {2003}),\ \href@noop {} {\emph {\bibinfo {title}
  {Universe in a helium droplet}}}\ (\bibinfo  {publisher} {Oxford University
  Press, Oxford UK})\BibitemShut {NoStop}%
\bibitem [{\citenamefont {Volovik}(2011)}]{Volovik3HeA}%
  \BibitemOpen
  \bibfield  {author} {\bibinfo {author} {\bibnamefont {Volovik}, \bibfnamefont
  {G.~E.}}} (\bibinfo {year} {2011}),\ \href@noop {} {\bibfield  {journal}
  {\bibinfo  {journal} {JETP Lett.}\ }\textbf {\bibinfo {volume} {93}},\
  \bibinfo {pages} {66}}\BibitemShut {NoStop}%
\bibitem [{\citenamefont {Volovik}(2013)}]{volovikLectNotes13}%
  \BibitemOpen
  \bibfield  {author} {\bibinfo {author} {\bibnamefont {Volovik}, \bibfnamefont
  {G.~E.}}} (\bibinfo {year} {2013}),\ \href@noop {} {\emph {\bibinfo {title}
  {Topology of quantum vacuum}}},\ \bibinfo {series} {Lecture Notes in
  Physics}, Vol.\ \bibinfo {volume} {870}\ (\bibinfo  {publisher} {Springer
  Berlin})\BibitemShut {NoStop}%
\bibitem [{\citenamefont {Ho\ifmmode~\check{r}\else
  \v{r}\fi{}ava}(2005)}]{HoravaPRL05}%
  \BibitemOpen
  \bibfield  {author} {\bibinfo {author} {\bibnamefont
  {Ho\ifmmode~\check{r}\else \v{r}\fi{}ava}, \bibfnamefont {P.}}} (\bibinfo
  {year} {2005}),\ \href {\doibase 10.1103/PhysRevLett.95.016405} {\bibfield
  {journal} {\bibinfo  {journal} {Phys. Rev. Lett.}\ }\textbf {\bibinfo
  {volume} {95}},\ \bibinfo {pages} {016405}}\BibitemShut {NoStop}%
\bibitem [{\citenamefont {Wada}\ \emph {et~al.}(2008)\citenamefont {Wada},
  \citenamefont {Murakawa}, \citenamefont {Tamura}, \citenamefont {Saitoh},
  \citenamefont {Aoki}, \citenamefont {Nomura},\ and\ \citenamefont
  {Okuda}}]{Wada:2008ly}%
  \BibitemOpen
  \bibfield  {author} {\bibinfo {author} {\bibnamefont {Wada}, \bibfnamefont
  {Y.}}, \bibinfo {author} {\bibfnamefont {S.}~\bibnamefont {Murakawa}},
  \bibinfo {author} {\bibfnamefont {Y.}~\bibnamefont {Tamura}}, \bibinfo
  {author} {\bibfnamefont {M.}~\bibnamefont {Saitoh}}, \bibinfo {author}
  {\bibfnamefont {Y.}~\bibnamefont {Aoki}}, \bibinfo {author} {\bibfnamefont
  {R.}~\bibnamefont {Nomura}}, \ and\ \bibinfo {author} {\bibfnamefont
  {Y.}~\bibnamefont {Okuda}}} (\bibinfo {year} {2008}),\ \href {\doibase
  10.1103/PhysRevB.78.214516} {\bibfield  {journal} {\bibinfo  {journal} {Phys.
  Rev. B}\ }\textbf {\bibinfo {volume} {78}},\ \bibinfo {pages}
  {214516}}\BibitemShut {NoStop}%
\bibitem [{\citenamefont {Wan}\ \emph {et~al.}(2011)\citenamefont {Wan},
  \citenamefont {Turner}, \citenamefont {Vishwanath},\ and\ \citenamefont
  {Savrasov}}]{WanVishwanathSavrasovPRB11}%
  \BibitemOpen
  \bibfield  {author} {\bibinfo {author} {\bibnamefont {Wan}, \bibfnamefont
  {X.}}, \bibinfo {author} {\bibfnamefont {A.~M.}\ \bibnamefont {Turner}},
  \bibinfo {author} {\bibfnamefont {A.}~\bibnamefont {Vishwanath}}, \ and\
  \bibinfo {author} {\bibfnamefont {S.~Y.}\ \bibnamefont {Savrasov}}} (\bibinfo
  {year} {2011}),\ \href {\doibase 10.1103/PhysRevB.83.205101} {\bibfield
  {journal} {\bibinfo  {journal} {Phys. Rev. B}\ }\textbf {\bibinfo {volume}
  {83}},\ \bibinfo {pages} {205101}}\BibitemShut {NoStop}%
\bibitem [{\citenamefont {Wang}\ and\ \citenamefont
  {Levin}(2014)}]{WangLevin2014}%
  \BibitemOpen
  \bibfield  {author} {\bibinfo {author} {\bibnamefont {Wang}, \bibfnamefont
  {C.}}, \ and\ \bibinfo {author} {\bibfnamefont {M.}~\bibnamefont {Levin}}}
  (\bibinfo {year} {2014}),\ \href {\doibase 10.1103/PhysRevLett.113.080403}
  {\bibfield  {journal} {\bibinfo  {journal} {Phys. Rev. Lett.}\ }\textbf
  {\bibinfo {volume} {113}},\ \bibinfo {pages} {080403}}\BibitemShut {NoStop}%
\bibitem [{\citenamefont {Wang}\ \emph
  {et~al.}(2013{\natexlab{a}})\citenamefont {Wang}, \citenamefont {Potter},\
  and\ \citenamefont {Senthil}}]{WangPotterSenthilgapTI2013}%
  \BibitemOpen
  \bibfield  {author} {\bibinfo {author} {\bibnamefont {Wang}, \bibfnamefont
  {C.}}, \bibinfo {author} {\bibfnamefont {A.~C.}\ \bibnamefont {Potter}}, \
  and\ \bibinfo {author} {\bibfnamefont {T.}~\bibnamefont {Senthil}}} (\bibinfo
  {year} {2013}{\natexlab{a}}),\ \href {\doibase 10.1103/PhysRevB.88.115137}
  {\bibfield  {journal} {\bibinfo  {journal} {Phys. Rev. B}\ }\textbf {\bibinfo
  {volume} {88}},\ \bibinfo {pages} {115137}}\BibitemShut {NoStop}%
\bibitem [{\citenamefont {{Wang}}\ \emph {et~al.}(2014)\citenamefont {{Wang}},
  \citenamefont {{Potter}},\ and\ \citenamefont
  {{Senthil}}}]{WangPotterSenthil2014}%
  \BibitemOpen
  \bibfield  {author} {\bibinfo {author} {\bibnamefont {{Wang}}, \bibfnamefont
  {C.}}, \bibinfo {author} {\bibfnamefont {A.~C.}\ \bibnamefont {{Potter}}}, \
  and\ \bibinfo {author} {\bibfnamefont {T.}~\bibnamefont {{Senthil}}}}
  (\bibinfo {year} {2014}),\ \href {\doibase 10.1126/science.1243326}
  {\bibfield  {journal} {\bibinfo  {journal} {Science}\ }\textbf {\bibinfo
  {volume} {343}},\ \bibinfo {pages} {629}}\BibitemShut {NoStop}%
\bibitem [{\citenamefont {{Wang}}\ and\ \citenamefont
  {{Senthil}}(2013)}]{WangSenthil2013}%
  \BibitemOpen
  \bibfield  {author} {\bibinfo {author} {\bibnamefont {{Wang}}, \bibfnamefont
  {C.}}, \ and\ \bibinfo {author} {\bibfnamefont {T.}~\bibnamefont
  {{Senthil}}}} (\bibinfo {year} {2013}),\ \href {\doibase
  10.1103/PhysRevB.87.235122} {\bibfield  {journal} {\bibinfo  {journal}
  {\prb}\ }\textbf {\bibinfo {volume} {87}},\ \bibinfo {eid}
  {235122}}\BibitemShut {NoStop}%
\bibitem [{\citenamefont {{Wang}}\ and\ \citenamefont
  {{Senthil}}(2014)}]{Wang2014}%
  \BibitemOpen
  \bibfield  {author} {\bibinfo {author} {\bibnamefont {{Wang}}, \bibfnamefont
  {C.}}, \ and\ \bibinfo {author} {\bibfnamefont {T.}~\bibnamefont
  {{Senthil}}}} (\bibinfo {year} {2014}),\ \href {\doibase
  10.1103/PhysRevB.89.195124} {\bibfield  {journal} {\bibinfo  {journal}
  {\prb}\ }\textbf {\bibinfo {volume} {89}},\ \bibinfo {eid}
  {195124}}\BibitemShut {NoStop}%
\bibitem [{\citenamefont {Wang}\ \emph
  {et~al.}(2013{\natexlab{b}})\citenamefont {Wang}, \citenamefont {Ding},
  \citenamefont {Fedorov}, \citenamefont {Yao}, \citenamefont {Li},
  \citenamefont {Lv}, \citenamefont {Zhao}, \citenamefont {Zhang},
  \citenamefont {Xu}, \citenamefont {Schneeloch}, \citenamefont {Zhong},
  \citenamefont {Ji}, \citenamefont {Wang}, \citenamefont {He}, \citenamefont
  {Ma}, \citenamefont {Gu}, \citenamefont {Yao}, \citenamefont {Xue},
  \citenamefont {Chen},\ and\ \citenamefont
  {Zhou}}]{wang_d_wave_Bi2Se3_NatPhys13}%
  \BibitemOpen
  \bibfield  {author} {\bibinfo {author} {\bibnamefont {Wang}, \bibfnamefont
  {E.}}, \bibinfo {author} {\bibfnamefont {H.}~\bibnamefont {Ding}}, \bibinfo
  {author} {\bibfnamefont {A.~V.}\ \bibnamefont {Fedorov}}, \bibinfo {author}
  {\bibfnamefont {W.}~\bibnamefont {Yao}}, \bibinfo {author} {\bibfnamefont
  {Z.}~\bibnamefont {Li}}, \bibinfo {author} {\bibfnamefont {Y.-F.}\
  \bibnamefont {Lv}}, \bibinfo {author} {\bibfnamefont {K.}~\bibnamefont
  {Zhao}}, \bibinfo {author} {\bibfnamefont {L.-G.}\ \bibnamefont {Zhang}},
  \bibinfo {author} {\bibfnamefont {Z.}~\bibnamefont {Xu}}, \bibinfo {author}
  {\bibfnamefont {J.}~\bibnamefont {Schneeloch}}, \bibinfo {author}
  {\bibfnamefont {R.}~\bibnamefont {Zhong}}, \bibinfo {author} {\bibfnamefont
  {S.-H.}\ \bibnamefont {Ji}}, \bibinfo {author} {\bibfnamefont
  {L.}~\bibnamefont {Wang}}, \bibinfo {author} {\bibfnamefont {K.}~\bibnamefont
  {He}}, \bibinfo {author} {\bibfnamefont {X.}~\bibnamefont {Ma}}, \bibinfo
  {author} {\bibfnamefont {G.}~\bibnamefont {Gu}}, \bibinfo {author}
  {\bibfnamefont {H.}~\bibnamefont {Yao}}, \bibinfo {author} {\bibfnamefont
  {Q.-K.}\ \bibnamefont {Xue}}, \bibinfo {author} {\bibfnamefont
  {X.}~\bibnamefont {Chen}}, \ and\ \bibinfo {author} {\bibfnamefont
  {S.}~\bibnamefont {Zhou}}} (\bibinfo {year} {2013}{\natexlab{b}}),\ \href
  {http://dx.doi.org/10.1038/nphys2744} {\bibfield  {journal} {\bibinfo
  {journal} {Nat Phys}\ }\textbf {\bibinfo {volume} {9}}~(\bibinfo {number}
  {10}),\ \bibinfo {pages} {621}}\BibitemShut {NoStop}%
\bibitem [{\citenamefont {Wang}\ \emph
  {et~al.}(2014{\natexlab{a}})\citenamefont {Wang}, \citenamefont {Liu},
  \citenamefont {Xu}, \citenamefont {Wu}, \citenamefont {Gu},\ and\
  \citenamefont {Duan}}]{Duan_TCI_surface}%
  \BibitemOpen
  \bibfield  {author} {\bibinfo {author} {\bibnamefont {Wang}, \bibfnamefont
  {J.}}, \bibinfo {author} {\bibfnamefont {J.}~\bibnamefont {Liu}}, \bibinfo
  {author} {\bibfnamefont {Y.}~\bibnamefont {Xu}}, \bibinfo {author}
  {\bibfnamefont {J.}~\bibnamefont {Wu}}, \bibinfo {author} {\bibfnamefont
  {B.-L.}\ \bibnamefont {Gu}}, \ and\ \bibinfo {author} {\bibfnamefont
  {W.}~\bibnamefont {Duan}}} (\bibinfo {year} {2014}{\natexlab{a}}),\ \href
  {\doibase 10.1103/PhysRevB.89.125308} {\bibfield  {journal} {\bibinfo
  {journal} {Phys. Rev. B}\ }\textbf {\bibinfo {volume} {89}},\ \bibinfo
  {pages} {125308}}\BibitemShut {NoStop}%
\bibitem [{\citenamefont {{Wang}}\ and\ \citenamefont
  {{Wen}}(2013)}]{WangWen2013}%
  \BibitemOpen
  \bibfield  {author} {\bibinfo {author} {\bibnamefont {{Wang}}, \bibfnamefont
  {J.}}, \ and\ \bibinfo {author} {\bibfnamefont {X.-G.}\ \bibnamefont
  {{Wen}}}} (\bibinfo {year} {2013}),\ \href@noop {} {\ }\Eprint
  {http://arxiv.org/abs/1307.7480} {arXiv:1307.7480} \BibitemShut {NoStop}%
\bibitem [{\citenamefont {Wang}\ \emph
  {et~al.}(2015{\natexlab{a}})\citenamefont {Wang}, \citenamefont {Gu},\ and\
  \citenamefont {Wen}}]{Wang2014c}%
  \BibitemOpen
  \bibfield  {author} {\bibinfo {author} {\bibnamefont {Wang}, \bibfnamefont
  {J.~C.}}, \bibinfo {author} {\bibfnamefont {Z.-C.}\ \bibnamefont {Gu}}, \
  and\ \bibinfo {author} {\bibfnamefont {X.-G.}\ \bibnamefont {Wen}}} (\bibinfo
  {year} {2015}{\natexlab{a}}),\ \href {\doibase
  10.1103/PhysRevLett.114.031601} {\bibfield  {journal} {\bibinfo  {journal}
  {Phys. Rev. Lett.}\ }\textbf {\bibinfo {volume} {114}},\ \bibinfo {pages}
  {031601}}\BibitemShut {NoStop}%
\bibitem [{\citenamefont {Wang}\ \emph
  {et~al.}(2015{\natexlab{b}})\citenamefont {Wang}, \citenamefont {Santos},\
  and\ \citenamefont {Wen}}]{Wang2014b}%
  \BibitemOpen
  \bibfield  {author} {\bibinfo {author} {\bibnamefont {Wang}, \bibfnamefont
  {J.~C.}}, \bibinfo {author} {\bibfnamefont {L.~H.}\ \bibnamefont {Santos}}, \
  and\ \bibinfo {author} {\bibfnamefont {X.-G.}\ \bibnamefont {Wen}}} (\bibinfo
  {year} {2015}{\natexlab{b}}),\ \href {\doibase 10.1103/PhysRevB.91.195134}
  {\bibfield  {journal} {\bibinfo  {journal} {Phys. Rev. B}\ }\textbf {\bibinfo
  {volume} {91}},\ \bibinfo {pages} {195134}}\BibitemShut {NoStop}%
\bibitem [{\citenamefont {Wang}\ and\ \citenamefont {Wen}(2015)}]{WangWen2014}%
  \BibitemOpen
  \bibfield  {author} {\bibinfo {author} {\bibnamefont {Wang}, \bibfnamefont
  {J.~C.}}, \ and\ \bibinfo {author} {\bibfnamefont {X.-G.}\ \bibnamefont
  {Wen}}} (\bibinfo {year} {2015}),\ \href {\doibase
  10.1103/PhysRevB.91.035134} {\bibfield  {journal} {\bibinfo  {journal} {Phys.
  Rev. B}\ }\textbf {\bibinfo {volume} {91}},\ \bibinfo {pages}
  {035134}}\BibitemShut {NoStop}%
\bibitem [{\citenamefont {Wang}\ \emph
  {et~al.}(2015{\natexlab{c}})\citenamefont {Wang}, \citenamefont {Essin},
  \citenamefont {Hermele},\ and\ \citenamefont
  {Motrunich}}]{wang_essin_hermele_motrunich_PRB_15}%
  \BibitemOpen
  \bibfield  {author} {\bibinfo {author} {\bibnamefont {Wang}, \bibfnamefont
  {L.}}, \bibinfo {author} {\bibfnamefont {A.}~\bibnamefont {Essin}}, \bibinfo
  {author} {\bibfnamefont {M.}~\bibnamefont {Hermele}}, \ and\ \bibinfo
  {author} {\bibfnamefont {O.}~\bibnamefont {Motrunich}}} (\bibinfo {year}
  {2015}{\natexlab{c}}),\ \href {\doibase 10.1103/PhysRevB.91.121103}
  {\bibfield  {journal} {\bibinfo  {journal} {Phys. Rev. B}\ }\textbf {\bibinfo
  {volume} {91}},\ \bibinfo {pages} {121103}}\BibitemShut {NoStop}%
\bibitem [{\citenamefont {Wang}\ \emph {et~al.}(2012)\citenamefont {Wang},
  \citenamefont {Liu}, \citenamefont {Xu}, \citenamefont {Yang}, \citenamefont
  {Miao}, \citenamefont {Yao}, \citenamefont {Gao}, \citenamefont {Shen},
  \citenamefont {Ma}, \citenamefont {Chen}, \citenamefont {Xu}, \citenamefont
  {Liu}, \citenamefont {Zhang}, \citenamefont {Qian}, \citenamefont {Jia},\
  and\ \citenamefont {Xue}}]{wang_Xue_science_12}%
  \BibitemOpen
  \bibfield  {author} {\bibinfo {author} {\bibnamefont {Wang}, \bibfnamefont
  {M.-X.}}, \bibinfo {author} {\bibfnamefont {C.}~\bibnamefont {Liu}}, \bibinfo
  {author} {\bibfnamefont {J.-P.}\ \bibnamefont {Xu}}, \bibinfo {author}
  {\bibfnamefont {F.}~\bibnamefont {Yang}}, \bibinfo {author} {\bibfnamefont
  {L.}~\bibnamefont {Miao}}, \bibinfo {author} {\bibfnamefont {M.-Y.}\
  \bibnamefont {Yao}}, \bibinfo {author} {\bibfnamefont {C.~L.}\ \bibnamefont
  {Gao}}, \bibinfo {author} {\bibfnamefont {C.}~\bibnamefont {Shen}}, \bibinfo
  {author} {\bibfnamefont {X.}~\bibnamefont {Ma}}, \bibinfo {author}
  {\bibfnamefont {X.}~\bibnamefont {Chen}}, \bibinfo {author} {\bibfnamefont
  {Z.-A.}\ \bibnamefont {Xu}}, \bibinfo {author} {\bibfnamefont
  {Y.}~\bibnamefont {Liu}}, \bibinfo {author} {\bibfnamefont {S.-C.}\
  \bibnamefont {Zhang}}, \bibinfo {author} {\bibfnamefont {D.}~\bibnamefont
  {Qian}}, \bibinfo {author} {\bibfnamefont {J.-F.}\ \bibnamefont {Jia}}, \
  and\ \bibinfo {author} {\bibfnamefont {Q.-K.}\ \bibnamefont {Xue}}} (\bibinfo
  {year} {2012}),\ \href {\doibase 10.1126/science.1216466} {\bibfield
  {journal} {\bibinfo  {journal} {Science}\ }\textbf {\bibinfo {volume}
  {336}}~(\bibinfo {number} {6077}),\ \bibinfo {pages} {52}}\BibitemShut
  {NoStop}%
\bibitem [{\citenamefont {Wang}\ \emph
  {et~al.}(2014{\natexlab{b}})\citenamefont {Wang}, \citenamefont {Liu},
  \citenamefont {Zhang}, \citenamefont {Samarth}, \citenamefont {Zhang},\ and\
  \citenamefont {Liu}}]{WangLiu14}%
  \BibitemOpen
  \bibfield  {author} {\bibinfo {author} {\bibnamefont {Wang}, \bibfnamefont
  {Q.-Z.}}, \bibinfo {author} {\bibfnamefont {X.}~\bibnamefont {Liu}}, \bibinfo
  {author} {\bibfnamefont {H.-J.}\ \bibnamefont {Zhang}}, \bibinfo {author}
  {\bibfnamefont {N.}~\bibnamefont {Samarth}}, \bibinfo {author} {\bibfnamefont
  {S.-C.}\ \bibnamefont {Zhang}}, \ and\ \bibinfo {author} {\bibfnamefont
  {C.-X.}\ \bibnamefont {Liu}}} (\bibinfo {year} {2014}{\natexlab{b}}),\ \href
  {\doibase 10.1103/PhysRevLett.113.147201} {\bibfield  {journal} {\bibinfo
  {journal} {Phys. Rev. Lett.}\ }\textbf {\bibinfo {volume} {113}},\ \bibinfo
  {pages} {147201}}\BibitemShut {NoStop}%
\bibitem [{\citenamefont {Wang}\ \emph
  {et~al.}(2015{\natexlab{d}})\citenamefont {Wang}, \citenamefont {Deng},
  \citenamefont {Moore}, \citenamefont {Sun},\ and\ \citenamefont
  {Duan}}]{Deng_chiral}%
  \BibitemOpen
  \bibfield  {author} {\bibinfo {author} {\bibnamefont {Wang}, \bibfnamefont
  {S.-T.}}, \bibinfo {author} {\bibfnamefont {D.-L.}\ \bibnamefont {Deng}},
  \bibinfo {author} {\bibfnamefont {J.~E.}\ \bibnamefont {Moore}}, \bibinfo
  {author} {\bibfnamefont {K.}~\bibnamefont {Sun}}, \ and\ \bibinfo {author}
  {\bibfnamefont {L.-M.}\ \bibnamefont {Duan}}} (\bibinfo {year}
  {2015}{\natexlab{d}}),\ \href {\doibase 10.1103/PhysRevB.91.035108}
  {\bibfield  {journal} {\bibinfo  {journal} {Phys. Rev. B}\ }\textbf {\bibinfo
  {volume} {91}},\ \bibinfo {pages} {035108}}\BibitemShut {NoStop}%
\bibitem [{\citenamefont {Wang}\ \emph
  {et~al.}(2013{\natexlab{c}})\citenamefont {Wang}, \citenamefont {Steinberg},
  \citenamefont {Jarillo-Herrero},\ and\ \citenamefont {Gedik}}]{Wang453}%
  \BibitemOpen
  \bibfield  {author} {\bibinfo {author} {\bibnamefont {Wang}, \bibfnamefont
  {Y.~H.}}, \bibinfo {author} {\bibfnamefont {H.}~\bibnamefont {Steinberg}},
  \bibinfo {author} {\bibfnamefont {P.}~\bibnamefont {Jarillo-Herrero}}, \ and\
  \bibinfo {author} {\bibfnamefont {N.}~\bibnamefont {Gedik}}} (\bibinfo {year}
  {2013}{\natexlab{c}}),\ \href {\doibase 10.1126/science.1239834} {\bibfield
  {journal} {\bibinfo  {journal} {Science}\ }\textbf {\bibinfo {volume}
  {342}}~(\bibinfo {number} {6157}),\ \bibinfo {pages} {453}},\ \Eprint
  {http://arxiv.org/abs/http://science.sciencemag.org/content/342/6157/453.full.pdf}
  {http://science.sciencemag.org/content/342/6157/453.full.pdf} \BibitemShut
  {NoStop}%
\bibitem [{\citenamefont {{Wang}}\ \emph {et~al.}(2010)\citenamefont {{Wang}},
  \citenamefont {{Qi}},\ and\ \citenamefont {{Zhang}}}]{WangQiZhang2010}%
  \BibitemOpen
  \bibfield  {author} {\bibinfo {author} {\bibnamefont {{Wang}}, \bibfnamefont
  {Z.}}, \bibinfo {author} {\bibfnamefont {X.-L.}\ \bibnamefont {{Qi}}}, \ and\
  \bibinfo {author} {\bibfnamefont {S.-C.}\ \bibnamefont {{Zhang}}}} (\bibinfo
  {year} {2010}),\ \href {\doibase 10.1088/1367-2630/12/6/065007} {\bibfield
  {journal} {\bibinfo  {journal} {New Journal of Physics}\ }\textbf {\bibinfo
  {volume} {12}}~(\bibinfo {number} {6}),\ \bibinfo {eid} {065007}}\BibitemShut
  {NoStop}%
\bibitem [{\citenamefont {{Wang}}\ \emph {et~al.}(2011)\citenamefont {{Wang}},
  \citenamefont {{Qi}},\ and\ \citenamefont {{Zhang}}}]{WangQiZhang2011}%
  \BibitemOpen
  \bibfield  {author} {\bibinfo {author} {\bibnamefont {{Wang}}, \bibfnamefont
  {Z.}}, \bibinfo {author} {\bibfnamefont {X.-L.}\ \bibnamefont {{Qi}}}, \ and\
  \bibinfo {author} {\bibfnamefont {S.-C.}\ \bibnamefont {{Zhang}}}} (\bibinfo
  {year} {2011}),\ \href {\doibase 10.1103/PhysRevB.84.014527} {\bibfield
  {journal} {\bibinfo  {journal} {\prb}\ }\textbf {\bibinfo {volume} {84}},\
  \bibinfo {eid} {014527}}\BibitemShut {NoStop}%
\bibitem [{\citenamefont {{Wang}}\ \emph {et~al.}(2012)\citenamefont {{Wang}},
  \citenamefont {{Qi}},\ and\ \citenamefont {{Zhang}}}]{WangGreenFunction2012}%
  \BibitemOpen
  \bibfield  {author} {\bibinfo {author} {\bibnamefont {{Wang}}, \bibfnamefont
  {Z.}}, \bibinfo {author} {\bibfnamefont {X.-L.}\ \bibnamefont {{Qi}}}, \ and\
  \bibinfo {author} {\bibfnamefont {S.-C.}\ \bibnamefont {{Zhang}}}} (\bibinfo
  {year} {2012}),\ \href {\doibase 10.1103/PhysRevB.85.165126} {\bibfield
  {journal} {\bibinfo  {journal} {\prb}\ }\textbf {\bibinfo {volume}
  {85}}~(\bibinfo {number} {16}),\ \bibinfo {eid} {165126}}\BibitemShut
  {NoStop}%
\bibitem [{\citenamefont {Wang}\ \emph {et~al.}(2012)\citenamefont {Wang},
  \citenamefont {Sun}, \citenamefont {Chen}, \citenamefont {Franchini},
  \citenamefont {Xu}, \citenamefont {Weng}, \citenamefont {Dai},\ and\
  \citenamefont {Fang}}]{Dai_predition_Na3Bi}%
  \BibitemOpen
  \bibfield  {author} {\bibinfo {author} {\bibnamefont {Wang}, \bibfnamefont
  {Z.}}, \bibinfo {author} {\bibfnamefont {Y.}~\bibnamefont {Sun}}, \bibinfo
  {author} {\bibfnamefont {X.-Q.}\ \bibnamefont {Chen}}, \bibinfo {author}
  {\bibfnamefont {C.}~\bibnamefont {Franchini}}, \bibinfo {author}
  {\bibfnamefont {G.}~\bibnamefont {Xu}}, \bibinfo {author} {\bibfnamefont
  {H.}~\bibnamefont {Weng}}, \bibinfo {author} {\bibfnamefont {X.}~\bibnamefont
  {Dai}}, \ and\ \bibinfo {author} {\bibfnamefont {Z.}~\bibnamefont {Fang}}}
  (\bibinfo {year} {2012}),\ \href {\doibase 10.1103/PhysRevB.85.195320}
  {\bibfield  {journal} {\bibinfo  {journal} {Phys. Rev. B}\ }\textbf {\bibinfo
  {volume} {85}},\ \bibinfo {pages} {195320}}\BibitemShut {NoStop}%
\bibitem [{\citenamefont {Wang}\ \emph
  {et~al.}(2013{\natexlab{d}})\citenamefont {Wang}, \citenamefont {Weng},
  \citenamefont {Wu}, \citenamefont {Dai},\ and\ \citenamefont
  {Fang}}]{wangCd3As2PRB13}%
  \BibitemOpen
  \bibfield  {author} {\bibinfo {author} {\bibnamefont {Wang}, \bibfnamefont
  {Z.}}, \bibinfo {author} {\bibfnamefont {H.}~\bibnamefont {Weng}}, \bibinfo
  {author} {\bibfnamefont {Q.}~\bibnamefont {Wu}}, \bibinfo {author}
  {\bibfnamefont {X.}~\bibnamefont {Dai}}, \ and\ \bibinfo {author}
  {\bibfnamefont {Z.}~\bibnamefont {Fang}}} (\bibinfo {year}
  {2013}{\natexlab{d}}),\ \href {\doibase 10.1103/PhysRevB.88.125427}
  {\bibfield  {journal} {\bibinfo  {journal} {Phys. Rev. B}\ }\textbf {\bibinfo
  {volume} {88}},\ \bibinfo {pages} {125427}}\BibitemShut {NoStop}%
\bibitem [{\citenamefont {{Wang}}\ and\ \citenamefont
  {{Zhang}}(2012)}]{WangZhang2012}%
  \BibitemOpen
  \bibfield  {author} {\bibinfo {author} {\bibnamefont {{Wang}}, \bibfnamefont
  {Z.}}, \ and\ \bibinfo {author} {\bibfnamefont {S.-C.}\ \bibnamefont
  {{Zhang}}}} (\bibinfo {year} {2012}),\ \href {\doibase
  10.1103/PhysRevX.2.031008} {\bibfield  {journal} {\bibinfo  {journal} {Phys.
  Rev. X}\ }\textbf {\bibinfo {volume} {2}}~(\bibinfo {number} {3}),\ \bibinfo
  {eid} {031008}}\BibitemShut {NoStop}%
\bibitem [{\citenamefont {{Wang}}\ and\ \citenamefont
  {{Zhang}}(2014)}]{WangZhang2014}%
  \BibitemOpen
  \bibfield  {author} {\bibinfo {author} {\bibnamefont {{Wang}}, \bibfnamefont
  {Z.}}, \ and\ \bibinfo {author} {\bibfnamefont {S.-C.}\ \bibnamefont
  {{Zhang}}}} (\bibinfo {year} {2014}),\ \href {\doibase
  10.1103/PhysRevX.4.011006} {\bibfield  {journal} {\bibinfo  {journal} {Phys.
  Rev. X}\ }\textbf {\bibinfo {volume} {4}}~(\bibinfo {number} {1}),\ \bibinfo
  {eid} {011006}}\BibitemShut {NoStop}%
\bibitem [{\citenamefont {{Wegner}}(1979)}]{Wegner1979}%
  \BibitemOpen
  \bibfield  {author} {\bibinfo {author} {\bibnamefont {{Wegner}},
  \bibfnamefont {F.}}} (\bibinfo {year} {1979}),\ \href {\doibase
  10.1007/BF01319839} {\bibfield  {journal} {\bibinfo  {journal} {Zeitschrift
  fur Physik B Condensed Matter}\ }\textbf {\bibinfo {volume} {35}},\ \bibinfo
  {pages} {207}}\BibitemShut {NoStop}%
\bibitem [{\citenamefont {Weinberg}(1981)}]{Weinberg81}%
  \BibitemOpen
  \bibfield  {author} {\bibinfo {author} {\bibnamefont {Weinberg},
  \bibfnamefont {E.~J.}}} (\bibinfo {year} {1981}),\ \href {\doibase
  10.1103/PhysRevD.24.2669} {\bibfield  {journal} {\bibinfo  {journal} {Phys.
  Rev. D}\ }\textbf {\bibinfo {volume} {24}},\ \bibinfo {pages}
  {2669}}\BibitemShut {NoStop}%
\bibitem [{\citenamefont {Wen}(1990)}]{wenTopOrder1990}%
  \BibitemOpen
  \bibfield  {author} {\bibinfo {author} {\bibnamefont {Wen}, \bibfnamefont
  {X.~G.}}} (\bibinfo {year} {1990}),\ \href {\doibase
  10.1142/S0217979290000139} {\bibfield  {journal} {\bibinfo  {journal}
  {International Journal of Modern Physics B}\ }\textbf {\bibinfo {volume}
  {04}}~(\bibinfo {number} {02}),\ \bibinfo {pages} {239}}\BibitemShut
  {NoStop}%
\bibitem [{\citenamefont {{Wen}}(2002)}]{Wen2002}%
  \BibitemOpen
  \bibfield  {author} {\bibinfo {author} {\bibnamefont {{Wen}}, \bibfnamefont
  {X.-G.}}} (\bibinfo {year} {2002}),\ \href {\doibase
  10.1103/PhysRevB.65.165113} {\bibfield  {journal} {\bibinfo  {journal}
  {\prb}\ }\textbf {\bibinfo {volume} {65}}~(\bibinfo {number} {16}),\ \bibinfo
  {eid} {165113}}\BibitemShut {NoStop}%
\bibitem [{\citenamefont {Wen}(2012)}]{xiaogang_noninteract}%
  \BibitemOpen
  \bibfield  {author} {\bibinfo {author} {\bibnamefont {Wen}, \bibfnamefont
  {X.-G.}}} (\bibinfo {year} {2012}),\ \href {\doibase
  10.1103/PhysRevB.85.085103} {\bibfield  {journal} {\bibinfo  {journal} {Phys.
  Rev. B}\ }\textbf {\bibinfo {volume} {85}},\ \bibinfo {pages}
  {085103}}\BibitemShut {NoStop}%
\bibitem [{\citenamefont {Wen}(2013)}]{Xiaogang_anomalies}%
  \BibitemOpen
  \bibfield  {author} {\bibinfo {author} {\bibnamefont {Wen}, \bibfnamefont
  {X.-G.}}} (\bibinfo {year} {2013}),\ \href {\doibase
  10.1103/PhysRevD.88.045013} {\bibfield  {journal} {\bibinfo  {journal} {Phys.
  Rev. D}\ }\textbf {\bibinfo {volume} {88}},\ \bibinfo {pages}
  {045013}}\BibitemShut {NoStop}%
\bibitem [{\citenamefont {{Wen}}(2014)}]{Wen2014symmetry-protected}%
  \BibitemOpen
  \bibfield  {author} {\bibinfo {author} {\bibnamefont {{Wen}}, \bibfnamefont
  {X.-G.}}} (\bibinfo {year} {2014}),\ \href {\doibase
  10.1103/PhysRevB.89.035147} {\bibfield  {journal} {\bibinfo  {journal}
  {\prb}\ }\textbf {\bibinfo {volume} {89}}~(\bibinfo {number} {3}),\ \bibinfo
  {eid} {035147}}\BibitemShut {NoStop}%
\bibitem [{\citenamefont {Wen}(2015)}]{WenNLSMs2014}%
  \BibitemOpen
  \bibfield  {author} {\bibinfo {author} {\bibnamefont {Wen}, \bibfnamefont
  {X.-G.}}} (\bibinfo {year} {2015}),\ \href {\doibase
  10.1103/PhysRevB.91.205101} {\bibfield  {journal} {\bibinfo  {journal} {Phys.
  Rev. B}\ }\textbf {\bibinfo {volume} {91}},\ \bibinfo {pages}
  {205101}}\BibitemShut {NoStop}%
\bibitem [{\citenamefont {Wen}\ and\ \citenamefont {Niu}(1990)}]{WenNiu90}%
  \BibitemOpen
  \bibfield  {author} {\bibinfo {author} {\bibnamefont {Wen}, \bibfnamefont
  {X.-G.}}, \ and\ \bibinfo {author} {\bibfnamefont {Q.}~\bibnamefont {Niu}}}
  (\bibinfo {year} {1990}),\ \href {\doibase 10.1103/PhysRevB.41.9377}
  {\bibfield  {journal} {\bibinfo  {journal} {Phys. Rev. B}\ }\textbf {\bibinfo
  {volume} {41}},\ \bibinfo {pages} {9377}}\BibitemShut {NoStop}%
\bibitem [{\citenamefont {Weng}\ \emph
  {et~al.}(2015{\natexlab{a}})\citenamefont {Weng}, \citenamefont {Fang},
  \citenamefont {Fang}, \citenamefont {Bernevig},\ and\ \citenamefont
  {Dai}}]{TaAs_Weng}%
  \BibitemOpen
  \bibfield  {author} {\bibinfo {author} {\bibnamefont {Weng}, \bibfnamefont
  {H.}}, \bibinfo {author} {\bibfnamefont {C.}~\bibnamefont {Fang}}, \bibinfo
  {author} {\bibfnamefont {Z.}~\bibnamefont {Fang}}, \bibinfo {author}
  {\bibfnamefont {B.~A.}\ \bibnamefont {Bernevig}}, \ and\ \bibinfo {author}
  {\bibfnamefont {X.}~\bibnamefont {Dai}}} (\bibinfo {year}
  {2015}{\natexlab{a}}),\ \href {\doibase 10.1103/PhysRevX.5.011029} {\bibfield
   {journal} {\bibinfo  {journal} {Phys. Rev. X}\ }\textbf {\bibinfo {volume}
  {5}},\ \bibinfo {pages} {011029}}\BibitemShut {NoStop}%
\bibitem [{\citenamefont {Weng}\ \emph
  {et~al.}(2015{\natexlab{b}})\citenamefont {Weng}, \citenamefont {Liang},
  \citenamefont {Xu}, \citenamefont {Yu}, \citenamefont {Fang}, \citenamefont
  {Dai},\ and\ \citenamefont {Kawazoe}}]{MTC_nodal_line}%
  \BibitemOpen
  \bibfield  {author} {\bibinfo {author} {\bibnamefont {Weng}, \bibfnamefont
  {H.}}, \bibinfo {author} {\bibfnamefont {Y.}~\bibnamefont {Liang}}, \bibinfo
  {author} {\bibfnamefont {Q.}~\bibnamefont {Xu}}, \bibinfo {author}
  {\bibfnamefont {R.}~\bibnamefont {Yu}}, \bibinfo {author} {\bibfnamefont
  {Z.}~\bibnamefont {Fang}}, \bibinfo {author} {\bibfnamefont {X.}~\bibnamefont
  {Dai}}, \ and\ \bibinfo {author} {\bibfnamefont {Y.}~\bibnamefont {Kawazoe}}}
  (\bibinfo {year} {2015}{\natexlab{b}}),\ \href {\doibase
  10.1103/PhysRevB.92.045108} {\bibfield  {journal} {\bibinfo  {journal} {Phys.
  Rev. B}\ }\textbf {\bibinfo {volume} {92}},\ \bibinfo {pages}
  {045108}}\BibitemShut {NoStop}%
\bibitem [{\citenamefont {Williams}\ \emph {et~al.}(2012)\citenamefont
  {Williams}, \citenamefont {Bestwick}, \citenamefont {Gallagher},
  \citenamefont {Hong}, \citenamefont {Cui}, \citenamefont {Bleich},
  \citenamefont {Analytis}, \citenamefont {Fisher},\ and\ \citenamefont
  {Goldhaber-Gordon}}]{WilliamsGoldhaber12}%
  \BibitemOpen
  \bibfield  {author} {\bibinfo {author} {\bibnamefont {Williams},
  \bibfnamefont {J.~R.}}, \bibinfo {author} {\bibfnamefont {A.~J.}\
  \bibnamefont {Bestwick}}, \bibinfo {author} {\bibfnamefont {P.}~\bibnamefont
  {Gallagher}}, \bibinfo {author} {\bibfnamefont {S.~S.}\ \bibnamefont {Hong}},
  \bibinfo {author} {\bibfnamefont {Y.}~\bibnamefont {Cui}}, \bibinfo {author}
  {\bibfnamefont {A.~S.}\ \bibnamefont {Bleich}}, \bibinfo {author}
  {\bibfnamefont {J.~G.}\ \bibnamefont {Analytis}}, \bibinfo {author}
  {\bibfnamefont {I.~R.}\ \bibnamefont {Fisher}}, \ and\ \bibinfo {author}
  {\bibfnamefont {D.}~\bibnamefont {Goldhaber-Gordon}}} (\bibinfo {year}
  {2012}),\ \href {\doibase 10.1103/PhysRevLett.109.056803} {\bibfield
  {journal} {\bibinfo  {journal} {Phys. Rev. Lett.}\ }\textbf {\bibinfo
  {volume} {109}},\ \bibinfo {pages} {056803}}\BibitemShut {NoStop}%
\bibitem [{\citenamefont {Wilson}\ and\ \citenamefont
  {Kogut}(1974)}]{Wilson:1973jj}%
  \BibitemOpen
  \bibfield  {author} {\bibinfo {author} {\bibnamefont {Wilson}, \bibfnamefont
  {K.~G.}}, \ and\ \bibinfo {author} {\bibfnamefont {J.~B.}\ \bibnamefont
  {Kogut}}} (\bibinfo {year} {1974}),\ \href {\doibase
  10.1016/0370-1573(74)90023-4} {\bibfield  {journal} {\bibinfo  {journal}
  {Phys. Rept.}\ }\textbf {\bibinfo {volume} {12}},\ \bibinfo {pages}
  {75}}\BibitemShut {NoStop}%
\bibitem [{\citenamefont {Winterfeld}\ \emph {et~al.}(2013)\citenamefont
  {Winterfeld}, \citenamefont {Agapito}, \citenamefont {Li}, \citenamefont
  {Kioussis}, \citenamefont {Blaha},\ and\ \citenamefont
  {Chen}}]{winterfeld_PRB_13}%
  \BibitemOpen
  \bibfield  {author} {\bibinfo {author} {\bibnamefont {Winterfeld},
  \bibfnamefont {L.}}, \bibinfo {author} {\bibfnamefont {L.~A.}\ \bibnamefont
  {Agapito}}, \bibinfo {author} {\bibfnamefont {J.}~\bibnamefont {Li}},
  \bibinfo {author} {\bibfnamefont {N.}~\bibnamefont {Kioussis}}, \bibinfo
  {author} {\bibfnamefont {P.}~\bibnamefont {Blaha}}, \ and\ \bibinfo {author}
  {\bibfnamefont {Y.~P.}\ \bibnamefont {Chen}}} (\bibinfo {year} {2013}),\
  \href {\doibase 10.1103/PhysRevB.87.075143} {\bibfield  {journal} {\bibinfo
  {journal} {Phys. Rev. B}\ }\textbf {\bibinfo {volume} {87}},\ \bibinfo
  {pages} {075143}}\BibitemShut {NoStop}%
\bibitem [{\citenamefont {Witczak-Krempa}\ \emph {et~al.}(2014)\citenamefont
  {Witczak-Krempa}, \citenamefont {Chen}, \citenamefont {Kim},\ and\
  \citenamefont {Balents}}]{krempaBallentsAnnuRev2014}%
  \BibitemOpen
  \bibfield  {author} {\bibinfo {author} {\bibnamefont {Witczak-Krempa},
  \bibfnamefont {W.}}, \bibinfo {author} {\bibfnamefont {G.}~\bibnamefont
  {Chen}}, \bibinfo {author} {\bibfnamefont {Y.~B.}\ \bibnamefont {Kim}}, \
  and\ \bibinfo {author} {\bibfnamefont {L.}~\bibnamefont {Balents}}} (\bibinfo
  {year} {2014}),\ \href {\doibase 10.1146/annurev-conmatphys-020911-125138}
  {\bibfield  {journal} {\bibinfo  {journal} {Annual Review of Condensed Matter
  Physics}\ }\textbf {\bibinfo {volume} {5}}~(\bibinfo {number} {1}),\ \bibinfo
  {pages} {57}}\BibitemShut {NoStop}%
\bibitem [{\citenamefont {Witczak-Krempa}\ and\ \citenamefont
  {Kim}(2012)}]{witczak_kim_weyl_2012}%
  \BibitemOpen
  \bibfield  {author} {\bibinfo {author} {\bibnamefont {Witczak-Krempa},
  \bibfnamefont {W.}}, \ and\ \bibinfo {author} {\bibfnamefont {Y.~B.}\
  \bibnamefont {Kim}}} (\bibinfo {year} {2012}),\ \href@noop {} {\bibfield
  {journal} {\bibinfo  {journal} {Phys. Rev. B}\ }\textbf {\bibinfo {volume}
  {85}},\ \bibinfo {pages} {045124}}\BibitemShut {NoStop}%
\bibitem [{\citenamefont {Witten}(1982)}]{Witten1982}%
  \BibitemOpen
  \bibfield  {author} {\bibinfo {author} {\bibnamefont {Witten}, \bibfnamefont
  {E.}}} (\bibinfo {year} {1982}),\ \href {\doibase
  10.1016/0370-2693(82)90728-6} {\bibfield  {journal} {\bibinfo  {journal}
  {Phys.Lett.}\ }\textbf {\bibinfo {volume} {B117}},\ \bibinfo {pages}
  {324}}\BibitemShut {NoStop}%
\bibitem [{\citenamefont {Witten}(1985)}]{Witten:1985xe}%
  \BibitemOpen
  \bibfield  {author} {\bibinfo {author} {\bibnamefont {Witten}, \bibfnamefont
  {E.}}} (\bibinfo {year} {1985}),\ \href {\doibase 10.1007/BF01212448}
  {\bibfield  {journal} {\bibinfo  {journal} {Commun. Math. Phys.}\ }\textbf
  {\bibinfo {volume} {100}},\ \bibinfo {pages} {197}}\BibitemShut {NoStop}%
\bibitem [{\citenamefont {Wojek}\ \emph {et~al.}(2013)\citenamefont {Wojek},
  \citenamefont {Buczko}, \citenamefont {Safaei}, \citenamefont {Dziawa},
  \citenamefont {Kowalski}, \citenamefont {Berntsen}, \citenamefont
  {Balasubramanian}, \citenamefont {Leandersson}, \citenamefont {Szczerbakow},
  \citenamefont {Kacman}, \citenamefont {Story},\ and\ \citenamefont
  {Tjernberg}}]{Wojek_TCI_surface}%
  \BibitemOpen
  \bibfield  {author} {\bibinfo {author} {\bibnamefont {Wojek}, \bibfnamefont
  {B.~M.}}, \bibinfo {author} {\bibfnamefont {R.}~\bibnamefont {Buczko}},
  \bibinfo {author} {\bibfnamefont {S.}~\bibnamefont {Safaei}}, \bibinfo
  {author} {\bibfnamefont {P.}~\bibnamefont {Dziawa}}, \bibinfo {author}
  {\bibfnamefont {B.~J.}\ \bibnamefont {Kowalski}}, \bibinfo {author}
  {\bibfnamefont {M.~H.}\ \bibnamefont {Berntsen}}, \bibinfo {author}
  {\bibfnamefont {T.}~\bibnamefont {Balasubramanian}}, \bibinfo {author}
  {\bibfnamefont {M.}~\bibnamefont {Leandersson}}, \bibinfo {author}
  {\bibfnamefont {A.}~\bibnamefont {Szczerbakow}}, \bibinfo {author}
  {\bibfnamefont {P.}~\bibnamefont {Kacman}}, \bibinfo {author} {\bibfnamefont
  {T.}~\bibnamefont {Story}}, \ and\ \bibinfo {author} {\bibfnamefont
  {O.}~\bibnamefont {Tjernberg}}} (\bibinfo {year} {2013}),\ \href {\doibase
  10.1103/PhysRevB.87.115106} {\bibfield  {journal} {\bibinfo  {journal} {Phys.
  Rev. B}\ }\textbf {\bibinfo {volume} {87}},\ \bibinfo {pages}
  {115106}}\BibitemShut {NoStop}%
\bibitem [{\citenamefont {Wolgast}\ \emph {et~al.}(2013)\citenamefont
  {Wolgast}, \citenamefont {Kurdak}, \citenamefont {Sun}, \citenamefont
  {Allen}, \citenamefont {Kim},\ and\ \citenamefont
  {Fisk}}]{First_realization_TKI}%
  \BibitemOpen
  \bibfield  {author} {\bibinfo {author} {\bibnamefont {Wolgast}, \bibfnamefont
  {S.}}, \bibinfo {author} {\bibfnamefont {C.}~\bibnamefont {Kurdak}}, \bibinfo
  {author} {\bibfnamefont {K.}~\bibnamefont {Sun}}, \bibinfo {author}
  {\bibfnamefont {J.~W.}\ \bibnamefont {Allen}}, \bibinfo {author}
  {\bibfnamefont {D.-J.}\ \bibnamefont {Kim}}, \ and\ \bibinfo {author}
  {\bibfnamefont {Z.}~\bibnamefont {Fisk}}} (\bibinfo {year} {2013}),\ \href
  {\doibase 10.1103/PhysRevB.88.180405} {\bibfield  {journal} {\bibinfo
  {journal} {Phys. Rev. B}\ }\textbf {\bibinfo {volume} {88}},\ \bibinfo
  {pages} {180405}}\BibitemShut {NoStop}%
\bibitem [{\citenamefont {Wrasse}\ and\ \citenamefont
  {Schmidt}(2014)}]{2D_TCI}%
  \BibitemOpen
  \bibfield  {author} {\bibinfo {author} {\bibnamefont {Wrasse}, \bibfnamefont
  {E.~O.}}, \ and\ \bibinfo {author} {\bibfnamefont {T.~M.}\ \bibnamefont
  {Schmidt}}} (\bibinfo {year} {2014}),\ \href {\doibase 10.1021/nl502481f}
  {\bibfield  {journal} {\bibinfo  {journal} {Nano Letters}\ }\textbf {\bibinfo
  {volume} {14}}~(\bibinfo {number} {10}),\ \bibinfo {pages}
  {5717}}\BibitemShut {NoStop}%
\bibitem [{\citenamefont {Wray}\ \emph {et~al.}(2013)\citenamefont {Wray},
  \citenamefont {Xu}, \citenamefont {Neupane}, \citenamefont {Fedorov},
  \citenamefont {Hor}, \citenamefont {Cava},\ and\ \citenamefont
  {Hasan}}]{wray_hasan_JPCS_13}%
  \BibitemOpen
  \bibfield  {author} {\bibinfo {author} {\bibnamefont {Wray}, \bibfnamefont
  {L.~A.}}, \bibinfo {author} {\bibfnamefont {S.}~\bibnamefont {Xu}}, \bibinfo
  {author} {\bibfnamefont {M.}~\bibnamefont {Neupane}}, \bibinfo {author}
  {\bibfnamefont {A.~V.}\ \bibnamefont {Fedorov}}, \bibinfo {author}
  {\bibfnamefont {Y.~S.}\ \bibnamefont {Hor}}, \bibinfo {author} {\bibfnamefont
  {R.~J.}\ \bibnamefont {Cava}}, \ and\ \bibinfo {author} {\bibfnamefont
  {M.~Z.}\ \bibnamefont {Hasan}}} (\bibinfo {year} {2013}),\ \href
  {http://stacks.iop.org/1742-6596/449/i=1/a=012037} {\bibfield  {journal}
  {\bibinfo  {journal} {Journal of Physics: Conference Series}\ }\textbf
  {\bibinfo {volume} {449}}~(\bibinfo {number} {1}),\ \bibinfo {pages}
  {012037}}\BibitemShut {NoStop}%
\bibitem [{\citenamefont {Wray}\ \emph {et~al.}(2010)\citenamefont {Wray},
  \citenamefont {Xu}, \citenamefont {Xia}, \citenamefont {Hor}, \citenamefont
  {Qian}, \citenamefont {Fedorov}, \citenamefont {Lin}, \citenamefont {Bansil},
  \citenamefont {Cava},\ and\ \citenamefont {Hasan}}]{WrayCavaHasan}%
  \BibitemOpen
  \bibfield  {author} {\bibinfo {author} {\bibnamefont {Wray}, \bibfnamefont
  {L.~A.}}, \bibinfo {author} {\bibfnamefont {S.-Y.}\ \bibnamefont {Xu}},
  \bibinfo {author} {\bibfnamefont {Y.}~\bibnamefont {Xia}}, \bibinfo {author}
  {\bibfnamefont {Y.~S.}\ \bibnamefont {Hor}}, \bibinfo {author} {\bibfnamefont
  {D.}~\bibnamefont {Qian}}, \bibinfo {author} {\bibfnamefont {A.~V.}\
  \bibnamefont {Fedorov}}, \bibinfo {author} {\bibfnamefont {H.}~\bibnamefont
  {Lin}}, \bibinfo {author} {\bibfnamefont {A.}~\bibnamefont {Bansil}},
  \bibinfo {author} {\bibfnamefont {R.~J.}\ \bibnamefont {Cava}}, \ and\
  \bibinfo {author} {\bibfnamefont {M.~Z.}\ \bibnamefont {Hasan}}} (\bibinfo
  {year} {2010}),\ \href@noop {} {\bibfield  {journal} {\bibinfo  {journal}
  {Nat. Phys.}\ }\textbf {\bibinfo {volume} {6}},\ \bibinfo {pages}
  {855}}\BibitemShut {NoStop}%
\bibitem [{\citenamefont {Xia}\ \emph {et~al.}(2006)\citenamefont {Xia},
  \citenamefont {Maeno}, \citenamefont {Beyersdorf}, \citenamefont {Fejer},\
  and\ \citenamefont {Kapitulnik}}]{XiaKapitulnik06}%
  \BibitemOpen
  \bibfield  {author} {\bibinfo {author} {\bibnamefont {Xia}, \bibfnamefont
  {J.}}, \bibinfo {author} {\bibfnamefont {Y.}~\bibnamefont {Maeno}}, \bibinfo
  {author} {\bibfnamefont {P.~T.}\ \bibnamefont {Beyersdorf}}, \bibinfo
  {author} {\bibfnamefont {M.~M.}\ \bibnamefont {Fejer}}, \ and\ \bibinfo
  {author} {\bibfnamefont {A.}~\bibnamefont {Kapitulnik}}} (\bibinfo {year}
  {2006}),\ \href {\doibase 10.1103/PhysRevLett.97.167002} {\bibfield
  {journal} {\bibinfo  {journal} {Phys. Rev. Lett.}\ }\textbf {\bibinfo
  {volume} {97}},\ \bibinfo {pages} {167002}}\BibitemShut {NoStop}%
\bibitem [{\citenamefont {Xia}\ \emph {et~al.}(2009)\citenamefont {Xia},
  \citenamefont {Qian}, \citenamefont {Hsieh}, \citenamefont {Wray},
  \citenamefont {Pal}, \citenamefont {Lin}, \citenamefont {Bansil},
  \citenamefont {Grauer}, \citenamefont {Hor}, \citenamefont {Cava},\ and\
  \citenamefont {Hasan}}]{Xia:2009uq}%
  \BibitemOpen
  \bibfield  {author} {\bibinfo {author} {\bibnamefont {Xia}, \bibfnamefont
  {Y.}}, \bibinfo {author} {\bibfnamefont {D.}~\bibnamefont {Qian}}, \bibinfo
  {author} {\bibfnamefont {D.}~\bibnamefont {Hsieh}}, \bibinfo {author}
  {\bibfnamefont {L.}~\bibnamefont {Wray}}, \bibinfo {author} {\bibfnamefont
  {A.}~\bibnamefont {Pal}}, \bibinfo {author} {\bibfnamefont {H.}~\bibnamefont
  {Lin}}, \bibinfo {author} {\bibfnamefont {A.}~\bibnamefont {Bansil}},
  \bibinfo {author} {\bibfnamefont {D.}~\bibnamefont {Grauer}}, \bibinfo
  {author} {\bibfnamefont {Y.~S.}\ \bibnamefont {Hor}}, \bibinfo {author}
  {\bibfnamefont {R.~J.}\ \bibnamefont {Cava}}, \ and\ \bibinfo {author}
  {\bibfnamefont {M.~Z.}\ \bibnamefont {Hasan}}} (\bibinfo {year} {2009}),\
  \href@noop {} {\bibfield  {journal} {\bibinfo  {journal} {Nat. Phys.}\
  }\textbf {\bibinfo {volume} {5}},\ \bibinfo {pages} {398}}\BibitemShut
  {NoStop}%
\bibitem [{\citenamefont {Xiao}\ \emph {et~al.}(2009)\citenamefont {Xiao},
  \citenamefont {Shi}, \citenamefont {Clougherty},\ and\ \citenamefont
  {Niu}}]{xiaoPRL09}%
  \BibitemOpen
  \bibfield  {author} {\bibinfo {author} {\bibnamefont {Xiao}, \bibfnamefont
  {D.}}, \bibinfo {author} {\bibfnamefont {J.}~\bibnamefont {Shi}}, \bibinfo
  {author} {\bibfnamefont {D.~P.}\ \bibnamefont {Clougherty}}, \ and\ \bibinfo
  {author} {\bibfnamefont {Q.}~\bibnamefont {Niu}}} (\bibinfo {year} {2009}),\
  \href {\doibase 10.1103/PhysRevLett.102.087602} {\bibfield  {journal}
  {\bibinfo  {journal} {Phys. Rev. Lett.}\ }\textbf {\bibinfo {volume} {102}},\
  \bibinfo {pages} {087602}}\BibitemShut {NoStop}%
\bibitem [{\citenamefont {Xiao}\ \emph {et~al.}(2011)\citenamefont {Xiao},
  \citenamefont {Zhu}, \citenamefont {Ran}, \citenamefont {Nagaosa},\ and\
  \citenamefont {Okamoto}}]{okamotoNatComm2011}%
  \BibitemOpen
  \bibfield  {author} {\bibinfo {author} {\bibnamefont {Xiao}, \bibfnamefont
  {D.}}, \bibinfo {author} {\bibfnamefont {W.}~\bibnamefont {Zhu}}, \bibinfo
  {author} {\bibfnamefont {Y.}~\bibnamefont {Ran}}, \bibinfo {author}
  {\bibfnamefont {N.}~\bibnamefont {Nagaosa}}, \ and\ \bibinfo {author}
  {\bibfnamefont {S.}~\bibnamefont {Okamoto}}} (\bibinfo {year} {2011}),\
  \href@noop {} {\bibfield  {journal} {\bibinfo  {journal} {Nat Commun}\
  }\textbf {\bibinfo {volume} {2}},\ \bibinfo {pages} {596}}\BibitemShut
  {NoStop}%
\bibitem [{\citenamefont {{Xie}}\ \emph {et~al.}(2015)\citenamefont {{Xie}},
  \citenamefont {{Chou}},\ and\ \citenamefont
  {{Foster}}}]{Foster2015PhRvB..91b4203X}%
  \BibitemOpen
  \bibfield  {author} {\bibinfo {author} {\bibnamefont {{Xie}}, \bibfnamefont
  {H.-Y.}}, \bibinfo {author} {\bibfnamefont {Y.-Z.}\ \bibnamefont {{Chou}}}, \
  and\ \bibinfo {author} {\bibfnamefont {M.~S.}\ \bibnamefont {{Foster}}}}
  (\bibinfo {year} {2015}),\ \href {\doibase 10.1103/PhysRevB.91.024203}
  {\bibfield  {journal} {\bibinfo  {journal} {\prb}\ }\textbf {\bibinfo
  {volume} {91}}~(\bibinfo {number} {2}),\ \bibinfo {eid} {024203}}\BibitemShut
  {NoStop}%
\bibitem [{\citenamefont {Xie}\ \emph {et~al.}(2015)\citenamefont {Xie},
  \citenamefont {Schoop}, \citenamefont {Seibel}, \citenamefont {Gibson},
  \citenamefont {Xie},\ and\ \citenamefont {Cava}}]{Cava_CaP_ring}%
  \BibitemOpen
  \bibfield  {author} {\bibinfo {author} {\bibnamefont {Xie}, \bibfnamefont
  {L.~S.}}, \bibinfo {author} {\bibfnamefont {L.~M.}\ \bibnamefont {Schoop}},
  \bibinfo {author} {\bibfnamefont {E.~M.}\ \bibnamefont {Seibel}}, \bibinfo
  {author} {\bibfnamefont {Q.~D.}\ \bibnamefont {Gibson}}, \bibinfo {author}
  {\bibfnamefont {W.}~\bibnamefont {Xie}}, \ and\ \bibinfo {author}
  {\bibfnamefont {R.~J.}\ \bibnamefont {Cava}}} (\bibinfo {year} {2015}),\
  \href {\doibase http://dx.doi.org/10.1063/1.4926545} {\bibfield  {journal}
  {\bibinfo  {journal} {APL Materials}\ }\textbf {\bibinfo {volume}
  {3}}~(\bibinfo {number} {8}),\ \bibinfo {eid} {083602}}\BibitemShut {NoStop}%
\bibitem [{\citenamefont {Xu}\ and\ \citenamefont
  {Ludwig}(2013)}]{XuLudwig2011}%
  \BibitemOpen
  \bibfield  {author} {\bibinfo {author} {\bibnamefont {Xu}, \bibfnamefont
  {C.}}, \ and\ \bibinfo {author} {\bibfnamefont {A.~W.~W.}\ \bibnamefont
  {Ludwig}}} (\bibinfo {year} {2013}),\ \href {\doibase
  10.1103/PhysRevLett.110.200405} {\bibfield  {journal} {\bibinfo  {journal}
  {Phys. Rev. Lett.}\ }\textbf {\bibinfo {volume} {110}},\ \bibinfo {pages}
  {200405}}\BibitemShut {NoStop}%
\bibitem [{\citenamefont {Xu}\ \emph {et~al.}(2015{\natexlab{a}})\citenamefont
  {Xu}, \citenamefont {Wang}, \citenamefont {Felser}, \citenamefont {Qi},\ and\
  \citenamefont {Zhang}}]{XuZhang15}%
  \BibitemOpen
  \bibfield  {author} {\bibinfo {author} {\bibnamefont {Xu}, \bibfnamefont
  {G.}}, \bibinfo {author} {\bibfnamefont {J.}~\bibnamefont {Wang}}, \bibinfo
  {author} {\bibfnamefont {C.}~\bibnamefont {Felser}}, \bibinfo {author}
  {\bibfnamefont {X.-L.}\ \bibnamefont {Qi}}, \ and\ \bibinfo {author}
  {\bibfnamefont {S.-C.}\ \bibnamefont {Zhang}}} (\bibinfo {year}
  {2015}{\natexlab{a}}),\ \href {\doibase 10.1021/nl504871u} {\bibfield
  {journal} {\bibinfo  {journal} {Nano Letters}\ }\textbf {\bibinfo {volume}
  {15}}~(\bibinfo {number} {3}),\ \bibinfo {pages} {2019}}\BibitemShut
  {NoStop}%
\bibitem [{\citenamefont {Xu}\ \emph {et~al.}(2011)\citenamefont {Xu},
  \citenamefont {Weng}, \citenamefont {Wang}, \citenamefont {Dai},\ and\
  \citenamefont {Fang}}]{HgCrSe_Weyl_2011}%
  \BibitemOpen
  \bibfield  {author} {\bibinfo {author} {\bibnamefont {Xu}, \bibfnamefont
  {G.}}, \bibinfo {author} {\bibfnamefont {H.}~\bibnamefont {Weng}}, \bibinfo
  {author} {\bibfnamefont {Z.}~\bibnamefont {Wang}}, \bibinfo {author}
  {\bibfnamefont {X.}~\bibnamefont {Dai}}, \ and\ \bibinfo {author}
  {\bibfnamefont {Z.}~\bibnamefont {Fang}}} (\bibinfo {year} {2011}),\ \href
  {\doibase 10.1103/PhysRevLett.107.186806} {\bibfield  {journal} {\bibinfo
  {journal} {Phys. Rev. Lett.}\ }\textbf {\bibinfo {volume} {107}},\ \bibinfo
  {pages} {186806}}\BibitemShut {NoStop}%
\bibitem [{\citenamefont {Xu}\ \emph {et~al.}(2014{\natexlab{a}})\citenamefont
  {Xu}, \citenamefont {Liu}, \citenamefont {Wang}, \citenamefont {Ge},
  \citenamefont {Liu}, \citenamefont {Yang}, \citenamefont {Chen},
  \citenamefont {Liu}, \citenamefont {Xu}, \citenamefont {Gao}, \citenamefont
  {Qian}, \citenamefont {Zhang},\ and\ \citenamefont {Jia}}]{SC_Proximity_Jia}%
  \BibitemOpen
  \bibfield  {author} {\bibinfo {author} {\bibnamefont {Xu}, \bibfnamefont
  {J.-P.}}, \bibinfo {author} {\bibfnamefont {C.}~\bibnamefont {Liu}}, \bibinfo
  {author} {\bibfnamefont {M.-X.}\ \bibnamefont {Wang}}, \bibinfo {author}
  {\bibfnamefont {J.}~\bibnamefont {Ge}}, \bibinfo {author} {\bibfnamefont
  {Z.-L.}\ \bibnamefont {Liu}}, \bibinfo {author} {\bibfnamefont
  {X.}~\bibnamefont {Yang}}, \bibinfo {author} {\bibfnamefont {Y.}~\bibnamefont
  {Chen}}, \bibinfo {author} {\bibfnamefont {Y.}~\bibnamefont {Liu}}, \bibinfo
  {author} {\bibfnamefont {Z.-A.}\ \bibnamefont {Xu}}, \bibinfo {author}
  {\bibfnamefont {C.-L.}\ \bibnamefont {Gao}}, \bibinfo {author} {\bibfnamefont
  {D.}~\bibnamefont {Qian}}, \bibinfo {author} {\bibfnamefont {F.-C.}\
  \bibnamefont {Zhang}}, \ and\ \bibinfo {author} {\bibfnamefont {J.-F.}\
  \bibnamefont {Jia}}} (\bibinfo {year} {2014}{\natexlab{a}}),\ \href {\doibase
  10.1103/PhysRevLett.112.217001} {\bibfield  {journal} {\bibinfo  {journal}
  {Phys. Rev. Lett.}\ }\textbf {\bibinfo {volume} {112}},\ \bibinfo {pages}
  {217001}}\BibitemShut {NoStop}%
\bibitem [{\citenamefont {Xu}\ \emph {et~al.}(2014{\natexlab{b}})\citenamefont
  {Xu}, \citenamefont {Alidoust}, \citenamefont {Belopolski}, \citenamefont
  {Richardella}, \citenamefont {Liu}, \citenamefont {Neupane}, \citenamefont
  {Bian}, \citenamefont {Huang}, \citenamefont {Sankar}, \citenamefont {Fang},
  \citenamefont {Dellabetta}, \citenamefont {Dai}, \citenamefont {Li},
  \citenamefont {Gilbert}, \citenamefont {Chou}, \citenamefont {Samarth},\ and\
  \citenamefont {Hasan}}]{Xu_TI_SC}%
  \BibitemOpen
  \bibfield  {author} {\bibinfo {author} {\bibnamefont {Xu}, \bibfnamefont
  {S.-Y.}}, \bibinfo {author} {\bibfnamefont {N.}~\bibnamefont {Alidoust}},
  \bibinfo {author} {\bibfnamefont {I.}~\bibnamefont {Belopolski}}, \bibinfo
  {author} {\bibfnamefont {A.}~\bibnamefont {Richardella}}, \bibinfo {author}
  {\bibfnamefont {C.}~\bibnamefont {Liu}}, \bibinfo {author} {\bibfnamefont
  {M.}~\bibnamefont {Neupane}}, \bibinfo {author} {\bibfnamefont
  {G.}~\bibnamefont {Bian}}, \bibinfo {author} {\bibfnamefont {S.-H.}\
  \bibnamefont {Huang}}, \bibinfo {author} {\bibfnamefont {R.}~\bibnamefont
  {Sankar}}, \bibinfo {author} {\bibfnamefont {C.}~\bibnamefont {Fang}},
  \bibinfo {author} {\bibfnamefont {B.}~\bibnamefont {Dellabetta}}, \bibinfo
  {author} {\bibfnamefont {W.}~\bibnamefont {Dai}}, \bibinfo {author}
  {\bibfnamefont {Q.}~\bibnamefont {Li}}, \bibinfo {author} {\bibfnamefont
  {M.~J.}\ \bibnamefont {Gilbert}}, \bibinfo {author} {\bibfnamefont
  {F.}~\bibnamefont {Chou}}, \bibinfo {author} {\bibfnamefont {N.}~\bibnamefont
  {Samarth}}, \ and\ \bibinfo {author} {\bibfnamefont {M.~Z.}\ \bibnamefont
  {Hasan}}} (\bibinfo {year} {2014}{\natexlab{b}}),\ \href
  {http://dx.doi.org/10.1038/nphys3139} {\bibfield  {journal} {\bibinfo
  {journal} {Nat Phys}\ }\textbf {\bibinfo {volume} {10}}~(\bibinfo {number}
  {12}),\ \bibinfo {pages} {943}}\BibitemShut {NoStop}%
\bibitem [{\citenamefont {Xu}\ \emph {et~al.}(2015{\natexlab{b}})\citenamefont
  {Xu}, \citenamefont {Alidoust}, \citenamefont {Belopolski}, \citenamefont
  {Yuan}, \citenamefont {Bian}, \citenamefont {Chang}, \citenamefont {Zheng},
  \citenamefont {Strocov}, \citenamefont {Sanchez}, \citenamefont {Chang},
  \citenamefont {Zhang}, \citenamefont {Mou}, \citenamefont {Wu}, \citenamefont
  {Huang}, \citenamefont {Lee}, \citenamefont {Huang}, \citenamefont {Wang},
  \citenamefont {Bansil}, \citenamefont {Jeng}, \citenamefont {Neupert},
  \citenamefont {Kaminski}, \citenamefont {Lin}, \citenamefont {Jia},\ and\
  \citenamefont {Zahid~Hasan}}]{Weyl_NbAs_Xu}%
  \BibitemOpen
  \bibfield  {author} {\bibinfo {author} {\bibnamefont {Xu}, \bibfnamefont
  {S.-Y.}}, \bibinfo {author} {\bibfnamefont {N.}~\bibnamefont {Alidoust}},
  \bibinfo {author} {\bibfnamefont {I.}~\bibnamefont {Belopolski}}, \bibinfo
  {author} {\bibfnamefont {Z.}~\bibnamefont {Yuan}}, \bibinfo {author}
  {\bibfnamefont {G.}~\bibnamefont {Bian}}, \bibinfo {author} {\bibfnamefont
  {T.-R.}\ \bibnamefont {Chang}}, \bibinfo {author} {\bibfnamefont
  {H.}~\bibnamefont {Zheng}}, \bibinfo {author} {\bibfnamefont {V.~N.}\
  \bibnamefont {Strocov}}, \bibinfo {author} {\bibfnamefont {D.~S.}\
  \bibnamefont {Sanchez}}, \bibinfo {author} {\bibfnamefont {G.}~\bibnamefont
  {Chang}}, \bibinfo {author} {\bibfnamefont {C.}~\bibnamefont {Zhang}},
  \bibinfo {author} {\bibfnamefont {D.}~\bibnamefont {Mou}}, \bibinfo {author}
  {\bibfnamefont {Y.}~\bibnamefont {Wu}}, \bibinfo {author} {\bibfnamefont
  {L.}~\bibnamefont {Huang}}, \bibinfo {author} {\bibfnamefont {C.-C.}\
  \bibnamefont {Lee}}, \bibinfo {author} {\bibfnamefont {S.-M.}\ \bibnamefont
  {Huang}}, \bibinfo {author} {\bibfnamefont {B.}~\bibnamefont {Wang}},
  \bibinfo {author} {\bibfnamefont {A.}~\bibnamefont {Bansil}}, \bibinfo
  {author} {\bibfnamefont {H.-T.}\ \bibnamefont {Jeng}}, \bibinfo {author}
  {\bibfnamefont {T.}~\bibnamefont {Neupert}}, \bibinfo {author} {\bibfnamefont
  {A.}~\bibnamefont {Kaminski}}, \bibinfo {author} {\bibfnamefont
  {H.}~\bibnamefont {Lin}}, \bibinfo {author} {\bibfnamefont {S.}~\bibnamefont
  {Jia}}, \ and\ \bibinfo {author} {\bibfnamefont {M.}~\bibnamefont
  {Zahid~Hasan}}} (\bibinfo {year} {2015}{\natexlab{b}}),\ \href
  {http://dx.doi.org/10.1038/nphys3437} {\bibfield  {journal} {\bibinfo
  {journal} {Nat Phys}\ }\textbf {\bibinfo {volume} {11}}~(\bibinfo {number}
  {9}),\ \bibinfo {pages} {748}}\BibitemShut {NoStop}%
\bibitem [{\citenamefont {Xu}\ \emph {et~al.}(2015{\natexlab{c}})\citenamefont
  {Xu}, \citenamefont {Belopolski}, \citenamefont {Alidoust}, \citenamefont
  {Neupane}, \citenamefont {Bian}, \citenamefont {Zhang}, \citenamefont
  {Sankar}, \citenamefont {Chang}, \citenamefont {Yuan}, \citenamefont {Lee},
  \citenamefont {Huang}, \citenamefont {Zheng}, \citenamefont {Ma},
  \citenamefont {Sanchez}, \citenamefont {Wang}, \citenamefont {Bansil},
  \citenamefont {Chou}, \citenamefont {Shibayev}, \citenamefont {Lin},
  \citenamefont {Jia},\ and\ \citenamefont {Hasan}}]{Xu_Weyl_2015_first}%
  \BibitemOpen
  \bibfield  {author} {\bibinfo {author} {\bibnamefont {Xu}, \bibfnamefont
  {S.-Y.}}, \bibinfo {author} {\bibfnamefont {I.}~\bibnamefont {Belopolski}},
  \bibinfo {author} {\bibfnamefont {N.}~\bibnamefont {Alidoust}}, \bibinfo
  {author} {\bibfnamefont {M.}~\bibnamefont {Neupane}}, \bibinfo {author}
  {\bibfnamefont {G.}~\bibnamefont {Bian}}, \bibinfo {author} {\bibfnamefont
  {C.}~\bibnamefont {Zhang}}, \bibinfo {author} {\bibfnamefont
  {R.}~\bibnamefont {Sankar}}, \bibinfo {author} {\bibfnamefont
  {G.}~\bibnamefont {Chang}}, \bibinfo {author} {\bibfnamefont
  {Z.}~\bibnamefont {Yuan}}, \bibinfo {author} {\bibfnamefont {C.-C.}\
  \bibnamefont {Lee}}, \bibinfo {author} {\bibfnamefont {S.-M.}\ \bibnamefont
  {Huang}}, \bibinfo {author} {\bibfnamefont {H.}~\bibnamefont {Zheng}},
  \bibinfo {author} {\bibfnamefont {J.}~\bibnamefont {Ma}}, \bibinfo {author}
  {\bibfnamefont {D.~S.}\ \bibnamefont {Sanchez}}, \bibinfo {author}
  {\bibfnamefont {B.}~\bibnamefont {Wang}}, \bibinfo {author} {\bibfnamefont
  {A.}~\bibnamefont {Bansil}}, \bibinfo {author} {\bibfnamefont
  {F.}~\bibnamefont {Chou}}, \bibinfo {author} {\bibfnamefont {P.~P.}\
  \bibnamefont {Shibayev}}, \bibinfo {author} {\bibfnamefont {H.}~\bibnamefont
  {Lin}}, \bibinfo {author} {\bibfnamefont {S.}~\bibnamefont {Jia}}, \ and\
  \bibinfo {author} {\bibfnamefont {M.~Z.}\ \bibnamefont {Hasan}}} (\bibinfo
  {year} {2015}{\natexlab{c}}),\ \href {\doibase 10.1126/science.aaa9297}
  {\bibfield  {journal} {\bibinfo  {journal} {Science}\ }\textbf {\bibinfo
  {volume} {349}}~(\bibinfo {number} {6248}),\ \bibinfo {pages}
  {613}}\BibitemShut {NoStop}%
\bibitem [{\citenamefont {Xu}\ \emph {et~al.}(2012)\citenamefont {Xu},
  \citenamefont {Liu}, \citenamefont {Alidoust}, \citenamefont {Neupane},
  \citenamefont {Qian}, \citenamefont {Belopolski}, \citenamefont {Denlinger},
  \citenamefont {Wang}, \citenamefont {Lin}, \citenamefont {Wray},
  \citenamefont {Landolt}, \citenamefont {Slomski}, \citenamefont {Dil},
  \citenamefont {Marcinkova}, \citenamefont {Morosan}, \citenamefont {Gibson},
  \citenamefont {Sankar}, \citenamefont {Chou}, \citenamefont {Cava},
  \citenamefont {Bansil},\ and\ \citenamefont {Hasan}}]{Xu2012}%
  \BibitemOpen
  \bibfield  {author} {\bibinfo {author} {\bibnamefont {Xu}, \bibfnamefont
  {S.-Y.}}, \bibinfo {author} {\bibfnamefont {C.}~\bibnamefont {Liu}}, \bibinfo
  {author} {\bibfnamefont {N.}~\bibnamefont {Alidoust}}, \bibinfo {author}
  {\bibfnamefont {M.}~\bibnamefont {Neupane}}, \bibinfo {author} {\bibfnamefont
  {D.}~\bibnamefont {Qian}}, \bibinfo {author} {\bibfnamefont {I.}~\bibnamefont
  {Belopolski}}, \bibinfo {author} {\bibfnamefont {J.~D.}\ \bibnamefont
  {Denlinger}}, \bibinfo {author} {\bibfnamefont {Y.~J.}\ \bibnamefont {Wang}},
  \bibinfo {author} {\bibfnamefont {H.}~\bibnamefont {Lin}}, \bibinfo {author}
  {\bibfnamefont {L.~A.}\ \bibnamefont {Wray}}, \bibinfo {author}
  {\bibfnamefont {G.}~\bibnamefont {Landolt}}, \bibinfo {author} {\bibfnamefont
  {B.}~\bibnamefont {Slomski}}, \bibinfo {author} {\bibfnamefont {J.~H.}\
  \bibnamefont {Dil}}, \bibinfo {author} {\bibfnamefont {A.}~\bibnamefont
  {Marcinkova}}, \bibinfo {author} {\bibfnamefont {E.}~\bibnamefont {Morosan}},
  \bibinfo {author} {\bibfnamefont {Q.}~\bibnamefont {Gibson}}, \bibinfo
  {author} {\bibfnamefont {R.}~\bibnamefont {Sankar}}, \bibinfo {author}
  {\bibfnamefont {F.~C.}\ \bibnamefont {Chou}}, \bibinfo {author}
  {\bibfnamefont {R.~J.}\ \bibnamefont {Cava}}, \bibinfo {author}
  {\bibfnamefont {A.}~\bibnamefont {Bansil}}, \ and\ \bibinfo {author}
  {\bibfnamefont {M.~Z.}\ \bibnamefont {Hasan}}} (\bibinfo {year} {2012}),\
  \href@noop {} {\bibfield  {journal} {\bibinfo  {journal} {Nat. Commun.}\
  }\textbf {\bibinfo {volume} {3}},\ \bibinfo {pages} {1192}}\BibitemShut
  {NoStop}%
\bibitem [{\citenamefont {Xu}\ \emph {et~al.}(2015{\natexlab{d}})\citenamefont
  {Xu}, \citenamefont {Liu}, \citenamefont {Kushwaha}, \citenamefont {Sankar},
  \citenamefont {Krizan}, \citenamefont {Belopolski}, \citenamefont {Neupane},
  \citenamefont {Bian}, \citenamefont {Alidoust}, \citenamefont {Chang},
  \citenamefont {Jeng}, \citenamefont {Huang}, \citenamefont {Tsai},
  \citenamefont {Lin}, \citenamefont {Shibayev}, \citenamefont {Chou},
  \citenamefont {Cava},\ and\ \citenamefont {Hasan}}]{Xu18122014}%
  \BibitemOpen
  \bibfield  {author} {\bibinfo {author} {\bibnamefont {Xu}, \bibfnamefont
  {S.-Y.}}, \bibinfo {author} {\bibfnamefont {C.}~\bibnamefont {Liu}}, \bibinfo
  {author} {\bibfnamefont {S.~K.}\ \bibnamefont {Kushwaha}}, \bibinfo {author}
  {\bibfnamefont {R.}~\bibnamefont {Sankar}}, \bibinfo {author} {\bibfnamefont
  {J.~W.}\ \bibnamefont {Krizan}}, \bibinfo {author} {\bibfnamefont
  {I.}~\bibnamefont {Belopolski}}, \bibinfo {author} {\bibfnamefont
  {M.}~\bibnamefont {Neupane}}, \bibinfo {author} {\bibfnamefont
  {G.}~\bibnamefont {Bian}}, \bibinfo {author} {\bibfnamefont {N.}~\bibnamefont
  {Alidoust}}, \bibinfo {author} {\bibfnamefont {T.-R.}\ \bibnamefont {Chang}},
  \bibinfo {author} {\bibfnamefont {H.-T.}\ \bibnamefont {Jeng}}, \bibinfo
  {author} {\bibfnamefont {C.-Y.}\ \bibnamefont {Huang}}, \bibinfo {author}
  {\bibfnamefont {W.-F.}\ \bibnamefont {Tsai}}, \bibinfo {author}
  {\bibfnamefont {H.}~\bibnamefont {Lin}}, \bibinfo {author} {\bibfnamefont
  {P.~P.}\ \bibnamefont {Shibayev}}, \bibinfo {author} {\bibfnamefont {F.-C.}\
  \bibnamefont {Chou}}, \bibinfo {author} {\bibfnamefont {R.~J.}\ \bibnamefont
  {Cava}}, \ and\ \bibinfo {author} {\bibfnamefont {M.~Z.}\ \bibnamefont
  {Hasan}}} (\bibinfo {year} {2015}{\natexlab{d}}),\ \href {\doibase
  10.1126/science.1256742} {\bibfield  {journal} {\bibinfo  {journal}
  {Science}\ }\textbf {\bibinfo {volume} {347}}~(\bibinfo {number} {6219}),\
  \bibinfo {pages} {294}}\BibitemShut {NoStop}%
\bibitem [{\citenamefont {Yamakage}\ \emph {et~al.}(2011)\citenamefont
  {Yamakage}, \citenamefont {Nomura}, \citenamefont {Imura},\ and\
  \citenamefont {Kuramoto}}]{doi:10.1143/JPSJ.80.053703}%
  \BibitemOpen
  \bibfield  {author} {\bibinfo {author} {\bibnamefont {Yamakage},
  \bibfnamefont {A.}}, \bibinfo {author} {\bibfnamefont {K.}~\bibnamefont
  {Nomura}}, \bibinfo {author} {\bibfnamefont {K.-I.}\ \bibnamefont {Imura}}, \
  and\ \bibinfo {author} {\bibfnamefont {Y.}~\bibnamefont {Kuramoto}}}
  (\bibinfo {year} {2011}),\ \href {\doibase 10.1143/JPSJ.80.053703} {\bibfield
   {journal} {\bibinfo  {journal} {Journal of the Physical Society of Japan}\
  }\textbf {\bibinfo {volume} {80}}~(\bibinfo {number} {5}),\ \bibinfo {pages}
  {053703}}\BibitemShut {NoStop}%
\bibitem [{\citenamefont {Yamakage}\ \emph
  {et~al.}(2013{\natexlab{a}})\citenamefont {Yamakage}, \citenamefont {Nomura},
  \citenamefont {Imura},\ and\ \citenamefont {Kuramoto}}]{PhysRevB.87.205141}%
  \BibitemOpen
  \bibfield  {author} {\bibinfo {author} {\bibnamefont {Yamakage},
  \bibfnamefont {A.}}, \bibinfo {author} {\bibfnamefont {K.}~\bibnamefont
  {Nomura}}, \bibinfo {author} {\bibfnamefont {K.-I.}\ \bibnamefont {Imura}}, \
  and\ \bibinfo {author} {\bibfnamefont {Y.}~\bibnamefont {Kuramoto}}}
  (\bibinfo {year} {2013}{\natexlab{a}}),\ \href {\doibase
  10.1103/PhysRevB.87.205141} {\bibfield  {journal} {\bibinfo  {journal} {Phys.
  Rev. B}\ }\textbf {\bibinfo {volume} {87}},\ \bibinfo {pages}
  {205141}}\BibitemShut {NoStop}%
\bibitem [{\citenamefont {Yamakage}\ \emph
  {et~al.}(2013{\natexlab{b}})\citenamefont {Yamakage}, \citenamefont {Sato},
  \citenamefont {Yada}, \citenamefont {Kashiwaya},\ and\ \citenamefont
  {Tanaka}}]{YamakageTanaka13}%
  \BibitemOpen
  \bibfield  {author} {\bibinfo {author} {\bibnamefont {Yamakage},
  \bibfnamefont {A.}}, \bibinfo {author} {\bibfnamefont {M.}~\bibnamefont
  {Sato}}, \bibinfo {author} {\bibfnamefont {K.}~\bibnamefont {Yada}}, \bibinfo
  {author} {\bibfnamefont {S.}~\bibnamefont {Kashiwaya}}, \ and\ \bibinfo
  {author} {\bibfnamefont {Y.}~\bibnamefont {Tanaka}}} (\bibinfo {year}
  {2013}{\natexlab{b}}),\ \href {\doibase 10.1103/PhysRevB.87.100510}
  {\bibfield  {journal} {\bibinfo  {journal} {Phys. Rev. B}\ }\textbf {\bibinfo
  {volume} {87}},\ \bibinfo {pages} {100510}}\BibitemShut {NoStop}%
\bibitem [{\citenamefont {{Yamakage}}\ \emph {et~al.}(2015)\citenamefont
  {{Yamakage}}, \citenamefont {{Yamakawa}}, \citenamefont {{Tanaka}},\ and\
  \citenamefont {{Okamoto}}}]{Nodal_line_Tanaka}%
  \BibitemOpen
  \bibfield  {author} {\bibinfo {author} {\bibnamefont {{Yamakage}},
  \bibfnamefont {A.}}, \bibinfo {author} {\bibfnamefont {Y.}~\bibnamefont
  {{Yamakawa}}}, \bibinfo {author} {\bibfnamefont {Y.}~\bibnamefont
  {{Tanaka}}}, \ and\ \bibinfo {author} {\bibfnamefont {Y.}~\bibnamefont
  {{Okamoto}}}} (\bibinfo {year} {2015}),\ \href@noop {} {\bibfield  {journal}
  {\bibinfo  {journal} {ArXiv e-prints}\ }}\Eprint
  {http://arxiv.org/abs/1510.00202} {arXiv:1510.00202 [cond-mat.mes-hall]}
  \BibitemShut {NoStop}%
\bibitem [{\citenamefont {Yang}\ \emph {et~al.}(2015)\citenamefont {Yang},
  \citenamefont {Morimoto},\ and\ \citenamefont
  {Furusaki}}]{Bulk_Dirac_cone_rotation_Akira}%
  \BibitemOpen
  \bibfield  {author} {\bibinfo {author} {\bibnamefont {Yang}, \bibfnamefont
  {B.-J.}}, \bibinfo {author} {\bibfnamefont {T.}~\bibnamefont {Morimoto}}, \
  and\ \bibinfo {author} {\bibfnamefont {A.}~\bibnamefont {Furusaki}}}
  (\bibinfo {year} {2015}),\ \href {\doibase 10.1103/PhysRevB.92.165120}
  {\bibfield  {journal} {\bibinfo  {journal} {Phys. Rev. B}\ }\textbf {\bibinfo
  {volume} {92}},\ \bibinfo {pages} {165120}}\BibitemShut {NoStop}%
\bibitem [{\citenamefont {Yang}\ and\ \citenamefont
  {Nagaosa}(2014)}]{bulk_Dirac_Yang:2014aa}%
  \BibitemOpen
  \bibfield  {author} {\bibinfo {author} {\bibnamefont {Yang}, \bibfnamefont
  {B.-J.}}, \ and\ \bibinfo {author} {\bibfnamefont {N.}~\bibnamefont
  {Nagaosa}}} (\bibinfo {year} {2014}),\ \href
  {http://dx.doi.org/10.1038/ncomms5898} {\bibfield  {journal} {\bibinfo
  {journal} {Nat Commun}\ }\textbf {\bibinfo {volume} {5}}~(\bibinfo {number}
  {4898})}\BibitemShut {NoStop}%
\bibitem [{\citenamefont {{Yao}}\ and\ \citenamefont
  {{Ryu}}(2013)}]{YaoRyu2013}%
  \BibitemOpen
  \bibfield  {author} {\bibinfo {author} {\bibnamefont {{Yao}}, \bibfnamefont
  {H.}}, \ and\ \bibinfo {author} {\bibfnamefont {S.}~\bibnamefont {{Ryu}}}}
  (\bibinfo {year} {2013}),\ \href {\doibase 10.1103/PhysRevB.88.064507}
  {\bibfield  {journal} {\bibinfo  {journal} {\prb}\ }\textbf {\bibinfo
  {volume} {88}}~(\bibinfo {number} {6}),\ \bibinfo {eid} {064507}}\BibitemShut
  {NoStop}%
\bibitem [{\citenamefont {{Ye}}\ \emph {et~al.}(2013)\citenamefont {{Ye}},
  \citenamefont {{Allen}},\ and\ \citenamefont
  {{Sun}}}]{KaiSun_crystalline_Kondo}%
  \BibitemOpen
  \bibfield  {author} {\bibinfo {author} {\bibnamefont {{Ye}}, \bibfnamefont
  {M.}}, \bibinfo {author} {\bibfnamefont {J.~W.}\ \bibnamefont {{Allen}}}, \
  and\ \bibinfo {author} {\bibfnamefont {K.}~\bibnamefont {{Sun}}}} (\bibinfo
  {year} {2013}),\ \href@noop {} {\ }\Eprint {http://arxiv.org/abs/1307.7191}
  {arXiv:1307.7191} \BibitemShut {NoStop}%
\bibitem [{\citenamefont {{You}}\ \emph {et~al.}(2014)\citenamefont {{You}},
  \citenamefont {{BenTov}},\ and\ \citenamefont
  {{Xu}}}]{you_bentov_xu_arXiv_14}%
  \BibitemOpen
  \bibfield  {author} {\bibinfo {author} {\bibnamefont {{You}}, \bibfnamefont
  {Y.-Z.}}, \bibinfo {author} {\bibfnamefont {Y.}~\bibnamefont {{BenTov}}}, \
  and\ \bibinfo {author} {\bibfnamefont {C.}~\bibnamefont {{Xu}}}} (\bibinfo
  {year} {2014}),\ \href@noop {} {\bibfield  {journal} {\bibinfo  {journal}
  {ArXiv e-prints}\ }}\Eprint {http://arxiv.org/abs/1402.4151} {arXiv:1402.4151
  [cond-mat.str-el]} \BibitemShut {NoStop}%
\bibitem [{\citenamefont {You}\ and\ \citenamefont {Xu}(2014)}]{YouXu2014}%
  \BibitemOpen
  \bibfield  {author} {\bibinfo {author} {\bibnamefont {You}, \bibfnamefont
  {Y.-Z.}}, \ and\ \bibinfo {author} {\bibfnamefont {C.}~\bibnamefont {Xu}}}
  (\bibinfo {year} {2014}),\ \href {\doibase 10.1103/PhysRevB.90.245120}
  {\bibfield  {journal} {\bibinfo  {journal} {Phys. Rev. B}\ }\textbf {\bibinfo
  {volume} {90}},\ \bibinfo {pages} {245120}}\BibitemShut {NoStop}%
\bibitem [{\citenamefont {Young}\ \emph {et~al.}(2008)\citenamefont {Young},
  \citenamefont {Lee},\ and\ \citenamefont {Kallin}}]{Kallin_FTI}%
  \BibitemOpen
  \bibfield  {author} {\bibinfo {author} {\bibnamefont {Young}, \bibfnamefont
  {M.~W.}}, \bibinfo {author} {\bibfnamefont {S.-S.}\ \bibnamefont {Lee}}, \
  and\ \bibinfo {author} {\bibfnamefont {C.}~\bibnamefont {Kallin}}} (\bibinfo
  {year} {2008}),\ \href {\doibase 10.1103/PhysRevB.78.125316} {\bibfield
  {journal} {\bibinfo  {journal} {Phys. Rev. B}\ }\textbf {\bibinfo {volume}
  {78}},\ \bibinfo {pages} {125316}}\BibitemShut {NoStop}%
\bibitem [{\citenamefont {Young}\ and\ \citenamefont
  {Kane}(2015)}]{nonsymmorphic_cone_Kane}%
  \BibitemOpen
  \bibfield  {author} {\bibinfo {author} {\bibnamefont {Young}, \bibfnamefont
  {S.~M.}}, \ and\ \bibinfo {author} {\bibfnamefont {C.~L.}\ \bibnamefont
  {Kane}}} (\bibinfo {year} {2015}),\ \href {\doibase
  10.1103/PhysRevLett.115.126803} {\bibfield  {journal} {\bibinfo  {journal}
  {Phys. Rev. Lett.}\ }\textbf {\bibinfo {volume} {115}},\ \bibinfo {pages}
  {126803}}\BibitemShut {NoStop}%
\bibitem [{\citenamefont {Young}\ \emph {et~al.}(2012)\citenamefont {Young},
  \citenamefont {Zaheer}, \citenamefont {Teo}, \citenamefont {Kane},
  \citenamefont {Mele},\ and\ \citenamefont {Rappe}}]{BiO3_Dirac_semimetal}%
  \BibitemOpen
  \bibfield  {author} {\bibinfo {author} {\bibnamefont {Young}, \bibfnamefont
  {S.~M.}}, \bibinfo {author} {\bibfnamefont {S.}~\bibnamefont {Zaheer}},
  \bibinfo {author} {\bibfnamefont {J.~C.~Y.}\ \bibnamefont {Teo}}, \bibinfo
  {author} {\bibfnamefont {C.~L.}\ \bibnamefont {Kane}}, \bibinfo {author}
  {\bibfnamefont {E.~J.}\ \bibnamefont {Mele}}, \ and\ \bibinfo {author}
  {\bibfnamefont {A.~M.}\ \bibnamefont {Rappe}}} (\bibinfo {year} {2012}),\
  \href {\doibase 10.1103/PhysRevLett.108.140405} {\bibfield  {journal}
  {\bibinfo  {journal} {Phys. Rev. Lett.}\ }\textbf {\bibinfo {volume} {108}},\
  \bibinfo {pages} {140405}}\BibitemShut {NoStop}%
\bibitem [{\citenamefont {{Yu}}\ \emph {et~al.}(2011)\citenamefont {{Yu}},
  \citenamefont {{Qi}}, \citenamefont {{Bernevig}}, \citenamefont {{Fang}},\
  and\ \citenamefont {{Dai}}}]{YuQiBernevig2011}%
  \BibitemOpen
  \bibfield  {author} {\bibinfo {author} {\bibnamefont {{Yu}}, \bibfnamefont
  {R.}}, \bibinfo {author} {\bibfnamefont {X.~L.}\ \bibnamefont {{Qi}}},
  \bibinfo {author} {\bibfnamefont {A.}~\bibnamefont {{Bernevig}}}, \bibinfo
  {author} {\bibfnamefont {Z.}~\bibnamefont {{Fang}}}, \ and\ \bibinfo {author}
  {\bibfnamefont {X.}~\bibnamefont {{Dai}}}} (\bibinfo {year} {2011}),\ \href
  {\doibase 10.1103/PhysRevB.84.075119} {\bibfield  {journal} {\bibinfo
  {journal} {\prb}\ }\textbf {\bibinfo {volume} {84}}~(\bibinfo {number} {7}),\
  \bibinfo {eid} {075119}}\BibitemShut {NoStop}%
\bibitem [{\citenamefont {Yu}\ \emph {et~al.}(2015)\citenamefont {Yu},
  \citenamefont {Weng}, \citenamefont {Fang}, \citenamefont {Dai},\ and\
  \citenamefont {Hu}}]{Dai_Cu3PdN_ring}%
  \BibitemOpen
  \bibfield  {author} {\bibinfo {author} {\bibnamefont {Yu}, \bibfnamefont
  {R.}}, \bibinfo {author} {\bibfnamefont {H.}~\bibnamefont {Weng}}, \bibinfo
  {author} {\bibfnamefont {Z.}~\bibnamefont {Fang}}, \bibinfo {author}
  {\bibfnamefont {X.}~\bibnamefont {Dai}}, \ and\ \bibinfo {author}
  {\bibfnamefont {X.}~\bibnamefont {Hu}}} (\bibinfo {year} {2015}),\ \href
  {\doibase 10.1103/PhysRevLett.115.036807} {\bibfield  {journal} {\bibinfo
  {journal} {Phys. Rev. Lett.}\ }\textbf {\bibinfo {volume} {115}},\ \bibinfo
  {pages} {036807}}\BibitemShut {NoStop}%
\bibitem [{\citenamefont {Yu}\ \emph {et~al.}(2010)\citenamefont {Yu},
  \citenamefont {Zhang}, \citenamefont {Zhang}, \citenamefont {Zhang},
  \citenamefont {Dai},\ and\ \citenamefont {Fang}}]{Yu_QAHE}%
  \BibitemOpen
  \bibfield  {author} {\bibinfo {author} {\bibnamefont {Yu}, \bibfnamefont
  {R.}}, \bibinfo {author} {\bibfnamefont {W.}~\bibnamefont {Zhang}}, \bibinfo
  {author} {\bibfnamefont {H.-J.}\ \bibnamefont {Zhang}}, \bibinfo {author}
  {\bibfnamefont {S.-C.}\ \bibnamefont {Zhang}}, \bibinfo {author}
  {\bibfnamefont {X.}~\bibnamefont {Dai}}, \ and\ \bibinfo {author}
  {\bibfnamefont {Z.}~\bibnamefont {Fang}}} (\bibinfo {year} {2010}),\ \href
  {\doibase 10.1126/science.1187485} {\bibfield  {journal} {\bibinfo  {journal}
  {Science}\ }\textbf {\bibinfo {volume} {329}}~(\bibinfo {number} {5987}),\
  \bibinfo {pages} {61}}\BibitemShut {NoStop}%
\bibitem [{\citenamefont {{Zahid Hasan}}\ \emph {et~al.}(2014)\citenamefont
  {{Zahid Hasan}}, \citenamefont {{Xu}},\ and\ \citenamefont
  {{Neupane}}}]{HasanXuNeupane2014}%
  \BibitemOpen
  \bibfield  {author} {\bibinfo {author} {\bibnamefont {{Zahid Hasan}},
  \bibfnamefont {M.}}, \bibinfo {author} {\bibfnamefont {S.-Y.}\ \bibnamefont
  {{Xu}}}, \ and\ \bibinfo {author} {\bibfnamefont {M.}~\bibnamefont
  {{Neupane}}}} (\bibinfo {year} {2014}),\ \href@noop {} {\ }\Eprint
  {http://arxiv.org/abs/1406.1040} {arXiv:1406.1040} \BibitemShut {NoStop}%
\bibitem [{\citenamefont {Zak}(1989)}]{Zak1989}%
  \BibitemOpen
  \bibfield  {author} {\bibinfo {author} {\bibnamefont {Zak}, \bibfnamefont
  {J.}}} (\bibinfo {year} {1989}),\ \href {\doibase
  10.1103/PhysRevLett.62.2747} {\bibfield  {journal} {\bibinfo  {journal}
  {Phys. Rev. Lett.}\ }\textbf {\bibinfo {volume} {62}},\ \bibinfo {pages}
  {2747}}\BibitemShut {NoStop}%
\bibitem [{\citenamefont {Zareapour}\ \emph {et~al.}(2012)\citenamefont
  {Zareapour}, \citenamefont {Hayat}, \citenamefont {Zhao}, \citenamefont
  {Kreshchuk}, \citenamefont {Jain}, \citenamefont {Kwok}, \citenamefont {Lee},
  \citenamefont {Cheong}, \citenamefont {Xu}, \citenamefont {Yang},
  \citenamefont {Gu}, \citenamefont {Jia}, \citenamefont {Cava},\ and\
  \citenamefont {Burch}}]{burch_natComm12}%
  \BibitemOpen
  \bibfield  {author} {\bibinfo {author} {\bibnamefont {Zareapour},
  \bibfnamefont {P.}}, \bibinfo {author} {\bibfnamefont {A.}~\bibnamefont
  {Hayat}}, \bibinfo {author} {\bibfnamefont {S.~Y.~F.}\ \bibnamefont {Zhao}},
  \bibinfo {author} {\bibfnamefont {M.}~\bibnamefont {Kreshchuk}}, \bibinfo
  {author} {\bibfnamefont {A.}~\bibnamefont {Jain}}, \bibinfo {author}
  {\bibfnamefont {D.~C.}\ \bibnamefont {Kwok}}, \bibinfo {author}
  {\bibfnamefont {N.}~\bibnamefont {Lee}}, \bibinfo {author} {\bibfnamefont
  {S.-W.}\ \bibnamefont {Cheong}}, \bibinfo {author} {\bibfnamefont
  {Z.}~\bibnamefont {Xu}}, \bibinfo {author} {\bibfnamefont {A.}~\bibnamefont
  {Yang}}, \bibinfo {author} {\bibfnamefont {G.~D.}\ \bibnamefont {Gu}},
  \bibinfo {author} {\bibfnamefont {S.}~\bibnamefont {Jia}}, \bibinfo {author}
  {\bibfnamefont {R.~J.}\ \bibnamefont {Cava}}, \ and\ \bibinfo {author}
  {\bibfnamefont {K.~S.}\ \bibnamefont {Burch}}} (\bibinfo {year} {2012}),\
  \href {http://dx.doi.org/10.1038/ncomms2042} {\bibfield  {journal} {\bibinfo
  {journal} {Nat Commun}\ }\textbf {\bibinfo {volume} {3}},\ \bibinfo {pages}
  {1056}}\BibitemShut {NoStop}%
\bibitem [{\citenamefont {{Zeng}}\ \emph {et~al.}(2015)\citenamefont {{Zeng}},
  \citenamefont {{Fang}}, \citenamefont {{Chang}}, \citenamefont {{Chen}},
  \citenamefont {{Hsieh}}, \citenamefont {{Bansil}}, \citenamefont {{Lin}},\
  and\ \citenamefont {{Fu}}}]{Fang_ring_point}%
  \BibitemOpen
  \bibfield  {author} {\bibinfo {author} {\bibnamefont {{Zeng}}, \bibfnamefont
  {M.}}, \bibinfo {author} {\bibfnamefont {C.}~\bibnamefont {{Fang}}}, \bibinfo
  {author} {\bibfnamefont {G.}~\bibnamefont {{Chang}}}, \bibinfo {author}
  {\bibfnamefont {Y.-A.}\ \bibnamefont {{Chen}}}, \bibinfo {author}
  {\bibfnamefont {T.}~\bibnamefont {{Hsieh}}}, \bibinfo {author} {\bibfnamefont
  {A.}~\bibnamefont {{Bansil}}}, \bibinfo {author} {\bibfnamefont
  {H.}~\bibnamefont {{Lin}}}, \ and\ \bibinfo {author} {\bibfnamefont
  {L.}~\bibnamefont {{Fu}}}} (\bibinfo {year} {2015}),\ \href@noop {} {\
  }\Eprint {http://arxiv.org/abs/1504.03492} {arXiv:1504.03492} \BibitemShut
  {NoStop}%
\bibitem [{\citenamefont {{Zhang}}\ \emph {et~al.}(2015)\citenamefont
  {{Zhang}}, \citenamefont {{Xu}}, \citenamefont {{Belopolski}}, \citenamefont
  {{Yuan}}, \citenamefont {{Lin}}, \citenamefont {{Tong}}, \citenamefont
  {{Alidoust}}, \citenamefont {{Lee}}, \citenamefont {{Huang}}, \citenamefont
  {{Lin}}, \citenamefont {{Neupane}}, \citenamefont {{Sanchez}}, \citenamefont
  {{Zheng}}, \citenamefont {{Bian}}, \citenamefont {{Wang}}, \citenamefont
  {{Zhang}}, \citenamefont {{Neupert}}, \citenamefont {{Zahid Hasan}},\ and\
  \citenamefont {{Jia}}}]{Zhang_anomaly_Weyl_2015}%
  \BibitemOpen
  \bibfield  {author} {\bibinfo {author} {\bibnamefont {{Zhang}}, \bibfnamefont
  {C.}}, \bibinfo {author} {\bibfnamefont {S.-Y.}\ \bibnamefont {{Xu}}},
  \bibinfo {author} {\bibfnamefont {I.}~\bibnamefont {{Belopolski}}}, \bibinfo
  {author} {\bibfnamefont {Z.}~\bibnamefont {{Yuan}}}, \bibinfo {author}
  {\bibfnamefont {Z.}~\bibnamefont {{Lin}}}, \bibinfo {author} {\bibfnamefont
  {B.}~\bibnamefont {{Tong}}}, \bibinfo {author} {\bibfnamefont
  {N.}~\bibnamefont {{Alidoust}}}, \bibinfo {author} {\bibfnamefont {C.-C.}\
  \bibnamefont {{Lee}}}, \bibinfo {author} {\bibfnamefont {S.-M.}\ \bibnamefont
  {{Huang}}}, \bibinfo {author} {\bibfnamefont {H.}~\bibnamefont {{Lin}}},
  \bibinfo {author} {\bibfnamefont {M.}~\bibnamefont {{Neupane}}}, \bibinfo
  {author} {\bibfnamefont {D.~S.}\ \bibnamefont {{Sanchez}}}, \bibinfo {author}
  {\bibfnamefont {H.}~\bibnamefont {{Zheng}}}, \bibinfo {author} {\bibfnamefont
  {G.}~\bibnamefont {{Bian}}}, \bibinfo {author} {\bibfnamefont
  {J.}~\bibnamefont {{Wang}}}, \bibinfo {author} {\bibfnamefont
  {C.}~\bibnamefont {{Zhang}}}, \bibinfo {author} {\bibfnamefont
  {T.}~\bibnamefont {{Neupert}}}, \bibinfo {author} {\bibfnamefont
  {M.}~\bibnamefont {{Zahid Hasan}}}, \ and\ \bibinfo {author} {\bibfnamefont
  {S.}~\bibnamefont {{Jia}}}} (\bibinfo {year} {2015}),\ \href@noop {} {\
  }\Eprint {http://arxiv.org/abs/1503.02630} {arXiv:1503.02630} \BibitemShut
  {NoStop}%
\bibitem [{\citenamefont {Zhang}\ and\ \citenamefont
  {Kane}(2014{\natexlab{a}})}]{ZhangKane14}%
  \BibitemOpen
  \bibfield  {author} {\bibinfo {author} {\bibnamefont {Zhang}, \bibfnamefont
  {F.}}, \ and\ \bibinfo {author} {\bibfnamefont {C.~L.}\ \bibnamefont {Kane}}}
  (\bibinfo {year} {2014}{\natexlab{a}}),\ \href {\doibase
  10.1103/PhysRevB.90.020501} {\bibfield  {journal} {\bibinfo  {journal} {Phys.
  Rev. B}\ }\textbf {\bibinfo {volume} {90}},\ \bibinfo {pages}
  {020501}}\BibitemShut {NoStop}%
\bibitem [{\citenamefont {Zhang}\ and\ \citenamefont
  {Kane}(2014{\natexlab{b}})}]{ZhangKane14PRL}%
  \BibitemOpen
  \bibfield  {author} {\bibinfo {author} {\bibnamefont {Zhang}, \bibfnamefont
  {F.}}, \ and\ \bibinfo {author} {\bibfnamefont {C.~L.}\ \bibnamefont {Kane}}}
  (\bibinfo {year} {2014}{\natexlab{b}}),\ \href {\doibase
  10.1103/PhysRevLett.113.036401} {\bibfield  {journal} {\bibinfo  {journal}
  {Phys. Rev. Lett.}\ }\textbf {\bibinfo {volume} {113}},\ \bibinfo {pages}
  {036401}}\BibitemShut {NoStop}%
\bibitem [{\citenamefont {Zhang}\ \emph {et~al.}(2013)\citenamefont {Zhang},
  \citenamefont {Kane},\ and\ \citenamefont {Mele}}]{Kane_Mirror}%
  \BibitemOpen
  \bibfield  {author} {\bibinfo {author} {\bibnamefont {Zhang}, \bibfnamefont
  {F.}}, \bibinfo {author} {\bibfnamefont {C.~L.}\ \bibnamefont {Kane}}, \ and\
  \bibinfo {author} {\bibfnamefont {E.~J.}\ \bibnamefont {Mele}}} (\bibinfo
  {year} {2013}),\ \href {\doibase 10.1103/PhysRevLett.111.056403} {\bibfield
  {journal} {\bibinfo  {journal} {Phys. Rev. Lett.}\ }\textbf {\bibinfo
  {volume} {111}},\ \bibinfo {pages} {056403}}\BibitemShut {NoStop}%
\bibitem [{\citenamefont {Zhang}\ \emph {et~al.}(2014)\citenamefont {Zhang},
  \citenamefont {Huang}, \citenamefont {Haule},\ and\ \citenamefont
  {Vanderbilt}}]{ZhangVanderbilt14}%
  \BibitemOpen
  \bibfield  {author} {\bibinfo {author} {\bibnamefont {Zhang}, \bibfnamefont
  {H.}}, \bibinfo {author} {\bibfnamefont {H.}~\bibnamefont {Huang}}, \bibinfo
  {author} {\bibfnamefont {K.}~\bibnamefont {Haule}}, \ and\ \bibinfo {author}
  {\bibfnamefont {D.}~\bibnamefont {Vanderbilt}}} (\bibinfo {year} {2014}),\
  \href {\doibase 10.1103/PhysRevB.90.165143} {\bibfield  {journal} {\bibinfo
  {journal} {Phys. Rev. B}\ }\textbf {\bibinfo {volume} {90}},\ \bibinfo
  {pages} {165143}}\BibitemShut {NoStop}%
\bibitem [{\citenamefont {Zhang}\ \emph {et~al.}(2015)\citenamefont {Zhang},
  \citenamefont {Cheng},\ and\ \citenamefont {Schwingenschlogl}}]{Zhang_TCI}%
  \BibitemOpen
  \bibfield  {author} {\bibinfo {author} {\bibnamefont {Zhang}, \bibfnamefont
  {Q.}}, \bibinfo {author} {\bibfnamefont {Y.}~\bibnamefont {Cheng}}, \ and\
  \bibinfo {author} {\bibfnamefont {U.}~\bibnamefont {Schwingenschlogl}}}
  (\bibinfo {year} {2015}),\ \href {http://dx.doi.org/10.1038/srep08379}
  {\bibfield  {journal} {\bibinfo  {journal} {Sci. Rep.}\ }\textbf {\bibinfo
  {volume} {5}}}\BibitemShut {NoStop}%
\bibitem [{\citenamefont {Zhang}\ and\ \citenamefont
  {Liu}(2015)}]{zhang_chao-xing_PRB_15}%
  \BibitemOpen
  \bibfield  {author} {\bibinfo {author} {\bibnamefont {Zhang}, \bibfnamefont
  {R.-X.}}, \ and\ \bibinfo {author} {\bibfnamefont {C.-X.}\ \bibnamefont
  {Liu}}} (\bibinfo {year} {2015}),\ \href {\doibase
  10.1103/PhysRevB.91.115317} {\bibfield  {journal} {\bibinfo  {journal} {Phys.
  Rev. B}\ }\textbf {\bibinfo {volume} {91}},\ \bibinfo {pages}
  {115317}}\BibitemShut {NoStop}%
\bibitem [{\citenamefont {Zhao}\ and\ \citenamefont
  {Wang}(2013)}]{ZhaoWangPRL13}%
  \BibitemOpen
  \bibfield  {author} {\bibinfo {author} {\bibnamefont {Zhao}, \bibfnamefont
  {Y.~X.}}, \ and\ \bibinfo {author} {\bibfnamefont {Z.~D.}\ \bibnamefont
  {Wang}}} (\bibinfo {year} {2013}),\ \href {\doibase
  10.1103/PhysRevLett.110.240404} {\bibfield  {journal} {\bibinfo  {journal}
  {Phys. Rev. Lett.}\ }\textbf {\bibinfo {volume} {110}},\ \bibinfo {pages}
  {240404}}\BibitemShut {NoStop}%
\bibitem [{\citenamefont {Zhao}\ and\ \citenamefont
  {Wang}(2014)}]{ZhaoWangPRB14}%
  \BibitemOpen
  \bibfield  {author} {\bibinfo {author} {\bibnamefont {Zhao}, \bibfnamefont
  {Y.~X.}}, \ and\ \bibinfo {author} {\bibfnamefont {Z.~D.}\ \bibnamefont
  {Wang}}} (\bibinfo {year} {2014}),\ \href {\doibase
  10.1103/PhysRevB.89.075111} {\bibfield  {journal} {\bibinfo  {journal} {Phys.
  Rev. B}\ }\textbf {\bibinfo {volume} {89}},\ \bibinfo {pages}
  {075111}}\BibitemShut {NoStop}%
\bibitem [{\citenamefont {{Zirnbauer}}(1992)}]{Zirnbauer1992}%
  \BibitemOpen
  \bibfield  {author} {\bibinfo {author} {\bibnamefont {{Zirnbauer}},
  \bibfnamefont {M.~R.}}} (\bibinfo {year} {1992}),\ \href {\doibase
  10.1103/PhysRevLett.69.1584} {\bibfield  {journal} {\bibinfo  {journal}
  {Phys. Rev. Lett.}\ }\textbf {\bibinfo {volume} {69}},\ \bibinfo {pages}
  {1584}}\BibitemShut {NoStop}%
\bibitem [{\citenamefont {Zirnbauer}(1996)}]{Zirnbauer:1996fk}%
  \BibitemOpen
  \bibfield  {author} {\bibinfo {author} {\bibnamefont {Zirnbauer},
  \bibfnamefont {M.~R.}}} (\bibinfo {year} {1996}),\ \href
  {http://link.aip.org/link/?JMP/37/4986/1} {\bibfield  {journal} {\bibinfo
  {journal} {Journal of Mathematical Physics}\ }\textbf {\bibinfo {volume}
  {37}}~(\bibinfo {number} {10}),\ \bibinfo {pages} {4986}}\BibitemShut
  {NoStop}%
\bibitem [{\citenamefont {Zyuzin}\ and\ \citenamefont
  {Burkov}(2012)}]{Burkov_Weyl_electromagnetic_2012}%
  \BibitemOpen
  \bibfield  {author} {\bibinfo {author} {\bibnamefont {Zyuzin}, \bibfnamefont
  {A.~A.}}, \ and\ \bibinfo {author} {\bibfnamefont {A.~A.}\ \bibnamefont
  {Burkov}}} (\bibinfo {year} {2012}),\ \href {\doibase
  10.1103/PhysRevB.86.115133} {\bibfield  {journal} {\bibinfo  {journal} {Phys.
  Rev. B}\ }\textbf {\bibinfo {volume} {86}},\ \bibinfo {pages}
  {115133}}\BibitemShut {NoStop}%
\end{thebibliography}%

\end{document}